\documentclass[10pt,b5paper]{book}	% B5: 250 x 176 mm, 9.84 x 6.93 inches.
\usepackage[latin1]{inputenc}
\usepackage[british]{babel}
\usepackage{fancyhdr}
\usepackage{graphicx}
\usepackage[dvipsnames]{color}
\usepackage{geometry}
\usepackage{amssymb}
\usepackage{makeidx}
\usepackage{setspace}
\usepackage{cancel}
\usepackage{ccaption}
\usepackage{natbib}

\captiontitlefont{\small\itshape}
\captionnamefont{\small\bfseries\itshape}

\headwidth=\textwidth
\newcommand{\dive} {\vec{\nabla}\cdot}
\newcommand{\rot} {\vec{\nabla}\times}
\newcommand{\grad} {\vec{\nabla}}
\newcommand{\derpar}[1] {\frac{\partial}{\partial #1}}
\newcommand{\derparn}[2] {\frac{\partial #2}{\partial #1}}
\newcommand{\curlBrel} {\vec{\nabla}\times(e^\nu\vec{B})}
\newcommand{\curlB} {\vec{\nabla}\times\vec{B}}
\newcommand{\omtau} {\omega_B\tau_e}
\newcommand{\de} {{\rm d}}
\newcommand{\gcc} {g~cm$^{-3}$}

\begin{document}

%%%%%%%%%%%%% COVER %%%%%%%%%%%%%%%%%
% \begin{titlepage}
% \setcounter{page}{0}
% \setlength{\topmargin}{-0.2in}
% \setlength{\evensidemargin}{-0.25in}	% Left margin on even numbered pages (left side of the book). 1 inch is added to this value.
% \setlength{\oddsidemargin}{0in}		% Left margin on odd numbered pages (right side of the book). 1 inch is added to this value.
% 
% \pagestyle{empty}
% \pagecolor{yellow}
% \begin{center}
% 
% % \vskip1cm
% \textbf
% {\Huge \color{blue}
% MAGNETIC FIELDS\\
% \vskip15pt
% IN NEUTRON STARS}
% 
% \vskip4cm
% 
% \includegraphics[width=.6\textwidth]{images/pulsar_art.eps}
% 
% \vskip3cm
% 
% \textbf{\Huge \color{blue} Daniele Vigan\`o}
% 
% \end{center}
% \end{titlepage}

%%%%%%%%%%%%% COUNTER-COVER %%%%%%%%%%%%%%%%%
\pagecolor{white}
\newpage\null\thispagestyle{empty}

\setlength{\topmargin}{-0.3in}
\setlength{\evensidemargin}{0in}	% Left margin for this page. 1 inch is added to this value.
\newpage\thispagestyle{empty}

\begin{center}

\begin{figure}[ht]
\centering
\includegraphics[width=.7\textwidth]{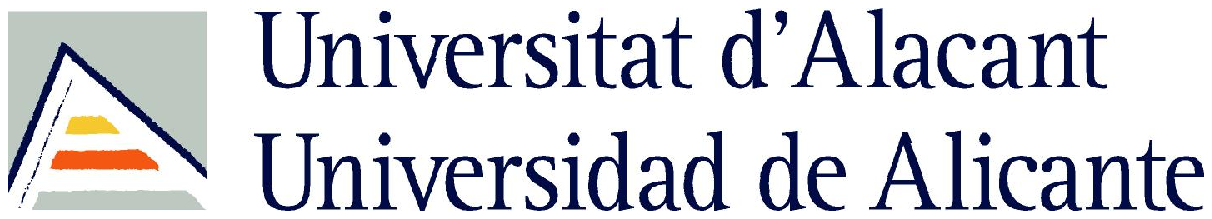}\\
\vskip1cm
{\large \textsc{Facultat de Ci\`encies - Departament de F\'isica Aplicada}}\\
{\large \textsc{Facultad de Ciencias - Departamento de F\'isica Aplicada}}\\
\end{figure}
\vskip0.8cm

\hrule

\vskip2cm

\textbf
{\Huge \color{blue} MAGNETIC FIELDS\\
\vskip15pt
IN NEUTRON STARS}

\vskip2cm

{\it \large PhD thesis by}\\
\vskip0.2cm
\textbf{\LARGE \color{blue} Daniele Vigan\`o}

\vskip0.6cm

\large \textit{Supervisors:}\\
\vskip0.2cm
\textbf{\large \color{blue}
Prof. Jos\'e Antonio Pons\\
Prof. Juan Antonio Miralles}

\vskip3cm

\hrule

\vskip0.8cm

\textsc{Alicante\\
September 2013}

\end{center}

\newpage\null\thispagestyle{empty}

\vfill
Print in July 2013, University of Alicante\\
Revised September 2013
\vskip0.5cm
\textit{Cover image} \copyright ~ by Antonella~Manzoni
\vskip3cm

\newpage\thispagestyle{empty}

\setlength{\topmargin}{-0.3in}
\setlength{\evensidemargin}{-0.5in}	% Left margin on even numbered pages (left side of the book). 1 inch is added to this value.
\setlength{\oddsidemargin}{0in}		% Left margin on odd numbered pages (right side of the book). 1 inch is added to this value.

%%%%%%%%%%%%% DEDICA %%%%%%%%%%%%%%%%%
\begin{flushright}
\null\vspace{\stretch{1}}
\large
\textit{To the wild Beauty, of any kind:\\
the Sky, the Earth, the free mind...}\\
\vspace{\stretch{2}}\null
\end{flushright}
%%%%%%%%%%%%%%%%%%%%%%%%%%%%%%%%%%%%%%%

\pagestyle{empty}
\setcounter{page}{0}
\pagenumbering{Roman}
\cleardoublepage

\pagestyle{fancy}
\tableofcontents
\cleardoublepage

\pagenumbering{arabic}
\setcounter{page}{1}

\pagestyle{empty}
\chapter*{Introduction}

This work aims at studying how magnetic fields affect the observational properties and the long-term evolution of isolated neutron stars, which are the strongest magnets in the universe. The extreme physical conditions met inside these astronomical sources complicate their theoretical study, but, thanks to the increasing wealth of radio and X-ray data, great advances have been made over the last years.

A neutron star is surrounded by magnetized plasma, the so-called magnetosphere. Modeling its global configuration is important to understand the observational properties of the most magnetized neutron stars, magnetars. On the other hand, magnetic fields in the interior are thought to evolve on long time-scales, from thousands to millions of years. The magnetic evolution is coupled to the thermal one, which has been the subject of study in the last decades. An important part of this thesis presents the state-of-the-art of the {\em magneto-thermal evolution} models of neutron stars during the first million of years, studied by means of detailed simulations. The numerical code here described is the first one to consistently consider the coupling of magnetic field and temperature, with the inclusion of both the Ohmic dissipation and the Hall drift in the crust.

The thesis is organized as follows. In chapter~\ref{ch:ns}, we give a general introduction to neutron stars. In chapter~\ref{ch:magnetosphere}, we focus on the magnetosphere, describing the analytical and numerical search for {\it force-free} configurations. We also discuss its imprint on the X-ray spectra. Chapter~\ref{ch:magnetic} describes the numerical method used for the magnetic field evolution. Chapter~\ref{ch:micro} reviews the microphysical processes and ingredients entering in the full magneto-thermal evolution code, the results of which are presented in chapter~\ref{ch:cooling}. In chapter~\ref{ch:unification}, we analyse observational X-ray data of isolated neutron stars, showing how their apparent diversity can be understood in the light of our theoretical results. In chapter~\ref{ch:timing}, we quantitatively discuss how the magnetic field evolution can contribute to explain the peculiar rotational properties observed in some neutron stars. In chapter~\ref{ch:conclusions} we summarize the main findings of this thesis.

Most of the original parts of our research (mostly contained in the second part of chapter~\ref{ch:magnetosphere}, and in chapters~\ref{ch:magnetic}, \ref{ch:cooling}, \ref{ch:unification}, \ref{ch:timing} and \ref{ch:conclusions}) have been published in the refereed papers listed below. We have extended and merged their contents in order to write a self-contained work including, when needed, overviews before entering into details of specific problems (e.g., chapters \ref{ch:ns} and \ref{ch:micro}).

We have also published on--line the results of our X-ray spectral analysis of 40 sources, including detailed references for every source, at the URL
\begin{center}
{\tt\bf http://www.neutronstarcooling.info/}
 \end{center}
We plan to update and extend periodically this freely accessible website.

\newpage
\section*{Publications directly related to this thesis.}

\subsection*{International refereed journals.}

\begin{itemize}
\item Vigan\`o D., Pons J.~A. \& Miralles J.~A. (2011),\\
{\it Force-free twisted magnetospheres of neutron stars}, A\&A, {\bf 533}, A125 
\item Vigan\`o D., Pons J.~A. \& Miralles J.~A. (2012),\\
{\it A new code for the Hall-driven magnetic evolution of neutron stars}, Comput. Phys. Comm., {\bf 183}, 2042
\item Vigan\`o D. \& Pons J.~A. (2012),\\
{\it Central compact objects and the hidden magnetic field scenario}, MNRAS, {\bf 425}, 2487 
\item Pons J.~A., Vigan\`o D. \& Geppert U. (2012),\\
{\it Pulsar timing irregularities and the imprint of magnetic field evolution}, A\&A, {\bf 547}, A9
\item Rea N., Israel G.~L., Pons J.~A., Turolla R., Vigan\`o D. \& 18 coauthors (2013),\\
{\it The outburst decay of the low magnetic field magnetar SGR 0418+5729}, ApJ, {\bf 770}, 65 
\item Pons J.~A., Vigan\`o D. \& Rea N. (2013),\\
{\it A highly resistive layer within the crust of X-ray pulsars limits their spin periods}, Nat. Phys., {\bf 9}, 431
\item Vigan\`o D., Rea N., Pons J.~A., Perna R., Aguilera D.~N. \& Miralles J.~A. (2013),\\
{\it Unifying the observational diversity of isolated neutron stars via magneto-thermal evolution models}, MNRAS, {\bf 434}, 123
\item Perna R., Vigan\`o D., Pons J.~A. \& Rea N. (2013),\\
{\it The imprint of the crustal magnetic field on the thermal spectra and pulse profiles of isolated neutron stars}, MNRAS, {\bf 434}, 2362
\end{itemize}

\subsection*{Proceedings of international conferences.}

\begin{itemize}
\item Vigan\`o D., Parkins N., Zane S., Turolla R., Pons J.~A. \& Miralles, J.~A. (2012),\\
{\it The influence of magnetic field geometry on magnetars X-ray spectra}, ``II iberian nuclear astrophysics meeting'', held in Salamanca (Spain), 22-23 September 2011
\item Geppert U., Gil J., Melikidze G., Pons J.~A. \& Vigan\`o D. (2012),\\
{\it Hall drift in the crust of neutron stars - Necessary for radio pulsar activity?}, ``Electromagnetic radiation from pulsars and magnetars'', held in Zielona G\'ora (Poland), 24-27 April 2012
\item Vigan\`o D., Pons J.~A. \& Perna R. (2013),\\
{\it Central compact objects in magnetic lethargy}, ``Thirteenth Marcel Grossman meeting on general relativity'', held in Stockholm (Sweden), 1-7 July 2012
\end{itemize}

\chapter*{Introducci\'on}

Esta tesis doctoral ha tenido como objetivo el estudio de c\'omo los campos magn\'eticos afectan a las propiedades observacionales y a la evoluci\'on a largo plazo de las estrellas de neutrones aisladas, que son los imanes m\'as potentes del universo. Las condiciones f\'isicas extremas dentro de estas estrellas complican su estudio te\'orico pero, gracias a la creciente cantidad de datos observaciones en rayos X y en radio, se han hecho grandes avances en los \'ultimos a\~{n}os.

Una estrella de neutrones est\'a tambi\'en rodeada por una regi\'on con plasma magnetizado, llamada magnetosfera. Modelizar su configuraci\'on global es importante para entender las estrellas de neutrones m\'as magnetizadas, los {\em magnetars}. Por otro lado, se cree que los campos magn\'eticos en el interior evolucionan en escalas de tiempo largas, de miles a millones de a\~{n}os. La evoluci\'on magn\'etica est\'a acoplada a la t\'ermica, que ha sido objeto de estudio en las \'ultimas d\'ecadas. Una parte importante de esta tesis presenta la vanguardia de los modelos de {\em evoluci\'on magnetot\'ermica} de las estrellas de neutrones durante el primer mill\'on de a\~{n}os de vida, estudiada mediante simulaciones num\'ericas detalladas. El c\'odigo num\'erico 
descrito en esta tesis es el primero capaz de calcular consistentemente el acoplamiento del campo magn\'etico y de la temperatura, incluyendo
los efectos del t\'ermino Hall en la corteza.

La tesis se organiza de la siguiente manera. En el cap.~\ref{ch:ns}, damos una introducci\'on general a las estrellas de neutrones. En el cap.~\ref{ch:magnetosphere}, nos centramos en la magnetosfera, describiendo distintas soluciones anal\'iticas y num\'ericas bajo ciertas aproximaciones. Tambi\'en discutimos su huella en el espectro de rayos X. El cap.~\ref{ch:magnetic} describe el m\'etodo num\'erico utilizado para la evoluci\'on del campo magn\'etico. El cap.~\ref{ch:micro} revisa los procesos e ingredientes microf\'isicos que entran en el c\'odigo completo de evoluci\'on magnetot\'ermica, cuyos resultados se presentan en el cap.~\ref{ch:cooling}. En el cap.~\ref{ch:unification}, se analizan los datos en rayos X de estrellas de neutrones aisladas, mostrando c\'omo su aparente diversidad puede entenderse a la luz de nuestros resultados te\'oricos. En el cap.~\ref{ch:timing}, discutimos cuantitativamente c\'omo la evoluci\'on del campo magn\'etico puede contribuir a explicar algunas propiedades peculiares observadas en algunas estrellas de neutrones. En el cap.~\ref{ch:conclusions} se resumen las principales conclusiones de esta tesis.

La mayor parte de los resultados originales de nuestra investigaci\'on (v\'ease la segunda parte del cap.~\ref{ch:magnetosphere}, y los cap.~\ref{ch:magnetic}, \ref{ch:cooling}, \ref{ch:unification}, \ref{ch:timing} y \ref{ch:conclusions}) se han publicado en revistas ci\'ent\'{\i}ficas internacionales de reconocido prestigio,
como listamos en la pagina anterior. Para completar esta tesis, hemos ampliado y fusionado el contenido de estos art\'{\i}culos ya publicados
con el fin de escribir una obra completa aut\'onoma, incluyendo, cuando ha sido necesario, una breve introducci\'on general antes 
de entrar en detalles t\'encinos sobre los problemas espec\'ificos (por ejemplo, los cap. \ref{ch:ns} y \ref{ch:micro}).

Tambi\'en hemos publicado on-line los resultados de nuestro an\'alisis del espectro de rayos X de 40 fuentes, incluyendo referencias detalladas de todas las fuentes, en la URL
\begin{center}
{\tt \bf http://www.neutronstarcooling.info/}
\end{center}
Tenemos la intenci\'on de actualizar y ampliar peri\'odicamente este sitio web, que es de acceso libre para la comunidad internacional.

% \newpage\null
% \newpage\pagestyle{fancy}
\newpage\pagestyle{fancy}
\chapter{Neutron stars}\label{ch:ns}

Supernovae are very bright and short-lasting stellar explosions. During the last two millennia, at least seven of them happened close enough to be detected by human eye and reported in historical records by different civilizations, mainly in China \citep{clark82}. Conversely, observations of their compact remnants, neutron stars, require advanced radio and X-ray telescopes, while the optical counterparts are usually barely detectable by space-based telescopes. This fact, together with the visionary ideas of a few physicists in the Thirties, explains why these objects are, until now, the only stars which existence and origin have been successfully predicted much before their discovery.

Soon after the experimental identification of the neutron \citep{chadwick32}, \cite{baade34} proposed the existence of neutron stars and their connection with supernovae, phenomena often observed at their workplace in the Mount Wilson Observatory. Note that Landau anticipated the basic idea of a giant nucleus star even before the neutron discovery \citep{yakovlev13}. In the Thirties, other works went in the direction of exploring the possibility of normal stars with a degenerate core \citep{chandrasekhar35}. In particular, \cite{oppenheimer39} proposed the equation of state for a relativistic gas of neutrons, preparing the basis to all successive works.

\begin{figure}
 \centering
\includegraphics[width=.7\textwidth]{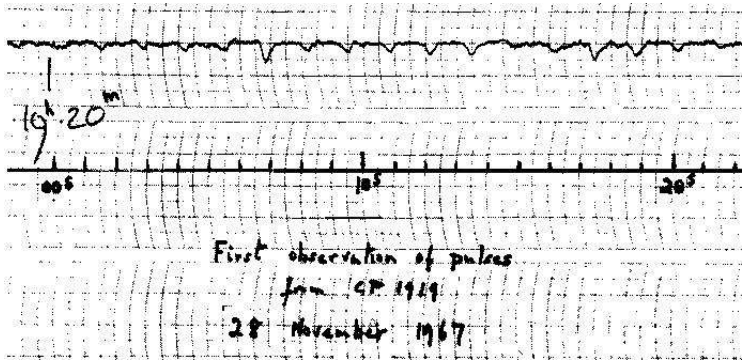}
\caption{Radio pulses of the first known pulsar CP1919 (taken from {\tt http://www.cv.nrao.edu}).} 
\label{fig:pulse_discovery}
\end{figure}

With the first rough X-ray data collected by rockets in the Sixties, neutron stars were among the candidates to explain the nature of the X-ray sources in the sky \citep{morton64}. The scientific novelties prospected by the brand new X-ray mission programs motivated several theoretical works about neutron star modeling \citep{tsuruta64}, with focus on the expected emission of thermal radiation \citep{chiu64,tsuruta65,pacini67} and gravitational waves \citep{chau67}. \cite{shkolvsky67} first correctly identified an astronomical source (Sco~X-1) as an accreting neutron star. In 1967, the notorious discovery of very regular radio pulses (\citealt{hewish68} and Fig.~\ref{fig:pulse_discovery}) was done at the Mullard Radio Astronomy Observatory. The nature of this object was identified as a rotating neutron star \citep{gold68}, dubbed {\it pulsar}, where magnetic field plays a key role \citep{gunn69,goldreich69,pacini69}. The doors to decades of intense theoretical and observational advances in the field were open.

Neutron stars are, until now, the only observed objects in the Universe where matter stably reaches so large values of pressure and density. The latter ranges from terrestrial values in the outermost layer, to $\rho\sim 10^{15}~$\gcc (several times the nuclear saturation density $\rho_0 \simeq 2.8\times 10^{14}~$\gcc) in the inner core. For this reason, their theoretical and observational study is a unique way to understand fundamental physics at regimes not achievable in terrestrial laboratories. Several branches of physics are involved. First of all, nuclear physics, because the fundamental nuclear interaction gives the equation of state that determines the structure of neutron stars. Thus, the uncertain properties of nuclear matter above the nuclear saturation density can be tested by astrophysical observations. Alternative models of compact stars, e.g. neutron stars containing also hyperons or kaon condensates, or even stars constituted by deconfined quark matter, are not excluded. However, until now there is no observational evidence favoring them with respect to the standard neutron star models.

Electrodynamics and plasma physics effects shape the observed electromagnetic radiation. Magnetic fields are about trillions of times more intense than in our Sun and have direct implications for observations. The diversity in the observed phenomenology needs accurate modeling of the electromagnetic processes taking place inside and outside the star.

Given their strong gravity, general relativistic effects are important. Neutron stars are also among the most promising candidates sources of gravitational waves. The forthcoming generation of gravitational wave detectors should be sensitive enough to frequently hear the emission from extragalactic mergers of compact objects (neutron stars, black holes, white dwarfs). Galactic neutron stars are also possible sources of gravitational waves during the first minutes after the supernova explosion. Note, however, that the birth of a neutron star is a rare event, being the galactic supernova rate $\sim 1$--$3$ per century \citep{diehl06,li11}.

\section{Origin.}

Stars spend most of their lives in a quiet balance between their own gravity and the energy released by the thermonuclear fusion of hydrogen into helium. This so-called main sequence stage lasts millions to tens of billions of years, depending on the mass of the star. When the hydrogen in the core is exhausted, the thermonuclear reactions stop. The core contracts under its own gravity while the external layers expand: the star becomes a giant. The internal temperature increases until it triggers a new chain of thermonuclear fusions that convert helium into carbon, neon and oxygen; the energy liberated in the fusion processes is able to keep high pressures that temporarily counteract gravity. This stage lasts much less than the main sequence stage, and produces a core mostly made of oxygen and carbon. The central density reaches values of the order of $\sim 10^6$--$10^{10}~$\gcc, and electrons are packed together so tightly that, due to the Pauli exclusion principle, their momenta are very large and the associated pressure, the {\em degeneracy pressure}, becomes important. If the star is light, like our Sun, the electron degeneracy pressure is able to counteract gravity, and the temperature never reaches the threshold to ignite new thermonuclear reactions. This is the endpoint expected for a star with $M\lesssim 10~M_\odot$ (where $M_\odot=1.99\times 10^{33}~$g is the mass of our Sun): a white dwarf.

In more massive stars the gravity is too strong to be balanced by the electron degeneracy pressure. Therefore, successive steps of core contractions increase the internal temperature, triggering the burning of carbon, neon, oxygen and silicon. These stages are increasingly faster and leave stratified shells of nuclear ashes, with iron and nickel in the core, which have the largest binding energy per nucleon: no energy can be released by further thermonuclear fusion of iron-like nuclei. Furthermore, the pressure provided by degenerate, relativistic electrons is still not enough to sustain the star, thus the core collapses. The high temperature reached, $T\gtrsim 3\times 10^9$ K, triggers two endothermic processes: the photo-disintegration of nuclei by thermal photons and neutrino emission. They waste part of the energy gained during the contraction and help the free-fall collapse of the core. The density increases up to the point that matter becomes composed by very heavy, neutron-rich nuclides, with a fraction of neutrons dripping out the nuclei and forming a degenerate fermionic gas. At densities $\gtrsim 10^{14}$ \gcc, the nuclei are completely dissolved into homogeneous nuclear matter mostly composed by neutrons. If the mass of the progenitor core is not too large, the degeneracy pressure of neutrons is finally enough to halt the collapse and triggers an outward shock, that reverses the in-fall of material, blowing up the star envelope and powering the supernova.

The compact remnant left behind is a hot ($T\sim 10^{11}~$K) {\em proto-neutron star}. It has a radius $R\sim 100~$km and is opaque to neutrinos, that diffuse out during the first tens of seconds. Subsequently, the star shrinks to the final size of the neutron star ($R_\star \sim 10$--$15$ km). It has a mass of $M\sim 1$--$2~M_\odot$, central densities $\sim 10^{15}~$\gcc, and spins $\sim 10$--$100$ times per second. At this point, the star becomes transparent to neutrinos. They are copiously produced in the interior and can freely escape away, draining energy from inside. The neutron star cools down to $\sim 10^9$ K within days. During the first weeks to months after birth, the outer layers of the star crystallize due to the repulsive Coulomb forces between ions. As temperature decreases, the neutrino emission processes become less efficient, but they still govern the cooling during the first $10^4$--$10^5$ yr (neutrino-cooling era). Only when the inner temperature goes below $\sim 10^8$ K, the photon emission from the surface becomes the main cause of cooling (photon-cooling era).

\section{Structure.}

\begin{figure}[t]
 \centering
\includegraphics[width=.7\textwidth]{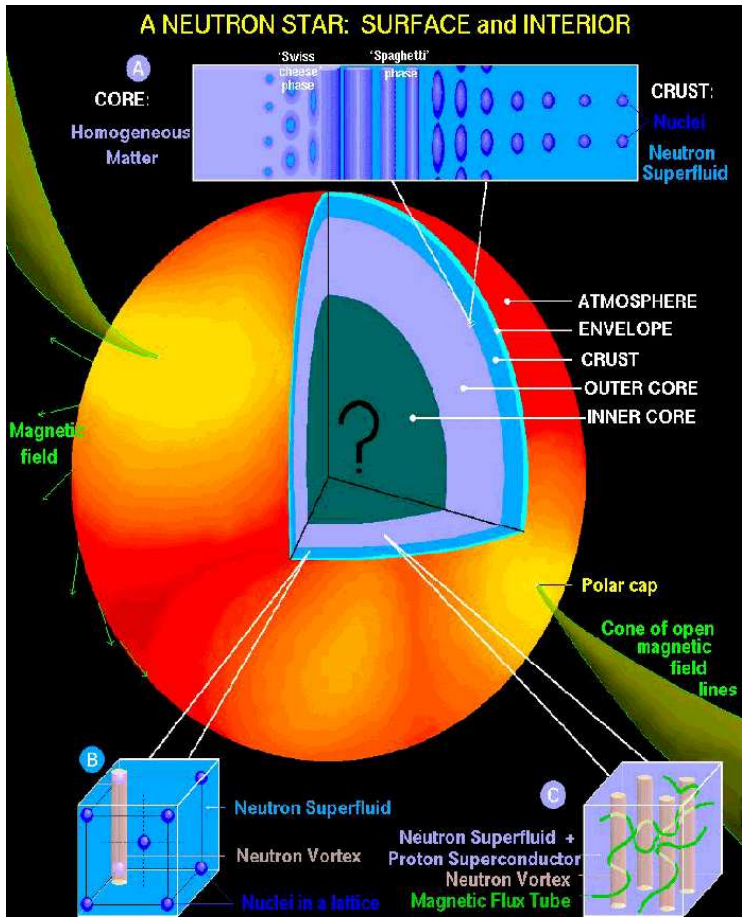}
\caption{Structure of a neutron star, taken from \cite{page06}.} 
\label{fig:ns_structure}
\end{figure}

The structure of a neutron stars is sketched in Fig.~\ref{fig:ns_structure}, taken from \citep{page06}. Below we briefly describe it.

\subsection{The envelope.}\label{sec:envelope}

The outermost ($\sim 100~$m) layer, called envelope or ocean, is liquid and has a relatively low density ($\rho \lesssim 10^9~$\gcc). It contains a negligible fraction of the total mass ($\sim 10^{-7}$--$10^{-8}~M_\odot$), and it is where the largest gradients of temperature, density and pressure are reached. In the classical picture, the density smoothly decreases down to values ($\rho \sim 1$ \gcc), and a gaseous atmosphere of a few centimeters may exist. An alternative scenario, suitable for neutron stars with large magnetic fields, is a condensed surface with a sudden drop of density from $\rho_s\sim 10^5$--$10^8~$\gcc~to the value of the surrounding environment (tens of orders of magnitude smaller).

In the outermost layer (atmosphere or condensed surface), the thermal radiation is released. The chemical composition is thought to leave an imprint on the observed spectra and depends on the particular history and environment: for instance, neutron stars in binary systems can accrete light elements from the companion.

\subsection{The crust.}\label{sec:intro_crust}

The crust is for the most part an elastic solid, comprising a Coulomb lattice of nuclei, a free gas of ultra-relativistic degenerate electrons, and coexisting with free neutrons in the inner part. The boundary between the crust and the envelope is defined by the density below which the matter is liquid. For the typical temperatures of observed neutron stars, the transition occurs at $\rho\sim 10^9$--$10^{10}~$\gcc. The crust extends up to $\rho\sim 10^{14}~$\gcc, with a thickness of about 1 km and comprising $\sim 1\%$ of the star mass. 

At these densities, matter can be described by the liquid compressible drop model, a semi-phenomenological estimate of the binding energy of nuclei, as a function of the atomic number $Z$ and the mass number $A$. The model takes into account the nuclear interaction, the surface energy, the Coulomb repulsion between protons, and the energy associated to the asymmetry in number of protons and neutrons. More sophisticated models include also the effects related to shell structure and nucleon pairing. The wealth of experimental data on nucleon-nucleon scattering allows to constrain the model parameters at low density, helping to infer the energetically favored nuclide as a function of density. 

The ion lattice is embedded in a dense gas of degenerate, ultra-relativistic electrons. Electrons are very energetic, therefore the electron capture process $e^- + p\rightarrow n + \nu_e$ is efficient and the ground state of matter becomes more and more neutron-rich with increasing density. For density $\rho > \rho_d \sim 2$--$4\times 10^{11}~$\gcc, neutrons are so abundant that they start to drip out the nuclei, forming a fermionic gas. The {\it neutron drip density} $\rho_d$ defines the boundary between the outer and inner crust. In the inner crust, heavy nuclei, electrons and free, degenerate neutrons coexist. The pressure in the outer crust and envelope is dominated by the contribution of degenerate electrons, while free neutrons become the main source of pressure for density $\rho\gtrsim 4\times 10^{12}~$\gcc.

The neutron gas is expected to undergo a phase transition when the temperature drops below a critical value $T_c\sim 10^9~$K. The onset of neutron superfluidity implies the formation of vortices, the dynamics of which is thought to be the driver of the frequently observed glitches, sudden increases in the star spin frequency.

Close to the crust-core interface, the spherical shape of nuclei is not energetically favored, due to the high energy cost given by Coulomb repulsion. For increasing density, cylindrical or planar forms are preferred \citep{ravenhall83}, because they minimize the energy by increasing the total attractive forces between electrons and protons, overcoming the surface energy. These phases are collectively named {\em nuclear pasta} (by analogy to the shape of {\em spaghetti}, {\em maccheroni} and {\em lasagne}), and they represent the transition to the uniform nuclear matter in the core.

The transport properties in the crust are expected to leave an imprint in various observed phenomena: glitches, quasi-periodic oscillations in Soft Gamma Repeaters, thermal relaxation of soft X-ray transients, long-term cooling, or magnetar behaviour. Chapters \ref{ch:magnetic}-\ref{ch:cooling} will deal with the physics of the crust involved in the long-term evolution of temperature and magnetic field.

\subsection{The core.}

With densities in the range $\rho\sim 10^{14}$--$10^{15}$ \gcc, and radius of $\sim 10~$km, the core accounts for $\sim 99\%$ of the total mass. It is classically described as a liquid, homogeneous mixture of $\sim 10^{57}$ baryons, $\sim 90\%$ of which are neutrons. The degenerate neutrons provide the bulk of pressure sustaining the star against gravity. As a consequence, the macroscopical properties of the star, most notably mass and radius, are determined by the physics at densities approaching and exceeding the nuclear saturation density.

However, the limited understanding of the fundamental nuclear interaction results in important theoretical uncertainties on the equation of state. At these densities, three-body interactions between nuclei are expected to become important, and, given the large Fermi momenta of degenerate electrons and nucleons, massive particles like muons and hyperons are expected to appear. Mass and radii inferred from observations are useful to constrain models. In particular, the recent discoveries of neutron stars with mass $\sim 2 M_\odot$ \citep{demorest10,antoniadis13} ruled out many equations of state containing strange or quark matter. Neutrons and protons are expected to form Cooper pairs below a certain critical temperature. Their superfluid and superconductive properties may significantly affect the magnetic and thermal evolution of the star.

\section{Magnetic fields.}

The magnetic field is directly related to several properties observed in neutron stars. In non degenerate stars, like our Sun, and in some white dwarfs, magnetic fields can be directly measured by the Zeeman splitting of spectral lines, or by polarization measurements. Another technique is the Doppler imaging, which consists in analysing the time-varying profiles of rotating stars, and allows to indirectly infer the presence of cold spots associated with the strongest magnetized regions of the surface. This direct measures are not possible in neutron stars.

The main signature of the magnetic field in neutron stars is the loss of rotational energy due to the electromagnetic torque. Thus, the rotational properties give an estimate of the large-scale dipolar magnetic field. From periods and period derivatives, typically $P\sim 0.001$--$12$ s and $\dot{P} \sim 10^{-16}$--$10^{-12}~$ss$^{-1}$, one can infer magnetic field intensities of $B \sim 10^{11}$--$10^{15}~$G: neutron stars are the strongest magnets in the Universe. In a few cases, X-ray spectra show hints for cyclotron lines, from which a value of the surface magnetic field can be estimated. 

The magnetic flux conservation from the progenitor and dynamo processes are the main candidates to explain the origin of such large magnetic fields. The magnetic configuration of a newly born neutron star is theoretically studied by numerical solutions of the magnetohydrodynamical equilibrium. However, these solutions are probably not unique, and it is not clear whether they are stable. Therefore, the initial configuration is an open question.

Another important issue is the long-term magnetic evolution. In the solid crust, the magnetic field evolves due to the Ohmic dissipation and the Hall drift, as explained in detail in chapter~\ref{ch:magnetic}. In the core, from hours to days after birth, protons undergo a transition to a type II superconducting phase, where magnetic field is confined to tiny flux tubes. The dynamics of flux tubes, likely coupled to the motion of superfluid neutron vortices, is a complex problem that makes the magnetic field evolution in the core formally difficult to face.

Seen from outside, a neutron star is naively seen as a spinning magnetic dipole, surrounded by the magnetosphere, a conducting region filled with magnetized plasma that ultimately shapes the emitted radiation across the electromagnetic spectrum. In localized regions of the magnetosphere, associated to the magnetic poles, a small fraction of the rotational energy loss is employed to accelerate charged particles, which in turn emit beamed electromagnetic radiation. Like a lighthouse, the electromagnetic beam can be detected as regular pulses if it periodically crosses our line of sight. The magnetospheric configuration and its observational imprints will be discussed in chapter~\ref{ch:magnetosphere}.

\section{Observations.}

In the last fifty years, observations in radio and X-ray bands have allowed to recognize a few thousands of astrophysical sources as neutron stars in different environments (isolated, in binary systems, with or without supernova remnants and/or wind nebulae...). More recently, new neutron stars and counterparts of known radio or X-ray pulsars have been detected also in the optical, infrared, ultraviolet and $\gamma$-ray bands. We now briefly summarize the neutron star phenomenology in different energy bands.

\subsection{Radio band.}

The most common manifestation of a neutron star is the detection of very regular pulses in the radio band. The individual pulses of a source have different shapes, but its period is very stable, challenging the most advanced atomic clocks built on Earth. Folding hundreds of pulses allows to measure it with an accuracy up to ten digits, very rarely obtained in astrophysics. Long follow-ups allow to measure accurately also the period derivative, that is the spin-down rate of the star. 

Despite the amazing regularity of periods, in most pulsars different kinds of timing anomalies are observed. Glitches are sudden spin-ups periodically observed in most pulsars and are thought to be associated with unpinning of superfluid neutron vortices in the crust \citep{anderson75}. They offer a variety of behaviours, in terms of spin-up amplitude and rotational properties observed during and after the post-glitch recovery. Other irregularities regard, in general, the so-called timing noise: non-negligible higher derivatives of spin periods. Other manifestations of pulse irregularity are the so-called mode changing, nulling, or intermittency. All of them are probably related to sudden changes in the magnetospheric radio emission. In particular, changes in average pulse shapes and amplitudes are likely associated with global magnetospheric reorganizations \citep{lyne10}.

The physical mechanism generating the coherent radio emission is not understood. Magnetospheric plasma effects leave an imprint on the observed polarization, and the location of the emitting region may be tested looking at the pulse profiles.

\subsection{X-rays.}

The great advances in X-ray observations during the last decades have improved our understanding of the physics of neutron stars. Most neutron stars shining in X-ray belong to binary systems, and their bright emission is powered by the accretion from the companion. X-ray binaries represents a very wide field of research, but in this work we will focus only on isolated neutron stars. More than 100 isolated neutron stars are seen in X-rays, and they are phenomenologically quite heterogeneous. Among them, the most puzzling sources are magnetars \citep{mereghetti08}: bright pulsating isolated neutron stars with relatively long periods (several seconds), showing sporadic bursts of X-ray radiation. The widely accepted view is that they are powered by their large magnetic fields, which are responsible for their bursting, timing and spectral properties. We reserve more details about X-ray observations of isolated neutron stars for chapter~\ref{ch:unification}.

\subsection{$\gamma$-ray.}

\begin{figure}[t]
 \centering
\includegraphics[width=.65\textwidth]{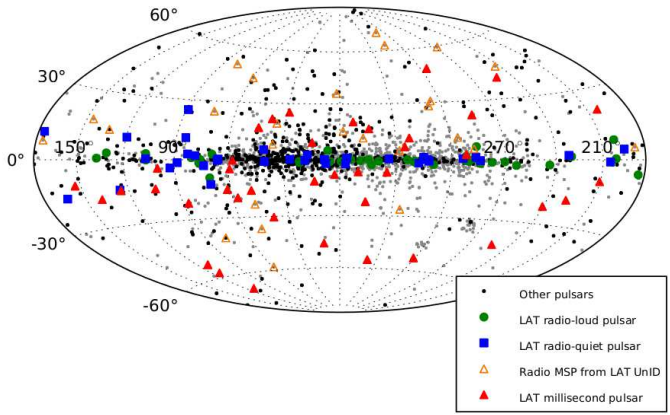}
\caption{Sky map of pulsars in Galactic coordinates (taken from \citealt{fermi2cat13}), including $117$ $\gamma$-ray pulsars and other pulsars (radio and X-ray).} 
 \label{fig:fermi_pulsars}
\end{figure}

The high energy emission of pulsars has been studied since the Seventies \citep{cheng77}, relying mostly on the data from two famous pulsars inside the Vela and Crab supernova remnants \citep{cheng86}. Later, the {\it Compton Gamma Ray Observatory} allowed to identify a few more $\gamma$-ray pulsars, one of which was the radio-quiet Geminga pulsar. The Large Area Telescope \citep{atwood09} on board of the {\it Fermi Gamma-Ray Space Telescope} mission, launched in 2008, has detected so far over one hundred pulsars \citep{fermi2cat13}, the sky map of which is shown in Fig.~\ref{fig:fermi_pulsars}. Within them, there are a comparable amount of radio-loud and radio-bright pulsars, with few of them shining also in X-rays. Furthermore, in the Very High Energy band (tens or hundreds of GeV), the most sensitive Cerenkov telescopes have detected emission from the Crab pulsar \citep{magic08,veritas11}.

The increasing statistics of high-energy pulsars helps to constrain the emission models, providing information about the region of magnetospheric emission and the involved electromagnetic processes, like curvature radiation and inverse Compton.

\subsection{Ultraviolet, optical and infrared bands.}

Neutron stars are intrinsically faint in the ultraviolet, optical and infrared wavelengths. However, technological advances recently led to the identification of counterparts of a few tens of pulsars (see \citealt{mignani12} and references within). As in the other energy bands, the analysis of the pulse profiles can give information about the location of the emitting region. Spectroscopy and polarization measurements, available for very few of them, can constrain the energy distribution and density of magnetospheric particles. Optical and ultraviolet observations can also help to infer the anisotropic thermal map of the surface and constrain the cooling model, because these bands include the bulk of thermal emission of cold stars, with ages $\gtrsim$ Myr. Optical and infrared observations are useful to test the presence of debris disks surrounding isolated neutron stars. Last, the good angular resolution of optical observations allows one to measure proper motions and parallaxes, getting fair estimates of distances.

\chapter{The magnetosphere}\label{ch:magnetosphere}

The magnetosphere is the region surrounding a star filled with magnetized plasma. To obtain a fully consistent magnetospheric model we should take into account, at the same time: the large-scale magnetic field configuration, the physical mechanisms producing electromagnetic radiation, the generation of plasma needed to sustain it, the interplay between radiation and plasma, and the short- and long-term evolution of the system. Given the complexity of the problem, one has to decide whether to focus on detailed microphysical processes of an approximate macrophysical solution or to study the large-scale electrodynamics with simplifying hypotheses about microphysics. The latter is the main subject of this chapter.

\begin{figure}[t]
 \centering
 \includegraphics[width=6cm]{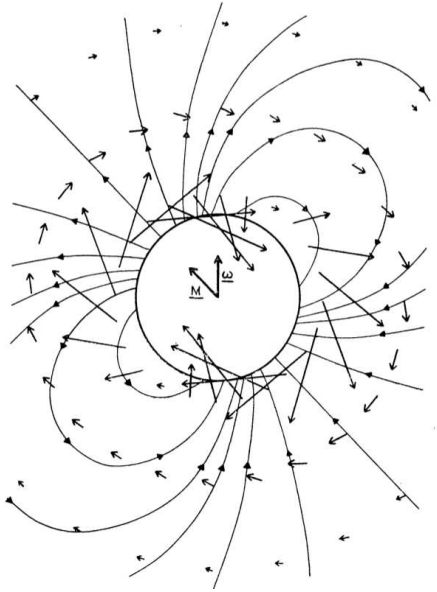}
 \caption{Picture of a rotating magnetized star {\it in vacuo} depicted by \cite{deutsch55}.}
 \label{fig:deutsch}
\end{figure}

The first global electrodynamics model around a rotating magnetized star can be traced back to well before the discovery of pulsars \citep{ferraro37}. \cite{deutsch55} first proposed the basic mechanism regulating the spin-down of a magnetized star {\it in vacuo} (see Fig.~\ref{fig:deutsch}). \cite{pacini67} proposed the first sketch of a vacuum oblique rotator model. \cite{goldreich69} described the pulsar electrodynamics in the simplest case of a rotating magnetic dipole, aligned with the rotational axis, surrounded by a charge-separated plasma.

Most of these pioneering models were based on the same underlying assumption: the magnetic field is arranged in a force-free configuration. Under this hypothesis, the electromagnetic forces, much stronger than any other force in the system, are exactly balanced to have no net force on a charge. Rotation induces currents that make the magnetic field deviate from vacuum, potential configurations. The resulting magnetic force must therefore be compensated by an electric field perpendicular to the magnetic field, which implies space-charge separation. The further assumption of axial symmetry leads to the so-called pulsar equation, the solutions of which describe the large-scale electrodynamics of an aligned rotator. The problem is analytically solvable only for a few simple, \textit{ad-hoc} choices of the current, but these solutions generally present unphysical features in some border regions. More than 30 years after the pulsar discovery, \cite{contopoulos99} presented the first numerical, realistic configuration for an aligned rotator, with a smooth matching between the inner and outer regions.

The dipolar aligned rotator was the first step before quantitatively considering the 3D effects of the misalignment between rotation and magnetic axes \citep{spitkovsky06}. The magnetic field inside a neutron star is thought to be more complex than in these models. Even in the aligned rotator model, the magnetic field has in general an azimuthal component, which twists the poloidal magnetic field lines. The internal configuration likely presents significant deviations from the dipolar geometry (see chapter~\ref{ch:magnetic}) and has a strong effect on the external magnetic field.

Note also that alternative models \citep{petri09} strongly question the stability of the large-scale force-free configurations \citep{krause85}. Working in axial symmetry, they proposed configurations with relatively small regions, called electrospheres, where a non-neutral plasma is confined and differentially rotates. Most of the surrounding space is a region depleted of plasma, where the electric field is not orthogonal to the magnetic field. These models, although interesting, are less developed than the long studied force-free magnetospheres: hereafter we consider only the latter.

Observationally, an interesting property of the persistent emission of most magnetars is that their X-ray spectra are well fitted by a thermal component with $k_bT_{bb} \sim 0.3$--$0.7~$keV (where $k_b=1.38\times 10^{-16}~$erg~K$^{-1}$ is the Boltzmann constant and $T_{bb}$ is the blackbody temperature), plus a hard non-thermal tail, described by a power-law with photon index $\beta\sim 2$--$4$. Some of them also exhibit a tail in the hard X-ray band, $E \sim 20$--$200~$keV \citep{enoto10}. These tails are commonly explained by the resonant Compton scattering of thermal photons coming from the surface by magnetospheric particles. This process is efficient in presence of a relatively dense and highly magnetized plasma. In this framework, the magnetosphere of standard radio pulsars is thought to be closer to the classical dipolar field geometry, while magnetar activity is compatible with a {\it twisted magnetosphere}, likely coming from the transfer of magnetic helicity from the crustal magnetic field \citep{thompson95}. 

The force-free condition cannot be accomplished in the whole magnetosphere: any suitable emission mechanism invokes the presence of regions ({\it gaps}) where the force-free condition is violated, and the electric field accelerates particles along magnetic field lines. Small deviations from force-free equilibrium are also required at a global scale in a twisted magnetosphere, in order to maintain a voltage $\sim 10^9~$V along magnetic field lines \citep{beloborodov07,beloborodov13}. In this scenario, electron-positron pairs are continuously produced and accelerated, thus they replenish the magnetosphere and are a source of radiation. The challenge is to model the coupling between the radiation field and the plasma dynamics, taking into account at the same time the force-free constraint. In Beloborodov's model, the plasma is immersed in an intense radiation field coming from the star surface and from back-scattered $\sim$ keV photons coming from the outer regions of the magnetosphere. The outflowing particles gradually slow down, by means of the emission of progressively less energetic photons, which populate a hard X-ray tail up to hundreds or thousands of keV. This mechanism self-regulates the deceleration of outflow, until the particles have radiated almost all the energy away. The radiation properties predicted by Beloborodov's model seem to be in good agreement with the outburst and persistent properties of magnetars \citep{beloborodov13,mori13}.

In this chapter, we are ultimately interested in studying the imprint of magnetic field geometry on the magnetar spectra. For this reason, we focus on the large-scale description of the magnetic field configuration. Resistive processes in the magnetosphere act on a typical timescale of years \citep{beloborodov09}, much longer than the typical response of the tenuous plasma, for which Alfv\'en velocity is close to the speed of light. Consequently, a reasonable approximation for the large-scale configuration is to consider stationary, force-free solutions, ignoring the small deviations from this condition. We then use the code by \cite{nobili08b} to compute the expected X-ray spectra under certain simplifying hypotheses.

\section{Force-free magnetospheres.}\label{sec:forcefree}

\subsection{Maxwell equations.}\label{sec:maxwell}

The evolution of the electromagnetic field is governed by the Maxwell equations, which, if relative to an observer at rest, are written in Gaussian units as

\begin{eqnarray}
 && \vec{\nabla}\cdot\vec{E}=4\pi \rho_q~, \label{eq:poisson_new}\\
 && \frac{1}{c}\frac{\partial \vec{E}}{\partial t} = \curlB - \frac{4\pi}{c}\vec{J}~, \label{eq:ampere_new}\\
 && \vec{\nabla}\cdot\vec{B}=0 \label{eq:divb_new}~, \\
 && \frac{1}{c}\frac{\partial \vec{B}}{\partial t} = - \vec{\nabla}\times\vec{E}~, \label{eq:induction_new}
\end{eqnarray}
where $c$ is the speed of light, $\vec{E}$ and $\vec{B}$ are the electric and magnetic fields, $\rho_q$ is the electric charge density, and $\vec{J}$ is the current density. General relativistic corrections are of the order of $\sim 20\%$ at the surface, and decrease linearly with distance. We neglect these corrections in this chapter, while they will be included for the evolution of the magnetic field inside the neutron star (chapter~\ref{ch:magnetic}).

Magnetohydrodynamics (MHD) studies a conducting, magnetized medium and its dynamics. Throughout this work, we will assume that the timescale of variation of the electromagnetic field is much larger than the typical timescale of collisions inside plasma. Therefore, in the Amp\`ere-Maxwell equation~(\ref{eq:ampere_new}), the displacement current, i.e. the left-hand side term, is negligible compared to the right-hand side terms. In other words, the conducting fluid is able to respond almost instantaneously to any variation of $\vec{B}$, in order to establish a current
\begin{equation}\label{eq:current_mhd_new}
 \vec{J} = \frac{c}{4\pi}(\curlB)~.
\end{equation}

\subsection{The unipolar induction.}

A perfect conductor rotating in a magnetic field undergoes the so-called {\it unipolar induction}. This principle was used by Faraday to build the first homopolar generator in 1831, in order to convert kinetic energy into electric voltage. The same mechanism is supposed to govern the highly conducting plasma surrounding pulsars, the rotational energy of which is converted in electromagnetic radiation. In absence of other forces, the charges co-rotating with the star feel the magnetic force orthogonal to both magnetic field and velocity. If  plasma particles can be freely supplied, the system is able to separate the electric charges so that the resulting electric force compensates the magnetic force:

\begin{equation}\label{eq:balance_em}
  \vec{E}+\frac{\vec{v}}{c}\times\vec{B}=0~.
\end{equation}
This implies that the accelerating electric fields along field lines is null, $\vec{E}\cdot\vec{B}=0$.

We can obtain the same result from a relativistic point of view, considering the star in the co-rotating frame, which will be denoted by primes. An inertial observer sees the plasma rotating with a velocity directed along the azimuthal direction $\hat{\varphi}$, and given by $c\vec{\beta}_{rot}=v_{rot}(r,\theta)\hat{\varphi}=\vec{\Omega}(r,\theta)\times \vec{r}$, where $\vec{\Omega}$ is the angular velocity vector. The relativistic Lorentz transformations for the electromagnetic field are \citep{jackson91}:

\begin{eqnarray}
  && \vec{E'}=\gamma_{rot}(\vec{E}+\vec{\beta}_{rot}\times\vec{B})-\frac{\gamma_{rot}^2}{\gamma_{rot}+1}\vec{\beta}_{rot}(\vec{\beta}_{rot}\cdot\vec{E})~, \label{eq:lorentz_e} \\
  && \vec{B}'=\gamma_{rot}(\vec{B}-\vec{\beta}_{rot}\times\vec{E})-\frac{\gamma_{rot}^2}{\gamma_{rot}+1}\vec{\beta}_{rot}(\vec{\beta}_{rot}\cdot\vec{B})~, \label{eq:lorentz_b} 
\end{eqnarray}
where $\gamma_{rot}\equiv(1-\beta_{rot}^2)^{-1/2}$ is the Lorentz factor. If the plasma surrounding the star is a perfect conductor, there is no electric field in the co-rotating frame: $\vec{E}'=0$. Multiplying eq.~(\ref{eq:lorentz_e}) by $\vec{\beta}_{rot}$, we obtain the condition $\vec{\beta}_{rot}\cdot\vec{E}=0$. An inertial observer sees an electric field perpendicular to the plane defined by the magnetic field and velocity:

\begin{equation}\label{eq:rotating_efield}
  \vec{E}=-\vec{\beta}_{rot}\times\vec{B}=-\frac{1}{c}(\vec{\Omega}\times \vec{r})\times\vec{B}~.
\end{equation}
In spherical coordinates $(r,\theta,\varphi)$, and indicating hereafter the partial derivatives respect to the coordinate $x$ with $\partial_x$, the curl of the electric field is\footnote{Remember that the derivatives of spherical unit vectors respect to $\varphi$ are not null.}: 

\begin{eqnarray}\label{eq:curl_ef}
  && \vec{\nabla}\times\vec{E}=(\vec{\beta}_{rot}\cdot\vec{\nabla})\vec{B}_{pol}-(\vec{B}_{pol}\cdot\vec{\nabla})\vec{\beta}_{rot}= \nonumber\\
  && \left[B_r\left(\frac{v_{rot}}{r}-\partial_r v_{rot}\right)+\frac{B_\theta}{r}\left(v_{rot}\frac{\cos\theta}{\sin\theta}-\partial_\theta v_{rot}\right)\right]\hat{\varphi}~,
\end{eqnarray}
where the ${pol}$ subscript indicates the poloidal projection, i.e., the meridional components, perpendicular to the azimuthal velocity (see Appendix~\ref{app:poloidal-toroidal} for definitions). The electric field is irrotational if and only if the plasma is rigidly rotating, $v_{rot}=\Omega r \sin\theta$, regardless of the magnetic field configuration. The Lorentz transformations for the magnetic field, eq.~(\ref{eq:lorentz_b}), combined with eq.~(\ref{eq:rotating_efield}), are:

\begin{eqnarray}
  && \vec{B}'_{pol}=\gamma_{rot}^{-1}\vec{B}_{pol}~,\\
  && \vec{B}'_{tor}=\vec{B}_{tor}~,
\end{eqnarray}
where we used $\vec{\beta}_{rot}\times(\vec{\beta}_{rot}\times\vec{B})=(\vec{\beta}_{rot}\cdot \vec{B})\vec{\beta}_{rot}-\beta_{rot}^2\vec{B}$. A co-moving observer sees a poloidal magnetic field smaller by a factor $\gamma_{rot}^{-1}$ than in the inertial frame, while the azimuthal component is the same. Note that, inside a neutron star, $\beta_{rot}\lesssim 10^{-3}-10^{-2}$ for the typical periods and radii, therefore $\gamma_{rot}\simeq 1$.

\subsection{Instability of vacuum surrounding a neutron star.}

The unipolar induction applies inside the highly conductive star, but what about the space surrounding the star? The seminal work by \cite{goldreich69} suggested that a rotating neutron star cannot be surrounded by vacuum. Below, we review their arguments.

Consider a perfectly conducting neutron star, rotating with angular velocity $\Omega$. The star is endowed with a magnetic dipole moment aligned with the the rotation axis and with value $|\vec{m}|=(B_pR_\star^3)/2$, where $R_\star$ is the star radius and $B_p$ is the magnetic field intensity at the pole. The components of the magnetic field are, in spherical coordinates:
\begin{eqnarray}\label{eq:mf_dipole}
 B_r &=& B_p \cos\theta \left(\frac{R_\star}{r}\right)^3~, \\
 B_\theta &=& \frac{B_p}{2} \sin\theta \left(\frac{R_\star}{r}\right)^3~.
\end{eqnarray}
The rotationally induced electric field, (eq.~\ref{eq:rotating_efield}), as seen by an observer in the inertial frame is:

\begin{equation}\label{gj_eint}
 \vec{E}_{int}=\frac{\vec{\Omega}\cdot\vec{m}}{cr^2}(\sin^2\theta \hat{r}-2\sin\theta\cos\theta\hat{\theta})~,
\end{equation}
corresponding to the electrostatic potential

\begin{equation}\label{eq:phi_int}
 \Phi_{int}(r,\theta)=\frac{\vec{\Omega}\cdot\vec{m}}{cr}\sin^2\theta + d~,
\end{equation}
where $d$ is an arbitrary constant. Inside the star the electric charge density is

\begin{equation}\label{eq:rhoint}
 \rho_q(r,\theta)=\frac{1}{4\pi}\vec{\nabla}\cdot\vec{E}_{int}= \frac{\vec{\Omega}\cdot\vec{m}}{2\pi c r^3}(3\cos^2\theta-1)~.
\end{equation}
The electric charge is necessary to sustain the rotation in a perfect conductor, in order to have a zero net Lorentz force. The effect is the separation of charges at different latitudes $\theta$, and the resulting distribution is quadrupolar.\footnote{The Gauss' theorem gives an unphysical non-zero net charge located in the origin $Q_c=(2/3c)\vec{\Omega}\cdot\vec{m}$. Note however that eq.~(\ref{eq:rhoint}) diverges in the origin. This is due to the idealized point magnetic dipole, which is not a physically consistent solution for the interior. In reality, the finite distribution of current inside the star will result in more realistic and complicated internal configurations of magnetic field (see chapter~\ref{ch:magnetic}).} The interior solution has to match with the electromagnetic field external to the star. If the star is surrounded by vacuum, the Laplace's equation $\nabla^2\Phi_{ext}=0$ holds, and the electrical potential can be described as a multipolar expansion:

\begin{equation}
 \Phi_{ext}(r,\theta)=\sum_l a_l P_l(\cos\theta)r^{-(l+1)}~,
\end{equation}
where $P_l$ are the Legendre polynomials (see Appendix~\ref{app:legendre}). The potential has to be continuous at the surface $r=R_\star$: looking at the eq.~(\ref{eq:phi_int}), the only multipoles permitted are the monopole $l=0$, associated to the net charge of the star, and the quadrupole $l=2$. Choosing the free parameter $d$ in eq.~(\ref{eq:phi_int}) in order to have no monopole term, the electric potential outside the star is: 

\begin{equation}
 \Phi_{ext}(r,\theta)=-\frac{\vec{\Omega}\cdot\vec{m}}{3 c}\frac{R_\star^2}{r^3}(3\cos^2\theta-1)~,
\end{equation}
which gives an external electric field

\begin{equation}\label{gj_eext}
 \vec{E}_{ext}=  -\frac{\vec{\Omega}\cdot\vec{m}}{3 c}\frac{R_\star^2}{r^4}\left[ (3\cos^2\theta-1)\,\hat{r} + 2\sin\theta\cos\theta\,\hat{\theta}  \right]~.
\end{equation}
Across the surface the radial electric field has a discontinuity which yields to a surface charge:

\begin{equation}\label{gj_surf_charge}
 \sigma_q(\theta)=\frac{\Delta E_r(r=R_\star)}{4\pi}=-\frac{\vec{\Omega}\cdot\vec{m}}{2\pi c R_\star^2}\cos^2\theta~.
\end{equation}
Integrating over the surface, the total charge is $Q_s=-(2/3c)\vec{\Omega}\cdot\vec{m}$, which is of the order of $10^{12}~$C, or few kilograms of electrons. The external quadrupolar electric field is very intense along the dipolar magnetic field lines:

\begin{equation}
 \vec{E}\cdot\vec{B}|_{ext}=E_rB_r+E_\theta B_\theta=-B_0\frac{2\vec{\Omega}\cdot\vec{m}}{c}\frac{R_\star^5}{r^7}\cos^3\theta~.
\end{equation}
Introducing $B_{12}=|\vec{B}|/10^{12}~$G and $R_6=R_\star/10^6~$cm, the strength of the electric field parallel to the magnetic field can be estimated as:

\begin{equation}
 E_\parallel \sim  10^{12}~\frac{R_6B_{12}}{P[\mbox{s}]}\left(\frac{R_\star}{r}\right)^4~\mbox{N~C}^{-1}~,
\end{equation}
where $P=2\pi/\Omega$ is the spin period. We can compare the electromagnetic forces with the other forces experienced, for instance, by an electron. The electric, gravitational and centrifugal accelerations are, respectively
\begin{eqnarray}
 a_{el} & \sim & \frac{eE_\parallel}{m_e} \sim 10^{25}~\frac{R_6B_{12}}{P[\mbox{s}]}\left(\frac{R_\star}{r}\right)^4  \mbox{~cm~s}^{-2}~,\\
 g & \sim & 1.3\times10^{14}~\frac{M}{M_\odot}\frac{1}{R_6^2}\left(\frac{R_\star}{r}\right)^2  \mbox{~cm~s}^{-2}~,\\
 \Omega^2 r & \sim & 3\times 10^7 ~\frac{R_6}{P[\mbox{s}]^2} \frac{r}{R_\star} \mbox{~cm~s}^{-2}~.
\end{eqnarray}
The ratios between them are:

\begin{eqnarray}
 && \frac{a_{el}}{\Omega^2 r}\sim 3\times 10^{17} ~ B_{12}P[\mbox{s}]  \left(\frac{R_\star}{r}\right)^5 ~,\label{eq:ratio_forces1}\\
 && \frac{a_{el}}{g}\sim 10^{11} ~ \frac{M_\odot}{M}\frac{B_{12}R_6^3}{P[\mbox{s}]}\left(\frac{R_\star}{r}\right)^2~.\label{eq:ratio_forces2}
\end{eqnarray}
Therefore we can safely neglect the centrifugal and gravitational terms, unless $r\gg R_\star$. Centrifugal and electromagnetic force would become comparable at $r\sim 10^3 ~R_\star$; the gravitational term is negligible.

Note that, in general, the net electric charge of the star could be non-zero, due, for instance, to stripped particles from the surface. In this case, a different choice of $d$ in eq.~(\ref{eq:phi_int}) leads to an additional monopole surface charge. In any case, if the star is surrounded by vacuum, a strong electric field parallel to $\vec{B}$ is expected at the surface. Thus, the enormous electric forces would pull out charged particles from the surface, unless the work function and cohesive forces are unrealistically large. There are several theoretical caveats about this naive picture, but the strong result is that a vacuum space surrounding a neutron star implies huge forces acting on the surface layer, that are avoided only if a conducting plasma surrounds the star.

\subsection{The co-rotating magnetosphere.}\label{sec:aligned_rotator}

\begin{figure}
 \centering
 \includegraphics[width=.7\textwidth]{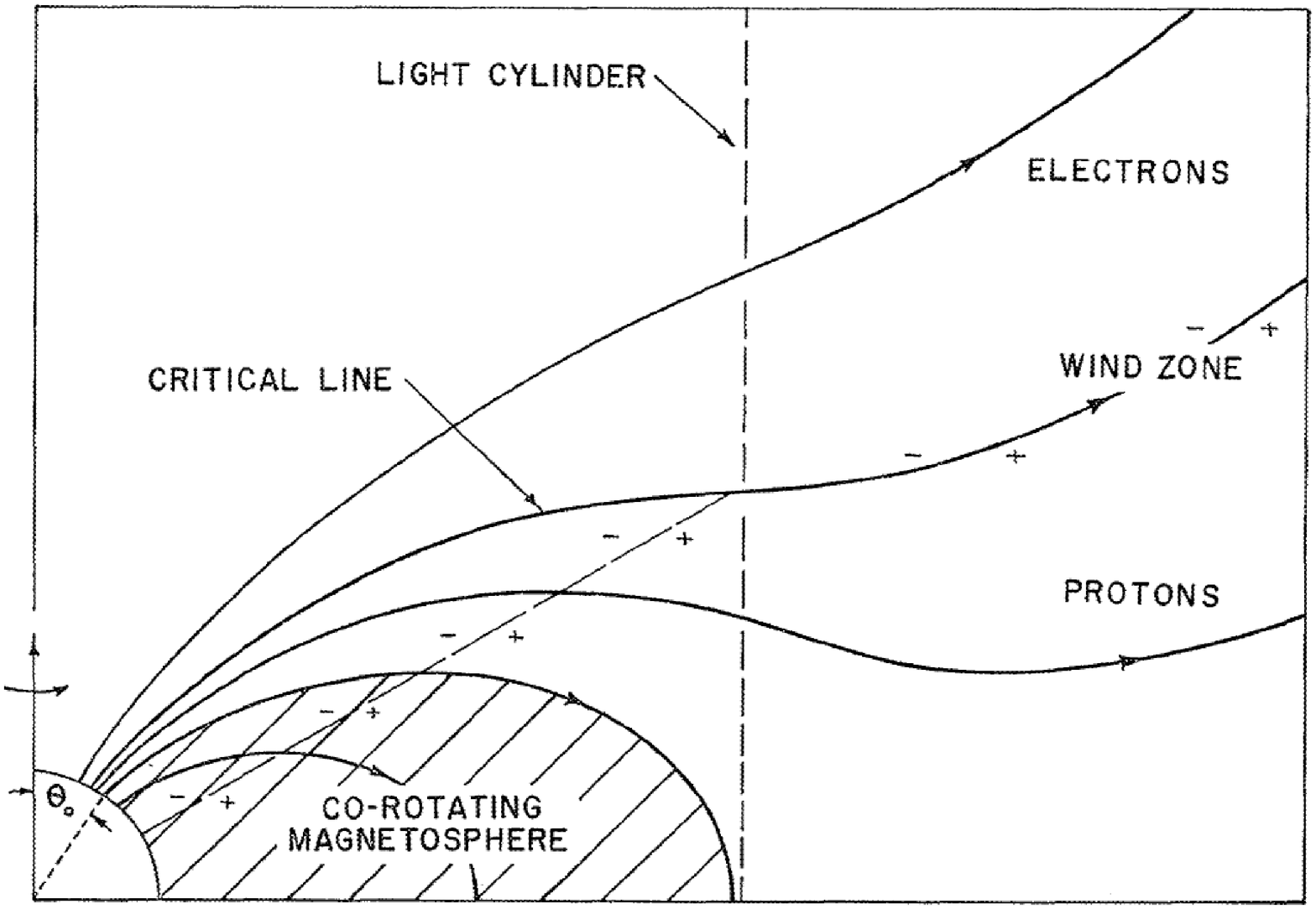}
 \caption{The famous picture of the aligned rotator model, taken from \cite{goldreich69}.}
 \label{fig:aligned_gj}
\end{figure}

\cite{goldreich69} also illustrated the global picture of pulsar electrodynamics, reproduced in Fig.~\ref{fig:aligned_gj} and summarized as follows. With the hypothesis of a free supply of plasma able to replenish the magnetosphere, consider an aligned rotator, i.e. a neutron star whose spin axis and magnetic moment are aligned. If the plasma is a perfect conductor, the unipolar induction drives the plasma to co-rotate rigidly with the star with rotational velocity $\vec{v}_{rot}=\vec{\Omega}\times\vec{r}$ and induces the separation of charges. The charge density is:

\begin{equation}\label{eq:charge_density_aligned}
  \rho_q(r,\theta) = -\frac{1}{4\pi c}\vec{B}\cdot[\vec{\nabla}\times(\vec{\Omega}\times\vec{r})]+\frac{\vec{\Omega}}{c^2}\cdot(\vec{r}\times\vec{J})
= -\frac{\vec{\Omega}\cdot\vec{B}}{2\pi c} + \frac{\Omega r\sin\theta J_\varphi}{c^2}~,
\end{equation}
where we have used eq.~(\ref{eq:current_mhd_new}) and $\vec{\nabla}\times(\vec{\Omega}\times\vec{r})=2\vec{\Omega}$. The co-rotation of the charge separated plasma induces a convection current:

\begin{equation}\label{eq:co-rotating_jphi}
  J_\varphi=\rho_q\Omega r \sin\theta~.
\end{equation}
In turn, this current is a source of magnetic field, that becomes more and more distorted with increasing distance, compared with a potential confguration (e.g., a dipole in vacuum). The {\em Goldreich-Julian density} is obtained combining eqs.~(\ref{eq:charge_density_aligned}) and (\ref{eq:co-rotating_jphi}):

\begin{equation}\label{eq:rho_gj}
  \rho_{gj}(r,\theta)=-\frac{\vec{\Omega}\cdot\vec{B}}{2\pi c(1-(\Omega r \sin\theta/c)^2)}~.
\end{equation}
For a magnetic dipole with $\vec{m}\cdot\vec{\Omega}>0$, the polar regions are negatively charged, while the equatorial region is positively charged. In the near zone where $\Omega r \sin\theta/c\ll 1$ the magnetic field is nearly potential and the charge density, for a dipolar magnetic field, is

\begin{equation}\label{eq:rho_gj_close}
  \rho_{gj}(r,\theta)\simeq-\frac{\Omega B_0 (1-(3/2)\sin\theta^2)}{2\pi c}~.
\end{equation}
In this case, the zero-charged surfaces (northern and southern hemispheres) are defined by $|\sin\theta_\pm|=\sqrt{2/3}$, which means $\theta_\pm=49^\circ,131^\circ$.  Assuming a completely separated charge plasma, and charges $Z=1$ (electrons, protons, positrons), the typical number density of particles, $n_{gj}=\rho_{gj}/e$, is

\begin{equation}\label{eq:n_gj}
  n_{gj} \sim 6.9\times10^{10}\frac{B_{12}}{P[\mbox{s}]} \mbox{ cm}^{-3}~.
\end{equation}
The largest inferred values of $B/P$ in observed pulsars is $\sim 5 \times 10^{14}$ G/s (PSR~J1808--2024), for which $n_{gj} \sim 3\times 10^{13}$ cm$^{-3}$ near the surface. Density quickly decreases with distance, due to the magnetic field dependence (for a vacuum dipole, $B\sim r^{-3}$). Note however, that, for increasing distance from the rotational axis, the magnetic field is distorted and eq.~(\ref{eq:rho_gj_close}) is not a good approximation.

\subsection{The light cylinder and the open field lines region.}

Plasma rigidly co-rotates with the star only in a region spatially limited by the finiteness of the speed of light. The {\em light cylinder} is the distance from the rotational axis at which a co-rotating particle would reach the speed of light:

\begin{equation}\label{eq:def_light_cyl}
  \varpi_l=\frac{c}{\Omega}=4.8\times 10^4~P[\mbox{s}]~\mbox{km}~,
\end{equation}
which typically corresponds to several hundreds to thousands of stellar radii. Some magnetic field lines close inside it, while those connected to the polar region cross it, and the particles moving along them cannot co-rotate. The separatrix is the line dividing the {\em co-rotating magnetosphere} from the open field lines region. The polar cap is the portion of the surface connected with the open field lines. For a dipolar magnetic field, a first estimate of its semi-opening angle is:

\begin{equation}\label{eq:critical_line}
  \sin\theta_0=\sqrt{\frac{R_\star}{\varpi_l}}=\sqrt{\frac{R_\star\Omega}{c}}~,
\end{equation}
which, for $R_\star = 10$ km, corresponds to

\begin{equation}\label{eq:polar_cap}
  \theta_0 \sim \frac{0.8^\circ}{\sqrt{P[\mbox{s}]}}~.
\end{equation}
The rotation has the effect to open the magnetic field lines and $\theta_0$ can be slightly larger. For the typical range of pulsar periods, $\theta_0$ can vary between $\sim 0.2^\circ$ and $\sim 20^\circ$. This has direct implications on the width of the emitted radio beam, biasing the radio detections against long period pulsars.

The electrodynamical description of the open field lines is not trivial. Along them, charged particles can flow to or from the so-called wind zone, providing currents which in turn modify the magnetic field configuration in the wind zone and beyond. In the regions beyond the light cylinder, the magnetic field is thought to be mainly radial and twisted. In general, building a self-consistent global solution which is smooth across the critical regions (light cylinder, separatrix, zero-charge surfaces) is not an easy task.

\subsection{Gaps and energy emission.}

The existence of a bunch of open magnetic field lines, connected to the outer space, raises questions about the mechanism of the electromagnetic emission. A related question is what the magnetospheric plasma is made of and how is it supplied. Two basic mechanisms can operate: extraction of particles from the neutron star surface and $e^--e^+$ pair production. The latter is possible in vacuum for magnetic fields $B\gtrsim 4\times 10^{13}$ G. The capability to extract particles depends on the structure and physical conditions of the outer layer (atmosphere, condensed surface...), which determines the cohesive energy of ions and the work function of electrons \citep{medin06a,medin06b}. 

Both mechanisms of plasma generation are related to the presence of gaps, regions depleted of plasma where the current needed to establish force-free conditions cannot be supplied. In these regions, the electric field is not screened and can accelerate particles along magnetic field lines. They consequently emit photons by curvature radiation or inverse Compton scattering. If the photons are energetic enough, they can trigger a cascade of pairs. There are different magnetospheric regions candidate to host the gaps.

The first models proposed an inner accelerator, just above the polar cap \citep{sturrock71,ruderman75,daugherty96}. In this vacuum region with typical height $\sim 10^4$ cm, a mechanism of continuous sparks discharging the voltage produces the pair cascade. Based on these models, more realistic scenarios describe the gap with a space-charge limited flow, with partial screening of electric field \citep{arons79}, and including pair production \citep{harding98}. \cite{gil03} consider a partially screened gap, whose efficiency in producing radiation depends on the local surface temperature and magnetic field \citep{medin07}. In this model, the particle bombardment heats up the surface, producing tiny hot spots. Similar models are the slot gaps: narrow regions located above the polar cap, elongated along the separatrix \citep{arons83,arons79,muslimov03}. Qualitatively different scenarios are the outer gap models \citep{cheng86,romani96}, in which the acceleration happens in the outer magnetosphere, along the separatrix or near the light cylinder.

A way to test a gap model is the analysis of the observed pulse profile at different energies (e.g., \citealt{venter12}). Most studies assume a location of the gap (e.g., above the polar cap), a magnetic field configuration (usually a vacuum dipole, but see \citealt{bai10} for more realistic force-free configurations), and a geometry for the radiation beam. Once the inclination and viewing angles are fixed, the theoretical pulse profile is obtained and can be compared with observations.

The pulse profiles in the high-energy band are in agreement with the prediction of the outer gap models, disfavoring the polar cap and slot gap models \citep{kerr13}. In radio, statistical studies of pulse widths indicate a typical altitude of $\sim 0.01-0.1 \varpi_l$, assuming the emission from the last open field line \citep{kijak03}. We stress that most gap models are phenomenological, in the sense that they study the possible geometric configurations for radiation beam, angles, and magnetic field, but the physical mechanism producing the radiation is not specified. In fact, it is still unknown in radio, while inverse Compton scattering and curvature radiation are the most promising candidates for high energy emission. The recent works by \citealt{timokhin13} and \citealt{chen13} try, at the same time, to solve the problem of electrodynamical self-consistency and to reproduce the radio and high-energy pulse profiles.

\section{Pulsar spin-down properties.}\label{sec:em_torque}

\subsection{Energy loss.}

Timing measurements of pulsars show a positive time derivative of their spin period, $\dot{P}$ (i.e., a negative time derivative of the angular velocity, $\dot{\Omega}$). The loss of rotational energy is
\begin{equation}\label{eq:erot}
  \dot{E}_{rot} \equiv I\Omega \dot{\Omega} = -3.95 \times 10^{46} ~ I_{45} \frac{\dot{P}}{P^3} \mbox{~erg~s}^{-1}~,
\end{equation}
where $I\equiv\int_V \rho(r) r^2 \de V$ is the moment of inertia of the star, and $I_{45}=I/(10^{45}~$g~cm$^2)$. An important issue regards the physical mechanisms responsible for the spin-down torque. Classical electrodynamics predicts that a time varying magnetic momentum $\vec{m}$ results in a loss of energy described by the Larmor formula $\dot{E}=(2/3)|\ddot{\vec{m}}|^2/c^3$ \citep{jackson91}. A magnetic dipole moment, forming an angle $\chi$ with the rotation axis, is written as
\begin{equation}
 \vec{m}=\frac{B_pR_\star^3}{2}\sin\chi ~e^{i\Omega t}~\hat{z}~,
\end{equation}
where $\hat{z}$ is the unit vector aligned to the spin vector. Then the Larmor formula,
\begin{equation}\label{eq:en_loss_vacuum}
\dot{E}_{vac} = - \frac{B_p^2R_\star^6\Omega^4}{6c^3} \sin^2\chi~,
\end{equation}
predicts that the energy is lost in form of Poynting flux with extremely low frequency $\Omega\sim$ Hz-kHz. When the effect of the magnetosphere is taken into account in the force-free models, the energy losses can be well approximated by \citep{spitkovsky06}:
\begin{equation}\label{eq:en_loss_forcefree}
\dot{E}_{ff} = - \frac{B_p^2R_\star^6\Omega^4}{4 c^3} (1+\sin^2\chi)\;.
\end{equation}
The power in this case is a factor $1.5(1+\sin^2\chi)$ larger than for the vacuum orthogonal rotator, eq.~(\ref{eq:en_loss_vacuum}), and it is non-zero even in the aligned case ($\chi=0$). This difference comes from the proper inclusion of the rotationally induced electric field. For higher order multipoles, the magnetic field decreases faster with distance, therefore they have significantly lower torques compared with the dipole: their contribution can be safely neglected.

The energy balance equation between radiation and rotational energy losses, eq.~(\ref{eq:erot}), reads
\begin{equation}\label{eq:spindown_forcefree}
I\Omega \dot{\Omega} =  \frac{B_p^2R_\star^6\Omega^4}{6 c^3}f_\chi~.
\end{equation}
Differences in the radiation mechanism are included in the factor $f_\chi$: magnetic dipole radiation losses scale as $\sin^2\chi$ for vacuum (eq.~\ref{eq:en_loss_vacuum}) or $1.5(1+\sin^2\chi)$ for force-free magnetospheres (eq.~\ref{eq:en_loss_forcefree}), while magnetospheric current losses scale as $\cos^2\chi$ for small $\chi$ \citep{beskin07}. The most recent resistive solutions for pulsar magnetospheres \citep{li12a} fit the spin-down luminosity with a pre-factor of the order of unity that also depends on the maximum potential drop along field lines in the co-rotating frame. In general, the previous balance can be written in terms of the spin period $P=\Omega/2\pi$ and its derivative:
\begin{equation}\label{eq:ppdot_spindown} 
  P \dot{P} = K B_p^2~,
\end{equation} 
with
\begin{equation}\label{eq:k_spindown}
  K=f_\chi\frac{2\pi^2}{3}\frac{R_\star^6}{Ic^3}=  2.44\times 10^{-40}~ f_\chi \frac{R_6^6}{I_{45}} \mbox{~s G}^{-2}~,
\end{equation}
where $R_6=R_\star/10^6~$cm. In literature, the fiducial values $I_{45}=1$, $R_6=1$ and $f_\chi=1$ (vacuum orthogonal rotator) are commonly used. 

Throughout this work, we will obtain the long-term evolution of $B_p(t)$ from simulations for a given neutron star model (fixed $R_\star$, $I$ and $f_\chi$). Integrating in time eq.~(\ref{eq:ppdot_spindown}), we will obtain the corresponding evolution of timing properties. We will always consider $K$ constant, ignoring the possible time variation of two quantities in eq.~(\ref{eq:k_spindown}): the angle, $f_\chi=f_\chi(t)$, and the {\it effective} moment of inertia, $I=I(t)$. The latter could vary during the early stages, with the growth of the superfluid region in the core, rotationally decoupled from the exterior \citep{glampedakis11c,ho12}. We will also neglect other possible mechanisms to the spin-down, like strong particle winds \citep{tong13}. We also neglect the spin-down by gravitational radiation, because it can be efficient only during the first minutes or hours of a neutron star life, when rotation is sufficiently fast and the mass quadrupole moment large enough (see e.g., \citealt{haskell06} and references therein).

Model dependencies enter in the spin-down factor $K$, eq.~(\ref{eq:k_spindown}), as $f_\chi R_6^6 /I_{45}$, where $R_6$ and $I_{45}$ depend on equation of state and star mass. To quantify such variations, \cite{lattimer01} considered the moments of inertia resulting from many equations of state. \cite{bejger02} revised it, finding a correlation between $I$, $M$ and $R_\star$, expressed by the following fit:
\begin{eqnarray}\label{eq:moment_inertia}
 && I=a(x)MR_\star^2~, \nonumber\\
 && a(x) = \left\{
 \begin{array}{l r}
  x/(0.1+2x) & x\le 0.1 \\
  2(1+5x)/9  & x > 0.1
 \end{array}
\right.~,\\
\end{eqnarray}
where $x$ is the dimensionless compactness parameter:
\begin{equation}
 x=\frac{M}{M_\odot}\frac{\mbox{km}}{R_\star}~.
\end{equation}
The value $a(x)=0.4$ corresponds to a constant density sphere, but realistic stars are expected to have a lower value, since the mass is concentrated towards the center. Most equations of state predict radii in the range $8$--$15$ km, and masses $1$--$2 M_\odot$. Considering the possible range of $x\sim 0.1$--$0.2$, the corresponding values are $a(x)\sim 0.1$--$0.2$ (see Fig.~1 in \citealt{bejger02}).

\subsection{Inferred surface magnetic field.}

Observations of pulsars provide their timing properties, $P$ and $\dot{P}$. Thus, inverting eq.~(\ref{eq:ppdot_spindown}) allows to infer the strength (at the pole) of the {\em dipolar component of the surface magnetic field}:
\begin{equation}\label{eq:inferred_bpole}
 B_p = C_{sd} \sqrt{P\mbox{[s]}~\dot{P}}~,
\end{equation}
where, according to eq.~(\ref{eq:k_spindown}):
\begin{equation}\label{eq:factor_binferred}
 C_{sd} = 6.4 \times 10^{19} ~ \sqrt{\frac{I_{45}}{f_\chi R_6^6}}~\mbox{G}~.
\end{equation}
In literature, the half-value is often found, and it corresponds to the value of the dipolar magnetic field at the equator, $B_e=B_p/2$. Considering the uncertainties in $R_\star$, $I$ and $f_\chi$ discussed above, the estimate of $B_p$ is reliable within a factor of $\sim 2$.

\subsection{Characteristic age.}\label{sec:chage}

The {\em characteristic age} (or spin-down age) is defined as
\begin{equation}\label{eq:chage}
 \tau_c=\frac{P}{2\dot{P}}~.
\end{equation}
This is a good estimate of the real age only if two hypotheses hold: the natal period, $P_0$, was much smaller than the present one, $P$, and the torque factor $KB_p^2$ in eq.~(\ref{eq:ppdot_spindown}), has been constant during the entire star life. Otherwise, the real age $t$ significantly deviates from $\tau_c$, as we demonstrate now. Integrating both sides of eq.~(\ref{eq:ppdot_spindown}) in time, from the birth, $t=0$, to the present age, $t=t_{real}$ (when we measure $P$ and $\dot{P}$, and their product $(KB_p^2)_{now}$), we have:
\begin{equation}
 \frac{1}{2}(P^2 - P_0^2) = \int_0^{t_{real}} (KB_p^2)(t') \de t'~.
\end{equation}
Dividing by $P\dot{P}$, and using the identity $t_{now}=[\int_0^{t_{real}} (KB_p^2)_{now}]/(KB_p^2)_{now}$, we obtain:
\begin{equation}\label{eq:treal_chage}
 t_{real} = \tau_c - \frac{P_0^2}{2P\dot{P}} - \frac{\int_0^{t_{real}} [(KB_p^2)(t') - (KB_p^2)_{now}] \de t'}{(KB_p^2)_{now}}~.
\end{equation}
If the initial period was close to the present value, the characteristic age is overestimating the age of the object. The same happens if $KB_p^2$ was on average larger in the past than today (for instance, if the magnetic field has decayed). On the other hand, if $KB_p^2$ in the past was on average smaller than today, the characteristic age underestimates the real age.

\subsection{Braking index.}\label{sec:braking_index_def}

A measurable quantity, closely related to the rotational evolution of pulsars, is the braking index $n$, defined as
\begin{equation}\label{eq:BIstandard}
 n = \frac{\ddot{\Omega}\Omega}{\dot{\Omega}^2} = 2 - \frac{\ddot{P}P}{\dot{P}^2} ~.
\end{equation} 
Using eq.~(\ref{eq:ppdot_spindown}), we have

\begin{equation}\label{eq:def_braking_index_mf}
 n=3-2\frac{P}{\dot{P}}\frac{1}{KB_p^2}\frac{d(KB_p^2)}{dt}~.
\end{equation}
If $KB_p^2$ is constant in time, the braking index is $n=3$. Most pulsars show strong deviations from $n=3$, implying a time variation of $KB_p^2$ (see chapter~\ref{ch:timing}).

\section{The pulsar equation.}\label{sec:pulsar_eq}

\begin{figure}
 \centering
 \includegraphics[width=.5\textwidth]{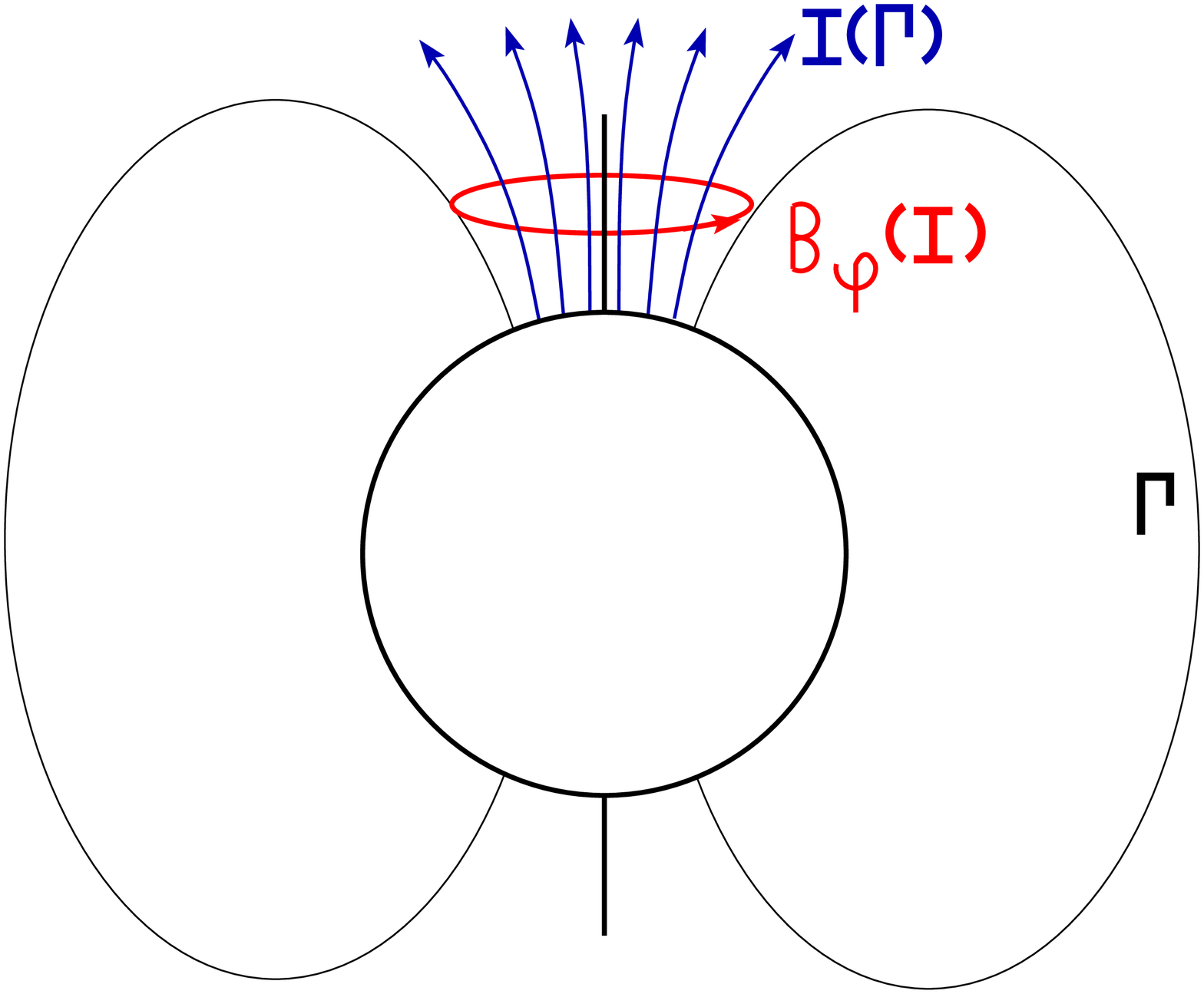}
 \caption{Schematic representation of the relation between the magnetic flux function $\Gamma$, the enclosed current function $I(\Gamma)$, and the toroidal magnetic field $B_\varphi$, in axial symmetry.}
 \label{fig:axisym}
\end{figure}

Since in the magnetospheric plasma the electromagnetic forces are much larger than the centrifugal, collisional and gravitational terms, eqs.~(\ref{eq:ratio_forces1}) and (\ref{eq:ratio_forces2}), the force equilibrium can be written as:

\begin{equation}\label{eq:forcefree_gen}
  \rho_q \vec{E}+\frac{1}{c}\vec{J}\times\vec{B}=0~,
\end{equation}
where $\vec{J}$ is defined by eq.~(\ref{eq:current_mhd_new}). With the hypothesis of completely separated charges, the relation $\vec{J}=\rho_q\vec{v}$ holds, where the velocity of plasma includes both the rotational contribution and the velocity along magnetic field lines. In axial symmetry, it is useful to split the magnetic field in poloidal and toroidal components (see Appendix~\ref{app:poloidal-toroidal} for more details about formalism). The poloidal magnetic field is expressed in terms of the {\it magnetic flux function} $\Gamma(r,\theta)$ as

\begin{equation}\label{eq:def_poloidal}
  \vec{B}_{pol}=\frac{\vec{\nabla} \Gamma(r,\theta) \times \hat{\varphi}}{r \sin\theta}~.
\end{equation}
In Fig.~\ref{fig:axisym} we schematically show the formalism employed. The function $\Gamma$ is constant along a magnetic field line ($\vec{B}_{pol}\cdot\vec{\nabla}\Gamma=0$). Thus its value labels the axisymmetric surface $S_\Gamma$ given by the azimuthal rotation of one field line. The magnetic flux flowing in between two surfaces, $S_{\Gamma_a}$ and $S_{\Gamma_b}$, is

\begin{equation}
  {\cal F}_B(\Gamma)\equiv 2\pi\int_{\theta_b}^{\theta_a} B_r(r,\theta)r^2\sin\theta\de \theta=2\pi\int\frac{d\Gamma}{d\theta}d\theta=2\pi(\Gamma_a-\Gamma_b)~,
\end{equation}
where $\theta_a$ and $\theta_b$ are the polar angles of the magnetic line footprints. The surface associated to the axis reduces to zero, which implies $\Gamma(\mu=\pm 1)=0$. As a consequence, $2\pi\Gamma$ is the magnetic flux within the surface $S_\Gamma$, and it is conserved by definition. In a configuration which is (anti)symmetric respect to the equatorial plane, $\Gamma$ is (anti)symmetric. In particular, for a dipole, the value of $\Gamma$ at the surface reaches the maximum value at the equatorial plane, where it represents the total magnetic flux flowing in the northern hemisphere. Note also that the continuity equation $\partial_t \rho_e + \vec{\nabla}\cdot \vec{J}=0$ for static solutions ($\partial_t\rho_e=0$) is automatically satisfied:

\begin{eqnarray}
  \vec{\nabla}\cdot \vec{J} && = \frac{c}{4\pi}\left(\alpha(\Gamma)\vec{\nabla}\cdot\vec{B}_{pol} + \frac{\partial\alpha}{\partial\Gamma}\vec{B}_{pol}\cdot\vec{\nabla}\Gamma\right)=\nonumber\\
  && = \frac{c}{4\pi r\sin\theta}\frac{\partial\alpha}{\partial\Gamma}(\vec{\nabla}\Gamma(r,\theta) \times\hat{\varphi})\cdot\vec{\nabla}\Gamma=0 ~.
\end{eqnarray}

The function $\Gamma(r,\theta)$ is the analogous of the Stokes' function that describes streamlines and velocity fluxes for three-dimensional, incompressible flows. In this case, the conditions $\vec{\nabla}\cdot\vec{v}=0$ and the definition of vorticity $\vec{\omega}\equiv\vec{\nabla}\times \vec{v}$ are analogous to $\vec{\nabla}\cdot\vec{B}=0$ and the definition of current, eq.~(\ref{eq:current_mhd_new}).

A toroidal component of the magnetic field, $\vec{B}_{tor}=B_\varphi \hat{\varphi}$, is present if sustained by poloidal currents $\vec{J}_{pol}$. An axisymmetric static solution only allows non-vanishing poloidal components of the electric field, otherwise the circulation of the electric field along a toroidal loop would be non-zero (integrating eq.~\ref{eq:induction_new}). As a consequence, $\vec{J}_{pol}\times\vec{B}_{pol}=0$, which means
\begin{equation}\label{eq:nablabtor}
 \curlB_{tor}=\frac{4\pi}{c}\vec{J}_{pol}=\alpha(r,\theta)\vec{B}_{pol}~,
\end{equation} 
where, {\em a priori}, the proportionality factor $\alpha$ is a generic function of the position. $\vec{J}$ and $\vec{B}$ are misaligned only if $\vec{B}_{tor}\neq 0$, but their projections on the meridional plane are always parallel everywhere: the force-free condition implies that currents flow parallel to the surfaces $S_\Gamma$. This also means that the current flowing within $S_\Gamma$ (see blue arrows in Fig.~\ref{fig:axisym}), has to be a function of only $\Gamma$. Integrating the Amp\`ere's law along a ring $l$ perpendicular to the axis, with length $2\pi r\sin\theta$, and enclosed surface $\vec{S}_l$, we have

\begin{equation}
 I(\Gamma) \equiv \int_{S_l} \vec{J}_{pol}\cdot \hat{n}\de S_l=\frac{c}{4\pi}\oint_l B_\varphi \de l ~,  
\end{equation}
where $\hat{n}$ is the normal to $S_l$. Therefore we obtain:

\begin{equation}\label{eq:bphi_definition}
 B_\varphi=\frac{2}{cr\sin\theta}I(\Gamma)~.
\end{equation}
Another way to see this fundamental relation between the toroidal and poloidal components is to use eq.~(\ref{eq:def_poloidal}) and integrate eq.~(\ref{eq:nablabtor}):

\begin{eqnarray}\label{eq:bphi_integrability}
  \partial_\theta (B_\varphi\sin\theta) = \frac{\alpha}{r}\partial_\theta \Gamma &\Rightarrow& B_\varphi(r,\theta)=\frac{1}{r\sin\theta}\int_0^\theta \alpha\partial_{\theta'} \Gamma\de \theta'~,  \\
  -\partial_r (B_\varphi r) = -\frac{\alpha}{\sin\theta}\partial_r \Gamma &\Rightarrow& B_\varphi(r,\theta)=\frac{1}{r\sin\theta}\int_{r_0}^r \alpha \partial_{r'}\Gamma \de r'~,
\end{eqnarray}
where we have assumed that $B_\varphi\sin\theta \rightarrow 0$ for $\theta\rightarrow 0$, and $rB_\varphi = 0$ at some radius $r_0$. The comparison of the two expressions above for $B_\varphi$ provides the integrability condition $\alpha=\alpha(\Gamma)$ and
\begin{equation}\label{eq:bphi_definition_alpha}
 B_\varphi=\frac{1}{cr\sin\theta}\int_0^\Gamma \alpha(\Gamma')\de\Gamma'~.
\end{equation}
Eq.~(\ref{eq:bphi_definition}) is recovered with the definition of the relation between $\alpha$ and the enclosed current function:

\begin{eqnarray}
 && I(\Gamma)=\frac{c}{2}\int_0^\Gamma \alpha(\Gamma')\de\Gamma'~, \label{eq:current_alpha}\\
 && \alpha(\Gamma) = \frac{2}{c}\frac{d I}{d\Gamma}~.\label{eq:alpha_current}
\end{eqnarray}
These relations tell us that the toroidal magnetic field (and the enclosed current) depends only of the poloidal magnetic flux, but its functional form is unconstrained. We have the freedom of choosing $\alpha(\Gamma)$, or, equivalently, $I(\Gamma)$ with the only requirement, given by eq.~(\ref{eq:current_alpha}), that

\begin{equation}
I(\Gamma=0)=0~.
\end{equation}
With the definitions of light cylinder $\varpi_l$ (\ref{eq:def_light_cyl}) and poloidal magnetic field (\ref{eq:def_poloidal}), the poloidal electric field (\ref{eq:balance_em}) and the charge density (\ref{eq:poisson_new}) are written as:

\begin{eqnarray}
 && \vec{E}_{pol} = -\frac{\vec{v}_{rot}\times(\vec{\nabla} \Gamma\times \hat{\varphi})}{c r \sin\theta} = -\frac{\vec{\nabla} \Gamma}{\varpi_l}~, \label{eq:ef_pulsar_equation}\\
 && \rho_q = -\frac{1}{4\pi \varpi_l}\nabla^2\Gamma ~. \label{eq:rhoe_pulsar_equation}
\end{eqnarray}
The poloidal components of the eq.~(\ref{eq:forcefree_gen}), $c\rho_q \vec{E}_{pol}+\vec{J}_{pol}\times \vec{B}_{tor} - \vec{B}_{pol}\times \vec{J}_{tor}=0$, reduce to

\begin{equation}
 \left(-\rho_q \vec{v}_{rot} - \frac{c}{4\pi}\alpha\vec{B}_{tor} + \vec{J}_{tor}\right)\times\vec{B}_{pol}=0~,
\end{equation}
where we have used eq.~(\ref{eq:nablabtor}). Vectors within parenthesis have azimuthal direction and their sum has to be zero. The consequent scalar equation is called {\it pulsar equation}. It is actually the Grad-Shafranov equation for force-free fields: the equilibrium equation in ideal MHD for a plasma in axisymmetric configurations. In general, it has to be solved numerically. Note that even for trivial choices of configurations without toroidal magnetic field ($\alpha = 0$), the co-rotation velocity of the charged particles represents a toroidal current that distorts the poloidal magnetic field, resulting in a deviation from any potential configurations $\curlB_{pol}=0$.

To solve the pulsar equation, we have to consider eqs.~(\ref{eq:bphi_definition_alpha}), (\ref{eq:rhoe_pulsar_equation}), and the toroidal current

\begin{equation}
 \vec{J}_{tor}=J_\varphi\hat{\varphi} = \frac{c}{4\pi}(\vec{\nabla}\times\vec{B}_{pol}) ~.
\end{equation}
In spherical coordinates, we have
\begin{eqnarray}\label{eq:pulsar_eq_terms_sph}
 && \vec{B}_{pol}= \frac{1}{r\sin\theta}(\partial_\theta\Gamma\hat{r}- \partial_\theta\Gamma\hat{\theta})~, \\
 && J_\varphi =-\frac{c}{4\pi r\sin\theta}\left(\partial_{rr}\Gamma+\frac{\partial_{\theta\theta}\Gamma}{r^2}\right)~,\\
 && \rho_q = \frac{1}{4\pi \varpi_l}\left(\partial_{rr}\Gamma+\frac{2\partial_r\Gamma}{r}+\frac{\partial_{\theta\theta}\Gamma}{r^2} + \frac{2\cos\theta}{r^2\sin\theta}\partial_\theta\Gamma\right) ~.
\end{eqnarray}
The pulsar equation in spherical coordinates is

\begin{eqnarray}\label{eq:pulsar_eq_sph}
 \left[1-\left(\frac{r\sin\theta}{\varpi_l}\right)^2\right]\left(\partial_{rr}\Gamma-\frac{\partial_\theta\Gamma\cos\theta}{r^2\sin\theta}+\frac{\partial_{\theta\theta}\Gamma}{r^2}\right) - \frac{2\sin{\theta}^2}{\varpi_l^2}\left(r\partial_r\Gamma+\frac{\partial_\theta\Gamma}{\sin\theta}\right) = && \nonumber\\
 = -\alpha\int \alpha(\Gamma)\de\Gamma~, &&
\end{eqnarray}
where the source term on the right side can be written also as $-4I(\de I/\de\Gamma)/c^2$.

In cylindrical coordinates $(\varpi,\varphi,z)$, where $\varpi=r\sin\theta$ and $z=r\cos\theta$, we have

\begin{eqnarray}\label{pulsar_eq:cyl_terms}
 && \vec{B}_{pol}= \frac{1}{\varpi}(-\partial_z\Gamma\hat{\varpi}+\partial_\varpi\Gamma\hat{z}) ~, \\
 && J_\varphi =-\frac{c}{4\pi\varpi}\left(\partial_{zz}\Gamma+\partial_{\varpi\varpi}\Gamma-\frac{\partial_{\varpi}\Gamma}{\varpi}\right)~,\\
 && \rho_q = -\frac{1}{4\pi \varpi_l}\left(\frac{\partial_\varpi\Gamma}{\varpi}+\partial_{\varpi\varpi}\Gamma+\partial_{zz}\Gamma\right)~,
\end{eqnarray}
and the pulsar equation is more compact:

\begin{equation}\label{pulsar_eq:cyl}
 \left(1-\frac{\varpi^2}{\varpi_l^2}\right)(\partial_{zz}\Gamma+\partial_{\varpi\varpi}\Gamma)-\frac{1}{\varpi}\left(1+\frac{\varpi^2}{\varpi_l^2}\right)\partial_\varpi\Gamma=-\alpha\int \alpha(\Gamma)\de\Gamma~.
\end{equation}

\subsection{Split monopole solution.}\label{sec:split_monopole}

A straightforward, fully analytical solution to the pulsar equation, including rotation and a smooth matching across the light cylinder is the {\em split monopole} presented by \cite{michel73b}. The form of the magnetic flux function is defined as

\begin{equation}
 \Gamma=\mp\Gamma_0 \cos\theta~,
\end{equation}
where each sign refers to a hemisphere, and $\Gamma_0$ is a normalization. The pulsar equation (\ref{eq:pulsar_eq_sph}) becomes

\begin{equation}
 \pm \frac{2\Gamma_0}{\varpi_l^2}\cos\theta \sin^2\theta = \frac{4}{c^2}I(\Gamma)\frac{d I}{d \Gamma}~.
\end{equation}
The choice of the enclosed current function satisfying the equation, and its derivative, are

\begin{eqnarray}\label{split_monopole_function}
 && I(\Gamma)=\frac{c}{2\varpi_l}\frac{\Gamma_0^2-\Gamma^2}{\Gamma_0} = \frac{c}{2\varpi_l}\Gamma_0\sin^2\theta ~,\\
 && \frac{d I}{d \Gamma}=-\frac{c\Gamma}{\varpi_l\Gamma_0} = \pm\frac{c}{\varpi_l}\cos\theta ~.
\end{eqnarray}
The magnetic field components are:

\begin{eqnarray}\label{split_monopole_components}
&& B_r=\pm\frac{\Gamma_0}{r^2} ~,\\
&& B_\theta=0 ~,\\
&& B_\varphi=\frac{\Omega}{c}\frac{\Gamma_0}{r}\sin\theta ~.
\end{eqnarray}
The magnetic field is directed outwards in one hemisphere and inwards in the other one, in order to preserve $\vec{\nabla}\cdot\vec{B}=0$. In the equatorial plane, the discontinuity of $B_r$ in the $\theta$-direction implies a toroidal current sheet $J_\varphi$. The magnetic field lines are twisted, and the angle between the toroidal and radial magnetic field components is $\arctan(r\sin\theta/\varpi_l)$. The charge density, eq.~(\ref{eq:rhoe_pulsar_equation}), is

\begin{equation}
 \rho_q=\pm \frac{\Gamma_0}{2\pi \varpi_l}\frac{\cos\theta}{r^2}~.
\end{equation}
The current is purely radial and proportional to the angular velocity $\Omega$:

\begin{equation}
 J_r=\frac{\Gamma_0 \Omega}{2\pi}\frac{\cos\theta}{r^2} = \pm \rho_q c~,
\end{equation}
which means that particles move only radially at the speed of light. To understand this result, consider the co-rotating frame, in which particles move only along the twisted field lines. Seen from an inertial observer, the toroidal component of this velocity is exactly compensated by the azimuthal drift of the magnetic field lines due to rigid rotation, thus $J_\phi=0$.

Despite its simplicity, this solution is thought to describe at least qualitatively the wind region, where the magnetic field is stretched outwards and there is an outflow of relativistic particles. A generalization for an oblique rotator has been found by \cite{bogovalov99} by means of a mathematical change of coordinates. It displays the same geometric features of the aligned split monopole model. The important difference is that, for an external observer, the current sheet oscillates around the rotational equatorial plane at the spin frequency.

\subsection{Numerical dipolar solutions.}

\begin{figure}
 \centering
 \includegraphics[width=.6\textwidth]{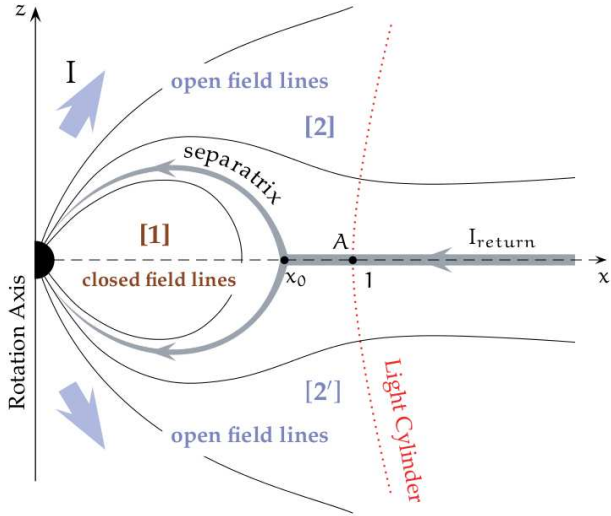}
 \caption{Numerical solution of the pulsar equation for an aligned rotator with a dipolar magnetic field, taken from \cite{timokhin10}.}
 \label{fig:dipole_cfl}
\end{figure}

Finding solutions to the pulsar equation is difficult also due to the mathematical singularity at the light cylinder. The intrinsic degeneracy of possible solutions could be partially removed considering the physical conditions in the outer region. As a matter of fact, the magnetic configuration of the surrounding environment (supernova remnant or interstellar medium), likely plays a determinant role \citep{goodwin04}.

A consistent numerical dipolar solution, smooth across the light cylinder, was first found numerically by \cite{contopoulos99} and later confirmed by several works \citep{goodwin04,contopoulos05,mckinney06,spitkovsky06}. The configuration of magnetic field lines is shown in Fig.~\ref{fig:dipole_cfl}. Compared with a non-rotating dipole, the magnetic field lines are inflated by the rotation, and a larger fraction (+36$\%$) of the magnetic flux is contained in the open field line region. In that region, the lines are twisted (i.e., the magnetic field has a non-zero toroidal component), and the pattern is remarkably similar to the split monopole discussed in \S~\ref{sec:split_monopole}, with slight deviation on the current distribution. The region beyond the light cylinder is smoothly connected with the open field lines, and consequently shows a radial configuration as well.

The characteristic feature of this solution is the so-called Y-point that defines the boundary between open and closed field lines. At the Y-point, the {\em separatrix line} displays a kink, and it lies on the equatorial plane, close to the light cylinder. Beyond the Y-point, an equatorial current sheet separates the northern and southern hemispheres, where radial and azimuthal fields point to opposite directions, like in the split-monopole configuration.

\cite{spitkovsky06} numerically solved the time-dependent oblique rotator in 3D. Similarly to the oblique split-monopole, the oblique dipole maintains the main features of the aligned version, with the main difference that the equatorial current sheet oscillates around the rotational equatorial plane. The Poynting flux across a sphere at infinity is increased compared with an orthogonal rotator in vacuum (see eq.~\ref{eq:spindown_forcefree}). These conclusions were confirmed by other simulations (e.g., \citealt{kalapotharakos09}). Recently, \cite{li12a,li12b} have also included the effect of finite conductivity of the magnetospheric plasma. This is expected also to cause magnetic reconnection events, which is important for the dynamics and evolution of the system.

\section{Non-rotating case: (semi-)analytical solutions.}\label{sec:semianalytical}

After having reviewed a few basic force-free solutions, we move to a simplified approach to the pulsar equation, in which no rotation is considered. This hypothesis is useful to model the co-rotating magnetosphere near the surface, where the effect of rotation can be safely neglected, unless the star spins very fast. In the pulsar equation~(\ref{eq:pulsar_eq_sph}), the terms proportional to $(1/\varpi_l)^2\equiv(\Omega/c)^2$ arise from the rotationally-induced electric field. Thus, the approximation works in the region well within the light cylinder, $r \ll \varpi_l$. In this limit, the electric field is negligible and the equilibrium equation is simply given by $\vec{J}\times\vec{B}=0$. The force-free condition simply requires the current to flow parallel to magnetic field lines everywhere. We can reformulate the problem with the simple equation
\begin{equation}\label{eq:rotB_nonrotating}
\curlB=\alpha(\Gamma)\vec{B}~,
\end{equation}
where the function $\alpha(\Gamma)$ is related to the enclosed current function by eq.~(\ref{eq:alpha_current}). The current density is simply given by
\begin{equation}\label{eq:current_force-free}
 \vec{J}=\frac{c}{4\pi}\alpha(\Gamma)\vec{B}~.
\end{equation}
Note the difference respect to the case with rotation: in eq.~(\ref{eq:nablabtor}), only the poloidal projections of current and magnetic field vectors were required to be parallel. The non-rotating pulsar equation in spherical coordinates is now expressed as

\begin{equation}\label{eq:gs_nonrotating}
\partial_{rr}\Gamma - \frac{\cos\theta}{\sin\theta}\frac{\partial_\theta \Gamma}{r^2} + \frac{\partial_{\theta\theta} \Gamma}{r^2}=-\alpha\int \alpha(\Gamma)\de\Gamma~.
\end{equation}
A usual approach is to expand the magnetic flux function in the orthonormal basis of Legendre polynomials, $P_l(\mu)$ (see also Appendix~\ref{app:legendre}):

\begin{equation}\label{eq:potential_gen}
  \Gamma(r,\mu)=\Gamma_0\sum_l \frac{r}{R_\star}a_l(r)(1-\mu^2)\frac{d P_l(\mu)}{d\mu},
\end{equation}
where $\mu=\cos\theta$, $a_l(r)$ is a dimensionless function giving the radius-dependent weight of the $l$-pole, and $\Gamma_0$ is a normalization that we can rewrite as
\begin{equation}\label{eq:gamma0}
  \Gamma_0=\frac{B_0R_\star^2}{2}~,
\end{equation}
where $B_0$ is the magnetic field strength at the pole in the case of a purely dipolar field with $a_1=1$. This leads to the following expression for the poloidal magnetic field components:

\begin{eqnarray}\label{eq:mf_leg}
  && B_r      = \frac{B_0}{2}\frac{R_\star}{r}\sum_l l(l+1)P_l(\mu)a_l(r) ~, \\
  && B_\theta = - \frac{B_0}{2}\frac{R_\star}{r}\sum_l \sqrt{1-\mu^2}\frac{d P_l(\mu)}{d\mu}\frac{d (ra_l(r))}{dr}~.
\end{eqnarray}
Note that the magnetic field is purely radial at the poles, as required by axial symmetry. We can obtain the governing differential equation from eq.~(\ref{eq:gs_nonrotating}):
\begin{equation}
  \frac{B_0R_\star}{2}(1-\mu^2)\sum_l \frac{d P_l(\mu)}{d\mu}\left[\frac{d^2 (ra_l(r))}{dr^2}-l(l+1)\frac{a_l(r)}{r}\right]= -\alpha\int \alpha(\Gamma)\de\Gamma~.
\end{equation}
Finally, by using the orthogonality relations of Legendre polynomials, eq.~(\ref{eq:leg_on2_app}), we have

\begin{equation}\label{eq:gs_ode_leg}
  B_0R_\star\frac{l(l+1)}{2l+1}\left[\frac{d^2 (ra_l(r))}{dr^2}-l(l+1)\frac{a_l(r)}{r}\right]= -\int_{-1}^1\frac{d P_l(\mu)}{d\mu}\left[\alpha\int \alpha(\Gamma)\de\Gamma\right]\de \mu~.
\end{equation}
The functional form of $\alpha(\Gamma)$ is responsible for the coupling between different multipoles. The rest of this section is devoted to reviewing some possible choices that make the problem (semi)analytically solvable. There are infinite possible choices of $\alpha(\Gamma)$; hereafter, we will explore the simple analytical functional form
\begin{equation}\label{eq:form_alpha}
 \alpha=\frac{k}{R_\star}\left(\frac{\Gamma}{\Gamma_0}\right)^q~,
\end{equation}
with $k$ being the dimensionless parameter related to the twist, and $q$ a free parameter. We discuss only a few simple choices of $q$ because of their simplicity.

\subsection{Potential solutions.}\label{sec:potential_solutions}

With the trivial choice $\alpha=0$, there is no toroidal magnetic field nor current. We recover the general potential solution (see also Appendix~\ref{app:vacuum_bc}):

\begin{eqnarray}
  && a_l(r)=c_l\frac{R_\star}{r}^{(l+1)}~,\\
  && B_r=\frac{B_0}{2}\sum_l c_l l(l+1)P_l(\mu)\frac{R_\star}{r}^{(l+2)}~,\\
  && B_\theta=\frac{B_0}{0}\sum_l c_l l\sqrt{1-\mu^2}\frac{d P_l(\mu)}{d\mu}\frac{R_\star}{r}^{(l+1)}~,\\
  && B_\varphi=0~,
\end{eqnarray}
where $c_l$ is the $l$-polar weight.

%%%%%%%%%%%%%%%%%%%%%%%%%%%%%%%%%%%%%%%%%%%%%%
\subsection{Spherical Bessel solutions.}\label{sec:bessel}

This choice of constant $\alpha=k/R_\star$, where $k$ is a dimensionless parameter, has been adopted in several studies of the solar corona \citep{chiu77,seehafer78} and of the open field line region of pulsar magnetospheres \citep{scharlemann73}. The right-hand side of eq.~(\ref{eq:gs_ode_leg}), due to the orthogonality properties of Legendre polynomials, does not couple different multipoles. For each $l$, we must independently solve the equation

\begin{equation}
  r^2\frac{d^2 a_l(r)}{dr^2}+2r\frac{d a_l(r)}{dr}+\left[\left(k\frac{r}{R_\star}\right)^2 - l(l+1)\right]a_l(r)=0~.
\end{equation}
The analytical solutions of this equation are the spherical Bessel functions of the first and second kind (see Appendix~\ref{app:bessel}). In this case, the physical solutions are represented by the functions of the second kind, $Y_l$, from which the potential solution $B_{p}\propto r^{-(l+1)}$ can be recovered in the limit $k\rightarrow 0$. The $Y_l$ functions are oscillatory at large distances, and the magnetic field components change sign as $r$ varies. The constant ratio $R_\star/k$ represents the length-scale of magnetic field variations. Note that for this family of analytical solutions, at large distances, all components (which have the same radial dependence for any $l$) decay slowly: $B_r\rightarrow r^{-2}, B_\theta\rightarrow r^{-1}, B_\varphi\rightarrow r^{-1}$. Thus this configuration cannot be a solution for the whole space, as it would imply infinite magnetic energy in an infinite volume. Also, these solutions cannot be continuously matched with vacuum, because it would require that, at the same radius $R_{out}$, $B_\varphi(R_{out})=0$ and $B_r(R_{out})\neq 0$, a condition that cannot be satisfied because those two components have the same radial dependence and the same zeros.

For numerical test purposes, we will employ in \S~\ref{sec:numerical_magnetosphere} the dipolar solution, $c_l=\delta_{l,1}$:
\begin{eqnarray}\label{bes_l1}
  && Y_1(x)=\frac{\cos x}{x^2} + \frac{\sin x}{x}~,\\
  && A_\varphi=B_0R_\star\sin\theta \left(\frac{\cos x}{x^2} + \frac{\sin x}{x}\right)~,\\
  && B_r=\frac{B_0R_\star}{r} \cos\theta \left(\frac{\cos x}{x^2} + \frac{\sin x}{x}\right)~,\\
  && B_\theta= \frac{B_0R_\star}{2r}\sin\theta\left(\frac{\cos x}{x^2} + \frac{\sin x}{x} +\cos x\right)~,\\
  && B_\varphi=\frac{kB_0}{2}\sin\theta \left(\frac{\cos x}{x^2} + \frac{\sin x}{x}\right)~,
\end{eqnarray}
where $x=k r/R_\star$.

%%%%%%%%%%%%%%%%%%%%%%%%%%%%%%%%%%%%%%%%%%%%%%
\subsection{Self-similar models.}\label{sec:selfsimilar}

\cite{low90}, \cite{wolfson95} and other authors, studied a particular class of self-similar solutions to describe the opening of the solar coronal magnetic field lines. \cite{thompson02} applied the same approach in the magnetar framework. Here we summarize their mathematical construction, that relies on a radial dependence $\alpha\propto 1/r$ and a radial power-law form for $\Gamma$:

\begin{eqnarray}
  && \Gamma=\Gamma_0 \left(\frac{R_\star}{r}\right)^pF(\mu)\label{eq:gamma_tlk} ~,\\
  && \alpha=\frac{k}{r}|F(\mu)|^{1/p}\label{alpha_tlk}=\frac{k}{R_\star}\left(\frac{|\Gamma|}{\Gamma_0}\right)^{1/p} ~,\\
  && I(\Gamma)=I_0\left|\frac{\Gamma}{\Gamma_0}\right|^{1+1/p}\label{eq:I_tlk} ~,
\end{eqnarray}
where

\begin{equation}\label{I0}
  I_0=\frac{kp}{4(p+1)}cB_0R_\star~.
\end{equation}
Eq.~(\ref{eq:gs_nonrotating}) becomes a second-order non-linear differential equation for the angular function $F(\mu)$:

\begin{equation}\label{eq:ode_tlk}
  (1-\mu^2)\frac{d^2 F}{d\mu^2}+p(p+1)F(\mu)=-C F(\mu)|F(\mu)|^{2/p}~,
\end{equation}
where $C=k^2p/(p+1)$. The magnetic field is given by

\begin{eqnarray}
 B_r       &=& -\frac{B_0}{2}\left(\frac{R_\star}{r}\right)^{(p+2)}\frac{dF}{d\mu}~, \label{eq:mf_tlk1}\\
 B_\theta  &=&  \frac{B_0}{2}\left(\frac{R_\star}{r}\right)^{(p+2)}p\frac{F(\mu)}{\sqrt{1-\mu^2}} ~,\label{eq:mf_tlk2}\\
 B_\varphi &=&  k\frac{B_0}{2}\left(\frac{R_\star}{r}\right)^{(p+2)}\frac{p}{p+1}\frac{F(\mu)|F(\mu)|^{1/p}}{\sqrt{1-\mu^2}}~.\label{eq:mf_tlk3}
\end{eqnarray}
The three components of the magnetic field have the same radial dependence: the configurations are self-similar. To solve eq.~(\ref{eq:ode_tlk}), we have to impose three physical requirements on the axis: $F(1)=F(-1)=0$ (only the radial field component is allowed), and $(\de F/\de\mu)(1)=-2$, which fixes the normalization $B_r(\mu=1)=B_0$. Note that an alternative boundary condition, $(dF/d\mu)(0)=0$, i.e. no radial field at the magnetic equator, implicitly assumes symmetry with respect to the equatorial plane and it can also be expressed requiring $\de F/\de\mu (-1)=\de F/\de\mu (1)$. 

\begin{figure}
\centering
\includegraphics[width=.46\textwidth]{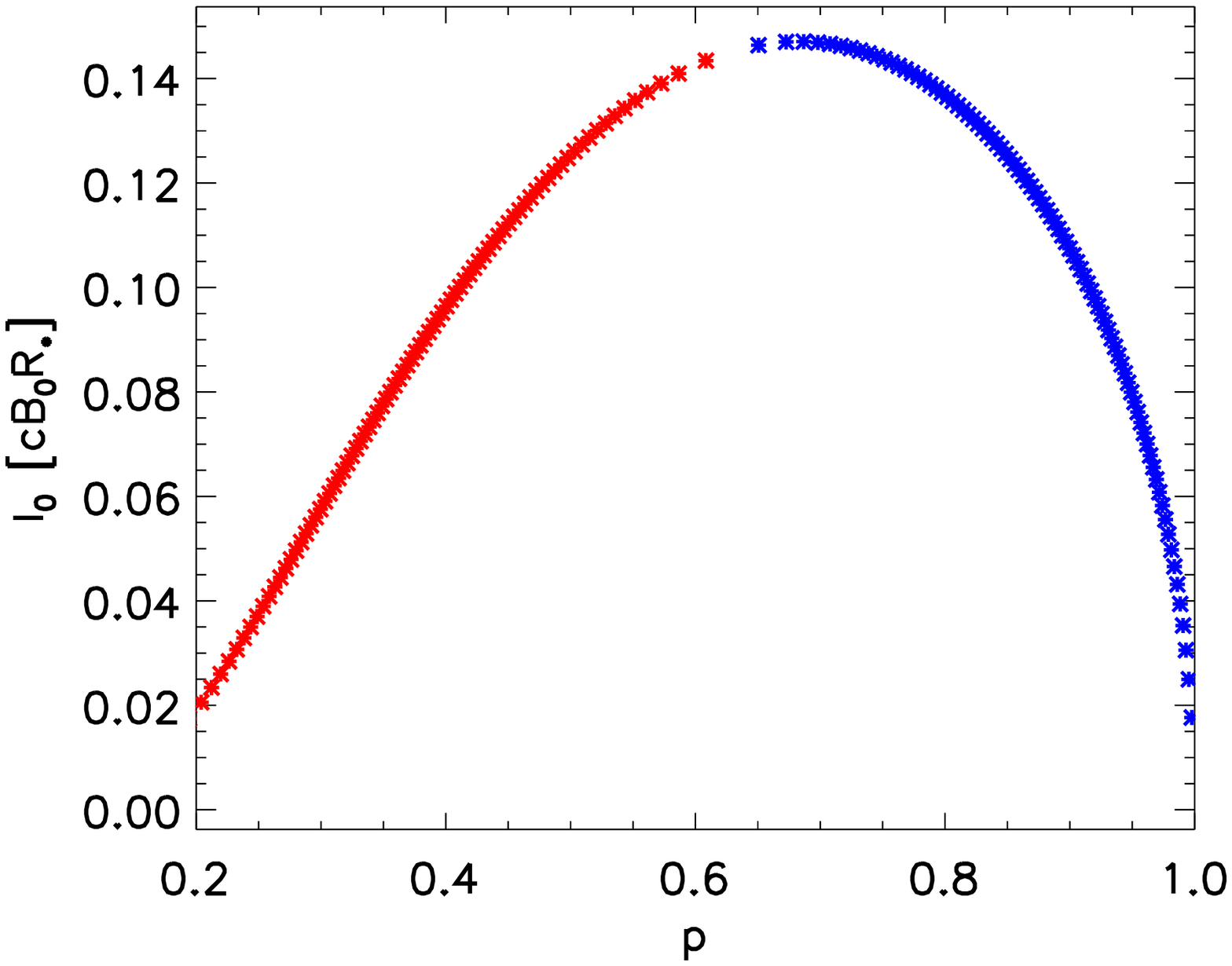}
\includegraphics[width=.46\textwidth]{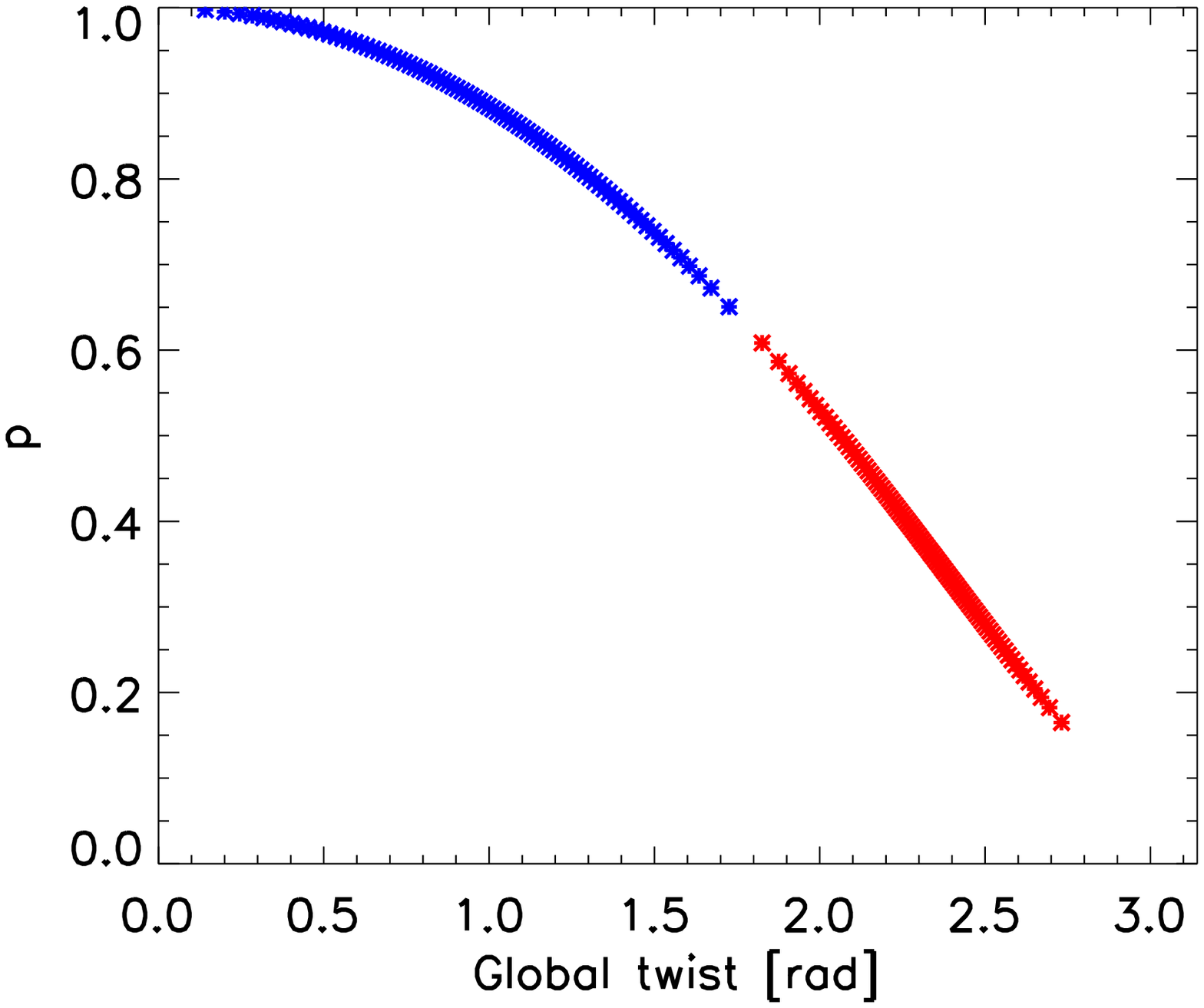}
\caption{Family of the self-similar twisted dipoles: curves of parameters $I_0(p)$ (left) and $p(\Delta\varphi_{ss})$ (right). Blue and red colors distinguish the two branches of solutions.} 
\label{fig:tlk_dipole}
\end{figure}

In a self-similar model, the local azimuthal displacement, $B_\varphi/B_\theta\sin\theta$, depends on $\theta$, thus the twist of a magnetic field line $l_\Gamma$ is simply (see the definition \ref{eq:def_twist_app}):
\begin{equation}
  \Delta \varphi_{tw}(\Gamma)=\frac{2k}{p+1}\int_0^{\mu_\Gamma} \frac{|F(\mu)|^{1/p}}{1-\mu^2} \de \mu~,
\end{equation}
where $\mu_\Gamma$ is the cosine of the polar angle $\theta$ at which the north footprint lies. $\Delta \varphi_{tw}(\Gamma)$ is a monotonic function of $\Gamma$, reaching its maximum value at the pole, $\mu_\Gamma = 1$. This means that the most twisted lines are those with footprints close to the pole ($\Gamma\rightarrow 0$). Their twist, hereafter called {\em global twisted}, is a parameter that uniquely describes the $l$-polar family of self-similar solutions:
\begin{equation}\label{eq:twist_tlk}
  \Delta \varphi_{ss}=\frac{2k}{p+1}\int_0^1 \frac{|F(\mu)|^{1/p}}{1-\mu^2} \de \mu~.
\end{equation}
We have numerically solved eq.~(\ref{eq:ode_tlk}) with a 4$^{th}$-order Runge-Kutta method, integrating $F(\mu)$ from $\mu=1$ and finding the value of $p$ that matches the condition $F(-1)=0$ through a shooting method. Given a value of $C$ (or $k$), there are an infinite number of solutions characterized by an eigenvalue $p$. Each solution represents a different multipole, except the dipole, for which there are two solutions. For each multipole $l$, a unique relation $k(p)$ (or $I_0(p)$ once the values of $B_0$ and $R_\star$ are fixed) defines the family of $l$-polar solutions.

\begin{table}[t]
\begin{center}
\begin{tabular}{c | c c | c c | c c}

\hline
\hline
& \multicolumn{2}{c|}{$C=0.5$} & \multicolumn{2}{c|}{$C=3$} & \multicolumn{2}{c}{$C=30$} \\
$l$ & $p$ & $\Delta\varphi_{ss}$ & $p$ & $\Delta\varphi_{ss}$ & $p$ & $\Delta\varphi_{ss}$ \\
\hline
1 & 0.39 & 2.28 & - & - & - & - \\
1 & 0.87 & 1.08 & - & - & - & - \\
2 & 1.97 & 0.68 & 1.82 & 1.66 & 0.79 & 4.58 \\
3 & 2.98 & 0.62 & 2.87 & 1.52 & 1.97 & 4.60 \\
4 & 3.98 & 0.57 & 3.89 & 1.41 & 3.03 & 4.51 \\
5 & 4.98 & 0.53 & 4.90 & 1.31 & 4.09 & 4.27 \\
\hline
\hline
\end{tabular}
\caption{Values of parameters in self-similar multipolar (up to $l=5$) configurations for three values of $C$.}
\label{tab:p_twist_multipole}
\end{center}
\end{table}

\begin{figure}[t]
\centering
\includegraphics[width=0.32\textwidth]{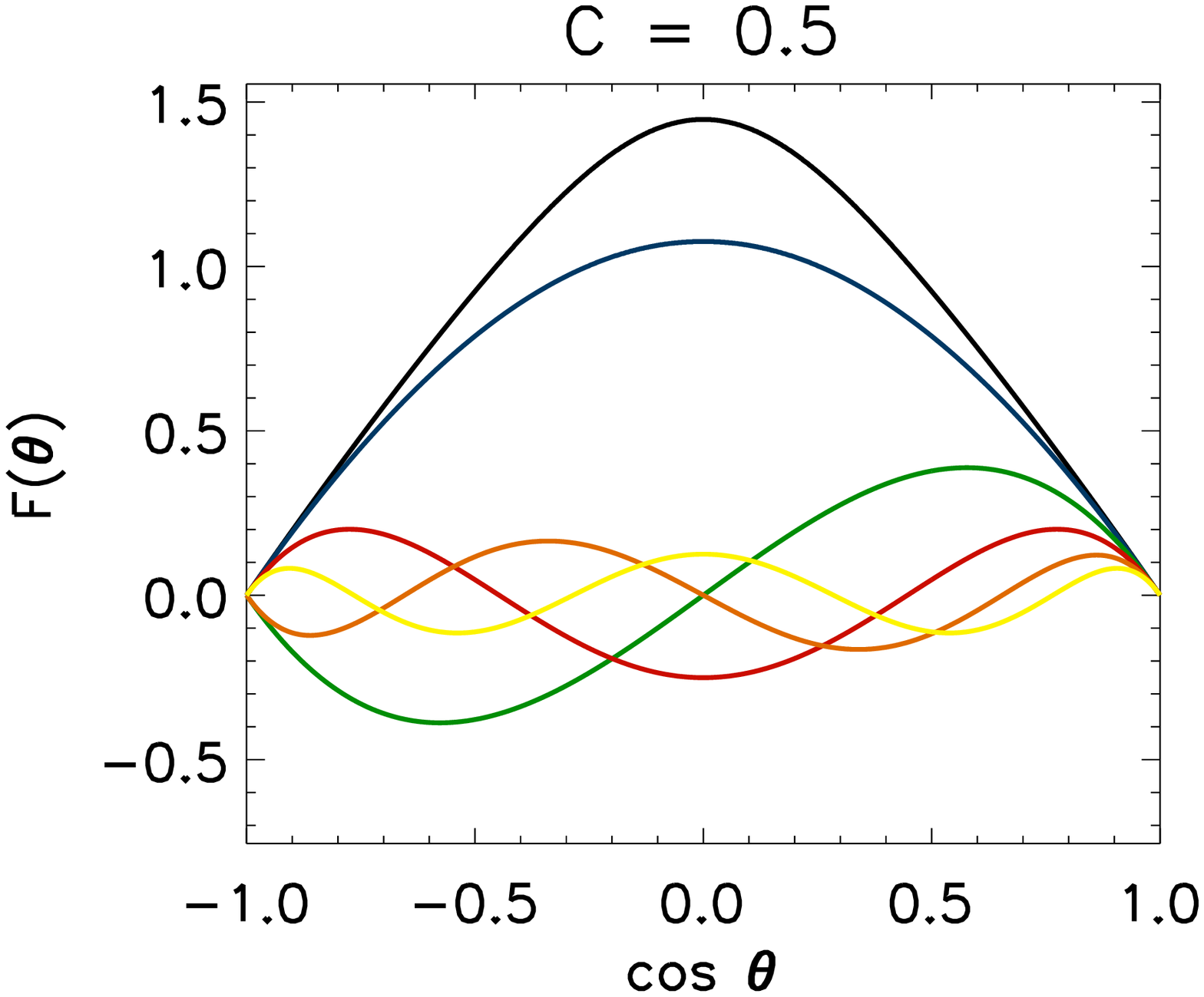}
\includegraphics[width=0.32\textwidth]{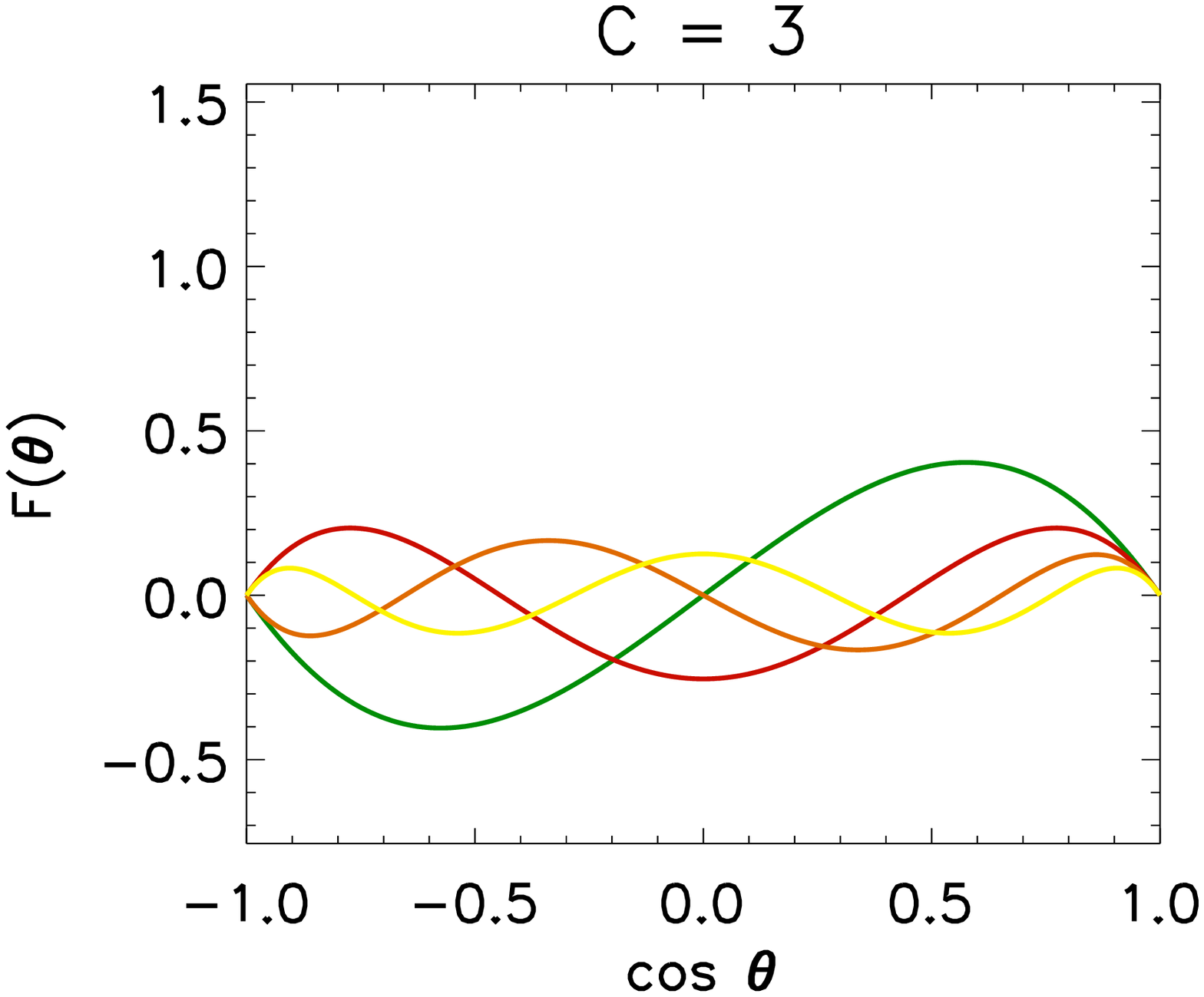}
\includegraphics[width=0.32\textwidth]{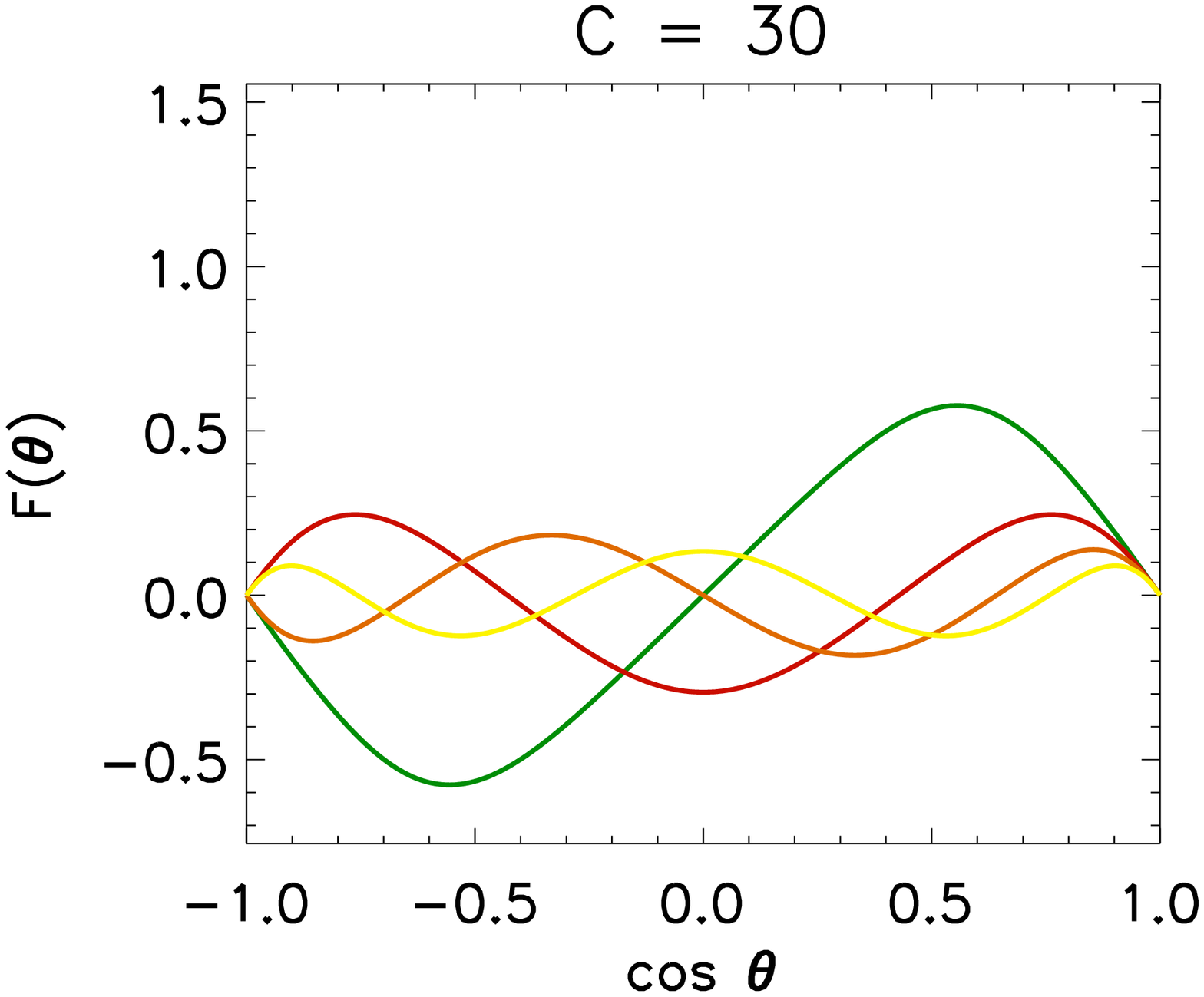}
\caption{Angular function $F(\mu)$ for the self-similar multipolar solutions of Table~\ref{tab:p_twist_multipole}. In each panel we show, if existing, the multipoles $l=1$ (black and blue lines), $l=2$ (green), $l=3$ (red), $l=4$ (orange), $l=5$ (yellow).} 
\label{fig:tlk_multipoles}
\end{figure}

In Fig.~\ref{fig:tlk_dipole} we show the curve $I_0(p)$ and $p(\Delta\varphi_{ss})$ for the dipolar family. For each value $0 < C \le 0.872$ there are two values of $p$ satisfying the equation. The first branch of solutions (blue points) recover the potential case, $F(\mu)=1-\mu^2$, in the limit $C = 0, p = 1$. The second branch (red points) goes towards $p=0, C=0$: this limit is the analytic split monopole solution discussed in \S~\ref{sec:split_monopole}, with $F(\mu)=1-|\mu|$ and twist $\Delta \varphi_{ss}=2\pi$.

Similarly, for higher multipoles, the allowed range of parameters is $p\in (0,l]$, $C\in [0,C_{max})$, $F(\mu)$ has a parity $(-1)^{l-1}$ and $(l+1)$ zeros. The potential $l$-polar solution is recovered if $p=l$, $C=0$, and $F(\mu)=(1-\mu^2)dP_p(\mu)/d\mu$. In Table~(\ref{tab:p_twist_multipole}) we show, for three values of $C$, the solutions corresponding to different multipoles, while the related eigen-functions $F(\mu)$ are shown in Fig.~(\ref{fig:tlk_multipoles}). Note that only in the first case (first column and left panel, $C=0.5$), dipolar solutions are found. For a fixed value of $C$, $\Delta\varphi_{ss}$ slightly decreases for higher multipoles. For increasing values of $C$, $F(\mu)$ deviates from sinusoidal forms, and shows cusps in the limit $C\rightarrow 0$, $p\rightarrow 0$.

Our results agree with previous dipolar \citep{thompson02} and multipolar \citep{pavan09} solutions. We will use these models in the numerical code presented in \S~\ref{sec:numerical_magnetosphere_code}, for testing purposes and for the sake of comparison with other numerical solutions.

Note however that all self-similar solutions are of limited generality, among other reasons, because a linear combination of solutions is not a solution itself, due to the nonlinear character of the problem. In other words, the fixed radial dependence in $\Gamma$ does not allow a combination of multipoles with different radial dependences. This also implies that all self-similar solutions have a defined symmetry with respect to the equator.

%%%%%%%%%%%%%%%%%%%%%%%%%%%%%%%%%%%%%%%%%%%%%%%%
\subsection{A multipole-coupling solution.}\label{sec:legendre_sol}

We explored a few more forms of $\alpha$ in eq.~(\ref{eq:form_alpha}), as extensively discussed in Appendix~\ref{app:forcefree}. An interesting choice is $\alpha=(k/R_\star)|\Gamma/\Gamma_0|^{1/2}$, with the condition $\Gamma\ge 0$ everywhere. Note that this is a generalization of the family of self-similar models with $p=2$. Eq.~(\ref{eq:gs_ode_leg}), with $\Gamma$ given by eq. (\ref{eq:potential_gen}), leads to:
\begin{equation}
  \frac{d^2 f_l}{d r^2}-l(l+1)\frac{f_l}{r^2}=-\left(\frac{k}{R_\star}\right)^2 \sum_{m,n=1}^\infty f_mf_n g_{lmn}~,
\end{equation}
where $f_l(r)=a_l(r)r/R_\star$, and the {\em Gaunt's factors} $g_{lmn}$ have been defined as
\begin{equation} 
g_{lmn}=\frac{2}{3}\frac{2l+1}{2m+1}\frac{m(m+1)}{l(l+1)}n(n+1) \left[\int P_lP_{m-1}P_n - \int P_lP_{m+1}P_n\right]~.\nonumber
\end{equation}
The integrals of the triple product of Legendre polynomials can be evaluated analytically (see Appendix~\ref{app:gaunt}). However, a strong limitation of the model is the condition $\Gamma \ge 0$ in its whole domain $(r,\mu)\in[1,r_{max}]\times[-1,1]$. Also due to this reason, we only find solutions with small values of twist, $k\lesssim 0.1$, close to be potential.

%%%%%%%%%%%%%%%%%%%%%%%%%%%%%%%%%%%%%%%%%%%%%%%%%%%%%%%
\section{Non-rotating case: numerical solutions.}\label{sec:numerical_magnetosphere_code}

In the previous section we have discussed some analytical and semi-analytical solutions that share the same drawback: the arbitrary choice of the enclosed current function $I(\Gamma)$ or, equivalently, $\alpha(\Gamma)$. Some of these solutions are nonphysical, in the sense that they can neither be extended to infinity nor matched to vacuum solutions. All these limitations make the (semi-)analytical approach insufficient for general purposes, because we have no physical argument for preferring one particular form of the current to another. Furthermore, \cite{beloborodov09} describes the time-dependent evolution of the twist due to the Ohmic dissipation. Globally twisted configurations, like the self-similar ones (\S~\ref{sec:selfsimilar}), are not expected. Long-lived currents flow in the largest loops, which footprints lie near the poles. These {\em j-bundles} are a likely configuration in magnetars.

The alternative is to find numerical solutions of the nonlinear, force-free equations describing a neutron star magnetosphere. We expect these solutions to be more general and in some cases very different from the semi-analytical ones.

%%%%%%%%%%%%%%%%%%%%%%%%%%%%%%%%%%%%%%%%%%%%%%%
\subsection{The magneto-frictional method.}\label{sec:magnetofrictional}

We assume that the slow-rotation approximation is valid up to an outer radius $R_{out}$, which reduces the problem to finding solutions of eq.~(\ref{eq:rotB_nonrotating}). In the {\em magneto-frictional method} \citep{yang86, roumeliotis94}, one begins with an initially non-force-free configuration and defines a fictitious velocity field proportional to the Lorentz force,

\begin{equation}
 \vec{v}_f=\nu_f \frac{\vec{J}\times\vec{B}}{B^2}~,
\end{equation}
where $\nu_f$ is a normalization constant, with units of length squared over time. The electric field, eq.~(\ref{eq:balance_em}), is written as

\begin{equation}\label{ef_code} 
\vec{E}_f  =\frac{\nu_f}{c}\left(\vec{J}-\frac{(\vec{J}\cdot \vec{B})\vec{B}}{B^2}\right)~.
\end{equation}
We evolve the magnetic field components directly by solving the induction equation 
\begin{equation}\label{eq:magnetofrictional_induction}
\frac{1}{c}\frac{\partial \vec{B}}{\partial t} =  -\vec{\nabla}\times \vec{E}_f~. \label{dbdt_code}
\end{equation}
$\vec{E}_f$ is a measure of the deviation from the force-free condition, because $\vec{J}\parallel \vec{B}$ is accomplished if and only if $\vec{E}_f\equiv 0$. It acts as a frictional term that forces the magnetic field to relax to a force-free configuration. Note that the time unit in eq.~(\ref{eq:magnetofrictional_induction}) is set by the value of $\nu_f$, and is not related to any physical evolution of the magnetosphere.

In the original method, \cite{roumeliotis94} write the magnetic field in terms of the magnetic flux function $\Gamma$ and another scalar function $\Theta$:
\begin{equation}\label{magnetofrictional_clebsch}
 \vec{B}=\vec{\nabla}\Gamma\times\vec{\nabla}\Theta~,
\end{equation}
in terms of which the induction equation becomes a system of two advection equations (see Appendix~\ref{app:poloidal-toroidal} for a derivation):
\begin{eqnarray}
 && \partial_t \Gamma+\vec{v}_f\cdot\vec{\nabla} \Gamma=0~, \label{eq:roum1}\\
 && \partial_t \Theta+\vec{v}_f\cdot\vec{\nabla} \Theta=0~. \label{eq:roum2}
\end{eqnarray}
The main reason for solving the induction equation for the magnetic field vector instead of eqs.~(\ref{eq:roum1}) and (\ref{eq:roum2}) is to allow for future extensions of the code by considering a real, rotationally-induced electric field. The disadvantage is that we have to be more careful when setting boundary conditions for the electric field, because we could converge to stationary solutions characterized by $\vec{\nabla}\times \vec{E}_f=0$, which are not necessarily force-free.

%%%%%%%%%%%%%%%%%%%%%%%%%%%%%%%%%%%%%%%%%%%%%%%%%%%%%%%%
\subsection{Linear analysis of the magneto-frictional method.}\label{sec:stability}
We now consider a background, uniform magnetic field $\vec{B}_0$ and a small perturbation
\begin{equation}
\delta \vec{B}=B_1 e^{i(\vec{k}\cdot\vec{r}-\omega t)}~. 
\end{equation}
In the linear regime, the equations read as

\begin{eqnarray}
 && \delta \vec{J} = \frac{c}{4\pi}i\vec{k}\times \vec{\delta B}~,\\
 && \delta \vec{E}_f=-\frac{\nu_f}{cB_0^2}[(\delta \vec{J}\times \vec{B}_0)\times \vec{B}_0]~,\\
 && \frac{1}{c}\frac{\partial \delta\vec{B}}{\partial t}=-i\frac{\omega}{c} \delta\vec{B}=-\vec{\nabla} \times \delta \vec{E}_f~.
\end{eqnarray}
Explicitly, the last equation can be written as
\begin{equation}\label{dispersion}
 -i \frac{4\pi\omega}{\nu_f}\delta\vec{B} = \frac{\vec{k}\times\vec{B}_0}{B_0^2}[\delta\vec{B}\cdot (\vec{k}\times\vec{B}_0)] -k^2 \delta \vec{B}~.
\end{equation}
If the perturbed current is orthogonal to the background magnetic field (either longitudinal perturbations $\delta\vec{B}\parallel \vec{B}_0$ or transverse perturbations with $\vec{k}\parallel\vec{B}_0$), the first term on the right-hand side of eq. (\ref{dispersion}) vanishes and the dispersion relation is purely dissipative:
\begin{equation}
 \omega = -i\frac{\nu_f}{4\pi}k^2~.
\end{equation}
Any perturbation of this type will be dissipated on a timescale $\propto k^{-2}$. In contrast, for transverse perturbations with both $\delta\vec{B}$ and $\vec{k}$ orthogonal to $\vec{B}_0$, the current is parallel to the background magnetic field, and the two terms in the right-hand side of eq. (\ref{dispersion}) cancel out, so that the perturbation does not evolve (a neutral mode with $\omega=0$).
Therefore, the magneto-frictional method is designed to dissipate all induced currents nonparallel to the magnetic field but allows for stationary solutions with currents parallel to the magnetic field. Since the largest length-scale in our problem is set by the size of the numerical domain, $\lambda_{max} \pi R_{out}$, the typical diffusion timescale on which we expect to converge to a force-free solution is $t_{dif}\propto R_{out}^2$.

%%%%%%%%%%%%%%%%%%%%%%%%%%%%%%%%%%%%%%%
\subsection{The numerical method.}\label{sec:method}

\begin{figure}
\centering
\includegraphics[width=6.5cm]{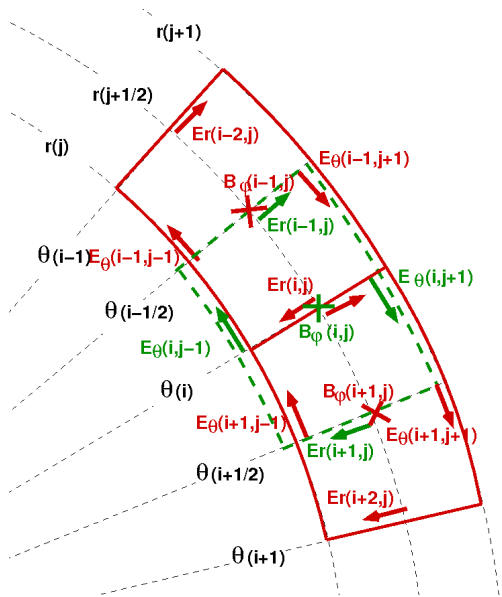}
\caption{Location of the variables in the numerical grid.}
\label{fig:grid}
\end{figure}

We work in spherical coordinates ($r,\theta,\varphi$) under the assumption of axial symmetry. We employ a fully explicit finite difference time domain method \citep{taflove75} with a numerical grid equally spaced in $\theta$ and logarithmic in the radial direction, unless the outer radius $r=R_{out}$ is very close to the surface of the star, $r=R_\star$, in which case we employ a linearly spaced radial grid. Our typical resolution varies between 30 and 200 points in the radial direction, and between 30 and 100 points in the angular direction. At each node $(\theta_i,r_j)$, we define all components of $\vec{B}^{(i,j)},\vec{B}^{(i,j)}$ and $\vec{B}^{(i,j)}$. In Fig.~\ref{fig:grid} we show the location of the variables needed for the time advance of $B_\varphi^{(i-1,j)}$, $B_\varphi^{(i+1,j)}$ (red), and $B_\varphi^{(i,j)}$ (green). The values of $\vec{J}^{(i,j)}$ and $\vec{B}^{(i,j)}$ directly provide $\vec{E}^{(i,j)}$. We advance the induction equation with the standard numerical definitions of geometric elements, fluxes and circulations, as described in Appendix~\ref{app:fdtd}. 

%%%%%%%%%%%%%%%%%
%%%%%%%%%%%%%%%%%%

We must mention that we also tried a method with a staggered grid \citep{yee66}, in which each $n$-component of the magnetic field is defined only at the center of the normal surface, $S_n^{(i,j)}$, while the electric field components are defined in the middle of its delimiting edges. Methods based on staggered grid are well-suited to solving Maxwell's equations, because they provide a natural way to time-advance one field by means of the circulation of the other one. However, we are not dealing with the true Maxwell's equations, but rather with an artificial electric field, eq. (\ref{ef_code}). Evaluating the dot product $\vec{J}\cdot \vec{B}$ requires the interpolation between two or four values of several of the six mutually displaced components. Considering the red components in Fig. \ref{fig:grid}, for instance, the calculation of $E_r^{(i+1,j)}$ also requires $B_\varphi^{(i+1,j)}$, which is not defined at the same location (see Appendix~\ref{app:couplings} for numerical couplings of the magneto-frictional method). The unavoidable interpolation errors prevent the code from completely relax to $\vec{E}_f=0$, except in the trivial case of untwisted configurations. For this reason, we decided to work with a standard grid.

%%%%%%%%%%%%%%%%%%%%%%%%%%%%%%%%%%%%%%%
\subsection{Boundary conditions.}\label{sec:boundary}

At the polar axis, we impose the vanishing of all angular components of magnetic field and currents: $B_\theta=B_\varphi=J_\theta=J_\varphi=E_\theta=E_\varphi=0~.$
At the surface, we have to fix the magnetic field components as provided by some interior solution. However, an arbitrary choice of  poloidal and toroidal magnetic fields may not be compatible with a force-free solution. We decided to impose $\vec{E}_f=0$, at the surface, which is equivalent to keep the value of the radial component fixed at the surface, $B_r(R_\star,\theta)$ and therefore to fix the angular dependence of the magnetic flux function, $\Gamma(R_\star,\theta)$. As a consequence, $B_\theta$ and $B_\varphi$ are allowed to evolve on the first radial grid point.

The external boundary is set at $r=R_{out}$. We have explored two different boundary conditions: $\vec{E}_f(r\ge R_{out})=0$ and the continuous matching to external vacuum solutions. The first choice is equivalent to fix the radial component $B_r(R_{out},\theta)$, while allowing for $B_\theta$ and $B_\varphi$ to evolve. Coupling to general vacuum solutions can be done if the radial field at the outer surface is known (see Appendix~\ref{app:vacuum_bc}). The vacuum region is characterized by $B_\varphi=0$ and the absence of currents or fictitious electric fields. This implies that a current sheet $J_\theta(R_{out})\ne 0$ is needed to ensure current conservation.

If we choose $\vec{E}_f=0$ as outer boundary condition, the code can actually converge to $\vec{E}_f\equiv 0$ at a round-off level, because mathematically this is the only solution compatible with $\vec{\nabla}\times \vec{E}_f=0$. The price to pay is a forced matching of the inner solution with a fixed value of $B_r(R_{out},\theta)$. In contrast, if we couple with vacuum, there is no guarantee that the final solution is $\vec{E}_f=0$ everywhere. We discuss below the influence of the different boundary conditions on the results.

%%%%%%%%%%%%%%%%%%%%%%%%%%%%%%%%%%%%%%%
\subsection{Convergence criterion and tests.}\label{sec:monitors}
Since the magneto-frictional method is based on the introduction of an artificial, viscous electric field that drives an arbitrary initial configuration into a force-free state, we need a convergence criterion to decide when our solutions are acceptable. For that purpose, we keep track of the following quantities during the run:
\begin{itemize}
 \item volume-integrated magnetic energy (both total and contribution from the toroidal magnetic field) 
 \begin{equation}
  {\cal E}_b=\int \frac{B^2}{2}\de V, \quad {\cal E}_{b\varphi}=\int \frac{B_\varphi^2}{2}\de V ~;
 \end{equation}
 \item volume-integrated energy stored in the fictitious electric field 
\begin{equation}
{\cal E}_e=\int \frac{E_f^2}{2}\de V ~;
\end{equation}
 \item total volume-integrated helicity (in Appendix~\ref{app:helicity} we discuss this definition)
\begin{equation}
 {\cal H}\equiv \int_V B_\varphi A_\varphi\de V ~;
\end{equation}
 \item volume-integrated absolute value of $\vec{\nabla}\cdot\vec{B}$ and $\vec{\nabla}\cdot\vec{J}$, which are expected to vanish at round-off level by construction;
 \item an average of the angle $\zeta$ between current and magnetic field\footnote{This average weighted with $J^2$ avoids numerical problems in regions where the numerical value of the current is very low and the angle $\zeta$ is numerically ill-defined.}
 \begin{equation}\label{eq:mean_angle}
  \sin^2\bar{\zeta}=\frac{c}{\nu_f}\frac{\sum  J^2\sin^2\zeta}{\sum J^2}=\frac{c}{\nu_f}\frac{\sum \vec{E}\cdot\vec{J}}{\sum J^2} ~,
 \end{equation}
  where the sum is performed over each node $(i,j)$;
 \item the consistency of the functions $I(\Gamma)$ and $\alpha(\Gamma)$ checking that: first, for each $n$-component, the three functions $\alpha_n(r,\theta)=4\pi J_n/cB_n(r,\theta)$ are the same; second, the eq.~(\ref{eq:current_alpha}) has to be satisfied.
\end{itemize}

%%%%%%%%%%%%%%%%%%%%%%%%%%%%%%%%%%%%%%%
\begin{table}[t]
\begin{center}
 \begin{tabular}{c c c c}
\hline
\hline
$R_{out} [R_\star]$ & $n_r$ & $n_\theta$ & $t_{dis}[R_\star^2/\nu_f]$  \\
\hline
5        & 30	& 30	   &  $41$ \\
5        & 50	& 100	   &  $35$ \\
10       & 30	& 30	   & $183$ \\
10       & 50	& 30	   & $162$ \\
100      & 50	& 30	   & $18000$ \\
\hline
\hline
\end{tabular}
\caption{Time needed to dissipate the numerical currents of the vacuum dipole.}
\label{tab_dip_test}
\end{center}
\end{table}
%%%%%%%%%%%%%%%%%%%%%%%%%%%%%%%%%%%%%%%

Hereafter, we show the electromagnetic quantities in units of $B_0$ (the magnetic field at the polar surface), the star radius $R_\star$, $c$, therefore the magnetic field $\vec{B}$ scales with $B_0$, the current density $\vec{J}$ with $cB_0/R_\star$, the enclosed current $I$ with $cB_0R_\star$, the magnetic flux function $\Gamma$ with $\Gamma_0=B_0 R_\star^2/2$, and the numerical time $t$ with $R_\star^2/\nu_f$. To test our code and to fix our convergence criteria for the realistic models, we performed a battery of tests. In the first basic test we considered the analytical vacuum dipole with $A_\varphi=B_0\sin\theta/2r^2$, $B_\varphi=0$ and checked the ability of the code to maintain this solution. Due to the discretization errors, a little numerical toroidal current appears, and consequently a non-vanishing toroidal fictitious electric field. These errors are small ($\Delta {\cal E}_b/{\cal E}_b\lesssim 1\%$), but it is interesting to see how long it takes to dissipate the perturbation to obtain ${\cal E}_e=0$ to machine accuracy. In Table \ref{tab_dip_test} we show the results with different resolutions and values of the outer radius, always keeping the time-step close to the maximum value allowed by the Courant condition. The expected behavior $t_{dis}\sim R_{out}^2$ is obtained, with the constant of proportionality depending on the grid resolution, which affects the strength of the initial numerical current.

The second test is provided by the dipolar spherical Bessel (\S~\ref{sec:bessel}) and self-similar solutions (\S~\ref{sec:selfsimilar}). Again we began with an initial model consisting in a known solution, and let the system dissipate the currents that come from discretization errors. To obtain the initial models, the self-similar solutions require the numerical resolution of the nonlinear ordinary differential equation (\ref{eq:ode_tlk}), while the analytical spherical Bessel solutions are directly implemented. We tried different parameters for the spherical Bessel solutions (varying $k$) and self-similar models (varying the multipole index and the global twist). For every model tested with analytical solutions, we observe a very slight numerical readjustment of the configuration and the code rapidly reaches the relaxed state, with relative changes in ${\cal E}_b$, ${\cal E}_{b\varphi}$, ${\cal H}$ less than $\sim 1\%$. We had to pay special attention to work with sufficient radial resolution in the case of highly twisted Bessel models $k\gtrsim 1$, due to their oscillatory radial dependence. For low resolution, the code may find, after a large scale reconfiguration, a completely different solution with smaller $k$, which is numerically more stable. If the resolution is high enough, all analytical solutions are found to be stable. 

\begin{figure}[t]
 \centering
 \includegraphics[width=.45\textwidth]{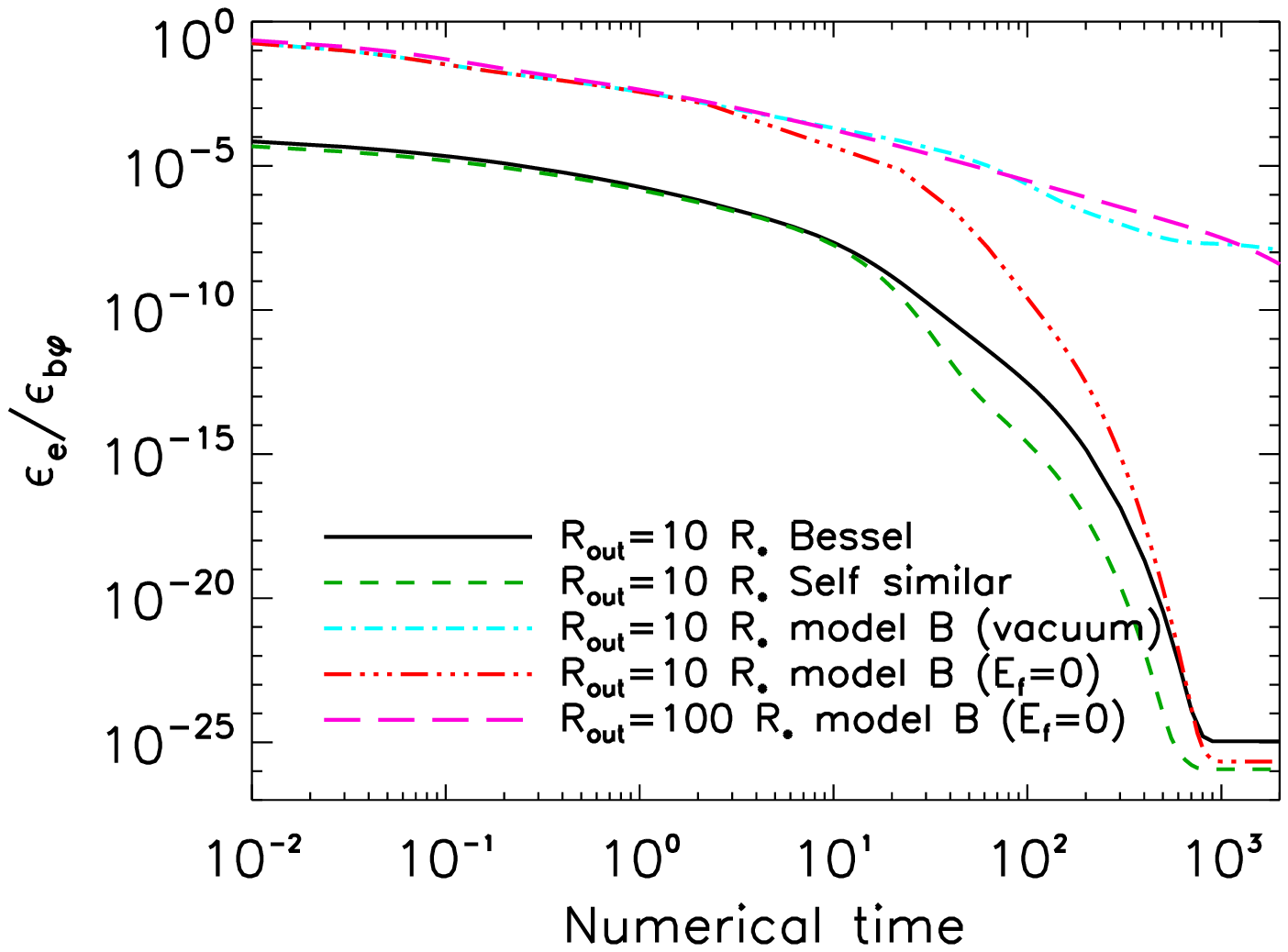}
 \includegraphics[width=.45\textwidth]{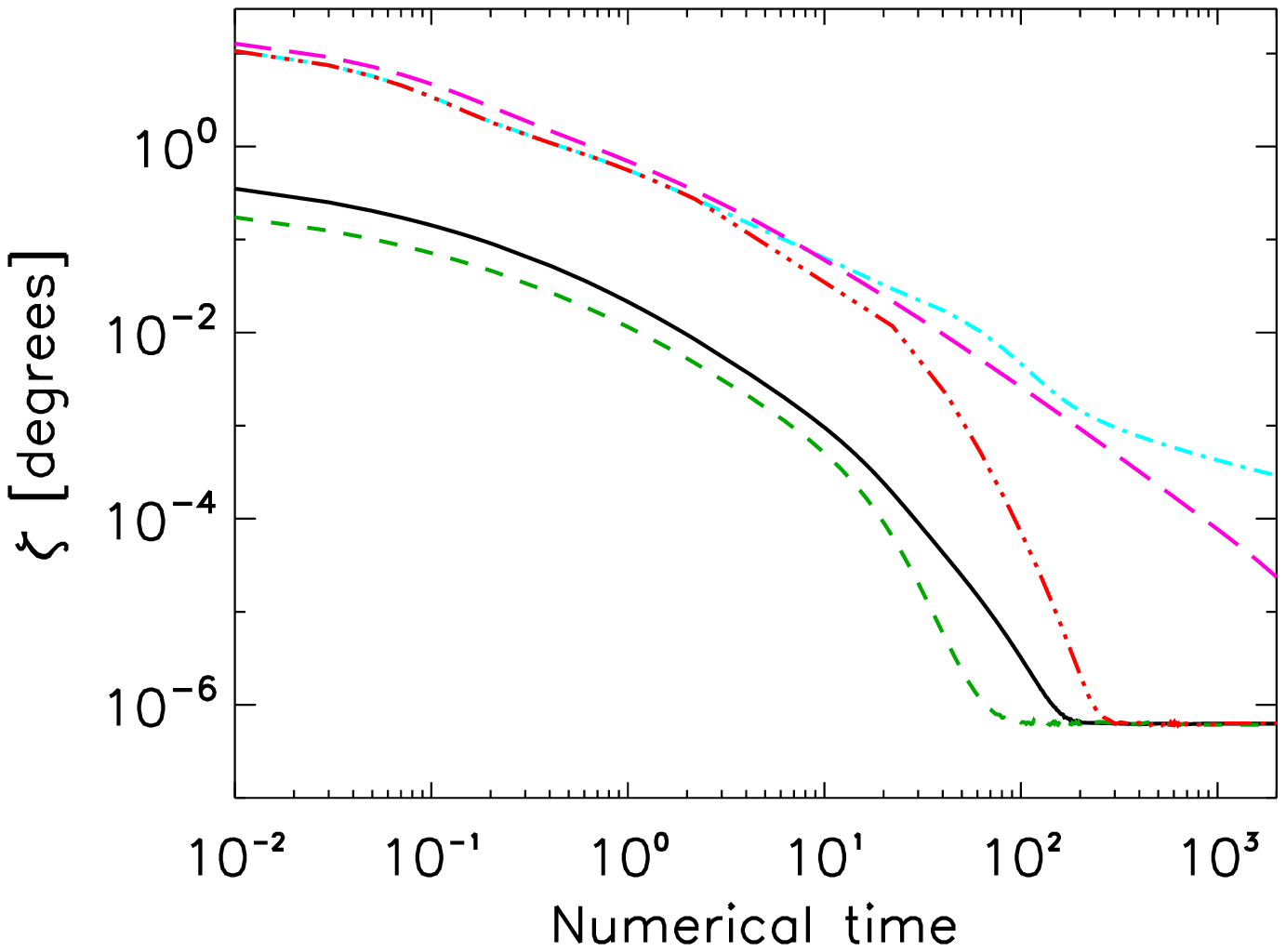}
 \caption{Influence of the outer boundary condition and the location of the external boundary on the evolution of the ratio ${\cal E}_e/{\cal E}_{b\varphi}$ (left) and $\bar{\zeta}$ (right) during the relaxation to a force-free configuration. For comparison, we show also an analytical spherical Bessel model and a twisted self-similar model.}
\label{fig:monitors}
\end{figure}

We have also studied the evolution of vacuum dipolar solutions with an additional toroidal magnetic field for different values of $R_{out}$. In this case the initial currents are due not only to discretization errors, but also to an inconsistency in the initial model. Moreover, the mean angle defined in eq.~(\ref{eq:mean_angle}) initially has a finite value $\bar{\zeta}_{in}$. In Fig.~\ref{fig:monitors} we show how some convergence monitors evolve as a function of time for three cases with an initial toroidal magnetic field of the form $B_\varphi=0.1B_0\sin\theta(R_\star/r)^3$ ($\bar{\zeta}_{in}=15.1^\circ$, model B in table~\ref{tab:initial_magnetosphere}), but different external boundary conditions: matching with vacuum at $R_{out}=10 R_\star$ or imposing $\vec{E}_f=0$ at  $R_{out}=10$ or 100 $R_\star$. For comparison, we also show results for two known dipolar solutions: a spherical Bessel and a twisted self-similar models, with the same helicity.

Finally, we tested a vacuum magnetic dipole perturbed by a weak toroidal component. This is a case of physical interest for quasi periodic oscillations of magnetars, as discussed in \cite{timokhin08} and \cite{gabler11}. Given a background poloidal magnetic field described by $\Gamma$, we chose an arbitrary functional form $I(\Gamma)$ and built the toroidal magnetic field according to eq. (\ref{eq:bphi_definition}). As expected, the perturbed configuration is stable and the stationary solution is rapidly reached after a small readjustment. Typically, for $\max(B_\varphi)=0.1\,B_0$, we have $\bar{\zeta}_{in}\sim O(1^\circ)$ and changes $\Delta B/B \sim 1\%$.

%%%%%%%%%%%%%%%%%%%%%%%%%%%%%%%%%%%%%%%
%%%%%%%%%%%%%%%%%%%%%%%%%%%%%%%%%%%%%%%

In general, the magnetic energy is not conserved, since the system has to dissipate part of the current to reach a force-free configuration. This effect is more evident for initial configurations with high helicity. When the outer boundary condition $\vec{E}_f=0$ is imposed, the helicity is conserved within a few percent, as expected (see Appendix~\ref{app:helicity} for the helicity conservation theorem), and both electric field and mean angle eventually vanish (to machine accuracy). However, when $R_{out}$ is large, or when vacuum boundary conditions are imposed, configurations with high initial helicity take a much longer time to relax (see Fig. \ref{fig:monitors}). In all cases, the relaxation process is faster near the surface, where the configuration of the magnetosphere is more important for our purposes. 

On the basis of these results, our convergence criteria for accepting that a configuration has reached a force-free state are hereafter ${\cal E}_e/{\cal E}_{b\varphi}< 10^{-8}$ and $\bar{\zeta}< 10^{-3}$ degrees, plus the requirement that both quantities are monotonically decreasing with time. Some short, initial relaxation phase, in which some large-scale reconfiguration occurs is possible. We chose to compare the electric energy to the magnetic energy contribution from the toroidal magnetic field, which is much more restrictive than simply the ratio of electric to magnetic energy, especially for low helicity.

%%%%%%%%%%%%%%%%%%%%%%%%%%%%%%%%%%%%%%%
\subsection{Results.}\label{sec:numerical_magnetosphere}

With the numerical code described above, we can obtain general solutions of force-free, twisted magnetospheres. We discuss separately the influence of the following relevant parameters:
\begin{itemize}
 \item the location of the outer radius $R_{out}$ and the external boundary condition;
 \item angular and radial dependence of the initial toroidal magnetic field;
 \item initial twist and helicity, fixed by the functional form and the strength of the initial toroidal magnetic field;
 \item the geometry of the initial poloidal magnetic field.
\end{itemize}

\begin{figure}[t]
 \centering
 \includegraphics[width=.23\textwidth]{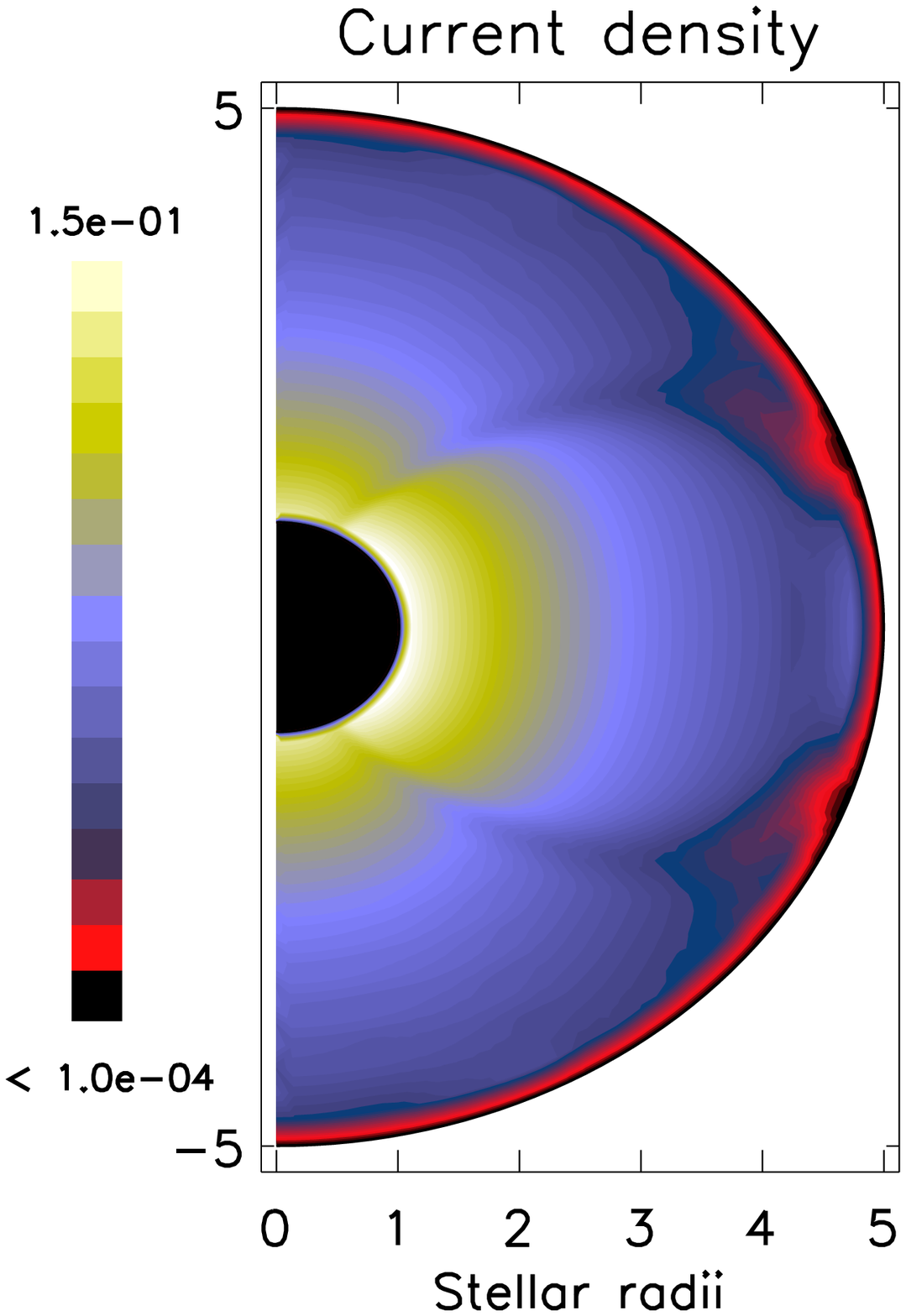}
 \includegraphics[width=.23\textwidth]{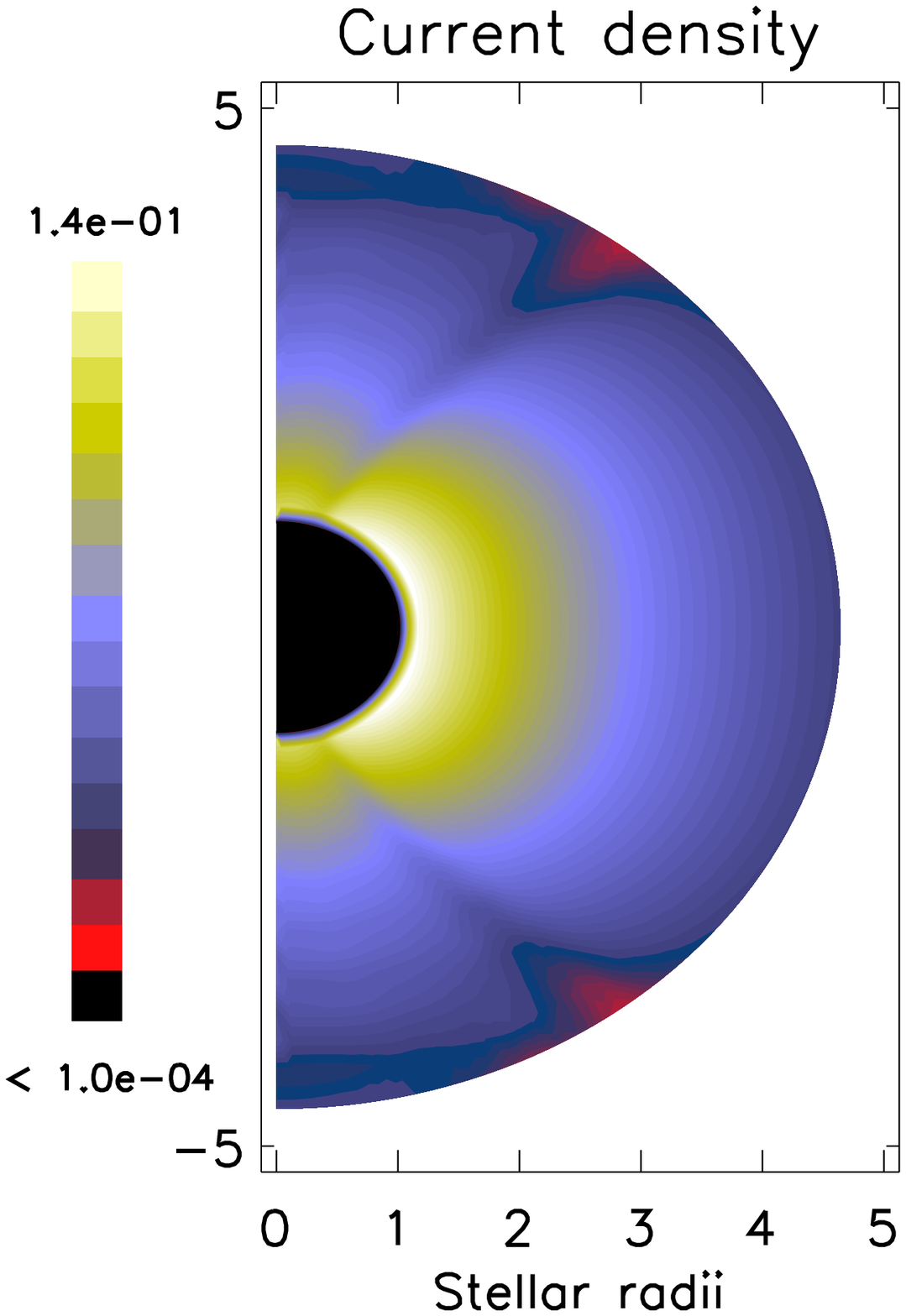}
 \includegraphics[width=.23\textwidth]{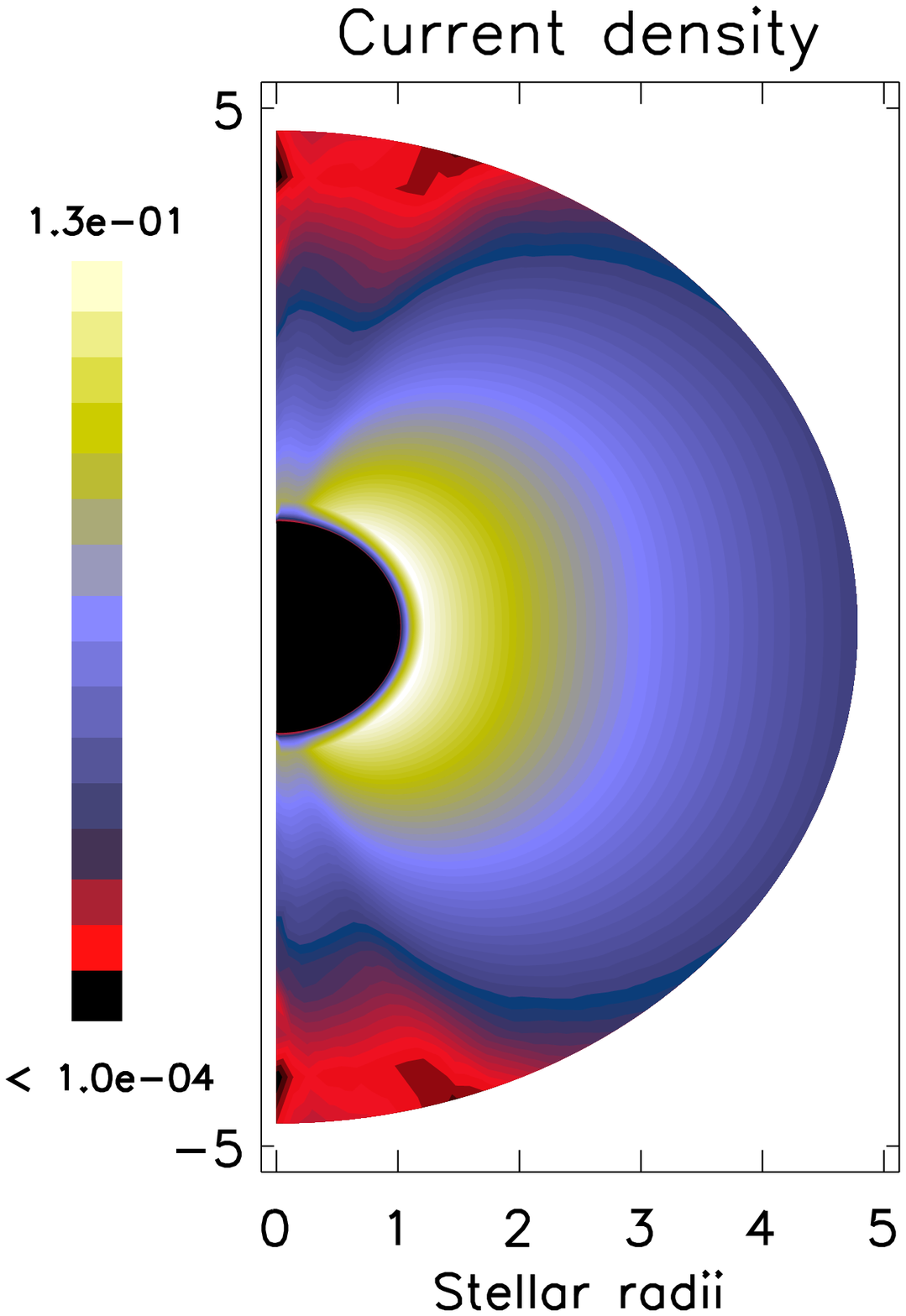}
 \includegraphics[width=.23\textwidth]{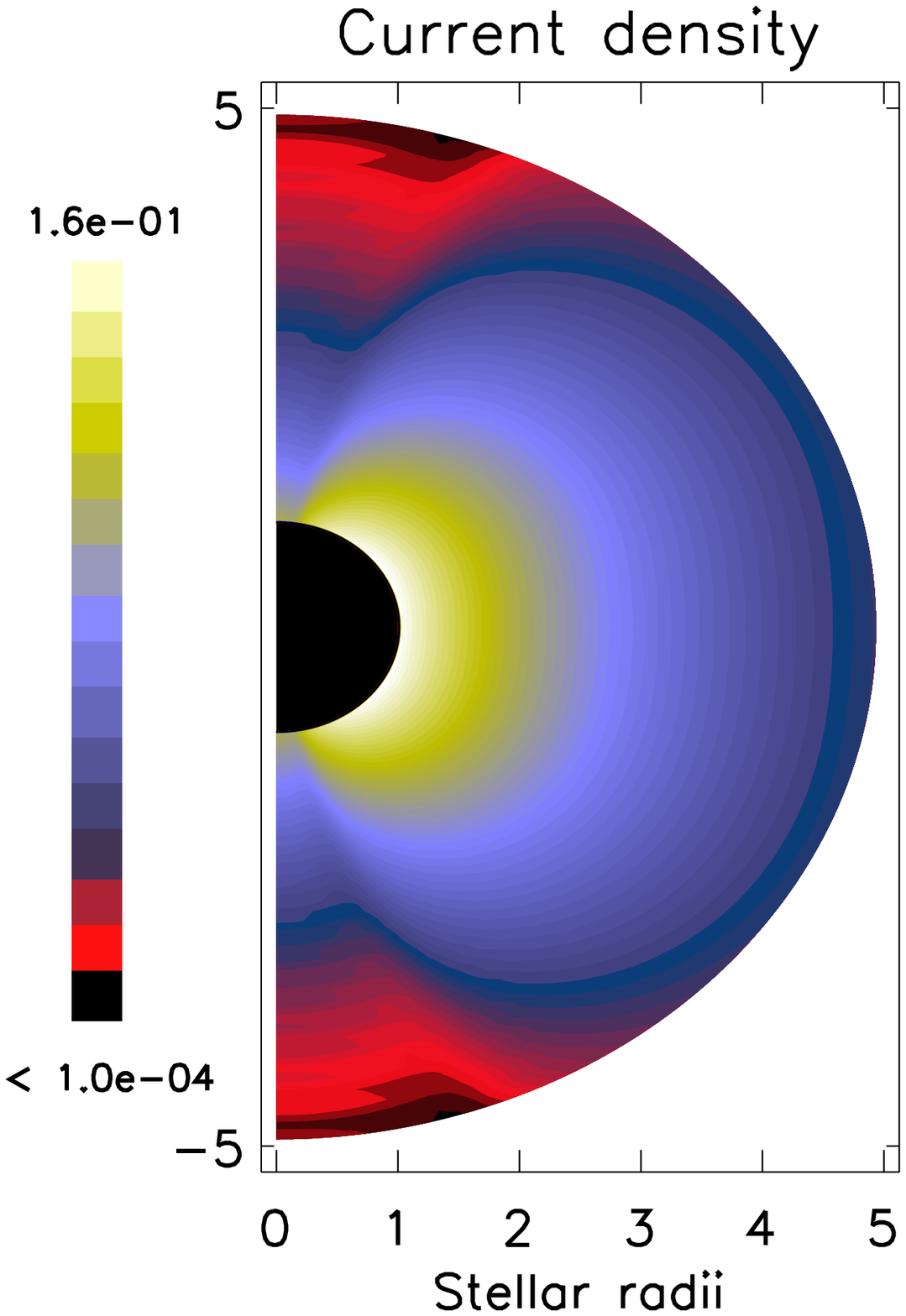}
 \caption{Current density distribution $|\vec{J}|$ in the near region $r\le 5 R_\star$ (colored logarithmic scale in units of $cB_0/R_\star$) for solutions obtained with the same initial data and boundary conditions, $\vec{E}_f=0$, but varying $R_{out}=5,10,50,100~R_\star$ (left to right).}
\label{fig:jrout}

 \includegraphics[width=.6\textwidth]{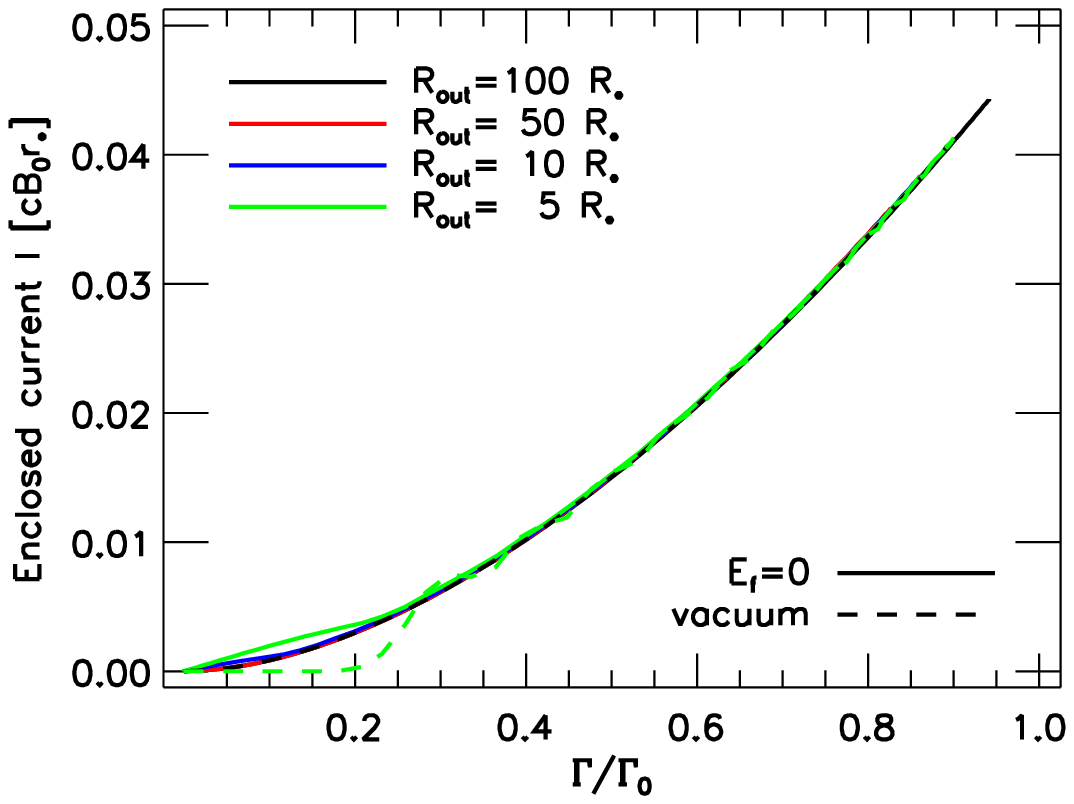}
 \caption{Enclosed current $I(\Gamma)$ for different boundary conditions: $\vec{E}_f=0$ (solid lines) or coupled with vacuum (dashes), at $R_{out}=100,50,10,$ or $5~R_\star$ (black, red, blue or green, respectively).}
\label{fig:igamma_bc}
\end{figure}
%%%%%%%%%%%%%%%%%%%%%%

In Fig. \ref{fig:jrout} we compare the distribution of currents in a solution obtained by imposing $\vec{E}_f=0$ at $R_{out}=5,10,50,$ or $100~R_\star$. In all cases the initial magnetic configuration is the same as the previous subsection (model B of Table~\ref{tab:initial_magnetosphere}). We observe that, near the axis, the solutions are clearly affected by the location of the external boundary, if it is not far enough from the surface ($R_{out}\lesssim 10~R_\star$). In such a case, the influence of the external boundary is important, and it introduces artificial features, although the current distribution in the equatorial region is similar in all cases. The final configurations become almost indistinguishable when $R_{out}=50$ or $100~R_\star$. Taking $R_{out} \gtrsim 100~R_\star$, we ensure that the numerical noise caused by the interaction with the external boundary is negligible. By neglecting the contribution from the open field lines, for which the twist is ill-defined, the maximum line twist (defined in eq.~\ref{eq:def_twist_app}) is similar in all models ($\sim 1.2$ rad).

The function $I(\Gamma)$ for the same four cases (solid lines) is shown in Fig. \ref{fig:igamma_bc}, together with two more cases corresponding to $R_{out}=5~R_\star$ and $R_{out}=100~R_\star$ but replacing the external boundary condition $\vec{E}_f=0$ by a smooth matching with vacuum solutions (dashes). The use of a different boundary condition affects the final solution only for low values of $R_{out}$. Matching with external vacuum implies that no currents can flow through the boundary, so that $I(\Gamma)=\alpha(\Gamma)=0$ along every field line crossing the outer boundary. As a consequence, a force-free configuration, coupled with a vacuum, will be characterized by a plateau $I(\Gamma)=0$ for $\Gamma<\Gamma_c$, with $\Gamma_c$ labeling the first closed field line. This means that in a bundle of open field lines the magnetic field is forced to be potential. The fraction of open field lines (the length of the plateau for low $\Gamma$) is large only for low values of $R_{out}$. For $\Gamma/\Gamma_0>0.2$ (equatorial region), all curves coincide. Increasing the outer boundary to $R_{out}>100~R_\star$ has no visible effect on the models with this helicity. For models with higher helicity, the interaction with the boundary becomes more important, and $R_{out}$ needs to be accordingly increased to minimize the boundary effects.

%%%%%%%%%%%%%%%%%%%%%%
\begin{table}[t]
 \begin{center}
  \begin{tabular}{c c c c c}
\hline
\hline
Model & $k_{tor}$ & $g_{in}(r)$ & $F_{in}(\theta)$ & $\bar{\zeta}_{in}$ [deg]\\
 \hline
A & 0.010  & $(R_\star/r)^{-3}$   					& $\sin\theta$ 					& 13.8 	\\
B & 0.100  & $(R_\star/r)^{-3}$   					& $\sin\theta$ 					& 15.1 	\\
C & 0.500  & $(R_\star/r)^{-3}$						& $\sin\theta$ 					& 29.9 	\\
D & 0.115  & $(R_\star/r)^{-3}$   					& $\sin^2\theta$ 				& 10.7 	\\
E & 0.200  & $(R_\star/r)^{-5}$   					& $\sin^2\theta$ 				& 25.1 	\\
F & 0.204  & $(R_\star/r)^{-3}e^{-[(r-R_\star)/0.5R_\star]^2}$ 		& $\sin^2\theta$ 				& 23.3 	\\
G & 0.113  & $(R_\star/r)^{-3}$   					& $\sin^2\theta$ 				& 34.2 	\\
H & 10     & $(R_\star/r)^{-3}$						& $\sin\theta e^{-[(\theta-\pi/8)/0.3]^2}$ 	& 63.7 	\\
J & 5      & $(R_\star/r)^{-3}(1+5)e^{-[(r-10R_\star)/5R_\star]^2}$	& $\sin\theta(1+4e^{-[(\theta-\pi/10)/0.2]^2}$	& 37.1	\\
\hline
\hline
 \end{tabular}
\caption{Parameters defining the initial toroidal magnetic field in our numerical models, as indicated in eq.~(\ref{eq:initial_tor_magnetosphere}).}
\label{tab:initial_magnetosphere}
\end{center}
\end{table}

The next step is to explore the influence of the strength and form of the initial toroidal magnetic field, by building different initial configurations. Models A to F, and model H are obtained starting from a dipolar poloidal component

\begin{equation}
 A_\varphi=\frac{B_0\sin\theta}{2}\left(\frac{R_\star}{r}\right)^2~.
\end{equation}
The initial configuration of model G is asymmetric with respect to the equator: the poloidal component is a superposition of a dipole and a quadrupole,

\begin{equation}
 A_\varphi=\frac{B_0\sin\theta}{2}\left(\frac{R_\star}{r}\right)^2+\frac{B_0\sin\theta\cos\theta}{2}\left(\frac{R_\star}{r}\right)^3~.
\end{equation}
Model J has a peculiar initial configuration with bunches of field lines near the north pole:

\begin{equation}
 A_\varphi(\theta,R_\star)=B_0\frac{\sin\theta}{2}\left(\frac{R_\star}{r}\right)^2\left[1+4e^{-\left(\frac{\theta-0.9\pi}{0.2}\right)^2}\right]~.
\end{equation}
Table~\ref{tab:initial_magnetosphere} summarizes the parameters of the initial models employed, the initial mean angle between $\vec{B}$ and $\vec{J}$ (eq.~\ref{eq:mean_angle}), and the form of the toroidal component parametrized as:
\begin{equation}\label{eq:initial_tor_magnetosphere}
 B_\varphi^{in}=k_{tor}B_0~g_{in}(r)f_{in}(\theta)~.
 \end{equation}
The angular part for models A to G is chosen to be of the form $f_{in}(\theta)=\sin^d\theta$, with $d$ being a positive integer, while in models H and J the initial toroidal magnetic field is confined to specific angles. The radial dependence is a power law, $g_{in}(r)=r^{-s}$, except in models F, and J for which we use rapidly decaying functions. We fix $R_{out}=100$ and the external boundary condition to $\vec{E}_f(R_{out})=0$ for all models.

\begin{table}[t]
 \begin{center}
 \begin{tabular}{c c c c c c}
\hline
\hline
Model & ${\cal H}$ & $\Delta\varphi_{max}$ & $\max(J)$  & $I_0$ & $p$\\
      & $[B_0^2R_\star^3]$ & [rad] & $[cB_0/R_\star]$ & $[cB_0R_\star]$ & \\
 \hline
S1 & 0.21 & 0.5     & 0.018 & 0.061 	& 0.97 \\
S2 & 1.11 & 1.6     & 0.081 & 0.15 	& 0.69 \\
A & 0.021 & 0.12    & 0.001 & 0.0050 	& 1.45 \\
B & 0.21  & 1.2     & 0.012 & 0.049 	& 1.40 \\
C & 1.11  & 6.5	    & 0.048 & $\ldots$  & $\ldots$ \\
D & 0.21  & 0.7     & 0.015 & 0.056  	& 1.14 \\
E & 0.21  & 0.4     & 0.042 & 0.106  	& 0.50 \\
F & 0.21  & 0.4     & 0.043 & $\ldots$  & $\ldots$ \\
G & 0.21  & 0.3     & 0.020 & 0.038  	& 2.20 \\
H & 0.80  & 2.3	    & 0.59  & $\ldots$	& $\ldots$ \\
J & 29.9  & 3.7	    & 29    & $\ldots$	& $\ldots$ \\
\hline
\hline
 \end{tabular}
\caption{Comparison between the properties of two self-similar solutions (S1-S2) and our numerical solutions (A-J).}
\label{tab:models_magnetosphere}
\end{center}
\end{table}
%%%%%%%%%%%%%%%%%%%%%%%

Table~\ref{tab:models_magnetosphere} shows the features of our final configurations: helicity, maximum line twist, maximum value of current density, and parameters of $I(\Gamma)=I_0(\Gamma/\Gamma_0)^{1+1/p}$.  We also include two self-similar solutions of similar helicity, models S1 and S2, where all components have the same radial dependence $B_i\sim r^{-(p+2)}$. In this case the current function is analytical, $I=I_0(\Gamma/\Gamma_0)^{1+1/p}$ with $(p,I_0)$ belonging to the family of solutions in Fig.~\ref{fig:tlk_dipole}. In the other cases $I_0$ and $p$ are obtained with fits to the numerical function. Note that not in all models $I(p)$ can be fit by a simple power law.

%%%%%%%%%%%%%%%%%%%%%%%%%%%%%%%%%%
\begin{figure}
\centering
\includegraphics[width=.24\textwidth]{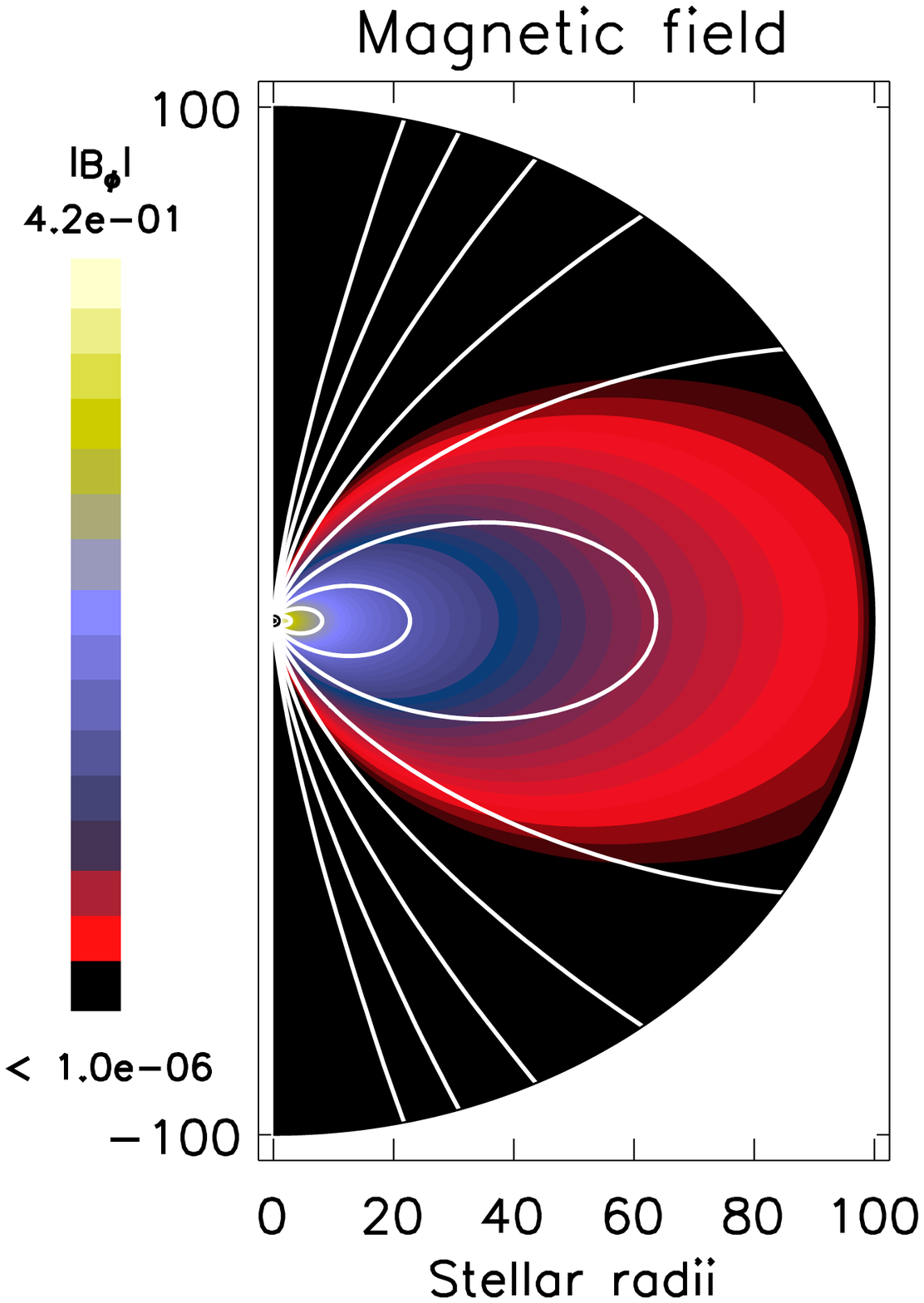}
\includegraphics[width=.24\textwidth]{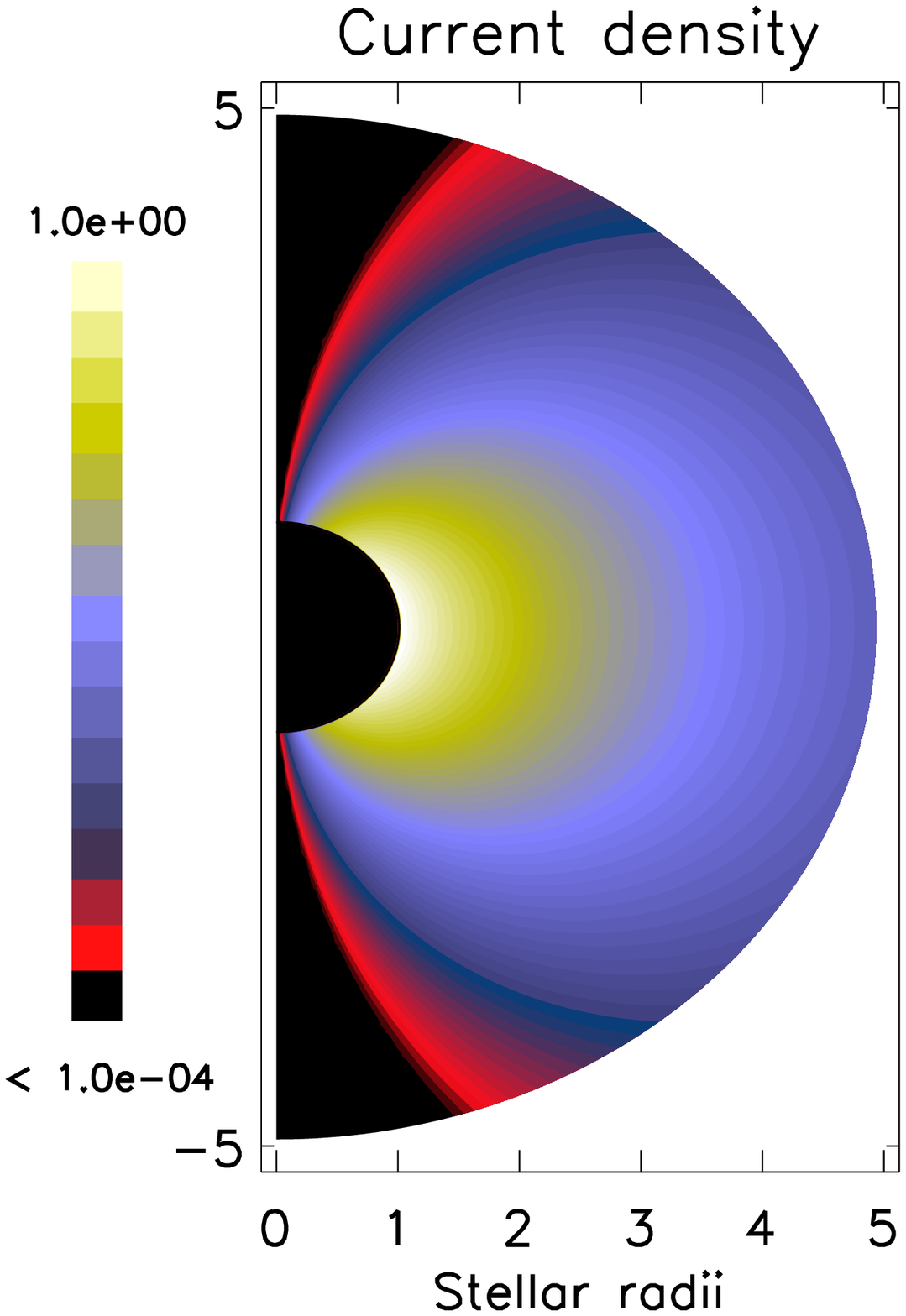}
\includegraphics[width=.24\textwidth]{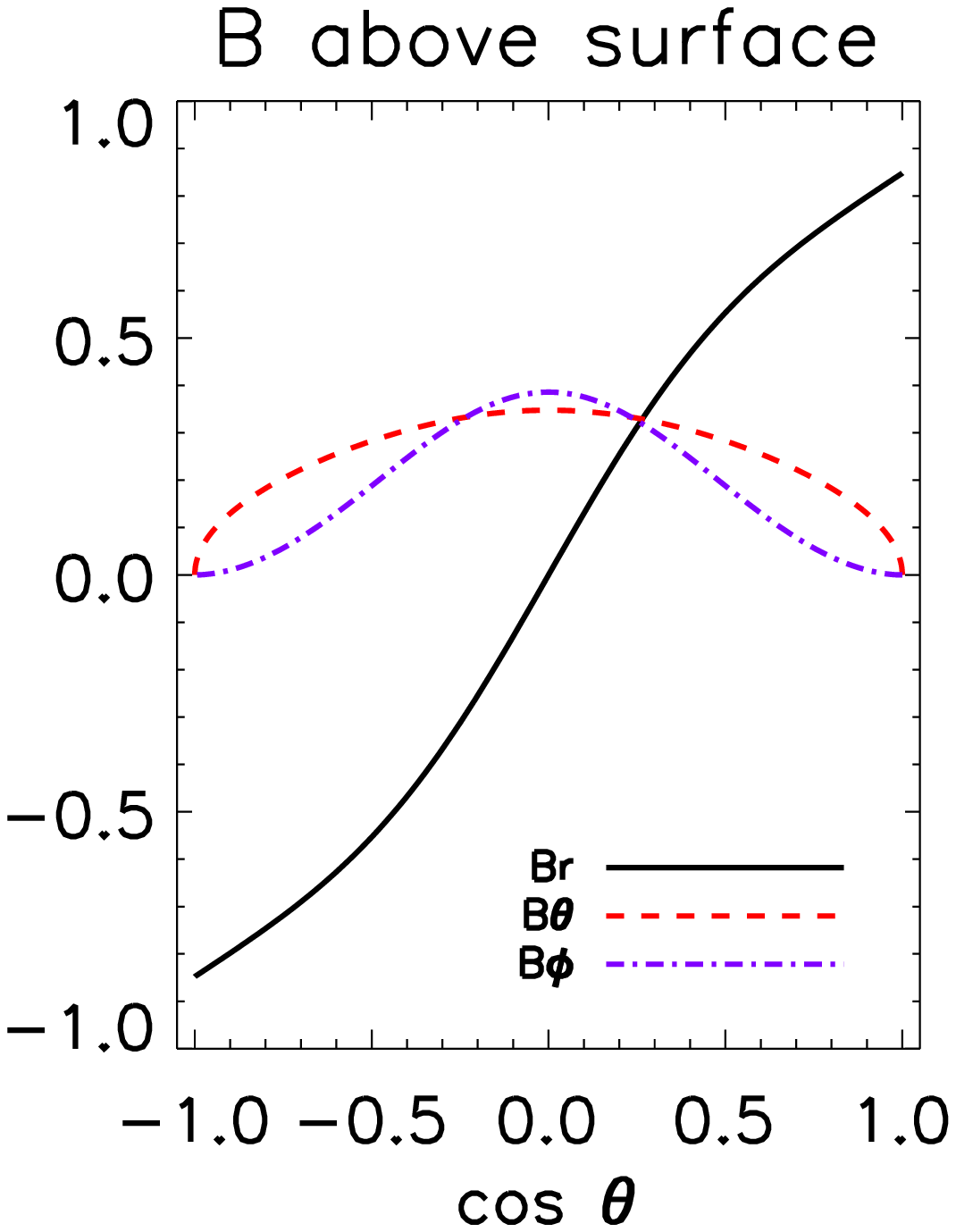}
\includegraphics[width=.24\textwidth]{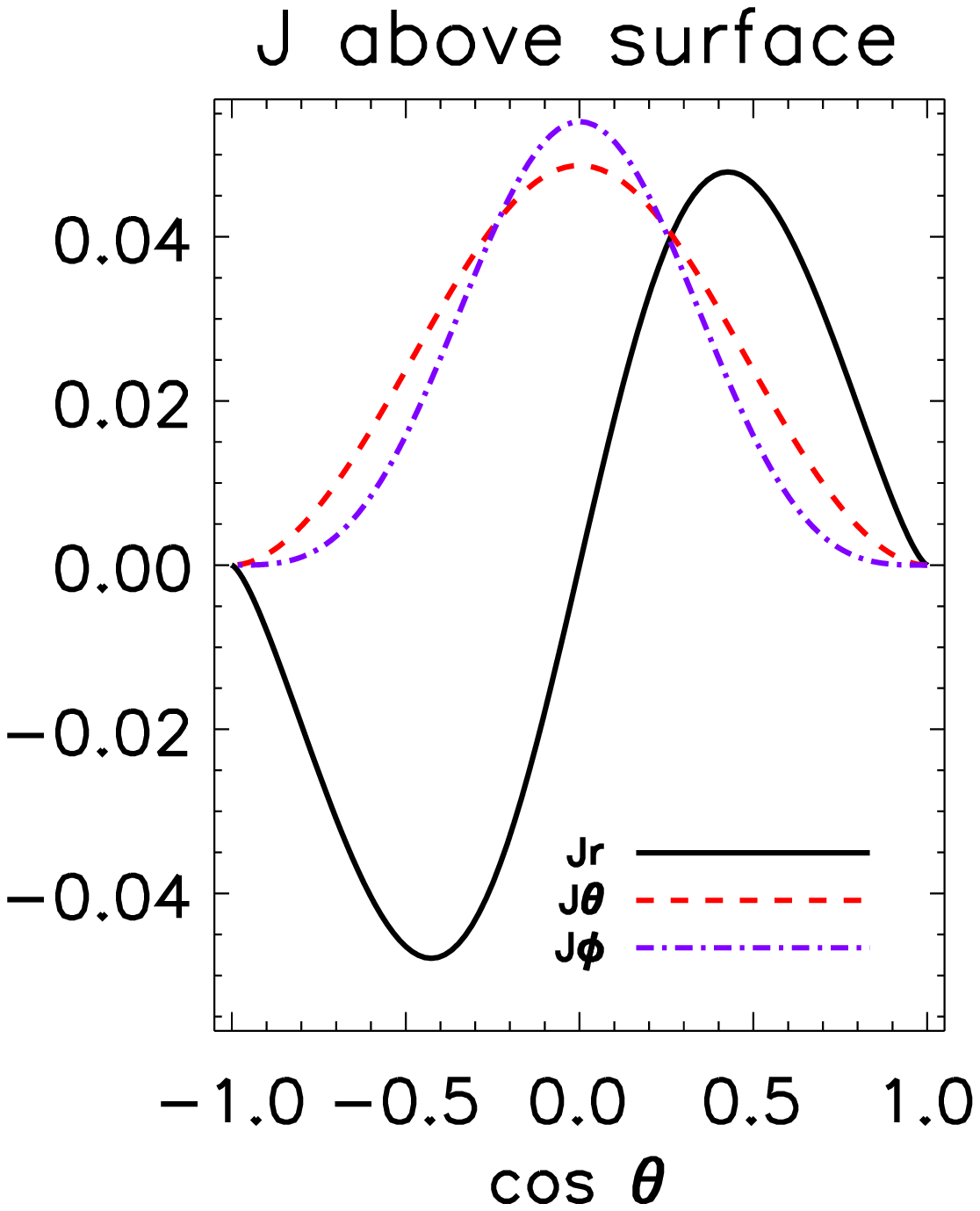}

\includegraphics[width=.24\textwidth]{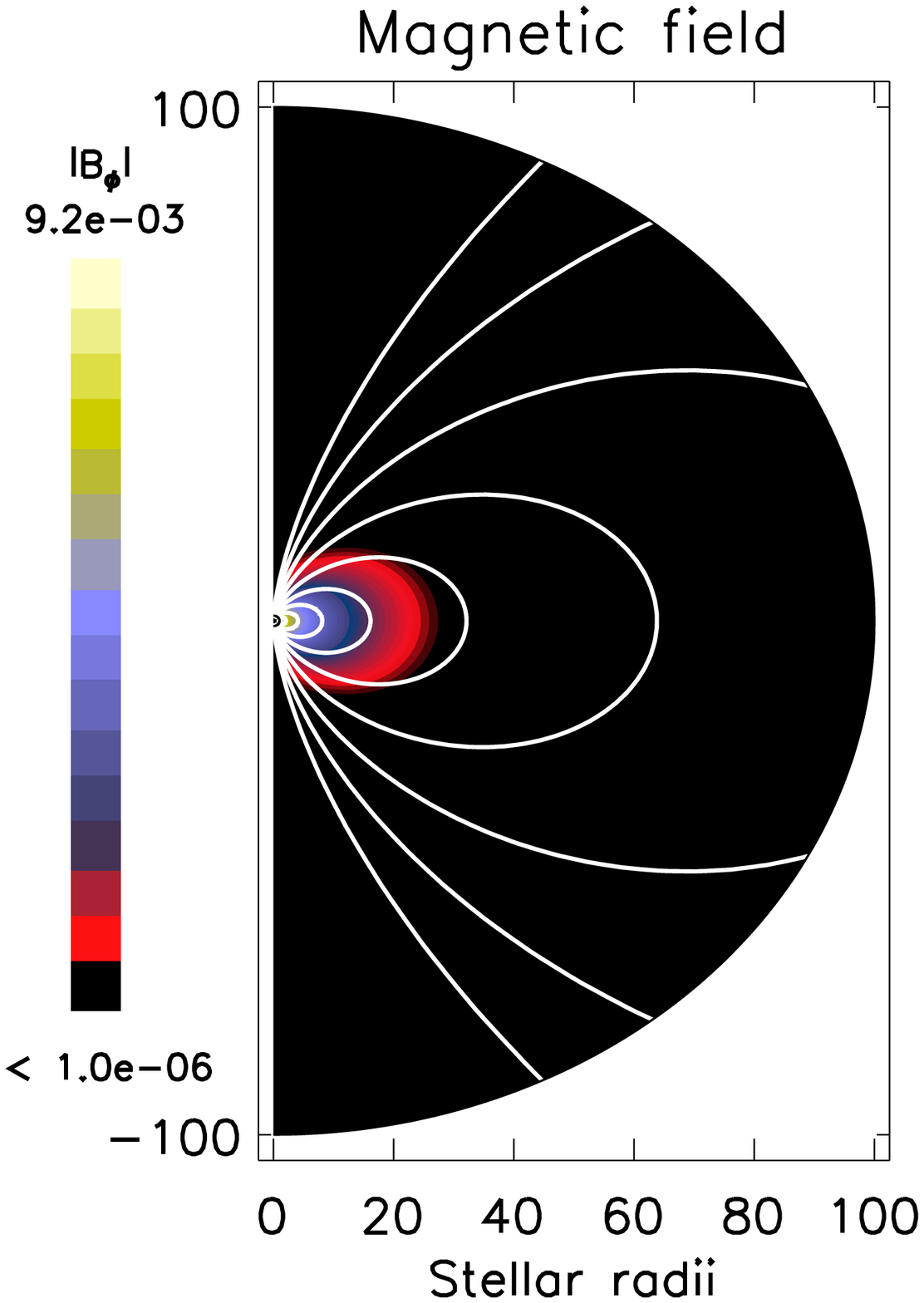}
\includegraphics[width=.24\textwidth]{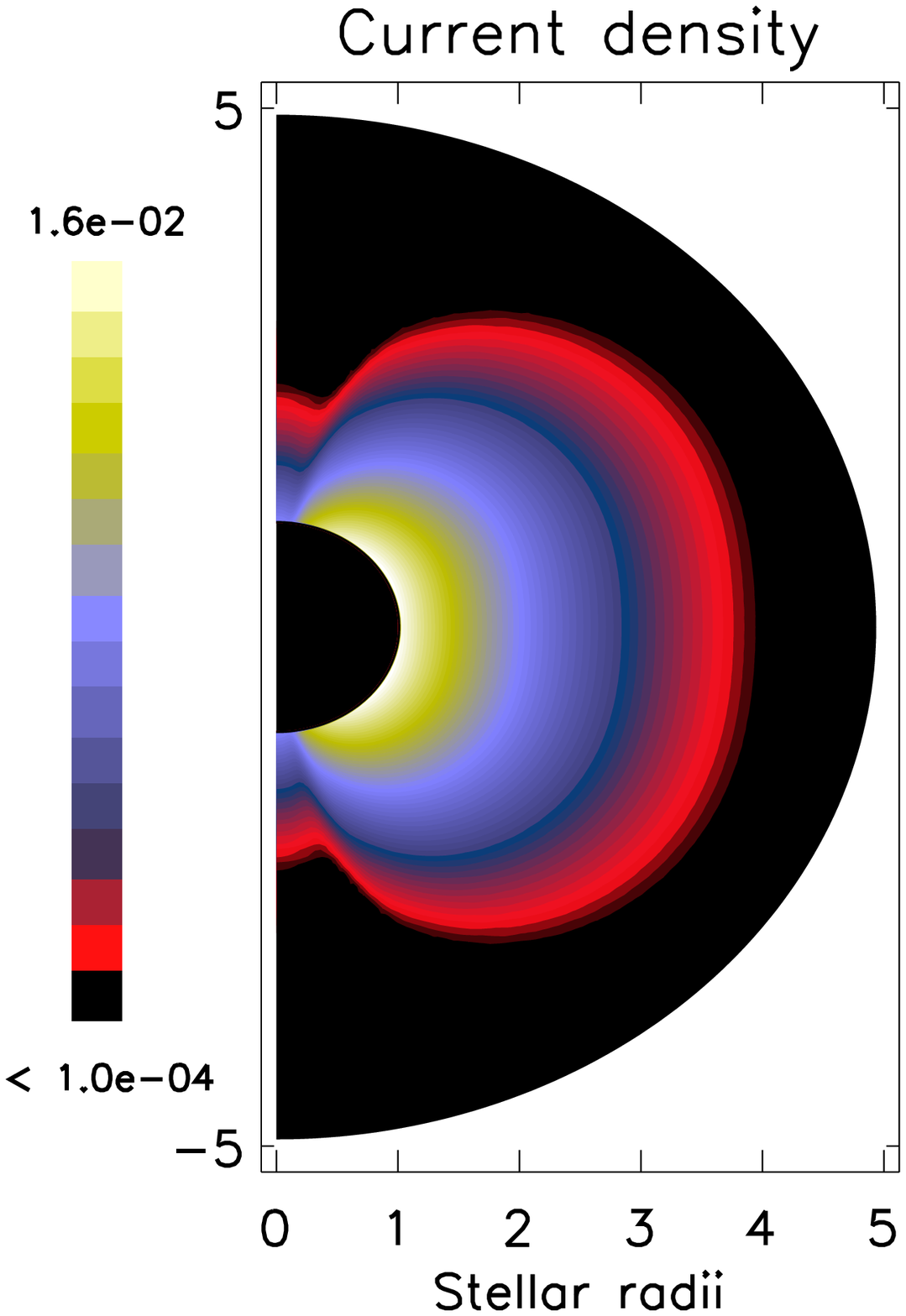}
\includegraphics[width=.24\textwidth]{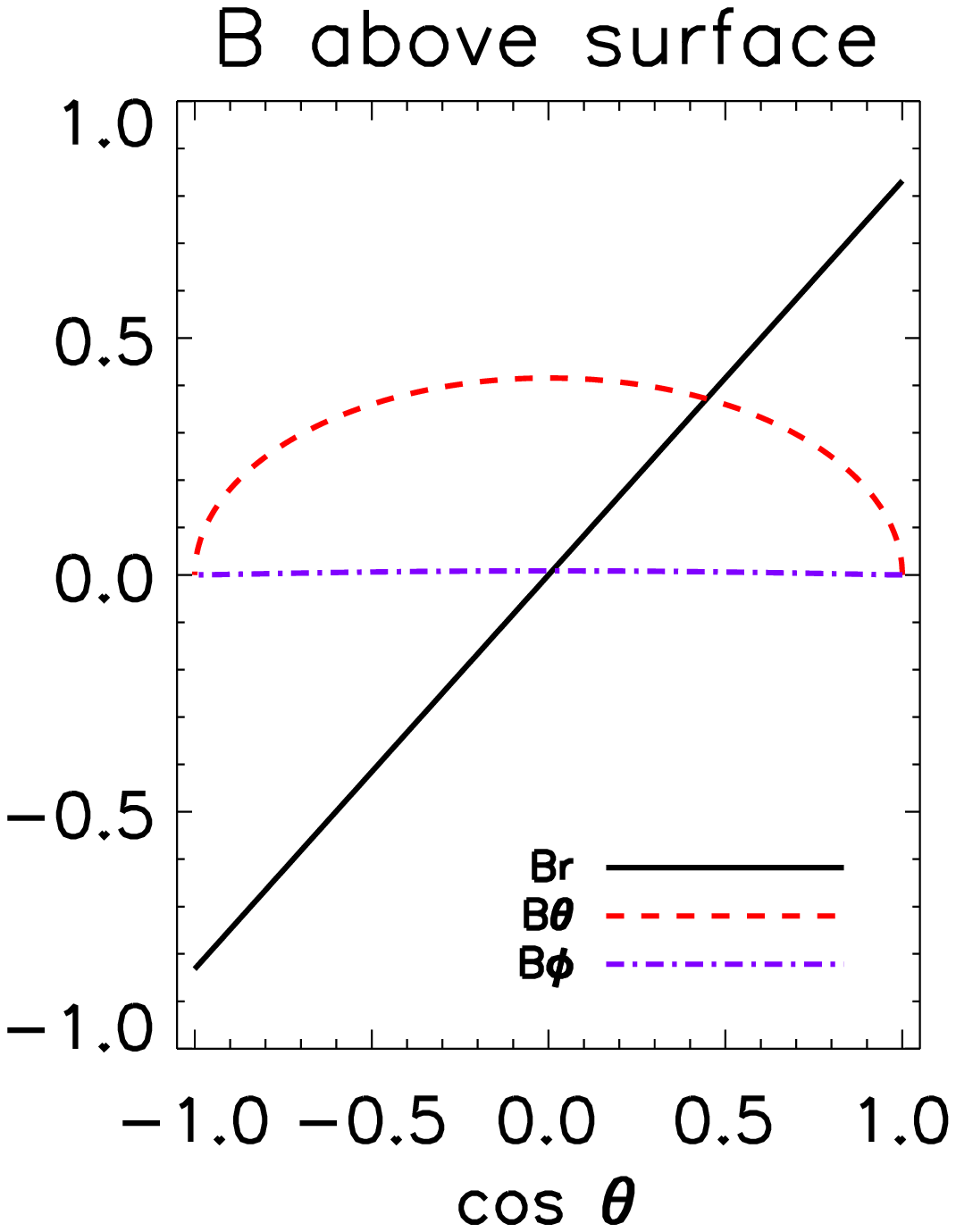}
\includegraphics[width=.24\textwidth]{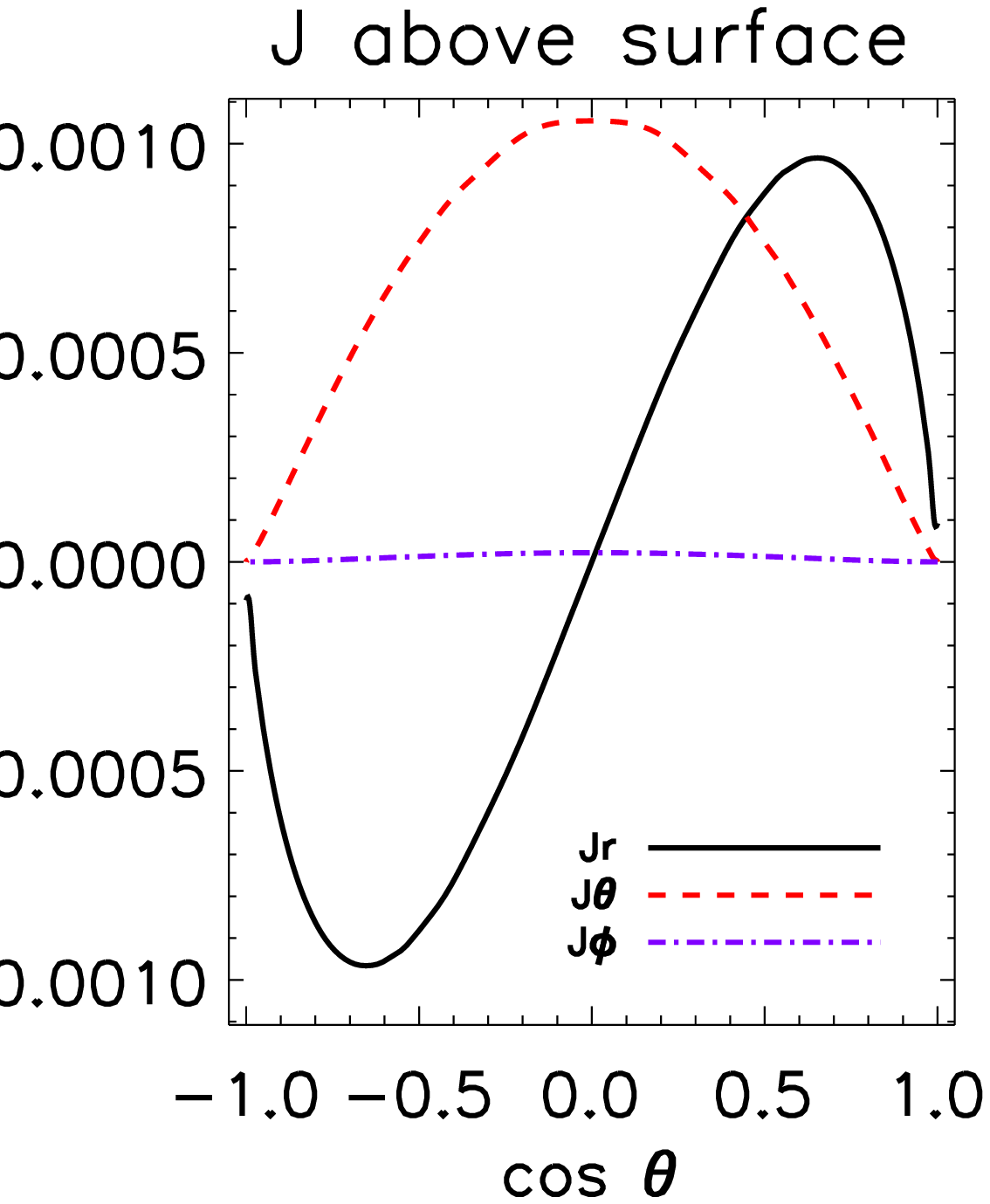}

\includegraphics[width=.24\textwidth]{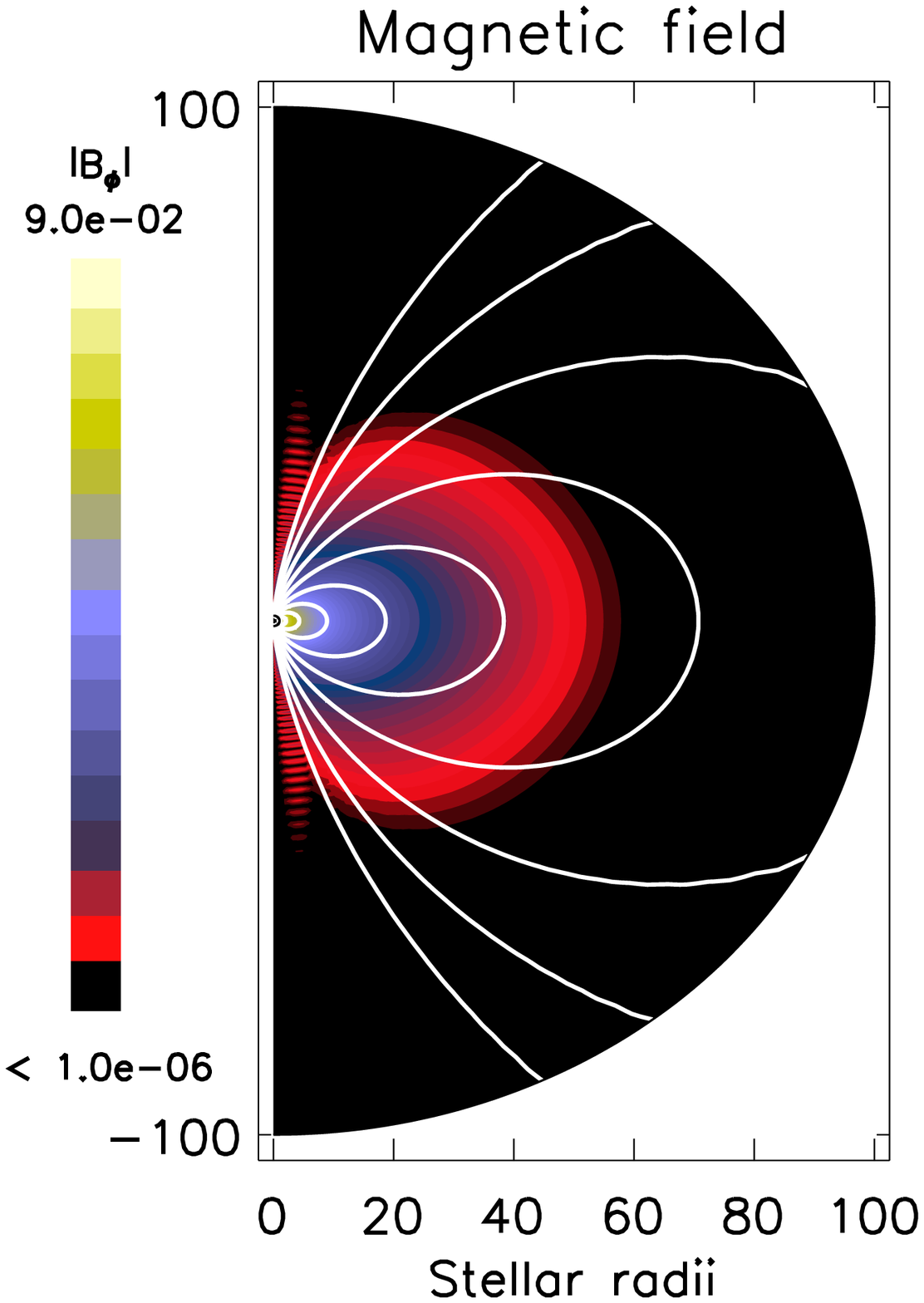}
\includegraphics[width=.24\textwidth]{images/jb.eps}
\includegraphics[width=.24\textwidth]{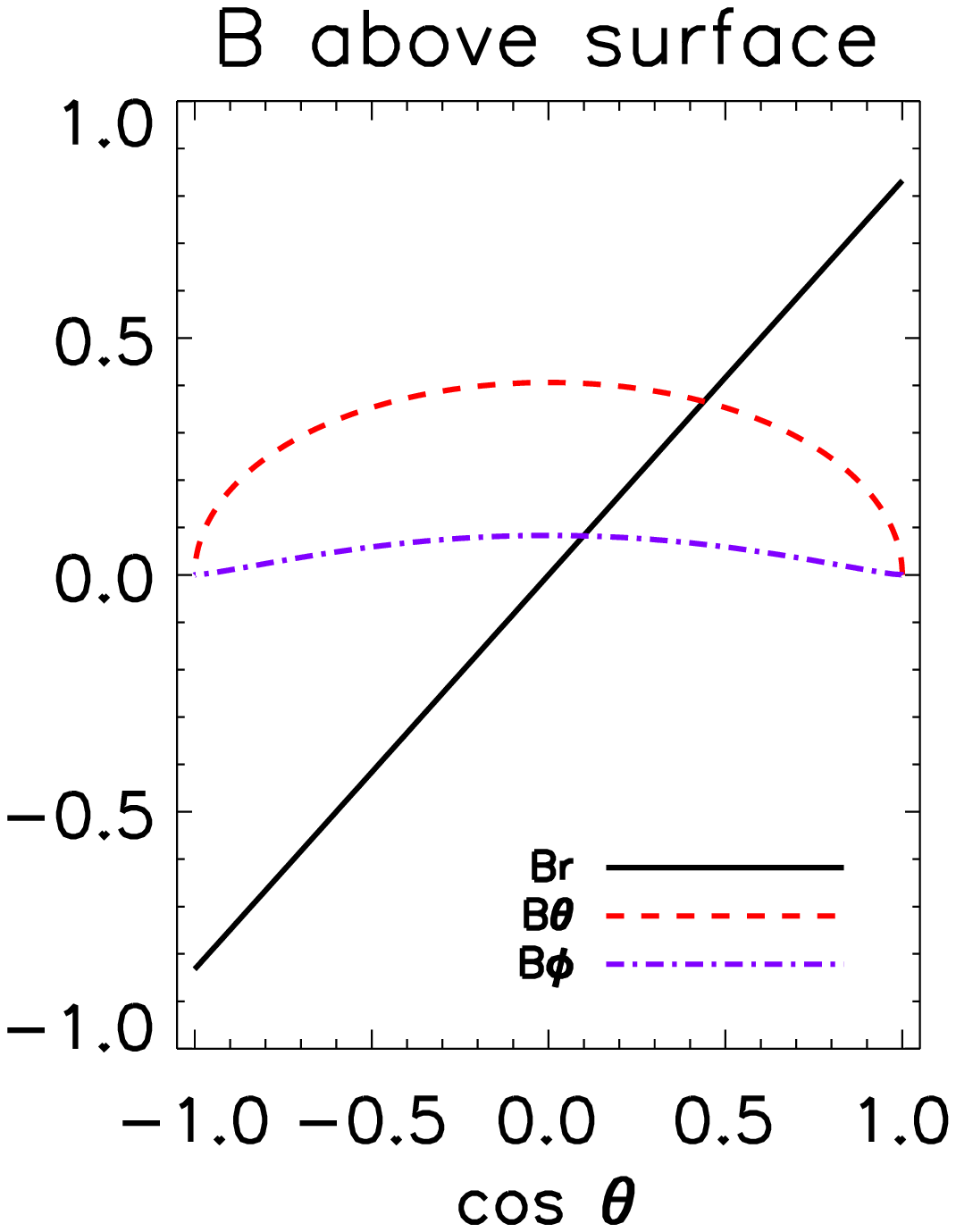}
\includegraphics[width=.24\textwidth]{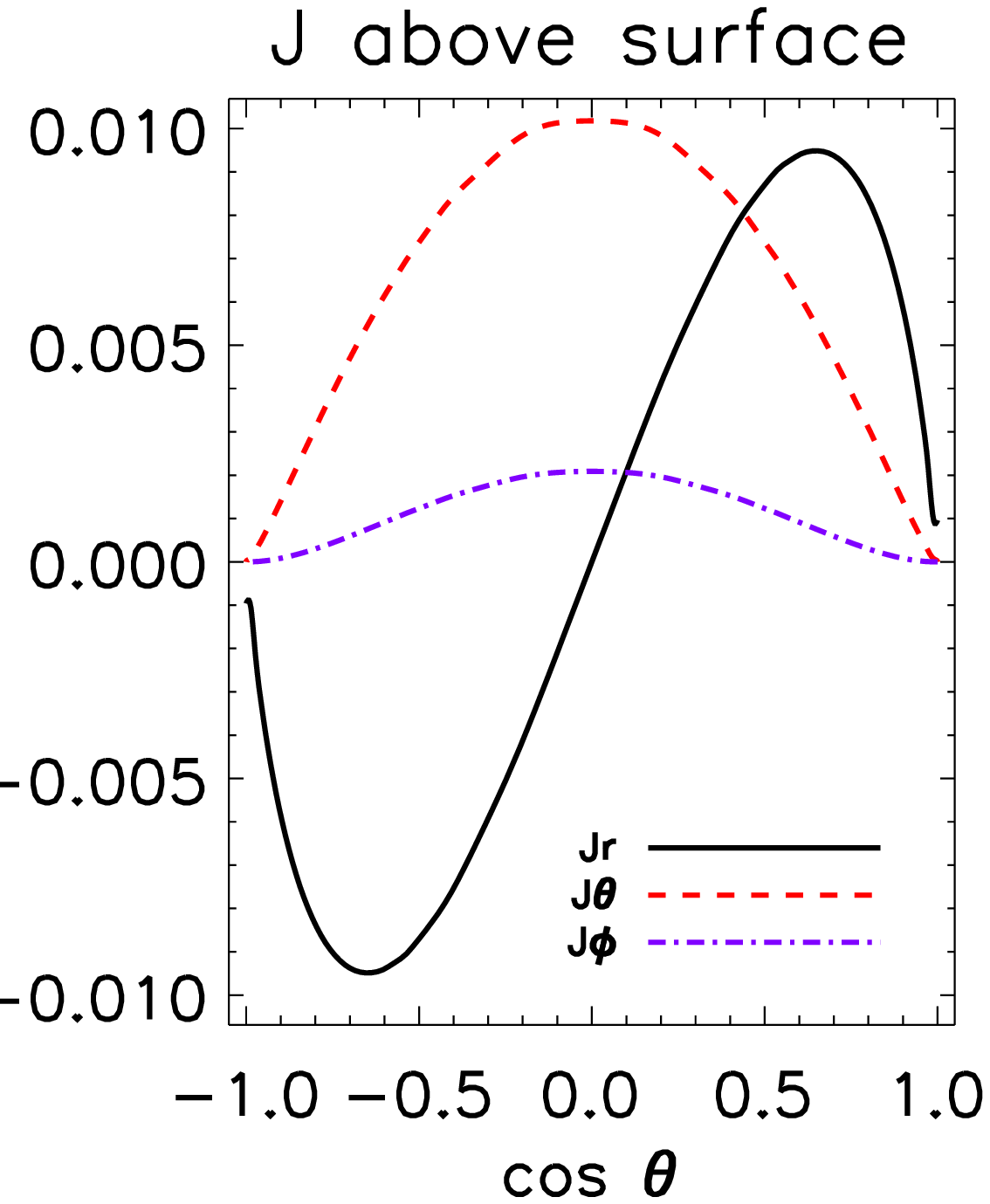}

\includegraphics[width=.24\textwidth]{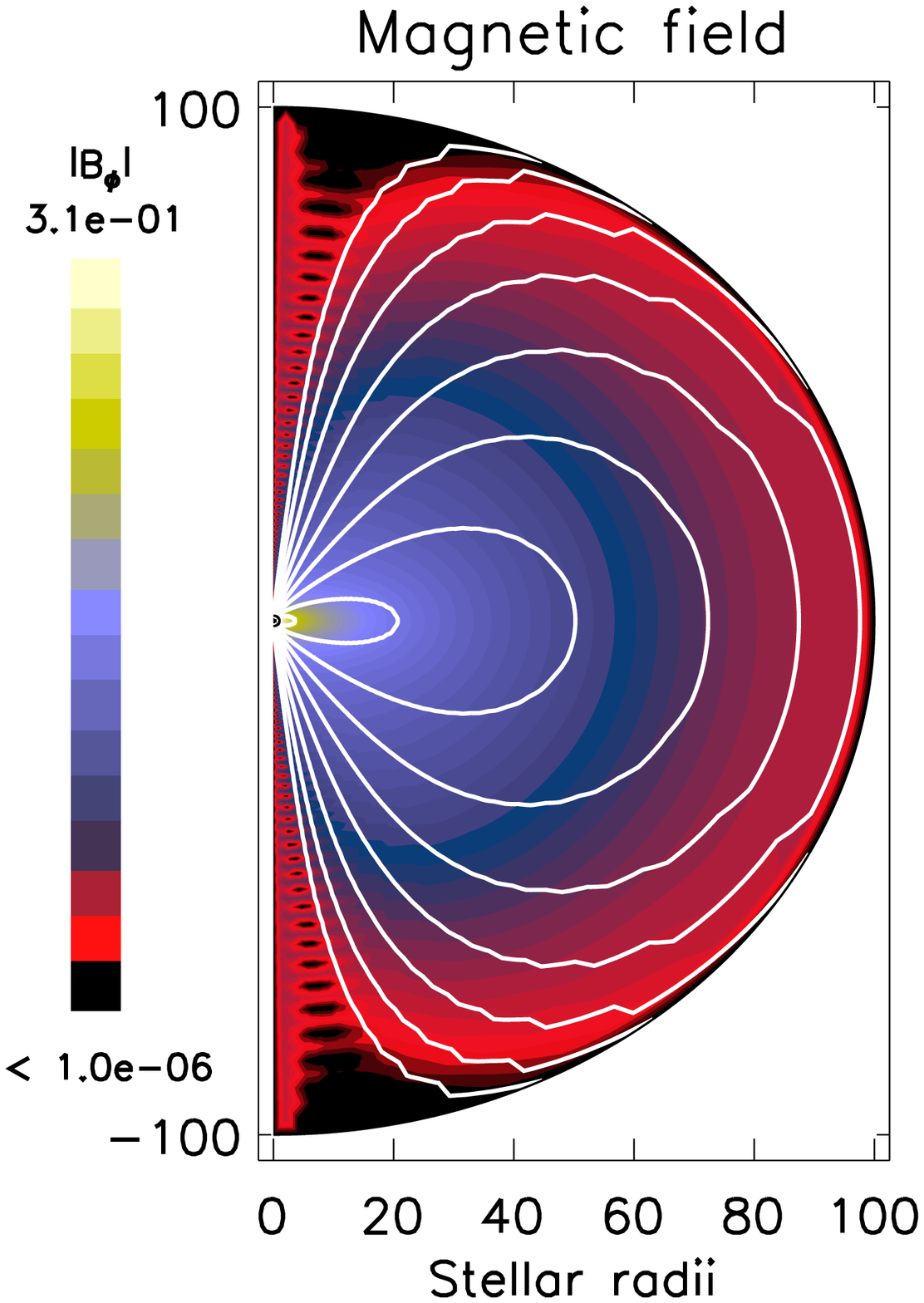}
\includegraphics[width=.24\textwidth]{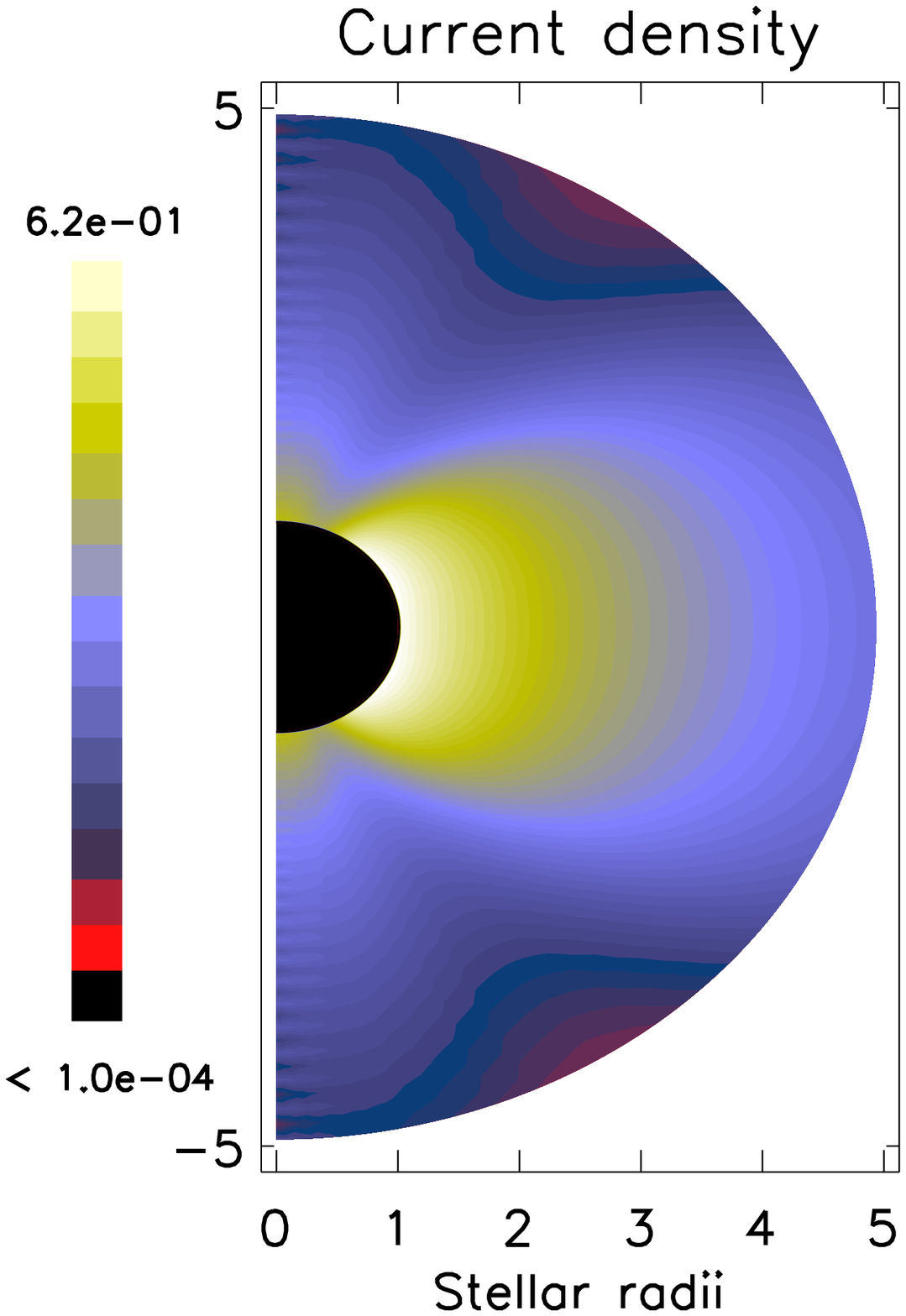}
\includegraphics[width=.24\textwidth]{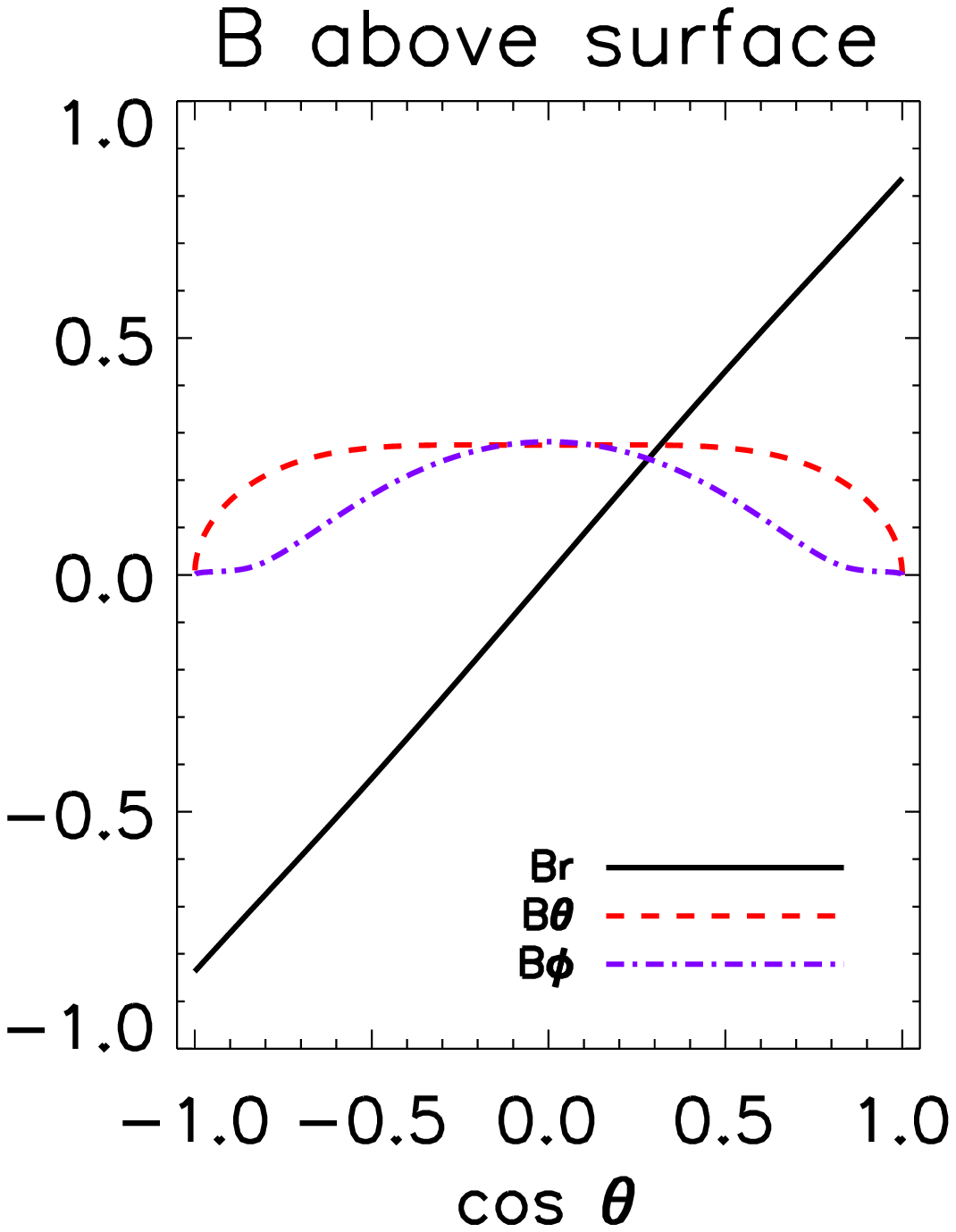}
\includegraphics[width=.24\textwidth]{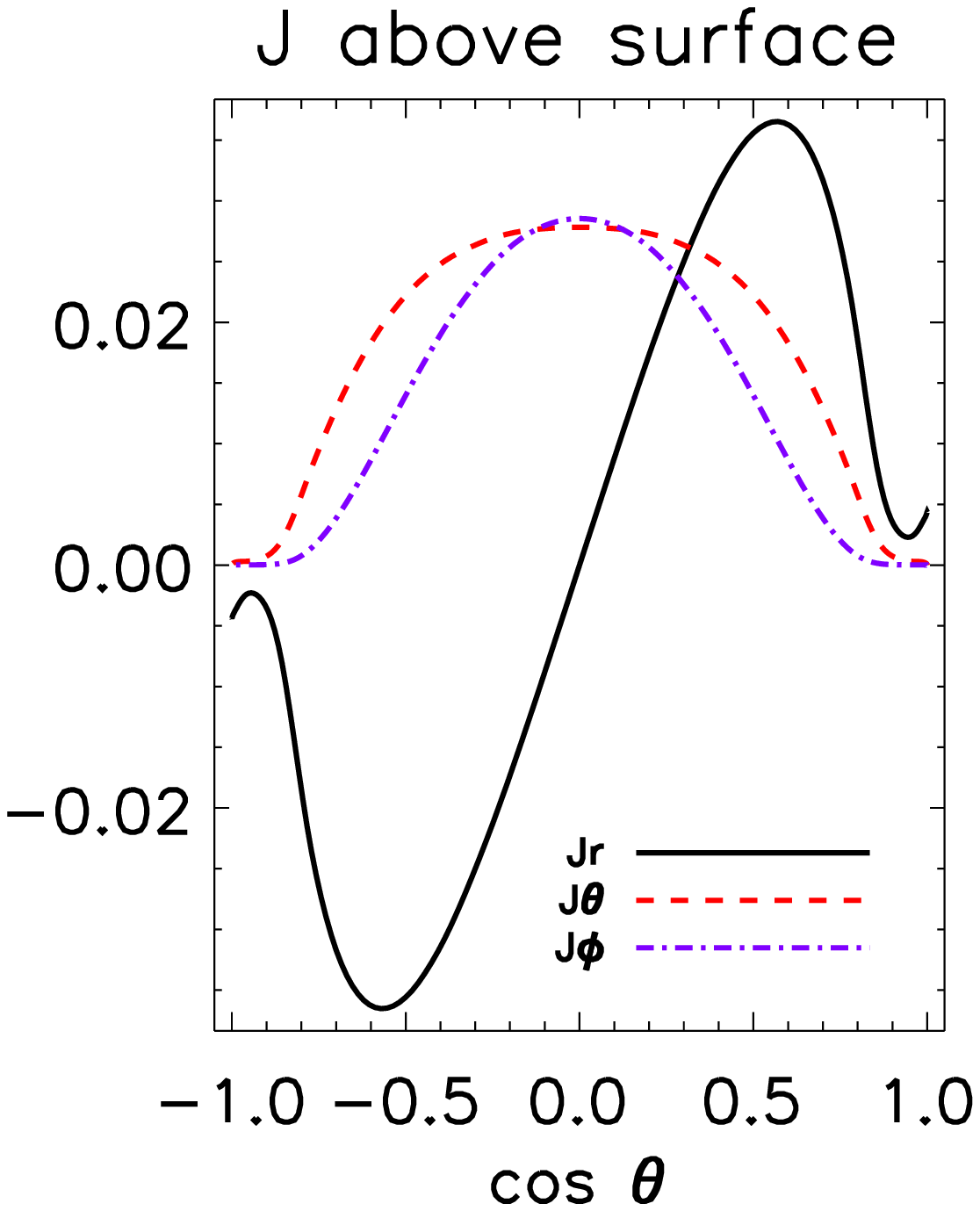}

\caption{Models S2 (self-similar), A, B and C (from top to bottom, respectively). From left to right: poloidal magnetic field lines (white) and strength of the toroidal component (colored logarithmic scale in units of $B_0$); current density distribution $|\vec{J}|$ in the near region $r\le 5~R_\star$ (colored logarithmic scale in units of $cB_0/R_\star$); angular profiles of the magnetic field components and of the current density components, both above the surface.}
\label{fig:model_s2-c}
\end{figure}

\begin{figure}
\centering

\includegraphics[width=.24\textwidth]{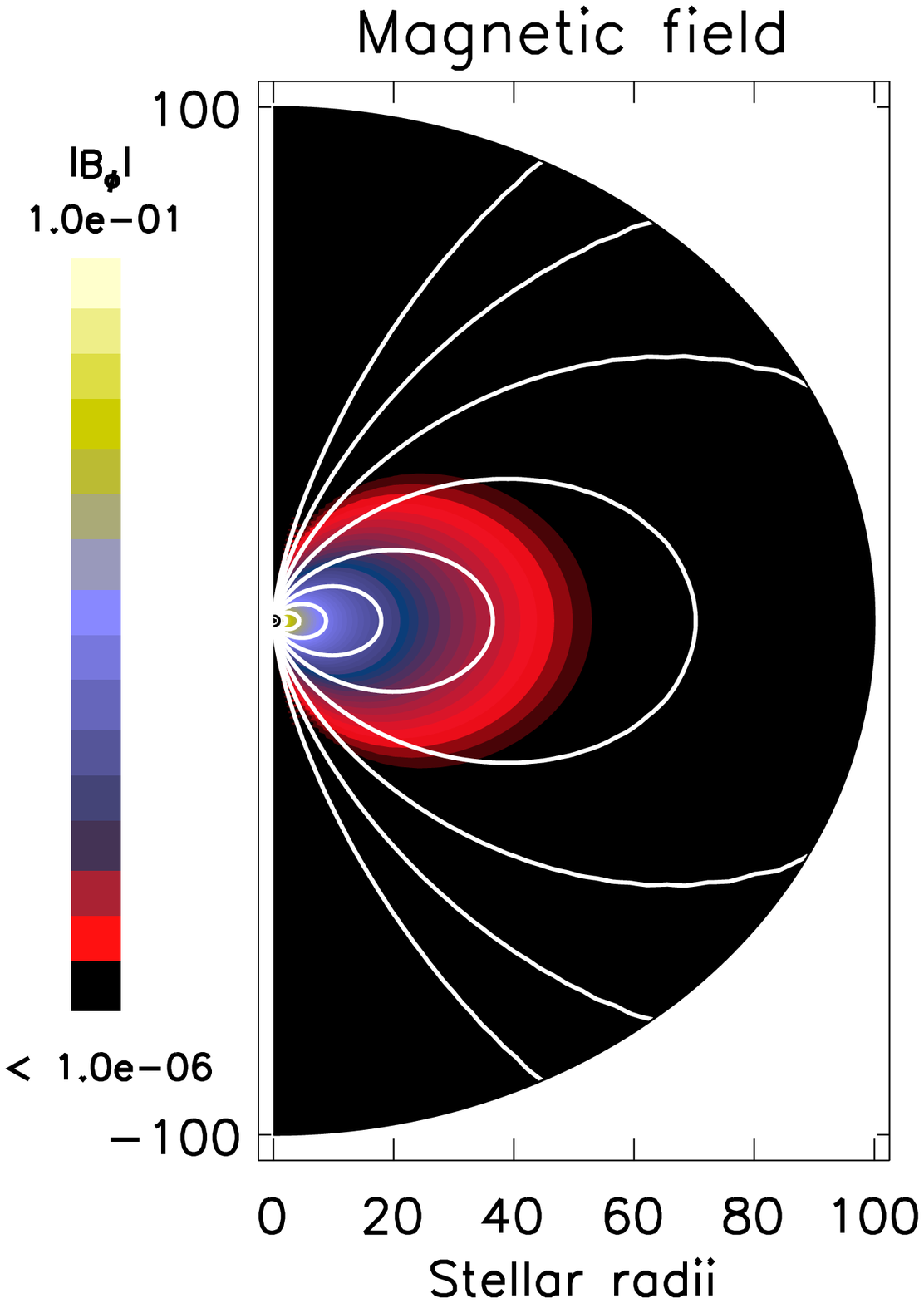}
\includegraphics[width=.24\textwidth]{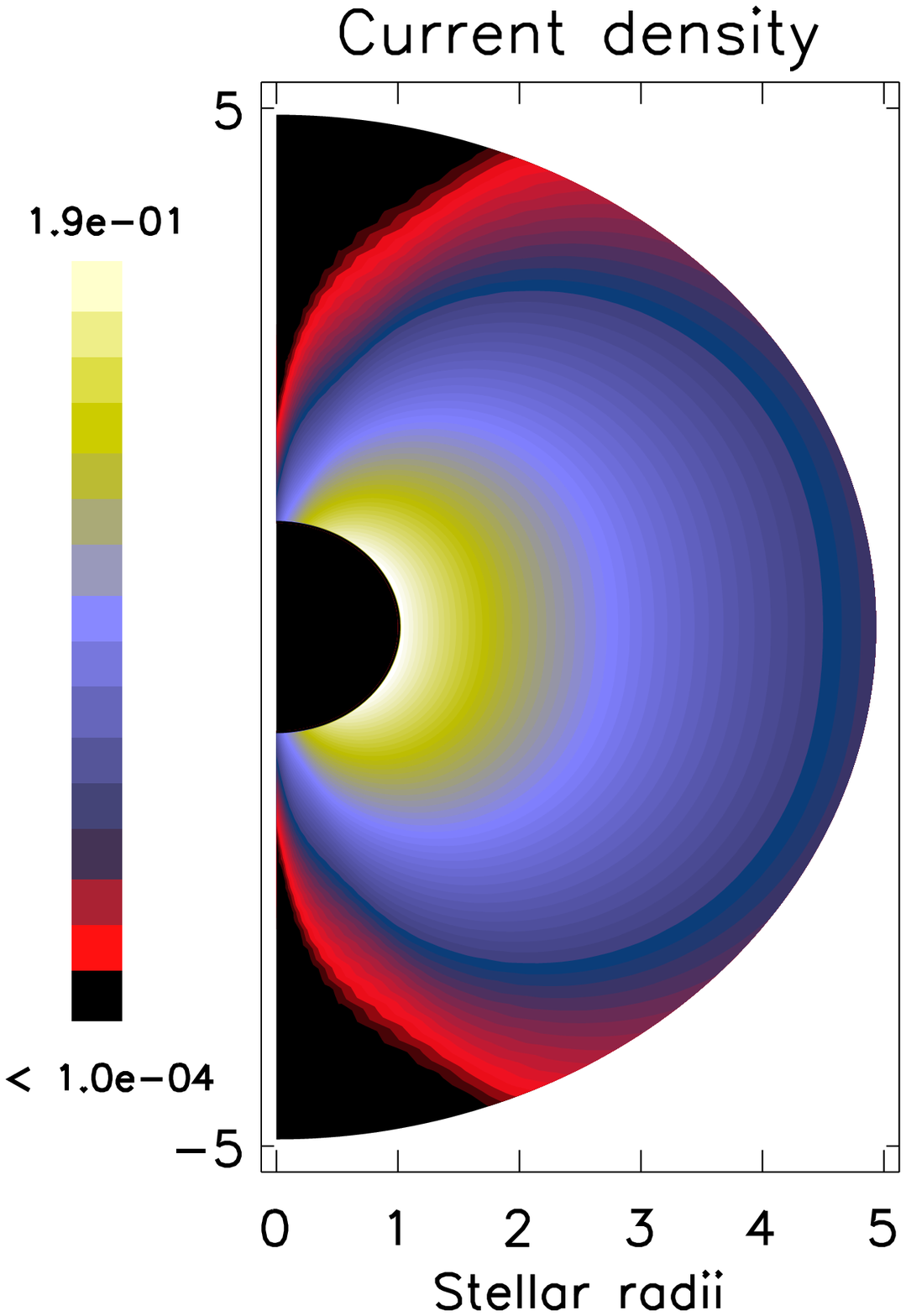}
\includegraphics[width=.24\textwidth]{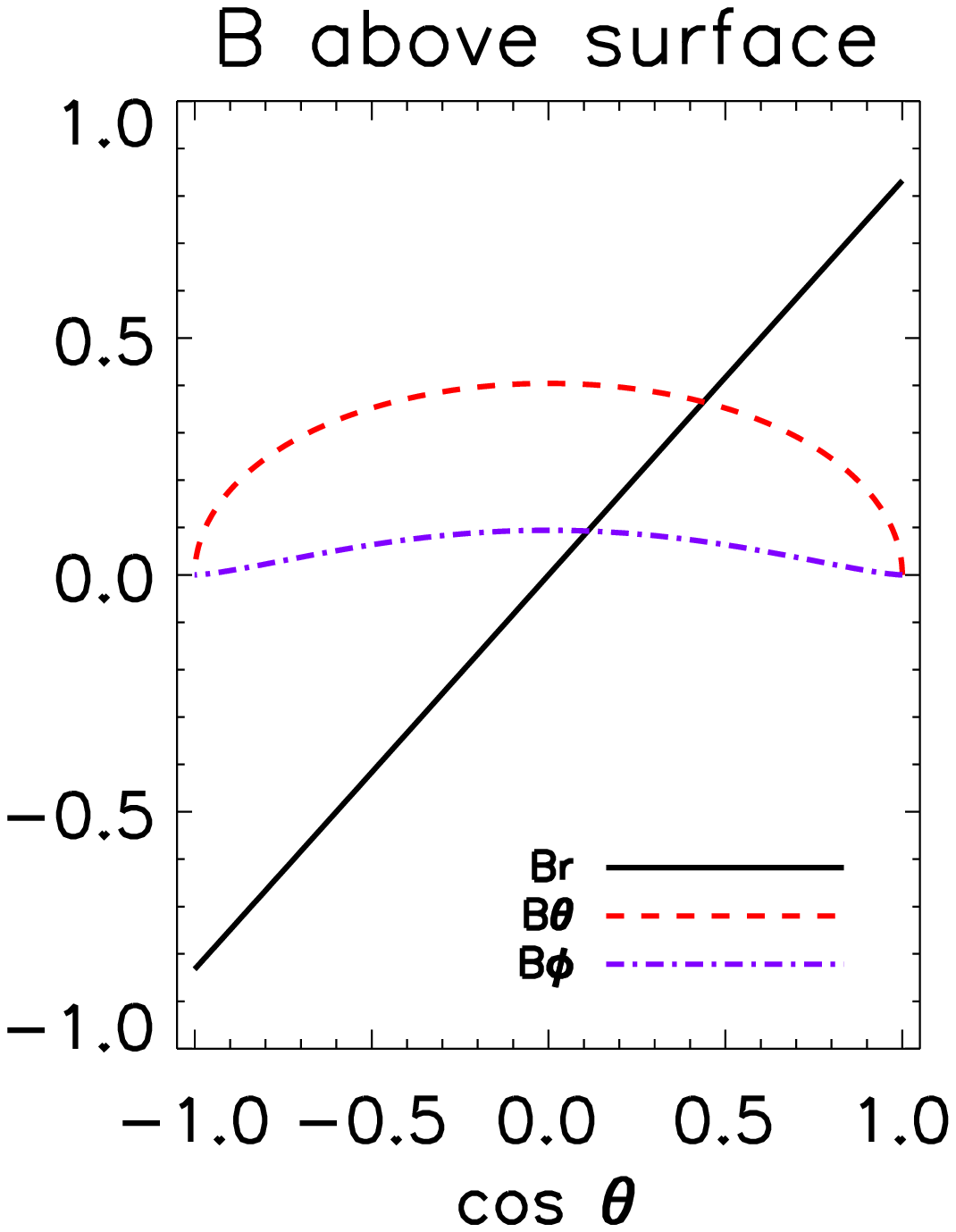}
\includegraphics[width=.24\textwidth]{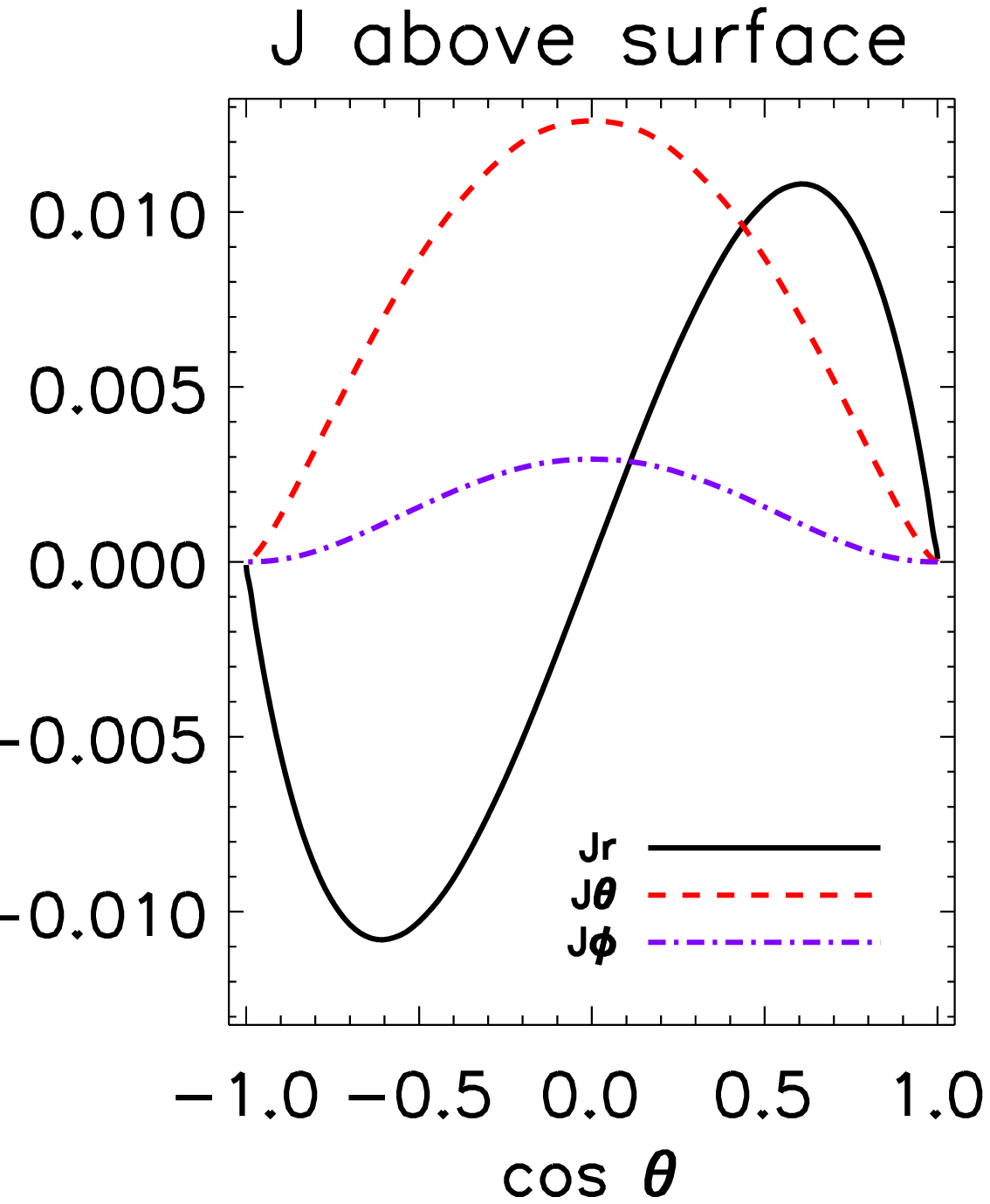}

\includegraphics[width=.24\textwidth]{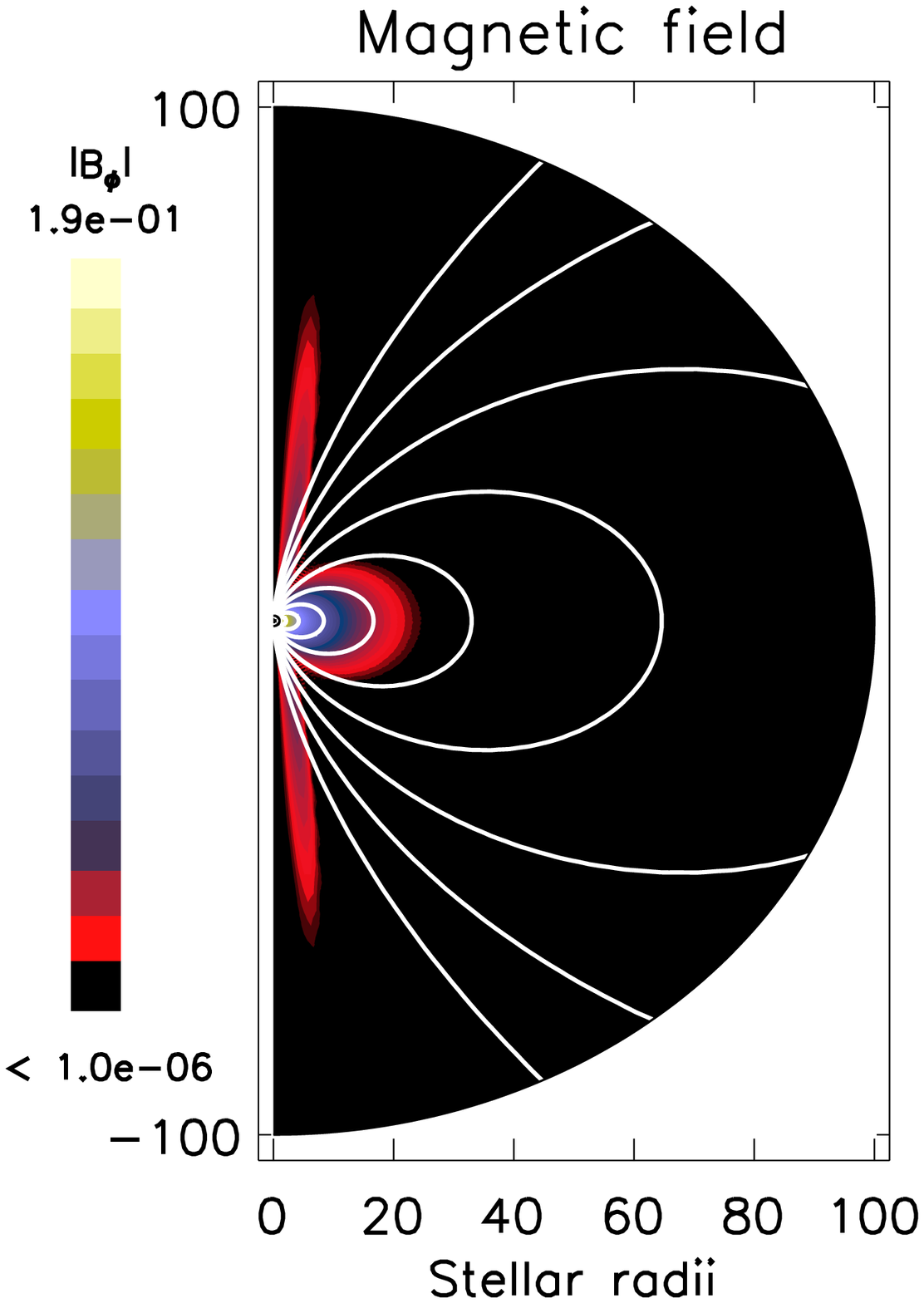}
\includegraphics[width=.24\textwidth]{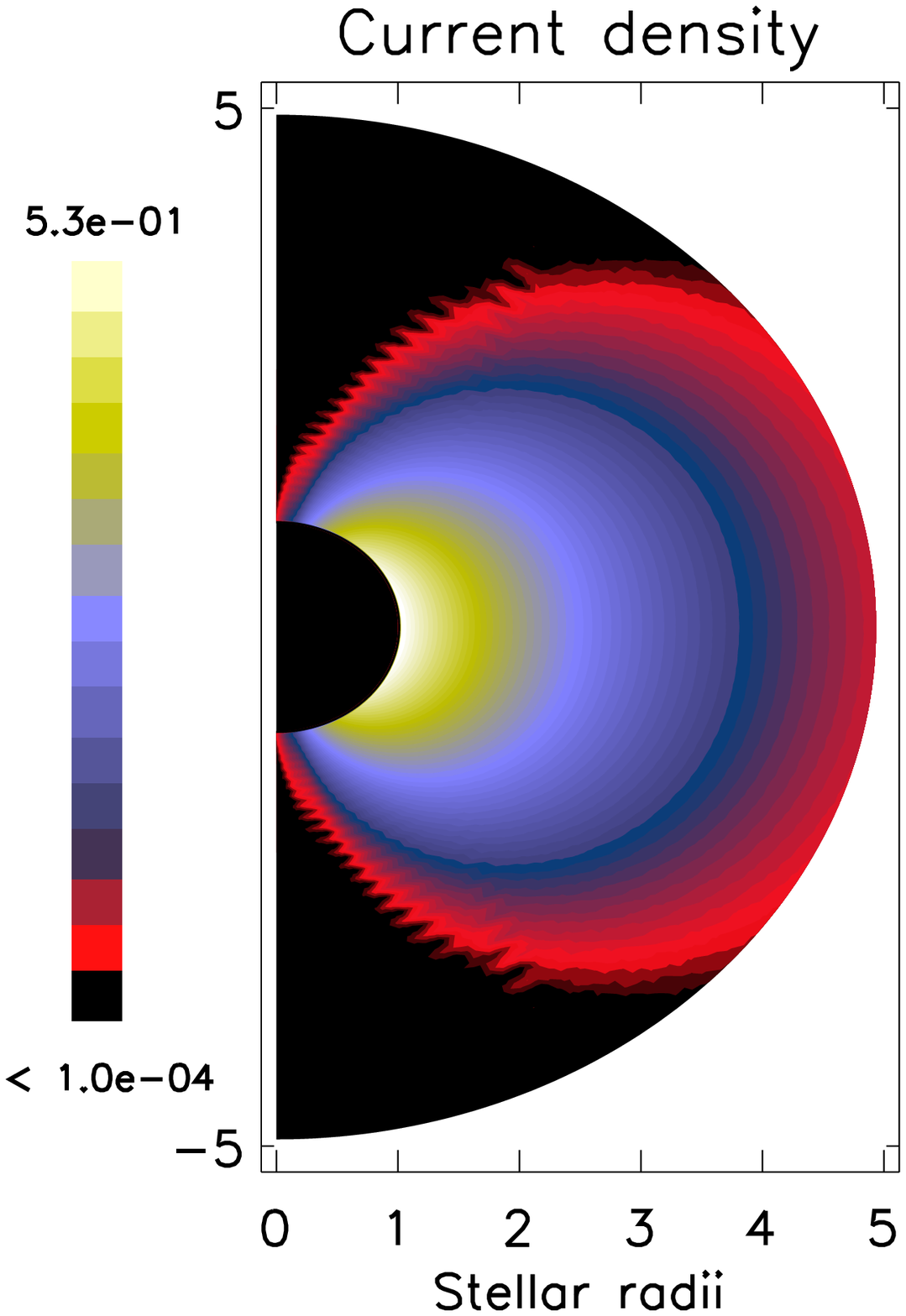}
\includegraphics[width=.24\textwidth]{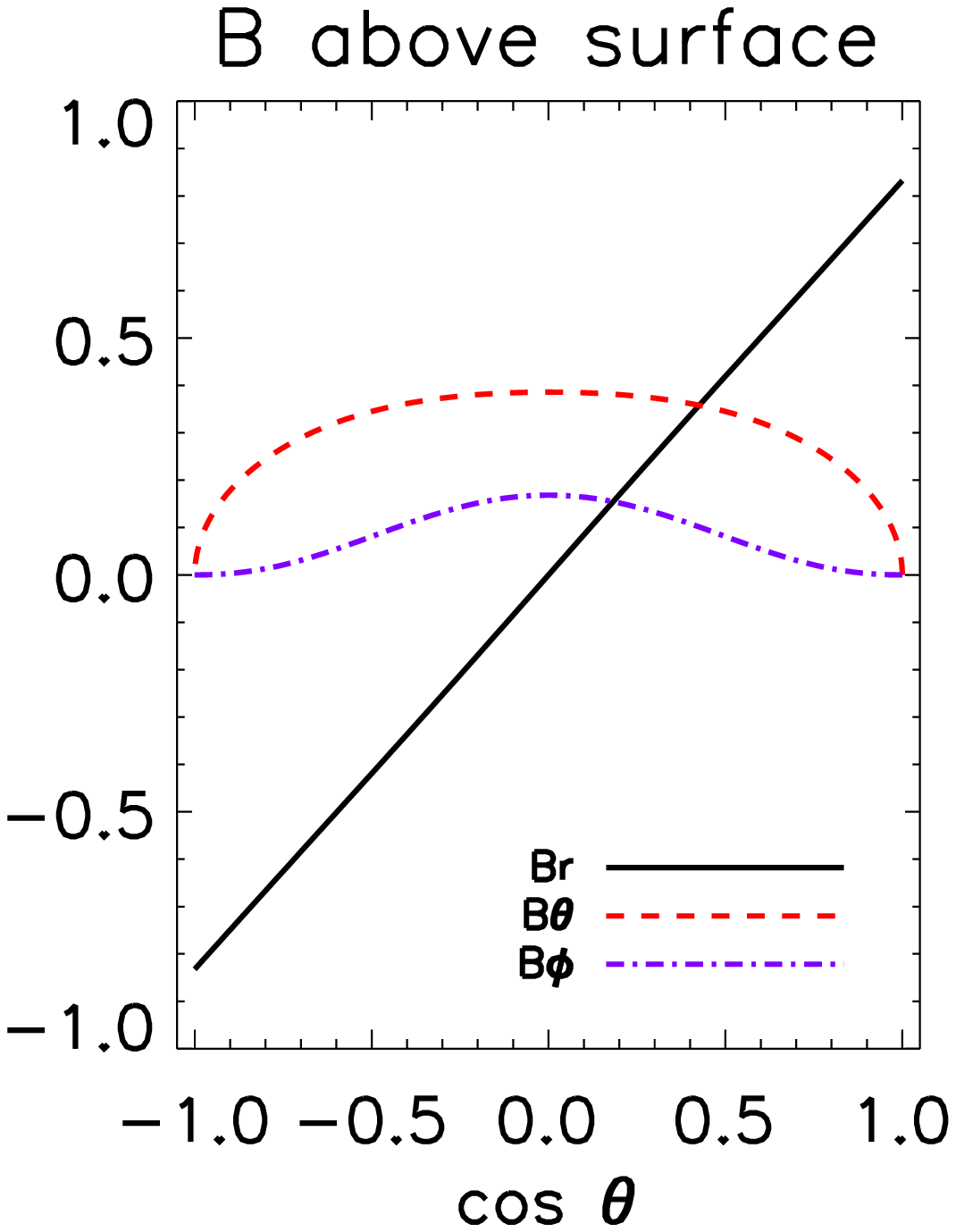}
\includegraphics[width=.24\textwidth]{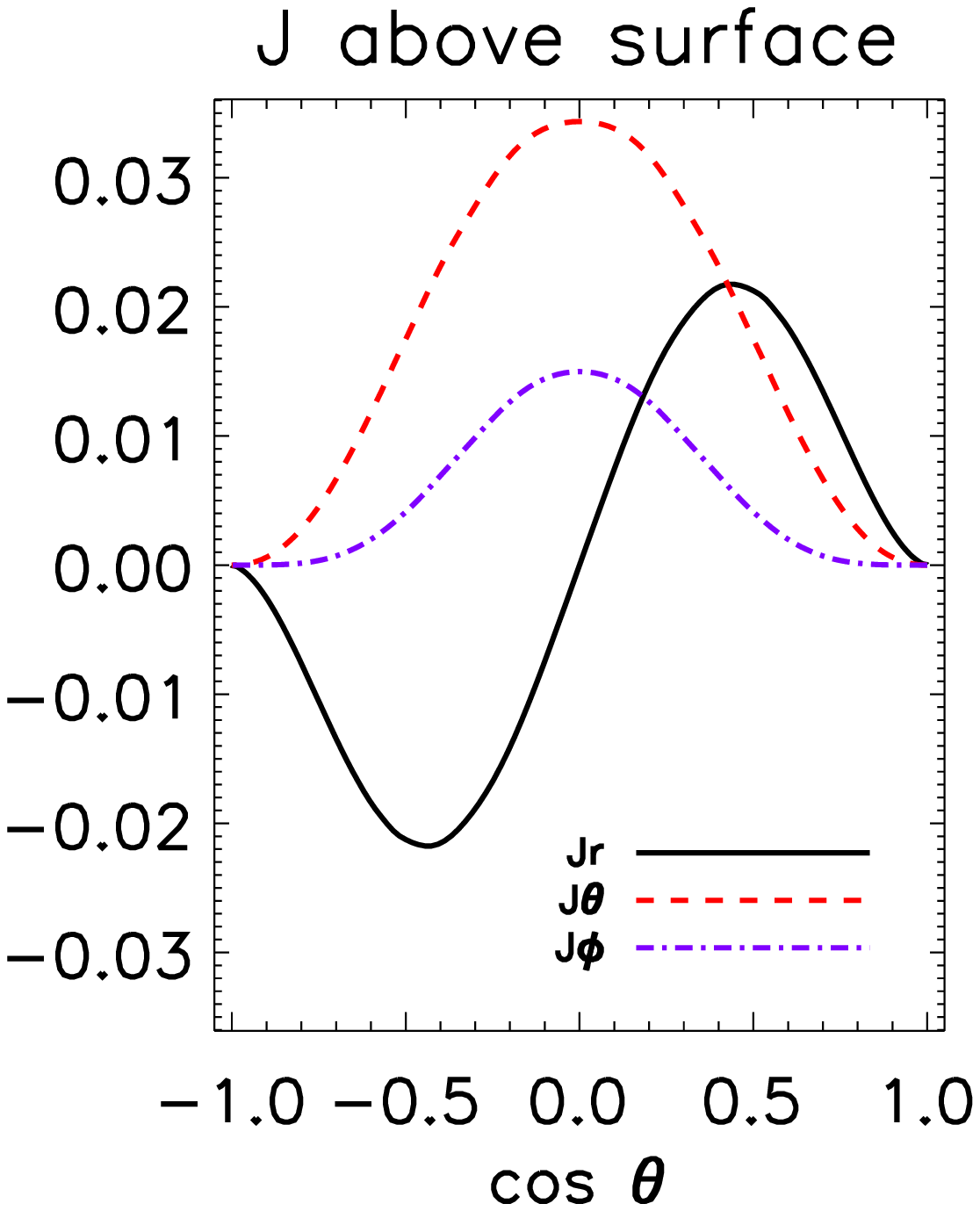}

\includegraphics[width=.24\textwidth]{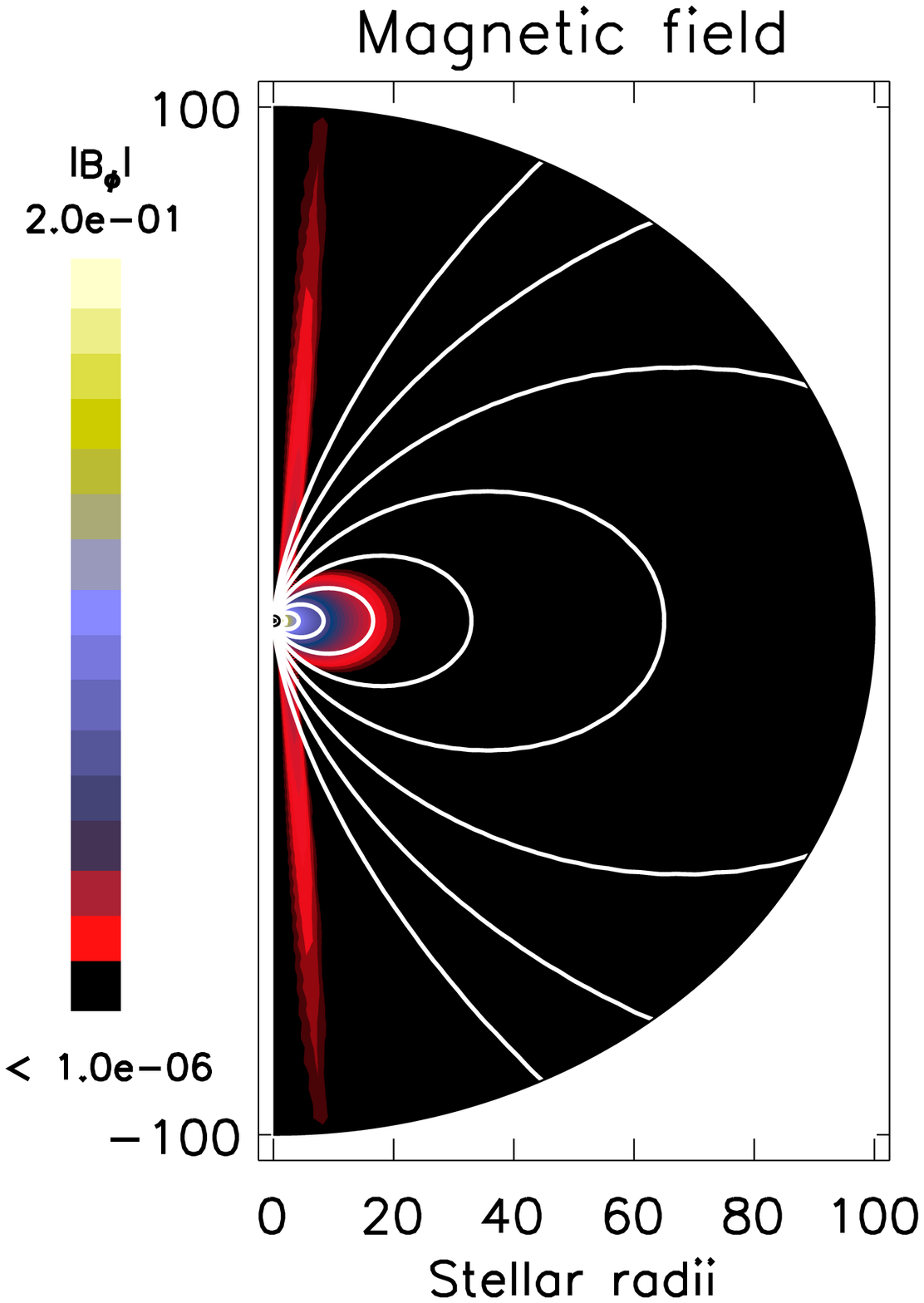}
\includegraphics[width=.24\textwidth]{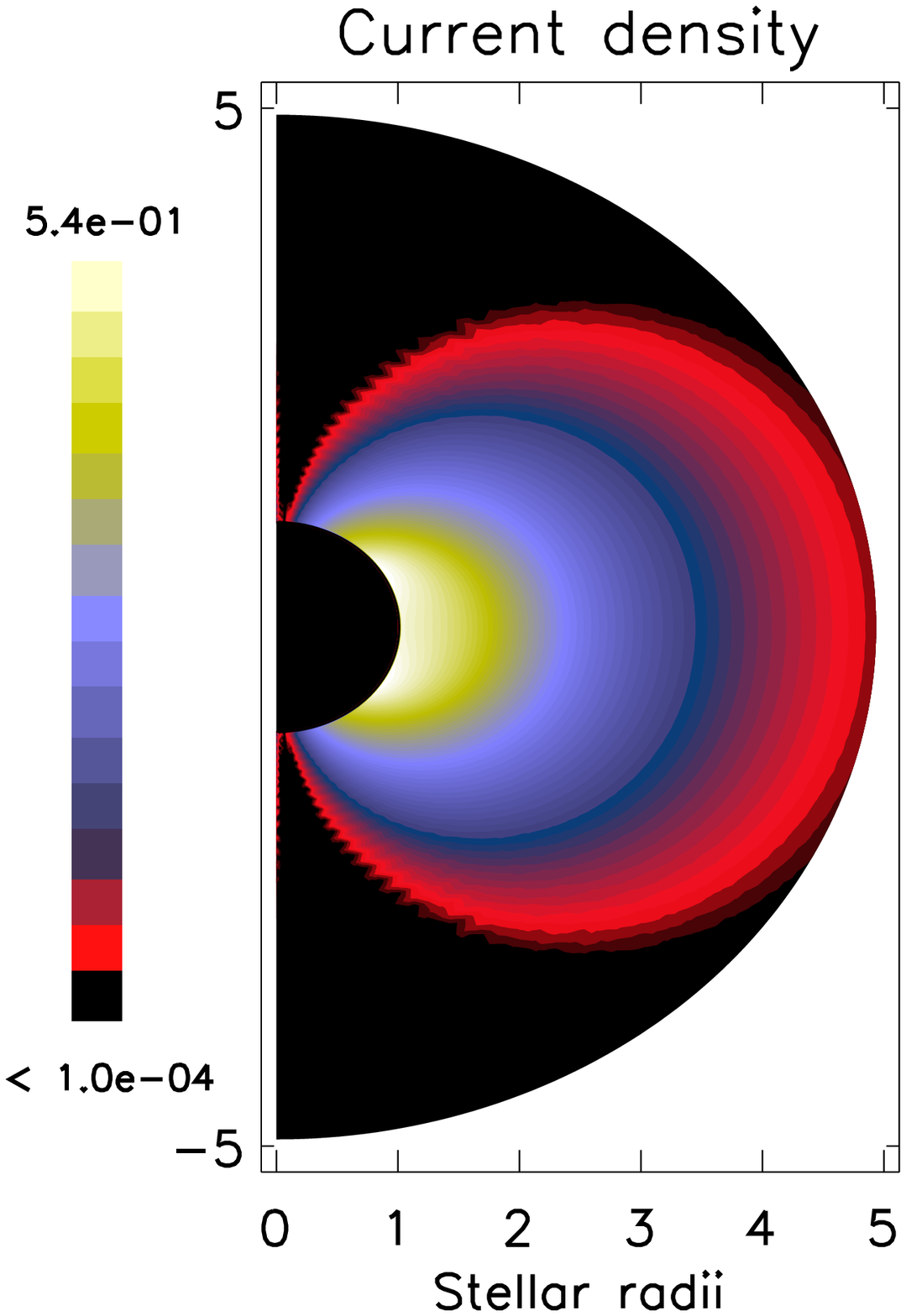}
\includegraphics[width=.24\textwidth]{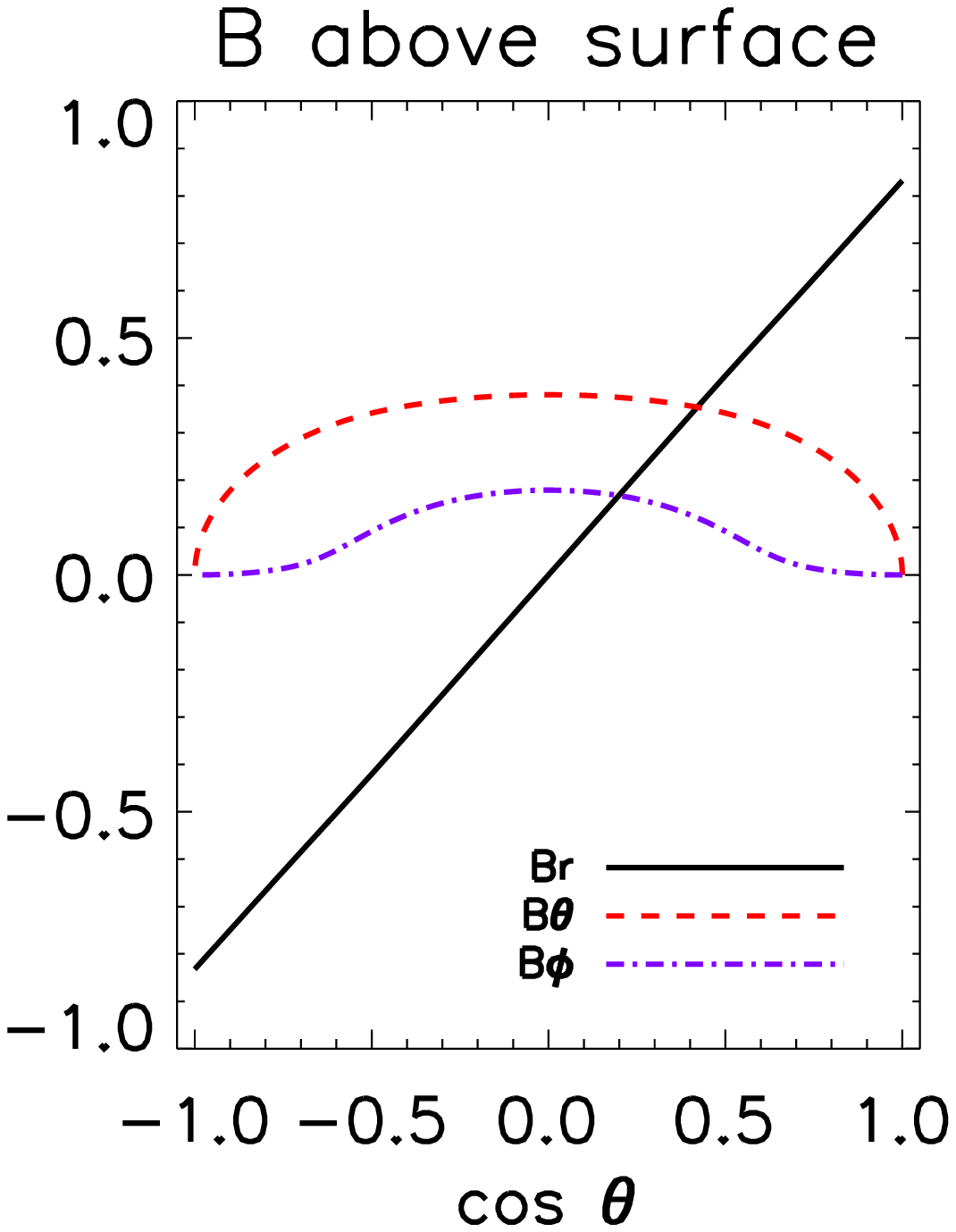}
\includegraphics[width=.24\textwidth]{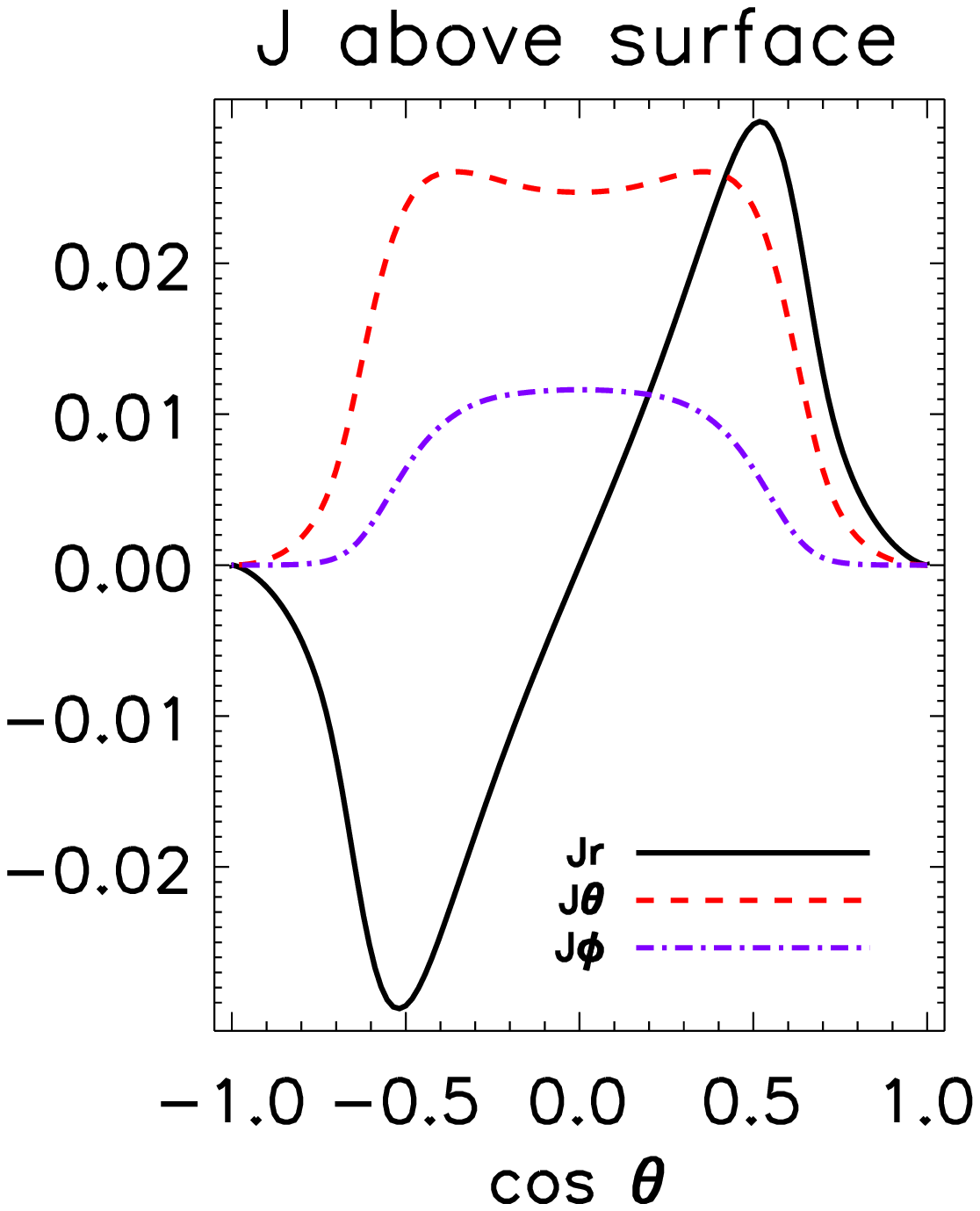}

\includegraphics[width=.24\textwidth]{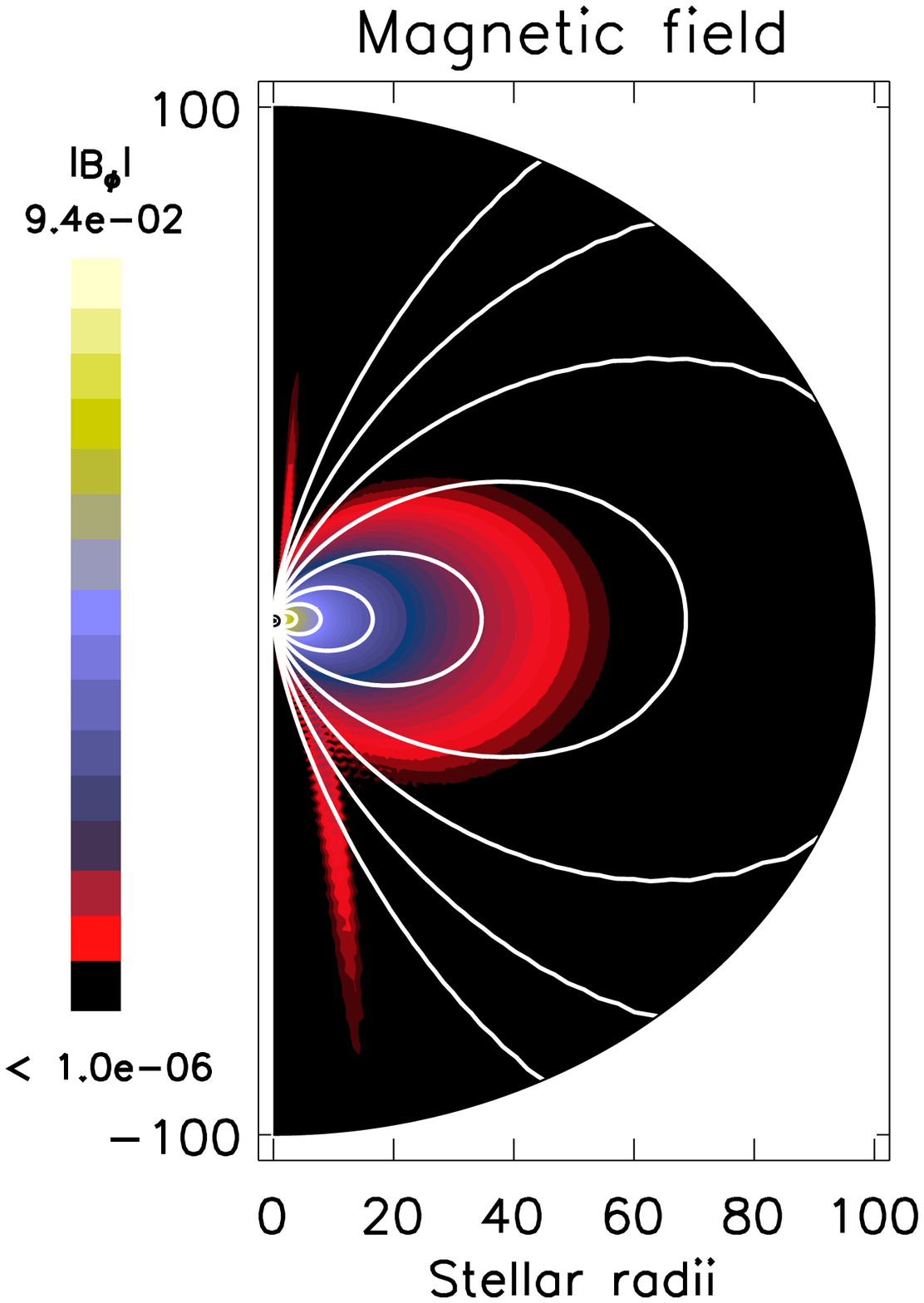}
\includegraphics[width=.24\textwidth]{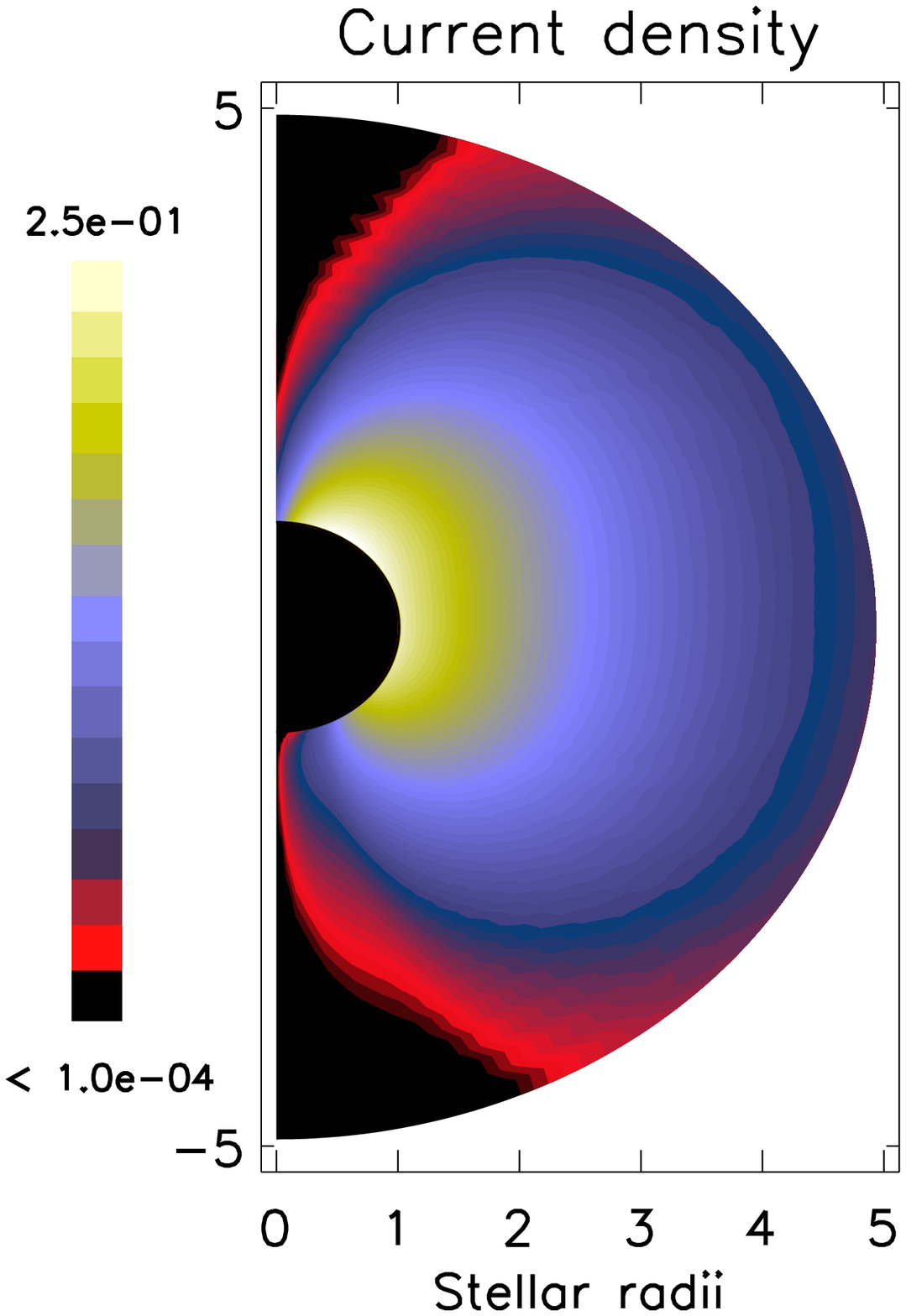}
\includegraphics[width=.24\textwidth]{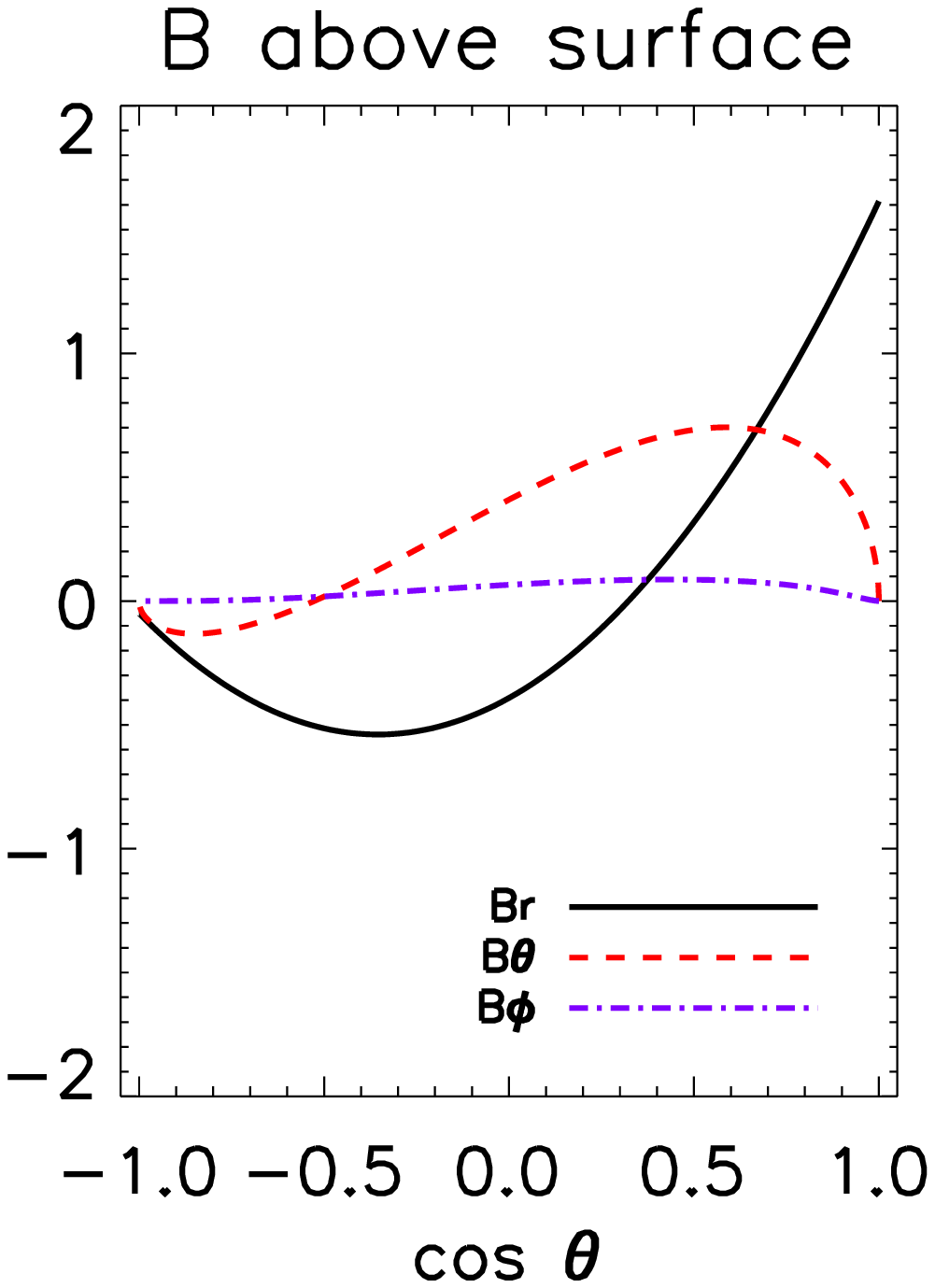}
\includegraphics[width=.24\textwidth]{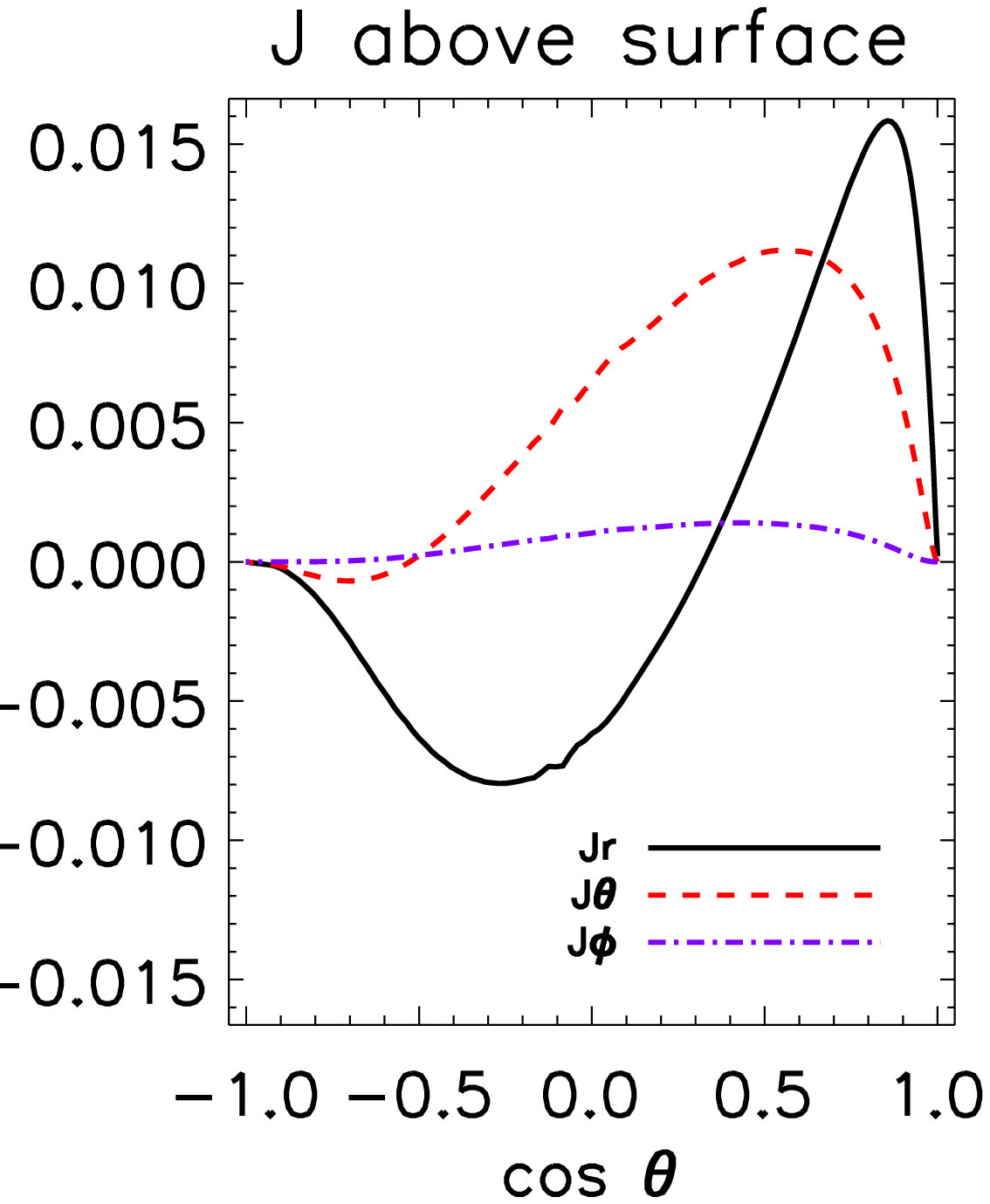}

\caption{Same as Fig. \ref{fig:model_s2-c} for numerical models D, E, F and G (from top to bottom). } 
\label{fig:models_d-g}
\end{figure}

\begin{figure}[t]
\centering

\includegraphics[width=.24\textwidth]{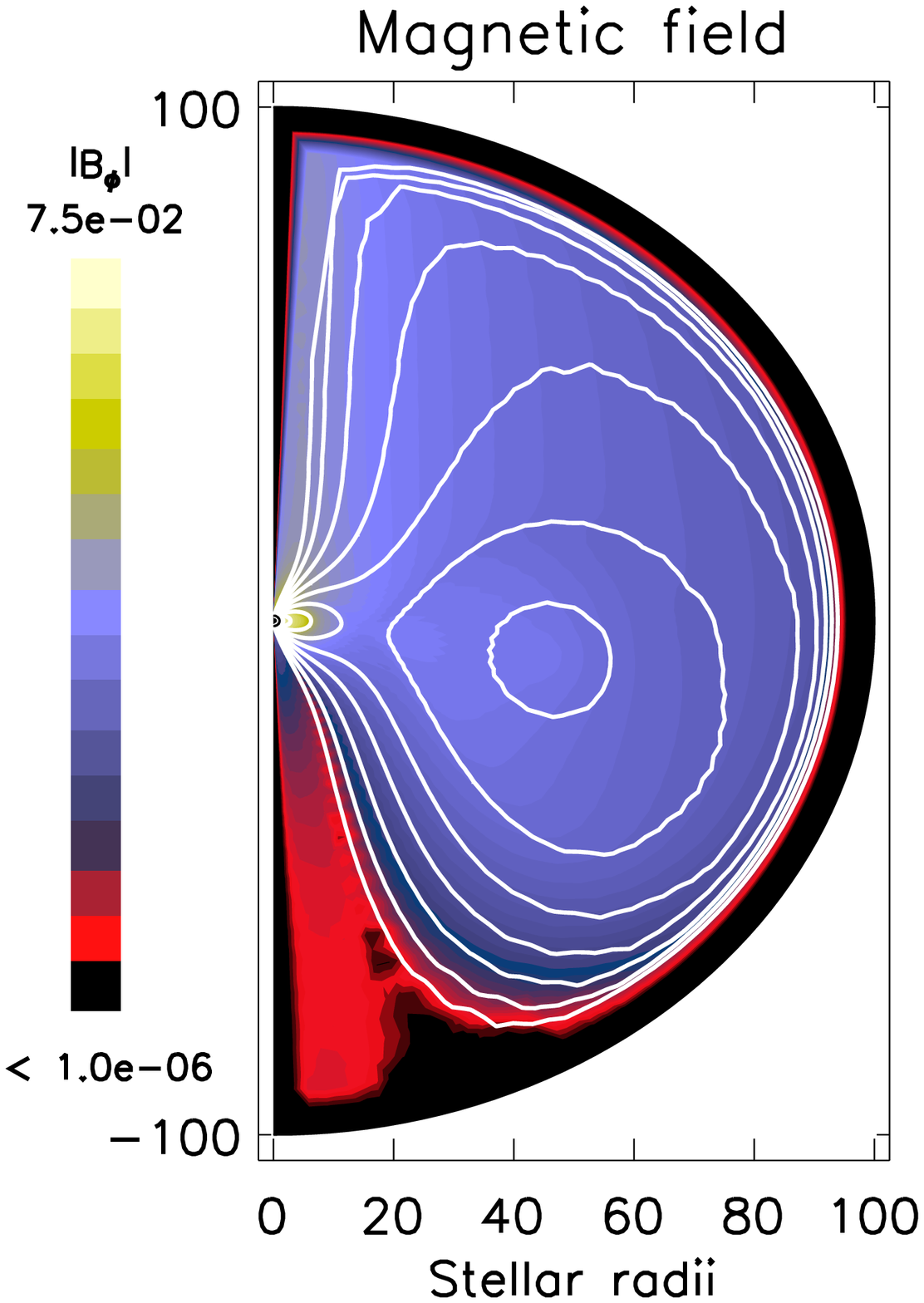}
\includegraphics[width=.24\textwidth]{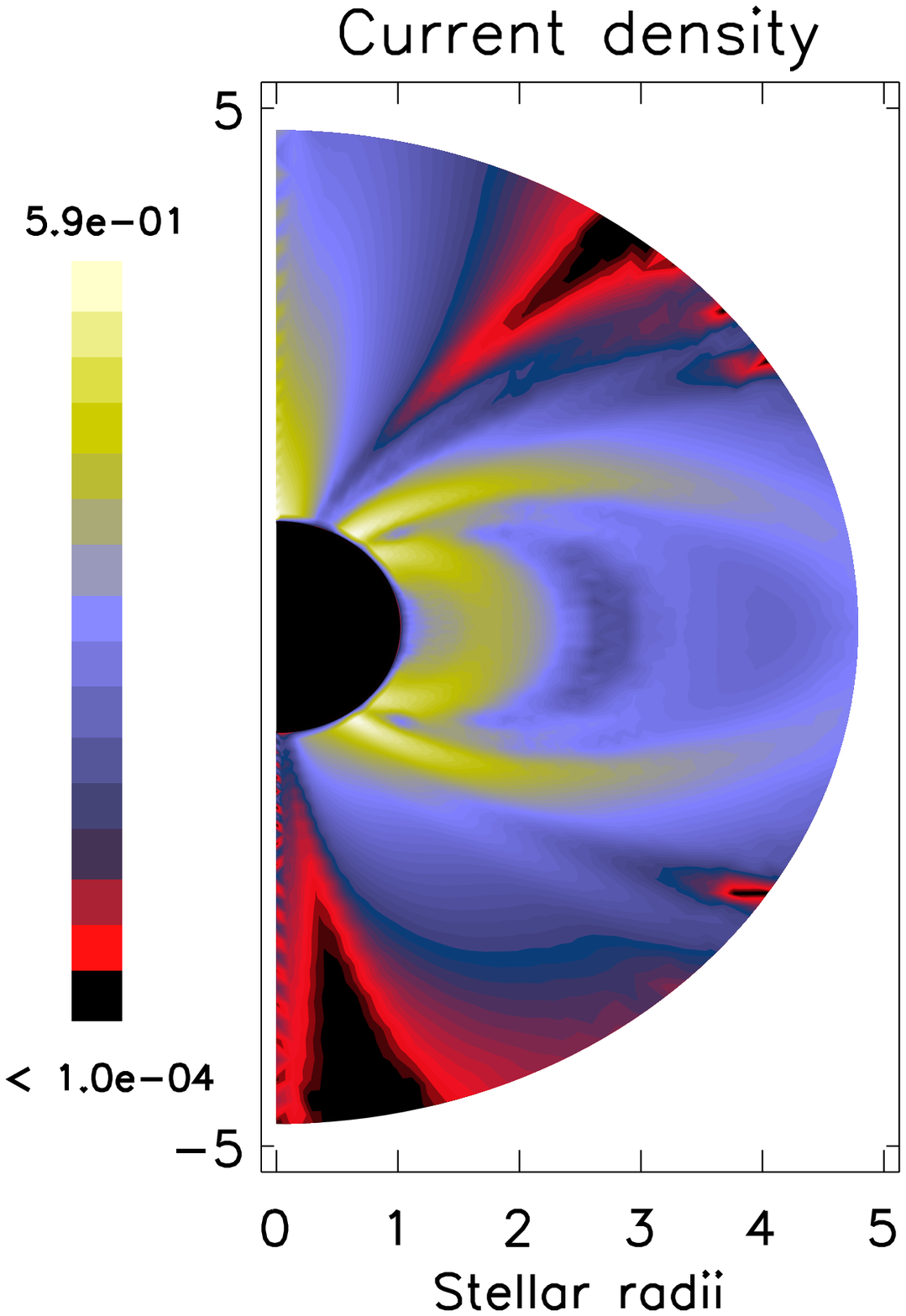}
\includegraphics[width=.24\textwidth]{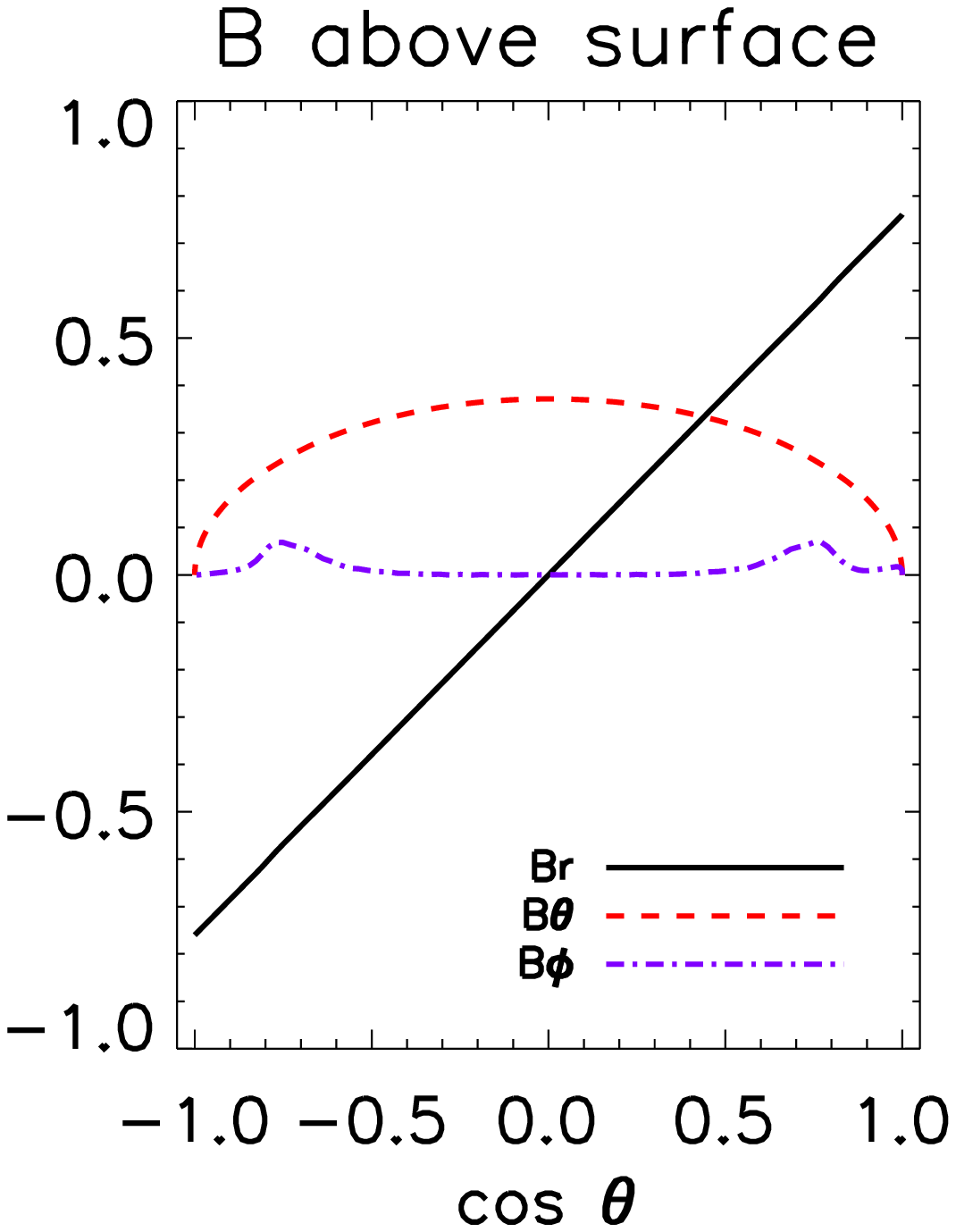}
\includegraphics[width=.24\textwidth]{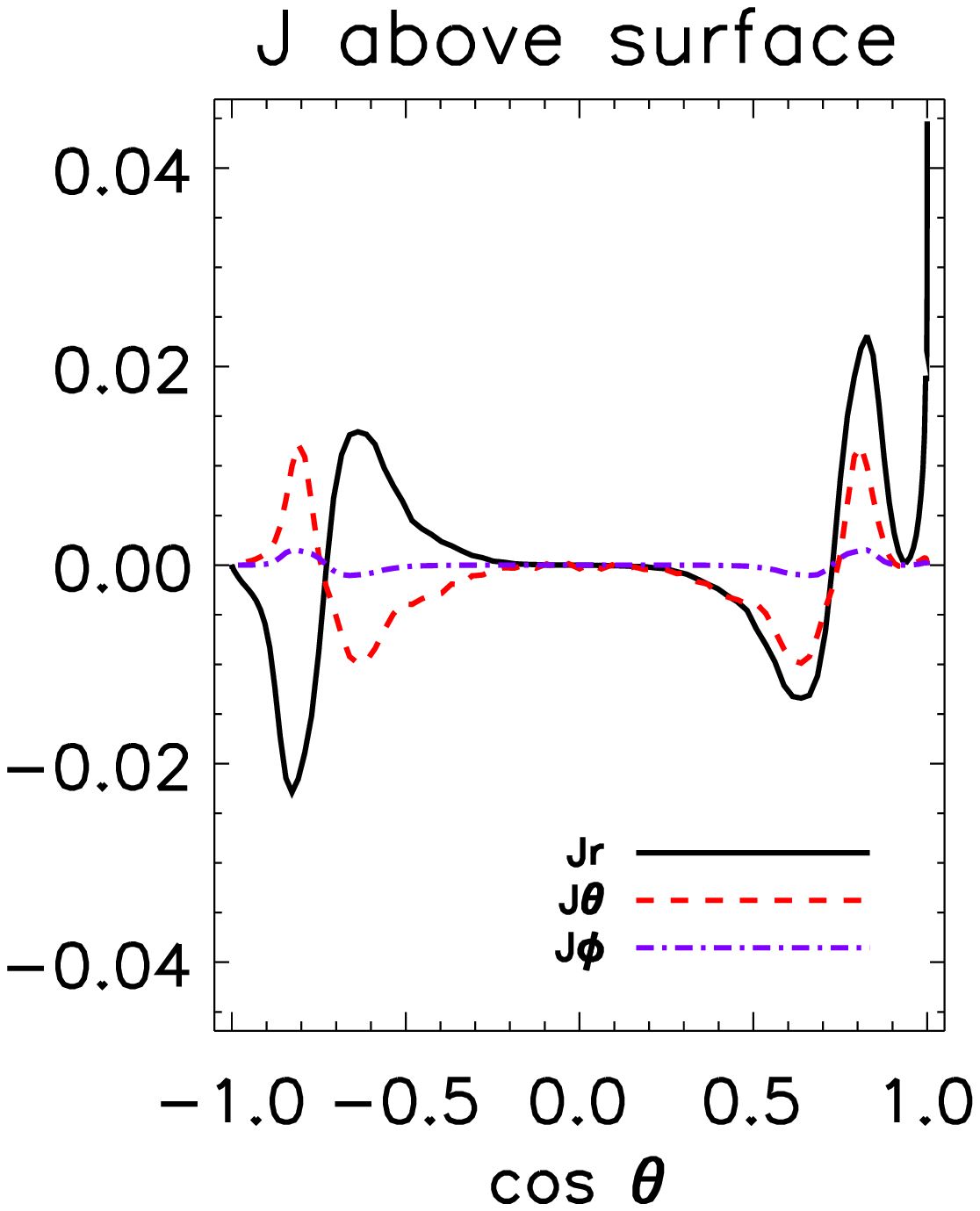}

\includegraphics[width=.24\textwidth]{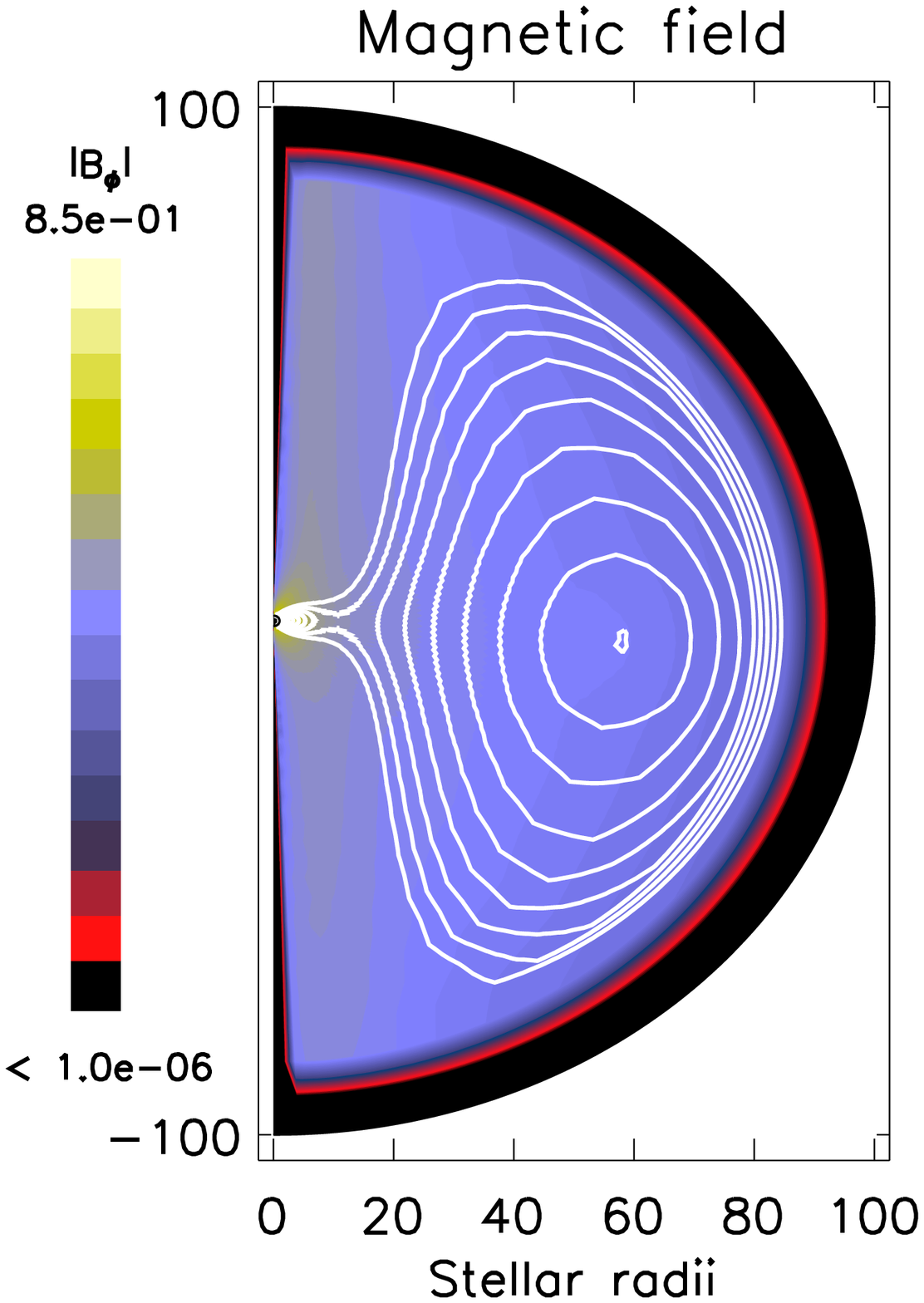}
\includegraphics[width=.24\textwidth]{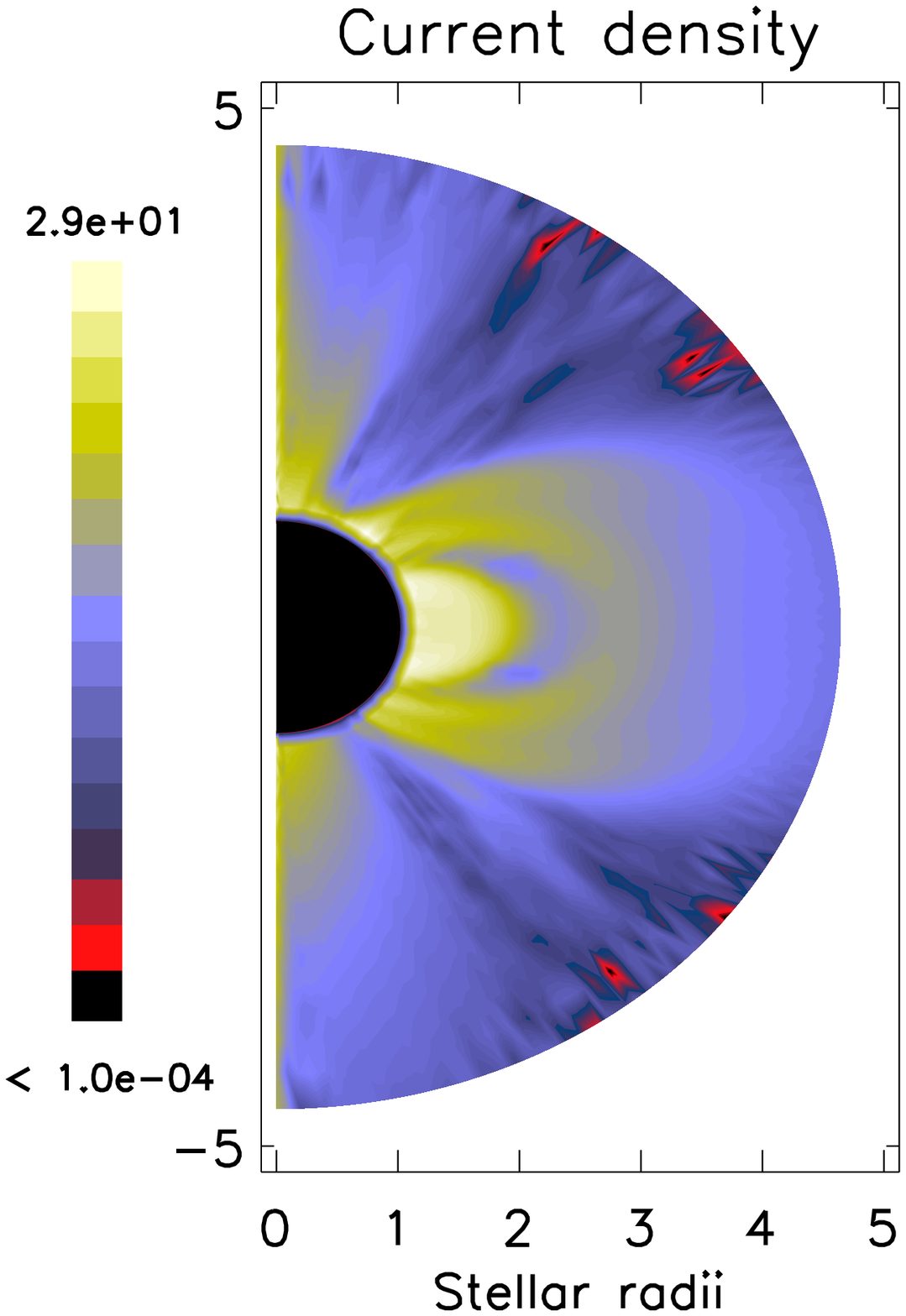}
\includegraphics[width=.24\textwidth]{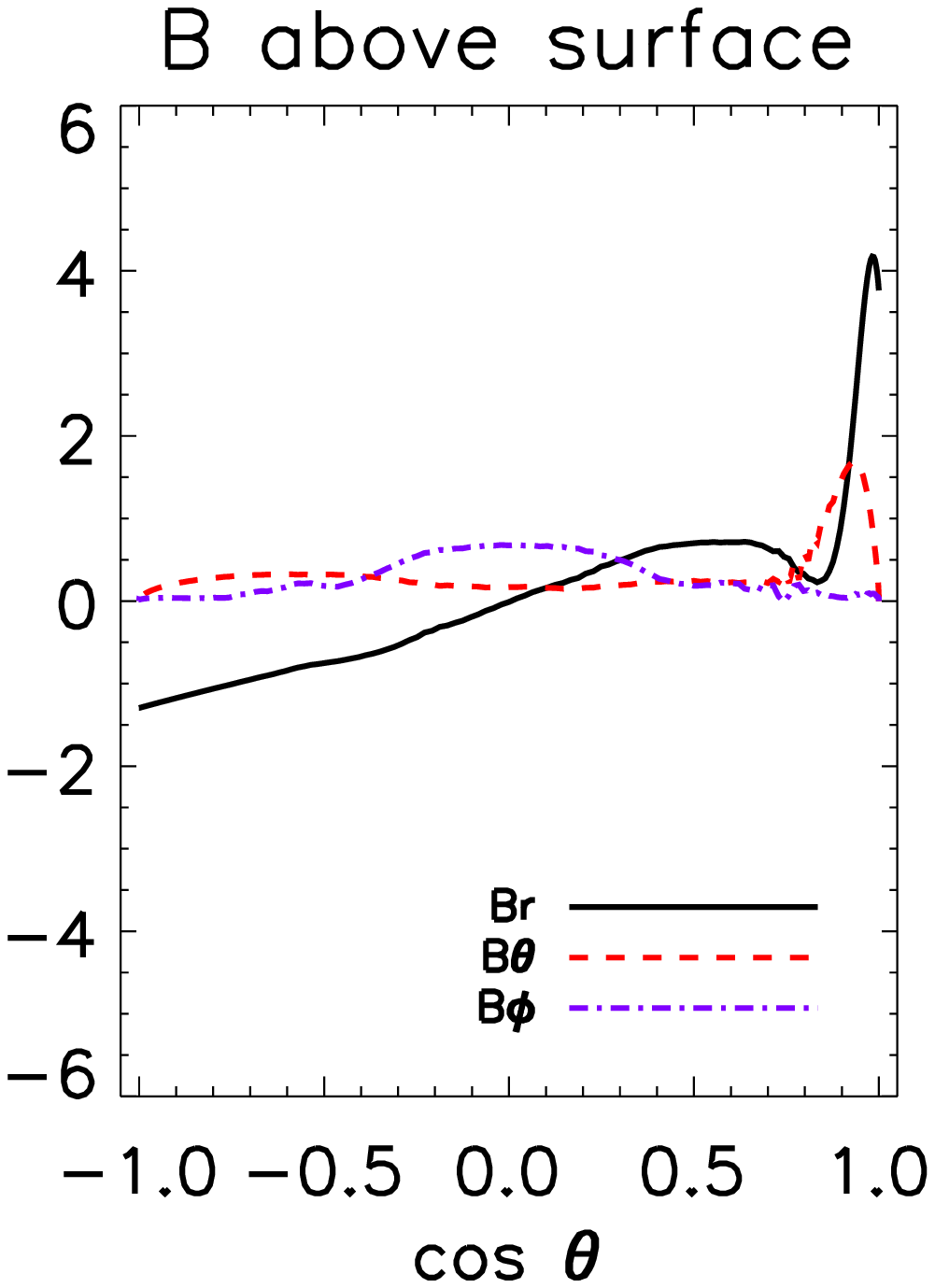}
\includegraphics[width=.24\textwidth]{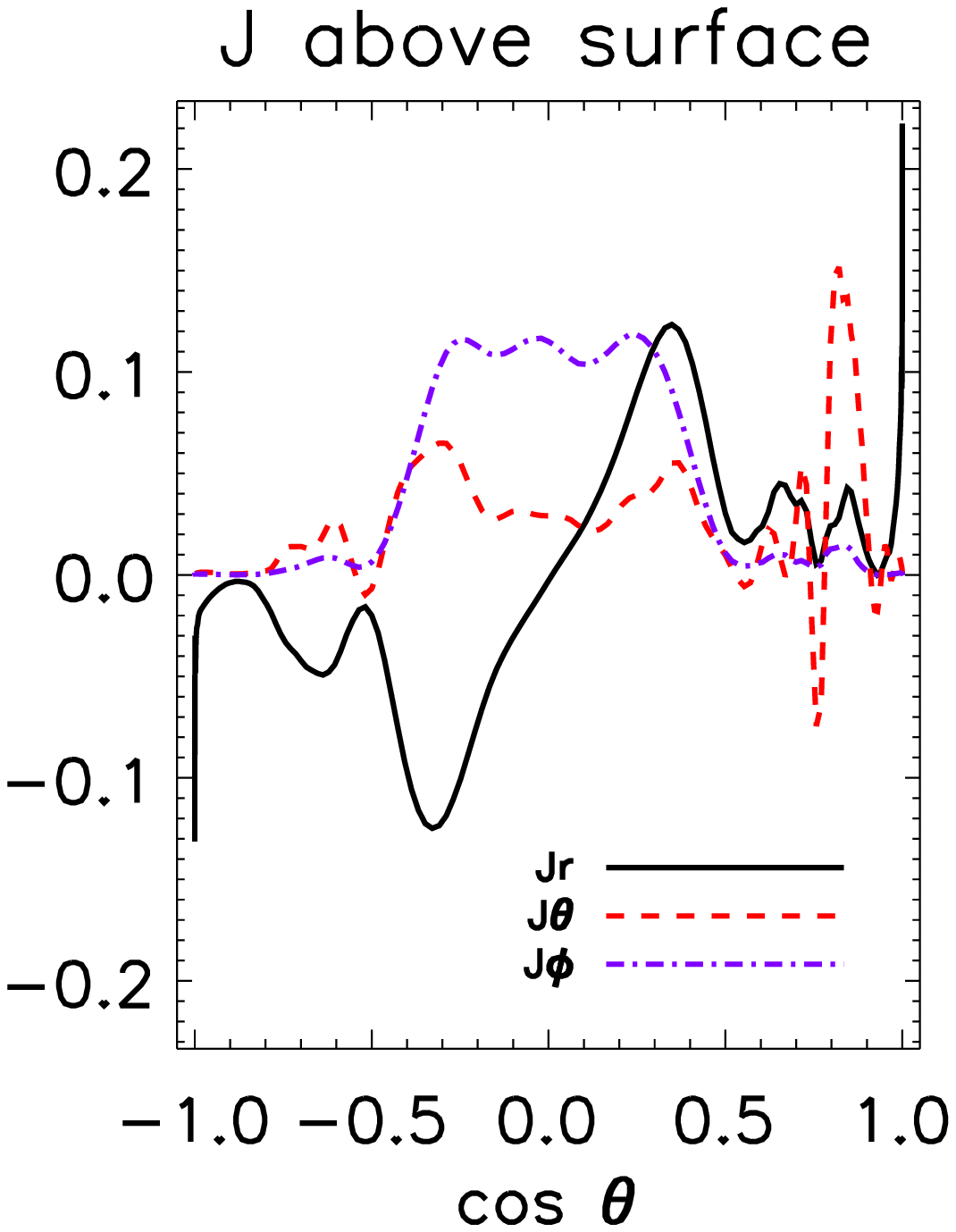}

\caption{Same as Fig. \ref{fig:model_s2-c} for numerical models H (top) and J (bottom). } 
\label{fig:models_h-j}
\end{figure}

The final geometry of the magnetic field and currents for all these models is shown in Figs. \ref{fig:model_s2-c} to \ref{fig:models_h-j}. For $k_{tor}\lesssim 0.1$ (models A and B), the initial poloidal magnetic field remains almost unaltered, and the behavior of the solutions is nearly linear. Toroidal field strength, helicity, current density $J$, enclosed current function $I(\Gamma)$, and global twist scale linearly with $k_{tor}$. In contrast, for $k_{tor}\gtrsim 0.1$, the high initial helicity results in a larger twist angle, up to several radians, which in turn corresponds to a highly deformed poloidal magnetic field. A direct comparison of models A, B and C, which differ from each other only in the strength of the initial toroidal component, illustrates this effect: models A and B have the same shape with just a different scale factor, but model C is qualitatively different. 

The general features in the low $k_{tor}$ models do not differ much from self-similar solutions, because the initial conditions were close to a slightly twisted dipole. In self-similar models, two characteristic features are the absence of radial currents on the axis and a higher concentration of currents around the equatorial plane. Conversely, in our numerical models, currents are more spread over the angular direction, and we allow for the existence of radial current on the axis. As a consequence, comparing numerical solutions A and B with self-similar models with comparable helicity, the former reach lower maximum values of current density with a higher global twist. We also note that in the most extreme case (model C) the angular dependence of the toroidal magnetic field and radial currents becomes steeper.\footnote{In some cases it approaches the formation of a current sheet near the equatorial plane, which introduces numerical noise that prevents us from converging to a smooth solution and accurately calculating the twist angle.} It is also interesting to compare models C and S2. Both have a similar helicity, but the global twist is larger in model C, while the maximum current density is larger in model S2.

Comparing models B and D, which only differ in the angular dependence of the initial data and the normalization (fixed to obtain the same helicity), we find that the converged solutions are very similar, except near the axis where model D has no radial currents. The effect of varying the initial radial dependence can be estimated by comparing model D to model E. In this case the final solution keeps memory of the initial model: the converged solution shows a toroidal magnetic field that decreases faster with distance for model E than for model D.

In model E a tiny toroidal magnetic field (note the color scale in the figures) appears near the axis. This is likely a numerical artifact that can be partly ascribed to the (narrow) bundle of lines that depart from polar region and interact with the outer boundary. As a matter of fact, these structures are stronger for low values of $R_{out}$, as already shown in  Fig. \ref{fig:jrout}. Moreover, the numerical dissipation of the current is slower near the axis and longer runs are needed to reach more restrictive convergence criteria. Note however that this axial current is much weaker than the current in the equatorial region.

\begin{figure}
\centering
\includegraphics[width=.24\textwidth]{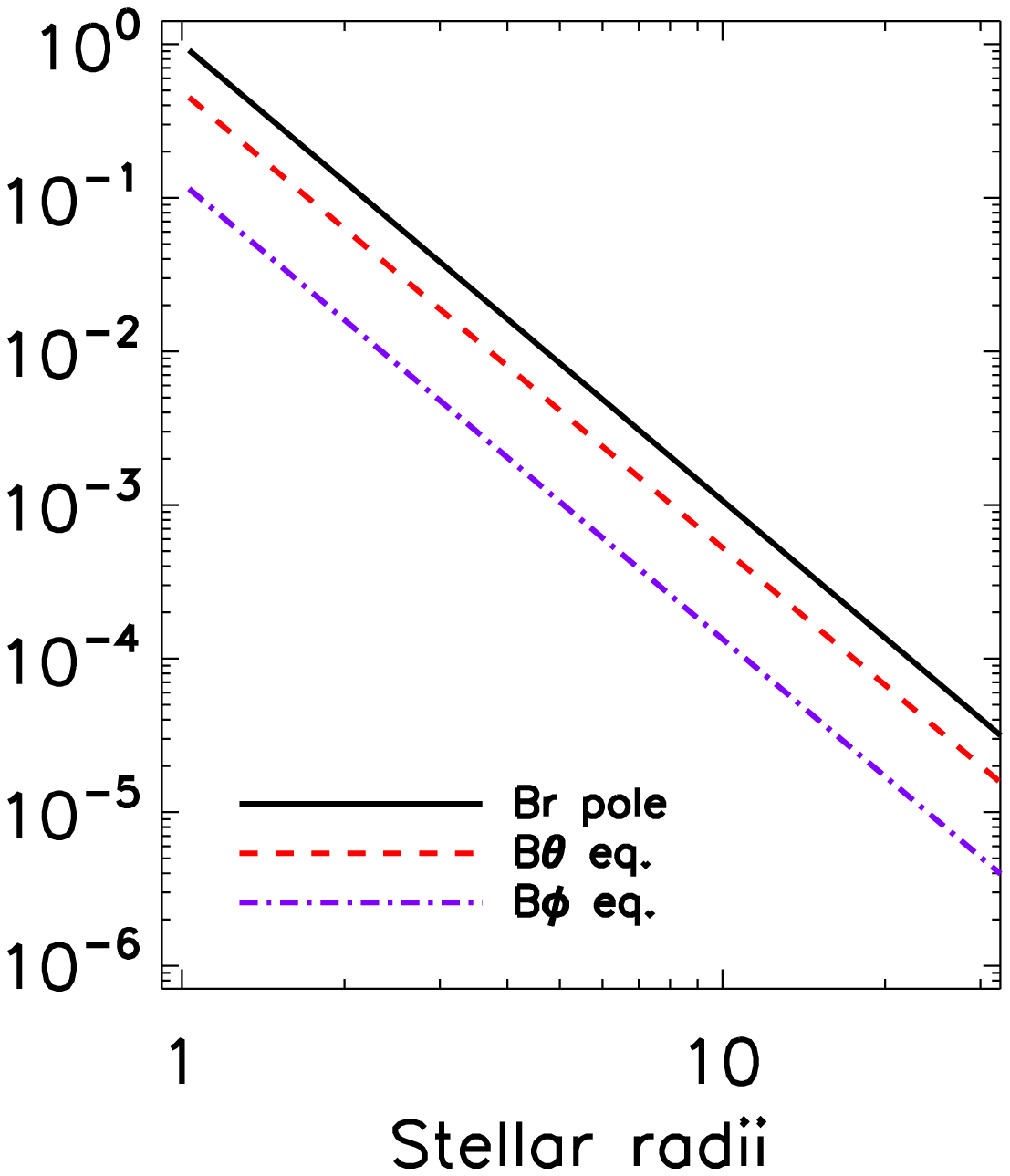}
\includegraphics[width=.24\textwidth]{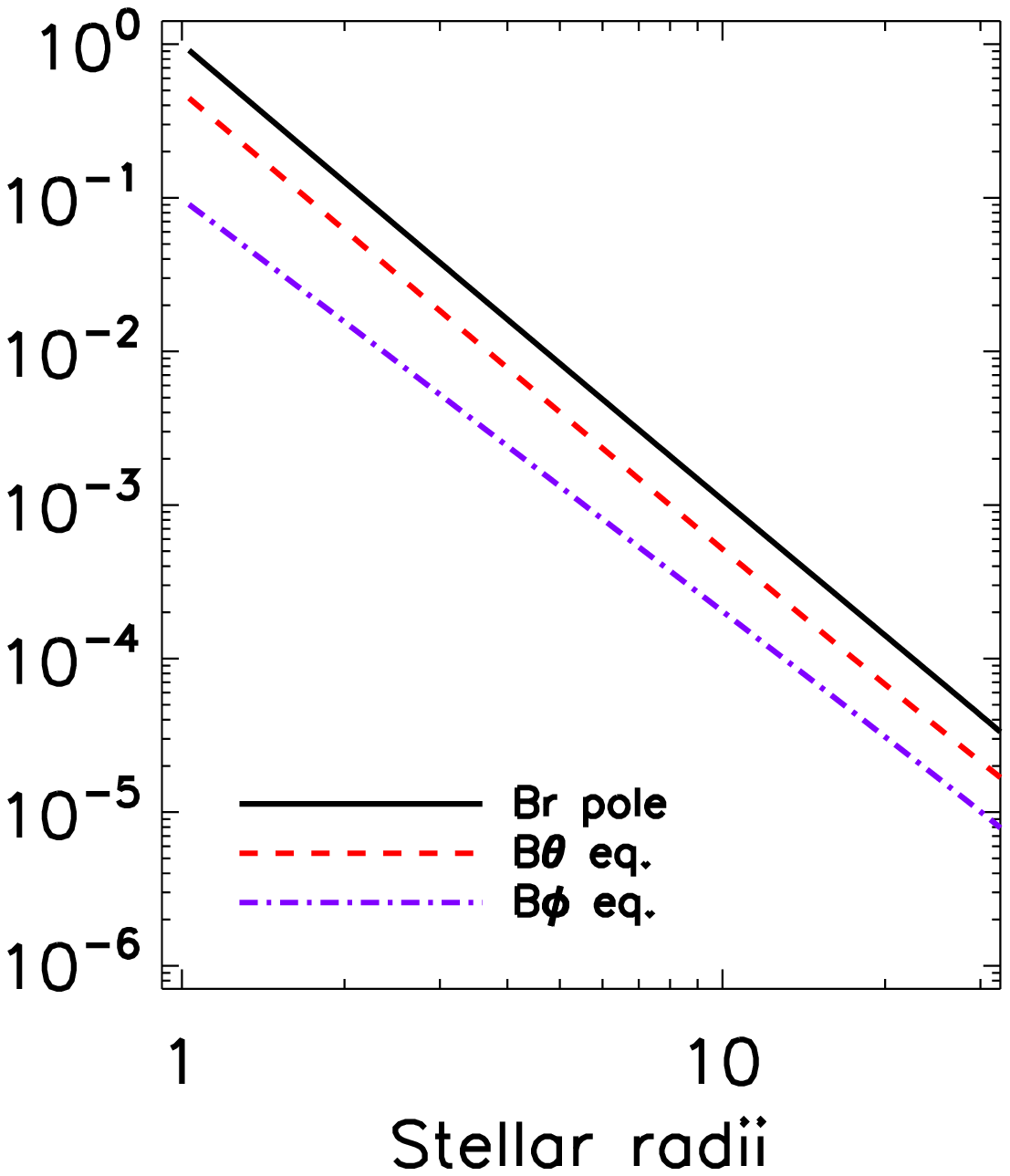}
\includegraphics[width=.24\textwidth]{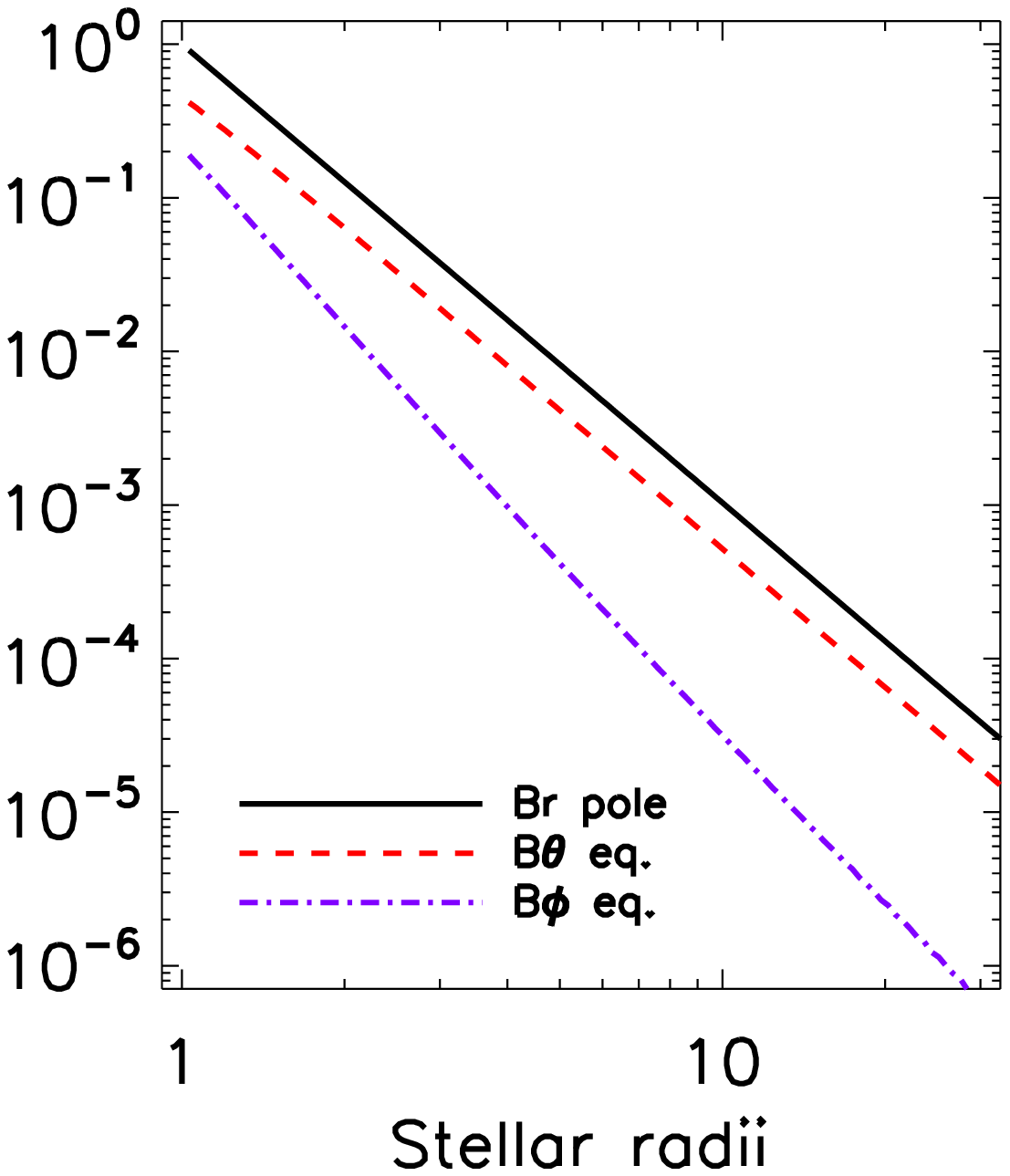}
\includegraphics[width=.24\textwidth]{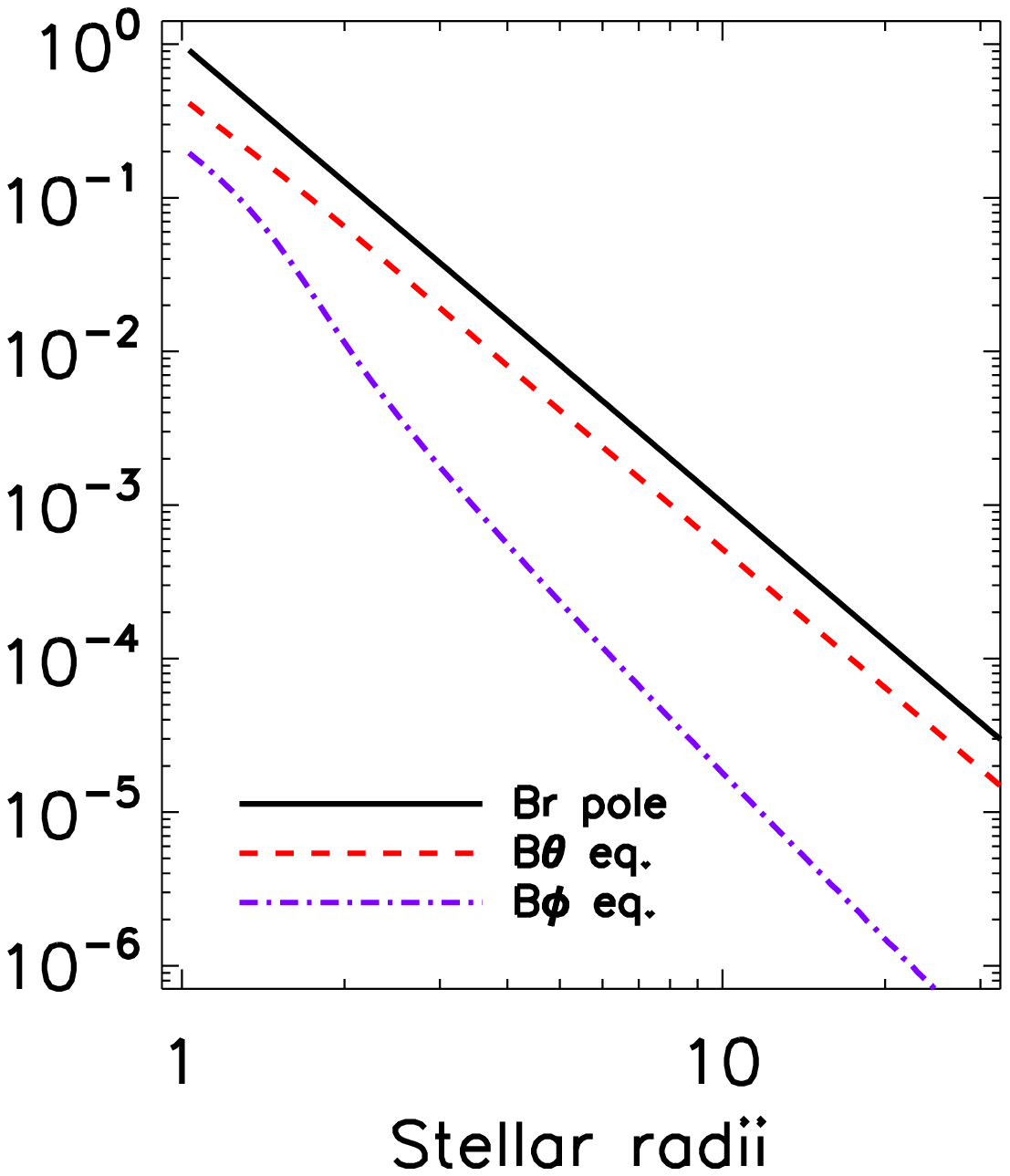}

\caption{Radial profiles of the magnetic field components, in units of $B_0$, for models S1, B, E and F (from left to right). We show $B_r$ on the axis and $B_\theta,B_\varphi$ at the equator for $r\in[1,30]~R_\star$.}
\label{fig:profr}
\end{figure}

The different radial dependences of the magnetic field components in models B, E and F, together with S1, are shown in Fig. \ref{fig:profr}. The self-similar solution has the same radial dependence $r^{-(p+2)}$ for the three components, but in the numerical solutions the toroidal magnetic field can decrease faster (model E and F) or slower (model B) than the poloidal components. Furthermore, the radial behavior may depend on $\theta$, too. In addition, the radial dependence of the poloidal components is close to the power law $r^{-3}$ when the twist is low, but it may significantly deviate from a power law for models with large twists.

Models G, H, and J are asymmetric configurations. In model G, most current is concentrated at a high latitude $\theta_m$. In model H the current and twist are concentrated in bundles near the equatorial region and near the southern semi-axis. Model J is more extreme: its helicity is much larger than for any other model. In both models H and J the currents are more localized, similarly to the expected configuration of a twisted magnetosphere \citep{beloborodov09}.

%%%%%%%%%%%%%%%%%%%%%%%
%%%%%%%%%%%%%%%%%%%%%%%

An interesting case (not shown) consists in an initial model with toroidal and poloidal magnetic fields of opposite parity (e.g. a dipolar poloidal magnetic field plus a quadrupolar toroidal magnetic field). In this case, the total initial helicity is zero and, since this is a conserved quantity, the current is dissipated and a potential solution is found by the numerical code. This is consistent with the fact that the numerical evolution converges towards the most trivial solution with the same helicity.

%%%%%%%%%%%%%%%%%%%%%%%
\begin{figure}
\centering
\includegraphics[width=.4\textwidth]{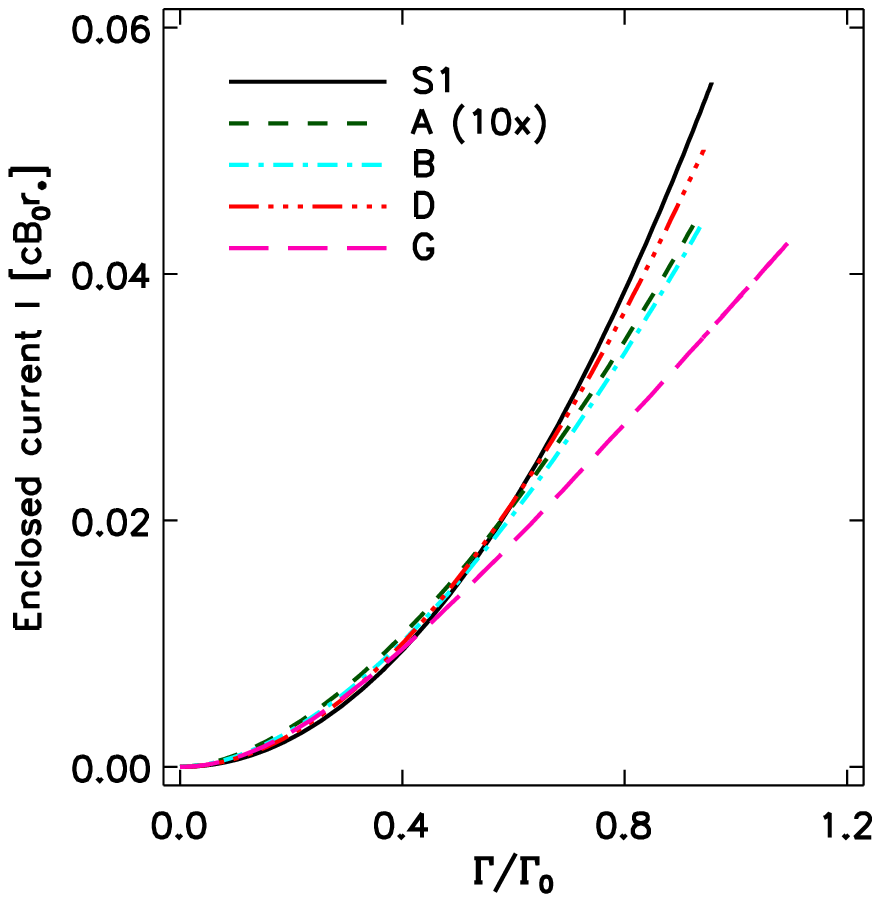}
\includegraphics[width=.4\textwidth]{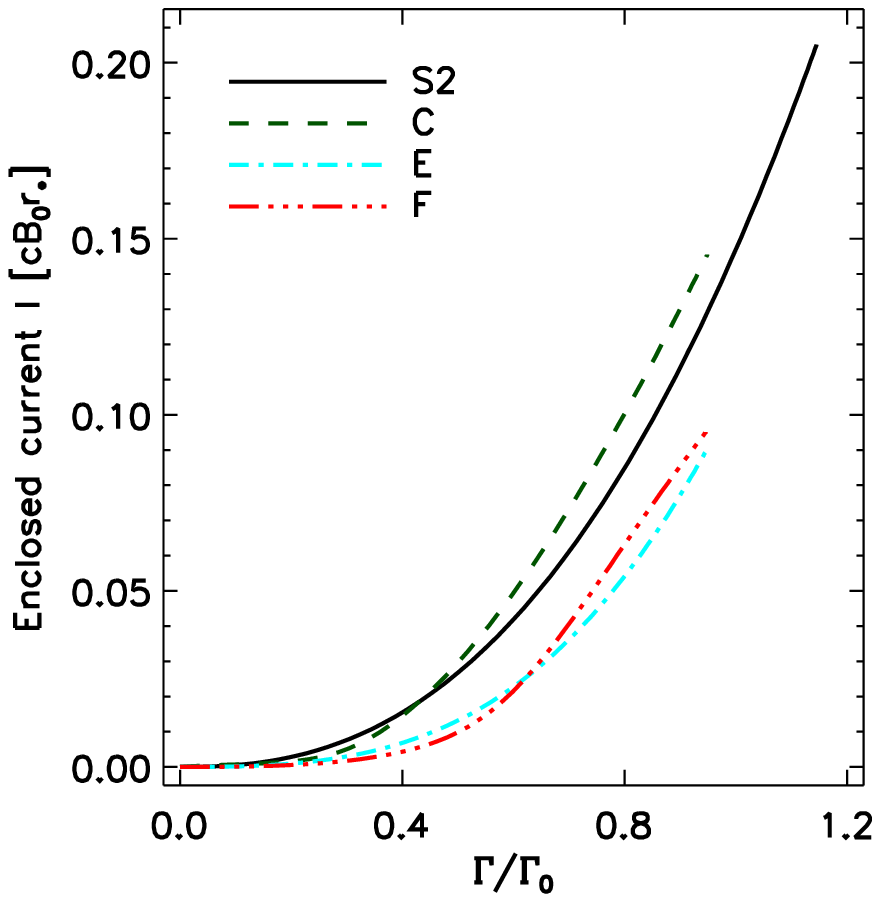}
\caption{$I(\Gamma)$ for different models. The curve of model A is magnified by a factor of 10.}
\label{fig:cfr_igamma}
\end{figure}
%%%%%%%%%%%%%%%%%%%%%%%

Finally, to better understand the differences among models, a comparison between the enclosed current function $I(\Gamma)$ is very helpful, as shown in Fig. \ref{fig:cfr_igamma}. In principle, if we know this function or a fit approximating the results of numerical simulations, the pulsar equation (eq. \ref{eq:pulsar_eq_sph}) can be solved and the magnetospheric structure can be determined. In our models A, B, D, E, and G, the current $I(\Gamma)$ is monotonic and can be well fitted by a single power law $I_0(\Gamma/\Gamma_0)^{1+1/p}$, but the values obtained for the parameter $p$ are not consistent with the value of $I_0$ describing the self-similar family of solutions (Fig. \ref{fig:tlk_dipole}), as expected. As a matter of fact, for models A, B, and D, the values of $p$ are greater than 1 (the self-similar dipolar family is described by $p\in[0,1]$). In contrast, model E lies quite close to the self-similar solution. The enclosed current of models C and E varies more rapidly, describing the concentration of currents in a smaller angular region. For some models, a power-law fit to $I(\Gamma)$ simply does not work. In model G, even if the symmetry with respect to the equatorial plane is broken, the resulting enclosed current $I(\Gamma)$ is not very different from models A, B, D, and S1 (after rescaling accordingly to the factor $k_{tor}$). The difference is the mapping between $\Gamma$ and the surface footprints. In all symmetric models, the maximum magnetic flux (proportional to $\Gamma$) is located at the equator, while it corresponds to $\theta_m\neq \pi/2$ for model G.

The geometric differences between magnetospheric models could leave an imprint in the formation of the output spectra, which is the subject of the next section.

\section{Magnetospheric resonant Compton scattering.}\label{sec:rcs}

\subsection{Magnetic Thomson scattering.}

The scattering between electrons and photons in absence of magnetic field can be described by classical electrodynamics. If the scattering electrons are non-relativistic, and the photon energy is $\hbar\omega\ll m_ec^2$, the process, known as Thomson scattering, is elastic. It can be naively seen in the following way: an incident electromagnetic wave exerts an electric force on the charged particle, which responds oscillating at the wave frequency. The oscillation, in turn, produces the emission of a photon at the same frequency. In absence of magnetic field, and in the elastic approximation, the Thomson cross section is given by (e.g., \citealt{jackson91}):

\begin{equation}\label{eq:thomson}
  \sigma_T = \frac{8\pi}{3}\left(\frac{e^2}{m_ec^2}\right)^2 = \frac{8\pi}{3}r_e^2 = 6.65\times10^{-25} \mbox{cm}^2~.
\end{equation}
where $r_e$ is the classical electron radius, at which the Coulomb energy equals its rest mass energy. The peculiarity of the Thomson cross-section is its independence on the photon frequency.

The propagation of electromagnetic waves in a plasma is affected by the presence of a strong magnetic field. In this case, charged particles describe helicoidal orbits around the magnetic field lines: they move freely along the magnetic field lines, while the motion perpendicular to them is restricted to circular orbits, with a fixed cyclotron frequency (or {\em gyro-frequency})
\begin{equation}\label{eq:def_gyrofrequency}
 \omega_B=\frac{|Ze|B}{mc}~,
\end{equation}
where $Ze$ and $m$ are the charge and mass of the particle. The corresponding energy is

\begin{equation}\label{eq:energy_gyrofrequency}
 \hbar\omega_B= 11.6~ |Z| B_{12}\frac{m_e}{m} \mbox{~keV}~.
\end{equation}
A free gas of charged particles with density $n$ is prone to collective modes, related to the dielectric function. The associated {\it plasma frequency} is defined as

\begin{figure}
\centering
\includegraphics[width=.6\textwidth]{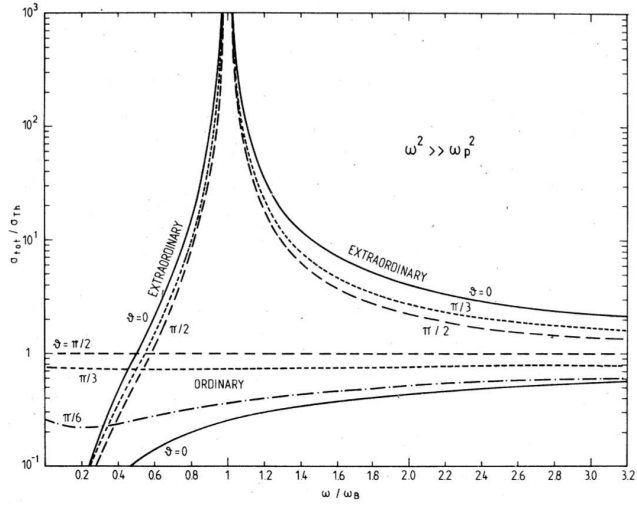}
\caption{Magnetic Thomson cross section of ordinary and extraordinary modes as a function of $\omega/\omega_B$ for different values of angle $\theta_{kB}$, taken from \cite{ventura79}.}
\label{fig:cross_section}
\end{figure}

\begin{equation}\label{eq:plasma_frequency}
 \omega_p=\sqrt{\frac{4\pi n (Ze)^2}{m}}~.
\end{equation}
Considering the typical Goldreich-Julian density (eq.~\ref{eq:n_gj}), the energy associated to the plasma frequency is
\begin{equation}
 \hbar\omega_{p}^{gj} \sim 10^{-5} ~ |Z|\sqrt{\frac{m_e}{m}}\sqrt{\frac{B_{12}}{P[\mbox{s}]}}~\mbox{eV}~,
\end{equation}
which, for any charge particle of interest (electrons, positrons, protons and heavy ions), is much smaller than the energies associated to cyclotron frequency (\ref{eq:def_gyrofrequency}) and X-ray photon frequencies, $\hbar\omega\sim 0.1-100~$keV.

Plasma and cyclotron frequencies enter in the polarization tensor, which is related to the wave propagation. Its treatment in the elastic limit has been considered in \cite{canuto71} and \cite{ventura79}. If $\omega_p\ll\omega_B$, the semi-transverse approximation applies, i.e. the polarization vector is orthogonal to the propagation direction $\hat{k}$. In this limit, the normal modes of propagation are in general elliptically polarized and the degree of ellipticity depends on the ratio $\omega/\omega_B$ and $\theta_{kB}$, defined as the the angle between $\hat{k}$ and $\vec{B}$. The normal modes given by the wave equations are called ordinary (O) and extraordinary (X). In the limit of propagation perpendicular to the magnetic field ($\theta_{kB}=\pi/2$), the modes are linearly polarized, with the polarization vector parallel (O-mode) or orthogonal (X-mode) to $\vec{B}$. For parallel propagation ($\theta_{kB}=0$) we recover the circularly polarized modes.

As shown in Fig.~\ref{fig:cross_section}, the magnetic Thomson cross section of the O and X modes depends in a non-trivial way on $\theta_{kB}$, and on the ratio $\omega/\omega_B$. The cross section of the O-mode is $\sigma=\sigma_T$ for $\theta_{kB}=\pi/2$, it scales with $\sin^2\theta_{kB}$ if $\omega\ll\omega_B$, and it does not depend dramatically on the frequency, unless $\theta_{kB}\rightarrow 0$. On the contrary, for low photon energies $\hbar \omega \ll \hbar\omega_B$, the cross section of the X-mode is strongly suppressed due to the reduced mobility of charged particles across magnetic field lines. At photon frequency $\omega=\omega_B$, the X-mode becomes resonant. In the limit of cold plasma, i.e. no motion of the charged particles, the X-mode resonant cross section diverges at the $E=\hbar \omega_B$. Physically, the divergence is cured by the non-vanishing damping term represented by the cyclotron radiation. Its power is described by the Larmor formula
\begin{equation}
W_L=\frac{2}{3}\frac{e^2|\dot{\vec{p}}|^2}{m^2c^3}~, 
\end{equation}
where $\dot{\vec{p}}$ is the time derivative of the relativistic momentum. The associated radiative de-excitation rate, for electrons, is \citep{ventura79}

\begin{equation}\label{eq:cyclotron_width}
 \Lambda_n=\frac{4}{3}\frac{e^2\omega^2}{m_ec^3}=\frac{4}{3}\frac{r_e\omega^2}{c} = 3.9\times 10^{15}~\left(\frac{\omega}{\omega_B}B_{12}\right)^2~\mbox{ s}^{-1}~.
\end{equation}
This rate is related with the width of the resonance, $\Lambda$, appearing in the resonant cross section for the X-mode:

\begin{equation}
  \sigma_{res} = \pi^2\frac{e^2}{m_e c}(1+\cos^2\theta_{kB})\frac{\Lambda/(2\pi)}{(\omega-\omega_B)^2+\Lambda^2/4}~,
\end{equation}
where the factor $(1+\cos^2\theta_{kB})$ arises from the assumption of unpolarized light. In general, $\Lambda$ is given by any process that introduces a damping force in the particle motion, such as Coulomb interactions that dissipate energy via Bremsstrahlung emission. However, in the typical magnetospheric conditions, the cyclotron radiation, eq.~(\ref{eq:cyclotron_width}), is by far the most dissipative process, thus $\Lambda=\Lambda_n$. The maximum value of the cross-section, reached at the resonance $\omega=\omega_B$, is:
\begin{equation}
  \sigma_{max}= \frac{3\pi}{2}\frac{c^2}{\omega_B^2}(1+\cos^2\theta_{bk})= 2.9\times 10^{-18}~\frac{1+\cos^2\theta_{bk}}{B_{12}^2}~\mbox{cm}^2~,
\end{equation}
which is more than six orders of magnitude larger than the non-resonant cross-section, $\sigma\sim \sigma_T$. Therefore, we can safely take the limit
\begin{equation}
 [\Lambda/(2\pi)]/[(\omega-\omega_B)^2+\Lambda^2/4] \rightarrow \delta(\omega-\omega_B)~,
\end{equation}
and the resonant cross section can be written as:
\begin{equation}\label{eq:sigma_res}
 \sigma_{res}=\pi^2(1+\cos^2\theta_{kB}) \frac{(Ze)^2}{mc}\delta(\omega-\omega_B)~.
\end{equation}

\subsection{Relativistic effects: Compton scattering.}

The elastic limit discussed in the previous subsection is not valid if the photon energy is comparable or greater than the rest energy of the scattering particle, or if the speed of the latter is relativistic. In the first case, photons transfer part of their energy to particles, but this is not the case for X-rays, for which $\hbar\omega \ll m_ec^2 \ll m_pc^2$. In the second case, particles can boost up photons to larger energies, and this inelastic process is known as inverse Compton scattering, or Compton up-scattering. It is common in astrophysics and is the primary candidate to convert soft thermal photons to high-energy photons in magnetars.

Consider an electron moving, in the inertial frame, with velocity $c\beta$. In the stellar frame, the resonant energy corresponds to the Doppler-shifted gyro-frequency:
\begin{equation}\label{eq:redshift_frequency}
\omega_D(B,\beta)=\frac{1}{\gamma(1-\beta\cos\theta_{\gamma e})}\frac{eB}{m_ec}~,
\end{equation}
where $\gamma\equiv(1-\beta^2)^{-1/2}$ is the Lorentz factor, and $\theta_{\gamma e}$ is the angle between the photon direction and the electron momentum. To derive the Compton cross-section, one has to consistently consider the conservation of momentum and energy (see, e.g., \citealt{nobili08b}). The spatial-dependent distribution of particle velocity strongly affects the processed spectra.

\subsection{Quantum electrodynamics effects.}

In presence of magnetic fields, quantum mechanics describes the orthogonal component of electronic motion as a bidimensional oscillator, which eigen-values of frequency are multiples of the fundamental gyro-frequency $\omega_B$, eq.~(\ref{eq:def_gyrofrequency}). The quantization has relevant effects for magnetic fields larger than the critical quantum electrodynamics magnetic field, defined as the value for which $\omega_B$ equals the rest mass of electrons:
\begin{equation}\label{eq:critcal_b}
 B_c=\frac{m_e^2c^3}{e\hbar}=4.4\times 10^{13} \mbox{~G}~,
\end{equation}
for which the largest possible gyro-radius would be given by $r_g=c/\Omega_B=\hbar/m_e c$. The proper quantum resonant cross-section taking into account the Landau quantization has been described by \cite{harding91} and \cite{nobili08b}. They include the various energy-dependent rates of the transitions between Landau levels, both for the excitation (photon absorption) and de-excitation (photon emission) to/from higher levels. The process of de-excitation to the ground level includes different possible transitions, including the generation of more than one photon. In the limit of $B\ll B_c$, the problem can be treated classically and the process of excitation and de-excitation can be described by classical electrodynamics as curvature radiation.

Another effect of strong magnetic fields is the vacuum birefringence \citep{schwinger51}, i.e., the different behaviour of the ordinary and extraordinary modes, due to the presence of virtual pairs even in absence of plasma. The quantum corrections enter in the wave equation with terms $\sim 10^{-4}(B/B_c)^2$ \citep{adler70}. When plasma and vacuum effects are both taken into account, the vacuum resonance strongly alters the polarization of radiation, allowing a conversion between X and O modes \citep{lai03}. This effect is important for densities $\rho\sim 1$ g~cm$^{-3}$, a condition met in the neutron star atmosphere. The magnetosphere is too sparse for the vacuum resonance to take place, but vacuum effects can be important in the regions where $\sin\theta_{kB}\rightarrow 0$. These processes are also thought to leave a strong polarization imprint in the radio band \citep{wang09}.

\subsection{Charge density in twisted magnetospheres.}

We now focus on the observable effects of having a twisted magnetosphere. First of all, we can estimate the contribution of the rotationally induced currents, eq.~(\ref{eq:co-rotating_jphi}), to the total current, eq.~(\ref{eq:current_force-free}):

\begin{equation}\label{eq:j_tw}
 \frac{J_{gj}}{J} \sim \left(\frac{r}{\varpi_l}\right)^2\frac{L_B}{r b_{tor}}~,
\end{equation}
where $b_{tor}\sim B_\varphi/B$, and $L_B$ is the typical length-scale of magnetic field variations. For moderately twisted magnetospheres, $b_{tor}\sim O(1)$, $r\sim L_B$, thus near the surface $J_{gj}/J\ll 1$: the twist strongly enhances the current density.

The space charge density, eq.~(\ref{eq:n_gj}), in absence of rotation, has to be zero. We assume that negative (electrons) and positive (protons, ions, positrons) charges co-exist, flowing at different velocity $v_j$ (where $j$ is the specie), and that they provide the current needed to sustain the twisted magnetic field:
\begin{equation}\label{eq:j_tw_def}
 J=\left|\sum_j n_j Z_je v_j\right| ~.
\end{equation}
However, we have no {\it a-priori} information about the composition of plasma and the velocities of its particles. These uncertainties can be parametrized by a factor $\kappa$, in terms of which the electron density is written in a compact form:

\begin{equation}\label{eq:n_tw_def}
 n_e = \frac{J}{\kappa c e}~.
\end{equation}
Comparing the last two equations, we have $\kappa=1$ if electrons flowing at velocity $c$ are the only carriers of current. Numerically, we estimate

\begin{eqnarray}\label{eq:n_tw}
 n_e \sim  \frac{b_{tor}B}{4\pi e L_B \kappa} = 10^{14}~\frac{b_{tor} B_{12}}{L_{B,6}\kappa}~\mbox{cm}^{-3}~.
\end{eqnarray}
where $L_{B,6}=L_B/10^6~$cm. The twist-induced electron density is several orders of magnitude larger than the charge-separated Goldreich-Julian density, eq.~(\ref{eq:n_gj}). These estimates hold in the co-rotating region of magnetars, in which the magnetic field is thought to deviate considerably from a dipole ($b_{tor} \gtrsim 1$), and the large periods result in $\varpi_l \gg L_B$. The relevant radiative processes occur for strong magnetic fields, that is, close to the surface. Here we can use the estimate of the electron density given by eq.~(\ref{eq:n_tw}).

\subsection{Thomson resonant optical depth.}

The scattering optical depth for a seed photon of energy $\hbar\omega$ depends on the particle density $n_Z$, the velocity of the scatterer and the intensity of the magnetic field. In the ideal case of no particle motion, a first rough estimate of the resonant optical depth is given by the integration of eq.~(\ref{eq:sigma_res}) along a given line of sight of constant $\theta$:

\begin{equation}\label{eq:tau_res}
 \tau_{res}(\theta,\omega)=\pi^2 n_ZZe(1+\cos^2\theta_{kB})\left|\frac{dB}{dr}\right|^{-1}~,
\end{equation}
where, and hereafter, all quantities are evaluated at the resonant radius $r_{res}(\omega,B)$, defined as the distance for which $\omega=\omega_B$. Note that the resonance broadening $\Lambda$ (\ref{eq:cyclotron_width}) implies that the resonant layer actually has a finite extent $\delta r_{res}=(\Lambda/\omega_B)L_B$ (for a dipole $L_B \simeq r_{res}/3$). For the natural width $\Lambda=\Lambda_n$ (eq.~\ref{eq:cyclotron_width}), $\delta r_{res}\sim 10^{-4}~ B_{12} r_{res}$, thus the resonant layer is very thin compared to the distance from the star.

The energy dependence of the optical depth, eq.~(\ref{eq:tau_res}), is given basically by the local ratio $(1+\cos^2\theta_{kB})n_Z/|dB/dr|$ (provided that $r_{res}$ lies above the star surface). Using the relation (\ref{eq:n_tw_def}), assuming $\kappa$ constant along the line of sight, and considering only the scattering between electrons and radially directed photons \citep{nobili08a}, eq.~(\ref{eq:tau_res}) becomes:

\begin{equation}\label{eq:tau_simple}
 \kappa\tau_{res}(\theta)=\pi^2 \frac{J_{tw}}{c}\left(1+\frac{B_r^2}{B^2}\right)\left|\frac{dB}{dr}\right|^{-1}~.
\end{equation}
If we consider self-similar models, $\tau_{res}$ is independent of $r_{res}$, that is independent on where the scattering occurs. This is because the local ratio $(1+\cos^2\theta_{kB})J/|dB/dr|$ is the same for each angle, since, for every component $i$, we have $B_i\propto r^{-p+2}$ and $J_i\propto r^{-p+3}$ (eqs.~\ref{eq:mf_tlk1} - \ref{eq:mf_tlk3}). Furthermore, the optical depth does not depend on the normalization of the magnetic field, $B_0$.

\begin{figure}[t]
\centering
\includegraphics[width=.24\textwidth]{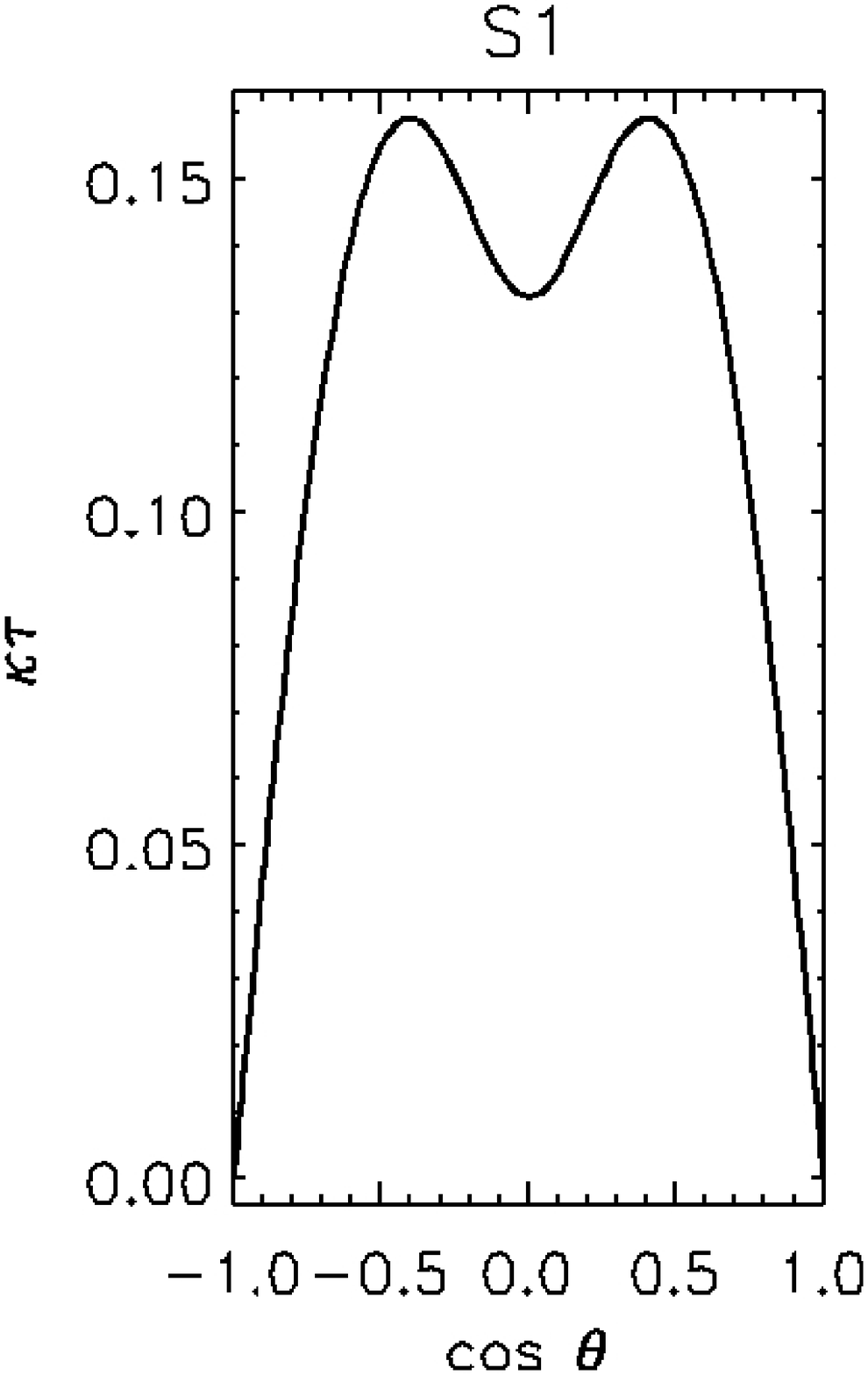}
\includegraphics[width=.24\textwidth]{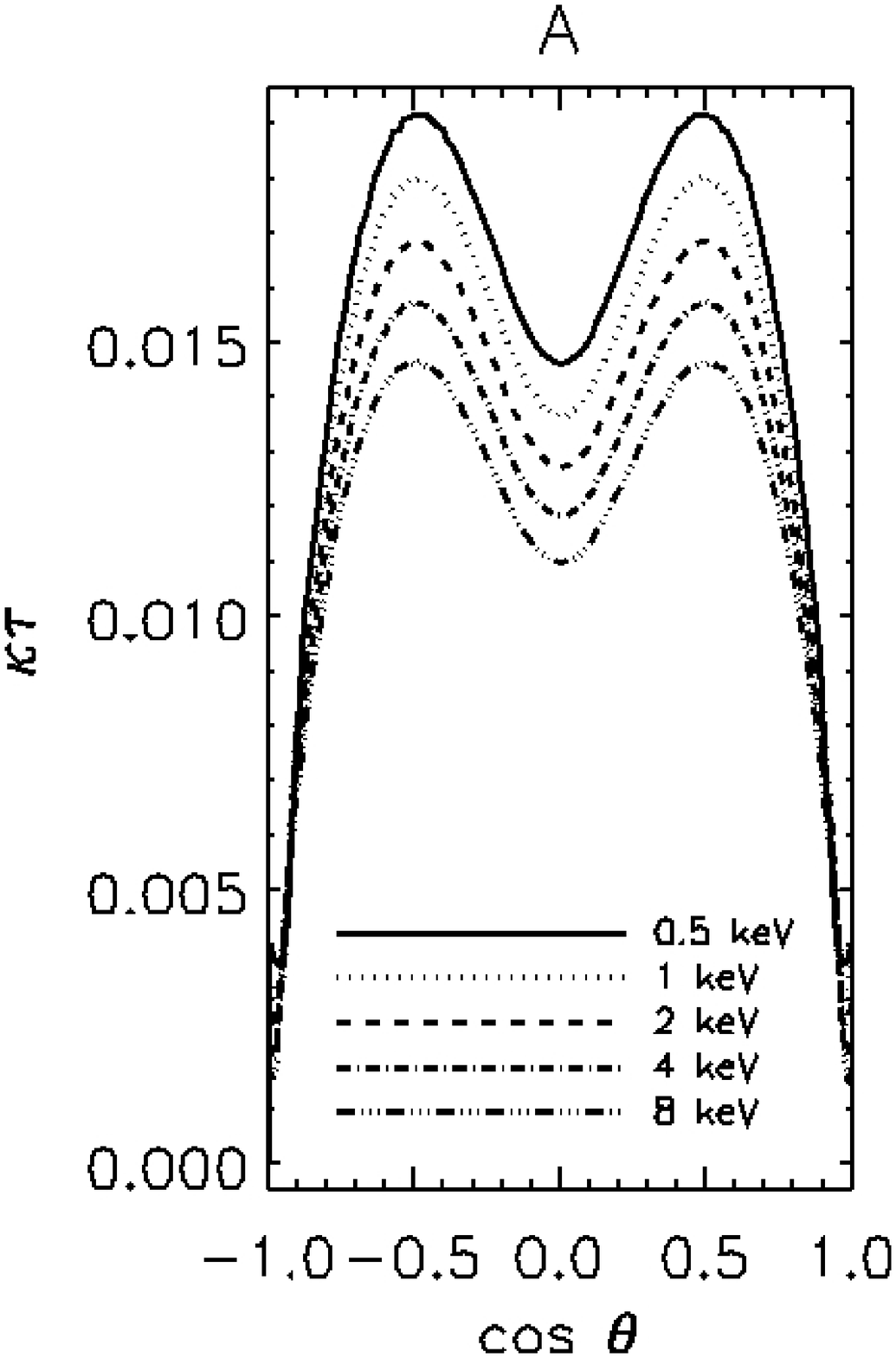}
\includegraphics[width=.24\textwidth]{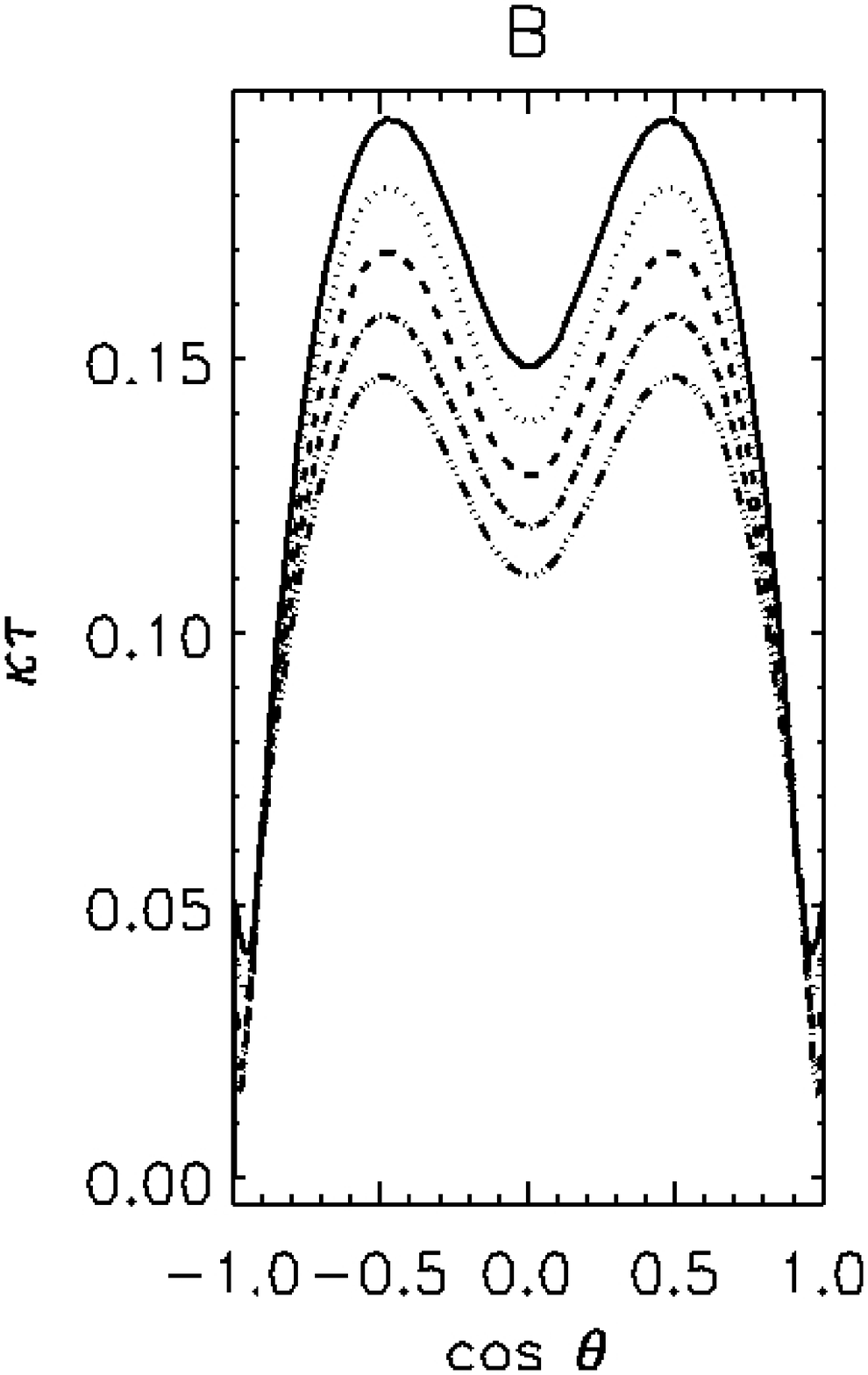}
\includegraphics[width=.24\textwidth]{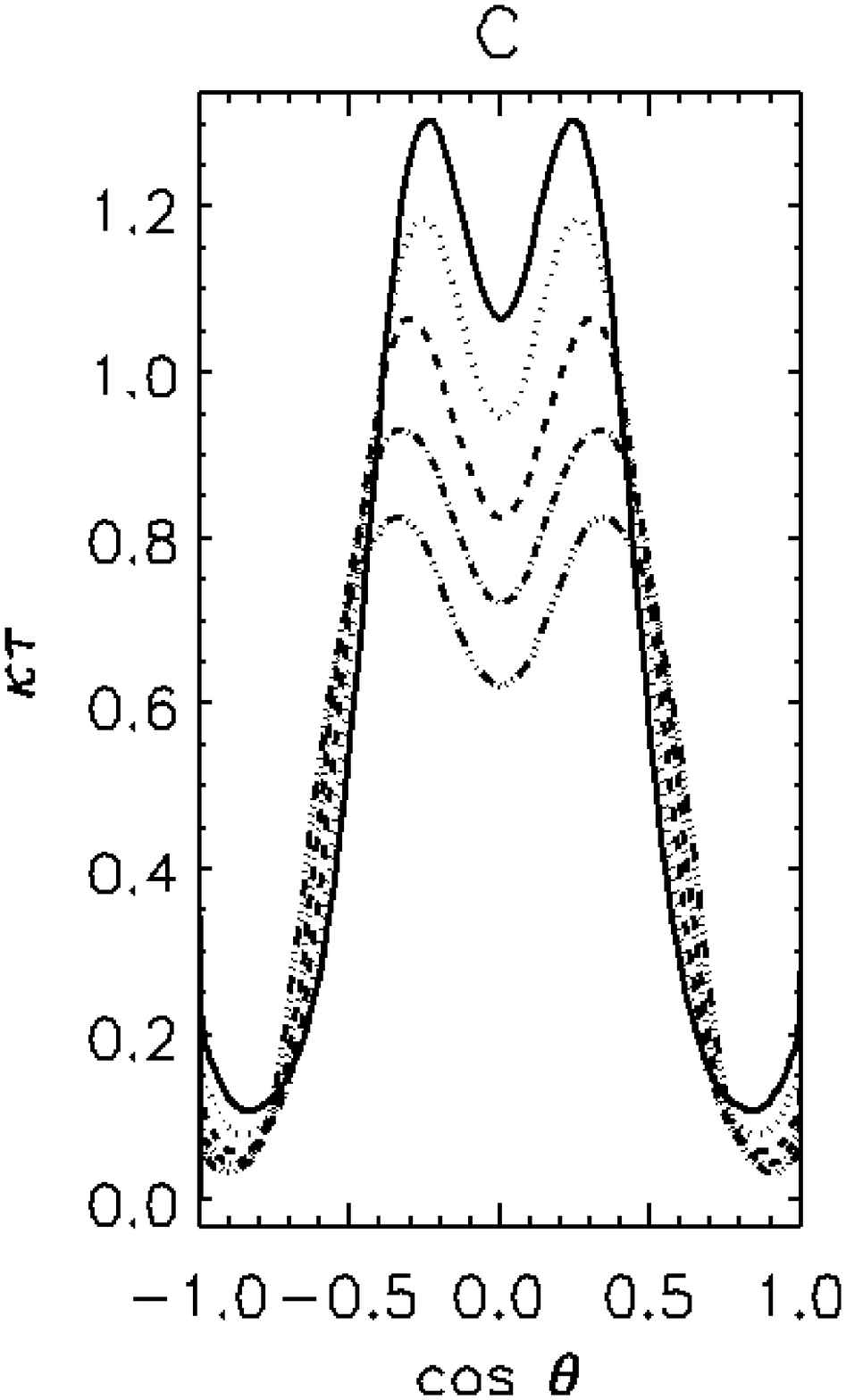}
\includegraphics[width=.24\textwidth]{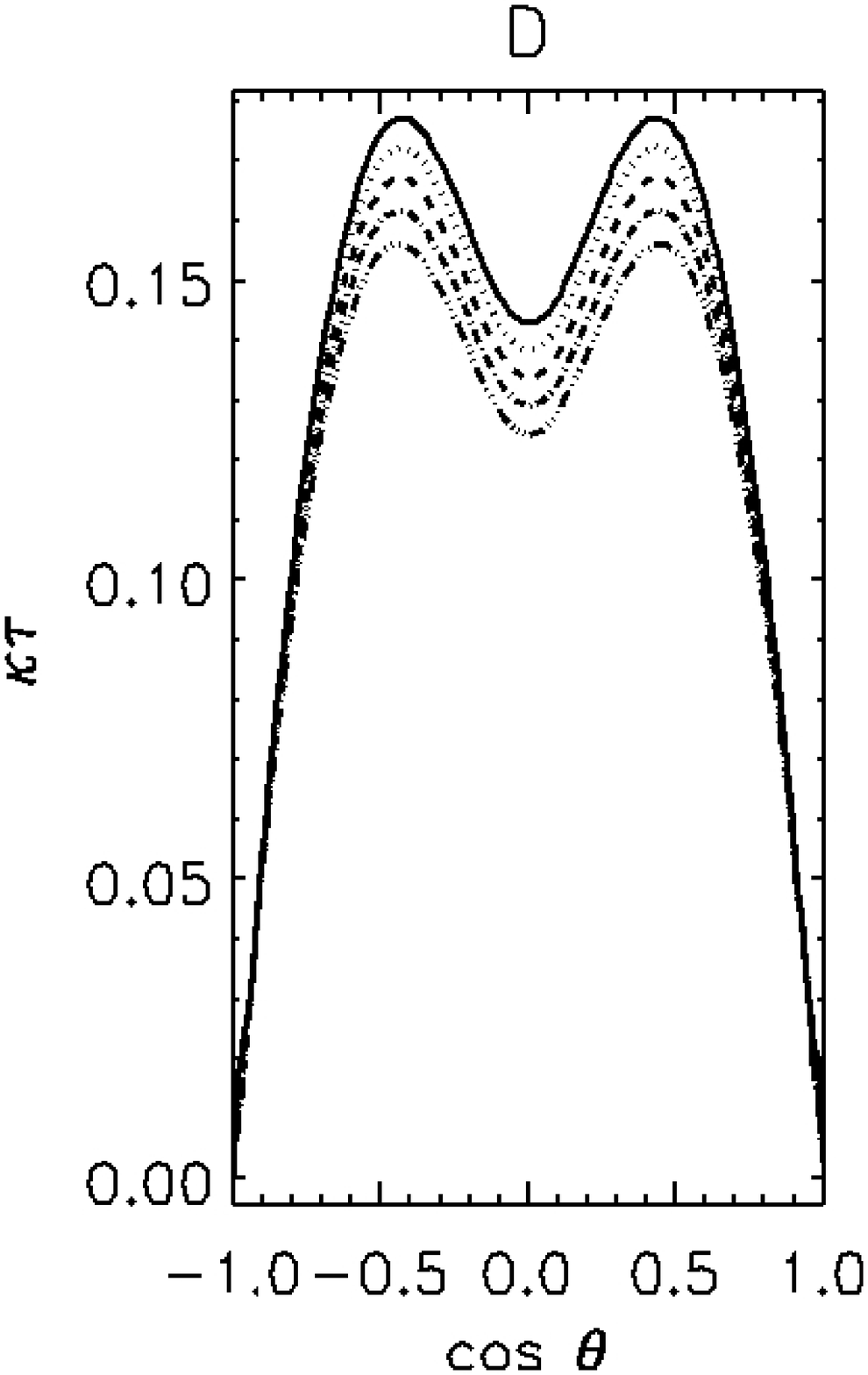}
\includegraphics[width=.24\textwidth]{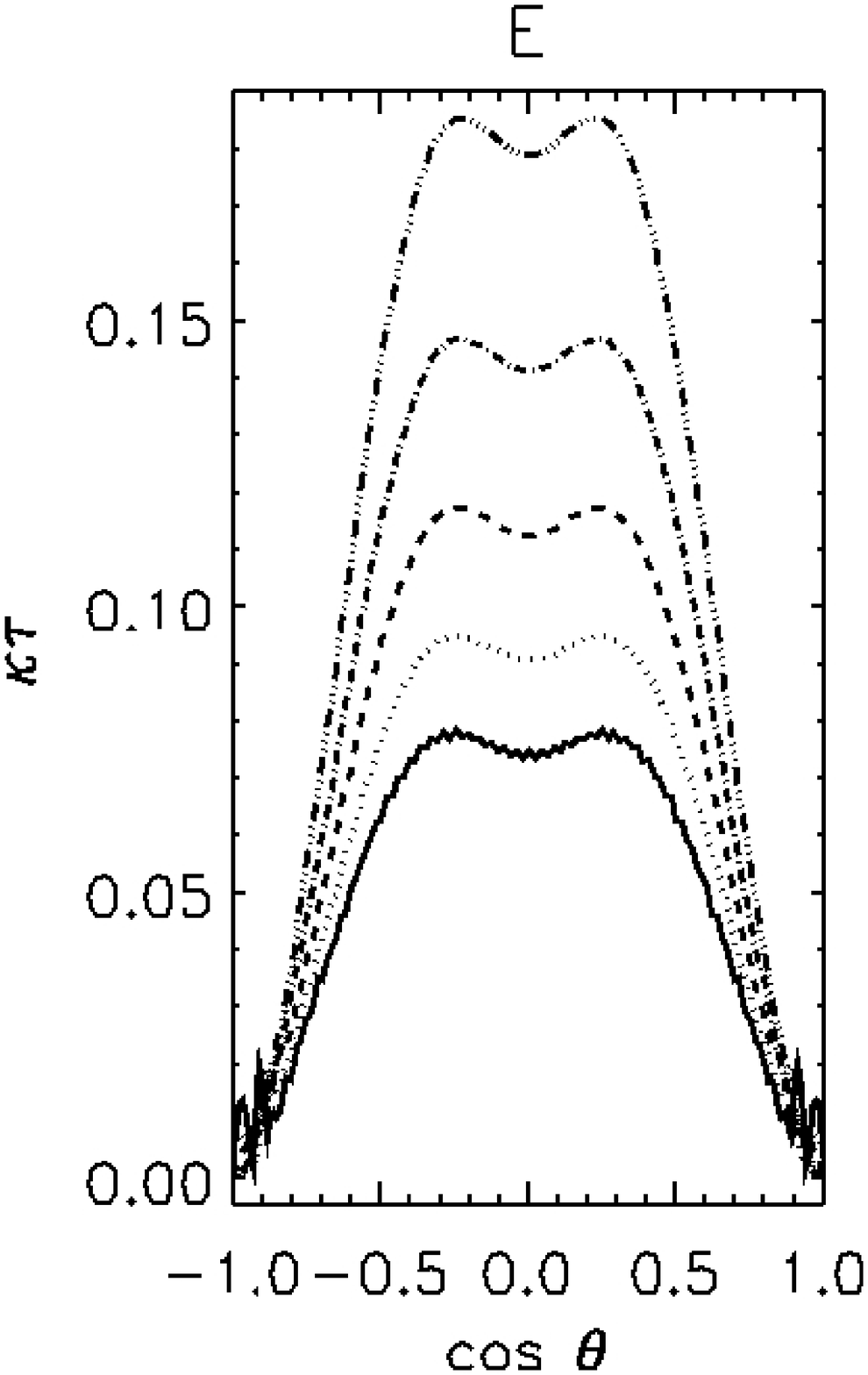}
\includegraphics[width=.24\textwidth]{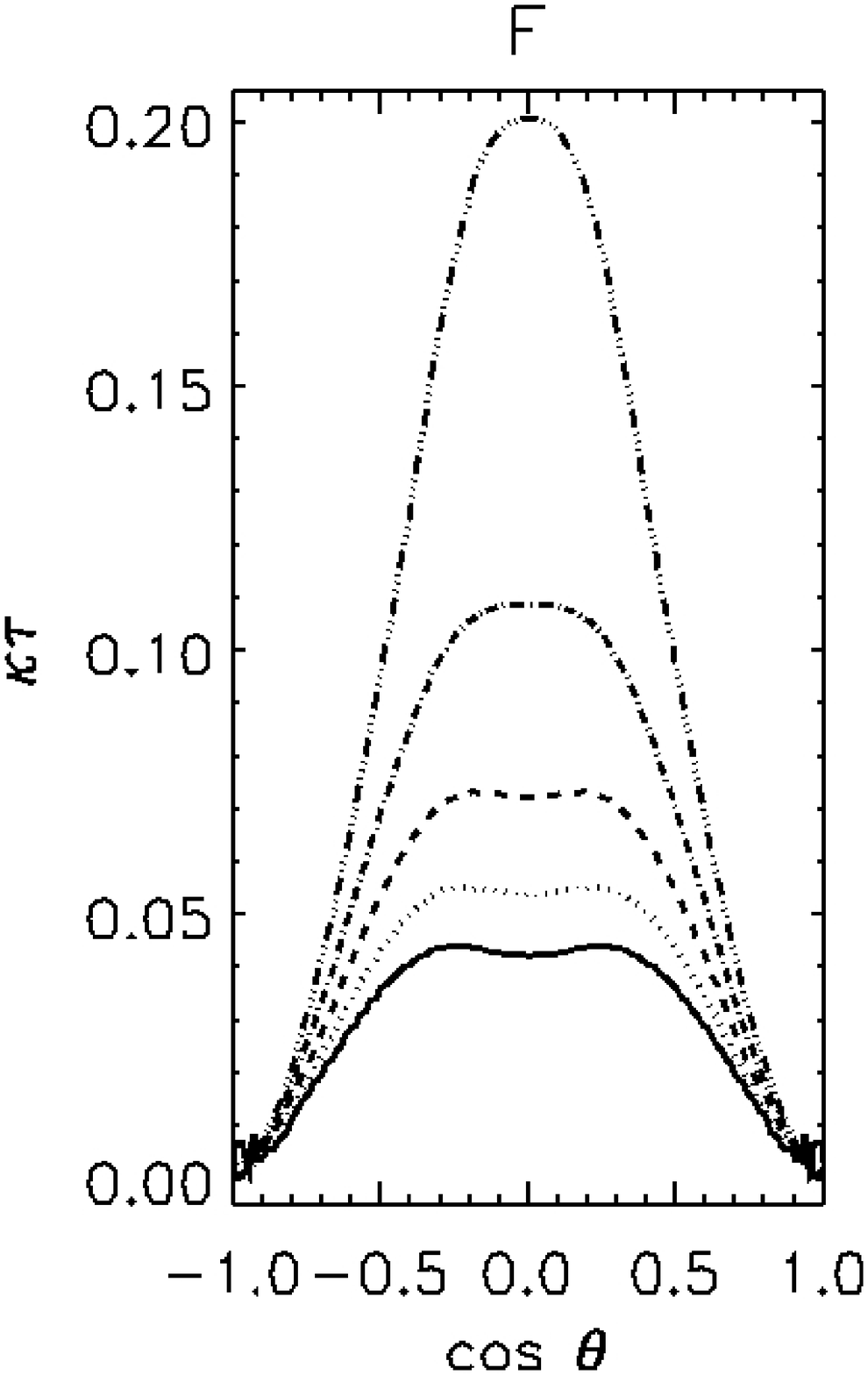}
\includegraphics[width=.24\textwidth]{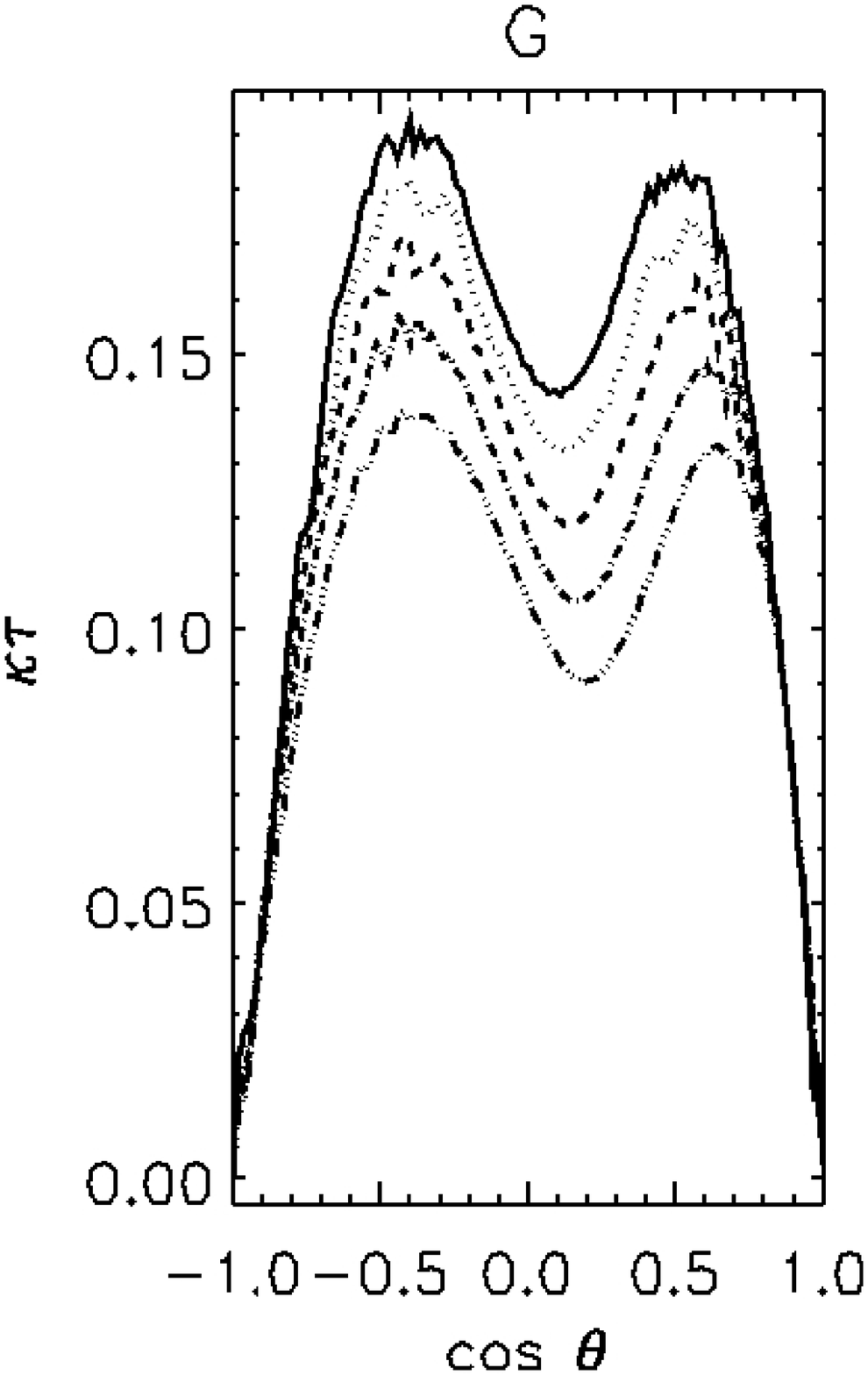}

\caption{Estimated Thomson resonant optical depth (multiplied by the microphysical factor $\kappa$) for radial photons as a function of the polar angle $\theta$. We show results for models S1, A, B, C (first row), D, E, F, G (second row) for different photon energies as indicated in the legend. We set $B_0=10^{13}$ G.}
\label{fig:tau}
\end{figure}

On the contrary, in our numerical models (\S~\ref{sec:numerical_magnetosphere}), $\tau_{res}$ depends on the photon energy, because the components of $\vec{J}$ and $\vec{B}$ cannot be described by the same power law. In Fig.~\ref{fig:tau} we show the estimated resonant optical depth, eq.~(\ref{eq:tau_simple}), for our models A-G, compared with the self-similar model S1. We plot $\kappa\tau_{res}$ as a function of the polar angle $\theta$, for different energies of the seed photons, $\hbar\omega=0.5,1,2,4,8$ keV, taking $B_0=10^{13}$ G. If the magnetic field is predominantly poloidal (small twist), the optical depth is roughly given by the ratio between toroidal and poloidal components at the resonant radius. Increasing the photon energy, the resonant radius will be closer to the surface, modifying the estimate of $J$ and $dB/dr$ which are involved in eq.~(\ref{eq:tau_simple}). Thus, if the toroidal magnetic field decays slower than the poloidal component, a higher photon energy implies a lower ratio $J/|dB/dr|$ at the resonant radius. Looking at the radial profiles (Fig.~\ref{fig:profr}), we can understand why for models E and F the optical depth increases with the photon energy while for other models  the optical depth is larger for soft photons. Due to the linear relation between $B$ and $\omega_B$, increasing the energy is equivalent to decrease the normalization $B_0$ by the same factor. A more precise prediction of how the spectrum is modified when using one or another magnetosphere model requires more elaborated calculations, as we discuss in the next subsection.

\subsection{Numerical simulations.}\label{sec:rcs_code}

To compute the expected spectrum processed by the resonant Compton scattering, we have to assume as inputs the surface radiation, the magnetospheric configuration of magnetic field and the (space-dependent) distribution of scatterers velocity and density. All these factors cause a spread of the resonant frequency, eq.~(\ref{eq:redshift_frequency}), and a detailed simulation of the processes in the whole magnetosphere is needed. The result is a continuum spectrum, with a strong deviation from the seed spectrum.

Several codes accounting for resonant Compton scattering have been built. The first attempts to consider quantitatively this process in magnetars were presented by \cite{lyutikov06}. They studied a simplified, semi-analytical 1D model by assuming that seed photons are radially emitted from the neutron star surface with a blackbody spectrum and the resonant scattering occurs in a thin, plane parallel magnetospheric slab. Assuming a certain bulk velocity of electrons, they found an average up-scattering for the transmitted radiation (forward scattered photons), while the mean energy of the reflected radiation remains the same. The model was extended by \cite{guver06} who relaxed the blackbody approximation for the surface radiation, including atmospheric effects.

Later, more accurate 3D Monte Carlo simulations have been performed for the non-relativistic \citep{fernandez07, nobili08a} and relativistic cases \citep{nobili08b}, accounting for polarization and quantum electrodynamics effects. The implementation in {\tt XSPEC} and the systematic fits to observational data have been successfully performed by \cite{rea08} and \cite{zane09} for the 1D and 3D models above, respectively. Below we employ the Monte Carlo code by \cite{nobili08b} and explore the dependence of the spectrum on the magnetosphere model by comparing results obtained with self-similar models and with two of our numerical configurations (models H and J of \S~\ref{sec:numerical_magnetosphere}).

Treating the problem on a large-scale allows to estimate the particle density, once the plasma velocity distribution is provided as input. The velocity distribution is assumed to be a one-dimensional (along the magnetic field line), relativistic Maxwellian distribution with a plasma temperature $T_e$ and a bulk velocity $\beta_{bulk}$, as described in \cite{nobili08a}. The strong simplification in this approach is that the velocity distribution does not depend on position, which reduces the microphysical inputs to two parameters: the temperature of the plasma $T_e$ and $\beta_{bulk}$. 

Given a seed spectrum for the photons emitted from the surface (assumed as a blackbody with  temperature $T_\star$), the Monte Carlo code follows the photon propagation through the scattering magnetosphere. When a photon enters in a parameter region where no more resonant scatterings are possible, the photon escapes to infinity, where its energy and direction are stored. The sky at infinity is divided into patches, so that viewing effects can be accounted for: if the line of sight is along the angles $\theta_s$, $\varphi_s$, only photons collected in the patch which contains that pair of angles are considered. The angle-averaged spectrum is obtained by averaging over all patches.

\subsection{Results.}\label{sec:rcs_results}

\begin{figure}
\centering
\includegraphics[width=.30\textwidth]{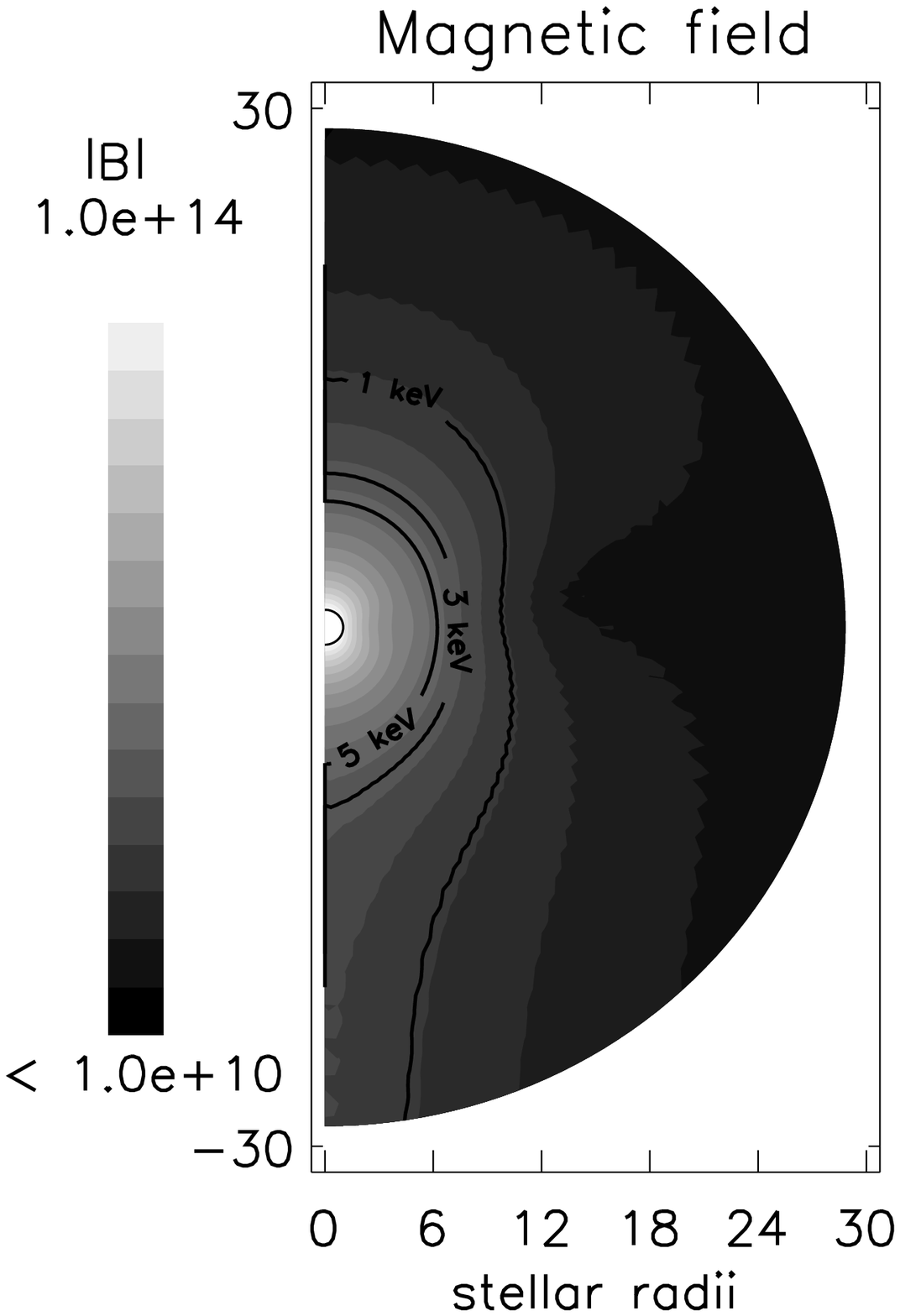}
\includegraphics[width=.30\textwidth]{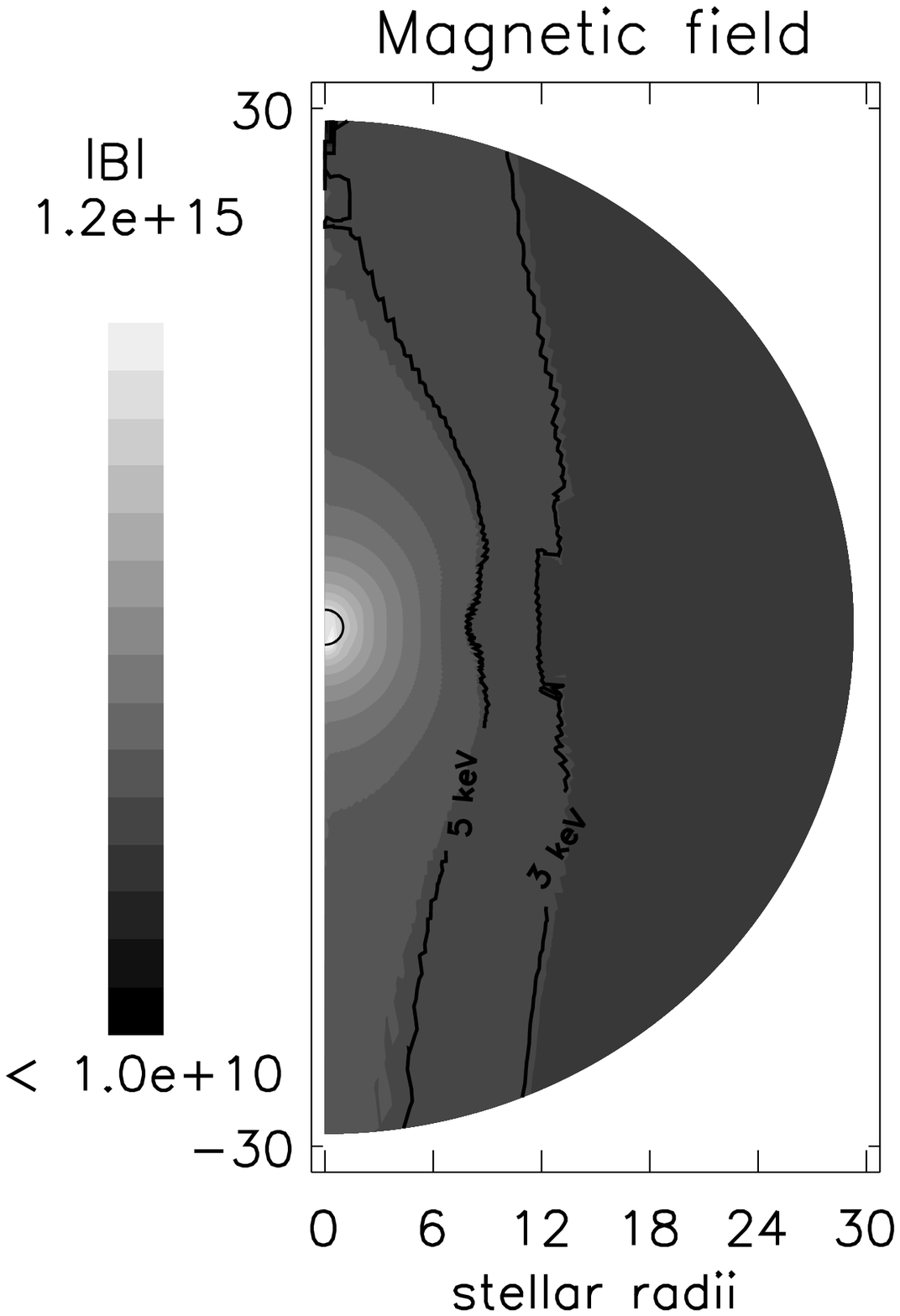}
\caption{Numerical model H (left) and model J (right): $|\vec B|$ (gray logarithmic scale) with scattering surfaces for photons of $1$, $3$ and $5~$keV (here $\beta_{bulk}=0.5$, $\cos\theta_{\gamma e}=0.5$; see text).}
\label{fig:scattering_surface}
\end{figure}

We present results comparing models with a magnetic field intensity at the pole of $B_p=10^{14}$~G. Fig.~\ref{fig:scattering_surface} shows the scattering surfaces in the two models H and J of \S~\ref{sec:numerical_magnetosphere}, for photons of 1, 3 and $5~$keV, assuming $\beta_{bulk}=0.5$ and $\cos\theta_{\gamma e}=0.5$. The surfaces are far from being spherically symmetric. In model J, which has the strongest helicity, the scattering surfaces of soft X-ray photons lie tens of stellar radii away from the surface, especially near the axis.

\begin{figure}[t]
\centering
\includegraphics[width=.45\textwidth]{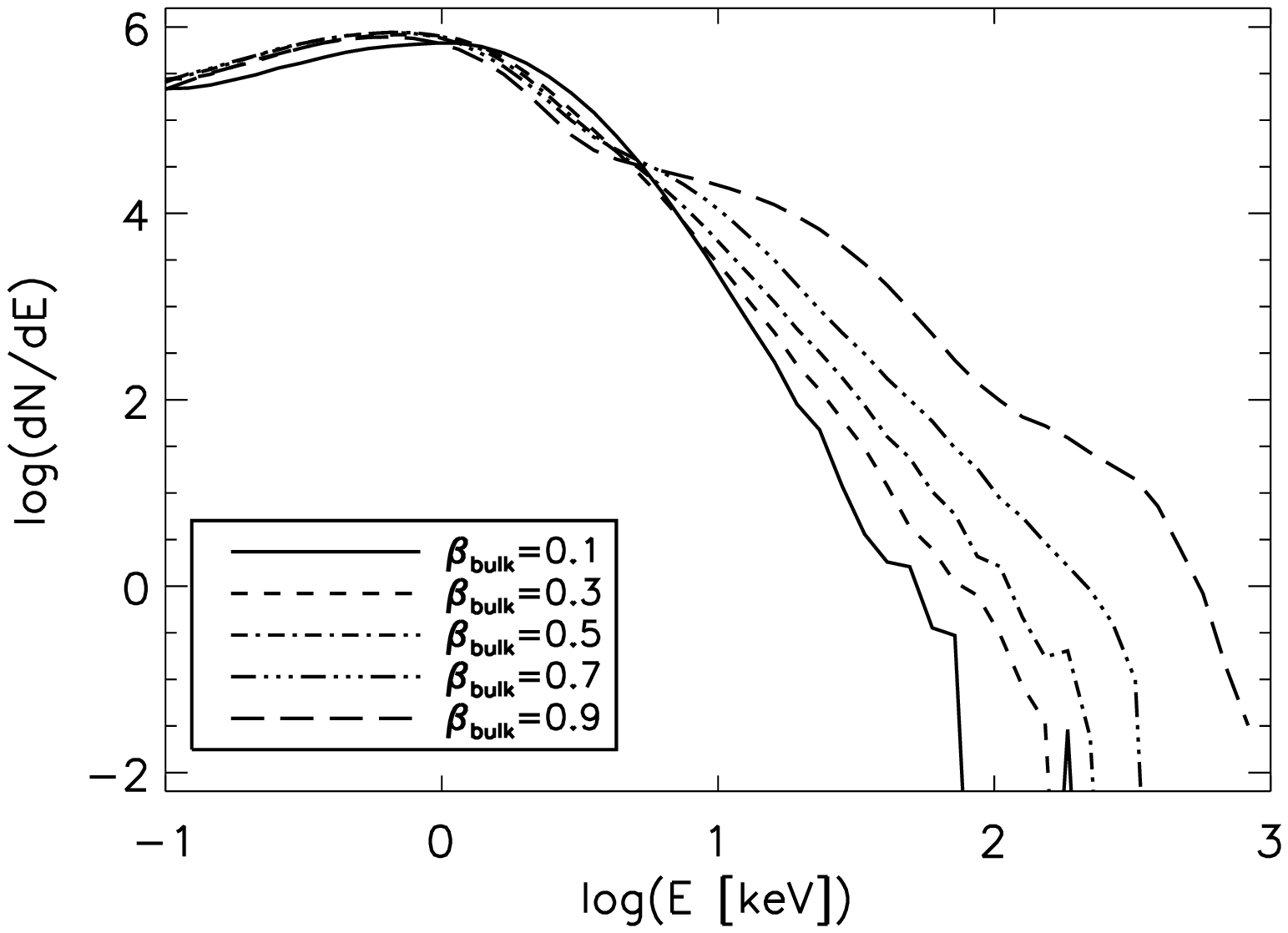}
\includegraphics[width=.45\textwidth]{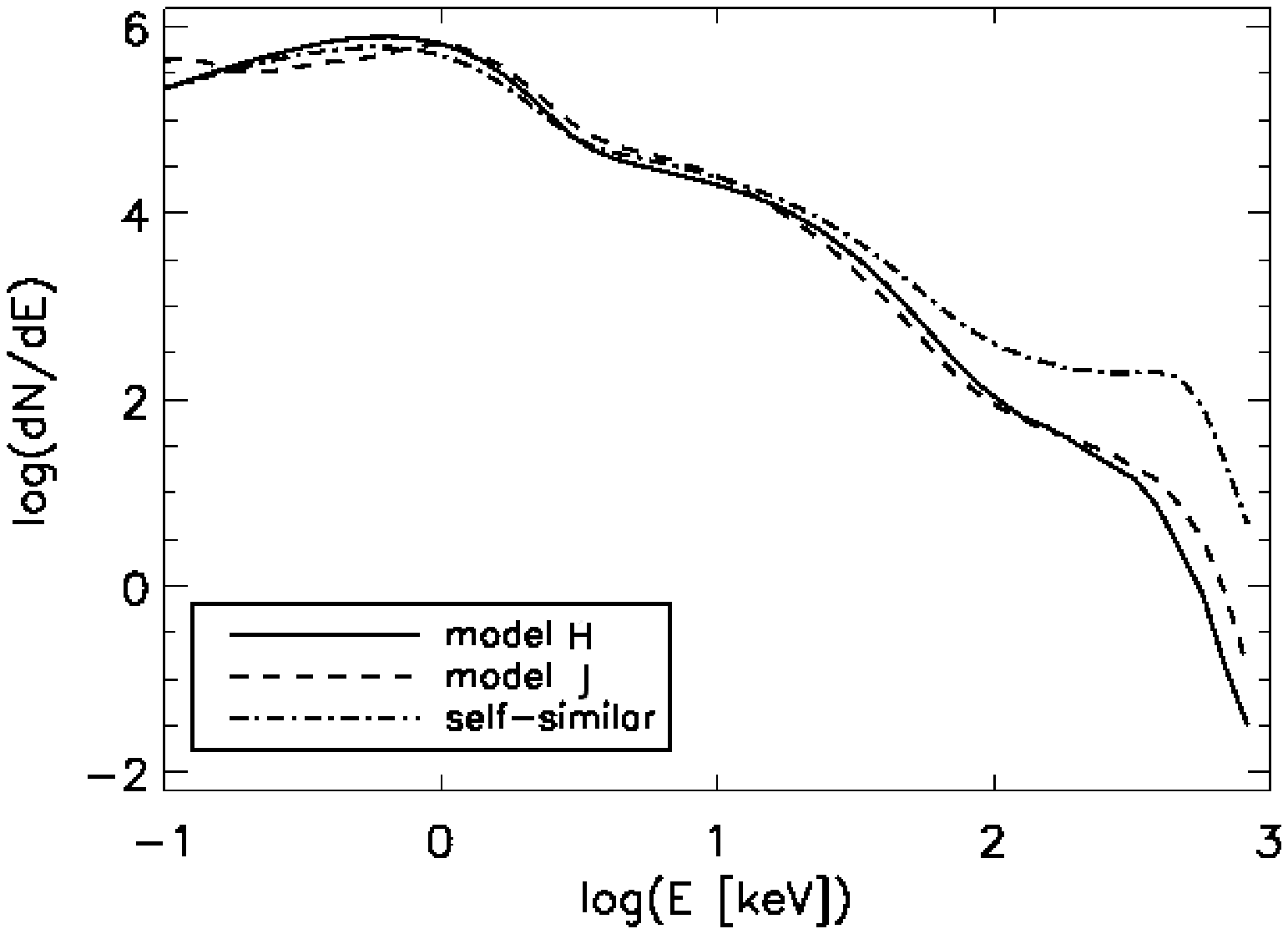}
\caption{Angle-averaged spectra for $k_bT_e=20~$keV. Left panel: model H, changing $\beta_{bulk}$. Right panel: $\beta_{bulk}=0.9$, changing magnetic field model (S2, H or J).}
\label{fig:average_spectra}
\end{figure}

The seed spectrum is assumed to be a $k_bT_\star=0.4~$keV Planckian isotropic distribution for both ordinary and extraordinary photons. In Fig.~\ref{fig:average_spectra} we plot the angle-averaged spectra, i.e., the distribution of all photons escaped to infinity, independently on the final direction. The left panel shows the effects produced by changes in the bulk velocity, keeping fixed the magnetic field configuration (model H), and the electron temperature ($k_bT_e=20~$keV). The main fact is that the spectrum in the region $E>10~$keV becomes harder as  $\beta_{bulk}$ is increased, while it is depleted of photons of energies in the $1$--$10~$keV range. A similar, but less pronounced effect, is obtained by increasing the electron temperature \citep{nobili08b}.

In the right panel we compare spectra obtained with fixed values of $k_bT_e=20~$keV and $\beta_{bulk}=0.9$, varying only the magnetic field topology. The three lines correspond to  model H (solid), model J (dashes) and a self-similar model (dash-dotted line) with $\Delta\varphi=1.36$, which approximately has the same total helicity of model H. We have chosen a high value of $\beta_{bulk}$ to show a case where the effects are larger, but our conclusions do not change qualitatively for lower values of $\beta_{bulk}$. We see that the major differences arising from changes in magnetic field topology appear in the hard tail of the spectra, which is clearly harder for the self-similar model. The thermal part of the spectrum is more depleted in model J.

\begin{figure}[t]
\centering
\includegraphics[width=.47\textwidth]{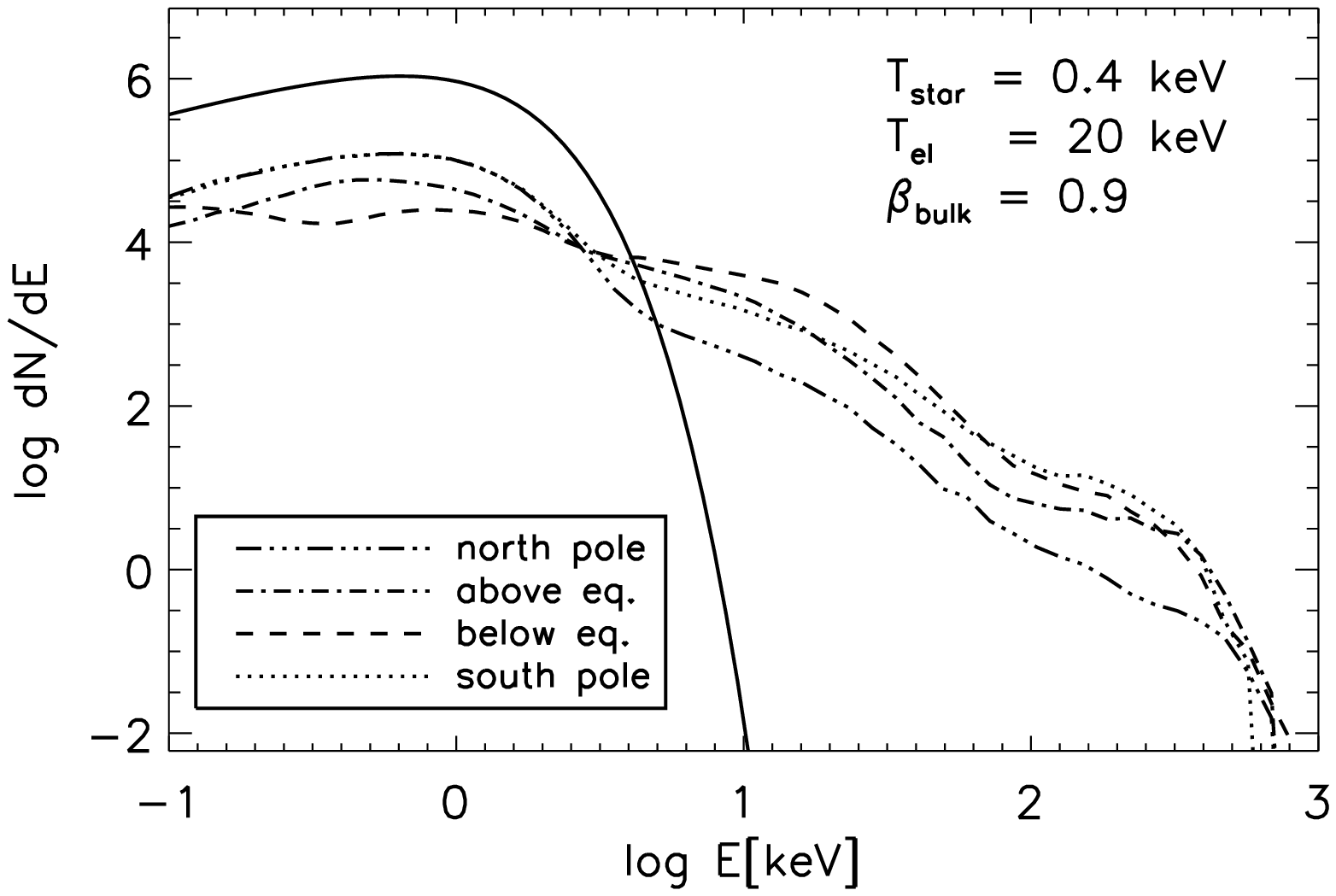}
\includegraphics[width=.47\textwidth]{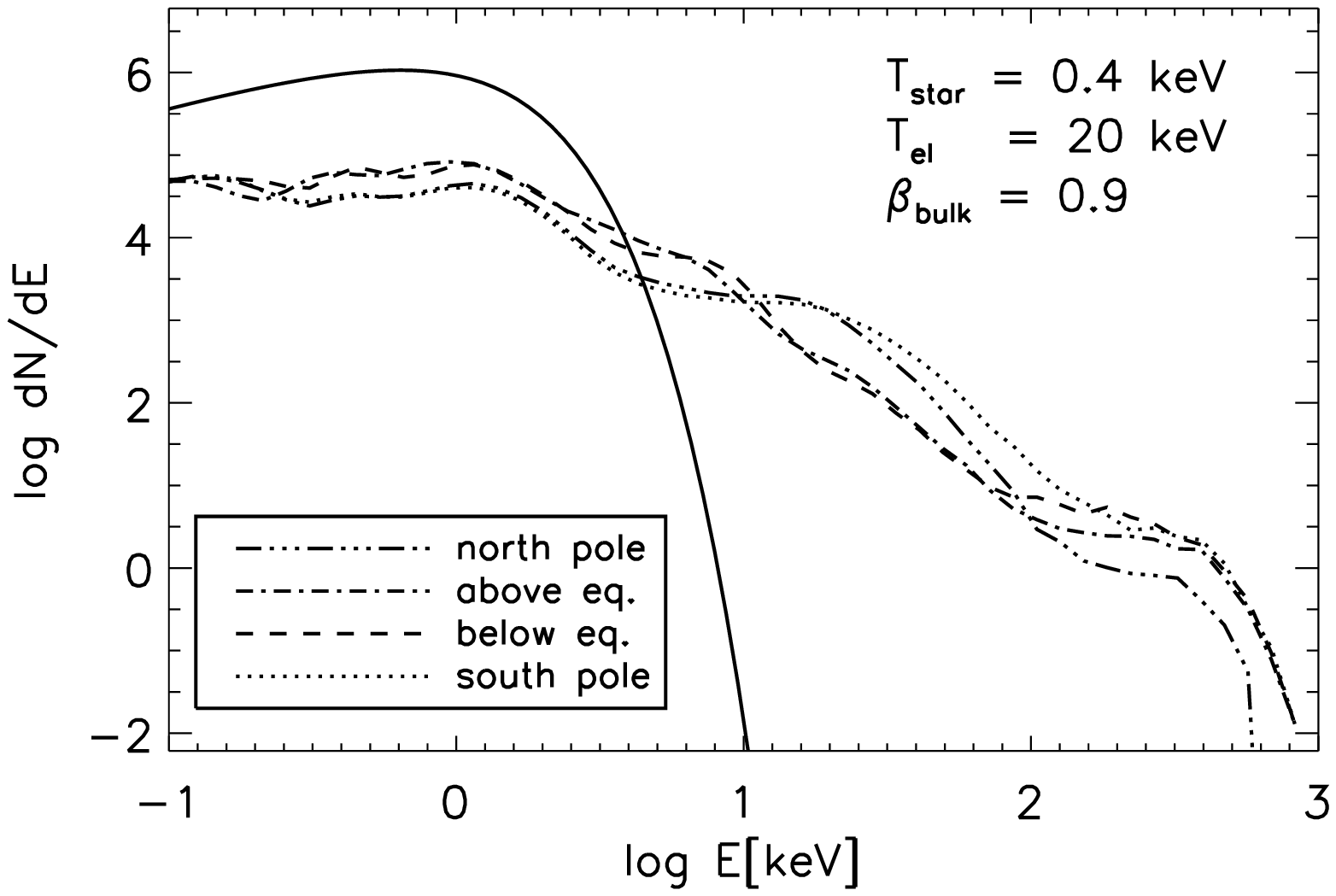}\\
\medskip
\includegraphics[width=.47\textwidth]{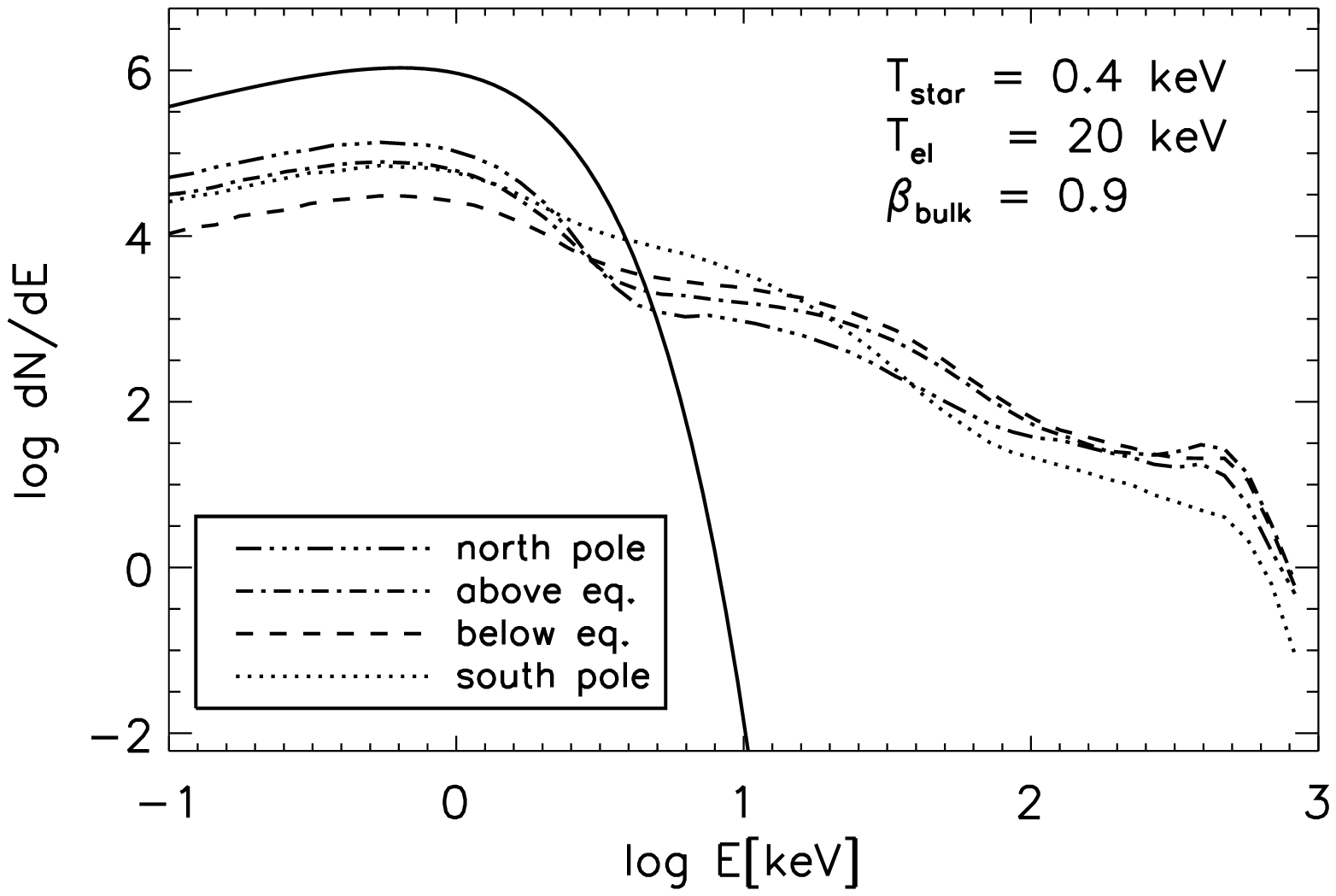}
\includegraphics[width=.47\textwidth]{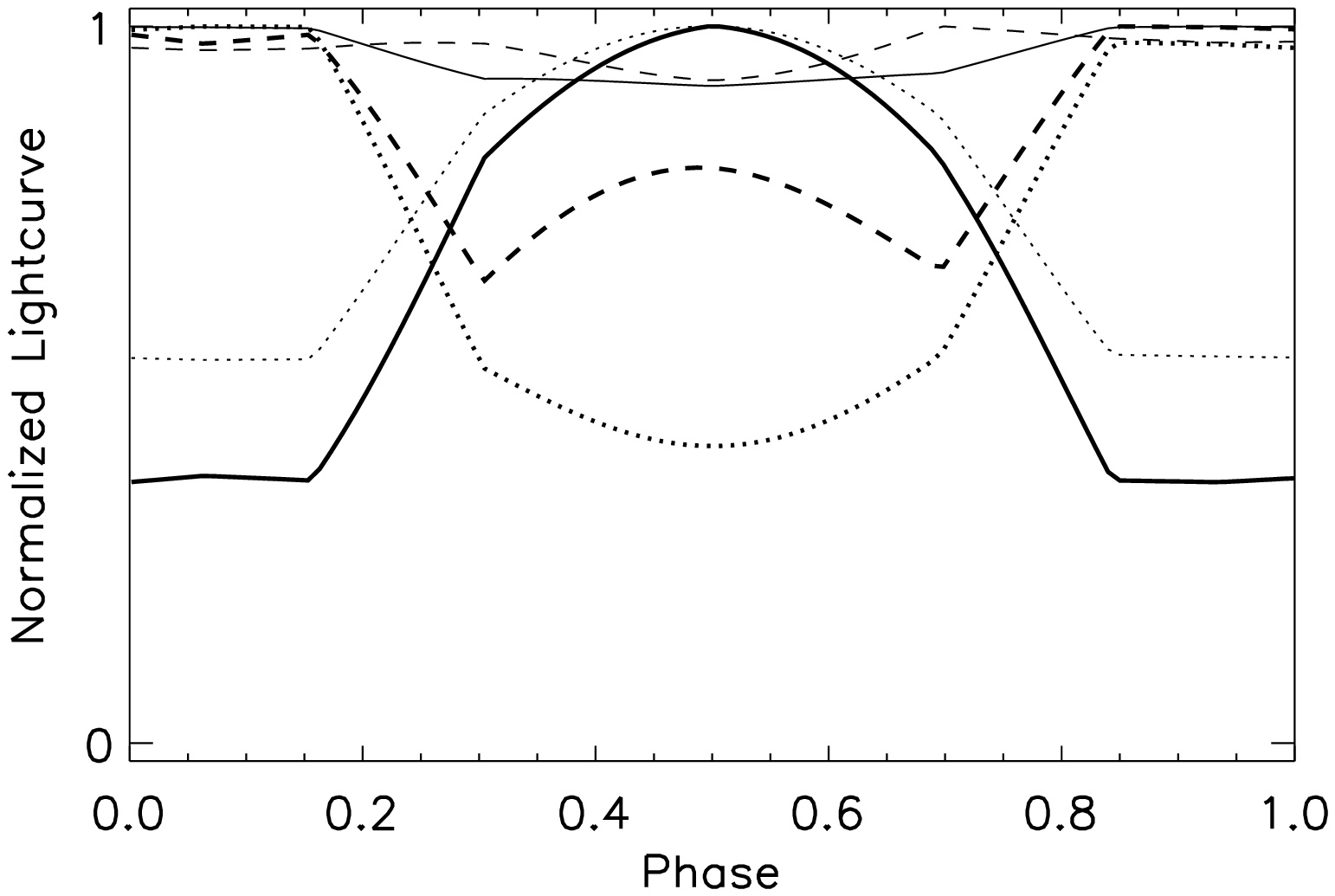}
\caption{Synthetic spectra for model H (upper left panel), model J (upper right), self-similar model (lower left) with the parameters indicated in the figures. Legends indicate the four viewing angles here considered: aligned with north or south pole, just above and just below the equator. The seed blackbody is shown for comparison (solid line). Fourth panel shows the pulse profiles obtained for model H (solid line), model J (dotted), self-similar (dashed), in the energy ranges $0.5$--$10~$keV (thin lines) and $20$--$200~$keV (thick lines), with $\xi=\chi=30^\circ$ (see text).}
\label{fig:dir}
\end{figure}

In the angle-averaged spectrum, the differences due to the magnetic field topology are partially smeared out, and the comparison between synthetic spectra as seen from different viewing angles (first three panels in Fig. \ref{fig:dir}) is more interesting. The comptonization degree reflects the inhomogeneous particle density distribution and therefore  the particular current distribution produces important differences between different viewing angles. Since neutron stars rotate, the study of pulse profiles and phase-resolved spectra can trace the geometric features of the scattering region (tens of stellar radii). In the lower right panel we show the light curves for the three magnetic field models, in bands $0.5$--$10~$keV and $20$--$200$~keV, for an oblique rotator with a certain line of sight ($\xi=\chi=30^\circ$ in \citealt{nobili08a}).

For self-similar models (lower left panel), the thermal part of the spectrum shows a smooth dependence with the viewing angle, which in turn translates into relatively regular light curves. In models H and J, differences in the spectra induced by viewing angles are more pronounced. In these cases, the spectrum is much more irregular and asymmetries are larger. In particular, model H (upper left) shows a softer spectrum when seen from northern latitudes, with important spectral differences. The pulsed fraction of model H is very high in the hard X-ray band, while it is comparable with the self-similar model in the soft range. Model J (upper right) has a more symmetric distribution of currents, and the comptonization degree at different angles depends on the energy band in a non trivial way. This results in large pulsed fraction for both energy ranges and in noteworthy differences between their pulse profiles.

The high energy tail, seen at different colatitudes, can vary by one order of magnitude or more. Investigating the geometry can help to recognize the different components seen in the hard tails via pulse phase spectroscopy. However, comparing the variabilities introduced by changes of $\beta_{bulk}$ (left panel of Fig.~\ref{fig:average_spectra}) and magnetic topology (right panel of Fig.~\ref{fig:average_spectra}, and Fig.~\ref{fig:dir}), the main conclusion is that the kinematic properties of plasma are the main factors affecting the spectrum.

Therefore, the macrophysical approach used in this work to obtain the magnetic field topology should be accompanied by the corresponding microphysical description to determine the velocity distribution of particles, here simply taken as a constant. \cite{beloborodov11,beloborodov13} worked in this direction, with a more consistent treatment of the coupled plasma dynamics and radiative transport in a 1D magnetic loop. The resulting velocity distribution is not trivial and strongly depends on position. A fully consistent, global description of both magnetic field geometry and the spatial and velocity distribution of particles is needed to advance in a more detailed interpretation of magnetar spectra.

\chapter{Magnetic field evolution in neutron stars}\label{ch:magnetic}

The magnetic field evolution in the interior of neutron stars has been extensively studied by a number of authors \citep{baym69,ewart75,sang87,chanmugam89,goldreich92,geppert02,hollerbach02,hollerbach04,cumming04,arras04,pons07b,pons09,gonzalez10,glampedakis11b}.

Soon after birth, a solid crust about 1 km thick is formed. Under these conditions, ions form a lattice and the electrical conduction is governed by the electrons. On the other side, the evolution in the liquid core is very uncertain. We will review the MHD equation for a general description of the problem, but we will mainly focus our attention on the well-understood crust. We will pay special attention to the effects of the Hall term.

\section{MHD of a two-fluid model.}\label{sec:mhd}

In the crust and the core of a neutron star, the hydrostatic forces (gravity and pressure) are much larger than the electromagnetic forces, which can modify only the structure of the envelope. However, the mass contained in this layer is a negligible fraction of the star mass. As a consequence, at first approximation the global structure of the star (i.e., the profiles of density, pressure and composition) is the result of the hydrostatic balance between the pressure, $p$, and the gravity:

\begin{equation}\label{eq:hydro_eq}
 \grad p + \rho\vec{\nabla}\Psi=0~,
\end{equation}
where $\Psi$ is the gravitational potential. In a magnetized plasma with different fluid components, the equation of motion for each $j$ specie (Euler equation) reads:

\begin{equation}\label{eq:motion}
  m_j^\star \left(\frac{\partial \vec{v}_j}{\partial t} + (\vec{v}_j\cdot\vec{\nabla})\vec{v}_j\right) + \frac{1}{n_j}\grad p_j = \vec{f}_j~,\label{eq:euler}\\
\end{equation}
where $m_j^\star,\vec{v}_j,p_j, n_j$ and $\rho_j$ are the effective mass, velocity, pressure, particle density and mass density of the $j$-particles, respectively, and $f_j$ is the sum of all external forces acting on each $j$ particle. Following \cite{goldreich92}, as external forces we have considered the gravity, the Lorentz force, and the frictional forces between the different fluid components, caused by collisions between particles:

\begin{equation}\label{eq:external_forces}
  \vec{f}_j = - m_j^\star \vec{\nabla}\Psi + Z_je \left(\vec{E} + \frac{\vec{v}_j}{c}\times\vec{B}\right) -  m_j^\star \sum_k \frac{\vec{v}_j - \vec{v}_k}{\tau_{jk}}~.
\end{equation}
The frictional forces are characterized by the {\it relaxation time} $\tau_{jk}$, defined as the inverse collision rate of the $j$ particles against $k$ particles. Typical $\tau_{jk}$ are much shorter than the time-scale of velocity variations. This, and the small velocity $\vec{v}_j$ typically considered, allow to neglect the acceleration terms in the left-hand side of the Euler equation (\ref{eq:euler}).

According to thermodynamics, the gradient of pressure in eq.~(\ref{eq:motion}) can be expressed in terms of the chemical potentials $\mu_j$ and the gradient of temperature (supposed to be the same for all species):

\begin{equation}\label{eq:gradp}
 \grad p_j = n_j\grad \mu_j + \left.\frac{\partial p_j}{\partial T}\right|_{\mu_j} \grad T~.
\end{equation}
Hereafter, we consider as fixed the background structure given by eq.~(\ref{eq:hydro_eq}), neglecting deviations from hydrostatic equilibrium ($m_j^\star\grad \Psi$ and $\grad p_j$ terms) in the equations of motion (\ref{eq:motion}). On top of the fixed background we consider the dynamics of two fluids: electrons and ions with electric charge $+Ze$ ($Z=1$ to consider protons). Their equations of motion become

\begin{eqnarray}
&& n_Z Z e \left(\vec{E} + \frac{\vec{v}_Z}{c}\times\vec{B}\right) - n_Z\frac{m_Z^\star}{\tau_{Zn}}\vec{v}_Z  - n_Z\frac{m_Z^\star}{\tau_{Ze}} (\vec{v}_Z-\vec{v}_e)=0~, \label{eq:euler_ions}\\
&& -n_e e \left(\vec{E} + \frac{\vec{v}_e}{c}\times\vec{B}\right) -  n_e\frac{m_e^\star}{\tau_{en}}\vec{v}_e  - n_e\frac{m_e^\star}{\tau_{eZ}} (\vec{v}_e-\vec{v}_Z)=0~. \label{eq:euler_electrons}
\end{eqnarray}
Inside the neutron star the large density implies charge neutrality, $\rho_q\equiv e (Zn_Z - n_e) = 0$, therefore the current density is microscopically defined as

\begin{equation}
 \vec{J} \equiv e (n_Z Z \vec{v}_Z - n_e\vec{v}_e) = en_e(\vec{v}_Z - \vec{v}_e) ~. \label{eq:current_micro}
\end{equation}
The conservation of momentum implies that

\begin{equation}
 n_Z \frac{m_Z^\star}{\tau_{Ze}} = n_e\frac{m_e^\star}{\tau_{eZ}}~.
\end{equation}
Thus, combining eqs.~(\ref{eq:euler_ions}) and (\ref{eq:euler_electrons}), and with the definition (\ref{eq:current_micro}), we obtain

\begin{equation}\label{eq:ambipolar_velocity}
 \frac{\vec{J}\times\vec{B}}{cn_e} = \frac{m_Z^\star}{\tau_{Zn}}\vec{v}_Z + \frac{m_e^\star}{\tau_{en}}\vec{v}_e \equiv \left( \frac{m_Z^\star}{\tau_{Zn}} + \frac{m_e^\star}{\tau_{en}} \right)\vec{v}_{amb}~,
\end{equation}
where we have defined the ambipolar velocity $\vec{v}_{amb}$, interpreted as a weighted mean velocity of the multi-component fluid \citep{goldreich92}. The interaction between electrons and neutrons is much weaker than between electrons and ions (or protons), or between ions (or protons) and neutrons. This implies that

\begin{eqnarray}
 && \frac{m_e^\star}{\tau_{en}}\ll \frac{m_Z^\star}{\tau_{Ze}} ~,\\
 && \frac{m_e^\star}{\tau_{en}}\ll \frac{m_Z^\star}{\tau_{Zn}} ~.
\end{eqnarray}
Thus, neglecting the electron-neutron interaction, we obtain that the ambipolar velocity, eq.~(\ref{eq:ambipolar_velocity}), is due only to the ion (proton) velocity $\vec{v}_{amb}=\vec{v}_Z$. From eqs.~(\ref{eq:euler_ions}) and (\ref{eq:euler_electrons}) we obtain the generalized Ohm's law relating electric field and current, that allows us to write the induction equation as:

\begin{equation}\label{eq:induction_general}
  \frac{\partial \vec{B}}{\partial t} =  - c\vec{\nabla}\times\left(\frac{\vec{J}}{\sigma_0}\right) + \vec{\nabla}\times(\vec{v}_{amb}\times\vec{B}) - \vec{\nabla}\times\left(\frac{\vec{J}\times\vec{B}}{n_e e}\right)~,
\end{equation}
where the we have defined the conductivity as

\begin{equation}\label{eq:sigma0}
 \sigma_0=\frac{n_ee^2}{m_e^\star}\tau_{eZ}~.
\end{equation}
The first term in eq.~(\ref{eq:induction_general}) is the resistive term, the second is the ambipolar diffusion term, and the third one is the Hall term. They are linear, cubic, and quadratic with $B$, respectively. Note that allowing deviations from hydrostatic equilibrium, eq.~(\ref{eq:hydro_eq}), results in the inclusion of additional terms $\propto \vec{\nabla}\mu_i$ and $\propto \vec{\nabla} T$ (see eq.~\ref{eq:gradp}) in eqs.~(\ref{eq:euler_ions}) and (\ref{eq:euler_electrons}). The terms $\propto \grad\mu_i$ are related with a chemical imbalance, the effects of which has been partially discussed, for instance, in \cite{goldreich92}. The term proportional to $\grad T$ provides the thermo-electric contribution \citep{urpin80b,blandford83}. Its effect has been studied in a series of analytical and numerical perturbative studies \citep{geppert91,geppert95,wiebicke91,wiebicke92,wiebicke95,wiebicke96}, and it is thought to be important in the envelope, where the gradient of temperature is strong enough. The study of these terms go beyond the purpose of this work. They are usually thought to be of limited importance for the long-term evolution.

\subsection{Magnetic evolution in the core.}

The dynamics of the magnetic field in the core is not understood. We can consider a core composed by a mixture of neutrons, protons and electrons, where all components are supposed to be completely degenerated and no superfluidity and superconductivity is taken into account. In this case, eq.~(\ref{eq:euler_ions}) with $Z=1$ holds and the fastest time-scale of the ambipolar term can be estimated taking into account the typical relaxation times (\citealt{goldreich92} and references within):
\begin{equation}
 t_{amb} \sim 3\times 10^9~\frac{T_8^2 L_{B,5}^2}{B_{12}^2}~\mbox{yr}~,
\end{equation}
where $T_8=T/10^8$ K, and $L_{B,5}=L_B/10^5$ cm, where $L_B$ is the typical length-scale of magnetic field variations. The conductivity of the core is very large, therefore the Ohmic time-scale is very long \citep{goldreich92}:

\begin{equation}
 t_{ohm}^{core} \sim 2\times 10^{11}~\frac{L_{B,5}^2}{T_8^2}\left(\frac{\rho}{\rho_{nuc}}\right)^3~\mbox{yr}~.
\end{equation}
If the Ohmic decay is assumed to be the only important process in the core, then the magnetic field is basically frozen on cooling time-scales (Myr). The Hall time-scale can be estimated as

\begin{equation}
 t_{hall}^{core} \sim 5\times 10^8~\frac{L_{B,5}^2}{B_{12}}\left(\frac{\rho}{\rho_{nuc}}\right)~\mbox{yr}~.
\end{equation}
A further complication is that protons in the core (or at least in a fraction of its volume) are thought to constitute a type II superconductor. In this picture, the magnetic field can thread the core only by being confined in tiny flux tubes. The dynamics of the fluid with free neutrons, neutron vortices, electrons, superconducting protons and flux tubes is very complicated and the first quantitative studies are on their way. The uncertainty on the coupling between the different components of the fluid makes very hard to guess how this can affect the evolution. In particular, the regime of vortex pinning requires a more sophisticated modeling of detailed vortex/flux tube dynamics \citep{ruderman98,glampedakis11b}. It has been suggested that the magnetic buoyancy would expel the magnetic field. However until now no detailed simulations have been performed.

The detailed study of these mechanisms in the core goes beyond the purpose of this work, in which we will include only Ohmic dissipation.

\section{The relativistic Hall induction equation.}\label{sec_induction}

In the crust, ions form the background reference frame, while electrons form an ideal degenerated gas. Therefore, $\vec{v}_Z=0$ and $\tau_{Zn}=0$ (equivalent to an infinite friction force) in eq.~(\ref{eq:euler_ions}), and the ambipolar velocity (\ref{eq:ambipolar_velocity}) is zero. Alternatively, the lattice of ions can be thought as a component with very large inertia $m_l^\star\rightarrow \infty$. This limit is known as electron MHD (EMHD), and it is apt to describe the solid crust, where electrons are the only moving particles and carry all the current, $\vec{J}=-en_e\vec{v}_e$.

The simplicity of the approach allows to introduce relativistic corrections without complicating the formalism. While in the magnetosphere they are of the order of $\sim 20\%$ at the surface but decay with $r$, inside the neutron star a general relativistic treatment is required. For our purposes, the small structural deformations induced by rotation and the magnetic field can be safely neglected, thus we consider the standard static metric
\begin{equation}\label{eq:metric}
 \de s^2 = - c^2 e^{2\nu(r)}\de t^2 + e^{2\lambda(r)}\de r^2 + r^2\de\theta^2 + r^2\sin^2\de\varphi^2~,
\end{equation}
where $e^{\lambda(r)} = (1 - 2Gm(r)/c^2r)^{-1/2}$ is the space curvature factor, $m(r)$ is the enclosed gravitational mass within radius $r$, and $e^\nu(r)$ is the lapse function that accounts for redshift corrections. The Maxwell equations, presented in \S~\ref{sec:maxwell}, are relativistically modified as follows:

\begin{eqnarray}
 && \vec{\nabla}\cdot\vec{E}=4\pi \rho_q~, \label{eq:poisson}\\
 && \frac{1}{c}\frac{\partial \vec{E}}{\partial t} = \vec{\nabla}\times(e^\nu \vec{B}) - \frac{4\pi}{c}e^\nu \vec{J}~, \label{eq:ampere}\\
 && \vec{\nabla}\cdot\vec{B}=0~, \label{eq:divb} \\
 && \frac{1}{c}\frac{\partial \vec{B}}{\partial t} = - \vec{\nabla}\times(e^\nu \vec{E})~, \label{eq:induction}
\end{eqnarray}
where, hereafter, the $\vec{\nabla}$ operators implicitly take into account the metric factors. Neglecting the displacement current, the relativistic Amp\`ere's law is: 

\begin{equation}\label{eq:current_mhd}
 \vec{J} = e^{-\nu}\frac{c}{4\pi}\curlBrel~,
\end{equation}
so that the electron velocity becomes
\begin{equation}\label{eq:hall_velocity}
 \vec{v}_e= - \frac{ce^{-\nu}}{4\pi e n_e} \curlBrel ~.
\end{equation}
The electron equation of motion (\ref{eq:euler_electrons}) becomes

\begin{equation}\label{eq:euler_electrons_hall}
  n_ee \left(\vec{E} + \frac{\vec{v}_e\times\vec{B}}{c}\right) + m_e^\star n_e\frac{\vec{v}_e}{\tau_e} = 0~.
\end{equation}
The last term is the frictional force per electron, where we have introduced the {\it electron relaxation time} $\tau_e$. In \S~\ref{sec:conductivity} we will explain its relation with the scattering rates with all the particles (ions, impurities, phonons, electrons themselves). We can rewrite eq.~(\ref{eq:euler_electrons_hall}) as a generalized Ohm's law:

\begin{equation}\label{ohm_general}
\vec{J} = \sigma\left(\vec{E}+\frac{\vec{v}_e}{c}\times\vec{B}\right)~,
\end{equation}
where eq.~(\ref{eq:sigma0}) reduces to the electrical conductivity

\begin{equation}\label{eq:electrical_conductivity}
 \sigma=e^2n_e\tau_e/m^\star_e~.
\end{equation}
Its inverse is the resistivity. The electric field is given by

\begin{equation}\label{eq:efield_hall}
 \vec{E} =  \frac{ce^{-\nu}}{4\pi \sigma} \curlBrel + \frac{e^{-\nu}}{4\pi e n_e} (\curlBrel)\times\vec{B}~,
\end{equation}
thus the {\it Hall induction equation} is:

\begin{equation}\label{eq:induction_hall}
 \frac{\partial \vec{B}}{\partial t} =  -\vec{\nabla}\times [\eta \curlBrel + f_h (\curlBrel)\times\vec{B}]~,
\end{equation}
where the {\it magnetic diffusivity} $\eta$, and the Hall pre-factor $f_h$ are defined as

\begin{eqnarray}
 && \eta=\frac{c^2}{4\pi \sigma}~, \label{eq:magnetic_diffusivity}\\
 && f_h= \frac{c}{4\pi e n_e}~. \label{eq:hall_coefficients}
\end{eqnarray}
The first term on the right-hand side of eq.~(\ref{eq:induction_hall}) accounts for Ohmic dissipation. The magnetic energy converted into heat per unit time and unit volume measured by the Eulerian observer is

\begin{equation}\label{eq:def_joule}
 {\cal Q}_j = \frac{J^2}{\sigma} = \frac{4\pi}{c^2}\eta J^2~. 
\end{equation}
This heat source is thought to maintain strongly magnetized neutron stars hot longer than the weakly magnetized ones \citep{haensel90,page00,tauris01,pons09}. In a neutron star crust, $\sigma$ is dominated by the electronic transport and depends on the electron density, the crustal temperature, and the impurity concentration within the crust (see Chapter~\ref{ch:cooling}). Since the electron density varies over about four orders of magnitude in the crust and the temperature decreases by about two to three orders of magnitude in a pulsar's lifetime, the electric conductivity may vary both in space and time by many orders of magnitude. Therefore, to assume a uniform Ohmic decay time, independent of the location of currents and the pulsar age is misleading: Ohmic decay in a neutron star crust cannot be described by a single exponential law (see e.g. \citealt{page00} for a qualitative discussion), and an important role is played by the resistivity of the region where currents are placed. 

The second term is the Hall term. When it dominates, electric currents are squeezed in a smaller volume and move through the crust, creating small scale structures and allowing the energy interchange between poloidal and toroidal components of the magnetic field. To compare the relative importance of the two terms, we can write the Hall induction equation (\ref{eq:induction_hall}) as follows:
\begin{equation}\label{eq:induction_omtau}
 \frac{\partial \vec{B}}{\partial t} = -\vec{\nabla}\times \left\{\eta \left[\curlBrel + \omtau\frac{(\curlBrel)\times\vec{B}}{B}  \right]\right\}~,
\end{equation}
where $\omega_B=eB/m^\star_ec$ is the gyro-frequency of electrons (see eq.~\ref{eq:def_gyrofrequency}), and the {\it magnetization parameter},

\begin{equation}\label{eq:omtau}
 \omtau=\frac{cB}{4\pi en_e \eta}=\frac{B\sigma}{en_e c}~,
\end{equation}
is an indicator of the relative importance between the two terms on the right-hand side. From a point of view of the helicoidal motion of electrons along magnetic field lines, $\omtau$ represents the average number of gyro-rotations between two collisions. If $\omtau\ll 1$, collisions dominate, the induction equation is dominated by the Ohmic diffusive term, and the effect of magnetic field on transport properties is negligible. If $\omtau\gg 1$, the mean free path of electrons in their helicoidal motion is large, and they transport current and heat anisotropically. 

The Hall term plays an important role in the evolution of neutron stars with large magnetic fields, $B\gtrsim 10^{14}$ G. The Hall-driven evolution is probably at the origin of the observed strong activity in young {\it magnetars}. Therefore, the interest in modeling the internal evolution of the magnetic field is continuously growing \citep{pons07b,hoyos08,pons09,shabaltas12}. For typical temperatures (a few $10^8$ K) in a middle-age neutron star of $10^3-10^5$ yr, the Hall term dominates over Ohmic dissipation when the magnetic field is $> 10^{14}$ G. However, as the star cools down and the Ohmic time-scale becomes very long (billions of years), the Hall term may dominate even for much lower values of the magnetic field strength.

A further generalization is the introduction of an advective term, for which the electron velocity, eq.~(\ref{eq:hall_velocity}), becomes $\vec{v}_e  \rightarrow \vec{v}_e+\vec{v}_a$. Thus, the induction equation becomes:

\begin{equation}\label{eq:induction_hall_adv}
 \frac{\partial \vec{B}}{\partial t} =  -\vec{\nabla}\times [\eta \curlBrel + f_h (\curlBrel)\times\vec{B} + e^\nu\vec{v}_a\times\vec{B}]~.
\end{equation}
Possible physical scenarios including an advective term are the magnetic buoyancy in the core, or accretion in the solid crust (see \S~\ref{sec:submergence}). Below, we analyse in detail the Hall term.

\subsection{Analysis of the non diffusive Hall induction equation.}

We consider the Hall part of the induction equation, setting $\eta=0$ in eq.~(\ref{eq:induction_hall}), and splitting it into poloidal and toroidal parts:

\begin{eqnarray}
 && \partial_t \vec{B}_{pol} =  -\vec{\nabla}\times [f_h \vec{j}_{pol}\times\vec{B}_{pol}]~, \label{eq:hall_part_pol}\\
 && \partial_t \vec{B}_{tor} =  -\vec{\nabla}\times [f_h (\vec{j}_{pol}\times\vec{B}_{tor}-\vec{j}_{tor}\times\vec{B}_{pol})]~, \label{eq:hall_part_tor}
\end{eqnarray}
where we have defined for conciseness

\begin{equation}
 \vec{j}=\curlBrel= \frac{4\pi e^\nu}{c} \vec{J}~.
\end{equation}
We can see immediately that the poloidal equation (\ref{eq:hall_part_pol}) contains only terms coupling both poloidal and toroidal, while the toroidal equation (\ref{eq:hall_part_tor}) contains a self-coupling term and a term $\propto B_{pol}^2$. This means that if we start with a pure toroidal magnetic field, no poloidal components will arise. On the contrary, with an initially pure poloidal magnetic field, a toroidal component will develop.

\subsubsection{Toroidal self-coupling term: Burgers-like behaviour.}

We begin to analyse the self-coupling term in eq.~(\ref{eq:hall_part_tor}) in the purely toroidal limit, $\vec{B}_{pol}=0$:

\begin{eqnarray}
 \partial_t \vec{B}_{tor} && = -\rot [f_h \rot(e^\nu\vec{B}_{tor})\times \vec{B}_{tor}]~, \\
 \vec{B}_{tor}\cdot \partial_t \vec{B}_{tor} && = -\vec{B}_{tor}\cdot\rot [f_h \vec{\nabla}\times(e^\nu\vec{B}_{tor})\times \vec{B}_{tor}] = \nonumber\\
 && = \dive\{ f_h \vec{B}_{tor} \times [(\rot(e^\nu\vec{B}_{tor}))\times \vec{B}_{tor}]\} + \nonumber\\
 && - f_h[(\rot(e^\nu\vec{B}_{tor}))\times\vec{B}_{tor}]\cdot (\rot\vec{B}_{tor}) = \nonumber\\
 && = (\rot (e^\nu\vec{B}_{tor})) \cdot \vec{\nabla}(f_h B_{tor}^2) + f_h B_{tor}^2 (\vec{\nabla}\times \vec{B}_{tor})\cdot\vec{\nabla}e^\nu~,
\end{eqnarray}
where we have used several standard vectorial identities. In spherical coordinates:\footnote{See \citealt{pons07b} for a more straightforward formulation in cylindrical coordinates, without relativistic corrections.}

\begin{eqnarray}
 \partial_t B_\varphi && = \frac{1}{B_\varphi}\left\{ \frac{1}{r\sin\theta}\derpar{\theta}(B_\varphi e^\nu \sin\theta)\frac{1}{e^\lambda}\derpar{r}(f_hB_\varphi^2) - \frac{1}{r^2e^\lambda}\derpar{r}(rB_\varphi e^\nu) \derpar{\theta}(B_\varphi^2 f_h)\right\} + \nonumber\\
 && + \frac{f_h}{e^\lambda}\derparn{r}{e^\nu}B_\varphi\frac{1}{r\sin\theta}\derpar{\theta}(B_\varphi\sin\theta) = \nonumber\\
 && = 2\frac{\cos\theta}{r\sin\theta}\frac{e^\nu}{e^\lambda}\derpar{r}\left(\frac{f_h B_\varphi^2}{2}\right) + \frac{1}{e^\lambda}\left( e^\nu\derparn{r}{f_h} - f_h\derparn{r}{e^\nu} - 2\frac{f_h}{r}e^\nu\right)\frac{1}{r}\derpar{\theta}\left(\frac{B_\varphi^2}{2}\right) \nonumber\\
 && + \frac{\cos\theta}{r \sin\theta}\frac{1}{e^\lambda}\derparn{r}{e^\nu}f_hB_\varphi^2 ~,
\end{eqnarray}
which can be written in the compact form

\begin{equation}
 \frac{\partial B_\varphi}{\partial t} = \frac{2\cos\theta}{r \sin\theta e^\lambda}\derpar{r}\left(\frac{f_h e^\nu B_\varphi^2}{2}\right) + \frac{r^2 e^{2\nu}}{e^\lambda}\derpar{r}\left( \frac{f_h}{e^\nu r^2}\right)\frac{1}{r}\derpar{\theta}\left(\frac{B_\varphi^2}{2}\right)~.
\end{equation}
We can write it in a compact form:
\begin{equation}\label{eq:induction_burgers}
 \frac{\partial B_\varphi}{\partial t} + \lambda_r\frac{1}{f_h e^\lambda}\derpar{r}\left(\frac{f_h e^\nu B_\varphi^2}{2}\right) + \lambda_\theta\frac{1}{r}\derpar{\theta}\left(\frac{B_\varphi^2}{2}\right) = 0~,
\end{equation}
with

\begin{eqnarray}
 && \lambda_r = - 2f_h \frac{\cot\theta}{r}~, \label{eq:induction_burgers_lambda1}\\
 && \lambda_\theta = - r^2\frac{e^{2\nu}}{e^\lambda}\derpar{r}\left( \frac{f_h}{e^\nu r^2}\right)~.\label{eq:induction_burgers_lambda2}
\end{eqnarray}
Eq.~(\ref{eq:induction_burgers}) is hyperbolic and resembles a multidimensional inviscid Burgers equation for a fluid with velocity $u$:

\begin{equation}
 \frac{\partial \vec{u}}{\partial t} + (\vec{u}\cdot\vec{\nabla})\vec{u} = 0~.
\end{equation}
In eq.~(\ref{eq:induction_burgers}), $f_h e^\nu B_\varphi^2/2$ (or simply $B_\varphi^2/2$ for the $\theta$-direction) is interpreted as the flux, and $\lambda_rB_\varphi$ and $\lambda_\theta B_\varphi$ have velocity dimensions.

A natural outcome of the presence of Burgers-like terms is the formation of discontinuities in the magnetic field components (analogously to shocks in hydrodynamics), which imply the formation of {\em current sheets}. The location where they form depend on the magnetic field geometry. When we consider a neutron star crust, the gradient of the term in $\lambda_\theta$ is dominated by the stratification. The role played by a charge density gradient ($f_h=f_h(r)$ in our case) in the Burgers-like term was already noted by \cite{vainshtein00} in planar geometry. Furthermore, even in the Newtonian case, with $e^\nu=e^\lambda=1$, and in absence of stratification (constant $f_h$), $\lambda_r$ and $\lambda_\theta$ do not vanish due to the spherical geometrical terms, as already pointed out by \cite{pons07b}.

\subsubsection{Poloidal-toroidal coupling.}

If $\vec{B}_{pol}\neq 0$, the toroidal magnetic field equation~(\ref{eq:hall_part_tor}) contains also the following coupling term:

\begin{eqnarray}
 && \vec{\nabla}\times [f_h (\vec{j}_{tor}\times\vec{B}_{pol})] = - f_h\rot(\vec{j}_{tor}\times\vec{B}_{pol})-\grad f_h\times(\vec{j}_{tor} \times \vec{B}_{pol}) = \nonumber\\
 && = - f_h[(\vec{B}_{pol}\cdot \grad)\vec{j}_{tor} - (\vec{j}_{tor}\cdot \grad)\vec{B}_{pol}] - \grad f_h\times (\vec{j}_{tor} \times \vec{B}_{pol}) = \nonumber \\
 && = - f_h\left[\frac{B_r}{e^\lambda}\derparn{r}{j_\varphi} + \frac{B_\theta}{r}\derparn{\theta}{j_\varphi} - \frac{j_\varphi}{r\sin\theta}(B_r\sin\theta+B_\theta\cos\theta)\right] - \derparn{r}{f_h}j_\varphi \frac{B_r}{e^\lambda} = \nonumber\\
 && = - f_h\left[\frac{B_r}{e^\lambda}\derparn{r}{j_\varphi} + \frac{B_\theta}{r}\derparn{\theta}{j_\varphi} - \frac{j_\varphi}{r}\cot\theta B_\theta)\right] - r\derpar{r}\left(\frac{f_h}{r}\right)j_\varphi \frac{B_r}{e^\lambda}~,
\end{eqnarray}
and eq.~(\ref{eq:hall_part_tor}) becomes:

\begin{eqnarray}
 && \frac{\partial B_\varphi}{\partial t} + \lambda_r\frac{1}{f_h e^\lambda}\derpar{r}\left(\frac{f_h e^\nu B_\varphi^2}{2}\right) + \lambda_\theta\frac{1}{r}\derpar{\theta}\left(\frac{B_\varphi^2}{2}\right) = \nonumber \\
 && - f_h\left[\frac{B_r}{e^\lambda}\derparn{r}{j_\varphi} + \frac{B_\theta}{r}\derparn{\theta}{j_\varphi} - \frac{j_\varphi}{r}\cot\theta B_\theta\right] - \frac{r}{e^\lambda}\derpar{r}\left(\frac{f_h}{r}\right)j_\varphi B_r~.
\end{eqnarray}
Here, the poloidal magnetic field acts as a quadratic source term (it appears in the $\vec{j}$ terms). Its evolution is more complicated, as it includes higher order derivatives in the non-linear terms. The poloidal equations (\ref{eq:hall_part_pol}) read
\begin{eqnarray}
 \frac{\partial \vec{B}_{pol}}{\partial t} & = & -f_h\rot[(\curlBrel_{tor})\times\vec{B}_{pol}]-\grad f_h\times[(\curlBrel_{tor})\times \vec{B}_{pol}] = \nonumber\\
 & = & - f_h[(\vec{B}_{pol}\cdot \grad)\vec{j}_{pol} - (\vec{j}_{pol}\cdot \grad)\vec{B}_{pol}] - \grad f_h\times [\vec{j}_{pol}\times \vec{B}_{pol}]~.
\end{eqnarray}
Note that $f_h(\vec{j}_{pol}\cdot \grad)\vec{B}_{pol}$ is an advective term for $\vec{B}_{pol}$. Developing further, we obtain:

\begin{eqnarray}
 \frac{\partial \vec{B}_{pol}}{\partial t} & = & - f_h\left[\left(\frac{B_r}{e^\lambda}\derparn{r}{j_r} + \frac{B_\theta}{r}\derparn{\theta}{j_r}\right)\hat{r} + \left(\frac{B_r}{e^\lambda}\derparn{r}{j_\theta} + \frac{B_\theta}{r}\derparn{\theta}{j_\theta}\right)\hat{\theta} + \frac{B_\theta}{r}j_r\hat{\theta} + \right. \nonumber\\
 && \left. - \left(\frac{j_r}{e^\lambda}\derparn{r}{B_r} + \frac{j_\theta}{r}\derparn{\theta}{B_r}\right)\hat{r} - \left(\frac{j_r}{e^\lambda}\derparn{r}{B_\theta} + \frac{j_\theta}{r}\derparn{\theta}{B_\theta}\right)\hat{\theta} - \frac{B_r}{r}j_\theta\hat{\theta}\right] + \nonumber\\
 && + e^{-\lambda}\derparn{r}{f_h}(j_r B_\theta - j_\theta B_r)\hat{\theta}~.
\end{eqnarray}
Therefore, we can write the two poloidal magnetic field equations in this form:

\begin{eqnarray}
 \derparn{t}{B_r} + v_r\derparn{r}{B_r} + \frac{v_\theta}{r}\derparn{\theta}{B_r} & = & -f_h\left[\frac{B_r}{e^\lambda}\derparn{r}{j_r} + \frac{B_\theta}{r}\derparn{\theta}{j_r} \right]~,\\
 \derparn{t}{B_\theta} + v_r\derparn{r}{B_\theta} + \frac{v_\theta}{r}\derparn{\theta}{B_\theta} & = & -f_h\left[\frac{B_r}{e^\lambda}\derparn{r}{j_\theta} + \frac{B_\theta}{r}\derparn{\theta}{j_\theta} \right] + \nonumber\\
 && + \frac{r}{e^\lambda}\derpar{r}\left(\frac{f_h}{r}\right)(j_r B_\theta - j_\theta B_r)~,
\end{eqnarray}
where

\begin{eqnarray}
 && v_r      = -f_h j_r      = -f_h[\rot(e^\nu \vec{B})]\cdot\hat{r}~, \\
 && v_\theta = -f_h j_\theta = -f_h[\rot(e^\nu \vec{B})]\cdot\hat{\theta}~.
\end{eqnarray}
We can rewrite the non-resistive Hall induction equation ($\eta=0$ in eq.~\ref{eq:induction_hall}) as follows:

\begin{equation}\label{eq:hall_general}
 \derparn{t}{B_i} + \frac{v_i^r}{e^\lambda}\derparn{r}{{\cal F}_i} + \frac{v_i^\theta}{r}\derparn{\theta}{{\cal F}_i} = - \frac{c}{4\pi e n_e}\left(\derparn{r}{j_i}B_r + \derparn{\theta}{j_i}\frac{B_\theta}{r}\right) + {\cal S}_i~,
\end{equation}
where:
\begin{eqnarray}\label{eq:hall_general_vel}
 && v_r^r = v_\theta^r = -\frac{c}{4\pi e n_e}j_r~,\\
 && v_r^\theta = v_\theta^\theta = -\frac{c}{4\pi e n_e}j_\theta~, \\
 && v_\varphi^r = - \frac{c}{2\pi e} \frac{\cot\theta}{r}~, \\
 && v_\varphi^\theta = - \frac{c n_e}{4\pi e} r^2 e^{\nu-\lambda}\derpar{r}\left( \frac{1}{n_e e^\nu r^2}\right)~,
\end{eqnarray}
and the flux terms ${\cal F}_i$ are Burgers-like for the toroidal component ($i=\varphi$), or advective terms with velocities $v^j_i$ in the $j$ direction for the poloidal components ($i=r,\theta$):

\begin{eqnarray}\label{eq:hall_general_fluxes}
 && {\cal F}_r=B_r ~,\\
 && {\cal F}_\theta=B_\theta~,\\
 && {\cal F}_\varphi=\frac{e^\nu B_\varphi^2}{2 n_e}~.
\end{eqnarray}
The source terms ${\cal S}_i$ are non-linear in $B$, being proportional to a poloidal component of the magnetic field, and involving the derivatives of currents, geometrical factors and density:
\begin{eqnarray}\label{eq:hall_general_sources}
 && {\cal S}_r=0 ~,\\
 && {\cal S}_\theta= \frac{c}{4\pi e}\frac{r}{e^\lambda}\derpar{r}\left(\frac{1}{n_e r}\right)(j_rB_\theta - j_\theta B_r)~, \\
 && {\cal S}_\varphi= \frac{c}{4\pi e n_e}\frac{j_\varphi}{r}\cot\theta B_\theta - \frac{c}{4\pi e}\frac{r}{e^\lambda}\derpar{r}\left(\frac{1}{n_e r}\right)j_\varphi B_r~.
\end{eqnarray}
While the Burgers and advective terms are easy to treat with upwind methods, the source terms on the right-hand side couples in a non-trivial way poloidal and toroidal magnetic fields. Numerically, it is the most difficult part to deal with.

From this analysis, we learn that, in the case of Hall-dominated evolution, the induction equation consists of a hyperbolic term, plus a doubly constrained multidimensional advection term because, in addition to the $\vec{\nabla}\cdot\vec{B}=0$ constraint, the velocity field is proportional to the current, i.e., to the derivatives of the magnetic field (eq.~\ref{eq:hall_velocity}). This implies non-linearity and coupling between poloidal and toroidal components. An initially pure toroidal magnetic field will not develop a poloidal component, but any poloidal magnetic field will induce the formation of toroidal components. When both components are present, magnetic energy can be transferred between them. Although the Hall term itself conserves energy, the plausible creation of small-scale structures may accelerate the Ohmic dissipation.

%%%%%%%%%%%%%%%%%%%%%%%%%%%%%%%%%%%%
\section{The magnetic evolution code.}\label{sec:magnetic_code}

After having introduced and analysed the Hall induction equation, we now deal with the applications to neutron stars. The numerical challenge is to be able to follow the evolution of a system in which important parameters (density, resistivity) vary several orders of magnitude across the spatial domain, but also during the temporal evolution. A multi-purpose code must be able to work in both the purely diffusive regime ($\omtau=0$) and in the limit of large $\omtau$ when the Hall term dominates. It must also be stable to follow the evolution for many diffusion time-scales (up to hundreds of Hall time-scales).  Historically, spectral methods had been used to solve the Hall induction equation in simplified constant density layers \citep{hollerbach02}, but such attempts were always restricted to a magnetization parameter not exceeding 200, since fully spectral codes systematically have unsurmountable problems dealing with structures where discontinuities or very large gradients of the variables appear, which is a natural consequence of the equations, as seen in the previous section. \cite{pons07b} presented a code solving the Hall induction equation in a realistic crust using an alternative approach (spectral in angles but finite differences in the radial direction). This was a significant improvement to previous work, but still could not work in the limit of vanishing electrical resistivity. For this reason, the only long-term fully coupled 2D magneto-thermal evolution simulations available up to now \citep{pons09} were restricted to the purely diffusive case. 

Here we present a new code based on upwind, finite difference schemes that can handle the Hall term in the induction equation for vanishing physical resistivity. These numerical methods are of very general use, from problems involving the simple 1D Burgers equation to high resolution shock capturing schemes successfully used in MHD problems \citep{anton06,giacomazzo07,cerdaduran08}. The formation of current sheets is properly modeled, and the code can follow the evolution of complex geometries with large gradients and discontinuities overcoming intrinsic problems of previous studies.

%%%%%%%%%%%%%%%
\begin{figure}
 \centering
\includegraphics[width=.5\textwidth]{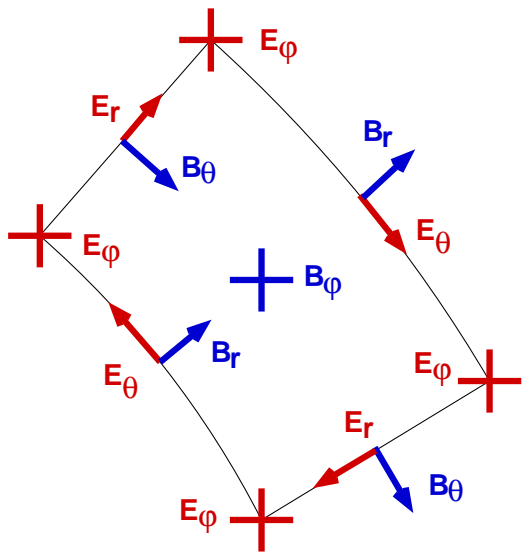}
\caption{Location of the variables on a staggered grid in spherical coordinates and for the axisymmetric case. Solid lines delimit the edges of the surface $S_\varphi$.} 
 \label{fig:staggered}
\end{figure}
%%%%%%%%%%%%%%%

%%%%%%%%%%%%%%%%%%%%%%%%%%%%%%
\subsection{The numerical staggered grid.}
Staggered grids \citep{yee66} are commonly used with finite difference time domain methods \citep{taflove75} to solve the Maxwell's equations. The magnetic field components are defined at the center, and normal to each face of the cell, while the components of electric field and current are defined along the edges. Our scheme ensures a correct propagation of waves and the preservation (by construction) of the divergence constraint (see Appendix~\ref{app:fdtd} for a derivation).

As an example, we show in Fig.~\ref{fig:staggered} the location of the variables in a numerical cell in spherical coordinates and assuming axial symmetry. In this case our grid maps a meridional section of the star in spherical coordinates, but below we also present tests in Cartesian coordinates.

%%%%%%%%%%%%%%%%%%%%%%%%%%%%%%
\subsection{Cell reconstruction and upwind method.}\label{sec:method_hall}

A simple centered difference scheme is not enough to resolve the current sheets that are typical of Hall MHD. For this reason, we decide to apply an upwind method together with the cell reconstruction.

The original Godunov's method is well known for its ability to capture discontinuous solutions, but it is only first-order accurate. This method can be easily extended to give second-order spatial accuracy on smooth solutions, but still avoiding non-physical oscillations near discontinuities. To achieve higher order accuracy, we can use a reconstruction procedure that improves the piecewise constant approximation. The simplest choice is a piecewise linear function in each cell. A very popular choice for the slopes of the linear reconstructed function is the {\it monotonized central-difference limiter} (MC; \citealt{vanleer77}). Given three consecutive points $x_{i-1},x_i,x_{i+1}$ on a numerical grid, and the numerical values of the function $f_{i-1},f_i,f_{i+1}$, the reconstructed function within the cell $i$ is given by $f(x)=f(x_i)+ \alpha (x-x_i)$, where the slope is
\begin{equation}
\alpha = {\rm minmod}\left( \frac{f_{i+1}-f_{i-1}}{x_{i+1}-x_{i-1}},2\frac{f_{i+1}-f_{i}}{x_{i+1}-x_{i}},
2\frac{f_{i}-f_{i-1}}{x_{i}-x_{i-1}}\right). 
\end{equation}
The ${\rm minmod}$ function of three arguments is defined by
\begin{equation}
{\rm minmod}(a,b,c) = \left\{ 
\begin{array}{cc}
{\rm min}(a,b,c) & {\rm if} ~a,b,c>0 \\
{\rm max}(a,b,c) & {\rm if} ~a,b,c<0 \\
0 & {\rm otherwise} 
\end{array}~.
\right.\nonumber
\end{equation}
This method has the advantage to behave like a normal centered difference where the function is smooth, while it approaches a purely upwind method when there is a strong discontinuity. We then reconstruct the magnetic field circulation elements, $C_k= B_k l_k$. From this, we directly obtain the components of $(\curlBrel)$ by means of Stokes' theorem and, when needed, we recover $B_x$ dividing the reconstructed circulation by the local length element.

Current and electric field components are always defined at the same location, but the Hall term includes products of tangential components of the magnetic field and velocity not always defined at the same edges. For such terms, we evaluate the needed components of the velocity by linear interpolation of the closest neighbor values and we take the {\it upwind} components $\vec{B}^w$ of the reconstructed value of magnetic field at each interface.

Therefore, at each interface, we look at the direction of velocity field to pick up the value at one side or the other of the interface. This value enters in the definition of the velocity induced electric field. For a given velocity field $\vec{v}$, we choose the corresponding value of $B_\varphi^w$ to calculate the electric field components at the $(i,j)$ node:
\begin{eqnarray}\label{wind_bphi_er}
&& E_r^{(i,j)}= - v_\theta^{(i,j)}B_\varphi^w + v_\varphi^{(i,j)}B_\theta^{(i,j)}~, \qquad \mbox{ with} \nonumber \\
&& B_\varphi^w=\left\{ 
\begin{array}{cc}
 B_\varphi^{(i+1,j)} & \mbox{ if} ~v_\theta^{(i,j)} < 0 \\
 B_\varphi^{(i-1,j)} & \mbox{ if} ~v_\theta^{(i,j)} > 0 \\
\end{array}\right. ~;
\end{eqnarray}

\begin{eqnarray}\label{wind_bphi_etheta}
&& E_\theta^{(i,j)}= v_r^{(i,j)}B_\varphi^w - v_\varphi^{(i,j)}B_r^{(i,j)} ~, \qquad \mbox{ with} \nonumber  \\
&& B_\varphi^w=\left\{ 
\begin{array}{cc}
 B_\varphi^{(i,j+1)} & \mbox{ if} ~v_r^{(i,j)} < 0 \\
 B_\varphi^{(i,j-1)} & \mbox{ if} ~v_r^{(i,j)} > 0 \\
\end{array}\right. ~;
\end{eqnarray}

\begin{eqnarray}\label{wind_etor}
&& E_\varphi^{(i,j)}= - v_r^{(i,j)}B_\theta^w + v_\theta^{(i,j)}B_r^w ~, \qquad \mbox{ with} \nonumber  \\
&& B_\theta^w=\left\{ 
\begin{array}{cc}
 B_\theta^{(i,j+1)} & \mbox{ if} ~v_r^{(i,j)} < 0 \\
 B_\theta^{(i,j-1)} & \mbox{ if} ~v_r^{(i,j)} > 0 \\
\end{array}\right. ~,\nonumber\\
&& B_r^w =\left\{ 
\begin{array}{cc}
 B_r^{(i+1,j)} & \mbox{ if} ~v_\theta^{(i,j)} < 0 \\
 B_r^{(i-1,j)} & \mbox{ if} ~v_\theta^{(i,j)} > 0 \\
\end{array}\right. ~.
\end{eqnarray}
While upwind methods work well for any independent velocity (e.g., advection velocity), in our case the velocity $\vec{v}=-f_h\curlBrel$ depends on the magnetic field itself. The key point is that, on the staggered grid, the electric field has to be calculated by the cross product between velocity \textit{in} the same point and the upwind value of $\vec{B}$.

\begin{figure}
 \centering
 \includegraphics[width=.45\textwidth]{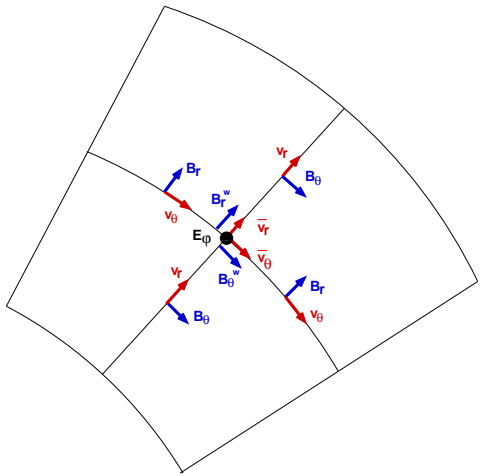}
\caption{Illustration of the procedure of calculation of the electric field: location of the 
components of velocity (red arrows) and magnetic field (blue) involved in the definition of the Hall term of $E_\varphi$ (black dot).}
 \label{fig:ewind}
\end{figure}

In Fig.~\ref{fig:ewind} we explicitly show the location, on the staggered grid, of $E_\varphi$ (black point) and the quantities needed for its evaluation. The average values of $\bar{v}_r$ and $\bar{v}_\theta$ are calculated taking the average of the two closest neighbors. In the example, both $\bar{v}_r$ and $\bar{v}_\theta$ are positive, therefore the upwind values of  $B_r^w$ and $B_\theta^w$ are the reconstructed values from the bottom and left sides, respectively.

%%%%%%%%%%%%%%%
%%%%%%%%%%%%%%%

%%%%%%%%%%%%%%%%%%%%%%%%%%%%%%
\subsection{Courant condition.}

We use an explicit, first-order time advance method with some intermediate corrections to improve the stability of the scheme. In explicit algorithms, the time-step is limited by the Courant condition, that avoids that any wave travels more than one cell length on each time-step. Since we want to evolve the system on long (Ohmic) time-scales, the Courant condition is particularly restrictive when $\omtau\gg 1$. At each time-step, we estimate the Courant time, $t_c$, as the minimum value, among all cells $(i,j)$, of 
\begin{equation}\label{estimate_courant_hall}
 t_c^{(i,j)} = \left.\frac{\Delta l}{f_h|\vec{\nabla} \times \vec{B}|}\right|_{(i,j)}, \end{equation}
where $\Delta l$ is the minimum length of the cell edges in any direction. We use a time-step
\begin{equation}\label{eq:timestep}
 \Delta t=k_c t_c~,
\end{equation}
where $k_c$ is a factor $< 1$. The time-step can vary by orders of magnitude during the evolution, becoming very small when the Hall term dominates and, in particular, where locally strong current sheets are formed. In typical situations with $B\gtrsim 10^{14}$ G, the Courant condition limits the time-step between $10^{-3}$ and 1 yr for a number of grid points of $50\times 50$.

\subsection{Time advance and hyper-resistivity}

After calculating all the line integrals of $\vec{E}$ along all edges as explained in Appendix~\ref{app:grid}, the time advance proceeds in two different steps. First we advance the toroidal component of the magnetic field, with a particular treatment of the quadratic term in $B_{tor}$ (see \S~\ref{sec_burgers}). With the updated value of this component, the electric field components are recalculated and then used to advance the poloidal components of the magnetic field. Schematically, the sequence of the time advance from $t_n$ to $t_{n+1}=t_n+\Delta t$ is the following:
\begin{itemize}
\item starting from $\vec{B}^n$, all currents and electric field components are calculated \\ 
$\vec{B}^n\rightarrow \vec{J}^n\rightarrow \vec{E}^n$;
\item $\vec{B}_{tor}^{n}$ is updated: $\vec{E}^n \rightarrow \vec{B}_{tor}^{n+1}$;
\item the new values $\vec{B}_{tor}^{n+1}$ are used to calculate the modified current components and $\vec{E}_{tor}$: \\
$\vec{B}_{tor}^{n+1} \rightarrow \vec{J}_{pol}^\star \rightarrow \vec{E}_{tor}^\star$;
\item finally, we use the values of $\vec{E}_{tor}^\star$ to update the remaining magnetic field components \\
$\vec{E}_{tor}^\star \rightarrow \vec{B}_{pol}^{n+1}$.
\end{itemize}
This two-step advance favors the stability of the method, as already pointed out for a 3D problem in Cartesian coordinates by \cite{osullivan06}. \cite{toth08} further discussed that the two-stage formulation is equivalent to introduce a fourth-order hyper-resistivity term. In our case, since the toroidal component is advanced explicitly, the hyper-resistive correction only acts on the evolution of the poloidal components. Now we derive this correction, neglecting for conciseness the relativistic factors. Given a magnetic field $\vec{B}^n$ at time $t_n$, we have:

\begin{eqnarray}
 && \vec{J}_{pol}^n = \vec{\nabla}\times\vec{B}_{tor}^n~, \nonumber\\
 && \vec{J}_{tor}^n = \vec{\nabla}\times\vec{B}_{pol}^n~, \nonumber\\
 && \vec{E}_{pol}^n = \eta\vec{J}_{pol}^n + f_h (\vec{J}_{tor}^n\times\vec{B}_{pol}^n - \vec{J}_{pol}^n\times\vec{B}_{tor}^n)~, \nonumber\\
 && \vec{B}_{tor}^{n+1} = \vec{B}_{tor}^n - c\Delta t(\vec{\nabla}\times\vec{E}_{pol}^n)~, \\
 && \vec{J}_{pol}^{n+1} = \vec{\nabla}\times\vec{B}_{tor}^{n+1}~, \nonumber\\
 && \vec{E}_{tor}^* = \eta\vec{J}_{tor}^n + f_h (\vec{J}_{pol}^{n+1}\times\vec{B}_{pol}^n)~, \nonumber\\
 && \vec{B}_{pol}^{n+1} = \vec{B}_{pol}^n - c\Delta t(\vec{\nabla}\times\vec{E}_{tor}^*)~.\nonumber
\end{eqnarray}
The intermediate toroidal electric field can be written considering the contribution $\delta\vec{E}_{tor}$ given by the updated current $\vec{J}_{pol}^{n+1}=\vec{J}_{pol}^n + \delta \vec{J}_{pol}^* = \vec{\nabla}\times(\vec{B}_{tor}^n + c\Delta t \partial_t\vec{B}_{tor})$:
\begin{equation}
 \vec{E}_{tor}^* = \vec{E}_{tor}^n + \delta\vec{E}_{tor}~,
\end{equation}
where
\begin{eqnarray}
 \vec{E}_{tor}^n & = & \eta\vec{J}_{tor}^n + f_h [(\vec{\nabla}\times\vec{B}_{tor}^n)\times\vec{B}_{pol}^n]~,\\
 \delta\vec{E}_{tor} & = & f_h (\delta\vec{J}_{pol}\times\vec{B}_{pol}^n) = f_h [(\vec{\nabla}\times\delta\vec{B}_{tor})\times\vec{B}_{pol}^n] = \nonumber\\
 & = & -c\Delta t\, f_h \{[\vec{\nabla}\times(\vec{\nabla}\times\vec{E}_{pol}^n)]\times\vec{B}_{pol}^n\} = \nonumber\\
 & = & -c\Delta t\, f_h \{[\vec{\nabla}\times(\vec{\nabla}\times\{\eta\vec{J}_{pol}^n + f_h  [\vec{J}_{tor}^n\times\vec{B}_{pol}^n - \vec{J}_{pol}\times\vec{B}_{tor}^n]\}]\times\vec{B}_{pol}^n\}~.
\end{eqnarray}
Therefore, the correction introduced by the intermediate step is:
\begin{eqnarray}
  \delta\vec{B}_{pol}^{n+1}=&&(c \Delta t)^2\,\vec{\nabla}\times\left\{ f_h  \{[\vec{\nabla}\times(\vec{\nabla}\times\{\eta(\curlB_{tor}^n)  \right. \nonumber\\
 && \left. +f_h  [(\curlB_{pol}^n)\times\vec{B}_{pol}^n - (\curlB_{tor}^n)\times\vec{B}_{tor}^n]\})]\times\vec{B}_{pol}^n\} \right\}~. 
\end{eqnarray}
From this expression, we can see that the additional correction given by $\delta\vec{E}_{tor}$ contains third and fourth-order spatial derivatives and scales with $(\Delta t)^2$, which is characteristic of hyper-resistive terms. We have found a significant improvement in the stability of the method compared with a fully explicit algorithm. 

The hyper-resistivity introduced by the two-steps method acts on the poloidal magnetic field alone. The method is even more stable if we also directly introduce a hyper-resistive term in the time advance of the toroidal component as follows:
\begin{equation}
 \vec{B}_{tor}^{n+1} = \vec{B}_{tor}^n - \Delta t [\vec{\nabla}\times\vec{E}_{pol}^n + \eta_{hyp}~(f_h B)~\nabla^4 \vec{B}_{tor}^n]~,
\end{equation}
where we set the dimensionless constant $\eta_{hyp}$ to ensure that the artificial hyper-resistivity is enough to stabilize the code. Following \cite{chacon03}, we set $\eta_{hyp}=0.2$. The term $\nabla^4 B_\varphi$ reads:
\begin{eqnarray}\label{eq:nabla4}
  && \nabla^4 B_\varphi = \frac{1}{r^4}\left[ \frac{\cos\theta}{\sin\theta}\left(2+\frac{1}{\sin^2\theta}\right)\partial_\theta B_\varphi  - \left(\frac{\cos\theta}{\sin\theta}\right)^2\partial_{\theta\theta}B_\varphi + 2\left(\frac{\cos\theta}{\sin\theta}\right)\partial_{\theta\theta\theta}B_\varphi +  \right.  \nonumber\\
  && \left. + \partial_{\theta\theta\theta\theta}B_\varphi + 2r^2\left(\frac{\cos\theta}{\sin\theta}\right) \partial_{rr\theta}B_\varphi + 2r^2 \partial_{r r \theta\theta} + 4r^3 \partial_{rrr}B_\varphi + r^4\partial_{rrrr}B_\varphi \right]~.
\end{eqnarray}
We have found that the method is always stable, if a sufficiently small time-step is used (typically $k_c \sim 10^{-2}$-$10^{-1}$ in eq.~\ref{eq:timestep}).

%%%%%%%%%%%%%%%%%%%%%%
\subsection{Energy balance.}\label{sec:en_balance}

The magnetic energy balance equation for Hall EMHD can be expressed as:
\begin{equation}\label{en_cons}
 \frac{\partial}{\partial t}\left(e^\nu\frac{B^2}{8\pi}\right)= - e^{2\nu}{\cal Q}_j - \vec{\nabla}\cdot(e^{2\nu} \vec{{\cal S}})~,
\end{equation}
where we have defined the Poynting vector
\begin{equation}
 \vec{{\cal S}}=\frac{c}{4\pi}\vec{E}\times\vec{B}~.
\end{equation}
During the evolution, the magnetic energy in a cell can only vary due to local Ohmic dissipation ${\cal Q}_j$ (eq.~\ref{eq:def_joule}) and by interchange between neighbor cells (Poynting flux).

Integrating eq.~(\ref{en_cons}) over the whole volume of the numerical domain, we define the total magnetic energy,
\begin{equation}
 {\cal E}_b=\int_V e^\nu\frac{B^2}{8\pi} \de V~,
\end{equation}
the total Joule dissipation rate,
\begin{equation}
 {\cal Q}_{tot}=\int_V e^{2\nu}{\cal Q}_j \de V~,
\end{equation}
and the Poynting flux through the boundaries $S$,
\begin{equation}
 {\cal S}_{tot}=\int_S e^{2\nu}\vec{{\cal S}}\cdot \hat{n} \de S~.
\end{equation}
The volume-integrated energy balance is 
\begin{equation}\label{integrated_balance}
 \frac{d}{d t}{\cal E}_b + {\cal Q}_{tot} + {\cal S}_{tot}=0~.
\end{equation}
A necessary test for any numerical code is to check the instantaneous (local and global) energy balance given by eqs.~(\ref{en_cons}) and (\ref{integrated_balance}). Any type of numerical instability usually results in the violation of the energy conservation. Therefore a careful monitoring of the energy balance is a powerful diagnostic test.

%%%%%%%%%%%%%%%%%%%%%%%%%%
\section{Numerical tests.}\label{sec_test}

In order to design a successful numerical algorithm, it is important to know the mathematical character of the equations and to identify the wave modes. In this section we show some illustrative examples of 2D Cartesian tests, reproducing whistler and Hall drift waves, and the Burgers flow. We will discuss the decay of Ohmic modes in a constant density sphere to compare against the analytic solutions for the purely resistive ($\omtau=0$) case. We pay special attention to the conservation of the total energy and the numerical viscosity of the method employed.

\subsection{2D tests in Cartesian coordinates.}

The Hall term introduces two wave modes into the system \citep{huba03}. In a constant density medium, the only modes of the Hall EMHD equation are the {\it whistler or helicon waves}, which are transverse field perturbations propagating along the magnetic field lines. In presence of a charge density gradient, additional {\it Hall drift waves} appear as transverse modes propagating in the $\vec{B} \times \vec{\nabla} n_e$ direction. 

%%%%%%%%%%%%%%%%%%%%%%%
\begin{figure}[t!]
 \centering
 \includegraphics[width=.5\textwidth]{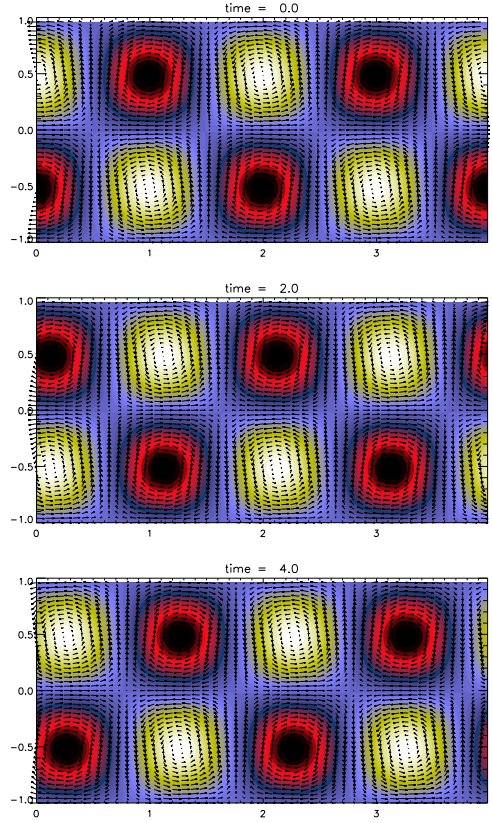}
\caption{Evolution of the initial configuration defined by eqs.~(\ref{inimod1}) with $B_0=10^3~B_1$ and $k_x L=\pi$ at three different times (in units of $\tau_0$). Arrows show the perturbed $B_x$ (subtracting $B_0$), and $B_z$-components, while the color scale represents the $B_y$ component (red/black positive, yellow/white negative).} 
\label{fig:whistler}
\end{figure}

\subsubsection{Whistler waves.}

The first test we have performed is to follow the correct propagation of whistler waves. Consider a two-dimensional slab, extending from $z=-L$ to $z=+L$ in the vertical direction. We impose periodic boundary conditions in the $x$-direction. All variables are independent of the $y$-coordinate. In the case of constant density, $n_e=n_0$, and $\eta=0$ (i.e. infinite magnetization parameter), the only modes present in the system are whistler waves. We consider the Hall induction equation for the following initial magnetic field
\begin{eqnarray}
\label{inimod1}
B_x&=& B_0 + B_1\cos(k_x z)\cos(k_x x)~, \nonumber \\
B_y&=& \sqrt{2}B_1 \sin(k_x z)\cos(k_x x)~, \\
B_z&=& B_1\sin(k_x z)\sin(k_x x)~, \nonumber
\end{eqnarray}
where $k_x = n \pi /L$, $n=1,2,...$, and $B_1\ll B_0$. We define the reference Hall time-scale as
\begin{equation}\label{tau0_wave}
\tau_0=\frac{4\pi e n_0 L^2}{c B_0}~.
\end{equation}
This case admits a pure wave solution confined in the vertical direction and traveling in the $x$-direction with speed 
\begin{equation}\label{vel_whistler}
v_w=-\frac{c}{4\pi e n_0}\sqrt{2}~k_x B_0=-\sqrt{2}\frac{L^2 k_x}{\tau_0}~.
\end{equation}
The evolution of the initial configuration defined by eqs.~(\ref{inimod1}) with $B_0=10^3~B_1$, and $k_x L=\pi$ at three different times (in units of $\tau_0$) is shown in Fig.~\ref{fig:whistler}, and for a $200 \times 50$ grid. The travel time to cross over the whole domain is $t=0.9~ \tau_0$, thus the perturbations have crossed through the horizontal domain several times, without apparent dissipation or shape change. The code runs for hundreds of Hall time-scales without any indication of instabilities, despite the fact that electrical resistivity is set to zero. The measured speed of the whistler waves is 0.9992 the analytical value for this particular resolution. We have also checked that the correct scaling of the propagation velocity (linear with $k_x$ and $B_0$) is recovered.

\subsubsection{Hall drift waves.}

%%%%%%%%%%%%%%%%%%%%%%%
\begin{figure}[t]
 \centering
 \includegraphics[width=.6\textwidth]{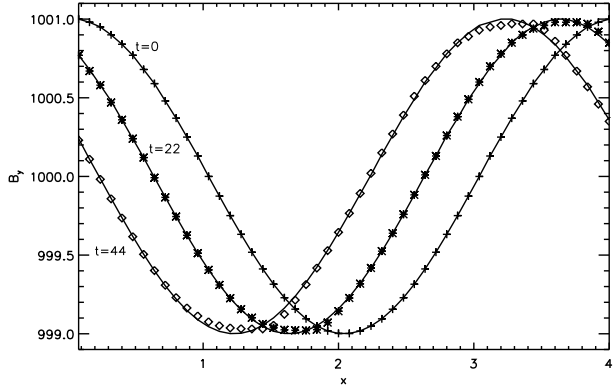}
\caption{Horizontal section ($z=0$) of the evolution of the initial configuration defined by eq.~(\ref{inimod2}) with $B_0=10^3~B_1$, $L=1$ and $k_x L=\pi/2$  at $t=0,22,$ and $44$ (in units of $\tau_0$). 
The perturbation crosses over the entire domain in $20 ~\tau_0$.} 
\label{fig:drift}
\end{figure}

As a second test in Cartesian coordinates, we set up the following initial configuration:
\begin{eqnarray}\label{inimod2}
B_x&=& 0~, \nonumber \\
B_y&=& B_0 + B_1\cos(k_x x) ~, \\
B_z&=& 0~, \nonumber
\end{eqnarray}
on a stratified background in the $z$-direction with 
\begin{equation}\label{n_wave}
n_e(z)= \frac{n_0}{1+\beta_L z}~,
\end{equation}
where $n_0$ is a reference density to which we associate the same Hall time-scale $\tau_0$ defined in eq.~(\ref{tau0_wave}), $z\in[-L,L]$, and $\beta_L$ is a parameter with dimensions of inverse length. We impose again periodic boundary conditions in the $x$-direction. In the $z$-direction, we copy the values of the magnetic field in the first and last cells ($z=\mp L$) on the first neighbor ghost cell, to simulate an infinite domain. For small perturbations ($B_1\ll B_0$), this configuration must induce Hall drift waves traveling in the $x$-direction with speed

\begin{equation}\label{vel_hd}
  v_{hd}=-\frac{\beta_L L^2}{\tau_0}~.
\end{equation}
For the particular model shown in Fig.~\ref{fig:drift}, with $B_0=10^3~B_1, k_x L=\pi/2$, and $\beta_L L=0.2$, this gives a horizontal drift velocity of $v_{hd}=-0.2 L/\tau_0$. We show the evolution of $B_y$ at three different times, when the perturbation has crossed over the numerical domain 1.1 and 2.2 times, respectively. The resolution used is $50\times50$. For the Hall drift modes, the propagation velocity scales linearly with $B_0$ and the gradient of $n_e^{-1}$, but it is independent of the wavenumber of the perturbation. All these properties are correctly reproduced.

%%%%%%%%%%%%%%%%%%%%%%%%%%%%%%%%%%%%%%
\subsubsection{The nonlinear regime and Burgers flows.}\label{sec_burgers}

With the two previous tests, we can conclude that the code is able to accurately propagate the two fundamental modes with the correct speeds. However, the Hall drift modes are a valid solution only in the linear regime. Let us consider more carefully the evolution of the $B_y$ component in a medium stratified in the $z$-direction. Assuming that $B_x=B_z=0$, it can be readily derived that the governing equation reduces to
\begin{equation}
 \frac{\partial B_y}{\partial t} + g(z) {B_y}\frac{\partial B_y}{\partial x}=0 ~.
\end{equation}
This is the Cartesian version of the Burgers equation (\ref{eq:induction_burgers}) in the $x$-direction with a coefficient that depends on the $z$ coordinate:

\begin{equation}
 g(z)=-\frac{d}{dz}\left(\frac{c}{4\pi e n_e}\right)~.
\end{equation}
Thus, the evolution of the $B_y$ component of the magnetic field is governed by independent Burgers-like equations with different propagation speeds at different heights. The solution of such equation in one dimension is well known and has been studied for decades. As a last test in Cartesian coordinates we consider the following initial configuration:
\begin{eqnarray}
\label{inimod3}
B_x&=& 0~, \nonumber \\
B_y&=& B_0 \cos(k_x x) ~,  \\
B_z&=& 0~,\nonumber
\end{eqnarray}
on the same stratified background as in the previous subsection, eq.~(\ref{n_wave}), which gives simply $g(z)=-\beta_L L^2/\tau_0 B_0$ and allows the direct comparison to the solution of the Burgers equation in one dimension.

Making use of the well-known numerical techniques applied to the Burgers equation, in our code we give an special treatment to the quadratic term in $B_y$ in the evolution equation for that component (we will proceed analogously for the $\varphi$-component in spherical coordinates). The key issue is to write the Burgers-like term in conservation form:
\begin{equation}
 \frac{\partial B_y}{\partial t} + g(z)\frac{\partial (B_y^2/2)}{\partial x}=0~,
\end{equation}
which can be discretized by an upwind conservative method
\begin{equation}
\frac{d B_y^j}{dt}= - \frac{\left( \hat{F}^{j+1/2}- \hat{F}^{j-1/2}\right)}{\Delta x}~,
\end{equation}
where the problem is reduced to calculate the upwind numerical fluxes at the interfaces $\hat{F}^{j\pm1/2}$. In this case the wave velocity determining the upwind direction is given by $\lambda = g(z)B_y$ and the flux $\hat{F}=g(z) B_y^2/2$. Using methods in flux-conservative form is particularly important when solving problems with shocks or other discontinuities (e.g. \citealt{toro09}), as a non-conservative method may give numerical solutions that look reasonable but are entirely wrong (incorrect propagation speed of discontinuous solutions). The particular treatment of the time advance of this term (different from the general method described before) has no impact on the divergence-preserving character of the method, since this is the component in the symmetry direction of the problem (the same argument applies for the toroidal magnetic field in  axial symmetry). 

%%%%%%%%%%%%%%%%%%%%%%%
\begin{figure}
 \centering
 \includegraphics[width=.6\textwidth]{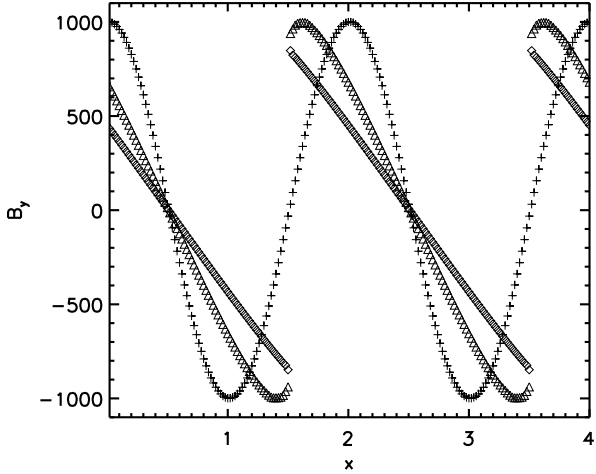}
\caption{Horizontal section of the evolution of the initial configuration defined by eq.~(\ref{inimod3}) with $B_0=10^3$ and $k_x L=\pi$ at $t=0,2,$ and $4$ (in units of $\tau_0$). The shock forms at $t=2$. The classical sawtooth shape developed during the evolution of the Burgers equation is evident.} 
\label{fig:burg1}
\end{figure}

In Fig.~\ref{fig:burg1} we show snapshots of the evolution of the initial conditions (\ref{inimod3}) with $k_x L=\pi$, $B_0=10^3$, and $\beta_L L=0.2$. It follows the typical Burgers evolution. The wave breaking and the formation of a shock at $t=2 \tau_0$ is clearly captured, as would happen in the solution of the inviscid Burgers equation. We must stress again that this test is done with zero physical resistivity, i.e., in the limit $\omtau \rightarrow \infty$, which is not reachable by spectral methods or standard centered difference schemes in non-conservative form.

We should also stress that the evolution of the energy spectrum of a Burgers-like problem leads to the scaling $k^{-2}$, as shown for example in \cite{tran10}. This is exactly the expected scaling, first discussed in the context of Hall MHD in neutron stars in \cite{goldreich92}, who pointed out the analogies and differences with turbulent velocity fields. The $k^{-2}$ scaling for the power spectrum has been discussed in the past in terms of a Hall cascade, transferring energy from larger scale to smaller scale modes, and there 
is some open debate about its interpretation as a global turbulent cascade or a local steepening of some magnetic field components.

%%%%%%%%%%%%%%%%%%%%%%%%%%%%%%%%%
\subsection{2D evolution in spherical coordinates: force-free solutions.}\label{sec:pure_ohmic_mode}

%%%%%%%%%%%%%%%%
\begin{figure}[t]
 \centering
 \includegraphics[width=.95\textwidth]{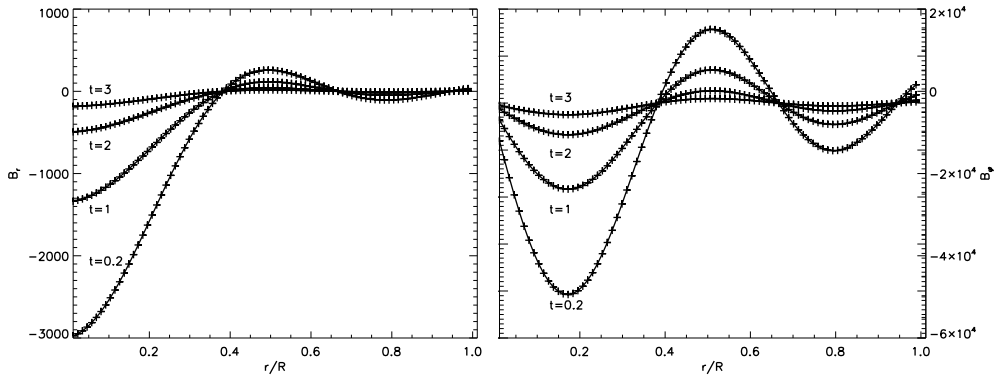}
\caption{Evolution of the purely Ohmic modes, eqs.~(\ref{eq:mf_ohmic1})-(\ref{eq:mf_ohmic3}), at $t=0.2, 1, 2,$ and $3$ diffusion times ($\tau_d$) for the model with $\alpha R_\star=11.5$ and a resolution of  $88\times60$ equally spaced grid points. In the figure we compare the 
analytical (lines) and numerical (crosses) radial profiles of $B_r$ (left) and $B_\varphi$ (right) at $\theta=\pi/3$.}
\label{fig:bessel_decay}
\end{figure}

In spherical coordinates, one of the few analytical solutions that can be used to confront numerical results is the evolution of pure Ohmic dissipation modes (i.e., in the limit $\omtau \rightarrow 0$). If the magnetic field satisfies $\curlB=\alpha\vec{B}$ everywhere, then we deal only with the resistive part. If $\alpha$ and the resistivity are both constant, and the metric is flat ($e^\nu=e^\lambda=1$), then, ignoring the Hall term in eq.~(\ref{eq:induction_hall}), we have:
\begin{equation}\label{eq:induction_ohmic_mode}
  \frac{\partial \vec{B}}{\partial t}=-\vec{\nabla}\times[\eta(\vec{\nabla}\times\vec{B})]= - \eta \alpha^2\vec{B}~.
\end{equation}
All components of the magnetic field decay exponentially as $\propto \exp{(-t/\tau_d)}$, where $\tau_d=(\eta \alpha^2)^{-1}$ is the diffusion time-scale. The family of solutions satisfying the above relation is described by radial parts involving the Bessel spherical functions and their derivatives (see Appendix~\ref{app:bessel}). We test the only solution with a regular behavior at the center, the dipolar $l=1$ function of the first kind, for which the magnetic field components are:

\begin{eqnarray}
  && B_r=\frac{B_0R_\star}{r}\cos\theta \left(\frac{\sin x}{x^2} - \frac{\cos x}{x}\right) ~, \label{eq:mf_ohmic1}\\
  && B_\theta=-\frac{B_0R_\star}{2r}\sin\theta\left(\frac{\sin x}{x^2}-\frac{\cos x}{x}-\sin x\right)~, \label{eq:mf_ohmic2}\\
  && B_\varphi=\frac{kB_0R_\star}{2}\sin\theta \left(\frac{\sin x}{x^2}-\frac{\cos x}{x}\right) ~, \label{eq:mf_ohmic3}
\end{eqnarray}
where $x=\alpha r$. Given this initial condition, we follow the evolution of the modes on several Ohmic time-scales, until the magnetic field has been almost completely dissipated.  As boundary conditions we impose the analytical solutions for $B_\theta$ and $B_\varphi$ at the central cell and the external ghost cell. Note also that the toroidal and poloidal components of the magnetic field are completely decoupled in the pure resistive limit, so that their evolution is totally independent.

In Fig.~\ref{fig:bessel_decay} we compare the evolution of the numerical (crosses) and analytical (solid lines) solutions of $B_r$ and $B_\varphi$ at different times, for a model with $\alpha R_\star=11.5$. They are indistinguishable in the graphic. To quantify the deviation, we have evaluated the $L^2$-norm of the deviation from the analytic solution $\Delta \vec{B}=(\vec{B}-\vec{B}_{an})$ (normalized to the initial magnetic field), defined as 
\begin{equation}\label{l2norm}
  L^2= \frac{\sum_{i,j} ||\Delta \vec{B}^{(i,j)}||}{\sum_{i,j} ||\vec{B}_{t=0}^{(i,j)}||}~,
\end{equation}
where the sum is performed over all grid cells $(i,j)$. The maximum of $L^2$ at any time for the same angular resolution (60 cells) and three different radial resolutions of 88, 173, and 346 grid points is $3.7\times10^{-4}$, $1.5\times10^{-4}$, and $1.0\times10^{-4}$ respectively.

\subsection{Evolution of a purely toroidal magnetic field in the crust.}\label{tor_sec}

We now consider the case of a purely toroidal magnetic field confined to the crust $R_{core}<r<R_\star$ of a realistic neutron star, with $R_{core}=10.8~$km and $R_\star=11.5~$km. For simplicity, we work in Newtonian case ($e^\nu=e^\lambda=1$), and we fix a constant temperature of $10^8$ K, which gives a magnetic diffusivity in the range $\eta\sim 0.01-10 $ km$^2$/Myr. The boundary conditions we impose in this case are vanishing magnetic field at both boundaries. According to the Hall induction equation, any initial toroidal configuration will remain purely toroidal during the evolution, but the shape and location of currents can be substantially modified. The initial magnetic field is given by:

\begin{equation}\label{btorq}
  B_\varphi=-B_0\frac{(R_\star-r)^2(r-r_c)^2\sin\theta\cos\theta}{r}~,
\end{equation}
where $B_0$ is the normalization adjusted to fix the initial maximum value of the toroidal magnetic field ($B_t^0$) to a specific value. This corresponds to an initial maximum value of magnetization parameter $\omega_B\tau_e \approx 10$ for $B_t^0=10^{14}$ G. We use this model to highlight a few important issues concerning energy conservation, numerical viscosity, and shock formation.

\begin{figure}[t]
 \centering
 \includegraphics[width=.6\textwidth]{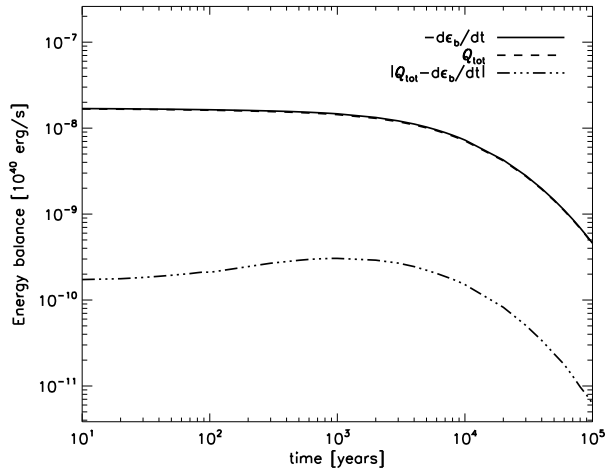}
 \caption{Volume-integrated energy balance as a function of time for a run on a 27$\times$60 grid. At each time-step loss of magnetic energy (solid line) is compensated by Joule dissipation rate (dashed), within the a small error (dot-dashed). The initial model is given by eq.~(\ref{btorq}) with $B_t^0=10^{14}$ G.} 
 \label{fig:bcons}
\end{figure}

%%%%%%%%%%%%%%%%%%%%%%%
\subsubsection{Energy conservation.}

Following the definitions of \S~\ref{sec:en_balance}, we check the instantaneous local conservation of energy, evaluating eq.~(\ref{en_cons}) in the whole volume numerical domain, during a time-step $\Delta t$:
\begin{equation}\label{balance}
- \sum \frac{{\bar{B^2}(t+\Delta t)}-{\bar{B^2}(t)}}{8\pi \Delta t} \Delta V = \sum \frac{4\pi \eta}{c^2} \bar{J^2} \Delta V~,
\end{equation}
where the sum is performed over the cells of the domain, and $\bar{B^2},\bar{J^2}$ are local averages inside each cell of volume $\Delta V$. We have used that the Poynting flux through the boundaries is zero, since the magnetic field vanishes at both the internal boundary and the star surface. In Fig.~\ref{fig:bcons} we show the two sides of eq.~(\ref{balance}) as a function of time (dashed and solid lines), for a model with $B_t^0=10^{14}$ G and with a coarse grid ($27\times 60$). Even for this low resolution, the error in the instantaneous energy balance (dot-dashed) is a couple of orders of magnitude less than the instantaneous magnetic energy loss (solid) and the Ohmic dissipation rate (dashed).

%%%%%%%%%%%%%%%%%%%%%%%
\subsubsection{Numerical viscosity.}

\begin{figure}[!t]
 \centering
 \includegraphics[width=.45\textwidth]{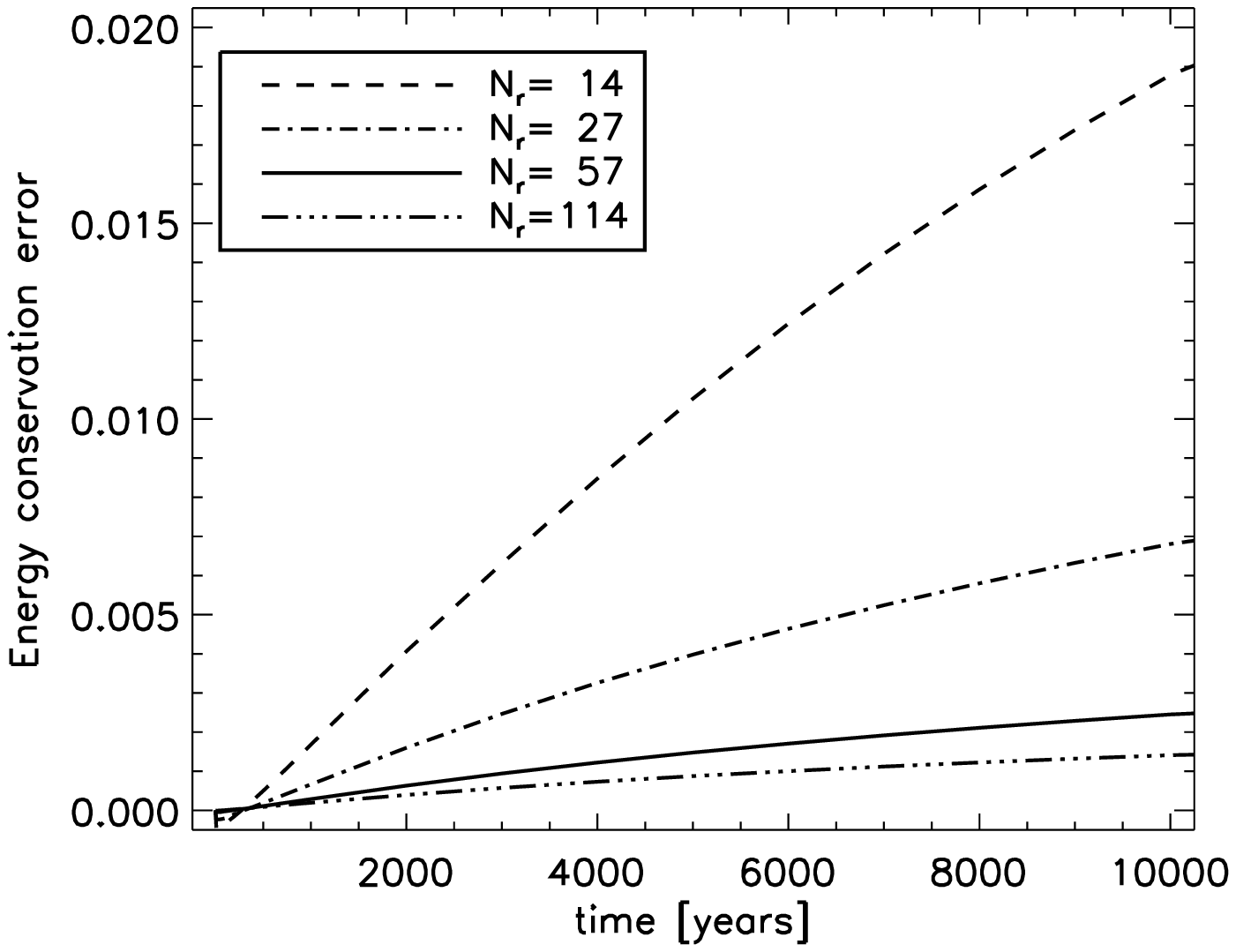}
 \includegraphics[width=.45\textwidth]{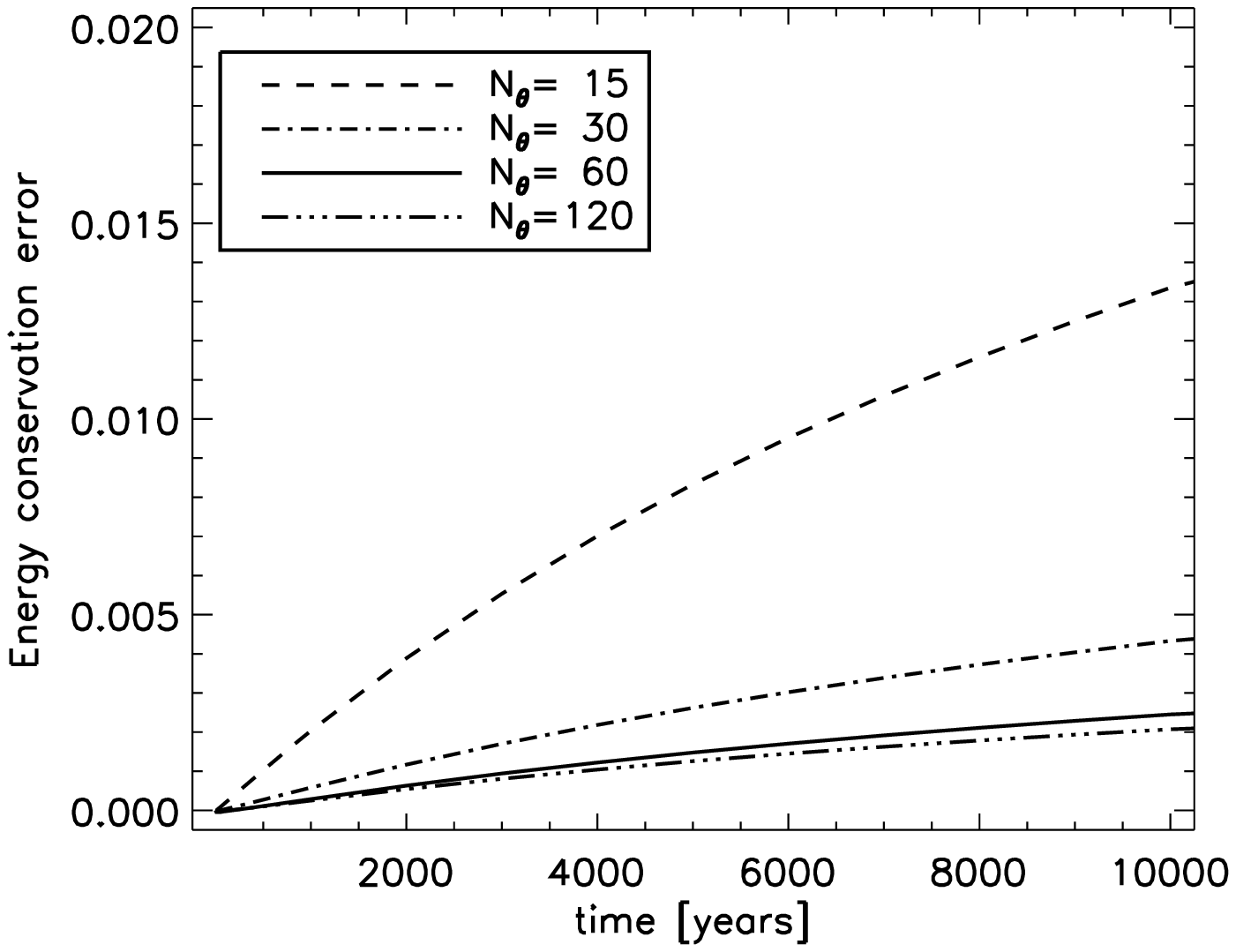}
 \caption{Accumulated relative error $\epsilon_{en}(t)$, eq.~(\ref{encons_error}), in the conservation of energy during the evolution of a purely toroidal magnetic field with $B_t^0=10^{14}$ G. We show $\epsilon_{en}(t)$ for different resolutions. Left: varying $N_r$, with fixed $N_\theta=60$; right: varying $N_\theta$, with fixed $N_r$.} 
 \label{fig:dissipation}
\end{figure}

It is well known that Godunov type methods provide stability in the most extreme conditions at the price of introducing numerical viscosity. Reconstruction procedures, such as the MC method we used for our code (see \S~\ref{sec:method_hall}), help decrease the numerical viscosity in smooth regions of the flow and increase the spatial order of the method. However, it is important to check that any numerical viscosity introduced by the method is well below the {\it physical} resistivity, in order not to affect the results.

\begin{figure}[t]
 \centering
 \includegraphics[width=.45\textwidth]{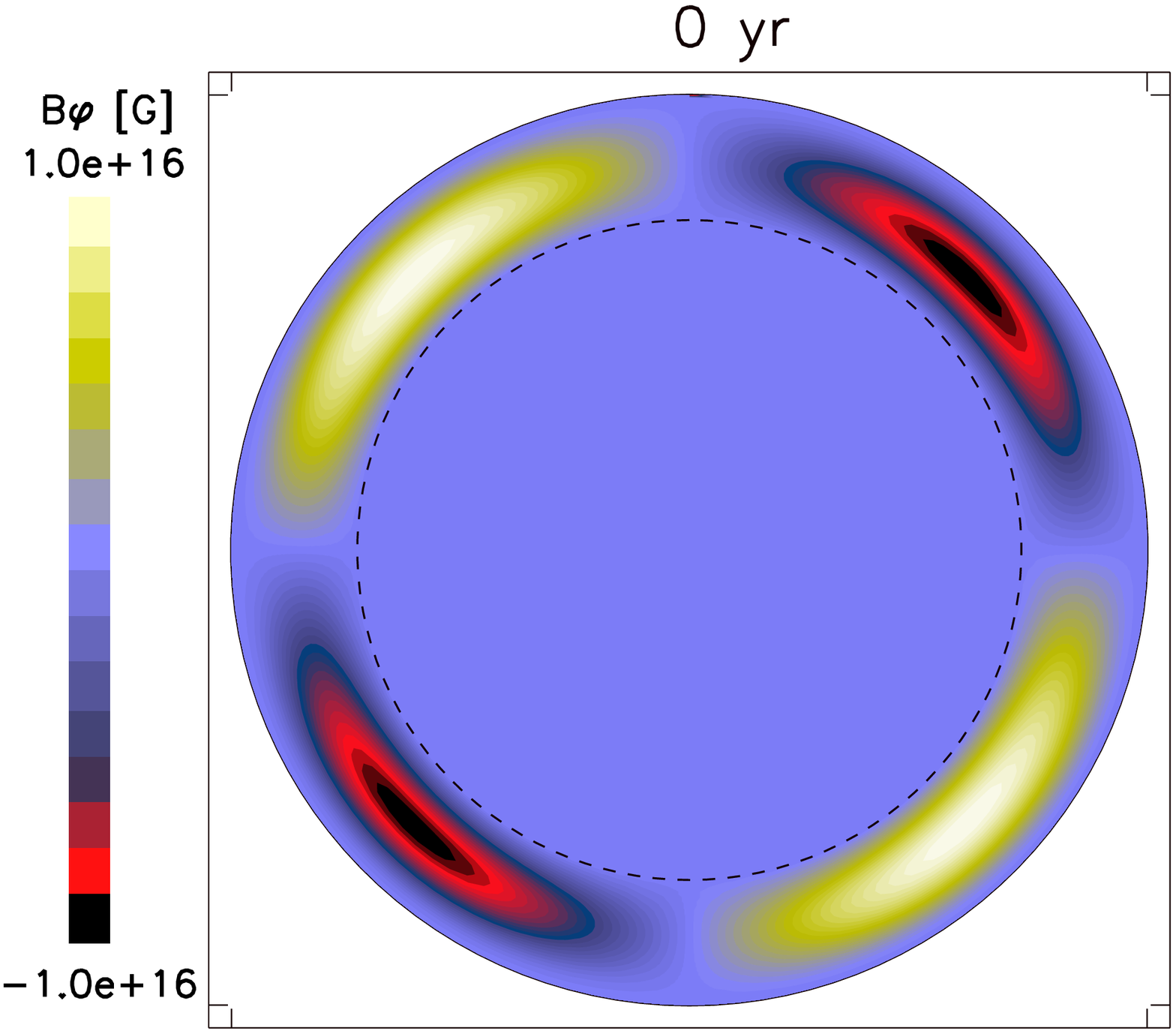}
 \includegraphics[width=.45\textwidth]{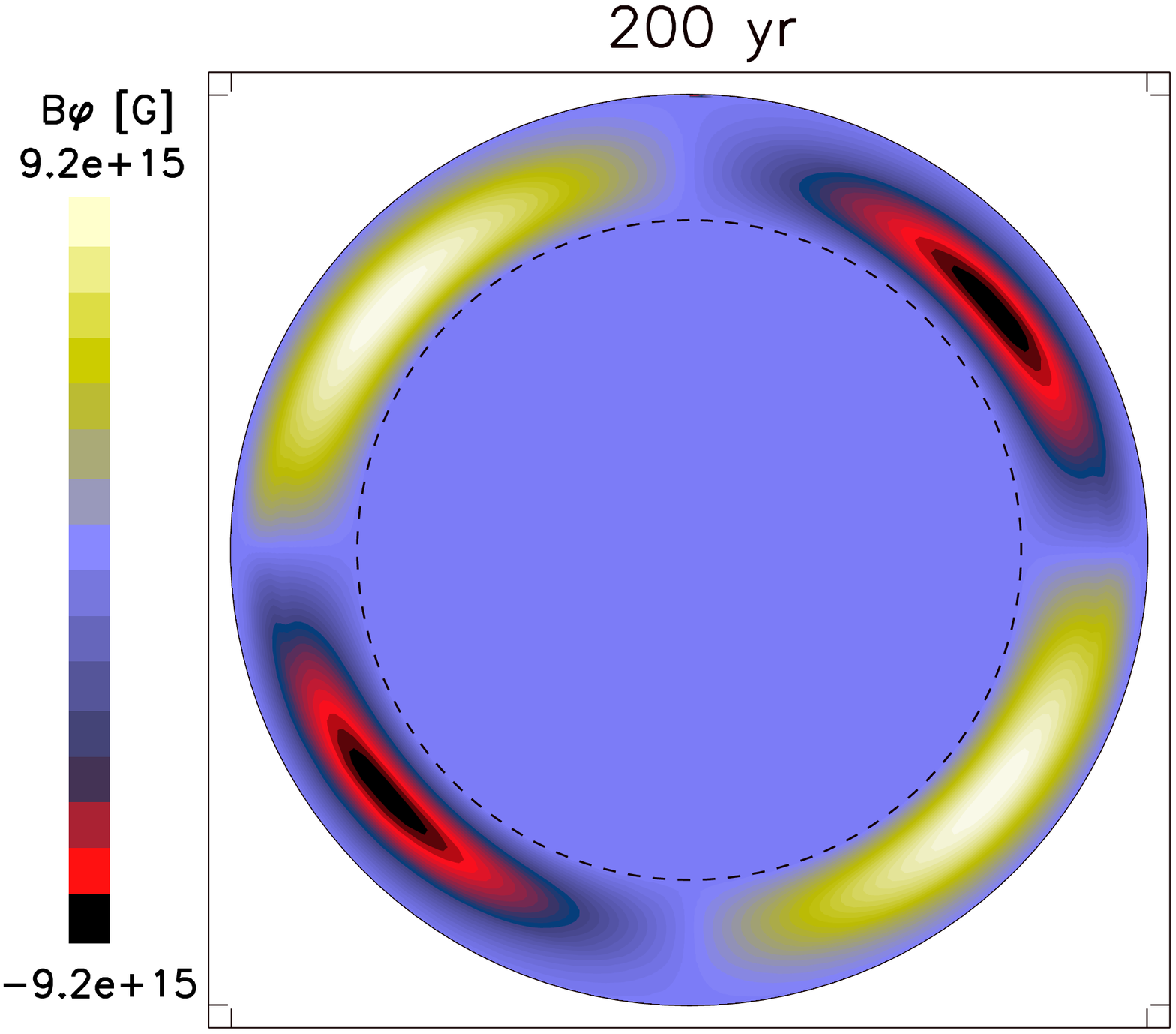}\\
 \includegraphics[width=.45\textwidth]{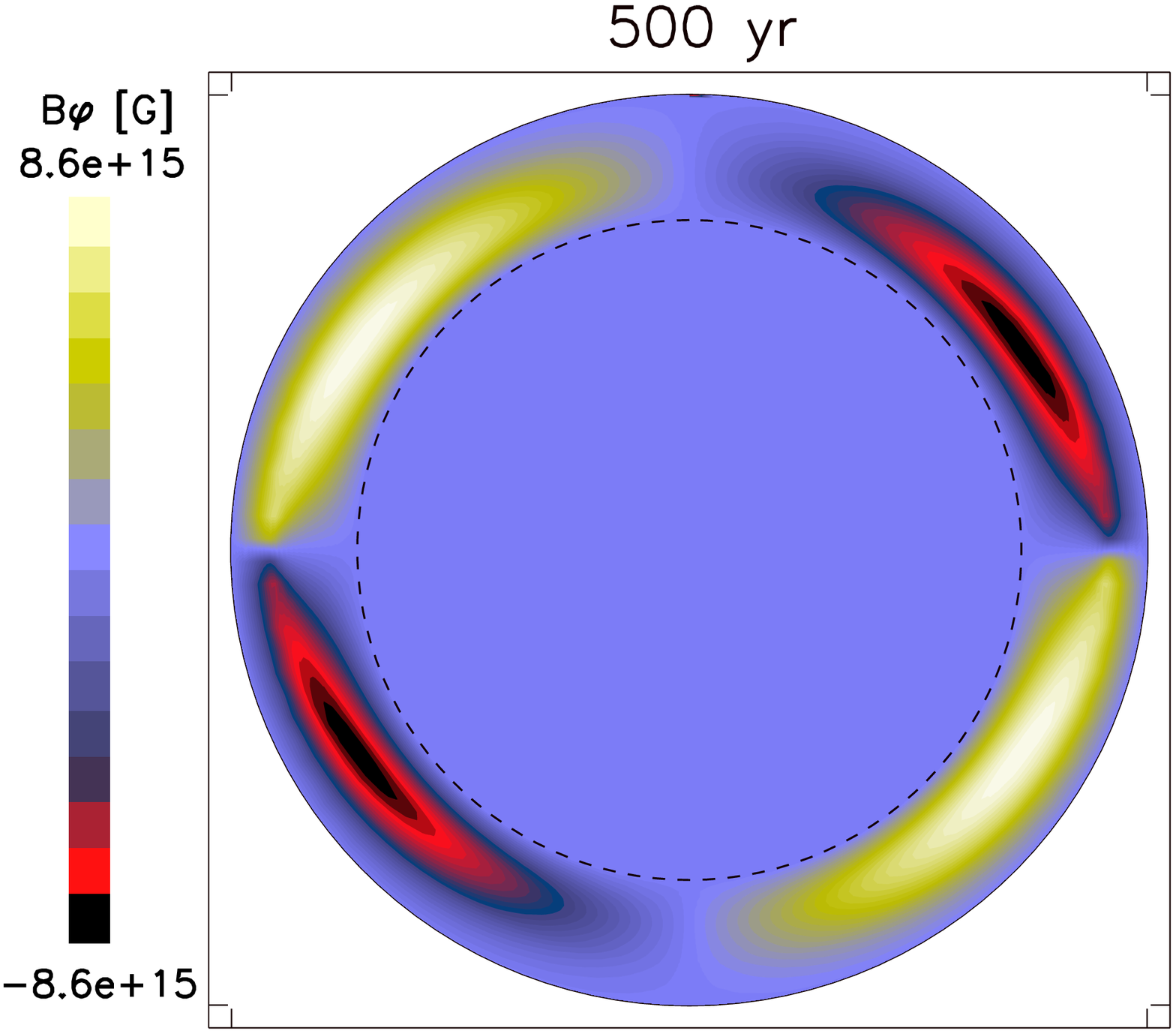}
 \includegraphics[width=.45\textwidth]{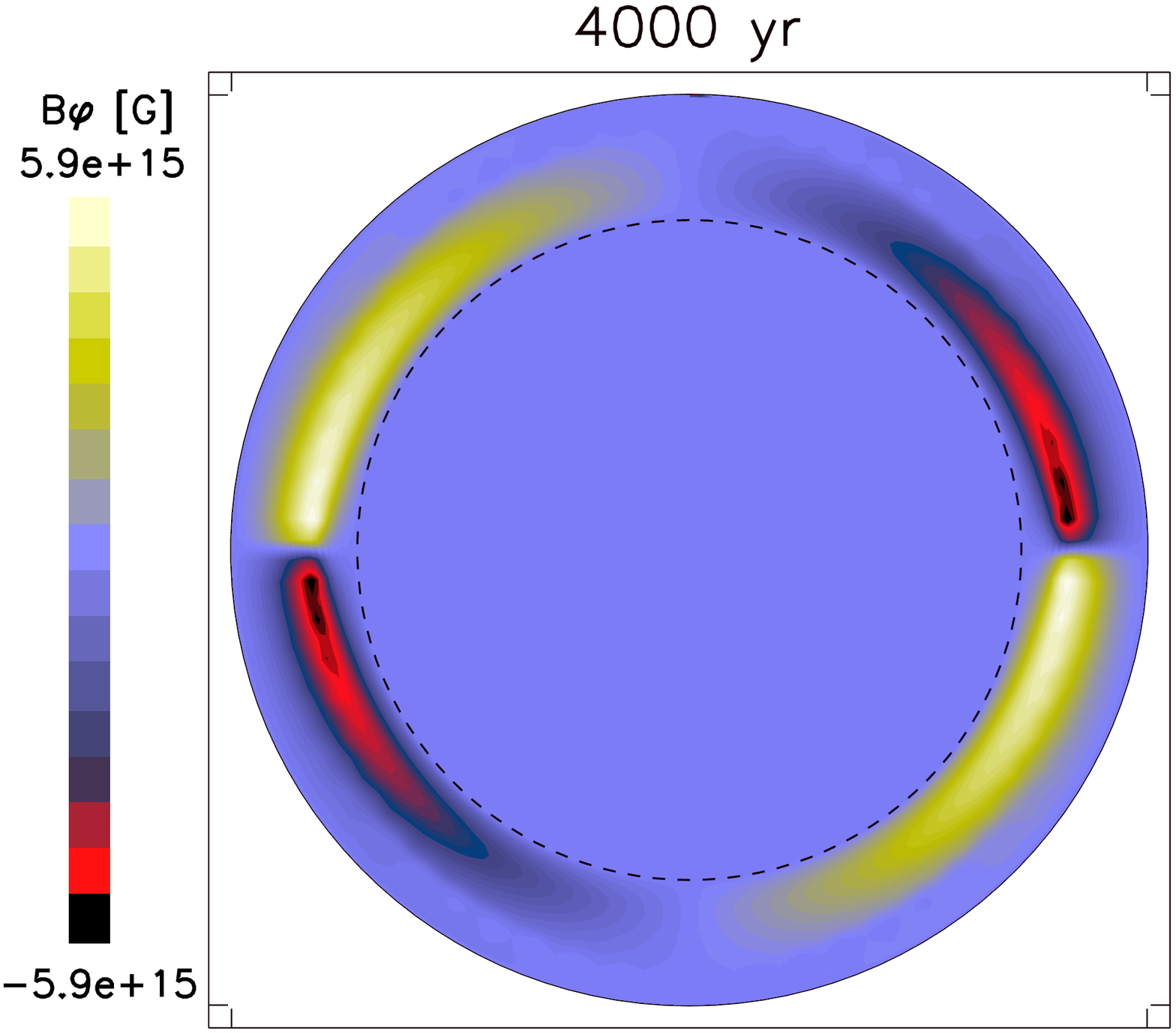}
\caption{Evolution of a purely quadrupolar toroidal magnetic field confined into the crust. We show snapshots at $t=0, 200, 500$ and $4000$ yr. The color scale indicates the toroidal magnetic field intensity (yellow: positive, red: negative). In the figure, the thickness of the crust has been stretched by a factor of 4 for clarity.} 
 \label{fig:bphi}
\end{figure}

In Fig.~\ref{fig:dissipation} we compare, for different resolutions, the relative error on the conservation of the total energy: 
\begin{equation}\label{encons_error}
  \epsilon_{en}(t)=1 - \frac{{\cal E}_{tot}(t)}{{\cal E}_{tot}(t=0)}~,
\end{equation}
where
\begin{equation}
{\cal E}_{tot}(t)= {\cal E}_b(t) + \int_0^t{\cal Q}_{tot}dt + \int_0^t{\cal S}_{tot}dt~. 
\end{equation}
The left panel shows $\epsilon_{en}(t)$ with fixed $N_\theta=60$ and $N_r=14,27,57,114$ (number of points in the crust) while the right panel shows results for $N_r=57$ and varying angular resolution $N_\theta=15,30,60,120$. The error in the energy conservation grows nearly linear with time in the first $10^4$ yr. Assuming that this deviation is entirely due to the numerical viscosity of the method, we can estimate the viscous time-scale of the numerical method by assuming that
\begin{equation}
  \epsilon_{en}(t)=1-\exp{(-t/\tau_{\rm num})}\simeq \frac{t}{\tau_{\rm num}}~,
\end{equation}
and extracting $\tau_{\rm num}$ from the results.

With a resolution of 57$\times$60 grid points (solid line in Fig.~\ref{fig:dissipation}), the error on energy conservation is of 0.25$\%$ at $10^4$ yr, which implies 
$\tau_{\rm num} \sim 4 $ Myr. Therefore we can estimate the order of magnitude of the numerical resistivity as
\begin{equation}
{\eta_{\rm num}} =\frac{\Delta r^2}{\tau_{\rm num}}~,
\end{equation}
where $\Delta r \sim (R_\star - R_{core})/N_r$ is the average radial size of a cell. For $N_r=57$, we obtain $\eta_{\rm num}\sim 10^{-4}$ km$^2$/Myr. Comparing this value with the typical physical resistivity of the crust at this temperature, $\eta\sim 10^{-2}-10^1$ km$^2$/Myr, we can assess that, even with a coarse grid of 57$\times$60 grid points, the numerical viscosity is at least two orders of magnitude less than the physical resistivity.

We have also studied the effect on the numerical viscosity of the Courant pre-factor $k_c$, defined in eq.~(\ref{eq:timestep}). We find that varying $k_c$ does not affect significantly the numerical dissipation. This effect is always much smaller than the dependence on the spatial resolution, as long as the evolution is stable (typically $k_c<0.1$).

%%%%%%%%%%%%%%%%%%%%
\subsubsection{Creation of current sheets.}

\begin{figure}[t]
 \centering
 \includegraphics[width=.45\textwidth]{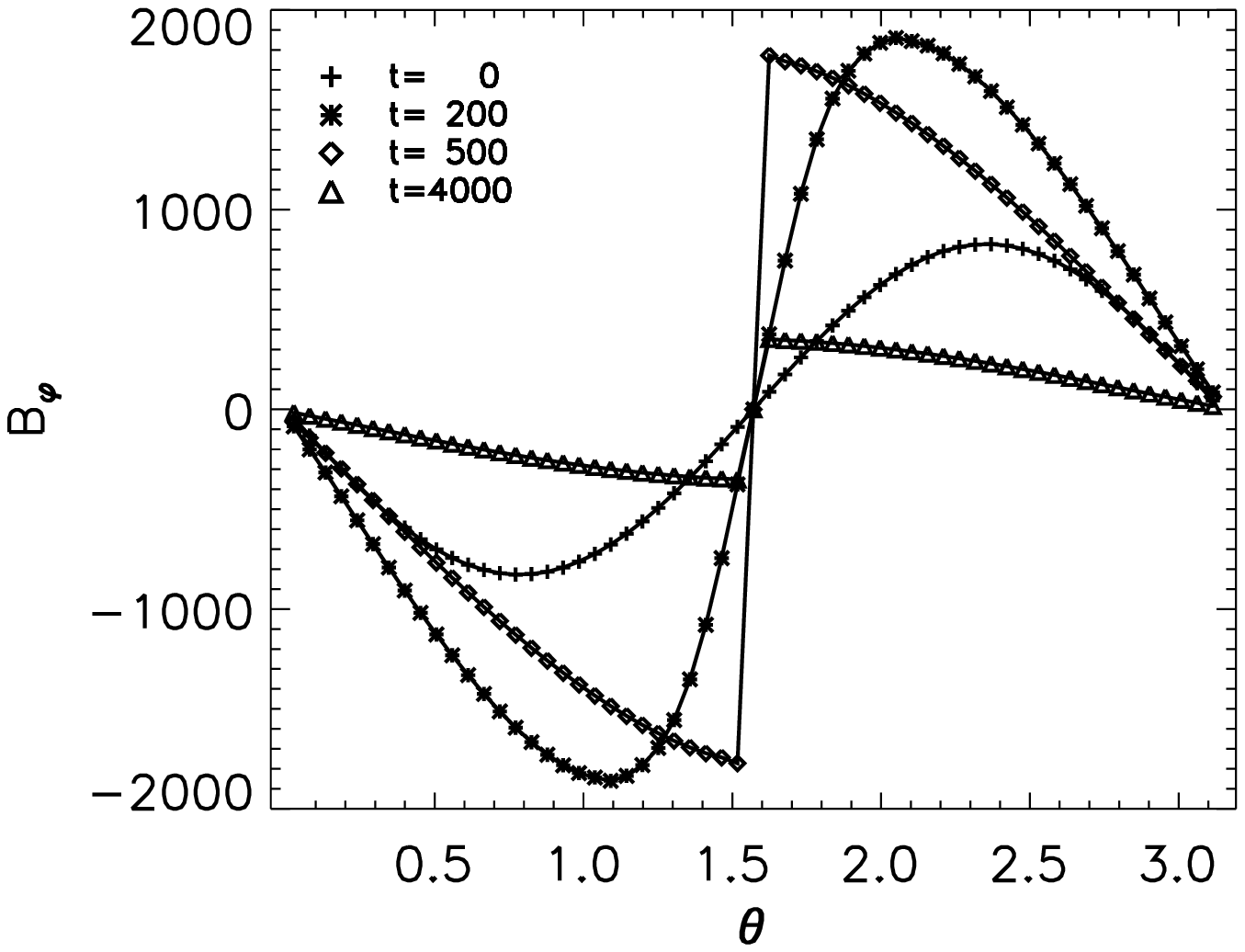}
 \includegraphics[width=.45\textwidth]{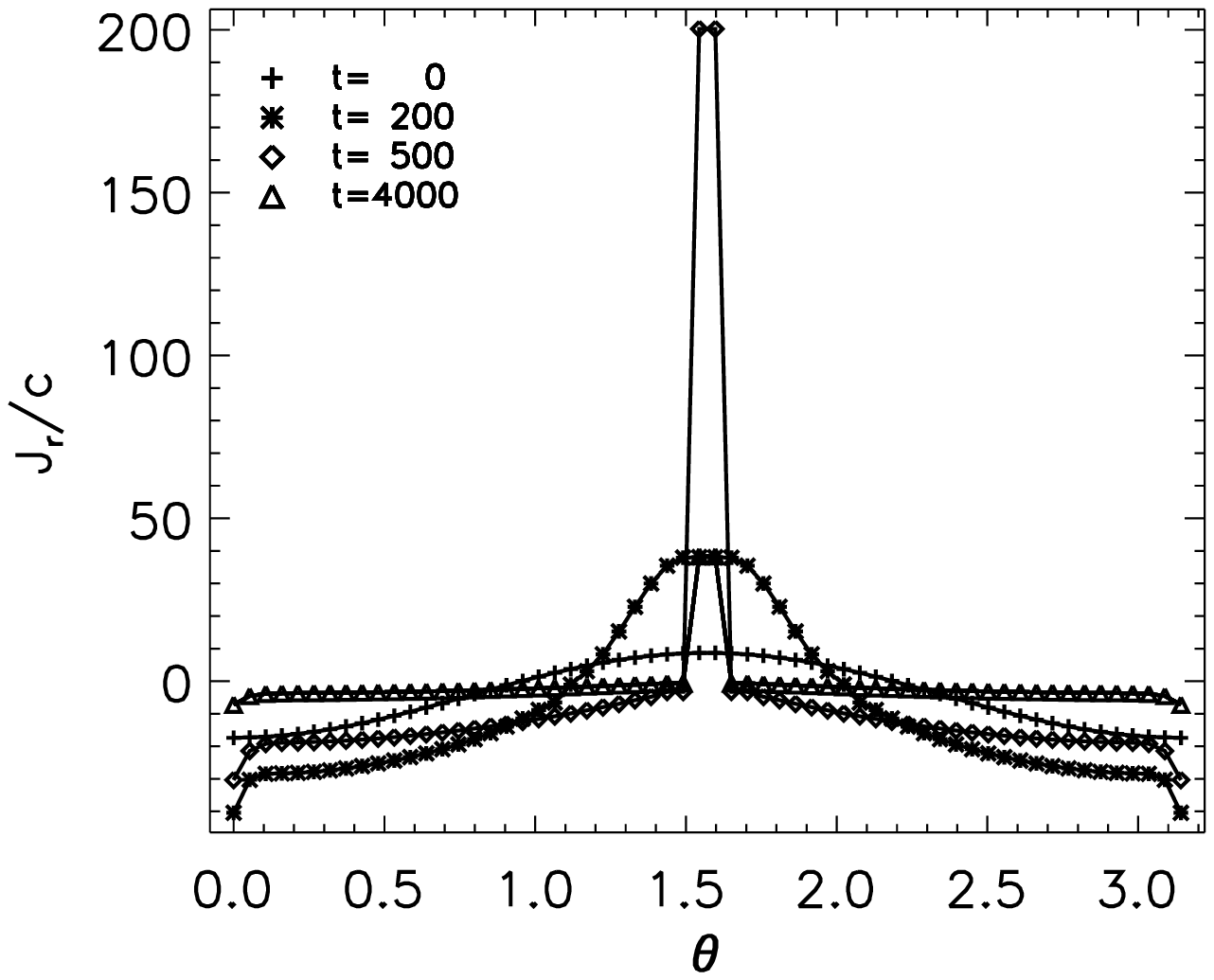}
\caption{Angular profiles of the same model as Fig.~\ref{fig:bphi}, just below the star surface, of $B_\varphi$ (in units of $10^{12}$ G) and $J_r/c$ (in units of $10^{12}$ G/km), at four different times $t=0, 200, 500,$ and $4000$ yr. The angular resolution used is $N_\theta=60$.} 
\label{fig:shock_up}
\end{figure}

\begin{figure}[t]
 \centering
 \includegraphics[width=.45\textwidth]{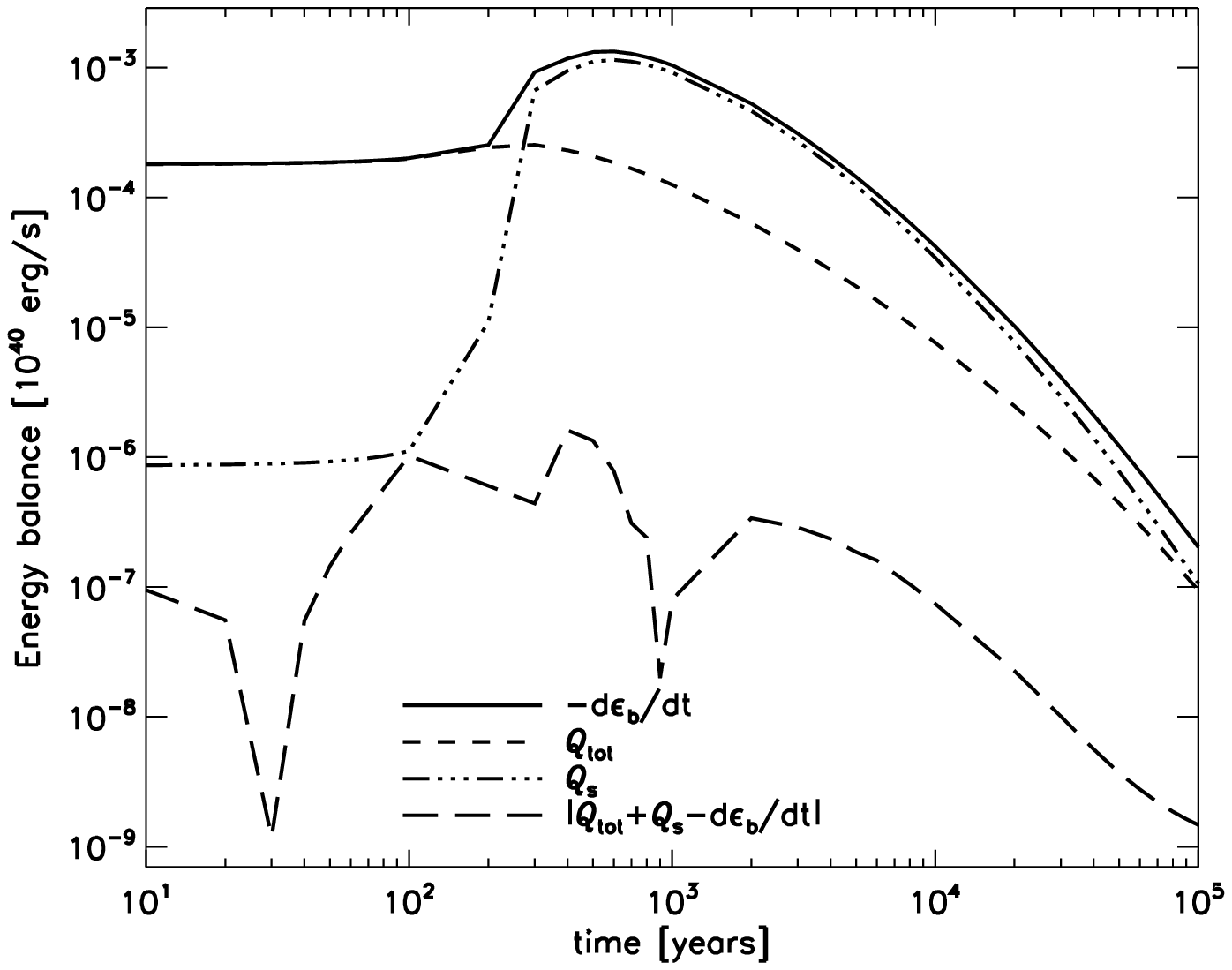}
 \includegraphics[width=.45\textwidth]{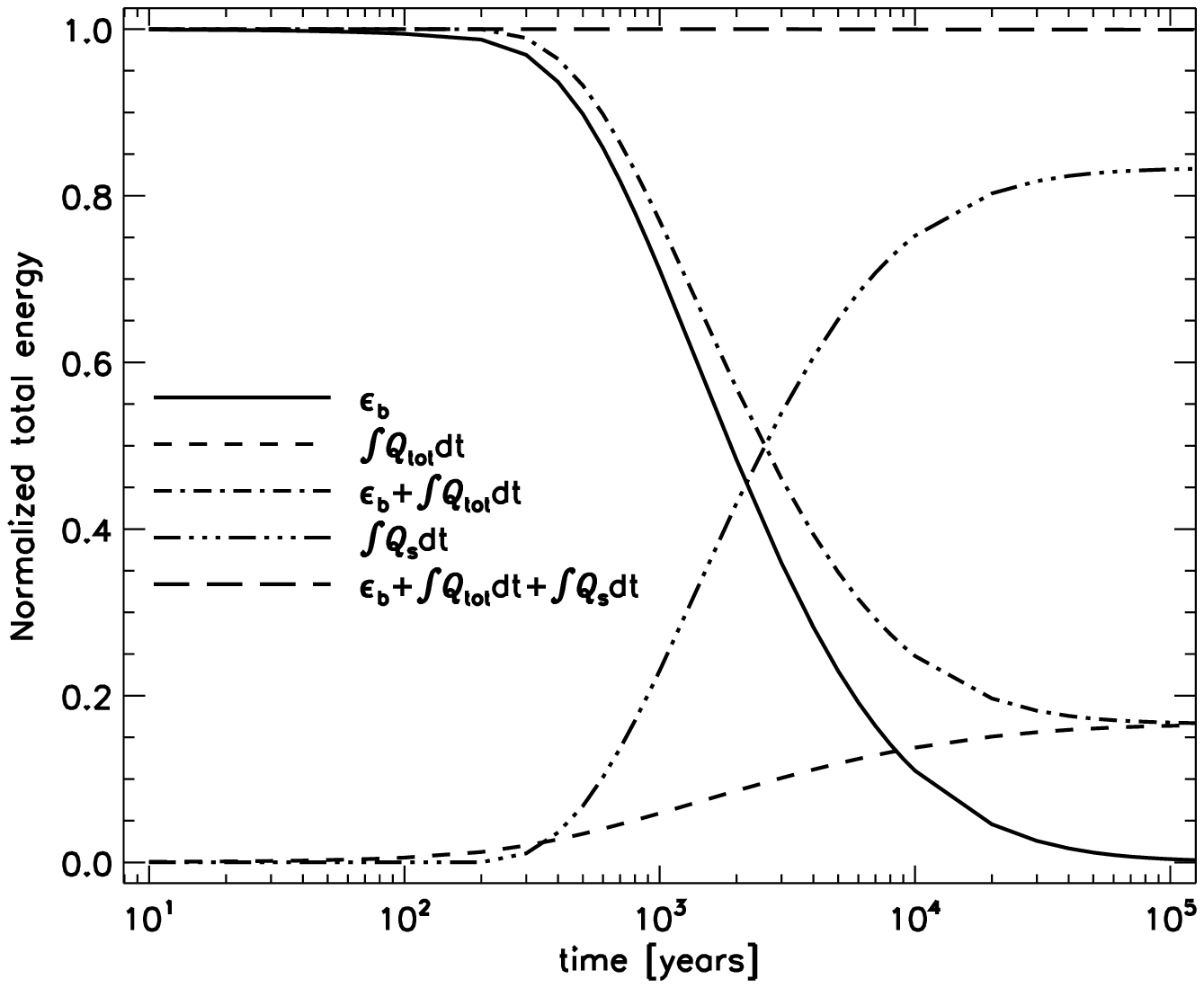}
\caption{Instantaneous (left) and time-integrated (right) energy balance for the same model as Fig.~\ref{fig:bphi}, with $114\times200$ crustal points. The variation of the magnetic energy (solid line), Joule dissipation rate (dashes) and dissipation rate at the shock (triple-dot-dash) are shown together with the total energy balance (long dashes). The right panel also shows the sum (dash-dots) of the magnetic energy and energy dissipated by Joule effect.}
 \label{fig:en_shock}
\end{figure}

The evolution equation for a purely toroidal magnetic field, neglecting resistivity, is described in spherical coordinates by eq.~(\ref{eq:induction_burgers}). In order to test this limit, we study the same model as in the previous section but with a larger field, taking $B_t^0=10^{16}$ G (implying $\omtau^{\rm max} \approx 1000$). Fig.~\ref{fig:bphi} shows four snapshots of the evolution of such configuration, where the dominant role of the Hall drift term leads to the fast formation of a current sheet in the equatorial plane. In Fig.~\ref{fig:shock_up} we show a sequence of angular profiles of $B_\varphi$ (left) and $J_r/c$ (right) close to the star surface. Our numerical code is able to follow the formation of the discontinuity and captures the thin current sheet with only two grid points.

In Fig.~\ref{fig:en_shock} we show the contribution of the different terms in the energy balance equation as a function of time. In the left panel, the sum of the magnetic energy (solid) and the energy dissipated by the Joule effect (short dashes) is shown by a dash-dotted line. The sum of these two terms is constant and equal to the initial magnetic energy for smooth solutions, as shown in the previous subsection. We can observe that now this is very accurately satisfied until $\approx 200$ yr, coinciding with the formation of the current sheet. After this point, the local energy balance calculated as above seems to be violated. About $90\%$ of the magnetic energy is dissipated in $10^4$ yr, but the energy dissipated by the Joule effect is only $13\%$ of the initial magnetic energy, yielding to an apparent loss of $77\%$ of the total energy. This seemingly incorrect result has a physical explanation: when a current sheet forms, there is energy dissipated at the discontinuity not accounted for by the standard numerical evaluation of the Joule dissipation rate. We find the same result with $N_\theta=60,120,$ or 200. However, we unavoidably underestimate the dissipation at the discontinuity, which affects the total energy balance, although the implied volume is small.

A rigorous mathematical study of the energy dissipation rate at the shock in one-dimensional Burgers flows, governed by $\partial_t u + u\partial_x u=\nu \partial_{xx}u$, shows that, in the inviscid limit $\nu=0$, the total energy dissipation rate goes as $2[u]^3/3$, where $[u]$ denotes the half-jump in the variable $u$ across the shock \citep{tran10}.  By analogy with the results obtained for one-dimensional Burgers flows, we propose an estimate of the energy dissipated at the shock across a surface $S_\alpha$:
\begin{equation}
 {\cal Q}_s=-\frac{1}{6\pi}\lambda_r[B_\varphi]^3 S_r~,
\end{equation}
if the discontinuity is in the radial direction, or
\begin{equation}\label{qs_ang}
 {\cal Q}_s=-\frac{1}{6\pi}\lambda_\theta[B_\varphi]^3 S_\theta~,
\end{equation}
if it is in the angular direction. This correction is accounting for the magnetic flux lost across the surface of the magnetic discontinuity. The coefficients $\lambda_r$, $\lambda_\theta$ are defined in eqs.~(\ref{eq:induction_burgers_lambda1}) and (\ref{eq:induction_burgers_lambda2}). Note that the correction has to be applied only at the interfaces where $\lambda_\alpha[B_\varphi]$ is negative, which is the necessary condition for having a shock. In Fig.~\ref{fig:en_shock} we show the effect of the above correction in the energy balance when a discontinuity in the toroidal magnetic field is detected. Initially, when the magnetic field is smooth, the extra term (triple-dot-dash) plays no role. It should vanish, but discretization errors result in a $\lesssim 1\%$ correction. However, when the current sheet is formed, it becomes the dominant contribution to explain the loss of magnetic energy. Applying this correction, the total energy balance is very well satisfied. After $10^5$ yr, when the magnetic field has been almost completely dissipated, the error in the total energy conservation is only $\epsilon_{en}\sim 7 \times 10^{-4}$.

Therefore, this estimation of the dissipated energy at the shock turns out to be an excellent approximation. Note that in this particular case (a purely toroidal magnetic field forming a current sheet exactly at the equator, with the discontinuity in the angular direction) the generalization of the planar exact result to spherical coordinates is a very good approximation, since each radial layer is governed by an independent one-dimensional Burgers-like equation in the vicinity of the equator. In more general cases (asymmetric with respect to the equator, or including poloidal components) this approximation  must be taken only as an estimate of the energy dissipation rate, within a factor of two. This term is important for realistic magneto-thermal evolution models, because it strongly enhances the local deposition of heat.

%%%%%%%%%%%%%%%%%%%%%%%%%%%%%%%%%
\section{Boundary conditions.}\label{sec:magnetic_bc}

\subsubsection{Axis.}

The axial symmetry requires reflecting boundary conditions on the axis for any vector field $\vec{F}$ (magnetic field, electric field, current density...): the magnetic field has to be purely radial, with $F_\theta=F_\varphi=0$. Indicating with $i_a$ the index labeling the axis, and considering the ghost cells just beyond the $\theta\in[0,\pi]$ domain, we have explicitly
\begin{eqnarray}
&& F_r^{(i_a-1)}=F_r^{(i_a+1)}~, \nonumber\\
&& F_\theta^{(i_a-1)}=-F_\theta^{(i_a+1)}~, \\
&& F_\varphi^{(i_a-1)}=-F_\varphi^{(i_a+1)}~. \nonumber
\end{eqnarray}

\subsubsection{Outer boundary condition: potential solution.}

The electron density drops over 20 orders of magnitude from the outer crust to the magnetosphere. As a consequence, the time-scale of current dissipation outside the star is much shorter than inside. Although magnetars are likely to have twisted, evolving magnetospheric configurations, for long-term evolution purposes one can assume that the magnetic field outside the star is a potential configuration. Therefore we match the crustal magnetic field to a potential solution at the surface $(r=R_\star)$. The vacuum boundary condition is equivalent to avoid current escaping (entering) from (into) the star. The non-vanishing Poynting flux across the outer boundary allows the (small) interchange of magnetic energy with the external field.

At each time-step, we reconstruct the value of $B_\theta$ in the first external ghost cell from the radial magnetic field $B_r(R_\star,\theta)$. In Appendix~(\ref{app:vacuum_bc}) we give details about the two equivalent numerical methods used (spectral decomposition or Green's method). We have verified that, for a given $B_r(R_\star,\theta)$, both methods provide the same function $B_\theta(R_\star,\theta)$, reproducing correctly the analytical cases (e.g., a dipole). By means of the spectral decomposition of $B_r(R_\star,\theta)$ (see Appendix~\ref{app:vacuum_spectral}), we also obtain the dipolar component of the external magnetic field

\begin{equation}\label{eq:def_bpole}
  B_p = \frac{3}{2}\int_{-1}^1  B_r(R_\star,\theta) \cos\theta~\de(\cos\theta)~.
\end{equation}
The time evolution of this value enters in the computation of the spin-down rate of the star, eq.~(\ref{eq:ppdot_spindown}).

\subsubsection{Inner boundary condition.}

The first choice for the inner boundary condition assumes that the magnetic field is confined to the crust. This could be the case if some mechanism was able to expel the magnetic field from the core, like the Meissner effect in type I superconductors. In this case, at the crust/core interface $r=R_{core}$, the radial magnetic field and the tangential electric field vanish, $B_r=E_\theta=E_\varphi=0$. Across the interface the Poynting flux is zero and no energy is allowed to flow into/from the superconducting core. Note that the tangential magnetic field, $B_\theta,B_\varphi$ and the radial electric field $E_r$ in general are non-zero.

If, instead, the magnetic field threads also the core of the star, then the mathematical singularity $r=0$ and the axial symmetry require the configuration to be regular. The dipolar solutions in the limit $r\rightarrow 0$, corresponds to a regular, homogeneous magnetic field aligned with the axis. As a boundary condition in the first radial point $r_1 > 0$, we apply:

\begin{eqnarray}
  && B_\theta^{(i,1)} =-B_r^{(i,1)}\tan\theta_i~, \label{bessel_center}\\
  && B_\varphi^{(i,1)} = 0~.
\end{eqnarray}

\chapter{Microphysics}\label{ch:micro}

In this chapter we revisit the main microphysical ingredients needed to simulate the long-term cooling of neutron stars. The aim is to give a brief overview, focusing mainly on the crustal properties. For more extended reviews, see, for instance, \cite{chamel08} and \cite{page12}.

\section{Equation of state.}\label{sec:eos}

%%%%%%%%%%%%%%%%%%%%%%%%%%%%%%%%%%%%%%
\begin{figure}
\centering
\includegraphics[width=0.7\textwidth]{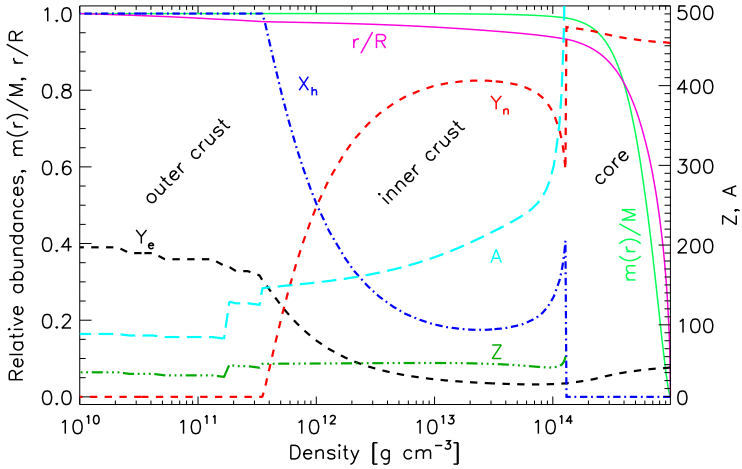}
\caption{Neutron star composition for the chosen equation of state. We indicate: the mass fraction inside the nuclei $X_h$ (blue dot-dashed line), the number of electrons per baryon $Y_e$ (black dashes) and free neutrons per baryon $Y_n$ (red dashes), the atomic number $Z$ (dark green triple dot-dashed), the mass number $A$ (cyan long dashes), the radius (normalized to $R_\star$) as a function of density (pink solid) and the corresponding enclosed mass (normalized to the star mass $M=1.4~M_\odot$, green solid).}
\label{fig:ns_profile}
\end{figure} 
%%%%%%%%%%%%%%%%%%%%%%%%%%%%%%%%%%%%%%%%%%%%%%%%%%%
\begin{table}
\begin{center}
 \begin{tabular}{c c c c c c c c}
\hline
\hline
$M$ 		& $\rho_{c,15}$ & $R_\star$ & $R_{core}$ & $e^{\lambda(R_\star)}$ & $g_{14}$ & $I_{45}$ & $C_{sd}(f_\chi=1)$ 	\\
$[M_\odot]$ 	&		&    [km]   &    [km]    &        		  & 	     &  	& [$10^{19}$ G] \\
\hline
1.10 & 0.79 & 11.82 & 10.54 &  1.174  &  1.04 & 1.24 & 4.31 \\
1.25 & 0.88 & 11.77 & 10.69 &  1.207  &  1.19 & 1.43 & 4.69 \\
1.32 & 0.93 & 11.74 & 10.75 &  1.223  &  1.27 & 1.52 & 4.87 \\
1.40 & 0.99 & 11.70 & 10.79 &  1.244  &  1.36 & 1.62 & 5.08 \\
1.48 & 1.05 & 11.64 & 10.82 &  1.265  &  1.45 & 1.71 & 5.30 \\
1.60 & 1.16 & 11.54 & 10.82 &  1.301  &  1.59 & 1.83 & 5.64 \\
1.70 & 1.28 & 11.42 & 10.79 &  1.336  &  1.73 & 1.92 & 5.95 \\
1.76 & 1.36 & 11.33 & 10.74 &  1.359  &  1.82 & 1.96 & 6.16 \\
\hline
\hline
\end{tabular}
\end{center}
\caption{Properties of the neutron star models considered in this work. We list, from left to right: the total mass, the central density (in units of $10^{15}$ \gcc), the radius of the star, the radius of the core, the space curvature factor at the surface (see eq.~\ref{eq:metric}), the surface gravity (normalized to $10^{14}$ cm s$^{-2}$), the moment of inertia (normalized to $10^{45}$ g~cm$^2$), and the pre-factor related to the inferred surface dipolar magnetic field (eq.~\ref{eq:inferred_bpole} with $f_\chi=1$).}
\label{tab:nsmasses}
\end{table}
%%%%%%%%%%%%%%%%%%%%%%%%%%%%%%%%%%%%%%%%

In this work, we employ a Skyrme-type equation of state at zero temperature to describe both the crust and the liquid core, based on the effective nuclear interaction SLy \citep{douchin01}, and ignoring the presence of muons. Below the neutron drip point, we use the standard BPS equation of state \citep{baym71}. In Fig.~\ref{fig:ns_profile}, we show, as a function of density, the profiles of the number of electrons and free neutrons per baryon ($Y_e$ and $Y_n$, respectively), the fraction of mass inside the nuclei $X_h$, the atomic number $Z$ and the mass number $A$. The appearance of free neutrons, $Y_n>0$, marks the neutron drip point at $\rho_d\simeq 3.5\times 10^{11}~$\gcc, which defines the boundary between outer and inner crust. The density at the crust-core interface, above which we have uniform nuclear matter, is $1.3\times 10^{14}~$\gcc, about half the value of nuclear saturation density. Close to the interface, $A$ and $Z$ reach values as large as 600 and 53, respectively. In the core, the fraction of protons is equal to $Y_e$ to ensure local charge neutrality. 

With this equation of state, we build neutron star models with eight different masses. The properties are listed in Table~\ref{tab:nsmasses}. The star radius slightly shrinks for increasing mass, but the thickness of the outer layer (crust plus envelope), $R_\star-R_{core}$, varies between 1.3 and 0.6 km. In Fig.~\ref{fig:ns_profile}, the pink solid line is the radius (normalized to $R_\star$) as a function of density, while the green line represents the fraction of the mass enclosed within this radius, normalized to the star mass $M=1.4~M_\odot$. Note that $\sim 99\%$ of the mass is contained in the core.

The pre-factor $C_{sd}$, defined in eq.~(\ref{eq:factor_binferred}), enters in the evaluation of the inferred magnetic dipolar field, $B_p = C_{sd} (P\mbox{[s]} \dot{P})^{(1/2)}$. In the classical fiducial model with $R_\star=10$ km, $I_{45}=1$, $f_\chi=1$, often found in literature, the pre-factor is $C_{sd}=6.4\times 10^{19}$ G, which is $5$--$30 \%$ larger than in our models (mainly due to the values of our radii, $R_\star>10$ km). If neutron stars were actually described by the chosen equation of state, then the magnetic field inferred from the fiducial values would be slightly overestimated. In other words, the magnetic spin-down for a given $B_p$ and $f_\chi$, eq.~(\ref{eq:k_spindown}), is faster in our models than in the commonly employed fiducial values.

It is not the purpose of this work to explore different models of the equation of state. For a recent review about this issue, including a discussion about measurements of masses and radii in neutron stars, see \cite{lattimer12}. 

\section{Plasma properties in the crust.}\label{sec:temperatures}

\begin{figure}
\centering
\includegraphics[width=0.7\textwidth]{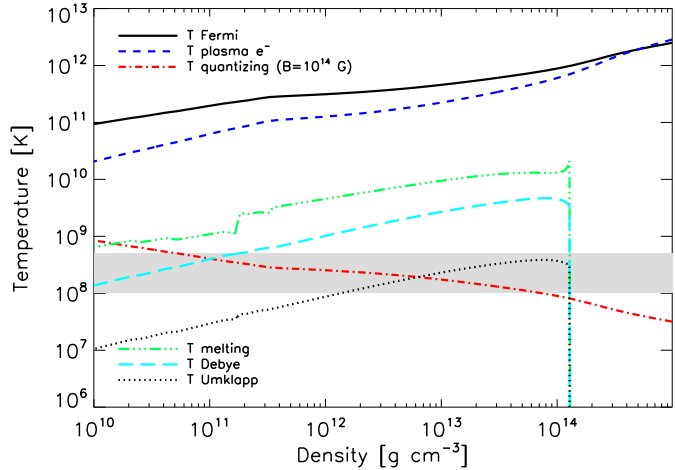}
\caption{Characteristic temperatures for the neutron star model of Fig.~\ref{fig:ns_profile} (see text for definitions). The quantizing temperature $T_{quant}$ is computed for $B=10^{14}$ G and scales linearly with $B$. The gray band indicates the typical internal temperature of detectable neutron stars.}
\label{fig:characteristic_temp}
\end{figure}

In order to understand the properties of matter in neutron star conditions, it is useful to introduce some characteristic temperatures. In Fig.~\ref{fig:characteristic_temp}, we show their profiles, as a function of density, above $\rho> 10^{10}$ \gcc (the envelope is not considered here). The typical range of crustal temperatures for detectable neutron star is $\sim 1-5 \times 10^8$ K, denoted by a gray band. Higher temperatures are reached only during the first years of a neutron star life, while neutron stars with lower temperatures are hardly detectable in X-rays.

%%%%%%%%%%%%%%%%%%%%%%%%%%%%%%%%%%%%%%
%%%%%%%%%%%%%%%%%%%%%%%%%%%%%%%%%%%%%%%%%%%%%%%%%%%

\subsubsection{Electron Fermi temperature.}
The Fermi momentum of electrons is defined by their density $n_e$:

\begin{equation}\label{eq:fermi_momentum}
p_F = \hbar(3\pi^2n_e)^{1/3} = \hbar(3\pi^2n_e)^{1/3}~.
\end{equation}
We define the dimensionless Fermi momentum as

\begin{equation}\label{eq:fermi_momentum_dimless}
 x_F \equiv \frac{p_F}{m_ec} = 0.01 \left(\frac{\rho}{\mbox{[\gcc]}}\frac{Z}{A}\right)^{1/3}~,
\end{equation}
where we have used $n_e=\rho Z/A m_u$, with $m_u=1.66\times 10^{-24}$ g (atomic mass unit). The electrons are ultra-relativistic when $x_F\gg 1$, which happens for $\rho \gg (A/Z)10^6$ \gcc. Their Fermi energy is
\begin{equation}
  \epsilon_F = m_ec^2\sqrt{1 + x_F^2} \equiv m_e^\star c^2~,
\end{equation}
where $m_e^\star$ is the effective mass of the electron. For comparison, we estimate the the potential energy related to the Coulomb attraction between electrons and ions as
\begin{equation}
  E_C = \frac{Ze^2}{a_i} \simeq 0.042~Z\left(\frac{A}{\rho_{10}}\right)^{1/3}~\mbox{MeV}~,
\end{equation}
where $\rho_{10}=\rho/10^{10}$ \gcc, and we have approximated the typical ion-electron distance to be equal to the radius of the ion sphere:
\begin{equation}\label{eq:ion_sphere}
 a_i = \left(\frac{3}{4 \pi n_i}\right)^{1/3} = \left(\frac{3 A m_u}{4 \pi \rho}\right)^{1/3} = 35 \left(\frac{A}{\rho_{10}}\right)^{1/3} \mbox{~fm}~.
\end{equation}
Since $\epsilon_F \gg E_C$ everywhere in the crust, matter is fully pressure-ionized and electrons can be considered an ideal gas.

The profile of the Fermi temperature, defined as
\begin{equation}
 T_F = \frac{m_ec^2}{k_b}\left(\sqrt{x_F^2+1}-1\right)~,
\end{equation}
is shown in Fig.~\ref{fig:characteristic_temp} with a black solid line. It is much larger than the star temperature during the whole lifetime of a neutron star. In summary, the electrons can be treated as an ideal, ultra-relativistic and strongly degenerated Fermi gas.

\subsubsection{Electron plasma temperature.}
The plasma frequency has already been defined in eq.~(\ref{eq:plasma_frequency}), in the context of the magnetosphere, as the frequency of the collective mode of a free gas composed by charged particles. Inside the star, the free gas of electrons can be associated to a plasma temperature

\begin{equation}\label{eq:t_plasma}
 T_{p,e}=\frac{\hbar}{k_b} \left(\frac{4\pi n_e e^2}{m_e}\right)^{1/2} = 3.3\times10^{10}~\left(\rho_{10}\frac{Z}{A}\right)^{1/2}~\mbox{K} ~.
\end{equation}
The plasma temperature is very large in the crust and in the core compared with the typical temperature (compare the blue dashes with the gray band in Fig.~\ref{fig:characteristic_temp}), therefore electrons in neutron stars are not prone to plasma effects, such as collective electron motions.

\subsubsection{Electron quantizing temperature.}
As we have seen in \S~\ref{sec:rcs}, magnetic fields affect the propagation of charged particles. The electron gyro-frequency, eq.~(\ref{eq:def_gyrofrequency}), is associated to the quantizing temperature 

\begin{equation}\label{eq:t_quantizing}
 T_{quant} = \frac{\hbar\omega_B}{k_b}\sim 1.34\times10^8~ \frac{B_{12}}{(1 + x_F^2)^{1/2}}~\mbox{K}~.
\end{equation}
Its profile for $B=10^{14}$ G is shown in Fig.~\ref{fig:characteristic_temp} with a red dot-dashed line. The density at which the Fermi energy reaches the lowest Landau level is

\begin{equation}\label{eq:rho_quantizing}
\rho_{quant} = \frac{A}{Z}\frac{m_u}{\pi^2 \sqrt{2}} \left(\frac{eB}{\hbar c}\right)^{3/2} \simeq 7.045\times 10^3 ~ \frac{A}{Z} B_{12}^{3/2} ~\mbox{\gcc}~.
\end{equation}
The magnetic field is strongly quantizing if $T\lesssim T_{quant}$ and $\rho \lesssim \rho_{quant}$. The latter condition is satisfied only in the outer envelope. In the crust, $T$ can be comparable with $T_{quant}$, but $\rho \gtrsim \rho_{quant}$. In this case the quantum oscillations are not very pronounced and occur around their classical value: the magnetic field is weakly quantizing. If $T \gg T_{quant}$ and $\rho \gg \rho_{quant}$, then many Landau levels are occupied, and the classical approximation holds. In \S~\ref{sec:conductivity} (see especially Fig.~\ref{fig:cond_class_quant}), we will consider the quantizing effects on the thermal and electrical conductivities.

\subsubsection{Ion melting temperature.}
The physical state of ions is described by the Coulomb parameter, that compares their Coulomb and thermal energies:
\begin{equation}\label{eq:coulomb_parameter}
  \Gamma_{coul} \equiv \frac{(Z e)^{2}}{k T a_i} \approx 4.77~\frac{Z^2}{T_8} \left(\frac{\rho_{10}}{A}\right)^{1/3}~,
\end{equation}
where $T_8$ is the temperature in units of $10^8$~K, and $a_i$ has been defined in eq.~(\ref{eq:ion_sphere}). When $\Gamma_{coul} < 1$, the ions form a Boltzmann gas, when $1 \leq \Gamma_{coul} < \Gamma_m$ their state is a coupled Coulomb liquid, and when $\Gamma_{coul} > \Gamma_m$ the liquid freezes into a Coulomb lattice. For a one component plasma (i.e., homogeneous composition), the melting value $\Gamma_m=175$ is found in molecular dynamics simulations of neutron star matter, for a body-centered cubic, pure lattice. For a multi-component plasma, i.e. with impurities in the lattice, $\Gamma_m$ can be $\sim 40\%$ larger \citep{horowitz07}. The corresponding melting temperature, shown with the green triple dot-dashed line in Fig.~\ref{fig:characteristic_temp}, is

\begin{equation}\label{eq:t_melting}
  T_m = \frac{(Z e)^{2}}{k_b \Gamma_m a_i} \approx 2.75\times10^6~\frac{Z^{2}}{(\Gamma_m/175)} \left( \frac{\rho_{10}}{A} \right)^{1/3}~\mbox{K}~.
\end{equation}
If the temperature in the star is lower than this critical value, ions form a strongly coupled Coulomb lattice, and the crust is a high-density version of a terrestrial metal. Cooling simulations show that matter at densities above the neutron drip point freezes very soon, from minutes to days after birth, while the layers with $\rho\sim 10^{10}$ \gcc~require a few years.

\subsubsection{Ion plasma and Debye temperature.}
Uncorrelated single particle motion for ions is relevant only if $T\gg T_{p,i}$, where $T_{p,i}$ is the ion plasma temperature

\begin{equation}\label{eq:t_plasma_ion}
 T_{p,i}=\frac{\hbar}{k_b} \left(\frac{4\pi n_i (Ze)^2}{m_i}\right)^{1/2} = 7.8\times10^{8}~\frac{Z}{A}\rho_{10}^{1/2} ~\mbox{ K} ~.
\end{equation}
In neutron stars, $T\lesssim T_{p,i}$, therefore the ions are subject to collective excitations, the phonons. The Debye temperature is closely related to $T_{p,i}$:

\begin{equation}\label{eq:t_debye}
  T_D\simeq 0.45 T_{p,i} = 3.5\times10^8~\frac{Z}{A}\rho_{10}^{1/2}~\mbox{~K} ~.
\end{equation}
When $T\gtrsim T_D$, phonons behave as a classical gas. If $T\lesssim T_D$, which usually holds in neutron stars (see cyan long dashes in Fig.~\ref{fig:characteristic_temp}), the number of thermal phonons is strongly reduced and their quantum nature becomes important. In this case one longitudinal and two transverse phonon modes are enough to describe the system. 

\subsubsection{Umklapp temperature.}
An important parameter related to the so-called Umklapp scattering between phonons and electrons is the Umklapp temperature:

\begin{equation}\label{eq:t_umklapp}
 T_U \simeq T_D \frac{Z^{1/3}e^2}{3 \hbar v_F} = 8.5 \times 10^6~\frac{Z^{4/3}}{A}\rho_{10}^{1/2}\frac{c}{v_F}~\mbox{K} ~,
 \end{equation}
where $v_F \simeq c$ is the Fermi velocity of the ultra-relativistic electrons. For temperatures $T<T_U$, the band structure of electrons cannot be neglected, and the approximation of free electrons does not hold anymore. The dots in Fig.~\ref{fig:characteristic_temp} show that $T_U$ is low compared with the typical temperature of the detectable neutron stars.

\subsection{Superfluidity and superconductivity.}

In neutron stars, nucleons can pair because of their attractive interactions, analogously to the electron pairing in low-temperature solids, described by the Bardeen-Cooper-Schreiffer theory \citep{bardeen57}. While superfluidity properties in Earth conditions are well-known from cold atoms experiments, pairing in nuclear matter is poorly known, and its study needs complicated theoretical calculations. In the crust, neutrons are expected to pair in $^1S_0$ waves, while in the core they pair in $^3P_2$ waves. Protons are expected to pair in $^1S_0$ waves in the core of a neutron star; as they are charged, the pairing implies a superconducting state.

Most models agree that the proton pairing ceases to exist at high densities $\rho \gtrsim 10^{15}$ \gcc. The situation for the pairing of neutrons in the core is very complicated, also because relativistic effects become important. Up to now, there is no conclusive approach to the problem. Most calculations agree that the nuclear pairing energy, or energy gap, is between $\Delta=0.1$ to $1$ MeV (see, e.g., \citealt{gezerlis10}), with much uncertainty about the dependence on density.

%%%%%%%%%%%%%%%%%%%%%%%%%%%%%%%%%%%
\begin{table}[t]
\begin{center}
\begin{tabular}{l c c c c c}
\hline
\hline
State & $D$ & $k_0$ & $k_1$ &$k_2$ &$k_3$ \\
      &   [MeV]    & [fm$^{-1}$] & [fm$^{-1}$] & [fm$^{-1}$] & [fm$^{-1}$]\\
\hline
$p$~$^1S_0$ & 120 & 0   & 9 & 1.3 & 1.8 \\
$n$~$^1S_0$ &  68 & 0.1 & 4 & 1.7 & 4   \\
$n$~$^3P_2$ & 0.15& 2   &0.1& 3.1 & 0.02\\
\hline
\hline
\end{tabular}
\end{center}
\caption{Parameters entering in the chosen functional form of the energy gap, eq.~(\ref{eq:gap_param}), taken from \cite{ho12}.}
\label{tab:gaps}
\end{table}

%%%%%%%%%%%%%%%%%%%%%%%%%%%%%%%%%%%%%%%%%%%%
\begin{figure}[t]
   \centering
   \includegraphics[width=0.6\textwidth]{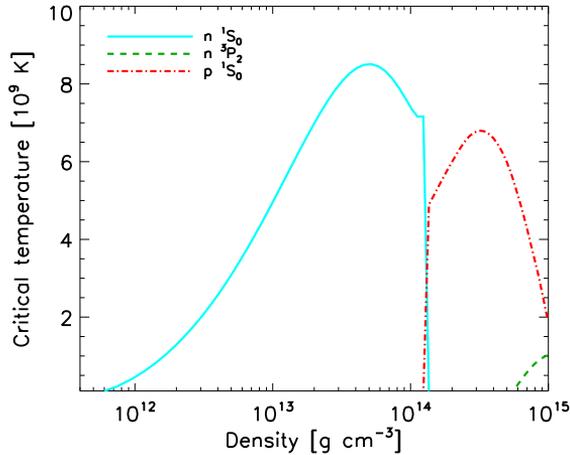}
\caption{Critical temperatures for the onset of superfluidity as a function of the density, for the chosen energy gap model (see text).}
\label{fig:sf_temp}
\end{figure}
%%%%%%%%%%%%%%%%%%%%%%%%%%%%%%%%%%%%%%%%%%%%%

We have employed the following phenomenological formula for the momentum dependence of the energy gap at zero temperature \citep{kaminker01}:
\begin{equation}\label{eq:gap_param}
 \Delta_0=D \frac{(k_F-k_0)^2}{(k_F-k_0)^2+k_1^2}\frac{(k_F-k_2)^2}{(k_F-k_2)^2+k_3^2}~,
\end{equation}
where $k_F=(3\pi^2n)^{1/3}$ is the Fermi wavenumber and $n$ is the particle density of each type of nucleons involved. The parameters $D$ and $k_i$, $i=1..4$ are listed in Table~\ref{tab:gaps}. This expression is valid for $k_0<k_F<k_2$, with vanishing $\Delta$ outside this range. Recently, the claimed observed temperature decrease of the neutron star in Cassiopeia A (see \citealt{elshamouty13} for a recent data analysis) has been attributed to the onset of superfluidity \citep{shternin11,page11}, although the interpretation of the observations has been strongly questioned \citep{posselt13}. Without loss of generality, we have arbitrarily adopted the parametrization shown in Table~\ref{tab:gaps}, employed by \cite{ho12} to explain the Cassiopeia A data.

The temperature dependence of the energy gap is given by \cite{levenfish94}:
\begin{equation}\label{eq:temperaturegaps}
\frac{\Delta(T)}{kT} \approx \sqrt{1- \frac{T}{T_c}}~\left(\alpha_g - \frac{\beta_g}{\sqrt{T/T_c}} +\frac{\gamma_g}{T/T_c}\right)~,
\end{equation}
where $\alpha_g=1.456$, $\beta_g=0.157$, and $\gamma_g=1.764$ for $^1S_0$, and $\alpha_g=0.789$, $\beta_g=0$, and $\gamma_g=1.188$ for $^3P_2$ states. The corresponding critical temperatures are
\begin{equation}
 T_c=\frac{\Delta_0}{k_b\gamma_g}~. 
\end{equation}
The density dependence of the critical temperatures is plotted in Fig.~\ref{fig:sf_temp}: the maximum critical temperature for crustal neutrons is $T^{\rm max}_{c,n} \simeq 8.5 \times 10^9$ K, while for protons in the core is $T^{\rm max}_{c,p} \simeq 7\times 10^9$ K. This implies that for these components the transition to the superfluid/superconducting phase starts soon after birth, with a  large part of the core becoming superconducting within the first year. Conversely, the epoch of the transition to a neutron superfluid phase in the core is very uncertain, likely happening tens to thousands of years later.

Superfluidity affects neutrino emissivities and the nucleon specific heat, without changing significantly the equation of state. Since the purpose of this work is not to explore the huge space of parameters included in superfluidity models, we remit detailed discussions about their imprint on cooling models to dedicated works (e.g., \citealt{page04}, \citealt{aguilera08b} and \citealt{page09}).

\section{Specific heat.}

%%%%%%%%%%%%%%%%%%%%%%%%%%%%%%%%%%%%%%%%%%%%%%%%%%%%%%%%%%%%%%%%%%%%%%%%%%
\begin{figure}[t]
\centering
\includegraphics[width=0.45\textwidth]{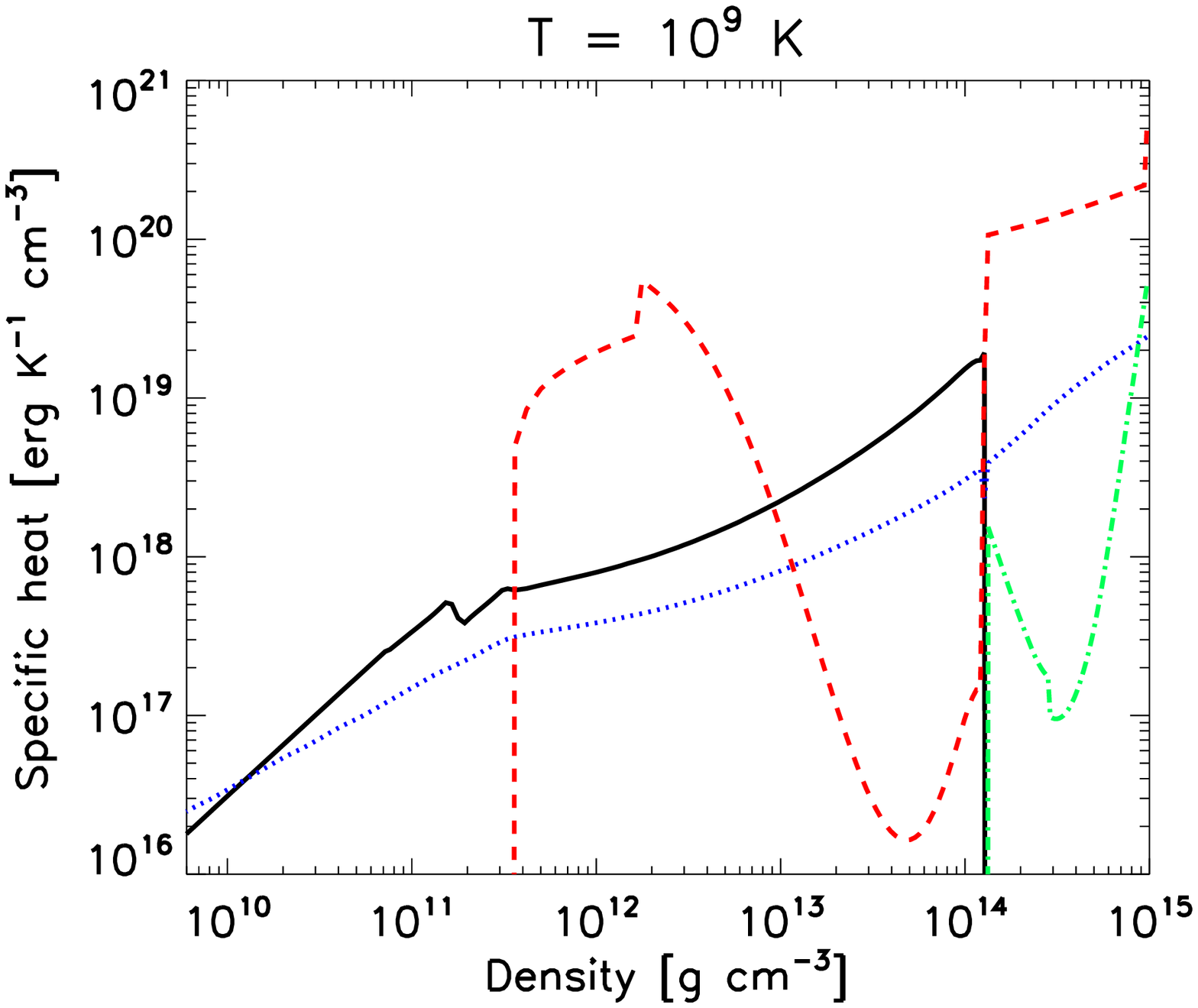}
\includegraphics[width=0.45\textwidth]{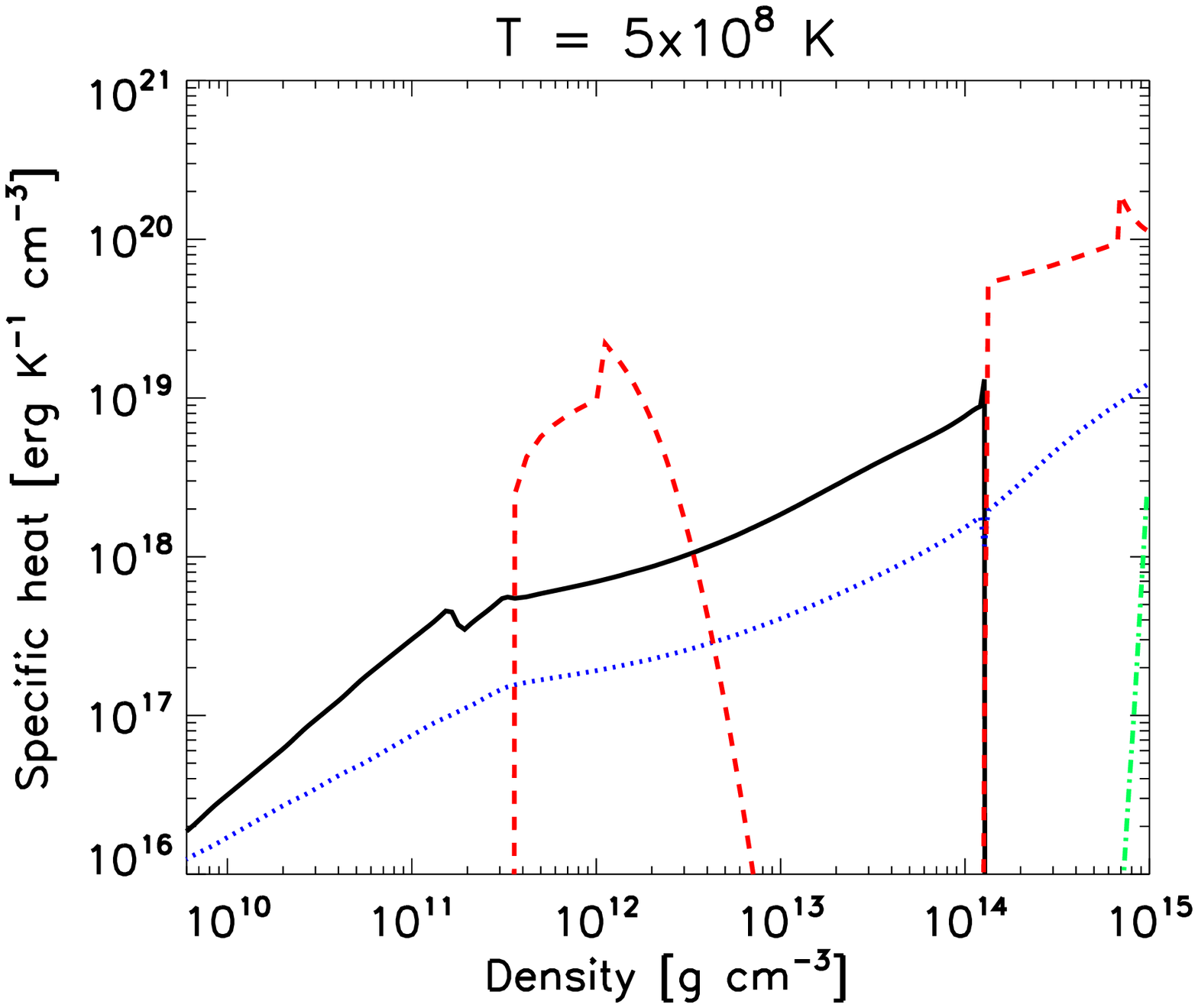}\\
\includegraphics[width=0.45\textwidth]{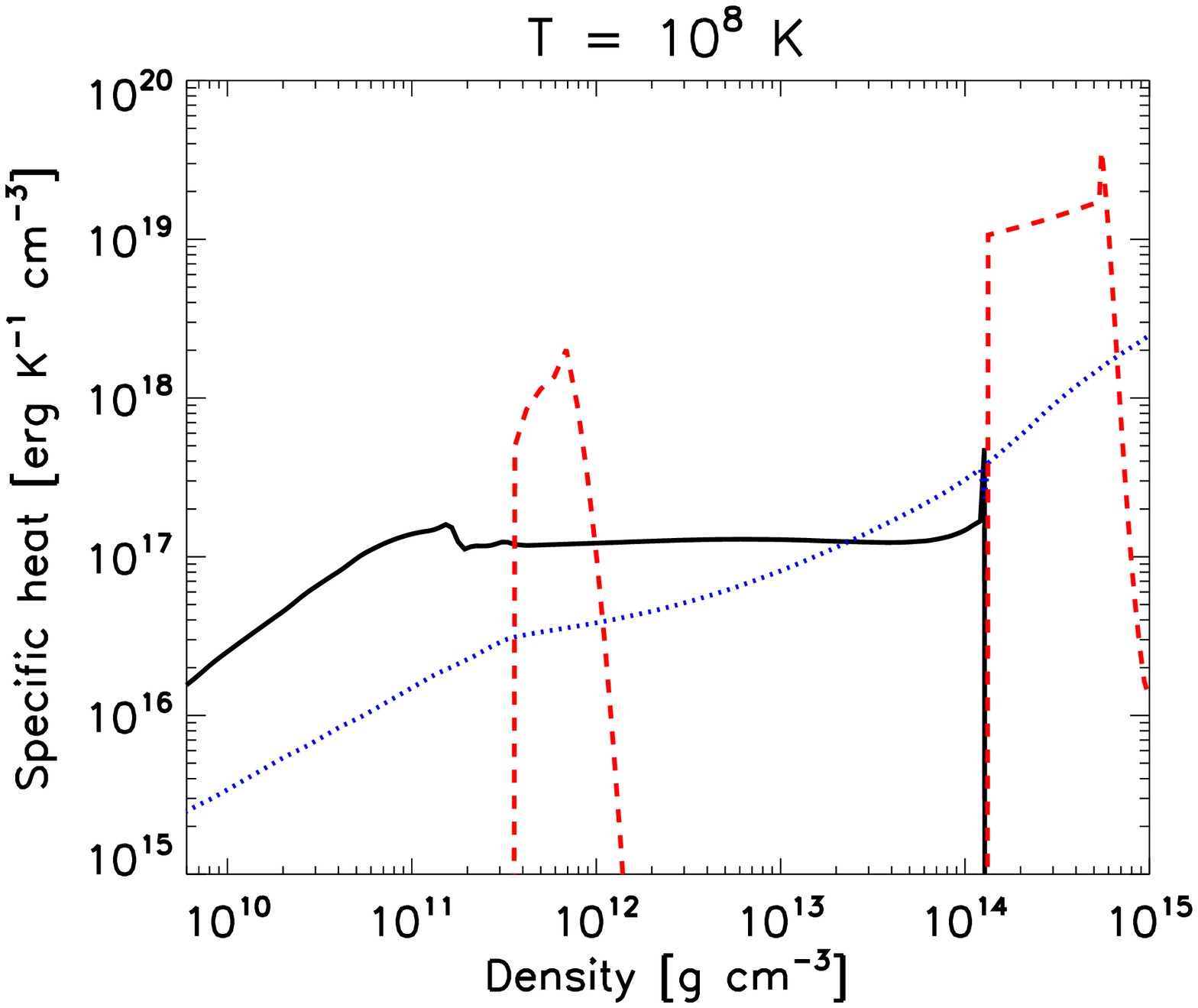}
\includegraphics[width=0.45\textwidth]{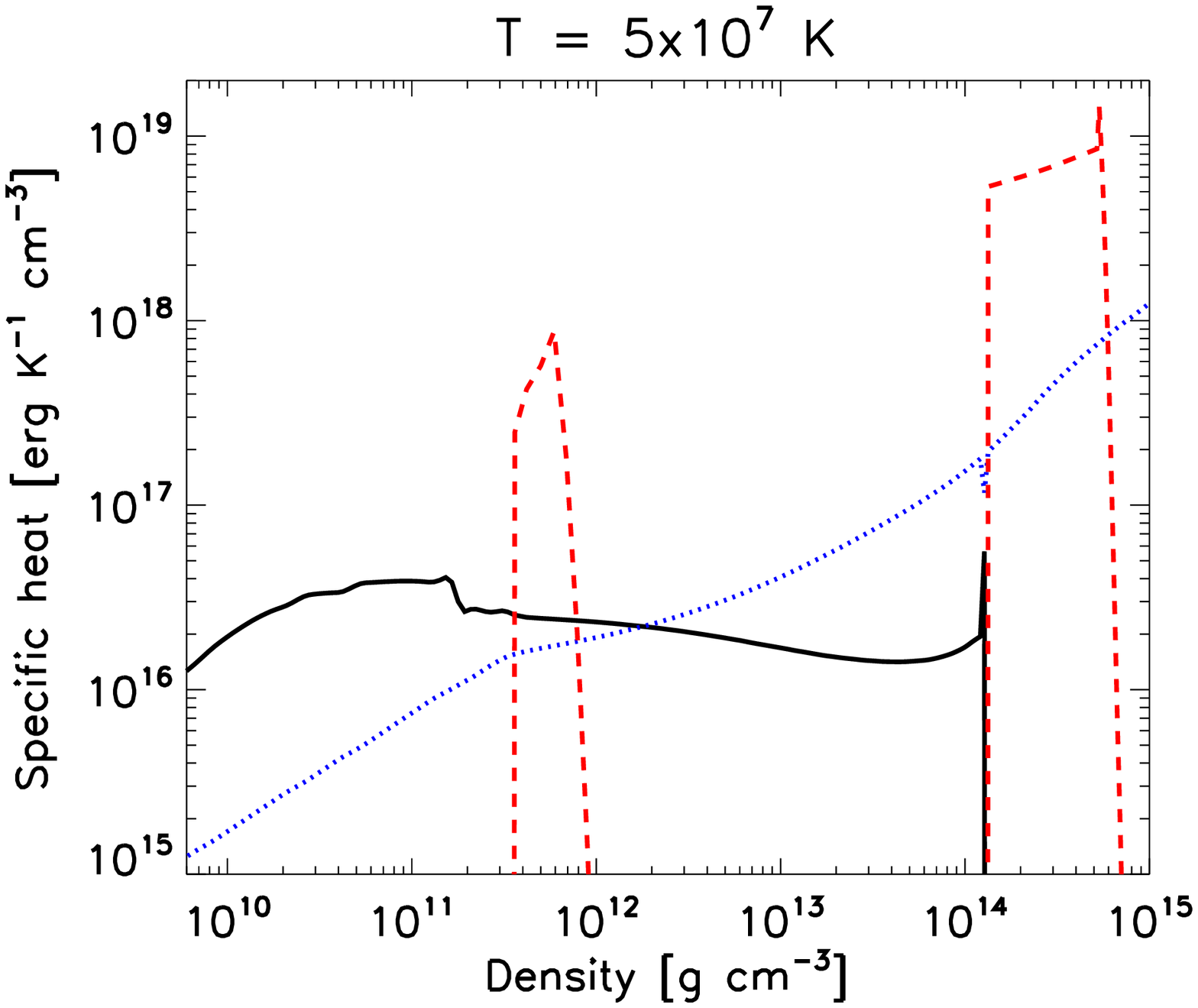}
\caption{Contribution to the specific heat as a function of density for different temperatures: neutrons (red dashes), protons (green dot-dashed), electrons (blue dots), and ions (black solid line).}
\label{fig:cv}
\end{figure}
%%%%%%%%%%%%%%%%%%%%%%%%%%%%%%%%%%%%%%%%%%%%%%%%%%%%%%%%%%%%%%

For degenerate fermions with Fermi momentum $k_F$ and number density $n$, the specific heat per unit volume is 
\begin{equation}\label{eq:specific_heat}
c_v=\pi^2\frac{n k_b^2T}{m c^2}\frac{\sqrt{x^2_F+1}}{x_F^2}~,
\end{equation}
where $x_F=\hbar k_F/mc$. The contribution of relativistic electrons is
\begin{equation}
c_{v,e}\simeq 5.4\times 10^{19}\, \left(\frac {n_e}{n_0}\right)^{2/3} T_9 \quad{\rm erg~cm}^{-3}{\rm K}^{-1}~,
\end{equation}
while for non-relativistic nucleons $N=n,p$ is
\begin{equation}
 c_{v,N}\simeq 1.6\times 10^{20}\,\frac{m^{\ast}_N}{m_N} \left(\frac {n_N}{n_0}\right)^{1/3} T_9 \,\mathcal R^{cv} \quad{\rm erg~cm}^{-3}{\rm K}^{-1}~,
\end{equation}
where $n_0 \simeq 0.16$ fm$^{-3}$ is the nuclear saturation density, and $\mathcal R^{cv}$ is a factor accounting for the exponential reduction in presence of superfluidity \citep{levenfish94}. Its value depends on the pairing state of the nucleons involved ($^1S_0$ or $^3P_2$).

In Fig.~\ref{fig:cv} we show the contribution to the specific heat by ions, electrons, protons and neutrons. The linear dependence of temperature is evident when comparing the four panels. The bulk of the total heat capacity of a neutron star is given by the core, where most of the mass is contained (see green line of Fig.~\ref{fig:ns_profile}). The regions with superfluid nucleons are visible as deep drops of the specific heat. The proton contribution is always negligible. Neutrons in the outer core are not superfluid, thus their contribution is dominant.

The crustal specific heat is given by the neutron gas, the degenerate electron gas and the nuclear lattice \citep{vanriper91}. For the exact computation of the specific heat in the crust, we use the latest version of the equation of state for a strongly magnetized, fully ionized electron-ion plasma \citep{potekhin10}\footnote{Code publicly available at {\tt http://www.ioffe.ru/astro/EIP/index.html}}. The specific heat of the lattice is generally the main contribution, except in parts of the inner crust where neutrons are not superfluid, or for temperatures $\lesssim 10^8$ K, when the electron contribution becomes dominant. In any case, the small volume of the crust implies that its heat capacity is small in comparison to the core contribution.

\section{Thermal and electrical conductivities.}\label{sec:conductivity}

The transport of heat is controlled by the thermal conductivity, while the electrical conductivity regulates the transport of momenta by charged particles. The conductivities are determined by the mean free path of the relevant carriers. The scattering processes of charged particles (electrons, protons, ions) will determine both the electrical and thermal conductivity, while neutron and phonon processes have influence only on the thermal conductivity. Radiative transport by photons is important only close to the surface and will not be treated here.

Giving details about the formal calculation of conductivities goes far beyond the purposes of this work: in this section we shortly review the main aspects.

\subsection{Electron conductivity.}

In presence of a magnetic field with direction $\hat{b}$, we express the electron contribution to the anisotropic flux of heat and electrical currents as \citep{urpin80b}:

\begin{eqnarray}
 && \vec{F}=-\hat{\kappa}_e\cdot\vec{\nabla}(e^\nu T) = - \kappa_\parallel \vec{\nabla}_\parallel (e^\nu T) - \kappa_\perp \vec{\nabla}_\perp (e^\nu T) - \kappa_H \hat{b}\times\vec{\nabla} (e^\nu T)~, \label{eq:heat_flux}\\
 && \vec{J}= \hat{\sigma}\cdot\vec{E} = \sigma_\parallel \vec{E}_\parallel + \sigma_\perp \vec{E}_\perp + \sigma_H \hat{b}\times\vec{E} ~, \label{eq:current_flux}
\end{eqnarray}
where we have neglected the thermoelectric coefficients. The symmetric part of the $\hat{\sigma}$ tensor is related to the components in the $\vec{E}-\vec{B}$ plane, explicitly, $\vec{E}_\parallel= \hat{b}(\vec{E}\cdot\hat{b})$, $\vec{E}_\perp=\hat{b}\times(\vec{E}\times\hat{b})$. The antisymmetric part is perpendicular to both $\hat{B}$ and $\vec{E}$. The same holds for $\hat{\kappa}$ and $\vec{\nabla} (e^\nu T)$ in eq.~(\ref{eq:current_flux}).

As the typical energy transferred during a collision is much smaller than $k_bT$, we use the relaxation time approximation, which allows to express the thermal conductivity tensor as

\begin{equation}\label{eq:thermal_conductivity_tensor}
 \hat{\kappa}_e = \frac{\pi^2 k_b^2 n_e T}{3 \epsilon_F} {\hat \tau}^{th}_e~,
\end{equation}
and the electrical conductivity tensor as

\begin{equation}\label{eq:electrical_conductivity_tensor}
 \hat{\sigma} = \frac{n_e c^2 e^2}{\epsilon_F} {\hat \tau}^{el}_e~.
\end{equation}
We have introduced the {\it electron relaxation time} tensors, the components of which are $(\sum_k \nu_k)^{-1}$, i.e. the inverse of all the collision frequencies $\nu_k$ of the relevant processes involved in the transport of heat or electrical current. In the strongly quantized regime (see \S~\ref{sec:temperatures}), the components of the tensors are non-trivial and need proper numerical calculations considering the limited number of occupied Landau levels.

If the quantizing effects are not relevant due to very large number of occupied Landau states, then the following classic relations hold \citep{urpin80b}:

\begin{eqnarray}\label{eq:ratio_cond}
  && \tau_\parallel=\tau_e~, \nonumber\\
  && \tau_\perp = \frac{\tau_e}{1 + (\omega_B\tau_e)^2}~, \\
  && \tau_H = \frac{\omega_B\tau_e^2}{1 + (\omega_B\tau_e)^2}~,\nonumber
\end{eqnarray}
where $\omega_B$ is the gyro-frequency, eq.~(\ref{eq:def_gyrofrequency}), and $\tau_e$ is the non-magnetic relaxation time. The conductivities across magnetic field lines are suppressed by a factor $1 + (\omega_B\tau_e)^2$. This results in an anisotropic transport of heat. Note also that in this limit the electric field can be expressed as

\begin{equation}
 \vec{E}= \hat{\sigma}^{-1}\vec{J}~,
\end{equation}
where $\hat{\cal R}\equiv\hat{\sigma}^{-1}$ is the electrical resistivity tensor. Explicitly, it can be shown that
\begin{equation}
\vec{E}= \frac{1}{\sigma_\parallel} \vec{J} + \frac{B}{cn_ee} \vec{J}\times\hat{b} ~, \label{eq:generalized_ohm_matrix}
\end{equation}
where the first part term represent the Ohmic resistivity in the induction equation~(\ref{eq:induction_hall}), with $\sigma=\sigma_\parallel$. The second term does not depend on $\tau_e$ and leads to the Hall term.

In the non-quantizing limit, and if the contribution from electron-electron processes is negligible, then ${\hat \tau}^{el}_e={\hat \tau}^{th}_e$ and the Wiedemann-Franz law is satisfied:
\begin{equation}
 \hat{\kappa}_e = \frac{\pi^2 k_b^2 T}{3 e^2} \hat{\sigma}_e~.
\end{equation}

\subsection{Electron-impurity processes.}

In the calculation of the electron relaxation times in the crust, the main processes to be considered are the electron-phonon scattering and the electron-impurity collisions \citep{flowers76}. For definition, electron-electron collisions cannot transport momentum and contribute only to the thermal conductivity. As already noted by \cite{urpin80a}, this contribution is unimportant if electrons are strongly degenerated.

Now we focus on the electron-impurity collisions, which are important for relatively low temperatures. They are Rayleigh scattering processes, thus their rate does not depend on temperature, at contrast with electron-phonon scattering. The impurity content of the lattice is parametrized by the {\it impurity parameter}, i.e. the mean quadratic deviation of the atomic number:

\begin{equation}\label{eq:qimp}
Q_{imp}= \sum_i Y_i ( Z_i^2 - \langle Z^2 \rangle ) ~.
\end{equation}
where $Y_i$ is the relative abundance of the nuclide with $Z_ie$ charge, and $\langle Z^2 \rangle$ is the average squared atomic number in the lattice. $Q_{imp}$ is a measure of the distribution of the nuclide charge numbers, and it tells how pure ($Q_{imp}\ll 1$) or impure ($Q_{imp}\gg 1$) is the crust. In accreting neutron stars in binary systems, the outer crust is being continuously replenished by newly processed nuclei and it is not expected to be pure. The thermal relaxation time-scales after an outburst in low-mass X-ray binaries is the most direct measurement of the thermal conductivity of the outer crust at low temperature. The comparison of cooling models with data favours $Q_{imp}\sim 5$ in the outer crust of these old, accreting systems \citep{shternin07,brown09}.

In isolated neutron stars, especially in magnetars which remain warmer for longer times, $Q_{imp}$ in the outer crust is expected to be low. The question is whether the system has reached the ground state or not. While weak reactions inside a nucleus are very fast in changing the atomic number $Z$, nuclear fusion is needed to change the mass number $A$. Fusion can be triggered by thermonuclear or pycnonuclear reactions. The first ones can only take place at the early stages after the formation of neutron star, when the temperature is high enough. Pycnonuclear reactions, on the other hand, are triggered by the quantum fluctuations at high pressure and can happen even at zero temperature \citep{salpeter69,yakovlev06}. However, they are much slower, with highly uncertain time-scales, indicatively hundreds or thousands of years. As a consequence, it cannot be excluded that some metastable nuclei populate the lattice also at late ages.

A particularly important issue is the problem of shell effects associated with unbound neutrons scattered by nuclear inhomogeneities \citep{magierski02}, which seem to result in the coexistence of several phases different from a body-centered cubic lattice. However, we note that there are also open issues about phase separation and recent molecular dynamics simulations seem to favor a regular body-centered cubic lattice, even in accreting systems, when a large number of impurities are present \citep{horowitz09a,horowitz09b}. Recent studies of diffusion (motion of defects) \citep{hughto11} suggest that the the crust of neutron stars will be crystalline and not amorphous, at least for the outer crust.

In the inner crust, the uncertainties on the structure of the lattice are even larger. The deepest layer of the crust is actually the most important region to determine the magnetic field dissipation time-scale, as we will see. Here the {\it nuclear pasta} is expected to appear (see \S~\ref{sec:intro_crust}). The relevant nuclear parameters that determine the range of densities at which pasta might appear are the symmetry energy and its density dependence close to the nuclear saturation density. These important parameters also determine global properties such as the radius and moment of inertia, and have been proposed to have a potential observational effect in the crust oscillation frequencies \cite{sotani11,gearheart11}. 

Detailed molecular dynamics simulations \citep{horowitz05,horowitz08} have shown that it may be unrealistic to predict the exact sizes and shapes of the pasta clusters, and that it is likely amorphous, with a very irregular distribution of charge. This lumpiness is expected to strongly limit the electrical conductivity. In the absence of more detailed microscopical calculations, we can use the impurity parameter formalism as a first approximation to the complex calculation of the conductivity.

A pasta region, by definition, is expected to have a large effective $Q_{imp}$ since it is expected that clustering happens in a very irregular manner. Even assuming spherical nuclei, the inner crust has also been proposed to be amorphous and heterogeneous in nuclear charge \citep{jones04a,jones04b}, with electrical and thermal conductivities much smaller than for a homogeneous body-centered cubic lattice. The expected values of $Q_{imp}$, coming from the crust formation process alone, could be of the order of a few tens, providing large values of resistivity. We also note that the conductivity of an amorphous crust calculated with molecular dynamics simulations is even lower by an order of magnitude than the estimates obtained using the impurity parameter formalism (as shown in \cite{daligault09} for the outer crust in the scenario of accreting neutron stars). The ``effective'' $Q_{imp}$ to be used in the impurity parameter formalism must then be high. We will set $Q_{imp}=0.1$ for densities below $10^{13}$ g/cm$^3$, and slowly increase it until the beginning of pasta region ($\rho>6\times10^{13}$ g/cm$^3$), where we will let it vary between $Q_{imp}=1-100$, to explore the sensitivity of our results to this important parameter.

\subsection{Phonon conductivity.}

The phonon thermal conductivity can be approximated by the expression
\begin{equation}
 \kappa_{ph} = \frac{1}{3}{c_v c_s \lambda_{ph}}~,
\end{equation}
where $c_s$ is the phonon wave speed, $c_v$ the specific heat, and $\lambda_{ph}$ the phonon mean free path in the lattice. The state-of-the-art calculations can be found in \cite{chugunov07}. The processes determining $\lambda_{ph}$ are the Umklapp scattering (i.e., electron-phonon collisions) and the phonon scattering on impurities. The electron-phonon conductivity is temperature-dependent, and it dominates when the star is warm, because, for $T > T_U$, $\lambda_{ph} \propto T^{-2}$.\footnote{At low temperatures, the calculation of the mean free path was classically modeled by an analytical formula including the exponential reduction of the scattering rate \citep{gnedin01}. \cite{chugunov12} shows that, at very low temperatures $T \lesssim 10^7$ K, such formula overestimates the electrical conductivity by several orders of magnitude. However, here we are interested in the evolution of a neutron star during the first $\sim$ Myr of its life, when the used approximation is valid.}

The mean free path of phonons is usually much shorter than for electrons. However, in presence of a strong magnetic field, the transverse electron transport is drastically suppressed, and the phonon contribution to the thermal conductivity may become dominant. 

In principle, superfluid neutrons may represent another important contribution to the thermal conductivity, by means of collective motions, the so-called superfluid phonons \citep{cirigliano11}. The most significant effect is to partially counteract the anisotropic electron conduction when the temperature is low $T\lesssim 10^7$ K \citep{aguilera09}. However, recent results indicate that entrainment between superfluid neutrons and nuclei is larger than expected, resulting in an increased effective mass of the nuclei in the inner crust \citep{chamel12,chamel13}, and a reduced mean free path due to the coupling with lattice phonons. For this reason, and in the absence of quantitative calculations, we have suppressed the superfluid phonon contribution.

%%%%%%%%%%%%%%%%%%%%%%%%%%%%%%%%%%%%%%%%%%%%%%%%%%%%%%%%%%%%%%%%%%%%%%%%%%
\begin{figure}[t]
\centering
\includegraphics[width=0.45\textwidth]{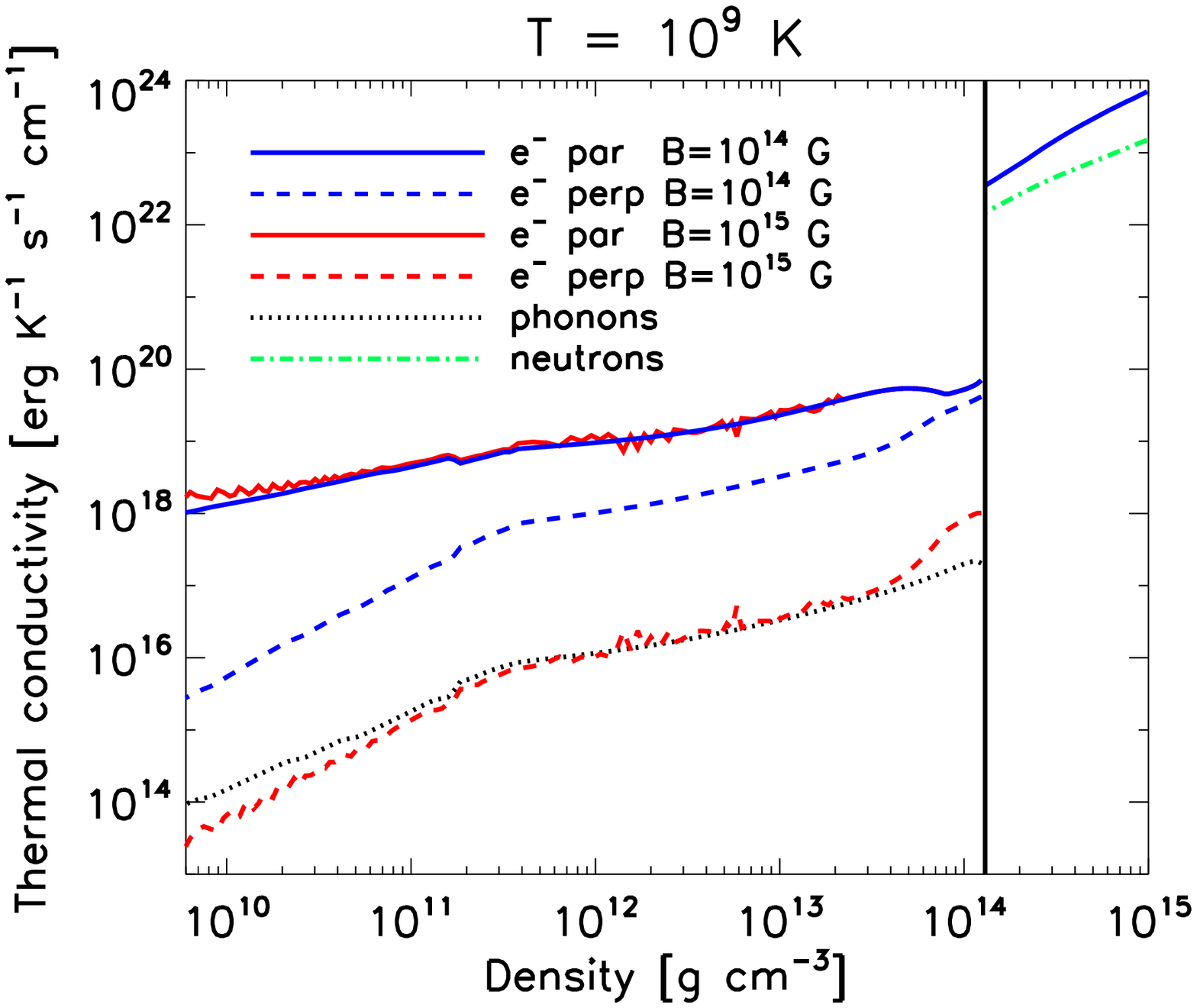}
\includegraphics[width=0.45\textwidth]{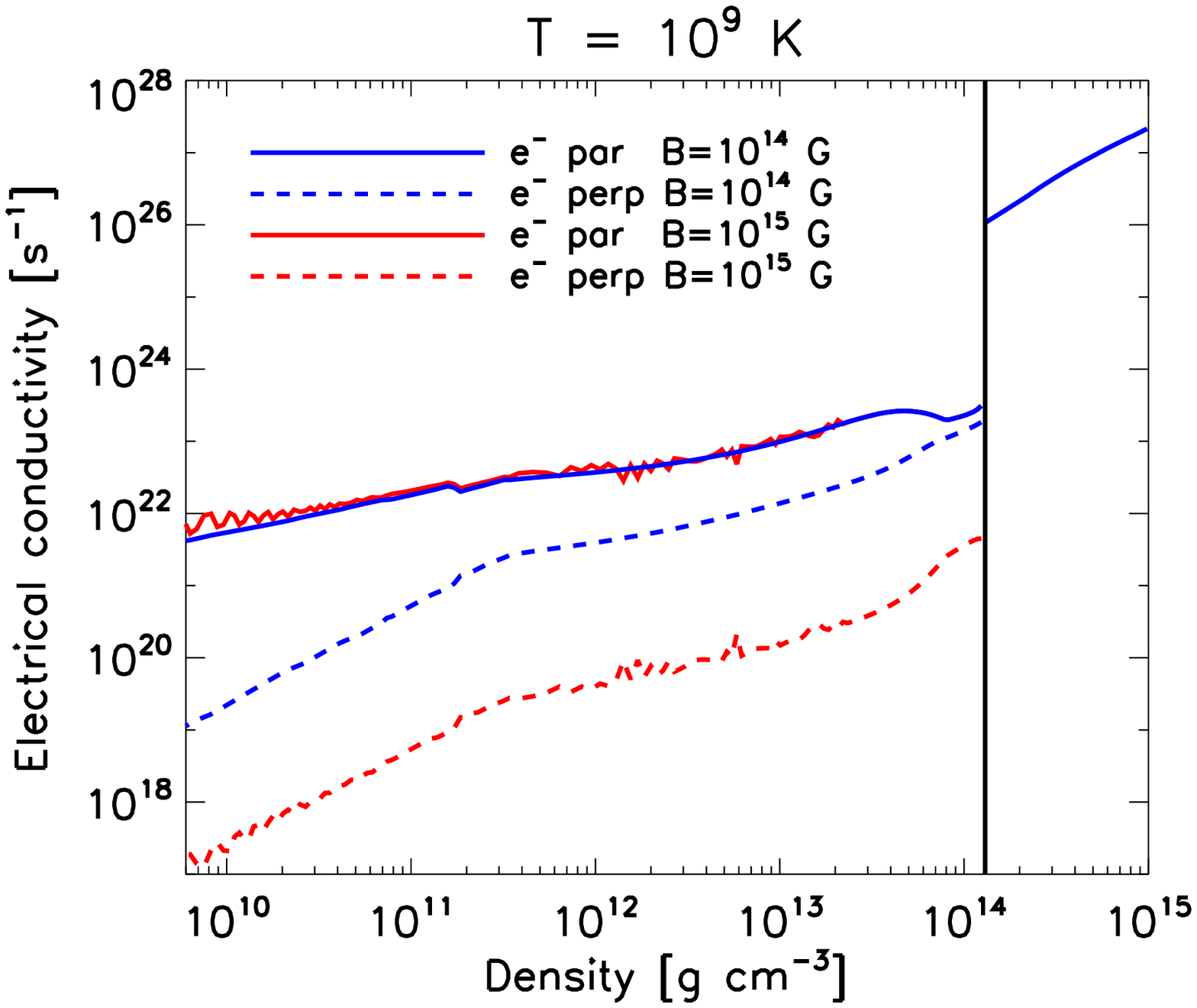}\\
\includegraphics[width=0.45\textwidth]{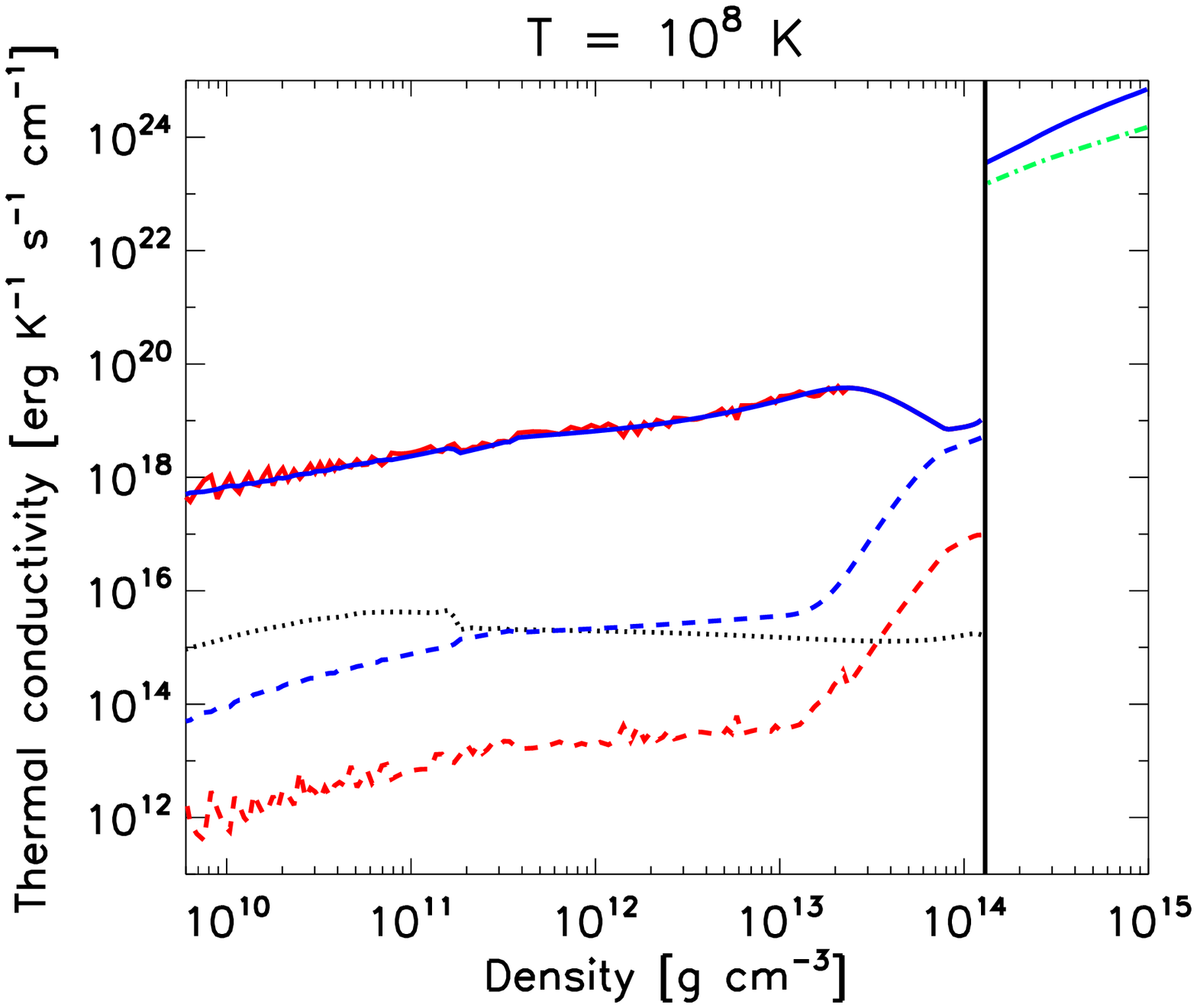}
\includegraphics[width=0.45\textwidth]{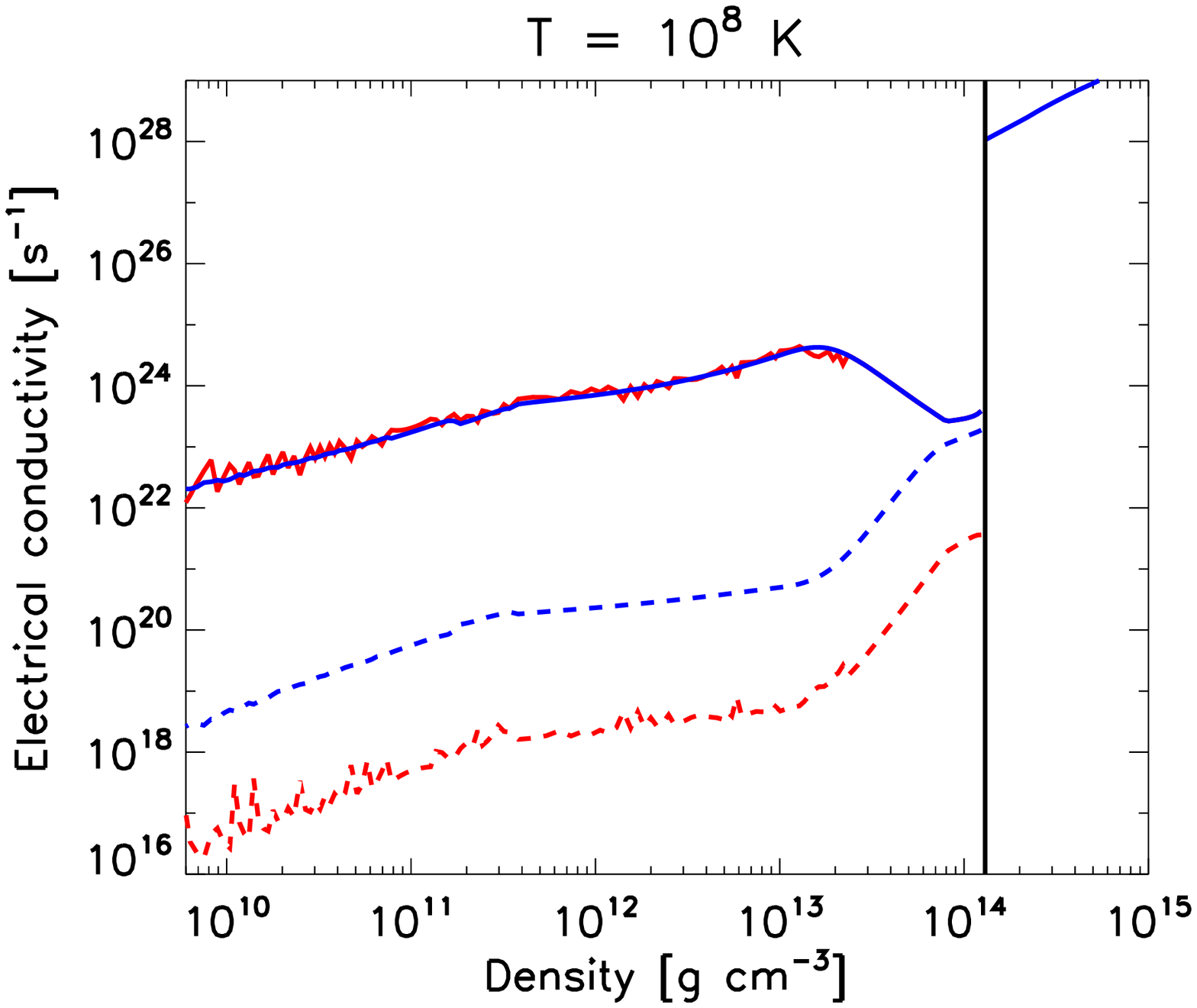}
\caption{Contributions to the thermal (left panels) and electrical (right panels) conductivities, at $T=10^9$ K (top) and $T=10^8$ K (bottom), without superfluidity.}
\label{fig:cond}
\end{figure}
%%%%%%%%%%%%%%%%%%%%%%%%%%%%%%%%%%%%%%%%%%%%%%%%%%%%%%%%%%%%%%

\subsection{Conductivity profiles in a realistic model.}

%%%%%%%%%%%%%%%%%%%%%%%%%%%%%%%%%%%%%%%%%%%%%%%%%%%%%%%%%%%%%%%%%%%%%%%%%%
\begin{figure}[ht]
\centering
\includegraphics[width=0.45\textwidth]{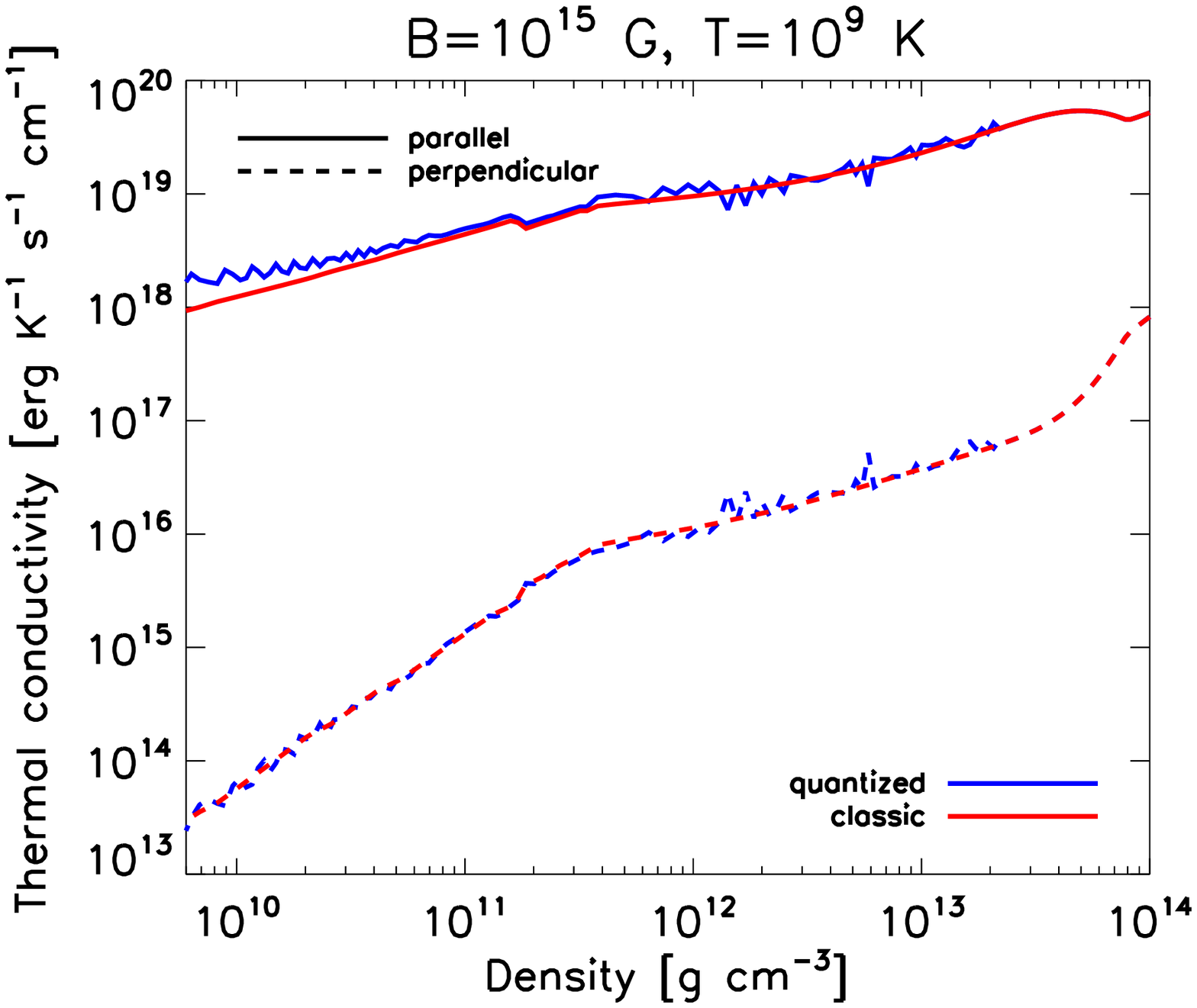}
\includegraphics[width=0.45\textwidth]{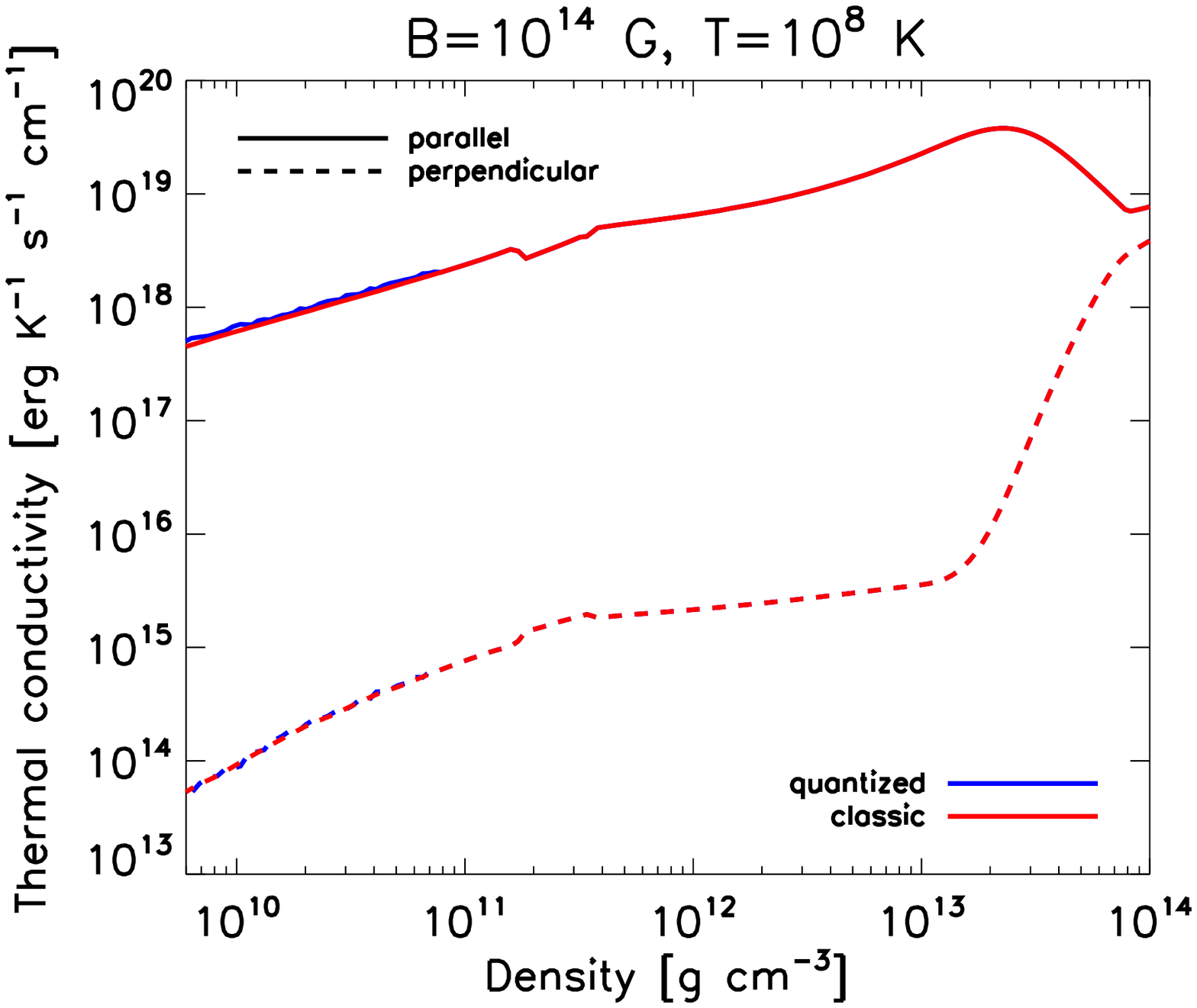}
\caption{Comparison between classical (red) and quantized (blue) thermal conductivity of electrons, in the direction parallel (solid lines) and perpendicular (dashes) to the magnetic field. We show the cases $T=10^9$ K, $B=10^{15}$ G (left panel) and $T=10^8$ K, $B=10^{14}$ G (right).}
\label{fig:cond_class_quant}
\end{figure}
%%%%%%%%%%%%%%%%%%%%%%%%%%%%%%%%%%%%%%%%%%%%%%%%%%%%%%%%%%%%%%

For the numerical implementation of conductivities, we use the latest {\tt FORTRAN} routines from the Saint~Petersburg group\footnote{{\tt http://www.ioffe.ru/astro/conduct/}}, which include all the relevant phonon and electronic contributions. The quantizing effects in presence of magnetic field (discussed above), the thermal averaging corrections accounting for partial degeneration of electrons (important only in the envelope), and the effect of the finite size of nuclei in the inner crust are consistently taken into account. The phonon-impurity scattering and the contribution by neutrons are ignored. Electron-electron scattering and some ionic quantum effects are considered in the limit $B=0$, since calculations are not available for the magnetized case.

The contributions to the conductivities by different carriers are shown in Fig.~\ref{fig:cond}, for a $M=1.40 M_\odot$ star, and $Q_{imp}=100$ in the pasta phase, and $Q_{imp}=0.1$ in the rest of the crust. We consider two temperatures: $T=10^9$ K (top panels) and $T=10^8$ K (bottom). The values of electrical conductivity are typically $\sigma \sim 10^{22}$--$10^{24}$ s$^{-1}$, orders of magnitude larger than in the most conductive terrestrial metals like silver at standard conditions of temperature and pressure $\sigma \sim 10^{17}$ s$^{-1}$, described by the band theory in solid state physics.

In the liquid core, electrons, neutrons and protons contribute to the conductivity \citep{gnedin95,baiko01}. Here we do not take account the poorly understood effects of the magnetic field; note that, due to the proton (type II) superconductivity, the magnetic field is likely confined into flux tubes that occupy a small fraction of its volume. We include contributions from electrons (blue solid lines) and neutrons (green dot-dashed). Since scattering rates are low, the very large thermal conductivity of matter, dominated by electrons, results in an isothermal core soon after birth, which implies that the precise value of the thermal conductivity is not important. The electrical conductivity of the core implies very long Ohmic time-scales.

In the crust, the conductivities are much lower. The dissipative processes responsible for the finite thermal conductivity include all the mutual interactions between electrons, lattice phonons (collective motion of ions in the solid phase), impurities (defects in the lattice), superfluid phonons (collective motion of superfluid neutrons) or normal neutrons, on the onset of superfluidity. The mean free path of free neutrons is limited by the interactions with the lattice. It is expected to be much shorter than for the electrons, but a consistent calculation has still to be done \citep{chamel08}. Ions trapped in the lattice (phonons) cannot transport net charge, therefore the electrical conductivity is determined by electron processes.

In Fig.~\ref{fig:cond}, we have considered the effects of $B=10^{14}$ G (blue) and $B=10^{15}$ G (red). The transport of heat and current flow in the direction parallel to the magnetic field is dominated by electrons (solid lines), while it is largely suppressed orthogonally to the magnetic field lines (dashed lines). A direct comparison between conductivities in the quantizing regime and in the classical approximation (eqs.~\ref{eq:ratio_cond}), is shown in Fig.~\ref{fig:cond_class_quant}. The quantizing effects are visible as little oscillations around classical values, corresponding to the gradual filling of Landau levels. They can be important only in the envelope, due to its lower density compared to the crust. Note also that we consider observables that depend on integrated quantities, therefore these little corrections are negligible in the crust. Nevertheless, we have included the quantizing effects in our simulations. 

Comparing the conductivities at different temperatures, it is evident that they generally increase, even by orders of magnitude, for decreasing temperatures (i.e., increasing age). Note, however, that the perpendicular electron conductivities (dashed lines) decrease, resulting in a progressive increase of the anisotropies. As a matter of fact, as the star cools down, the scattering processes are less frequent and the magnetization parameter $\omtau$, eq.~(\ref{eq:omtau}), becomes larger. An important exception occurs in regions where the electron-impurity processes dominate. In our model, they dominate in the pasta region, where the electron conductivity remains almost constant in time.

\section{Neutrino processes.}\label{sec:neutrino}

%%%%%%%%%%%%%%%%%%%%%%%%%BEGIN TABLE%%%%%%%%%%%%%%%%%%%%%%%%%%%%%%%%%%%

\begin{table}
\begin{center}
\begin{tabular}{l l l c}
\hline
\hline
{\bf Process} & $Q_\nu\,[{\rm erg\, cm^{-3}s^{-1}}]$  & {\bf Onset} & {\bf Ref}\\ 
\hline\noalign{\smallskip}
\multicolumn{4}{l}{{\it Core}}\\ 
\noalign{\smallskip}
\hline
Modified URCA ($n$-branch) & & & \\
 $nn\rightarrow pne\bar\nu_e$, $pne\rightarrow nn\nu_e$ & $8\times 10^{21}\,\mathcal R^{MU}_n \,n_p^{1/3}\,T^8_9$  & & 1 \\ 
Modified URCA ($p$-branch) & & & \\ 
 $np\rightarrow ppe\bar\nu_e$, $ppe\rightarrow np\nu_e$ & $8\times 10^{21} \,\mathcal R^{MU}_p \,n_p^{1/3}\,T^8_9$ & $Y_p^c=0.01$  & 1 \\ 
\hline
N-N Bremsstrahlung  & & &\\ 
$nn\rightarrow nn \nu \bar\nu$ & $7\times 10^{19}\,\mathcal R^{nn} \,n_n^{1/3}\,T^8_9$  & & 1 \\ 
$np\rightarrow np \nu \bar\nu$ & $1\times 10^{20}\,\mathcal R^{np} \,n_p^{1/3}\,T^8_9$  & & 1 \\ 
$pp\rightarrow pp \nu \bar\nu$ & $7\times 10^{19}\,\mathcal R^{pp} \,n_p^{1/3}\,T^8_9$  & & 1 \\ 
\hline
$e$-$p$ Bremsstrahlung & & &\\ 
$ep\rightarrow ep \nu \bar\nu $ & $2\times 10^{17}\,n_B^{-2/3}\,T^8_9$ & & 2 \\
\hline
Direct URCA & & &\\
$n\rightarrow pe\bar\nu_e, pe \rightarrow n\nu_e$& $4\times 10^{27}\,\mathcal R^{DU} \,n_e^{1/3}\,T_9^6$ & $Y_p^c=0.11$ & 3\\ 
$n\rightarrow p\mu\bar\nu_{\mu},p\mu \rightarrow n\nu_{\mu}$ & $4 \times 10^{27}\,\mathcal R^{DU} \,n_e^{1/3}\,T_9^6$ & $Y_p^c=0.14$ & 3\\
\hline\noalign{\smallskip}
\multicolumn{4}{l}{{\it Crust}}\\
\noalign{\smallskip}
\hline
Pair annihilation  & & &\\ 
$ee^+\rightarrow \nu \bar\nu$ & $9 \times 10^{20}\, F_{\rm pair}(n_e,n_{e^+})$ &  & 4 \\
\hline
Plasmon decay &&&\\
$\tilde e\rightarrow \tilde e\nu\bar\nu$ & $1\times 10^{20} \, I_{\rm pl} (T,y_e)$ & & 5\\
\hline
$e$-$A$ Bremsstrahlung &&&\\
$e (A,Z)  \rightarrow e (A,Z) \nu \bar\nu$ & $3 \times 10^{12}\, L_{eA}\,Z\,\rho_o\,n_e\,T^6_9$ & & 6\\
\hline 
$N$-$N$ Bremsstrahlung   &&&\\ 
$nn\rightarrow nn \nu \bar\nu$ & $7 \times 10^{19}\, \mathcal R^{nn}f_{\nu} \,n_n^{1/3}\,T^8_9$ & & 1  \\
\hline\noalign{\smallskip}
\multicolumn{4}{l}{{\it Core and crust}}\\ 
\noalign{\smallskip}
\hline
CPBF && &\\
$\tilde B + \tilde B \rightarrow \nu \bar \nu$ & $1\times 10^{21}\,n_N^{1/3} \,F_{A,B}\, T^7_9$ & & 7\\
\hline 
Neutrino synchrotron & & & \\  
$e \rightarrow (B) \rightarrow e \nu \bar\nu$  & $9\times10^{14}\,S_{AB,BC}\,B_{13}^2 \,T^5_9$ & & 8\\
\hline
\hline
\end{tabular}\\
\end{center}
\medskip
{\scriptsize Refs.: (1)~\cite{yakovlev95}; (2)~\cite{maxwell79}; (3)~\cite{lattimer91}; (4)~\cite{kaminker94}; (5)~\cite{yakovlev01}; (6)~\cite{haensel96,kaminker99}; (7)~\cite{yakovlev99}; (8)~\cite{bezchastnov97}}
\caption{Neutrino processes and their emissivities ${\cal Q}_\nu$ in the core and in the crust, taken from \citealt{aguilera08b}. The third column shows the onset for some processes to operate (critical proton fraction $Y_p^c$). We indicate the normalized temperature $T_9=T/10^9$ K; detailed functions and precise factors can be found in the references (last column).}
 \label{tab:neutrino} 
\end{table} 
%%%%%%%%%%%%%%%%%%%%%%%%%END TABLE%%%%%%%%%%%%%%%%%%%%%%%%%%%%%%%%%%%%

%%%%%%%%%%%%%%%%%%%%%%%%%%%%%%%%%%%%%%%%%%%%%%%%%%%%%%%%%%%%%%%%%%%%%%%%%
\begin{figure}[t]
   \centering
   \includegraphics[width=0.45\textwidth]{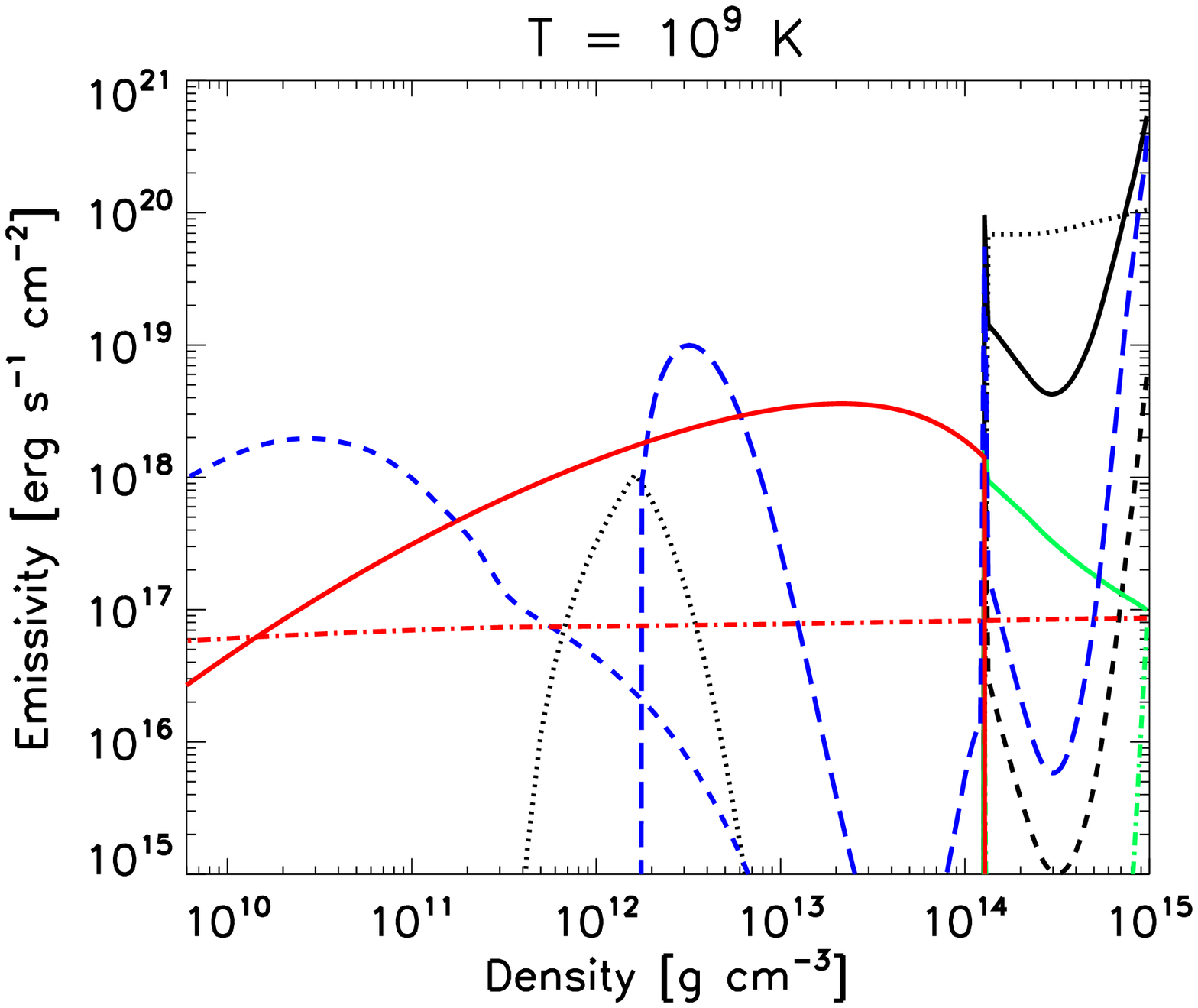}
   \includegraphics[width=0.45\textwidth]{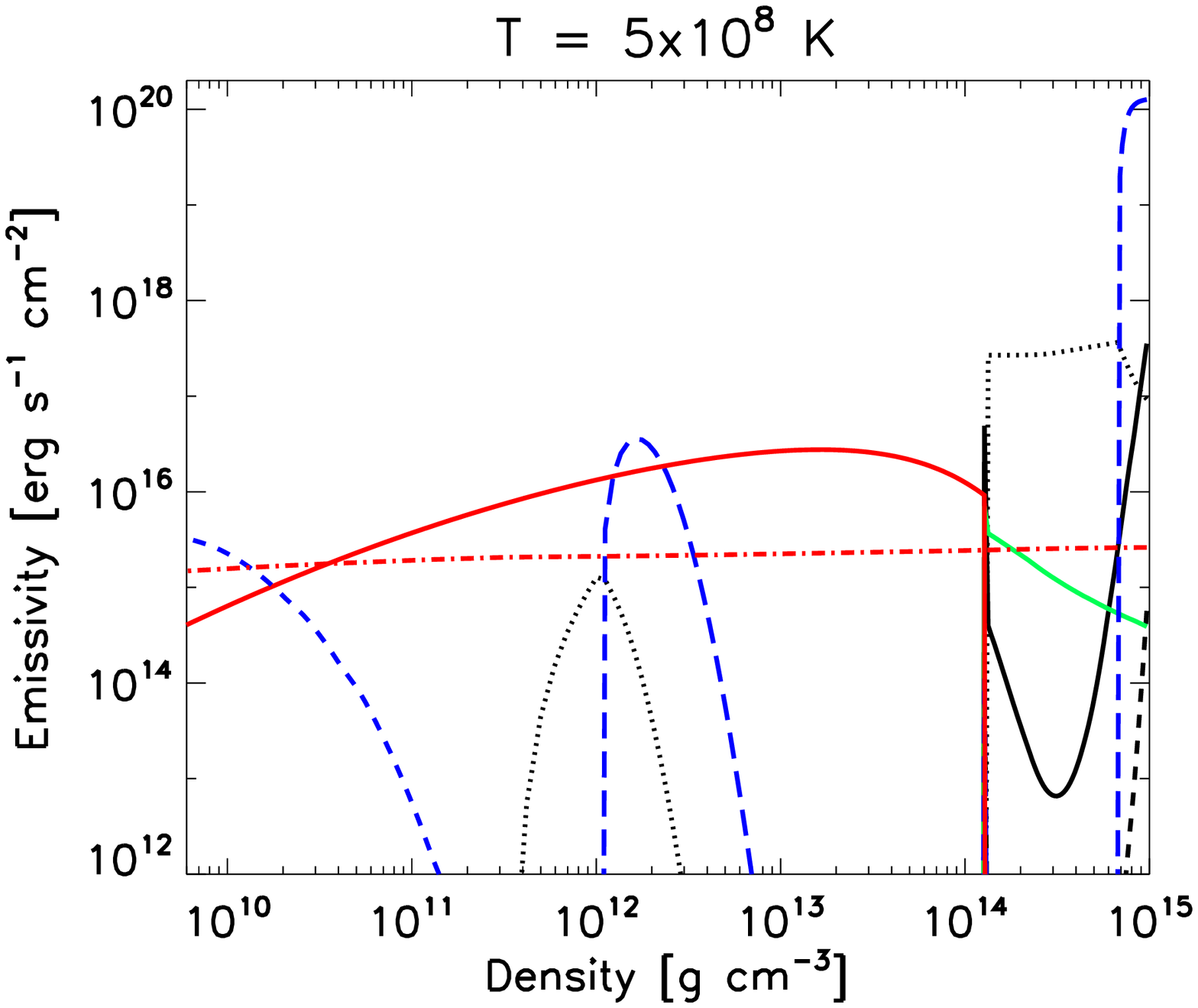}\\
   \includegraphics[width=0.45\textwidth]{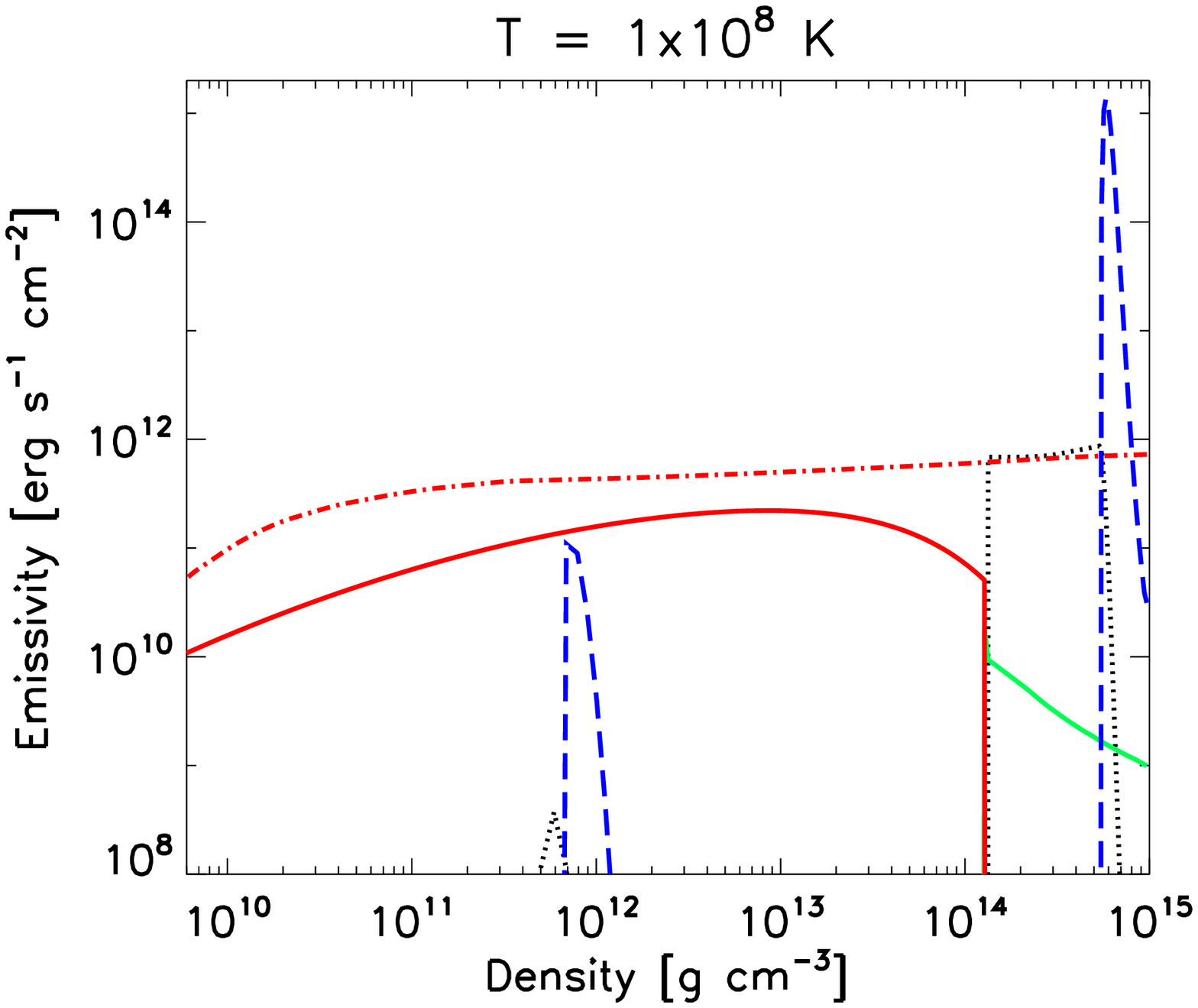}
   \includegraphics[width=0.45\textwidth]{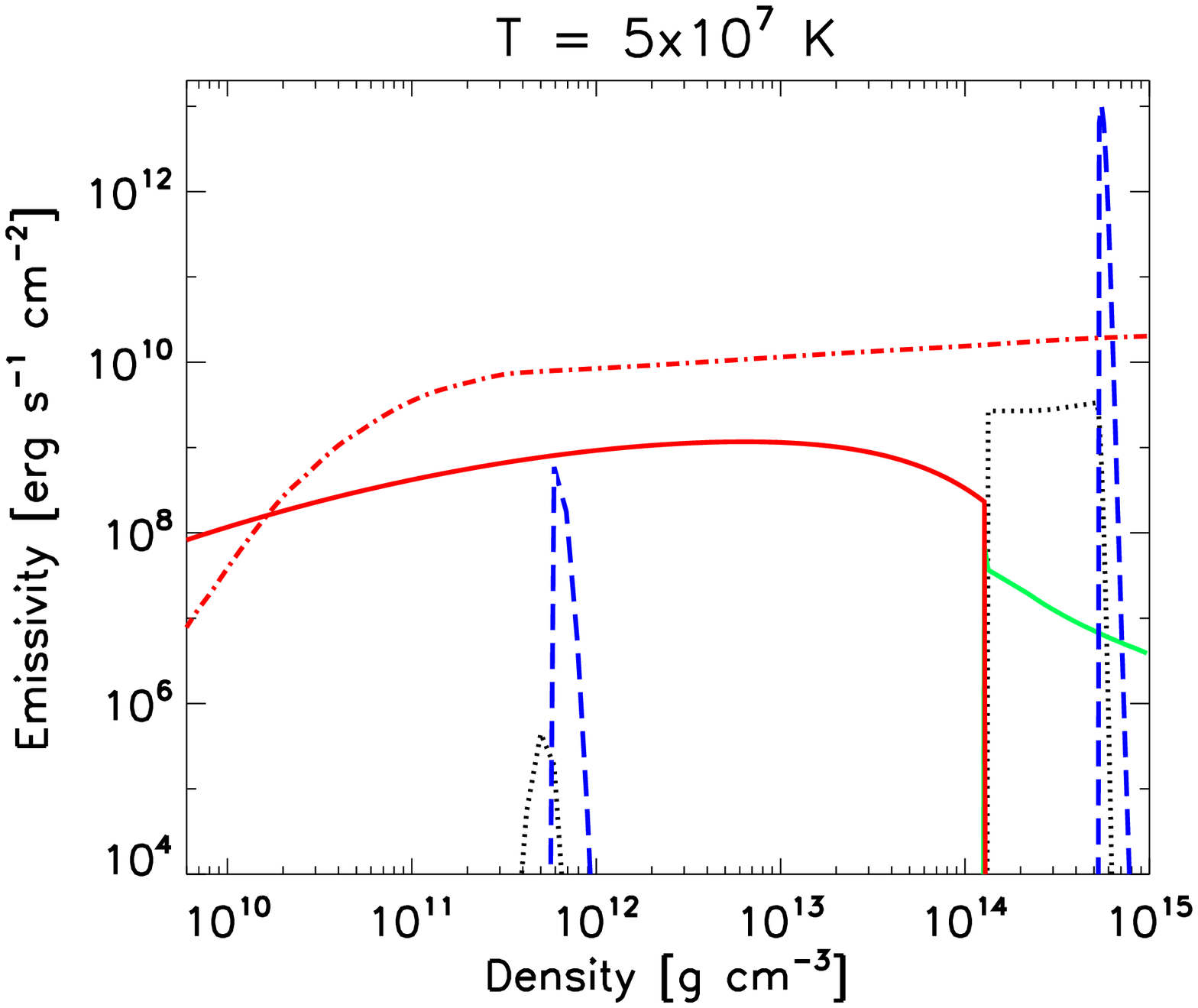}
\caption{Neutrino emissivities in the crust and in the core at the four indicated temperatures, with the chosen equation of state and superfluid gaps (see text), and mass $M=1.4~M_\odot$ (no direct URCA). Lines denote: modified URCA (black solid line), $n$-$n$ Bremsstrahlung (black dots), $n$-$p$ Bremsstrahlung (black dashes), $e$-$p$ Bremsstrahlung (green solid), $e$-$A$ Bremsstrahlung (red solid), plasmon decay (short blue dashes), CPBF (blue long dashes), and $\nu$-synchrotron for $B=10^{14}$ G (red dot-dashed line).}
\label{fig:neutrino}
\end{figure}
%%%%%%%%%%%%%%%%%%%%%%%%%%%%%%%%%%%%%%%%%%%%%%%%%%%%%%%%%%%%%%%%%%%%%%%%

Since the first minute after birth, neutrino mean free path is much longer than the star radius \citep{burrows86,pons99}. Therefore, when the star is still hot, the copious production of neutrinos carries energy away from the star. This is why the first $\approx 10^5$ yr are called neutrino-cooling era. Neutrino reactions occur in a wide range of matter compositions, which are very different in the solid crust and the liquid core. The theoretically calculated emissivities are complicated functions of the temperature and matter composition. In general, neutrino emissivities depend on a high power of temperature, and weakly on density: see the detailed review by \cite{yakovlev01} and the summary of relevant processes in Table~\ref{tab:neutrino}, taken from \cite{aguilera08b}. 

In the so-called {\it standard cooling} scenario, the total emissivity is dominated by slow processes in the core $\propto T^8$, such as modified URCA and the different Bremsstrahlung processes. In regions with high proton fraction $Y_p>Y_p^c \approx 0.11$ (reached only in the inner core of high mass neutron stars), the fast direct URCA process is allowed, and the early cooling is much faster: it is the {\it enhanced cooling} scenario, in which a temperature drop is predicted after tens of years. With the chosen equation of state, all our models have $Y_p<Y_p^c$ everywhere. Since we want to explore the effects of direct URCA, for practical purposes we artificially activate it above a threshold density of $\rho_c=10^{15}$ \gcc, reached in our models with $M > 1.4 M_\odot$ (see Table~\ref{tab:nsmasses}).

The {\it minimal cooling} model, in which pairing between nucleons and the effects of superfluidity are both included \citep{page04}, is more realistic. First, most of the effective neutrino processes, involving superfluid components (modified URCA, nucleon Bremsstrahlung), are exponentially suppressed by factors fitted by the mathematical functions $\mathcal R$ described in \cite{yakovlev01}. On the other hand, the onset of superfluidity opens new channels in the neutrino production: the Cooper Pair Breaking and Formation processes (CPBF). According to \cite{leinson06}, they are suppressed in the crust (neutron $^1S_0$ pairing), but they are active in the core (proton $^1S_0$ and neutron $^3P_2$ channels). The crustal superfluidity has a minor impact on the long-term evolution, as it affects only the early relaxation of the crust, a stage lasting about $100$ yr (as we will see in \S~\ref{sec:cooling_b0}), before the thermal coupling between crust and core is achieved \citep{gnedin01,page09}. The bulk of energy loss is regulated by the physics of the core, where most of the mass is contained. As a consequence, the total neutrino emissivity is strongly sensitive to the superfluid gaps.

A further complication arises from the presence of a strong magnetic field. In that case, the relativistic electrons have a different dispersion relation and they can directly emit neutrino pairs, analogously to the synchrotron emission of photons, with an emissivity proportional to $B$. A few processes, such as pair annihilation and photo-neutrino emissivity, are relevant only in the envelope.

In Fig.~\ref{fig:neutrino}, we show the emissivities of the relevant neutrino processes in our baseline model, for fixed temperatures of $10^9$~K (top left panel), $5\times 10^8$~K (top right), $10^8$~K (bottom left) and $5 \times 10^7$~K (bottom right). When the star is hot, the modified URCA and the nucleon-nucleon Bremsstrahlung are the most efficient processes. In regions with normal neutrons, $n$-$n$ Bremsstrahlung dominates. In the outer crust, the emissivity is dominated by electron-nuclei Bremsstrahlung, with localized important contribution by plasmon decay, CPFB, and synchrotron.

When the temperature has fallen to  $T=10^8$~K, neutrino losses are much less effective. The relevant neutrino processes are CPFB, the neutrino synchrotron if a magnetic field $B\gtrsim 10^{14}$ G is present, and the $e^-$-$A$ Bremsstrahlung, due to their weaker dependence on temperature. When temperature is even lower ($T\lesssim 5\times 10^7$ K), the neutrino emissivities become irrelevant, and photons radiated from the surface become the main cooling channel (photon-cooling era).

\chapter{Magneto-thermal evolution}\label{ch:cooling}

The theory of neutron star cooling was developed short after the first evidence of X-ray emission from the surface of neutron stars in the Sixties (e.g., \citealt{morton64,chiu64,tsuruta65}). During the following decades, microphysical inputs have been refined to improve 1D cooling models \citep{yakovlev81,nomoto86,page90,vanriper91,page92,pethick92}, but the effects of magnetic fields were barely taken into account \citep{miralles98,page00}. \cite{page04,page09} proposed the so-called minimal cooling scenario, in which all the necessary known microphysical ingredients are consistently included (superfluidity, neutrino emission processes), but fast cooling processes (direct URCA by nuclear or exotic matter) are excluded. The minimal cooling paradigm has been shown to be consistent with the luminosity of the weakly magnetized neutron stars ($B\lesssim 10^{13}$ G), but magnetars and some of the high-B pulsars show a relatively large X-ray luminosity, thus requiring the presence of additional heat sources, likely linked to the magnetic field evolution.

The presence of a strong magnetic field directly affects the microphysical processes that govern the thermal evolution of the crust, as the temperature in turn modifies the magnetic field evolution. For this reason, previous works \citep{aguilera08b,pons07b,pons09} paved the way for a fully coupled magneto-thermal evolution. \cite{aguilera08b} developed a 2D (axial symmetry) cooling code taking into account the interplay between temperature and magnetic field. However the decay of the latter was simulated by a phenomenological, analytical formula, without solving the induction equation. In \cite{pons07b}, a spectral code followed the evolution of the magnetic field, including both the resistive and the Hall terms, but the temperature evolution was described by an isotropic, analytical cooling law. The first 2D simulations of the fully coupled magneto-thermal evolution \citep{pons09} included only Ohmic dissipation, because of numerical limitations in the treatment of the Hall term.

Building on previous works \citep{page04,pons07b,aguilera08b,page09,pons09} we have updated and extended our magneto-thermal evolution code in two major ways: the proper treatment of the important Hall term in the induction equation describing the magnetic field evolution (see chapter~\ref{ch:magnetic}), and updated microphysics inputs (see chapter~\ref{ch:micro}). This allows us to follow the long-term evolution of magnetized neutron stars, a necessary step to understand timing and spectral properties of isolated neutron stars at different ages.

\section{The magneto-thermal evolution equations.} 

The thermal evolution in the star is described by a simple energy balance equation which, using the static metric (\ref{eq:metric}), reads
\begin{equation}\label{eq:heat_balance}
c_v e^\nu\frac{\partial T}{\partial t} + \vec{\nabla}\cdot(e^{2\nu}\vec{F}) = e^{2\nu}{\cal Q}~,
\end{equation}
where $c_v$ is the heat capacity per unit volume, $\vec{F}$ is the thermal flux, and ${\cal Q}$ is the sum of energy sources and losses per unit volume and time. In the diffusion limit, the thermal flux is given by
\begin{equation}\label{eq:flux_diffusion} 
\vec{F} = -e^{-\nu} \hat{\kappa} \cdot \vec{\nabla} (e^\nu T) ~,
\end{equation}
where $\hat \kappa$ is the thermal conductivity tensor. As source terms, we will consider the energy lost by neutrino emission ${\cal Q}_\nu$, and the Joule dissipation rate of the electrical currents, ${\cal Q}_j$, eq.~(\ref{eq:def_joule}). In summary, two fundamental equations govern the magneto-thermal evolution of the star: the heat balance equation and the induction equation (\ref{eq:induction_hall}), that we report again together:

\begin{eqnarray}
 && c_v e^\nu\frac{\partial T}{\partial t} - \vec{\nabla}\cdot[e^\nu \hat{\kappa} \cdot \vec{\nabla} (e^\nu T)]  = e^{2\nu}({\cal Q}_j - {\cal Q}_\nu) ~,  \label{eq:magnetothermal_t}\\
 && \frac{\partial \vec{B}}{\partial t} =  -\vec{\nabla}\times \left[\frac{c^2}{4\pi \sigma} \curlBrel + \frac{c}{4\pi e n_e}  (\curlBrel)\times\vec{B}\right]~. \label{eq:magnetothermal_b}
\end{eqnarray}
To solve these equations, we need a neutron star model, built by means of an equation of state, and the microphysical inputs: specific heat, thermal and electrical conductivities and neutrino emissivities. These ingredients depend in general on the local values of temperature, density, composition and magnetic field strength.

\section{The cooling code.}

The diffusion equation (\ref{eq:magnetothermal_t}) is discretized in a fully implicit way, building a linear system of equations described by a block tridiagonal matrix. The ``unknowns'' vector, formed by the temperatures in each cell, is advanced implicitly by inverting the matrix with standard numerical techniques, adapted to the particular case, as in \cite{aguilera08b}.

Values of temperature are defined at the center of each cell, where also the heating rate and the neutrino losses are evaluated. While the magnetic field evolution requires an explicit, upwind scheme with special care about the non-linear term, the evolution of temperature is much simpler. Its time-step is much larger than the magnetic one, and it is adjusted as the star evolves. Typically, $\Delta t_m=1$ year for the first 10 or 100 years, then it increases up to $\Delta t_m=100$ or 1000 years when the star is older and colder.

\subsection{Boundary conditions and the envelope model.}

The envelope is the liquid layer above the solid crust and below the atmosphere (see introduction), where the largest gradient of temperature is reached. Due to its relatively low density, its thermal relaxation time-scale is much shorter than that of the crust, which makes any attempt to perform cooling simulations in a numerical grid that comprises both regions computationally expensive. The widely used approach is to use fits to stationary envelope models to obtain a phenomenological relation between the temperature at the bottom of the envelope, $T_b$, and the effective temperature $T_s$. This $T_b$--$T_s$ relation depends on the surface gravity $g_s$, the envelope composition, and the magnetic field strength and orientation. Models assuming a non-magnetized envelope made of iron and iron-like nuclei show that the surface temperature is related to $T_b$ as follows \citep{gudmundsson83}:
\begin{equation}\label{eq:envelope_gudmu}
 T_{b,8}=1.288 \left[ \frac{T_{s,6}^4}{g_{14}}\right]^{\,0.455} ~,
\end{equation}
where $g_{14}$ is the surface gravity in units of $10^{14}$ cm~s$^{-2}$, $T_{b,8}$ is $T_b$ in $10^8$~K, and $T_{s,6}$ is $T_s$ in $10^6$~K.

Further 1D, plane-parallel simulations have followed the same approach, including neutrino processes and the presence of magnetic field, which affects the transport of heat, causing the heat flux to become anisotropic \citep{potekhin01,potekhin07}. Therefore, the geometry of the magnetic field is important, since regions permeated by radial magnetic field lines are thermally connected to the interior, while zones with tangential magnetic field (equator in the dipolar case) are thermally insulated. \cite{pons09} revised the envelope model with a 2D code, allowing for meridional transport of heat, finding that the most important effect is the attenuation of the anisotropy.

In a stationary envelope model, the surface temperature depends on $g$, $T_b$, $B$, and the angle $\varphi$ that the magnetic field forms with respect to the normal to the neutron star surface. The following functional form have been found to fit the $T_b$--$T_s$ relation \citep{potekhin01,pons09}:
\begin{equation}\label{PCY-iron} 
T_s(B,\varphi,g,T_b)\approx T_s^0(g,T_b) \,{\cal X}(B,\varphi,T_b)~,
\end{equation}
where
\begin{equation}
  {\cal X}(B,\varphi,T_b) = [ {\cal X}_\parallel^\iota(B,T_b)\cos^2\varphi + {\cal X}_\perp^\iota(B,T_b) \sin^2\varphi ]^{1/\iota}~,
  \label{fit3} 
\end{equation}
where the functional forms of ${\cal X}_\parallel$, ${\cal X}_\perp$ and the values of $T_s^0$ and $\iota$ are found by building a set of stationary envelope models with different values of $g$, $T_b$ and $B$, and for a fixed magnetic field geometry (i.e., different values of $\varphi$ at different latitudes). In particular, the value of the index has been found to be $\iota=4.5$ \citep{potekhin01} or $\iota=4$ \citep{greenstein83}.

For consistency, we have built our own envelope models using the microphysical inputs illustrated above and placing the outer boundary of our crust at $\rho=3\times 10^{10}$ g cm$^{-3}$. The fits to the obtained $T_b$--$T_s$ relations are then used as an external boundary condition in our simulations.

The {\it bolometric luminosity} seen by an observer at infinity is obtained by assuming blackbody emission at the temperature $T_s$ from each patch of the neutron star surface:
\begin{equation}\label{eq:bb_emission}
 L = 2\pi \int_0^\pi \sigma_{sb} R_\infty^2 T_\infty^4 \sin\theta~ \de \theta~,
\end{equation}
where $T_\infty=e^\nu T_s$, $R_\infty=e^{-\nu}R_\star$ are the redshifted surface temperature and the radius seen at infinity, and $\sigma_{sb}=5.67\times 10^{-5}$ erg cm$^{-2}$ s$^{-1}$ K$^{-4}$ is the Stefan-Boltzmann constant.

Other emission models have been proposed instead of a pure blackbody emission: magnetic atmospheres \citep{lloyd03,vanadelsberg06,ho07,ho08,suleimanov09,suleimanov12} or condensed surface \citep{vanadelsberg05,perezazorin05,medin07,potekhin12}. In the latter, in particular, the gaseous atmosphere and the outer envelope condensate to a solid state due to the cohesive interaction between ions caused by the magnetic field \citep{medin06a,medin06b}. In this case, the pressure vanishes at a finite density \citep{lai01}
\begin{equation}\label{eq:rho_condensation}
\rho_s \simeq 560 ~\frac{A}{Z^{3/5}}B_{12}^{6/5}~\mbox{\gcc} ~.
\end{equation}
Atmospheres and condensed surfaces have different emission properties and, consequently, the boundary condition has to be calculated differently, as for example in \cite{perezazorin06a}.

Overall, different emission models do not appreciably change the $T_b$--$T_s$ relation but shape the spectrum in different ways. Therefore, they can be important to interpret the observations, from which we infer the physical properties by means of a spectral fit to a particular emission model.

%%%%%%%%%%
\section{Cooling of weakly magnetized neutron stars.}\label{sec:cooling_b0}

\begin{figure}
 \centering
\includegraphics[width=.65\textwidth]{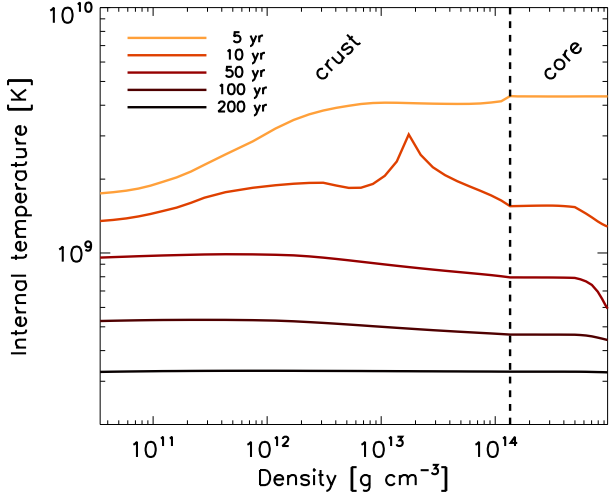}
\caption{Evolution of internal temperature in the early stages ($t=5,10,50,100$ and $200$ yr), starting with $T_{init}=10^{10}~$K, for a non-magnetized neutron star with $M=1.4~M_\odot$.}
 \label{fig:tint_b0}
\end{figure}

\begin{figure}
 \centering
\includegraphics[width=.46\textwidth]{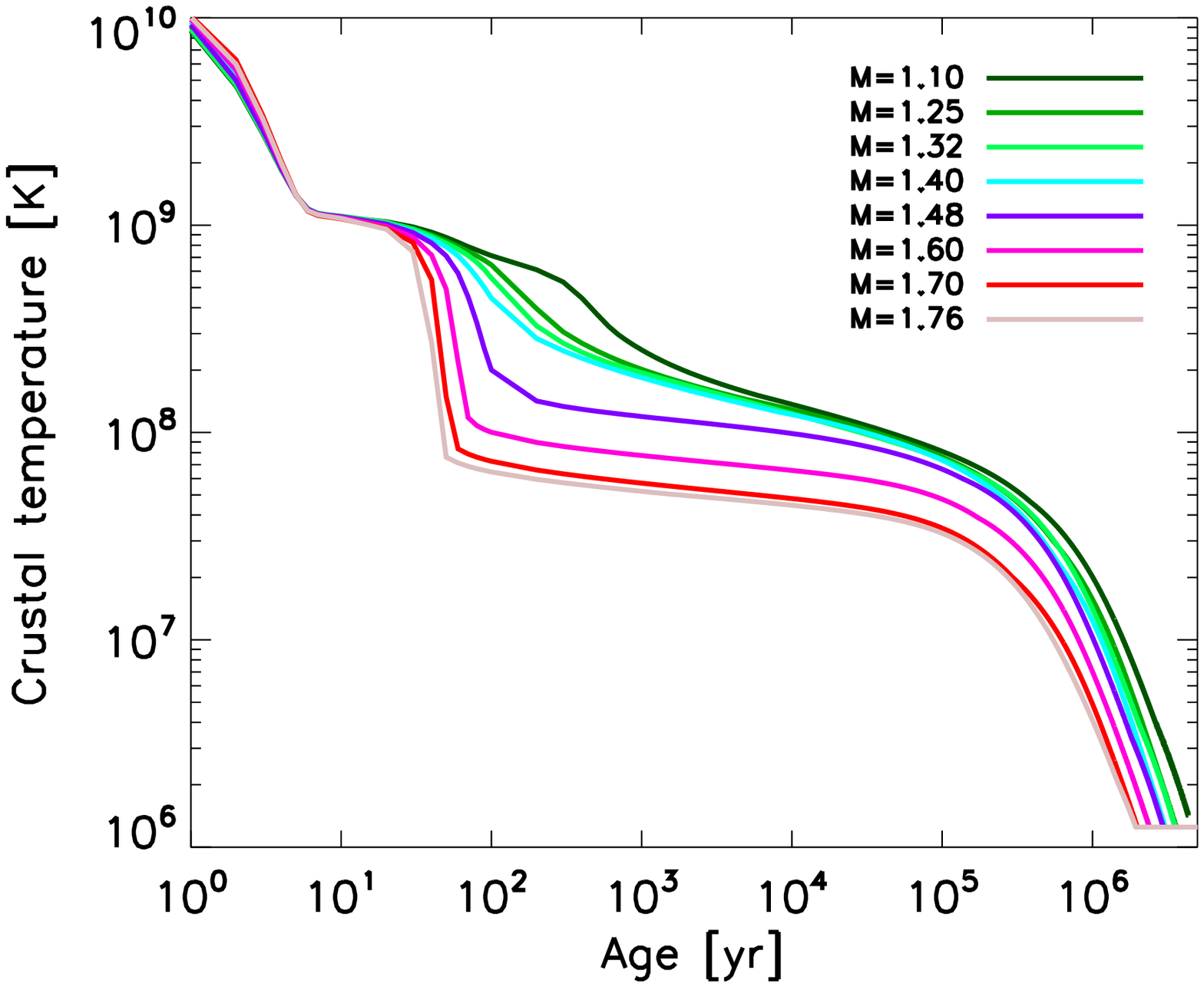}
\includegraphics[width=.46\textwidth]{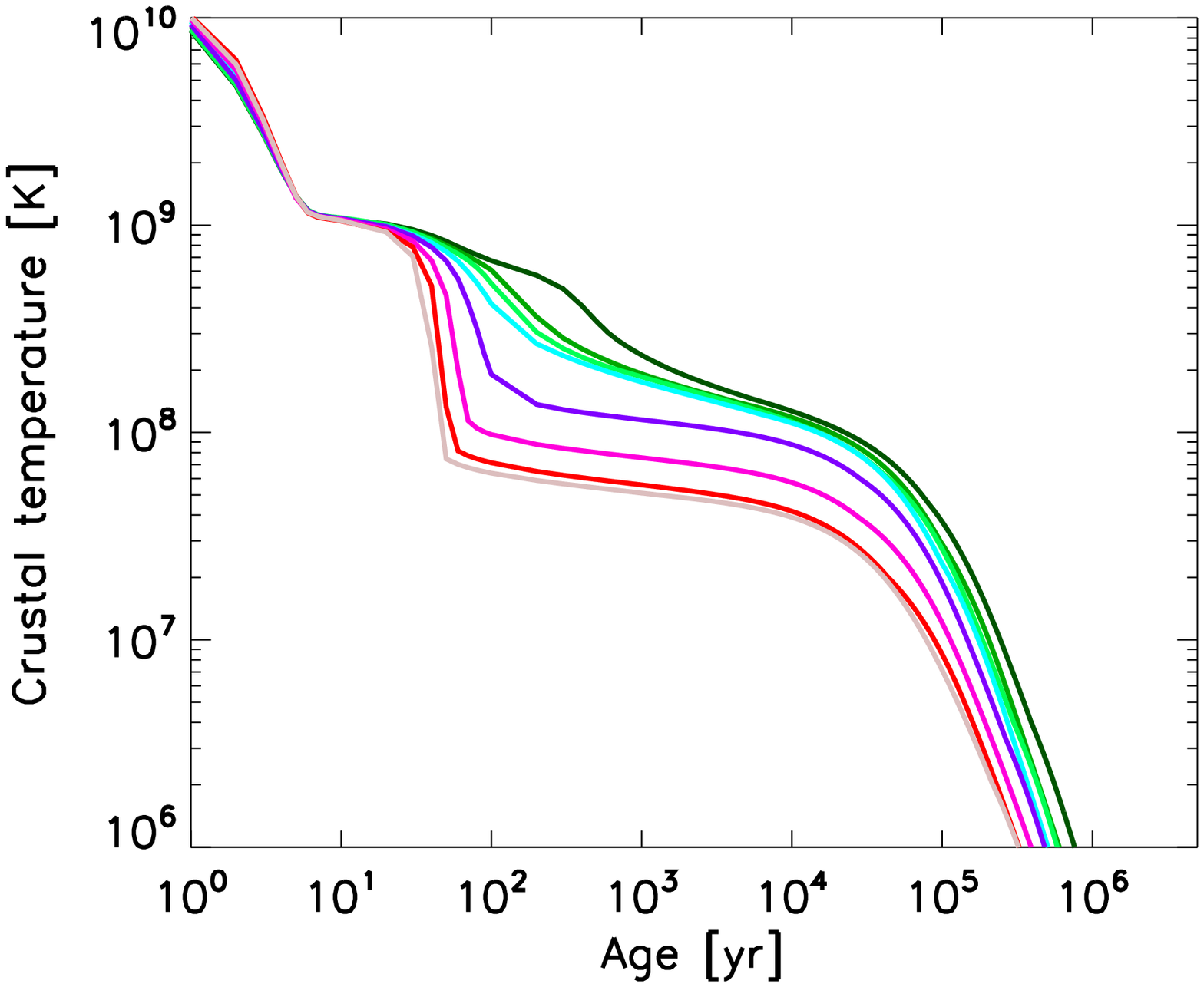}\\
\includegraphics[width=.46\textwidth]{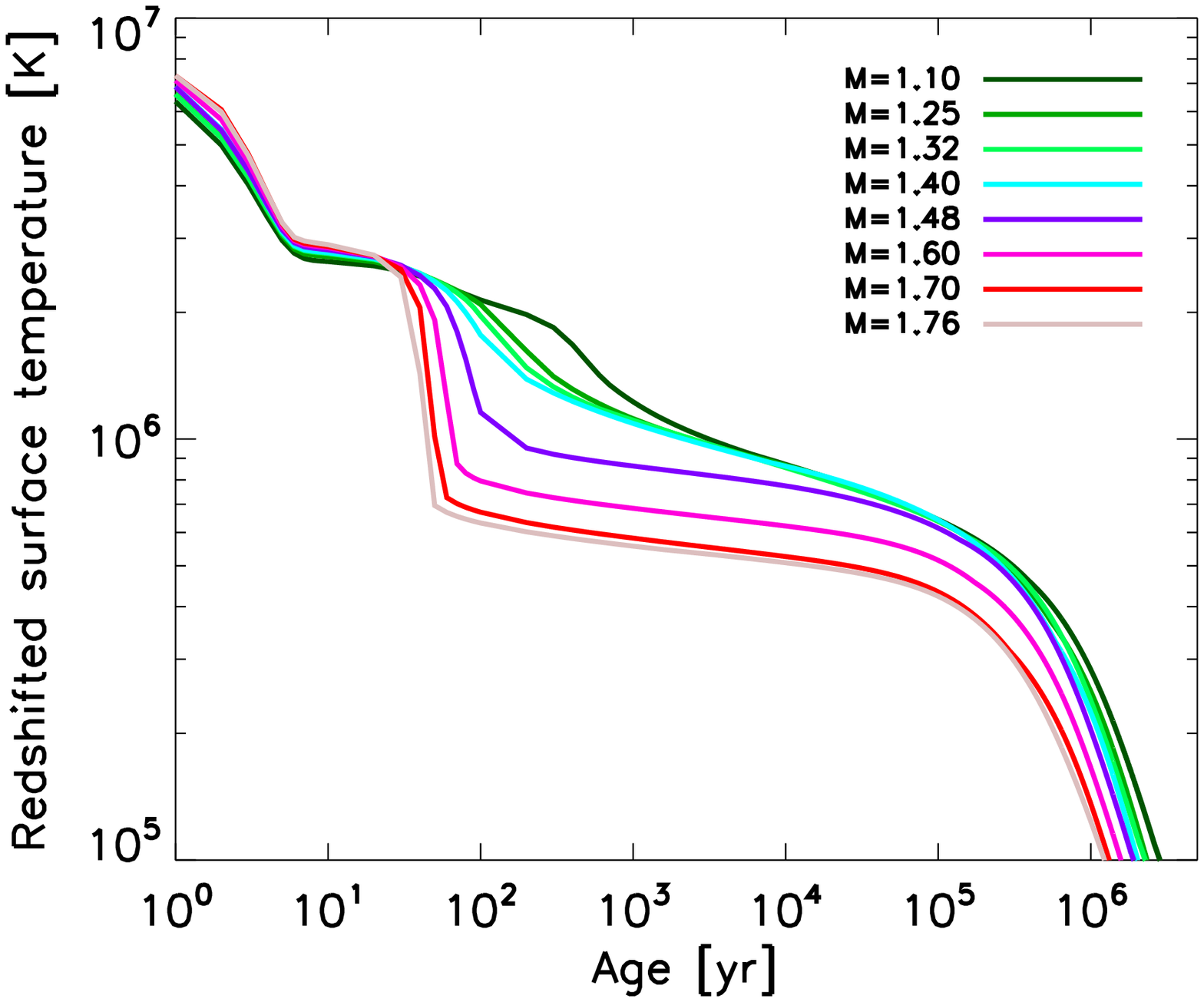}
\includegraphics[width=.46\textwidth]{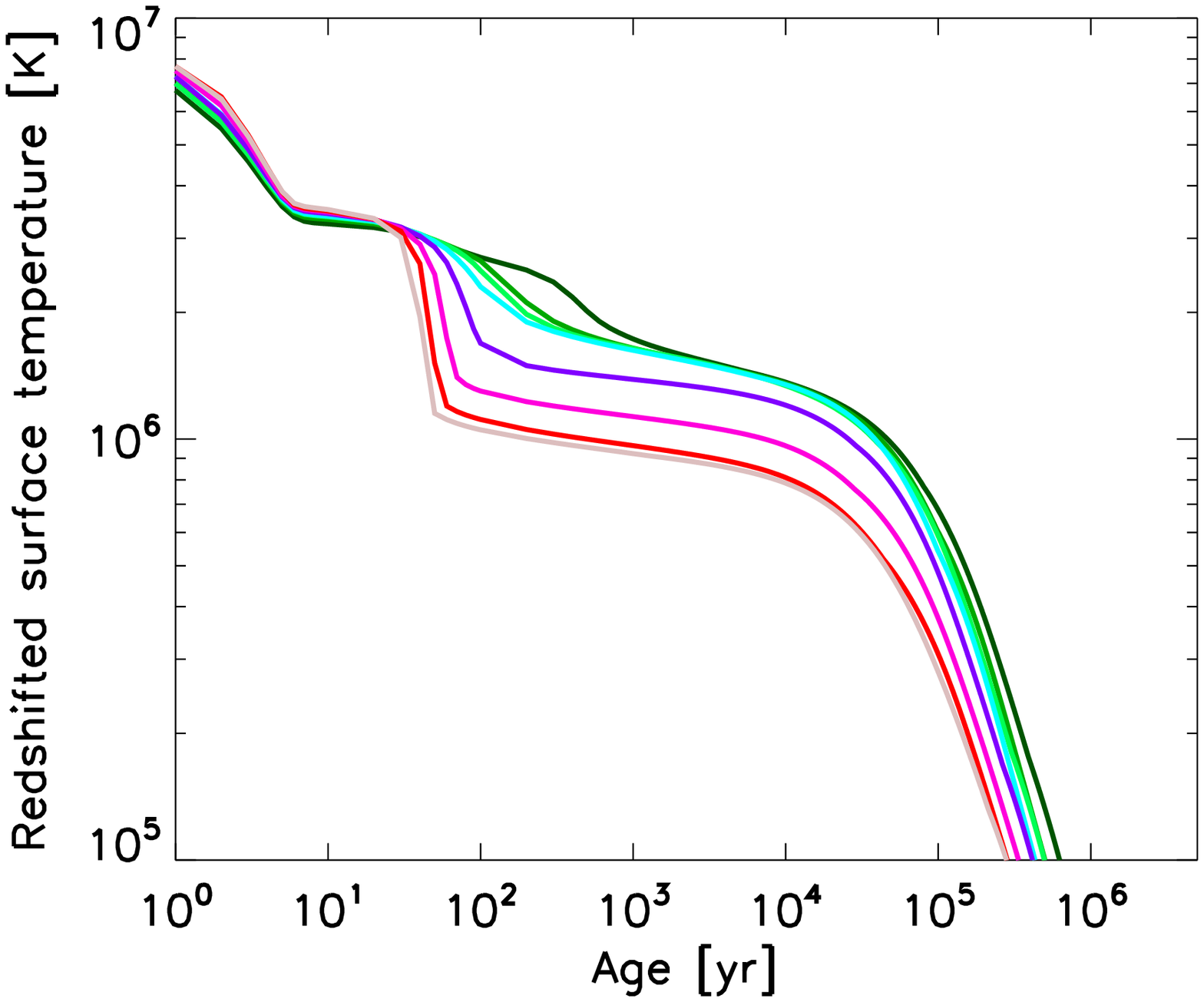}\\
\includegraphics[width=.46\textwidth]{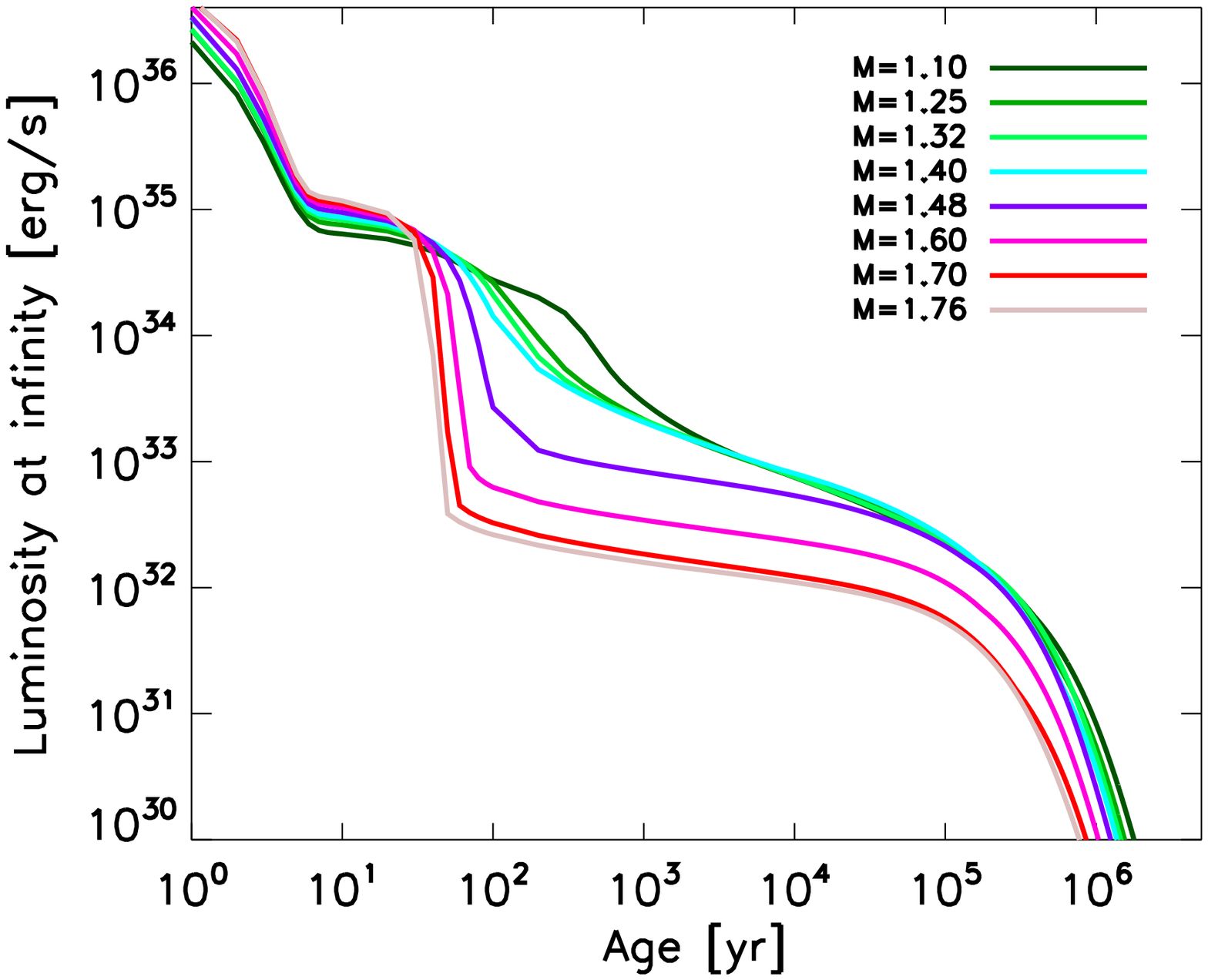}
\includegraphics[width=.46\textwidth]{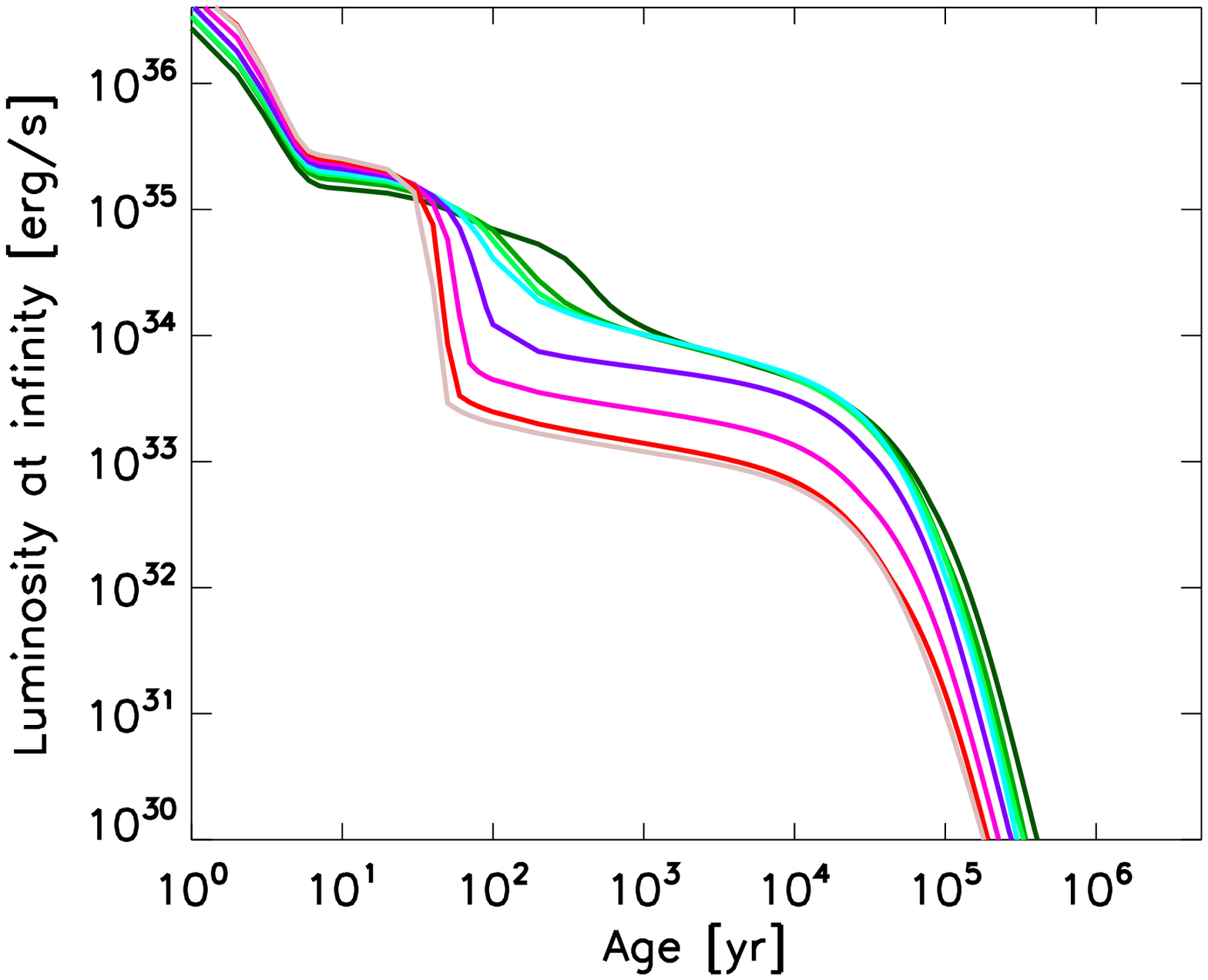}
\caption{Cooling curves for 8 masses ($M= 1.10, 1.25, 1.32, 1.40, 1.48, 1.60, 1.70$ and $1.76~M_\odot$): crustal temperatures (top panels), surface temperatures (middle) and bolometric luminosity (bottom). Left panels corresponds to models with iron envelopes and right panels to models with light-elements envelopes.}
 \label{fig:b0}
\end{figure}
%%%%%%%%%%%%%%%%%%%%%%%%%%%%%%%%%%%%%%%%%%

We now briefly revisit the cooling of non-magnetized neutron stars, with our microphysical setup. Comprehensive reviews can be found in \cite{page04} and \cite{yakovlev04}.

The profile of the internal temperature in the early stages is shown in Fig.~\ref{fig:tint_b0}, as a function of density. We have started with an initial temperature $T_{init}=10^{10}$ K. The strong neutrino emission in the inner core quickly lowers the temperature, that drops below $10^9$ K after a few decades, independently of the particular initial value $T_{init}$. The crust, with a larger thermal relaxation time-scale (a few years), follows the core cooling. This initial thermal relaxation takes at most a few centuries, after which, in absence of any additional interior heat (source positive terms ${\cal Q}>0$ in eq.~\ref{eq:heat_balance}), the star becomes isothermal, except in the envelope (not shown in the figure), where the large density gradient requires a strong temperature gradient at any age. Meanwhile, the temperature in the region $\rho\sim 10^{11}$--$10^{14}$ \gcc~ has become lower than the melting temperature (see Fig.~\ref{fig:characteristic_temp}) and the solid crust has been formed.

In Fig.~\ref{fig:b0}, we show the cooling curves for non-magnetized neutron stars with masses ranging between 1.10 and 1.76 $M_\odot$. We show the evolution in time of the crustal temperature $T_b$ (taken as the outermost layer of our crust, top panels), redshifted surface temperature $e^\nu T_s$ (middle) and bolometric luminosity (bottom), as defined by eq.~(\ref{eq:bb_emission}).

In general, after $\approx 100$ yr, low mass stars ($M \lesssim 1.4~M_\odot$) are hotter and brighter than high mass stars. For the high-mass family, $M\gtrsim 1.4~M_\odot$, the activated direct URCA processes\footnote{Artificially activated at $\rho>10^{15}~$\gcc for illustrative purposes, see \S~\ref{sec:neutrino}.} result in fast cooling before $\sim 100$ yr. Within the low-mass family, cooling curves are similar at early ages ($<100$ yr). The differences at $t\sim 10^2$--$10^3$ yr are due to the delayed transition of neutrons in the core to a superfluid state, which activates neutrino emission by means of CPFB (see \S~\ref{sec:neutrino}). This effect also depends on the equation of state employed: a stiffer equation of state results in lower densities and the transition to superfluidity is further delayed, assuming the same gap. After the effect of the transition to a superfluid core is finished, at $\gtrsim 10^3$ yr, cooling curves for low mass neutron stars tend to converge again, following the same curve independently of the mass. For models with heavy element envelopes (left panels), the luminosity drops below $10^{33}$ erg/s at most a few thousand years after birth, while light envelope models (right panels) predict luminosities around $10^{33}$--$10^{34}$ erg/s for quite a long period (up to several $10^4$ yr).

In young neutron stars, the evolution is dominated by neutrino cooling (flat part of the curves), while at late times, the emission of thermal photons from the star surface is the dominant factor (the steep decline of luminosity in the log-log plot). The turning point between the neutrino-dominated cooling era and the photon-dominated cooling era happens much earlier for light envelope models (a few $\times 10^4$ yr) than for heavy envelope ones ($\approx 10^5$ yr).  During the photon cooling, the evolution of the surface temperature of the star can be roughly described by a power law $T_s \propto t^{-1/8\xi}$, with $\xi\ll 1$, the exact value depending on the envelope model, which relates $T_s\propto T_b^{0.5 + \xi}$ \citep{page04}. For all weakly magnetized models, after $\sim 1$ Myr, the luminosity has dropped below $10^{31}$ erg/s, the surface temperature goes below $20$--$30$ eV and the star becomes invisible to X-ray observations.  Note also that in the photon cooling era neutron stars with light-elements envelopes are much cooler than those with iron envelopes. \cite{page11} suggested that a star could in principle have a light-elements envelope at the beginning of its life, residual of accretion of fall-back material after the supernova explosion, that is progressively converted to heavier elements by nuclear reactions.

As we will see in \S~\ref{ch:unification}, non-magnetized cooling curves cannot account for the high luminosities of highly magnetized objects. Thus it is necessary to add the effects of the magnetic field in the simulations, extending the four parameter family of classical theoretical cooling models (compositional differences in the core, superfluid properties, composition of the envelope, and mass) with one more, the magnetic field. This is the subject of the next sections.

%%%%%%%%%%
\section{Cooling of strongly magnetized neutron stars.}

\subsection{Initial magnetic field.}\label{sec:initial_b}

How to generate ultra-strong fields in neutron stars and what is the most probable field geometry in a newly born neutron star are still open questions. A possible mechanism to generate a strong magnetic field is the amplification of the progenitor magnetic field by a factor of $\sim (R_{prog}/R_{ns})^3\sim10^{10}$ \citep{woltjer64,ferrario06}  due to flux conservation. 
Alternatively, \cite{duncan92} and \cite{thompson93} proposed a dynamo process, acting in rapidly rotating proto-neutron stars with periods of a few milliseconds, as responsible for the amplification of magnetic field up to magnetar strengths ($10^{14}$--$10^{15}$ G). The magneto-rotational instability also could work in sub-magnetar classes, amplifying their magnetic field after the core collapse \citep{sawai13}. In all scenarios, the short dynamical timescales make plausible to expect that the proto-neutron star be able to reach a sort of MHD equilibrium very soon after birth.

\begin{figure}[t]
 \centering
\includegraphics[width=.6\textwidth]{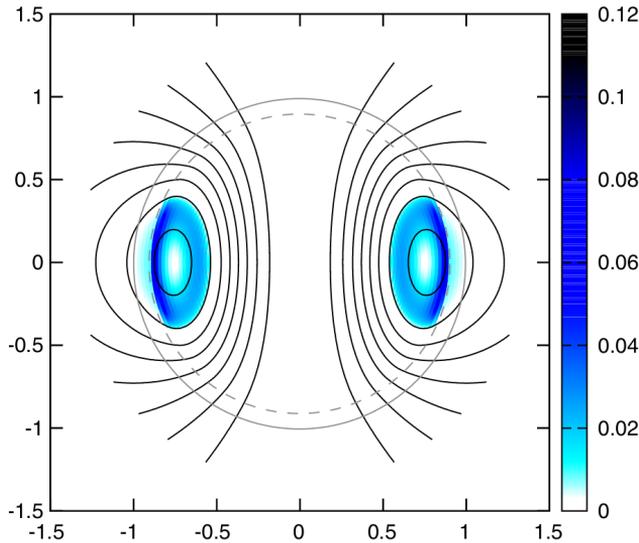}
\caption{Meridional cut of an MHD equilibrium solution in a star composed by a superconducting core and a crust composed solely by protons (interface with dashed line). Poloidal field lines are shown in black and the relative magnitude of the toroidal-field in color scale. Taken from \cite{lander13}.}
 \label{fig:mhd_lander}
\end{figure}

A classical approach to the study of the initial configuration is to build, or search for MHD equilibrium solutions, considering a fluid star \citep{braithwaite04}. These conditions are met in neutron stars before the crust freezes (hours to weeks after birth). The results depend on the equation of state and whether stratification is taken into account or not. However, in all cases the equilibrium consist in a large-scale poloidal magnetic field, either dipolar \citep{lander09}, or with higher multipoles \citep{ciolfi09}, with an equatorial torus where the lines are twisted. The mixture of poloidal and toroidal components is essential to have stability. \cite{lander13} considers a superconducting core coupled to a proton crust, obtaining a similar configuration, with the main difference that the toroidal magnetic field is tubular (see Fig.~\ref{fig:mhd_lander}, taken from that work). In most configurations, the strength of the toroidal component can be locally comparable with the poloidal one. \cite{glampedakis12} find that the inclusion of stratification and neutron superfluidity in the model leads to equilibria with a larger fraction of magnetic energy stored in the toroidal component. However, also in the latter case, most of the volume-integrated magnetic energy is provided by the large scale poloidal component. This is generally different from simulations for main-sequence stars, where the differential rotation induces larger toroidal components \citep{nelson13}. Note also that, according to \cite{lander12}, the boundary conditions play a major role in determining the fraction of magnetic energy stored in toroidal magnetic field. Vacuum boundary conditions applied to Grad-Shafranov equations lead, by definition, to toroidal magnetic fields confined within the closed poloidal magnetic field lines. Given the typical very fast MHD timescales, the system is supposed to reach the equilibrium, but it is not clear whether the complicated dynamics of the proto-neutron star actually leads to the kind of configuration described above, or whether they are stable (see \citealt{akgun13} for stability analyses).

%%%%%%%%%%%%%%%%%%%%%%%%%%%%%%%%%%%%%%%%%%
\begin{figure}[t]
 \centering
\includegraphics[width=.32\textwidth]{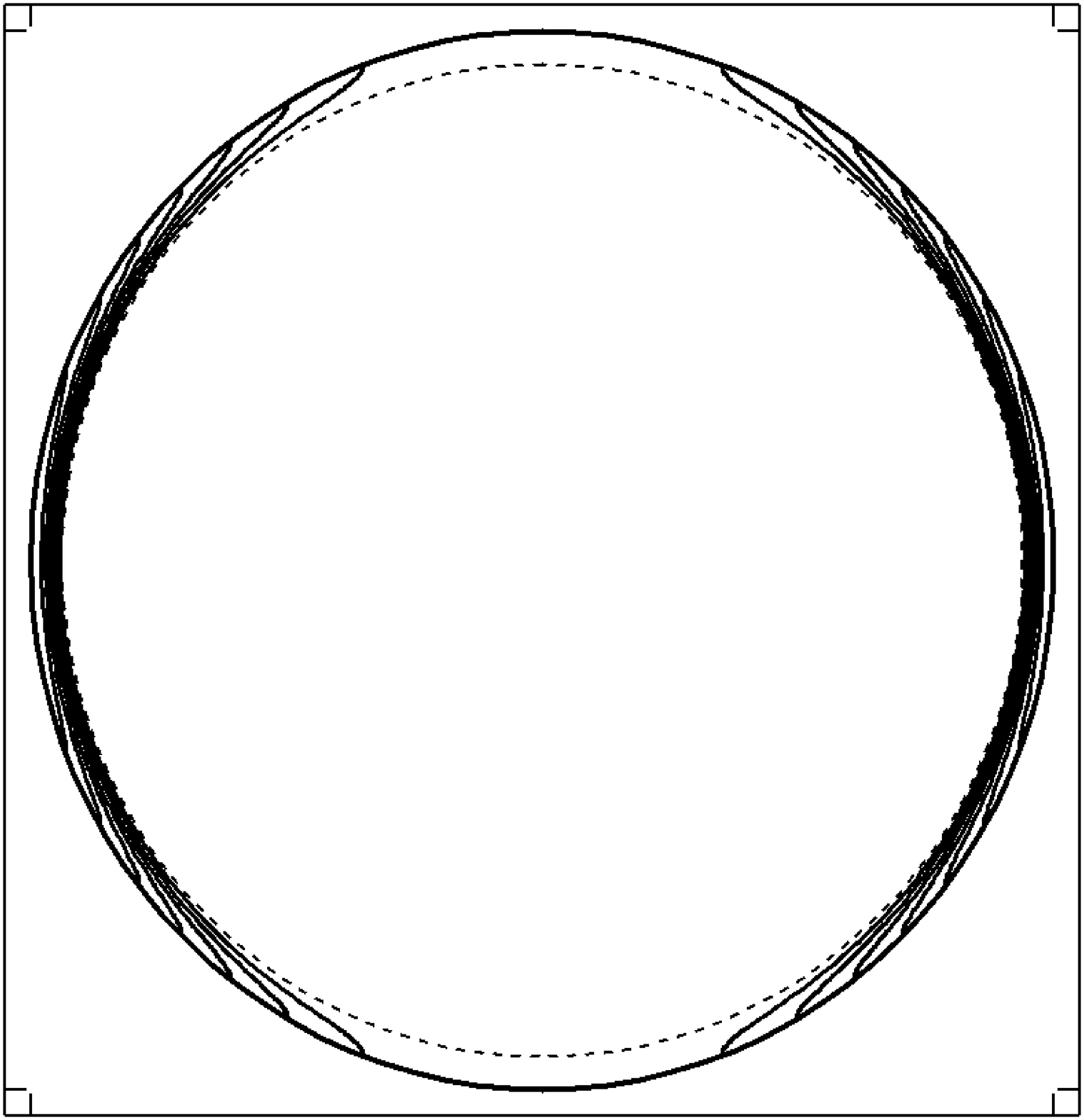}
\includegraphics[width=.32\textwidth]{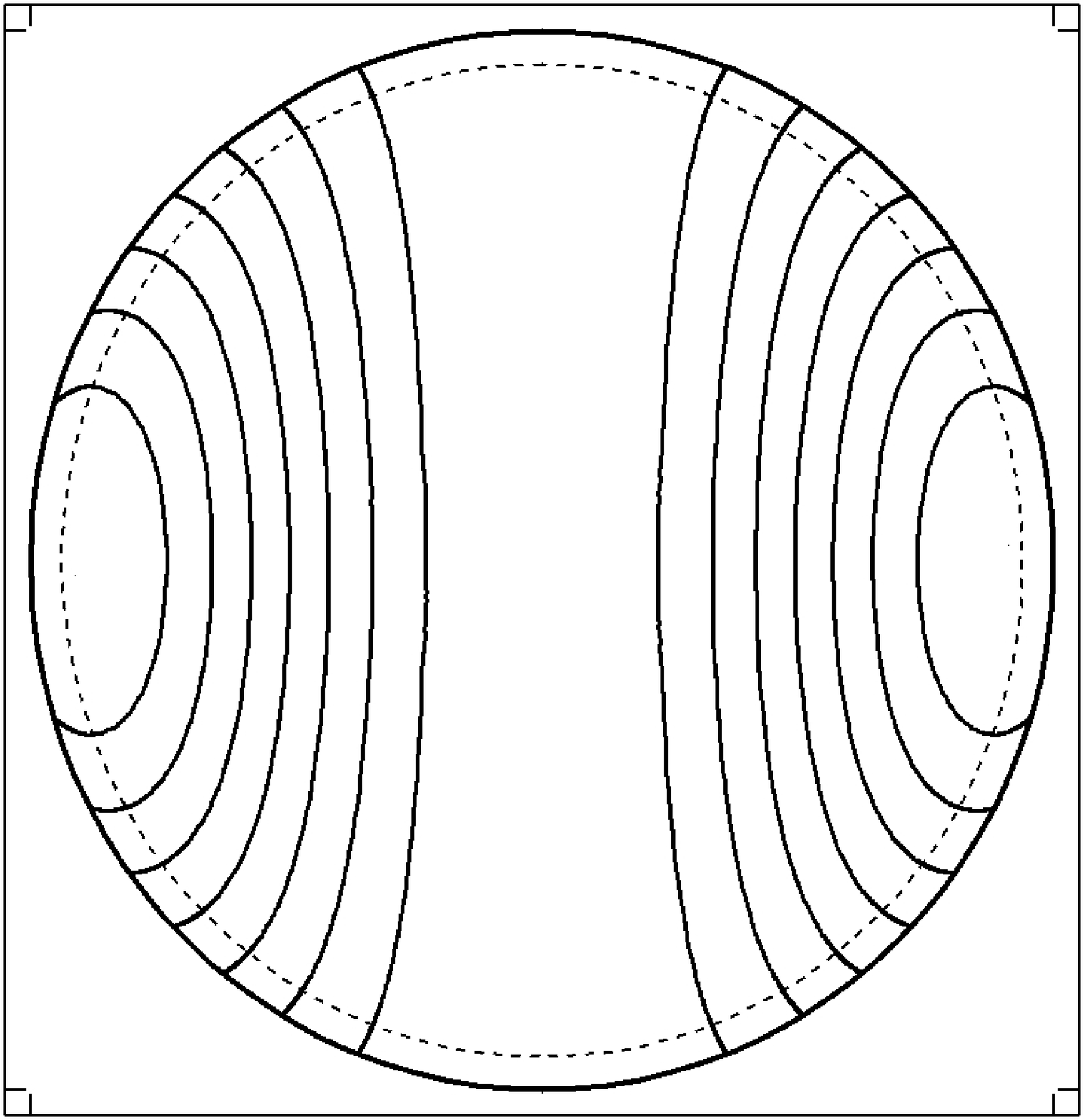}
\includegraphics[width=.32\textwidth]{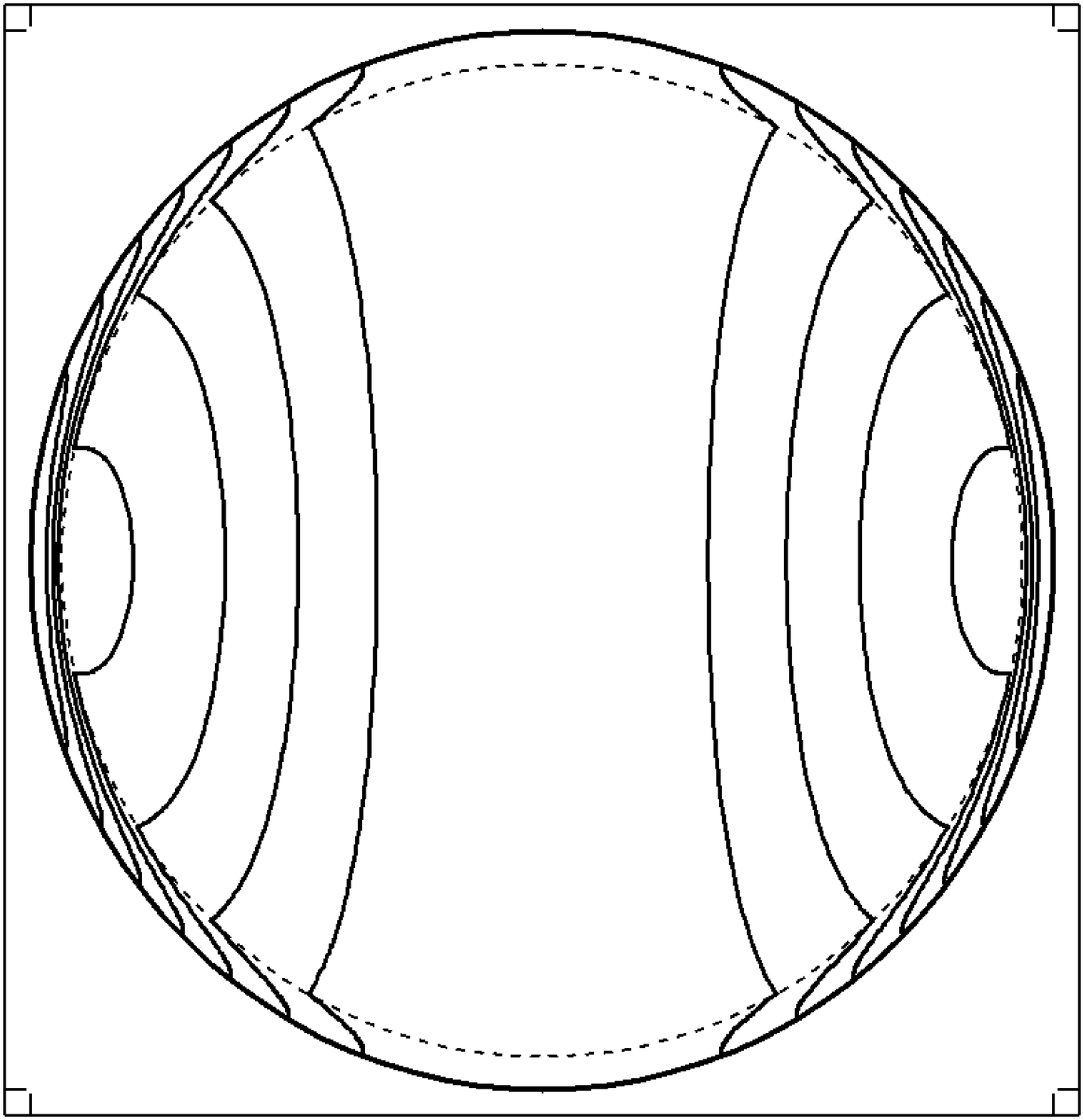}
\caption{Representative configurations of the initial magnetic field geometries. Left: type A, crustal confined field; center: type B, core-extended field; right: type C, hybrid. Solid lines show the poloidal magnetic field lines in a meridional plane.}
 \label{fig:initial_b}
\end{figure}
%%%%%%%%%%%%%%%%%%%%%%%%%%%%%%%%%%%%%%%%%%

Hereafter we denote by ``initial'' the configuration at the epoch when the crust is formed. Lacking robust arguments in favor of specific initial configurations, we consider three generic families with different geometries, illustrated in Fig.~\ref{fig:initial_b}. In all cases, the initial poloidal magnetic field matches an external vacuum dipolar solution at the surface, and it is built by means of a dipolar spherical Bessel configuration (see \S 2.1 of \citealt{aguilera08b} and our Appendix~\ref{app:bessel}).

In the ``crust-confined field'' (type A geometry, left panel), the magnetic field lines do not penetrate into the core ($B_r=0$ at the crust/core interface), and the currents are entirely contained in the crust. The toroidal magnetic field (not shown in the figure) is extended through the crust, and its form is arbitrarily chosen to be dipolar:
\begin{equation}
 B_\varphi=-k_{tor}(r-R_{core})^2(r-R_\star)^2\sin\theta/r~, 
\end{equation}
where $k_{tor}$ is a normalization factor. In type B configurations (central panel), the magnetic field threads the core, where the bulk of the current circulates. It mimics the typical twisted torus dipolar configuration discussed above. Type C geometry (right panel) is a hybrid case where a double system of currents supports a large-scale, core-extended dipole and an additional, stronger crustal field. Each family is parametrized by the intensity of the dipolar component at the pole, $B_p^0$, and the maximum intensity of the toroidal magnetic field at any point, $B_t^0$.  In Table~\ref{tab:models} we list the models studied in this work. We stress that what governs the evolution of the magnetic field is where the bulk of currents circulate, because their location determines the dissipation rate.

%%%%%%%%%%%%%%%%%%%%%%
\begin{table}
\begin{center}
\begin{tabular}{l c c c}
\hline
\hline
Model & current  & $B_p^0$ & $B_t^0$  \\
 & location  & [G] &  [G]  \\
\hline
A14	& crust	& $10^{14}$ & 0 	\\
A15	& crust	& $10^{15}$ & 0 	\\
A14T	& crust	& $10^{14}$ & $5\times10^{15}$ \\
B14	& core	& $10^{14}$ & 0 	\\
C14	& crust+core & $10^{14}$ & 0 \\
\hline
\hline
\end{tabular}
\end{center}
\caption{Summary of the initial magnetic field configurations.}
\label{tab:models}
\end{table}
%%%%%%%%%%%%%%%%%%%%%%%

\subsection{General evolution.}

The Ohmic dissipation time-scale can be estimated as
\begin{equation}
\tau_{d} \sim \frac{L_B^2}{\eta}~,
\end{equation} 
where $L_B$ is the typical length scale of the magnetic field variation. For the Hall term, several time-scales can be considered from the analysis of eq.~(\ref{eq:hall_general}): all share the important inverse dependence with magnetic field. Considering the advecting terms, which include the poloidal magnetic field evolution ($i=r,\theta$), or the source terms, the Hall time-scale is given by
\begin{equation}\label{eq:hall_time-scale_pol}
\tau_{h,pol} \sim \frac{4\pi e n_e}{c B} L_B^2 \sim \frac{\tau_{d}}{\omtau}~.
\end{equation} 
With these definitions, $\omtau$ gives the ratio between the Ohmic and Hall time-scales. From the toroidal magnetic field evolution we can estimate the typical time-scale for the formation of the current sheet and for the drift of the toroidal magnetic field in radial direction:
\begin{equation}\label{eq:hall_drift_time-scale}
 t_{h,drift} \simeq \frac{4 \pi e n_e}{cB_\varphi}\left[\frac{1}{R_\star}\left(\frac{1}{L_B} + \frac{1}{L_{n_e}}\right) + \frac{1}{L_B}\left(\frac{1}{L_{n_e}} + \frac{1}{R_\star}\right)\right]^{-1}~,
\end{equation}
where  $L_{n_e}$ is the typical length-scale of variation of electron density. In a realistic profile in the crust, $L_{n_e}\sim 0.1$ km, typically less than $L_B$ and $R_\star$. In general, we can estimate 
\begin{equation}\label{eq:hall_time-scale}
 t_{h,tor} \simeq \frac{4 \pi e n_e}{cB}L_BL_{min}~,
\end{equation}
where $L_{min}$ is the minimum length-scale between $L_B$ and $L_{n_e}$.

The Ohmic and Hall time-scales vary by orders of magnitude within the crust and during the evolution, depending strongly on density and magnetic field intensity and curvature. The Ohmic time-scale additionally depends on temperature. For this reason, rough analytical estimates can only help to understand realistic numerical simulations, which become a necessary
tool to understand the complex evolution.

\begin{figure}[t]
\centering
\includegraphics[width=.32\textwidth]{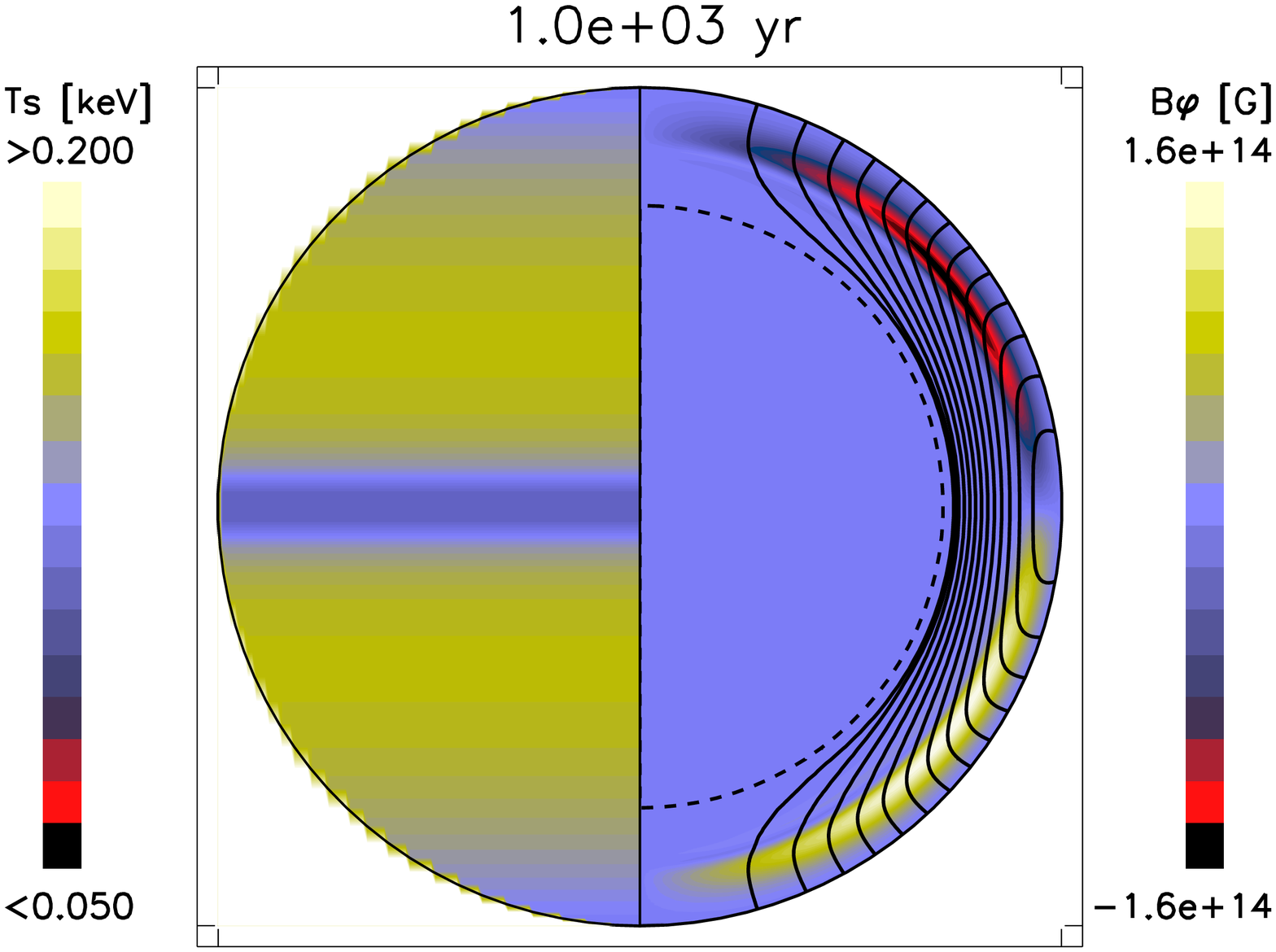}
\includegraphics[width=.32\textwidth]{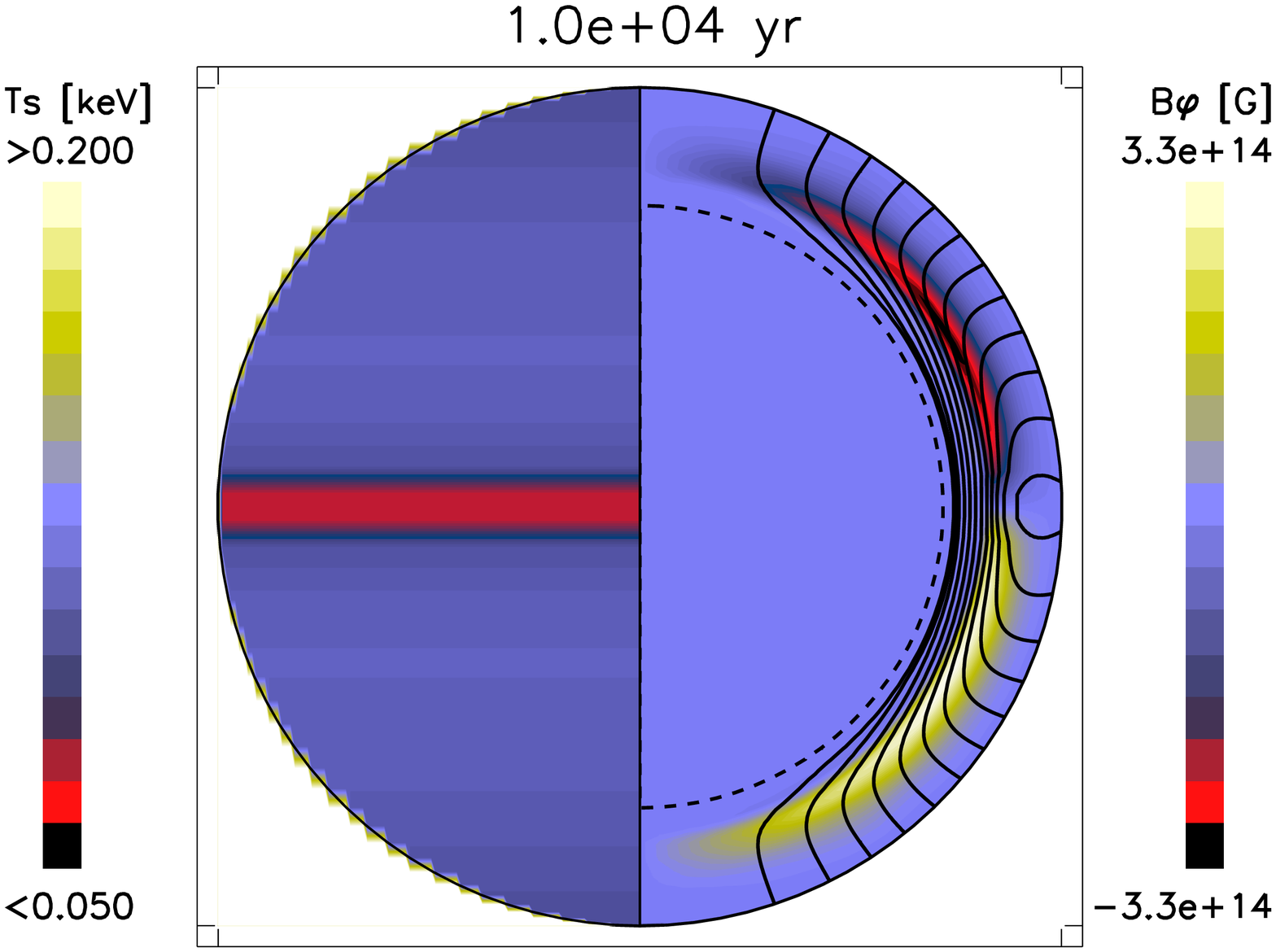}
\includegraphics[width=.32\textwidth]{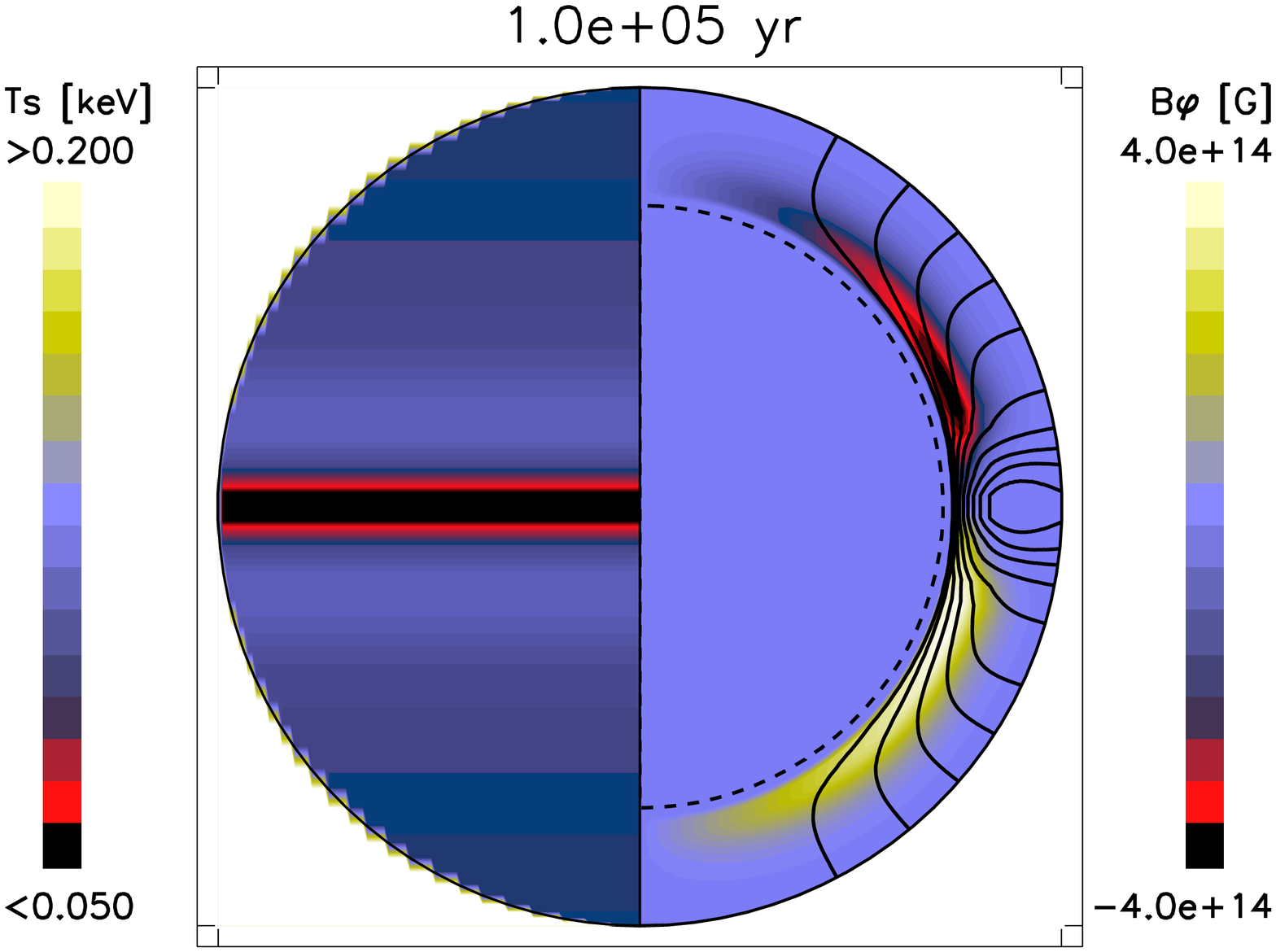}\\
\vskip0.2cm
\includegraphics[width=.32\textwidth]{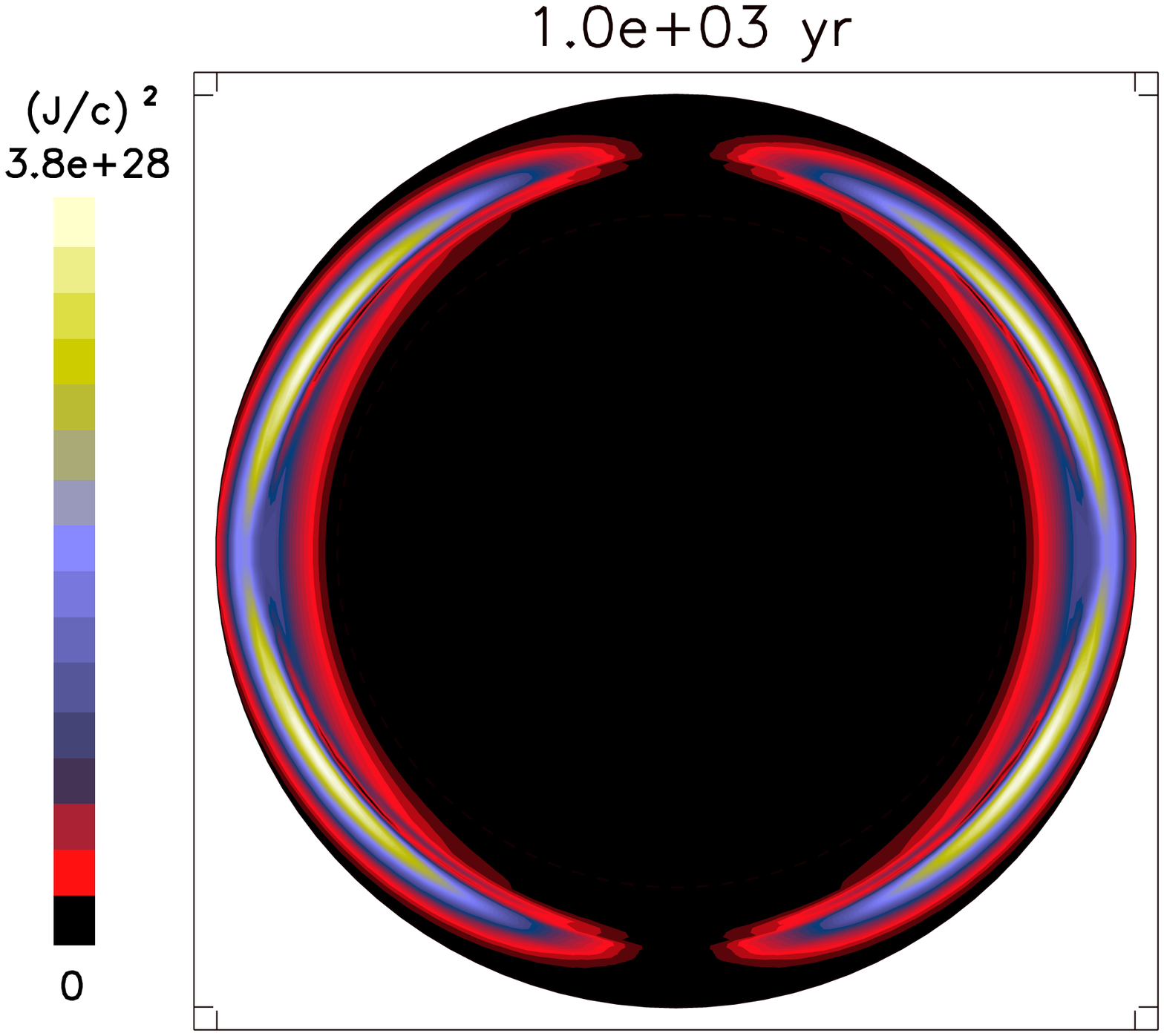}
\includegraphics[width=.32\textwidth]{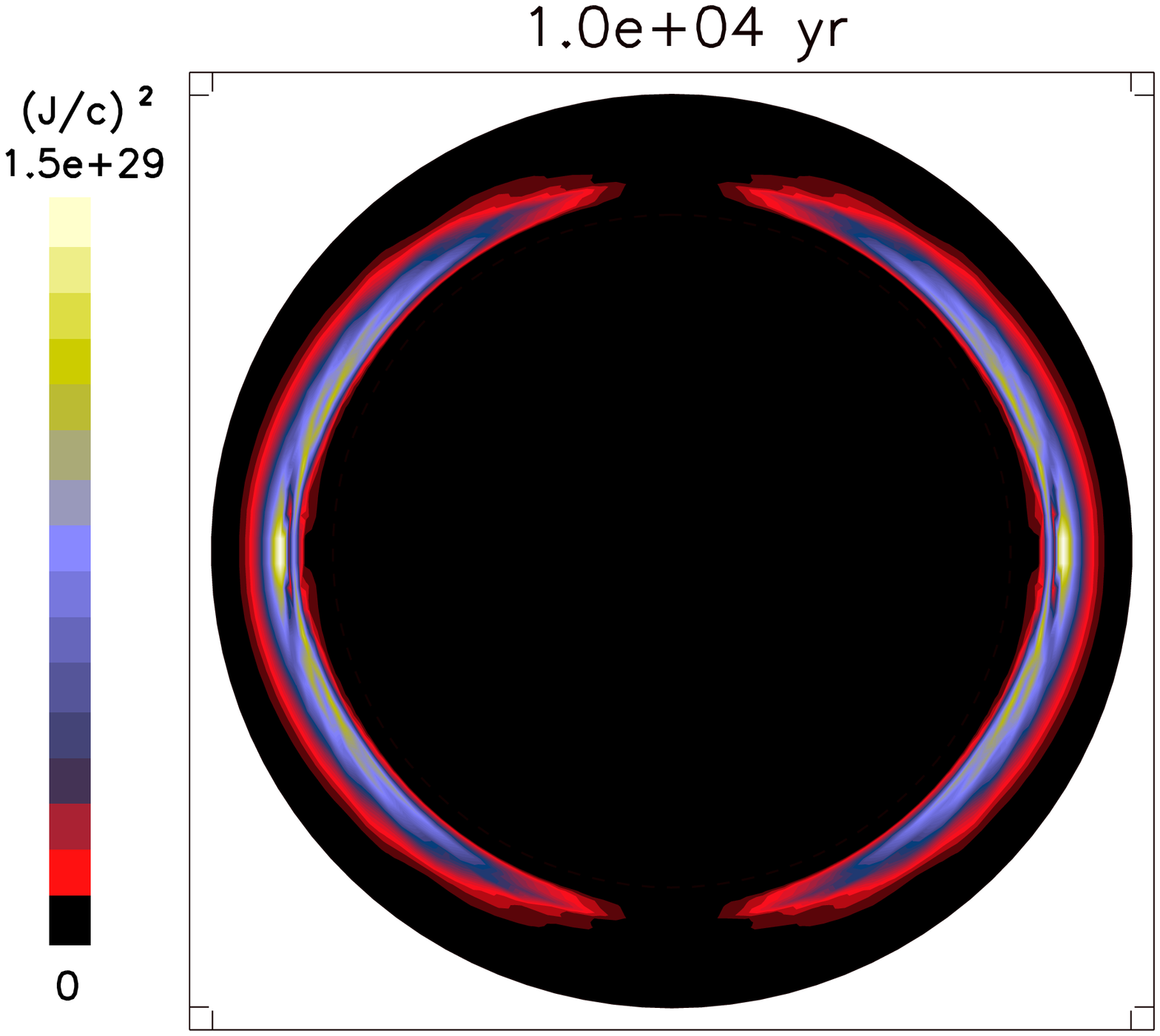}
\includegraphics[width=.32\textwidth]{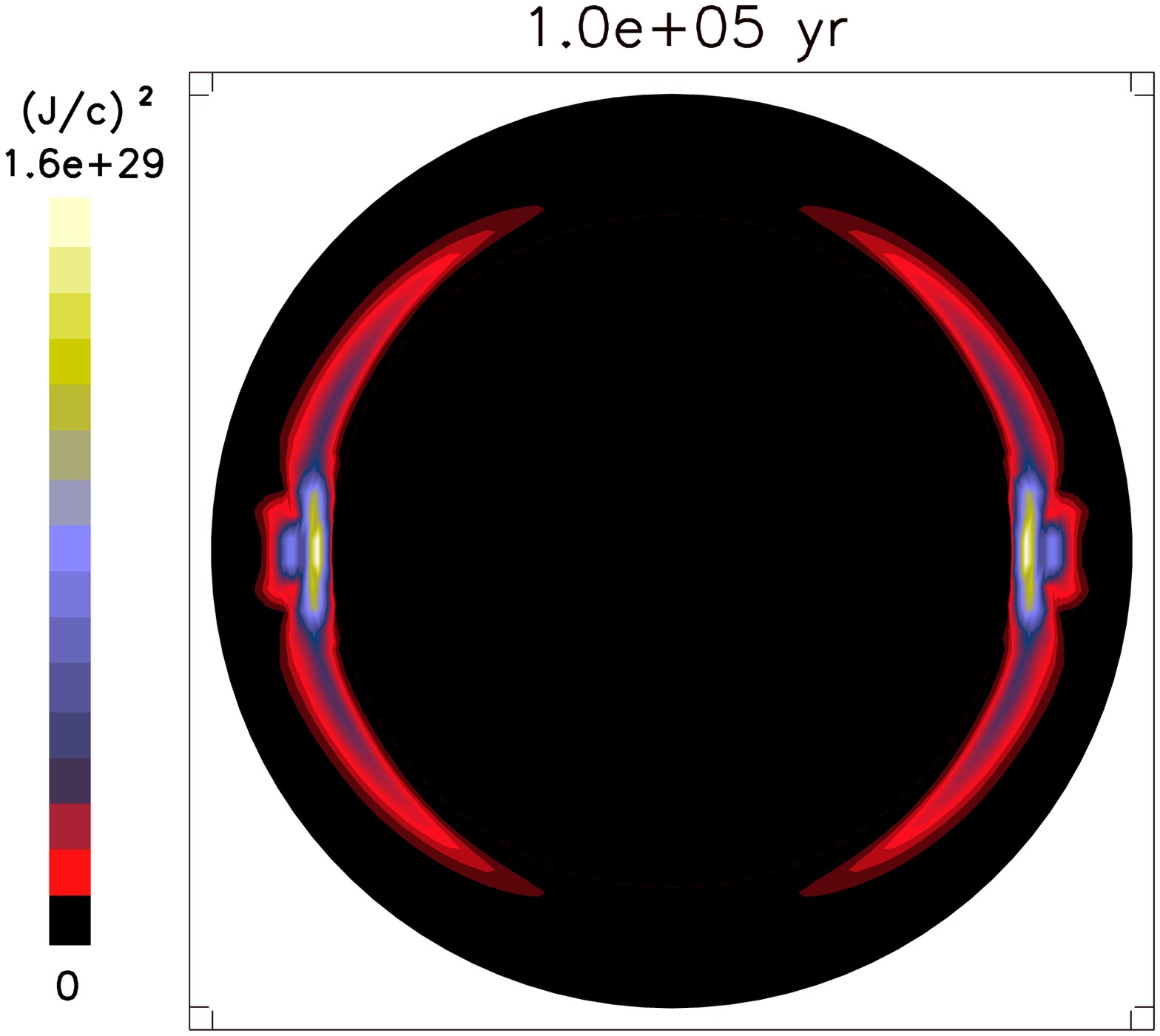}\\
\vskip0.2cm
\includegraphics[width=.32\textwidth]{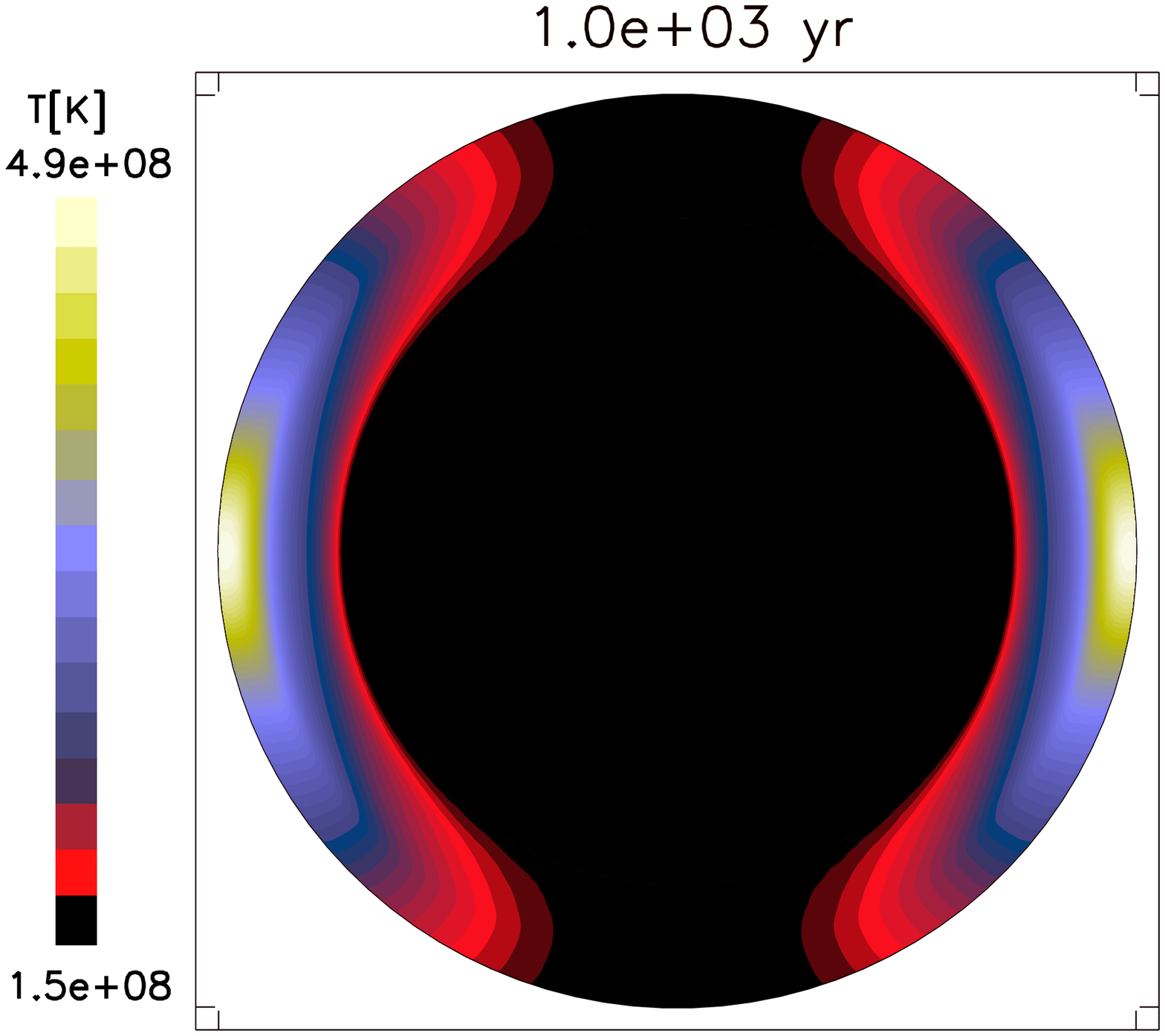}
\includegraphics[width=.32\textwidth]{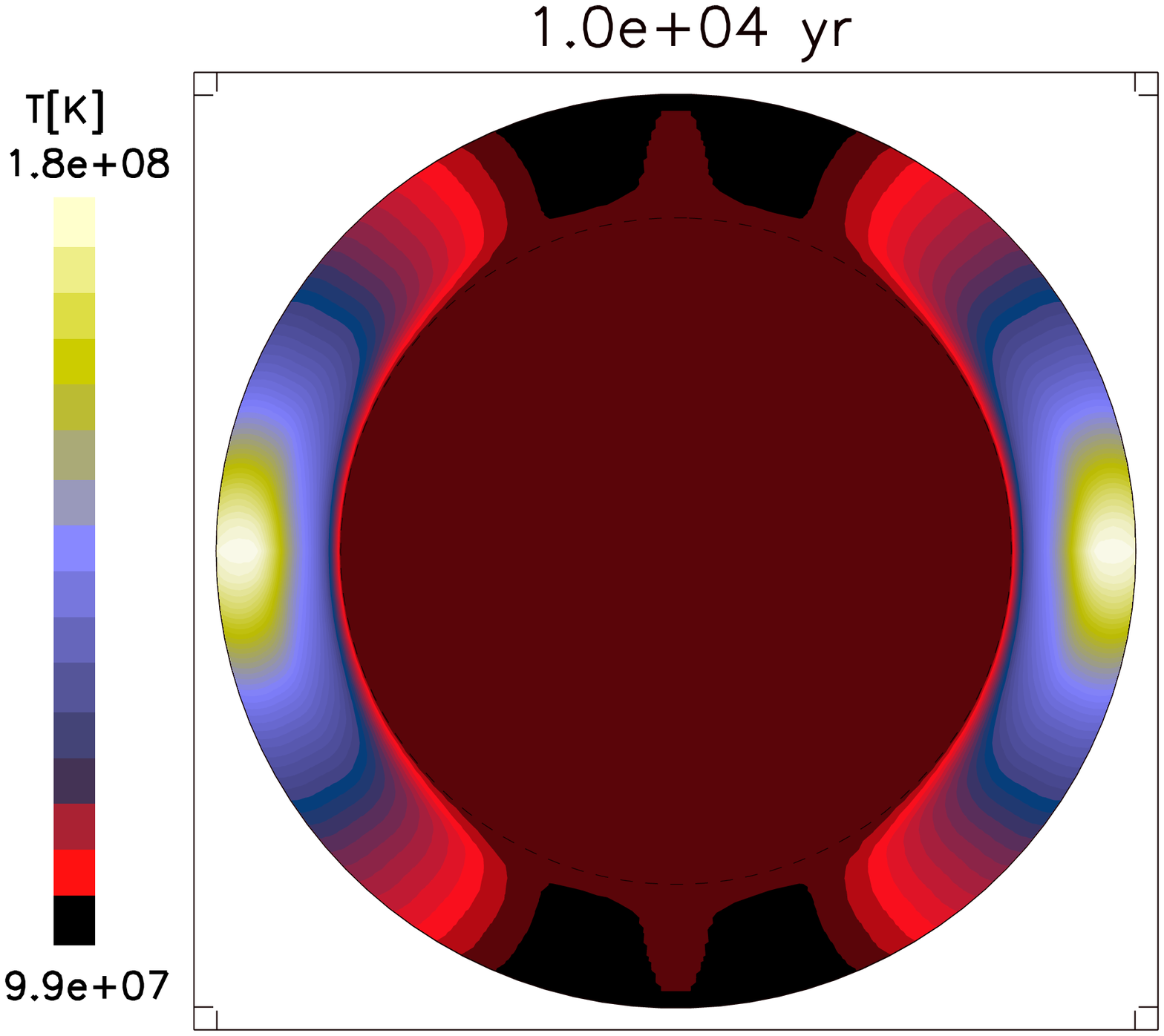}
\includegraphics[width=.32\textwidth]{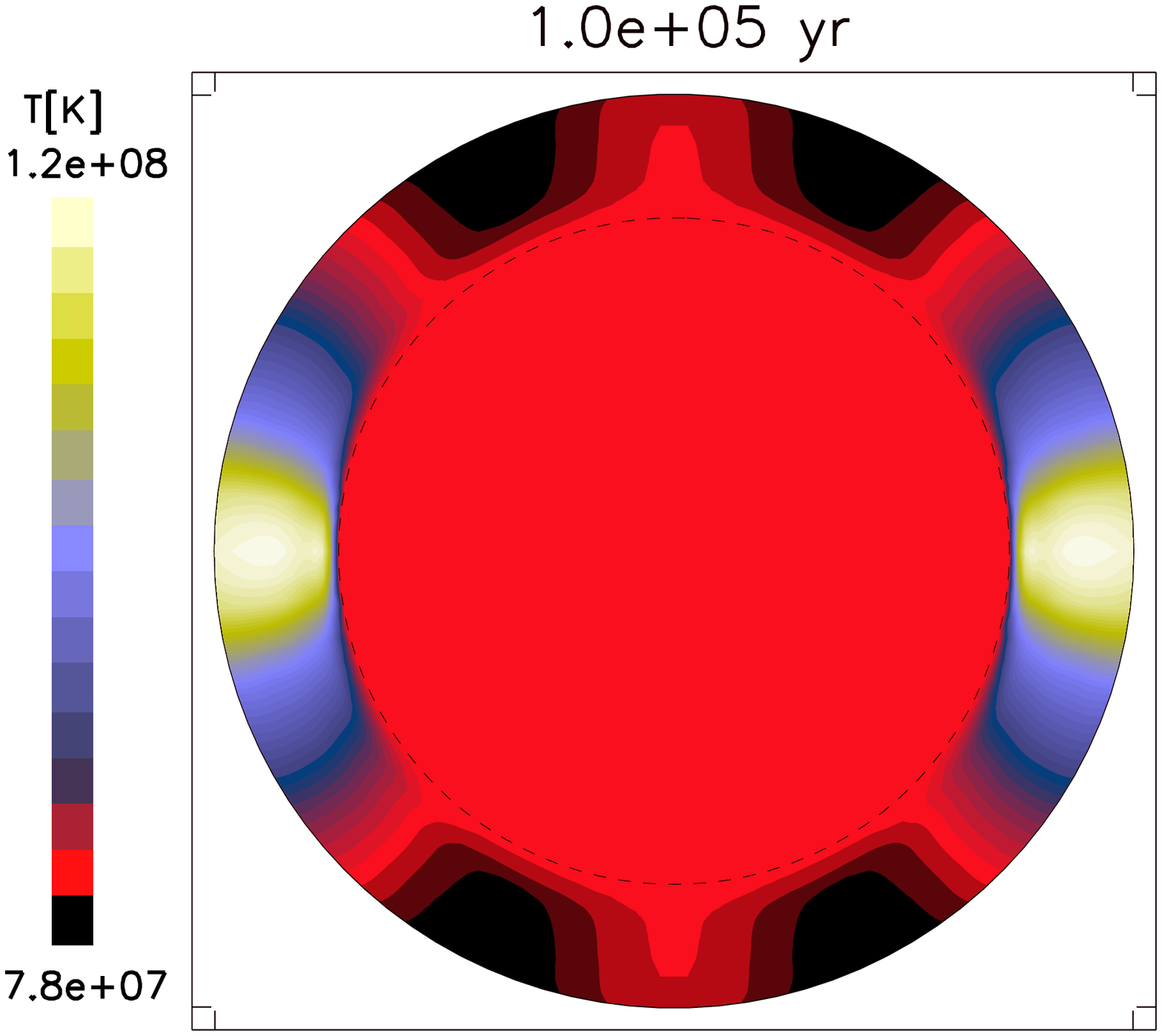}
\caption{Snapshots of the evolution of model A14 at $10^3, 10^4, 10^5$ yr, from left to right. Top panels: the left hemisphere shows in color scale the surface temperature, while the right hemisphere displays the magnetic configuration in the crust. Black lines are the poloidal magnetic field lines, while color scale indicates the toroidal magnetic field intensity (yellow: positive, red: negative). Middle panels: intensity of currents; the color scale indicates $J^2/c^2$, in units of $($G/km$)^2$. Bottom panels: temperature map inside the star. In all panels, the thickness of the crust has been enlarged by a factor of 4 for visualization purposes (see Fig.~\ref{fig:initial_b} for the real scale).}
 \label{fig:b14_evo}
\end{figure}

To gain some insight into the problem, we run simulations up to several Myr, monitoring the temperature and magnetic field inside the star and at its surface. We begin the discussion of our results by describing the general evolution of model A14. In Fig.~\ref{fig:b14_evo} we show three snapshots at $t=10^3, 10^4, 10^5$ yr. In the top panels, for each snapshot, we show in the left hemisphere the surface temperature distribution, and in the right hemisphere the evolution of the crustal magnetic field. We now concentrate on the latter.

%%%%%%%%%%%%%%%%%%%%%%%
%%%%%%%%%%%%%%%%%%%%%%%

The first effect of the Hall term in the induction equation is to couple the poloidal and toroidal magnetic field, so that even if the latter is zero at the beginning, it is soon created. After $\sim 10^3$ yr, the poloidal dipolar field has generated a mainly quadrupolar toroidal magnetic field, with a maximum strength of the same order of the poloidal magnetic field, with $B_\varphi$ being negative in the northern hemisphere and positive in the southern hemisphere. The Hall term is responsible for the partial conversion of energy from the dipolar component to higher multipoles, likely creating localized current sheets.

Thereafter, under the effect of the Hall drift, the toroidal magnetic field rules the evolution, dragging the currents into the inner crust (see middle panels), and compressing the magnetic field lines. These combined effects result in fast dissipation. When $L_B$ is so small than the Ohmic time-scale is comparable with Hall drift, the currents stop to drift inward. Thus, after $\sim 10^5$ yr, the toroidal magnetic field is mostly contained in the inner crust. In the outer crust, the magnetic field approaches a potential configuration that matches the external magnetic field.

After this phase, most of the current circulates close to the crust/core interface. Therefore, the dissipation of magnetic energy is regulated by the resistivity in this region. In this model, the highly resistive layer in the pasta region leads to a rapid decay of the magnetic field. At later ages, currents in the pasta regions have been dissipated and the magnetic field is sustained by the currents circulating in the immediately outer layer ($\rho \sim 10^{13}$ \gcc). The dissipation time-scale of the magnetic field reflects the resistivity of the regions where the currents circulate. This has a direct imprint on the observable rotational properties, by means of the evolution of the external dipolar component $B_p$ (see \S~\ref{sec:magnetic_micro}).

%%%%%%%%%%  Magnetic ENERGY
\begin{figure}
 \centering
\includegraphics[width=.45\textwidth]{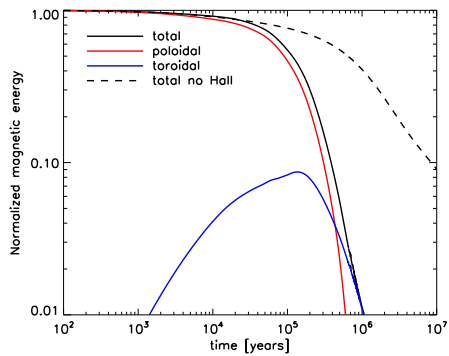}
\caption{Magnetic energy in the crust (normalized to the initial value) as a function of time, for model A14. The solid lines correspond to: total magnetic energy (black), energy in the poloidal (red) and toroidal (blue) components. The dashed line shows the effect of switching off the Hall term (only Ohmic dissipation).}
 \label{fig:entransf}
\end{figure}
%%%%%%%%%%%%%%%%%%%%%%%%%%%%%%%%

The Ohmic deposition of heat changes the map of the internal temperature. In the bottom panels of Fig.~\ref{fig:b14_evo} we show the crustal temperature evolution. At $t=10^3$ yr, the equator is hotter than the poles by a factor of 3. As the evolution proceeds and currents are dissipated, the temperature reflects the change of geometry of the poloidal magnetic field lines (see Fig.~\ref{fig:b14_evo}) and the anisotropy becomes weaker.

The presence of strong tangential components ($B_\theta$ and $B_\varphi$) insulates the surface from the interior. For a dipolar geometry, the magnetic field is nearly radial at the poles, and these are thermally connected with the interior, while the equatorial region is insulated by tangential magnetic field lines. This has a twofold effect: if the core is warmer than the crust, the polar regions will be warmer than the equatorial regions; however, if Ohmic dissipation heats up the equatorial regions, they will remain warmer than the poles. In addition, the insulating effect of the envelope  must be taken into account. Depending on the local conditions (temperature, field strength, angle of the magnetic field with the normal to the surface) it may invert the anisotropy: at the surface, the equator is cooler than the poles (see top panels of Fig.~\ref{fig:b14_evo}). 

In Fig.~\ref{fig:entransf} we show the evolution of the total magnetic energy, compared with the fraction stored in the different components.  We also compare our results to the purely resistive case, in which the Hall term has been switched off (and therefore there is no creation of toroidal magnetic field). When the Hall term is included, $\sim 99 \%$ of the initial magnetic energy is dissipated in the first $\sim 10^6$ yr, compared to only the $60\%$ of the purely resistive case. At the same time, a $\sim 10\%$ of the initial energy is transferred to the toroidal component in $10^5$ yr, before it begins to decrease. Note that the poloidal magnetic field, after $10^5$ yr, is dissipated faster than the toroidal magnetic field. The poloidal magnetic field is supported by toroidal currents concentrated in the inner, equatorial regions of the crust. Here the resistivity is high for two reasons: the effect of the pasta phase, and the higher temperature (see right bottom panel of Fig. \ref{fig:b14_evo}). Conversely, the toroidal magnetic field is supported by larger loops of poloidal currents that circulate in higher latitude and outer regions, where the resistivity is lower. This causes the more rapid dissipation of the poloidal magnetic field. As a result, at late times most of the magnetic energy is stored in the toroidal magnetic field.

%%%%%%%%%%%%%%%%%%%%%%%%%%%%%%%%%%%%%%%%%%%%%%%
\begin{figure}
 \centering
\includegraphics[width=.32\textwidth]{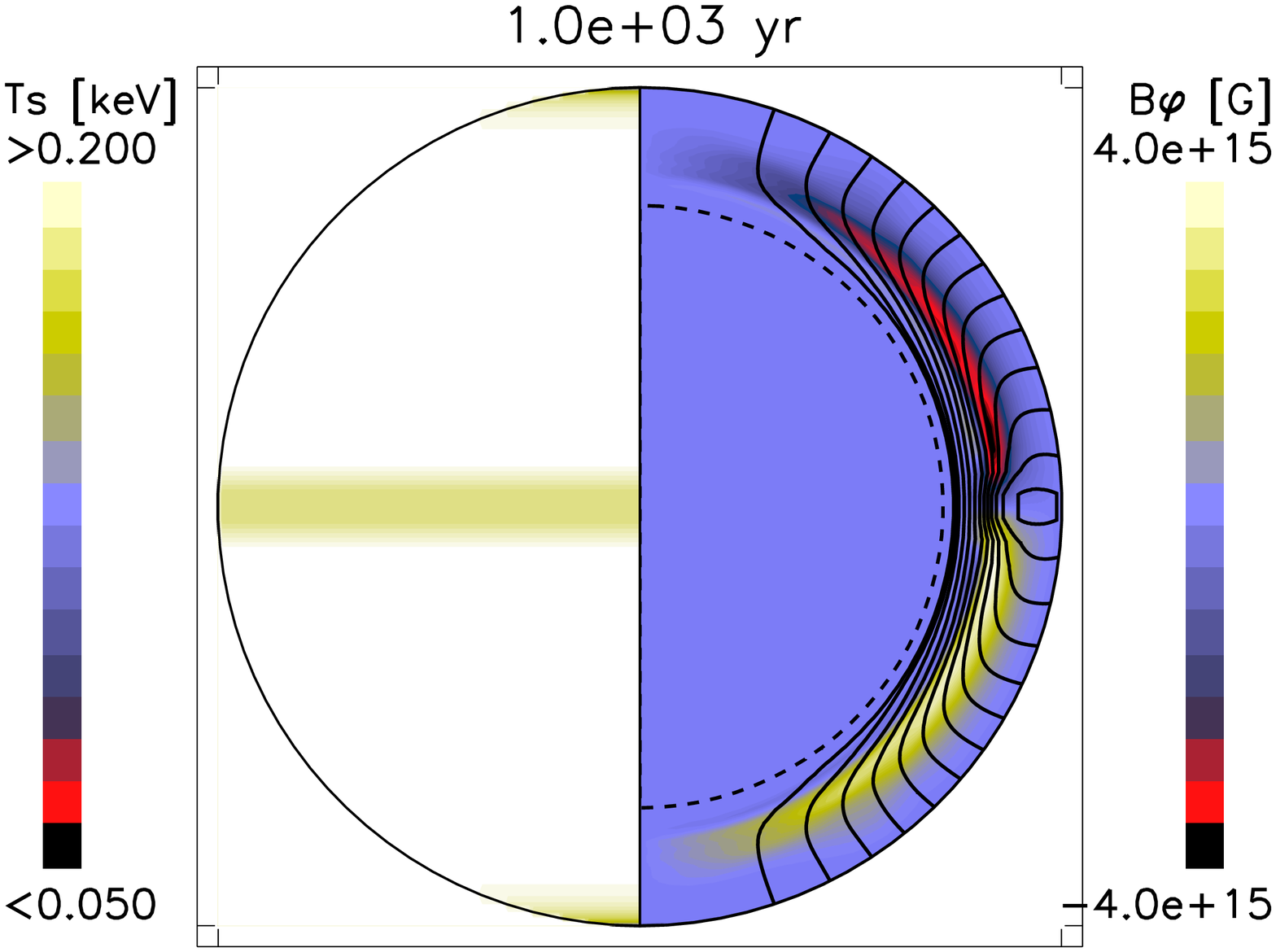}
\includegraphics[width=.32\textwidth]{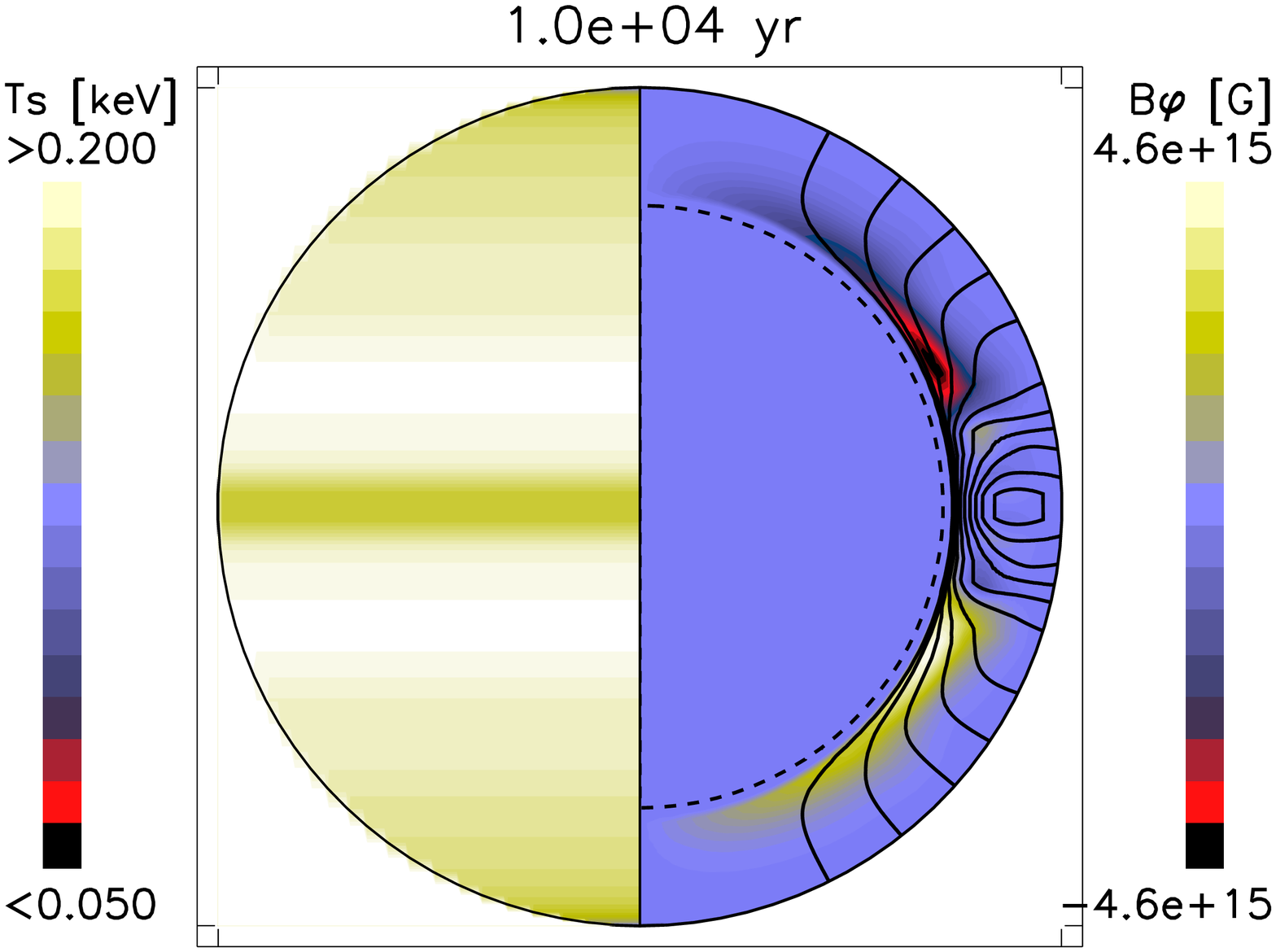}
\includegraphics[width=.32\textwidth]{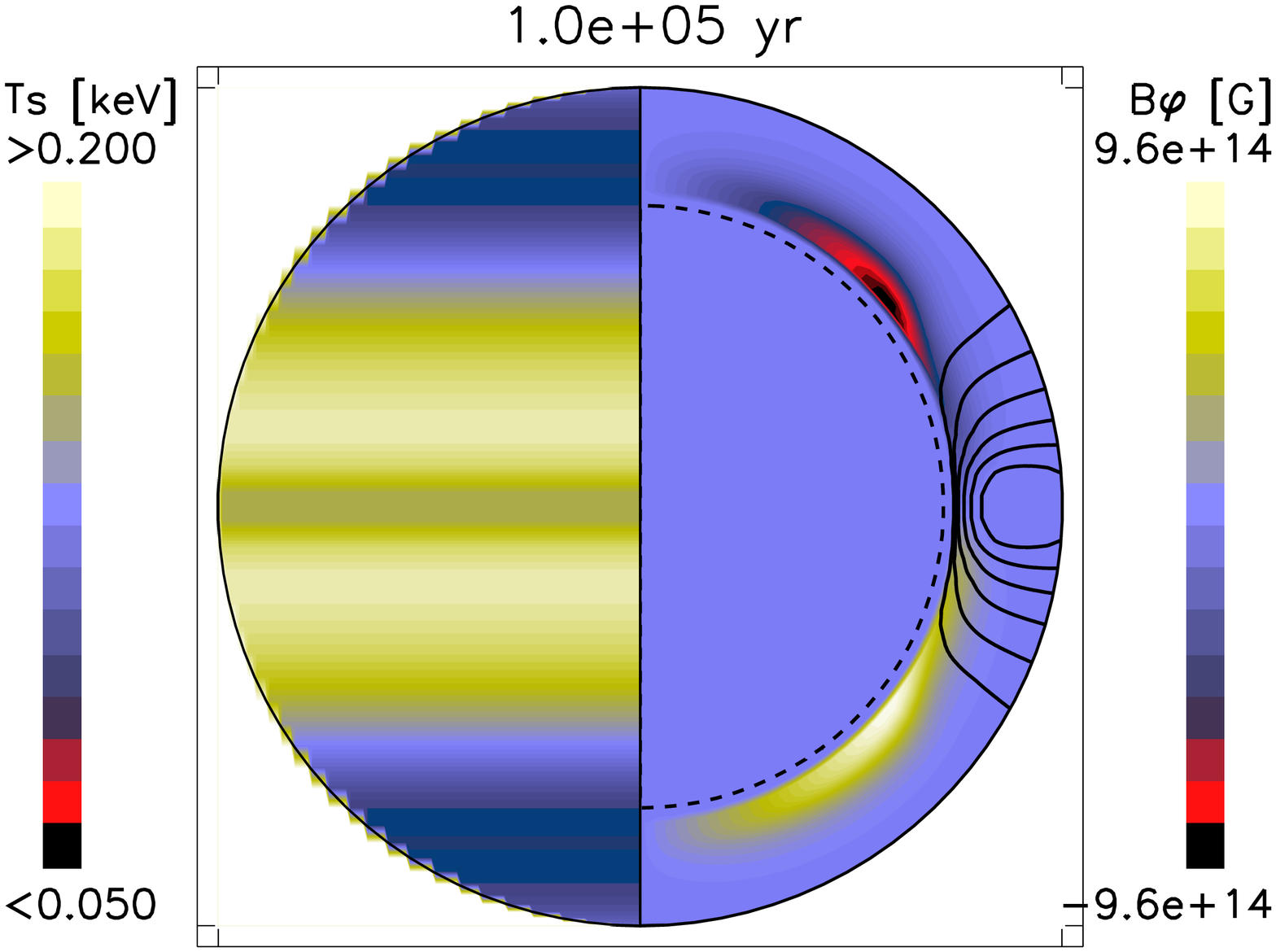}\\
\vskip0.5cm
\includegraphics[width=.32\textwidth]{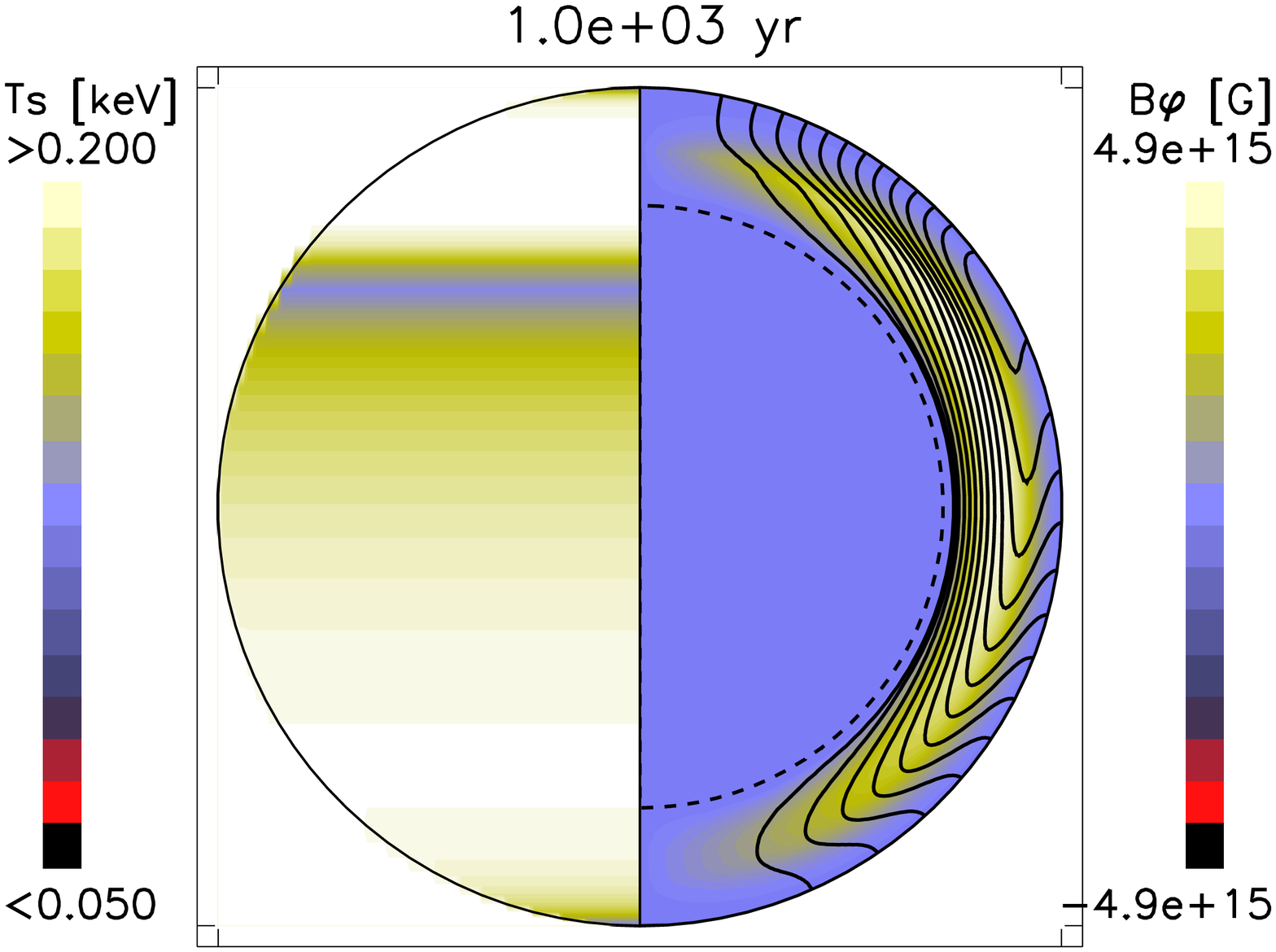}
\includegraphics[width=.32\textwidth]{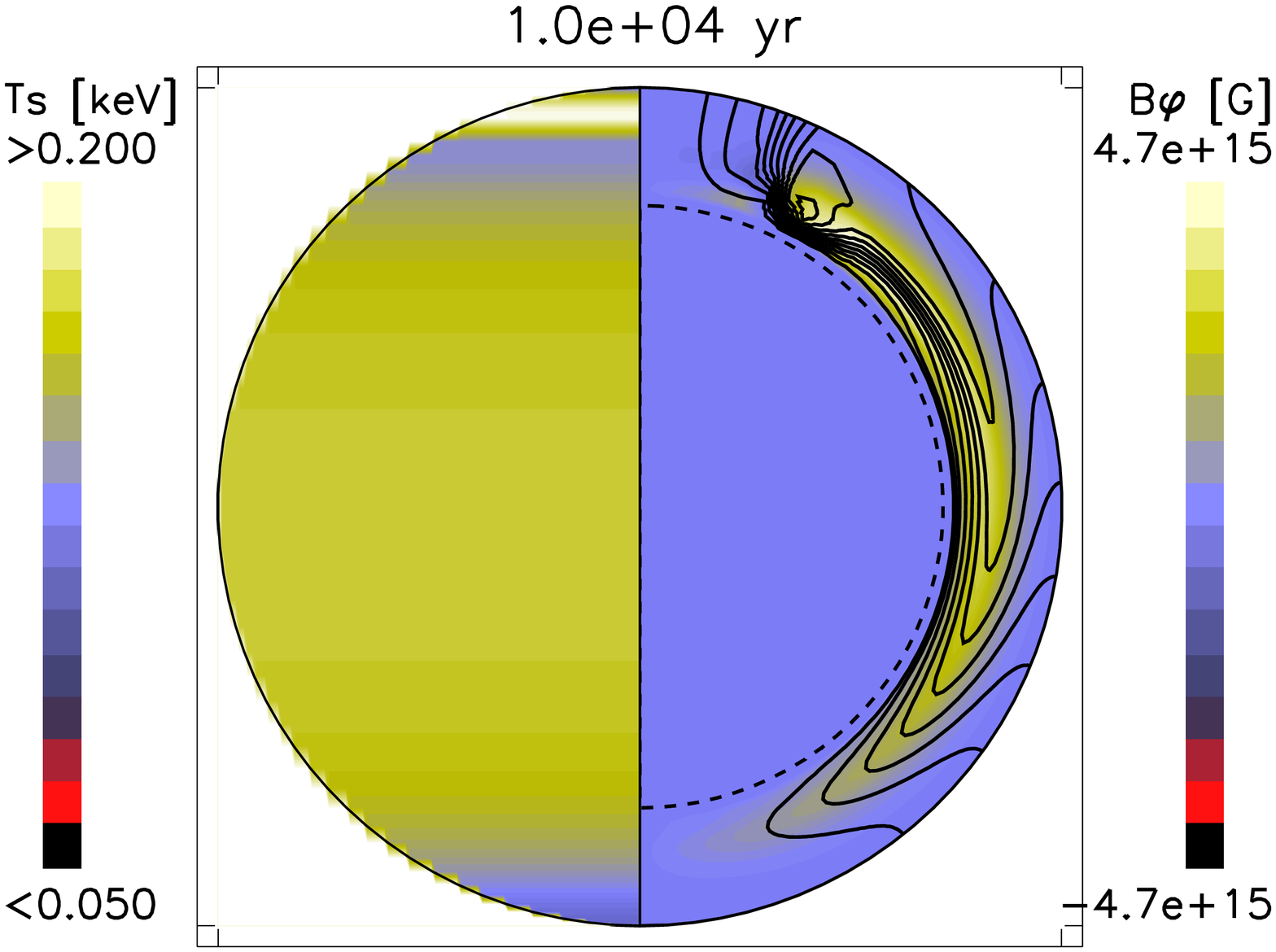}
\includegraphics[width=.32\textwidth]{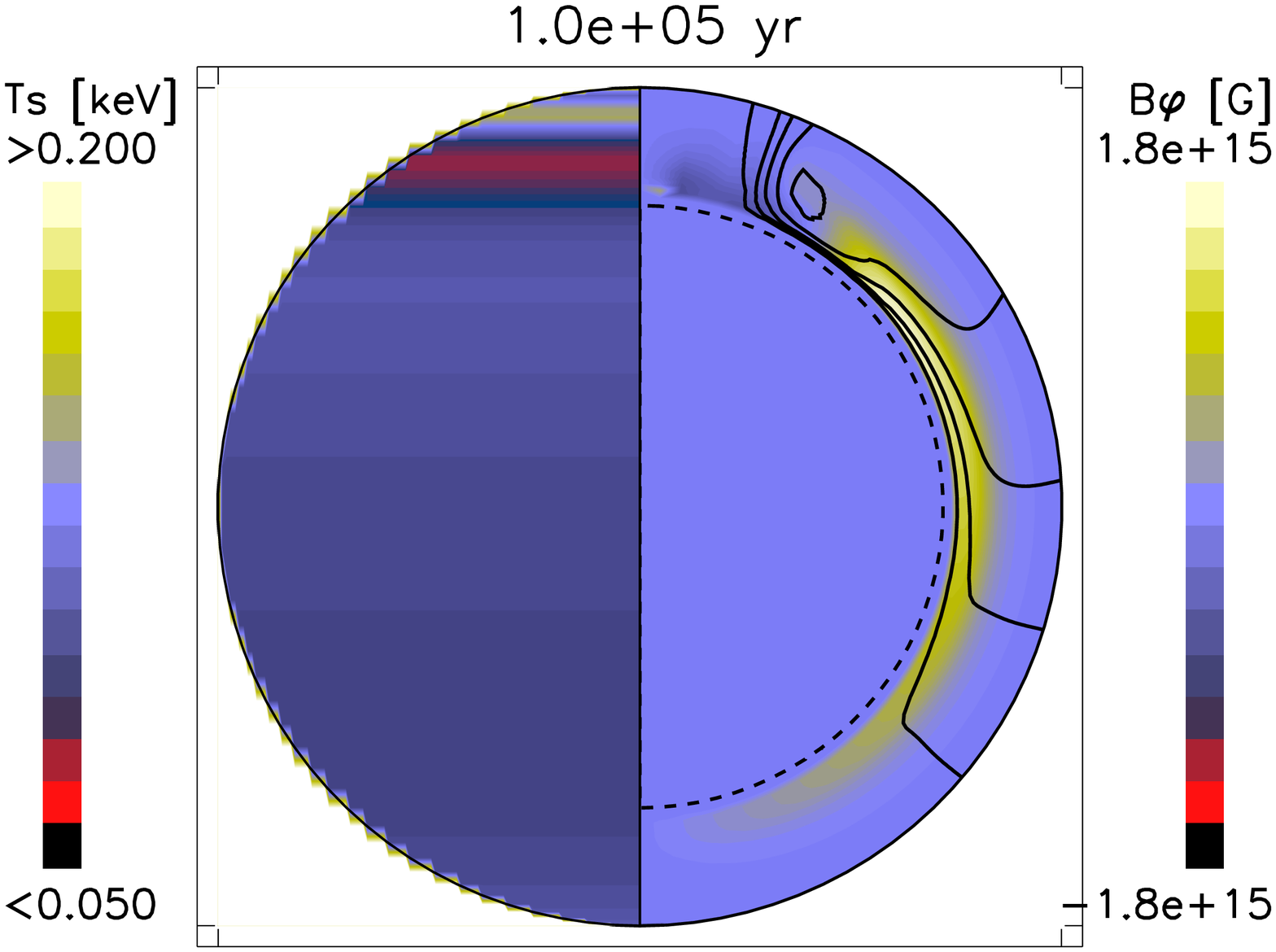}\\
\vskip0.5cm
\includegraphics[width=.32\textwidth]{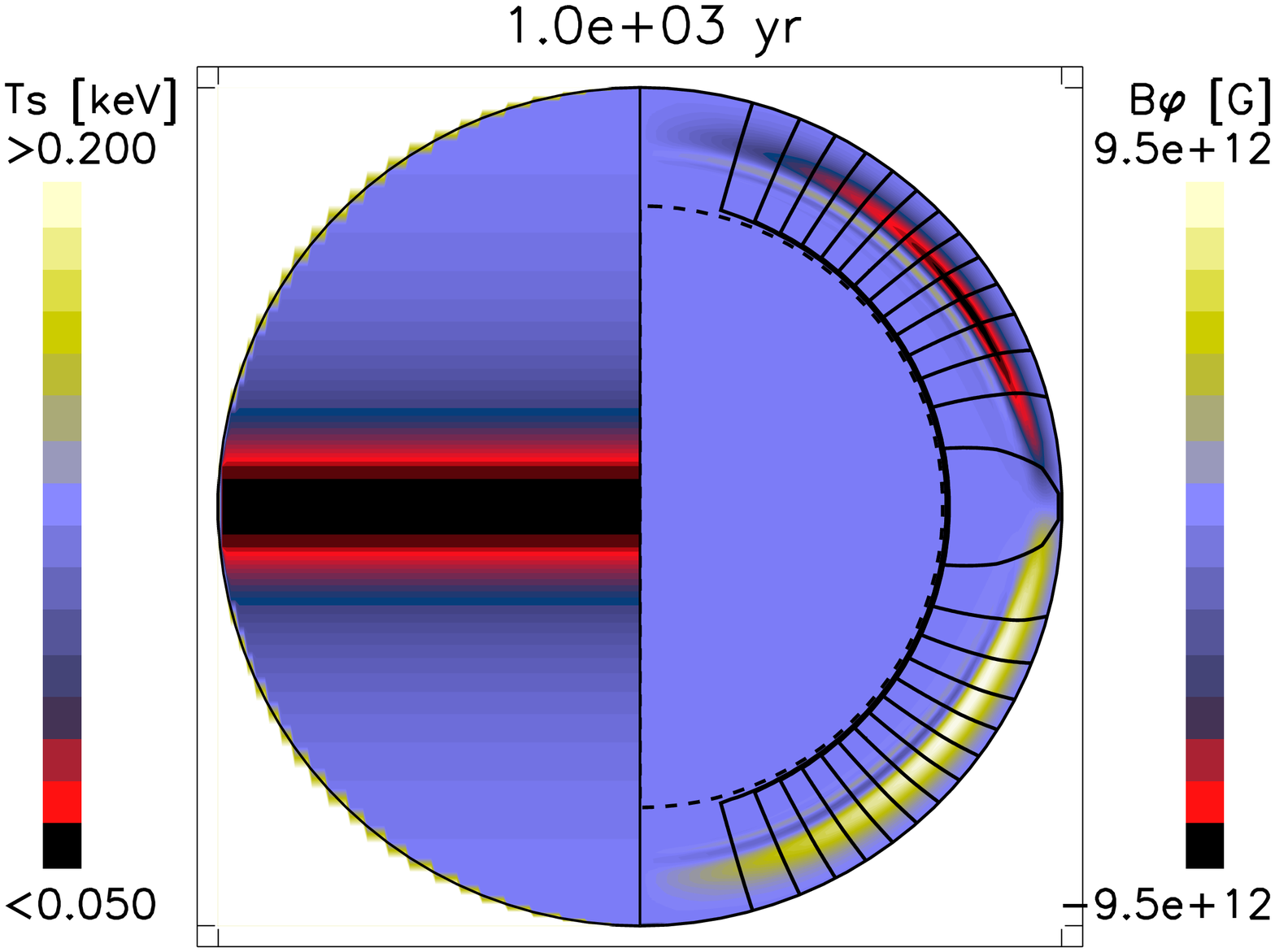}
\includegraphics[width=.32\textwidth]{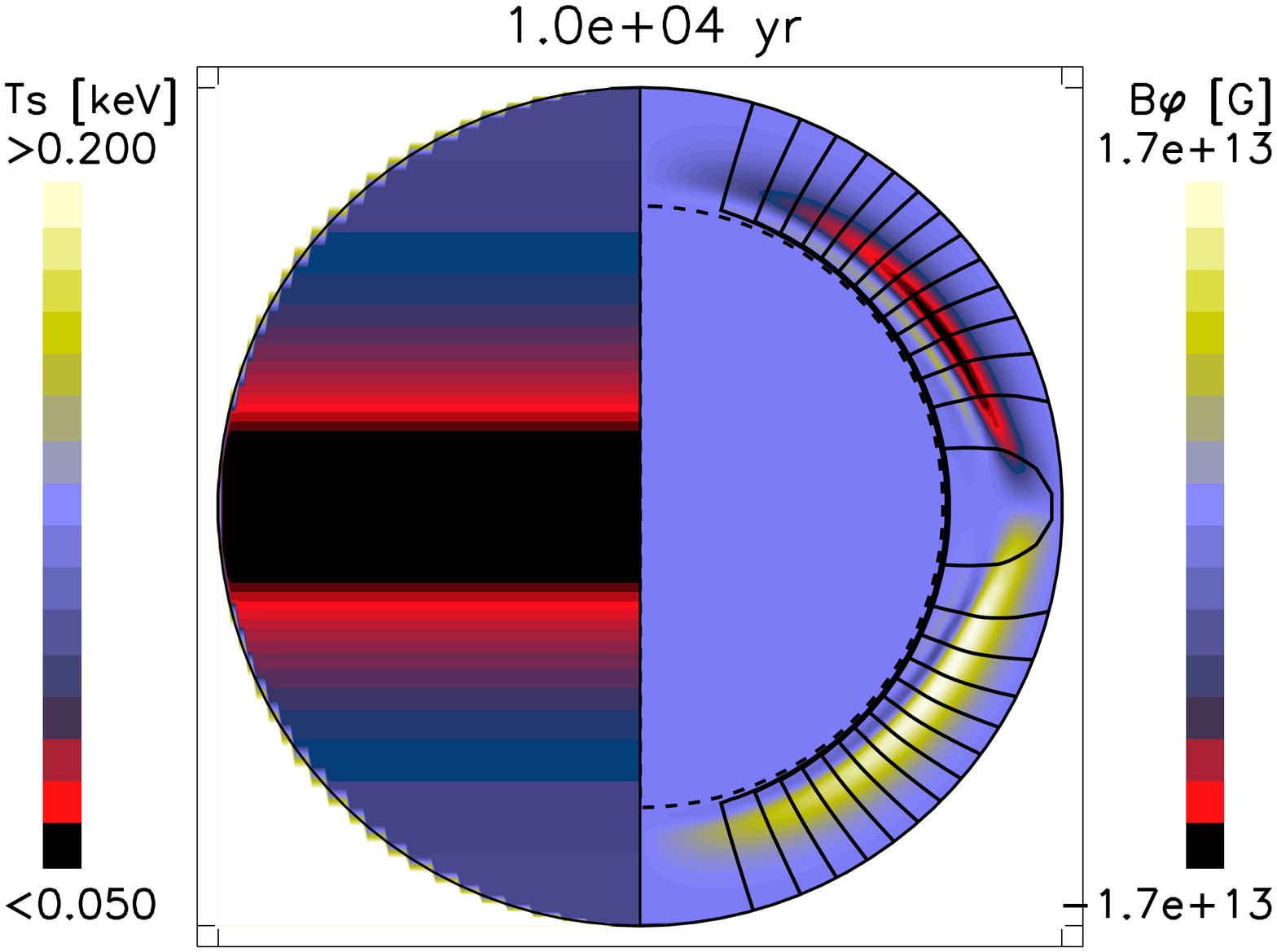}
\includegraphics[width=.32\textwidth]{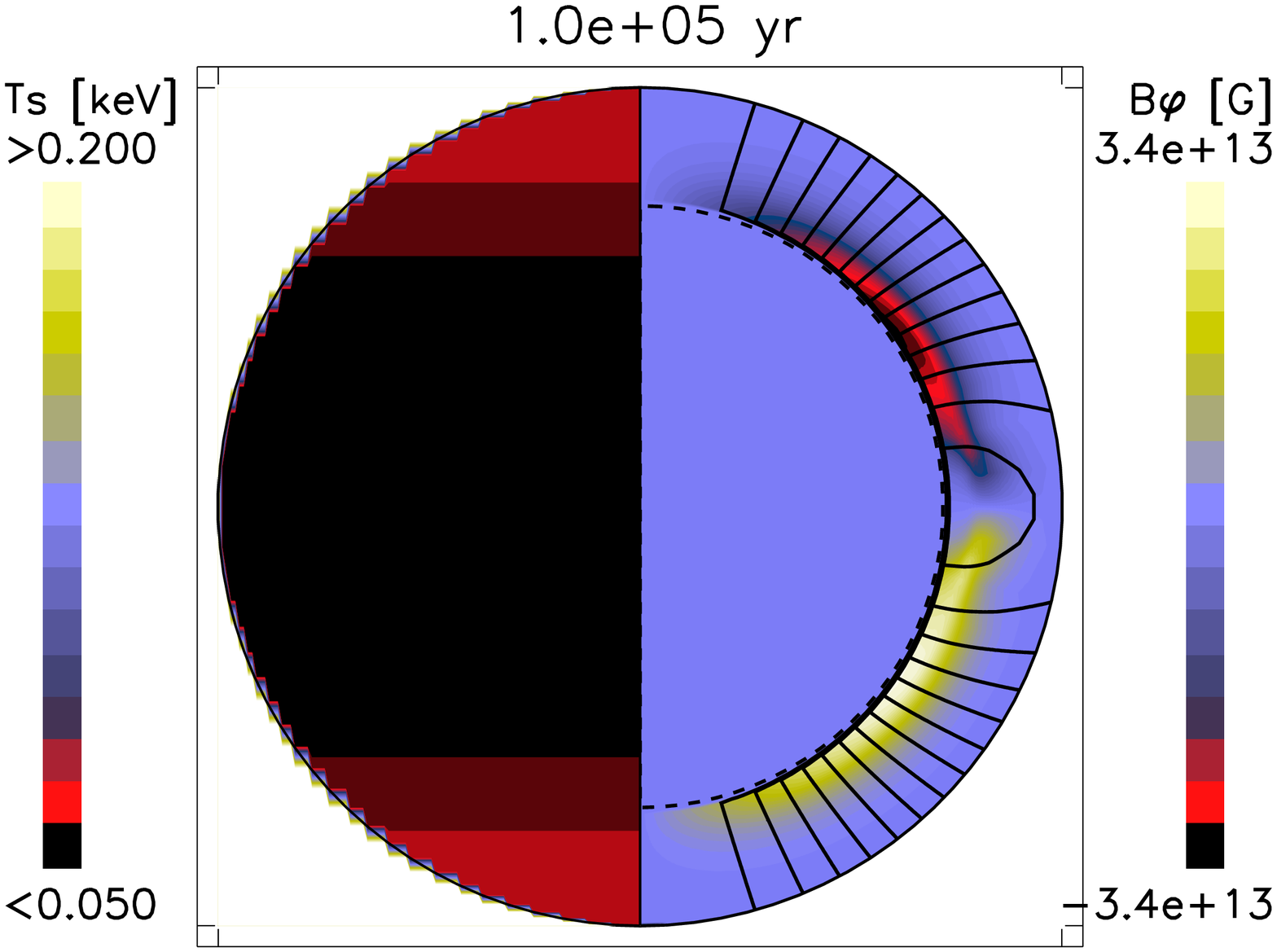}\\
\vskip0.5cm
\includegraphics[width=.32\textwidth]{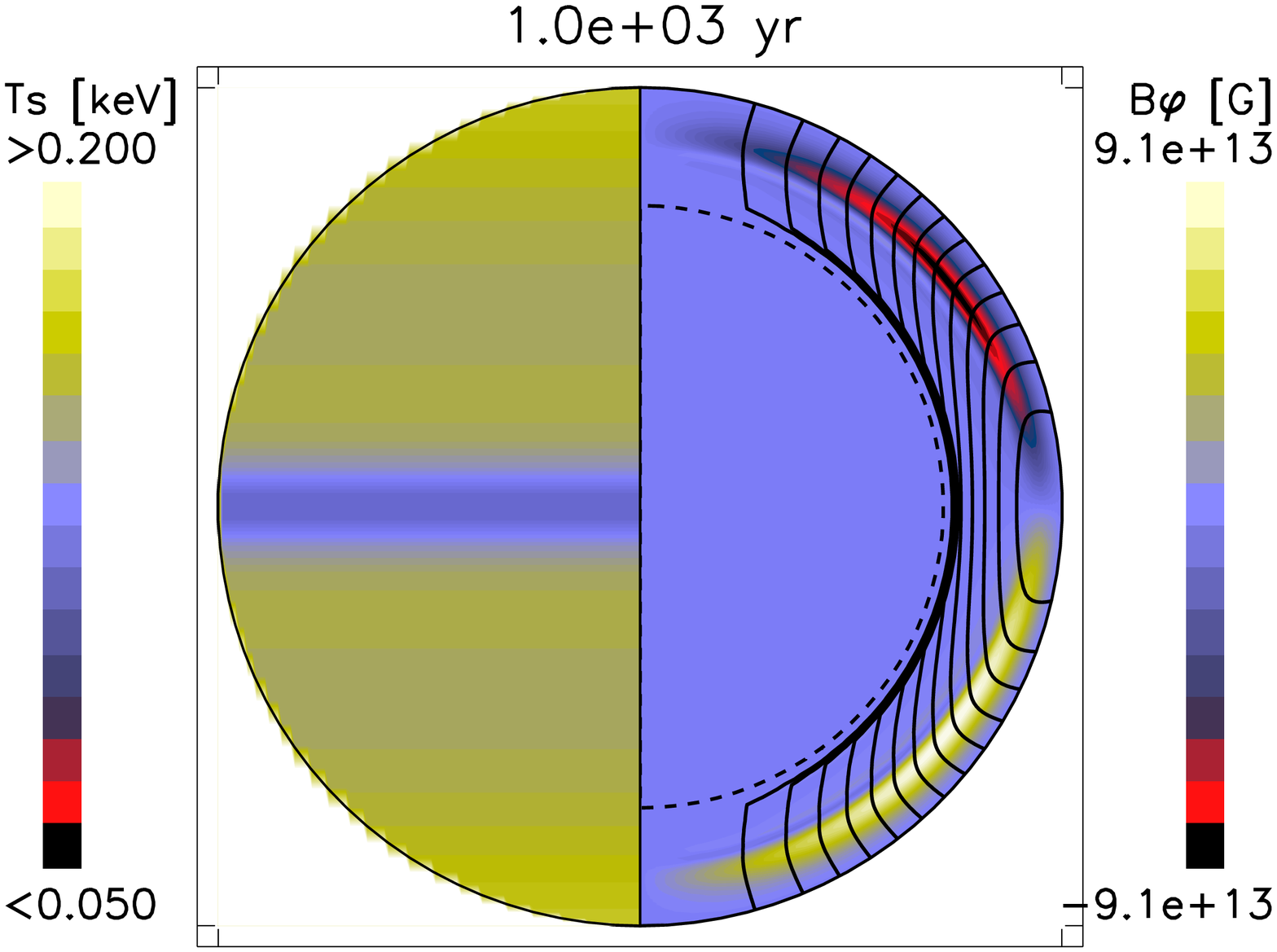}
\includegraphics[width=.32\textwidth]{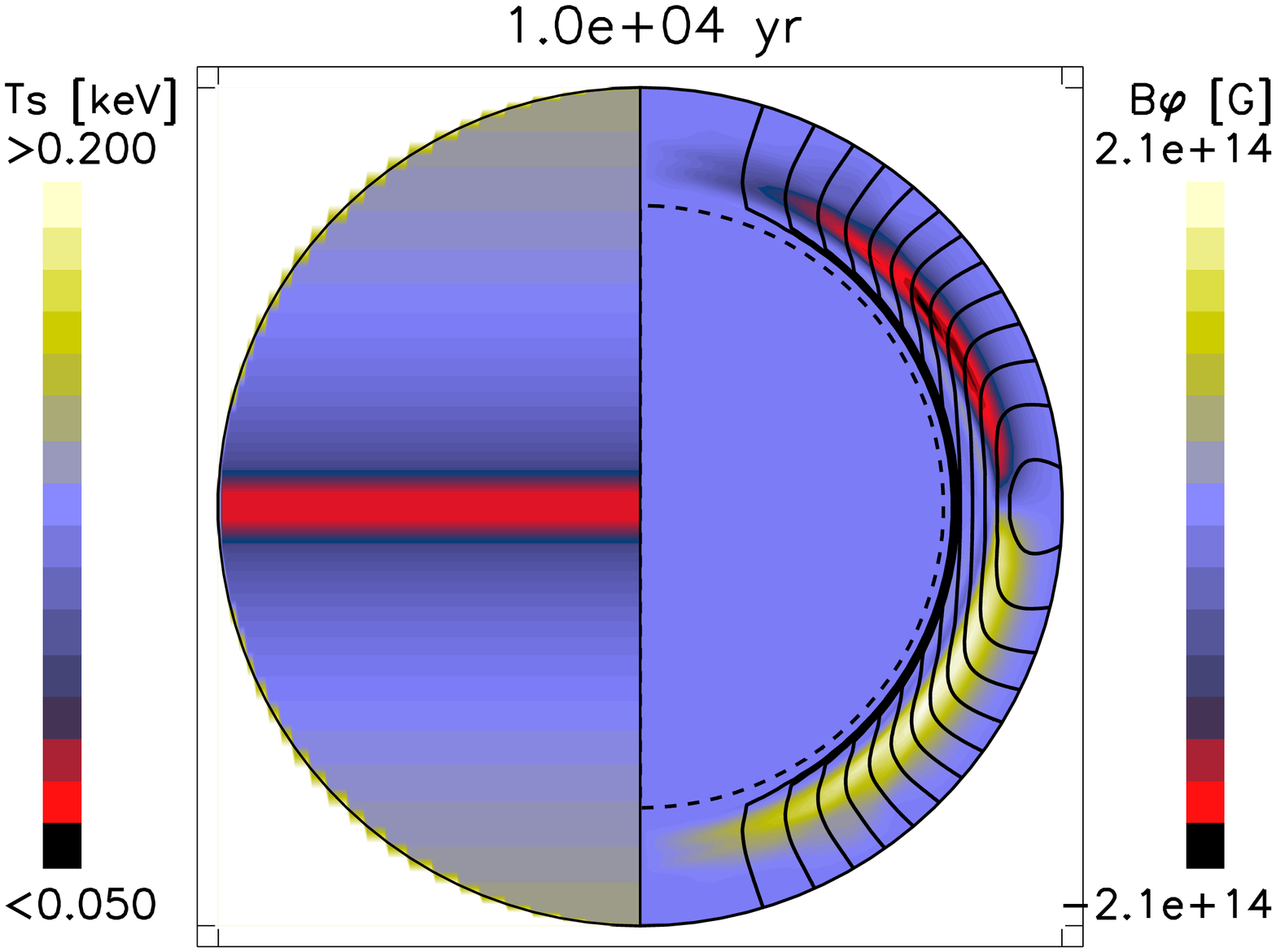}
\includegraphics[width=.32\textwidth]{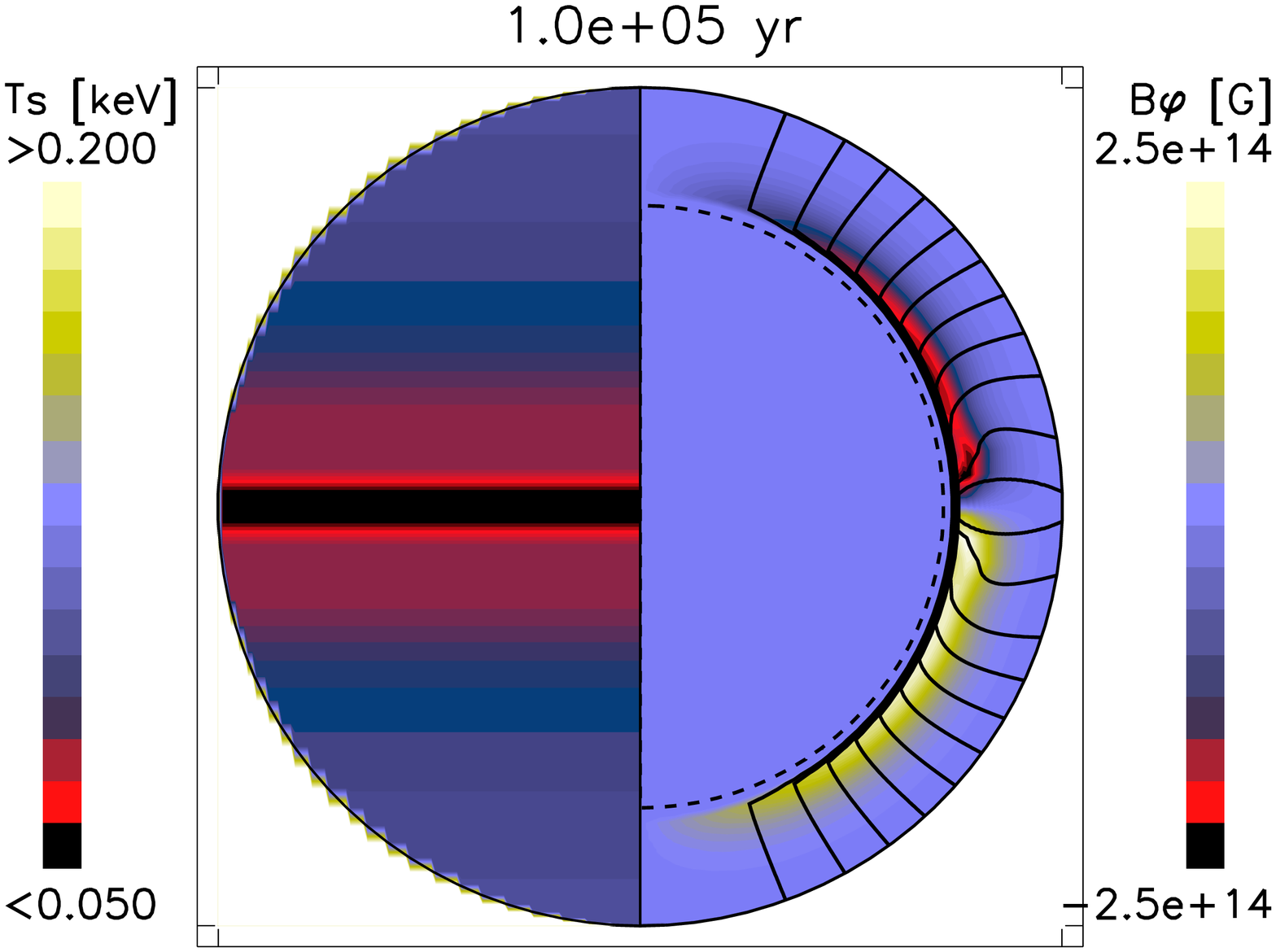}\\
\caption{Same as the upper panels of Fig.~\ref{fig:b14_evo}, but for models: A15 (first row), A14T (second), B14 (third) and C14 (fourth). In models B14 and C14, the magnetic field lines penetrate in the core, but they are not shown here.}
 \label{fig:other_evo}
\end{figure}
%%%%%%%%%%%%%%%%%%%%%%%%%%%%%%%%%%%%%%%%%%%%%%%%%%

%%%%%%%%%%%%%
\subsection{Dependence on initial field geometry and strength.}\label{sec:geo}

We continue the discussion by considering how the initial magnetic field configuration affects the evolution. In the top row of Fig.~\ref{fig:other_evo}, we present the results for the evolution of model A15.  The qualitative behaviour is the same as in model A14, but since the Hall time-scale goes as $\propto 1/B$, the dynamics is accelerated by a factor of ten. The more relevant role of the Hall term for stronger field leads to the formation of a discontinuity in $B_\varphi$.  In this particular case the current sheet is located at the equator (see the compressed lines and the ``colliding'' toroidal rings of opposite sign in the second panel), where the dissipation is strongly enhanced. Note that the average surface temperature is higher than for model A14 (Fig.~\ref{fig:b14_evo}).

The second row of Fig.~\ref{fig:other_evo} shows the evolution of model A14T. The magnetic energy initially stored in the toroidal component is $98\%$ of the total energy. Compared with model A14 (Fig.~\ref{fig:b14_evo}), there are significant differences. In particular, the symmetry with respect to the equator is broken because the vertical drift of the toroidal component acts towards the north pole. The initial poloidal magnetic field is distorted and bent at intermediate latitudes, with the formation of a small bundle of field lines, which has been proposed to be necessary for the radio pulsar activity \citep{geppert13}. This again locally increases the dissipation rate. At about $10^3$~yr, the northern hemisphere is, on average, warmer than the southern hemisphere, and a characteristic hot polar cap is observed. Later, the geometry of the magnetic field becomes more complicated, with the formation of localized bundles of nearly radial magnetic field lines. This results in a temperature pattern with hot and cool rings.

%%%%%%%%%%%%%%%%%%%%%%%%%%%%%%%%%%%%%%%%
\begin{figure}[t]
 \centering
\includegraphics[width=.47\textwidth]{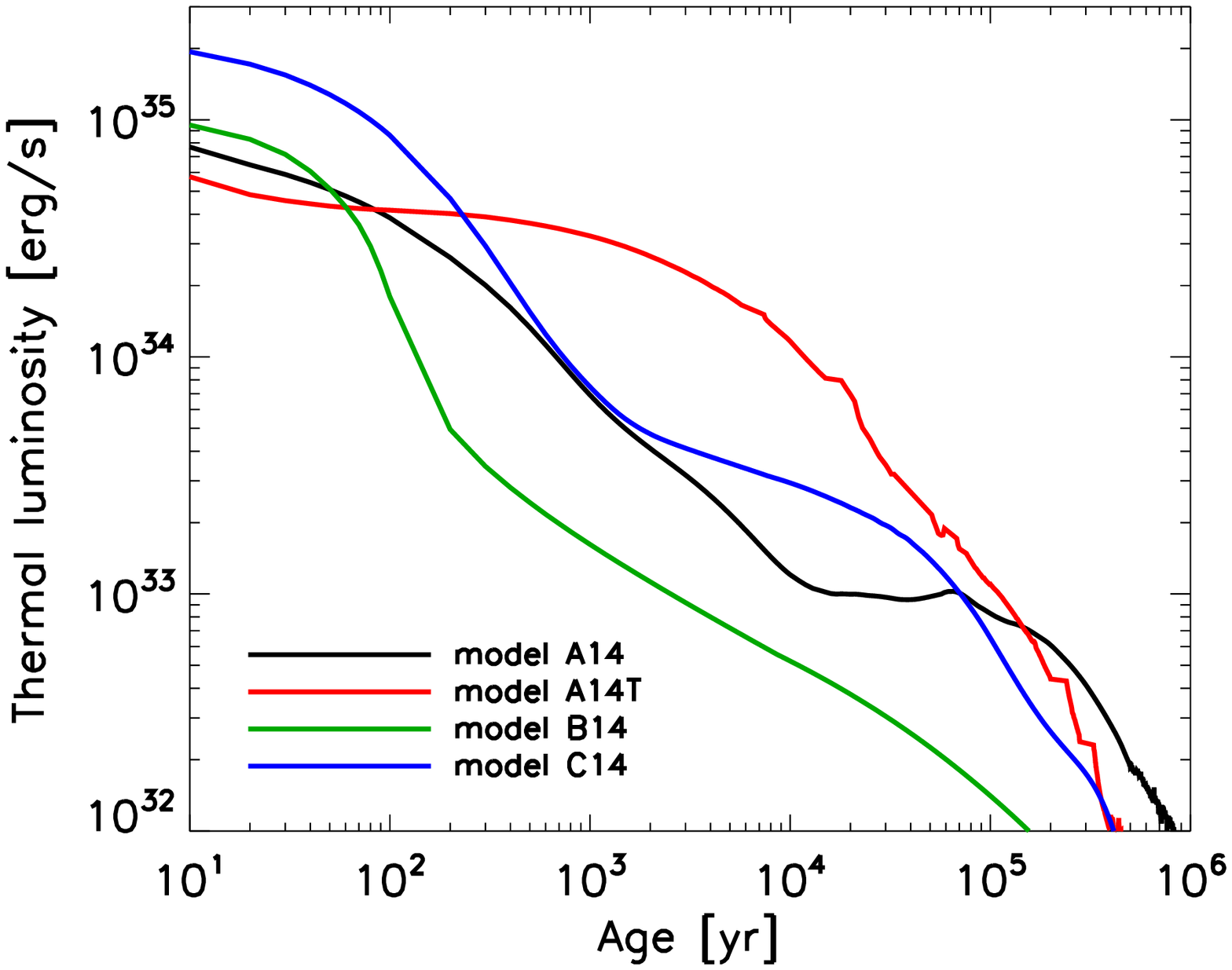}\\
\includegraphics[width=.47\textwidth]{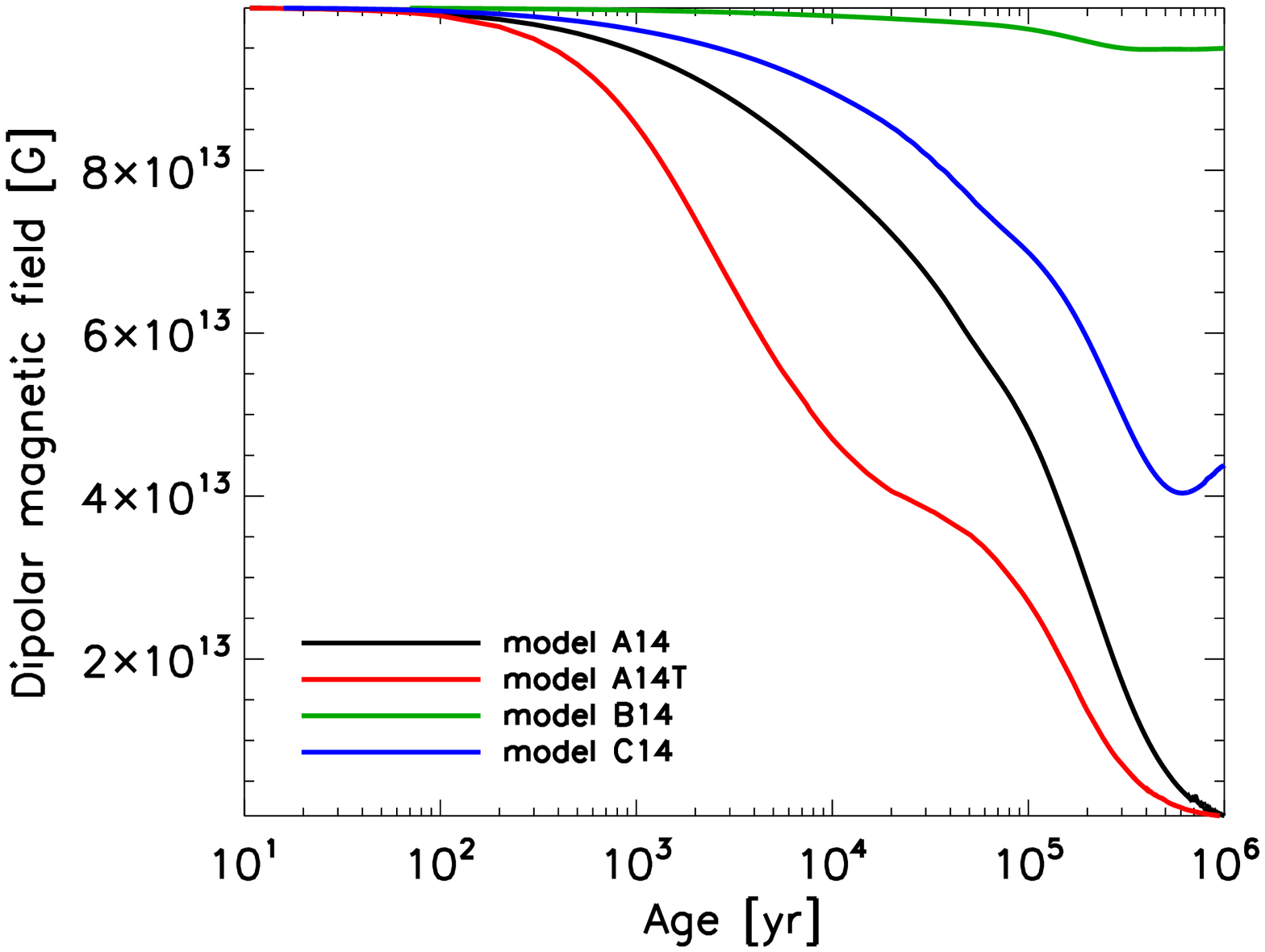}
\includegraphics[width=.47\textwidth]{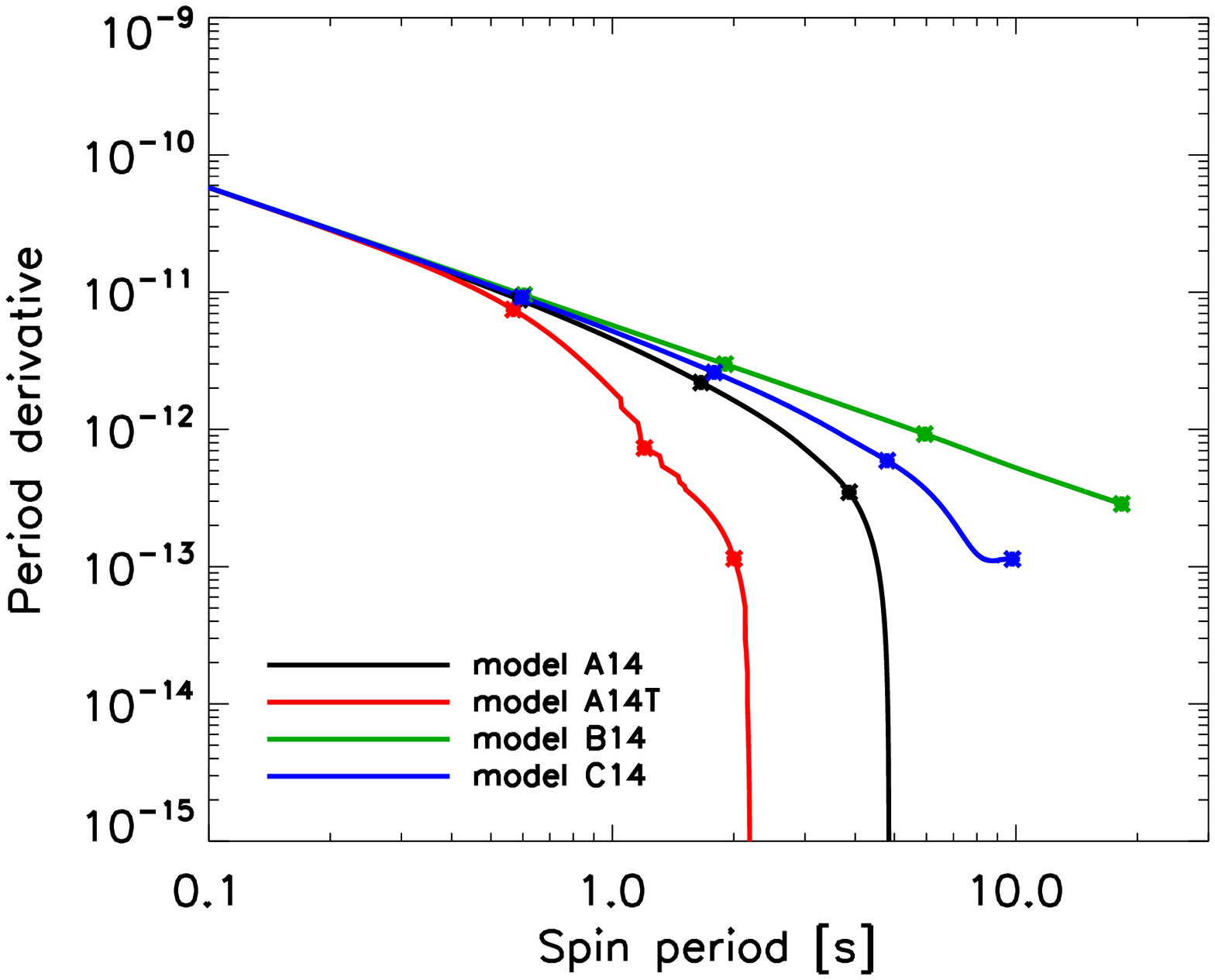}
\caption{Evolution of luminosity (top panel) and $B_p$ (bottom left), for different models: A14 (black), A14T (red), B14 (green) and C14 (blue). The bottom right panel shows the corresponding tracks in the $P$--$\dot{P}$ diagram, where, in each tracks, asterisks indicate the ages $t=10^3,10^4,10^5,10^6$ yr (from short to long periods).} 
 \label{fig:geometry}
\end{figure}
%%%%%%%%%%%%%%%%%%%%%%%%%%%%%%%%%%%%%%%%%%%

In the third row of Fig.~\ref{fig:other_evo}, the evolution of the core-extended configuration (model B14) is shown. The magnetic field lines are penetrating inside the core although, for clarity, the figure shows only the configuration in the crust (enlarged for visualization; see Fig. \ref{fig:initial_b} for the real scale). The field in the core is basically frozen due to the high electrical conductivity of the interior. Some weak Hall activity is developed at the bottom of the crust, but the maximum value of the toroidal magnetic field generated is about one order of magnitude weaker than in model A14, and the poloidal lines do not suffer any significant bending. In addition, the reduced heat deposition in the crust results in a much cooler surface compared to all the other models, and similar to the low field cooling models. 

Between the two extremes (mostly crustal currents versus mostly core currents), we also consider the intermediate case of model C14, shown in the bottom row of Fig.~\ref{fig:other_evo}. The presence of crustal currents activates the Hall dynamics and leads to a similar evolution as for model A14. The only relevant difference is that at very late times there will be a long-lasting magnetic field, supported by currents in the core.

In Fig.~\ref{fig:geometry} we compare the evolution of luminosity (top panel) and $B_p$ (bottom left) for four models, all with $B_p^0=10^{14}$ G. The difference between crustal-confined (models A14 and A14T) and the core-extended configuration (model B14) is evident. In the latter, the magnetic field is almost constant during 1 Myr. The configuration with an extremely strong initial toroidal magnetic field shows a larger luminosity (up to one order of magnitude at some age), due to the larger internal energy reservoir, but the higher temperatures also lead to a much faster decay of $B_p$ than for model A14. Moderate values of toroidal magnetic fields (e.g., with half of the magnetic energy stored in) result in a decay rate intermediate  between models A14 and A14T. In the hybrid case, model C14, $B_p$ the crustal currents dissipate before we can see oscillations around the value given by the currents circulating in the core. In the bottom panel of Fig.~\ref{fig:geometry}, we compare the luminosity for the same four models. Model B14 is similar to the non-magnetized case (Fig. \ref{fig:b0}), because most of the current is in the core and does not decay. On the other hand, when strong currents circulate in the crust, extra heat is deposited in the outer layers, resulting in higher luminosities.

The evolution of $B_p(t)$ allows us to calculate the evolution of the rotational properties by integrating eq.~(\ref{eq:ppdot_spindown}). In the bottom right panel of Fig.~\ref{fig:geometry} we show the corresponding evolutionary tracks in the $P$--$\dot{P}$ diagram, up to an age of 1 Myr (fixing $f_\chi=1.5$ in eq.~\ref{eq:k_spindown}), and an initial period $P_0=0.01$~s, low enough for the results not to be sensitive to the particular value of $P_0$. The difference in the magnetic field dissipation between models result in qualitative differences: in model C14, the track is straight, while for models A14 and A14T it shows a vertical drop. Comparing the timing properties at the same age (see asterisks), crust-confined models have much shorter periods.

One important general result is that the Hall term causes the self-regulation of the ratio of the toroidal and poloidal components. The differences due to particular initial ratios are much less important than in simulations including only the Ohmic term \citep{pons09}, except when the initial toroidal component is significantly larger than the poloidal magnetic field (like in model A14T).

%%%%%%%%%%%%%%%%%%%%%%%%%%%%%%%%%%%%%%%%
\begin{figure}
 \centering
\includegraphics[width=.6\textwidth]{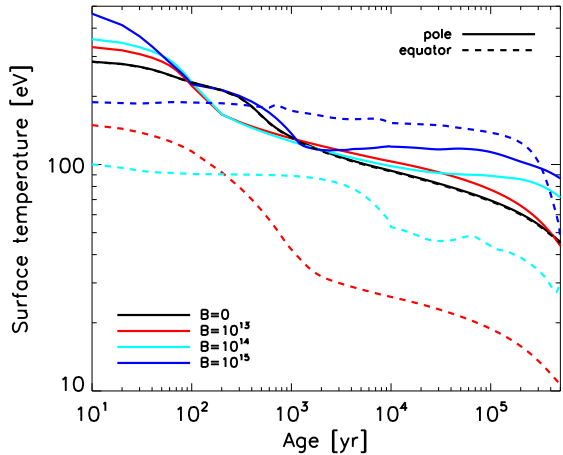}
\caption{Surface temperature at the pole (solid) and equator (dashed) for $B_p^0=0,10^{13},10^{14},10^{15}$ G (black, red, cyan and blue lines, respectively), in all cases for type A configurations.} 
 \label{fig:tevo_m140}
\end{figure}
%%%%%%%%%%%%%%%%%%%%%%%%%%%%%%%%%%%%%%%%

To conclude this subsection, we discuss the dependence of the surface temperature on the magnetic field strength. In Fig.~\ref{fig:tevo_m140}, we compare the surface temperatures at the pole and equator for $B_p^0=10^{13},10^{14},10^{15}$ G. At early times, little difference is seen in the evolution of the temperature at the pole, regardless of the magnetic field strength. For strong magnetic fields, the polar temperatures can be kept high for longer times. The largest differences, however, are seen in the equatorial temperatures. For low magnetic field models, the blanketing effect of the envelope leads to much cooler equatorial regions. Conversely, for the strongest magnetic field field model, the high energy deposition rate by field decay is able to compensate this effect and the equator is actually warmer than the pole after a few hundred years.

%%%%%%%%%
\subsection{Dependence on mass and relevant microphysical parameters.}\label{sec:magnetic_micro}

%%%%%%%%%%%%%%%%%%%%%%%%%%%%%%%%%%%%%%%%%%%%%%%
\begin{figure}[t]
 \centering
\includegraphics[width=.47\textwidth]{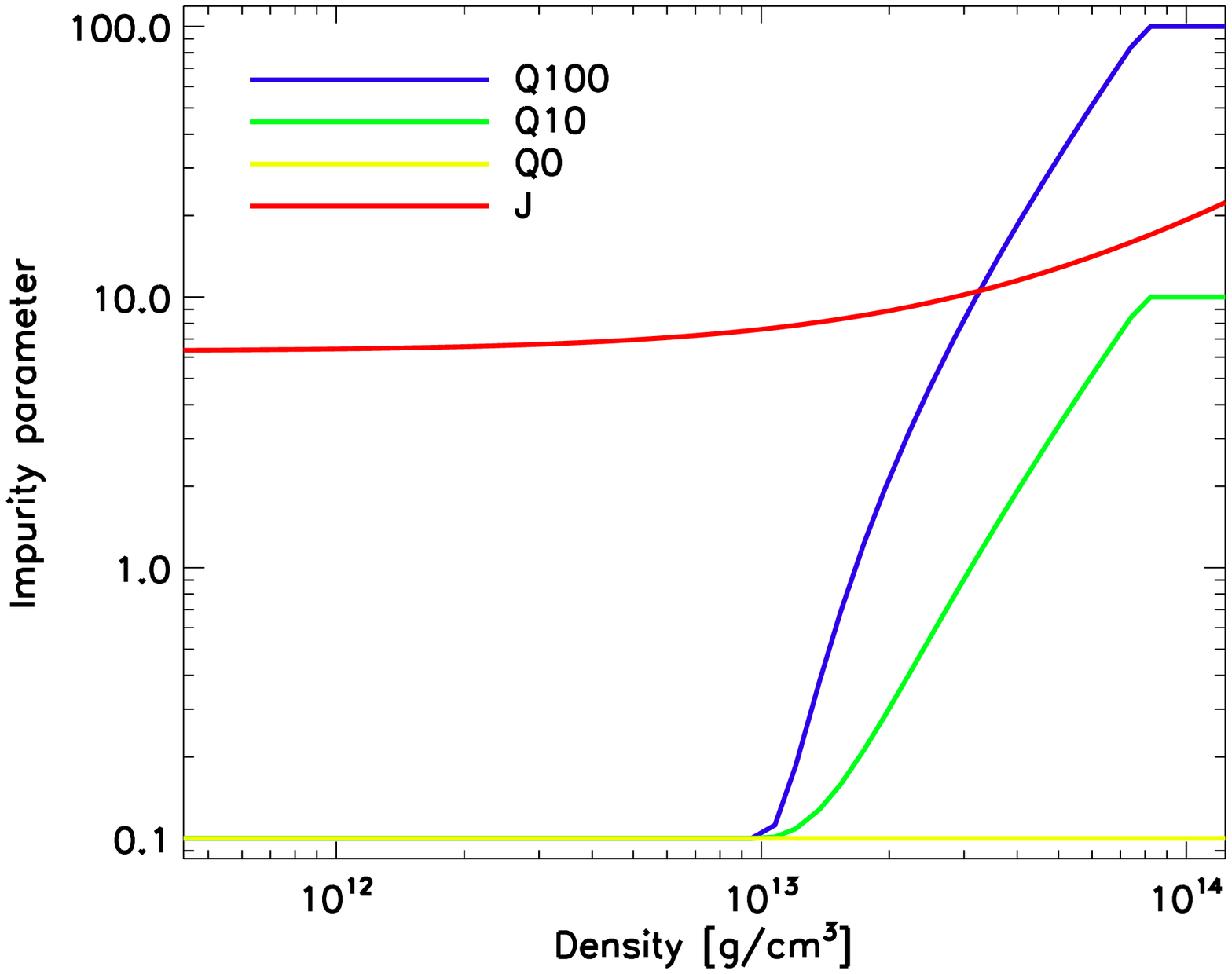}
\includegraphics[width=.47\textwidth]{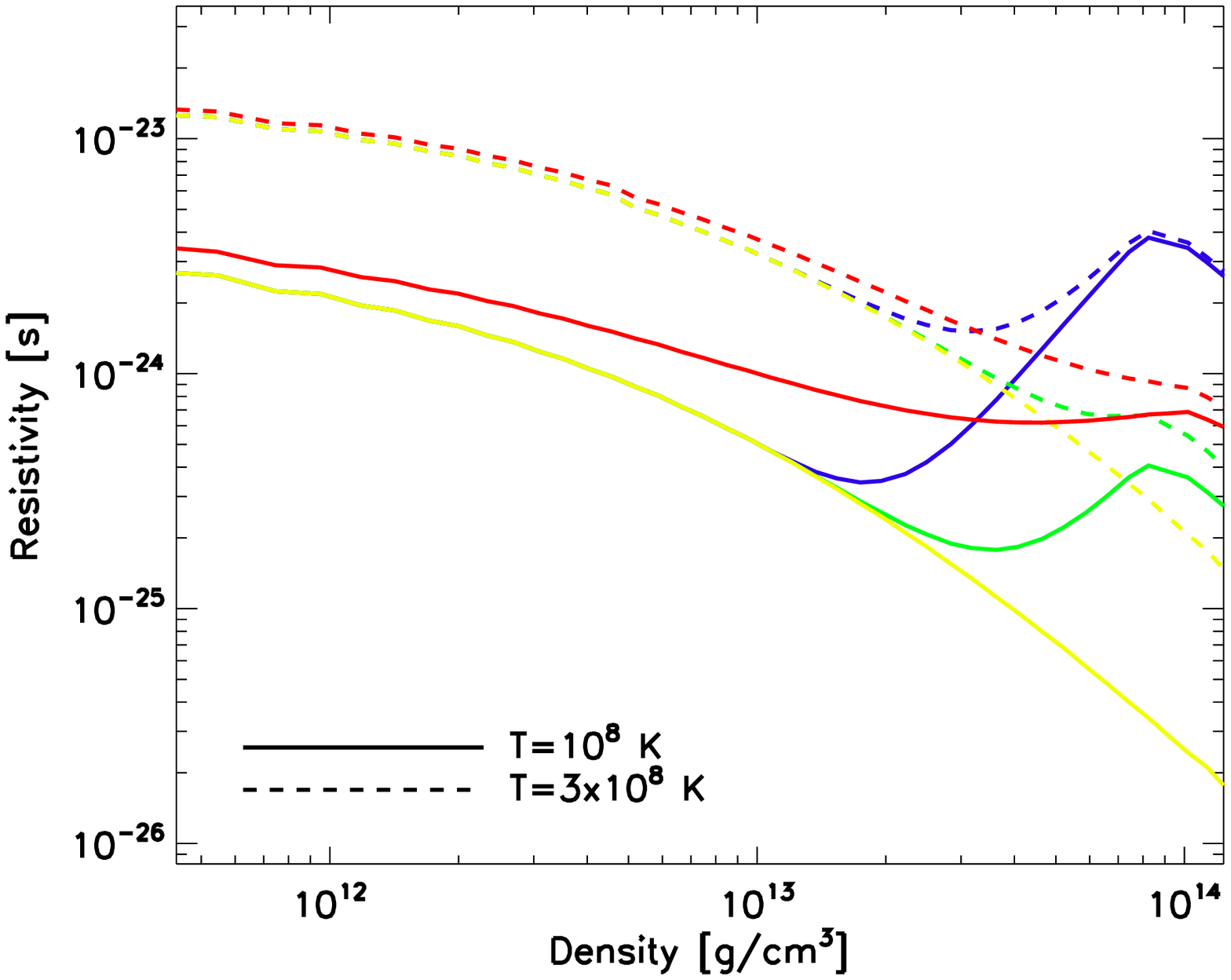}
\caption{Impurity parameter $Q_{imp}$ (left) and electrical resistivity (right) as a function of density for the four models with $M=1.4~M_\odot$. We plot the resistivity for two different temperatures to show more clearly the regions where the temperature-independent disorder resistivity dominates.}
 \label{fig:qimp}
\end{figure}
%%%%%%%%%%%%%%%%%%%%%%%%%%%%%%%%%%%%%%%%%%%%%%%

Now we briefly explore the dependence on the neutron star mass and some of the main microphysical parameters, such as the impurity content (see \S~\ref{sec:conductivity}). We use the same initial magnetic field (model A, with $B_p^0=3\times 10^{14}$ G), but with different neutron star masses, and different values of $Q_{imp}$. In particular, we consider, for $M=1.4~M_\odot$, two models (Q100, Q10) with a high $Q_{imp}$ only in the pasta region ($\rho > 10^{13}~$\gcc), but low $Q_{imp}$ in the rest of the inner crust, one model (Q0) with $Q_{imp}=0.1$ everywhere, and another model (J) corresponding to the extrapolation of the values calculated in \cite{jones04a,jones04b} at a few densities (a disordered crust). The profiles of $Q_{imp}$ as a function of density and the corresponding profiles of electrical resistivity (at two different temperatures) are shown in Fig.~\ref{fig:qimp}. When the resistivity is dominated by the electron-impurity scattering processes (large $Q_{imp}$, blue lines at high densities), there is very little dependence on temperature.

\begin{figure}[t]
 \centering
\includegraphics[width=.47\textwidth]{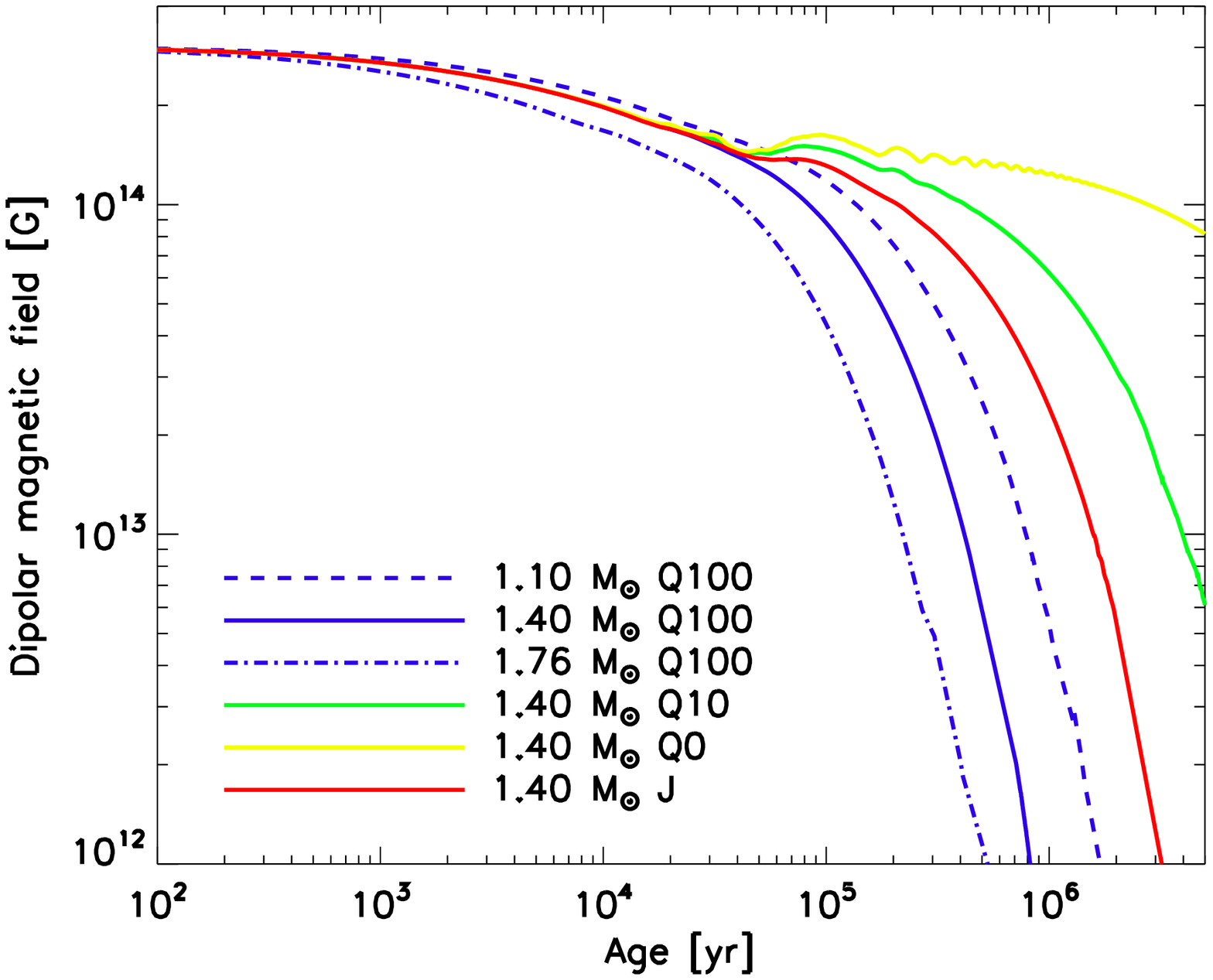}
\includegraphics[width=.47\textwidth]{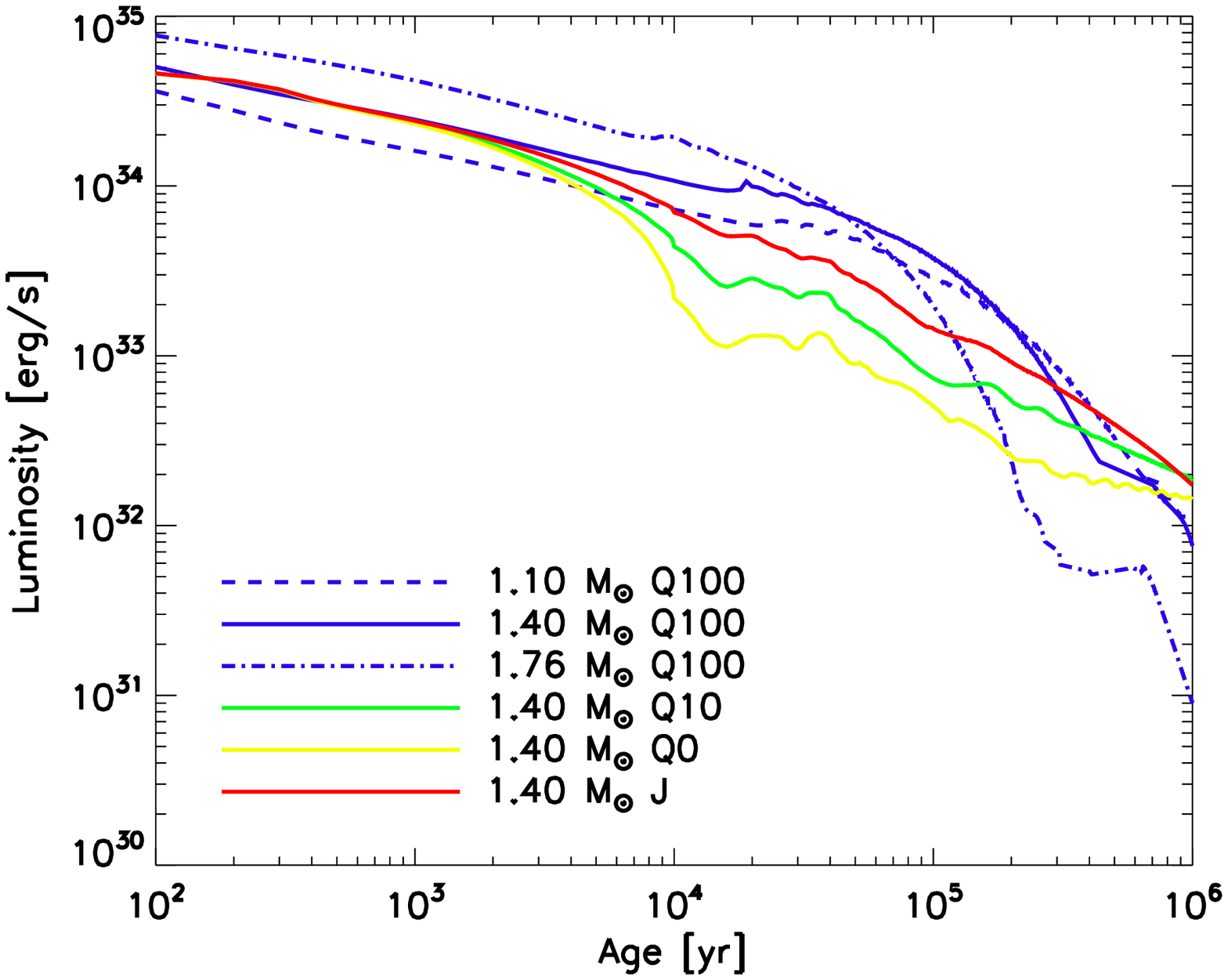}\\
\includegraphics[width=.47\textwidth]{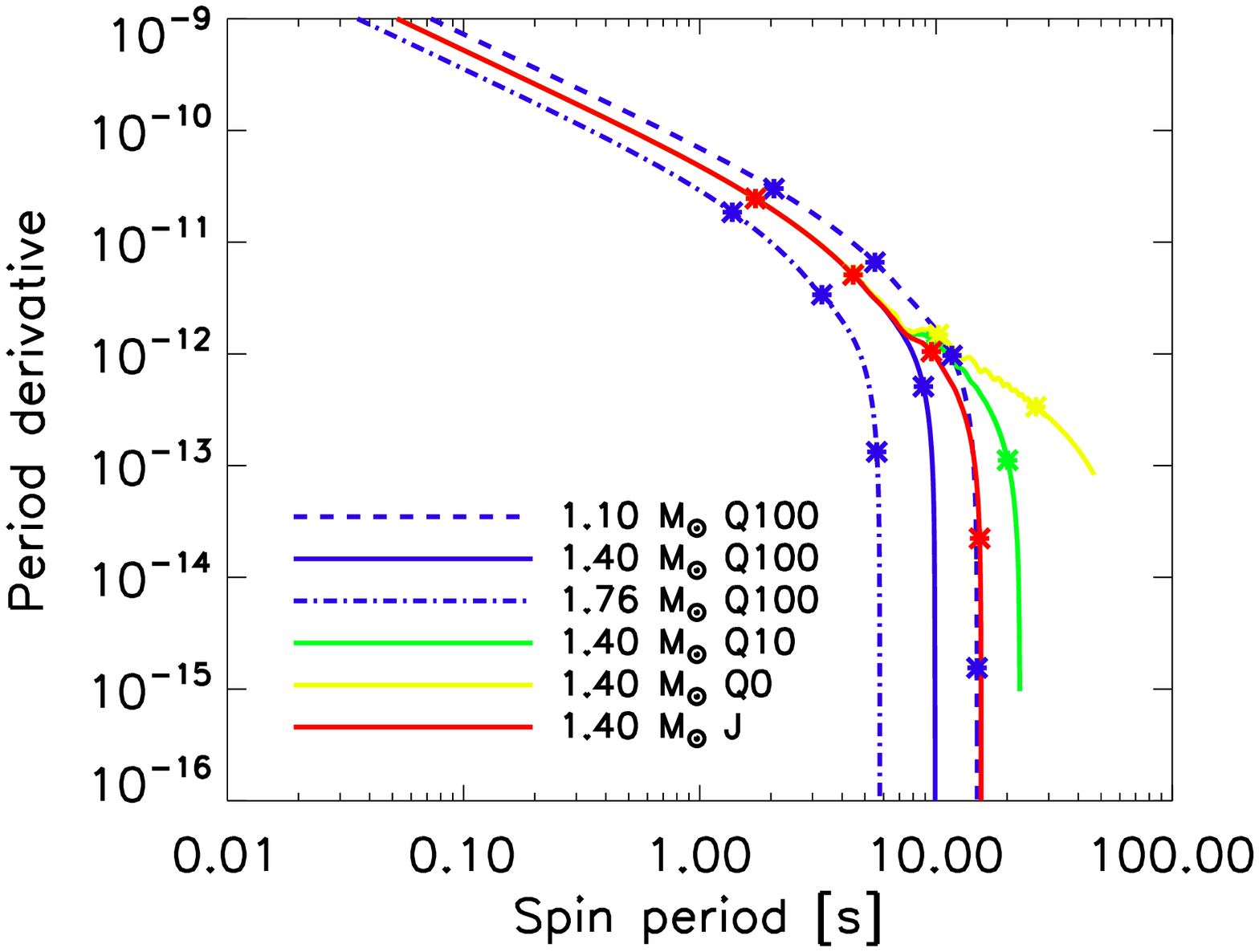}
\includegraphics[width=.47\textwidth]{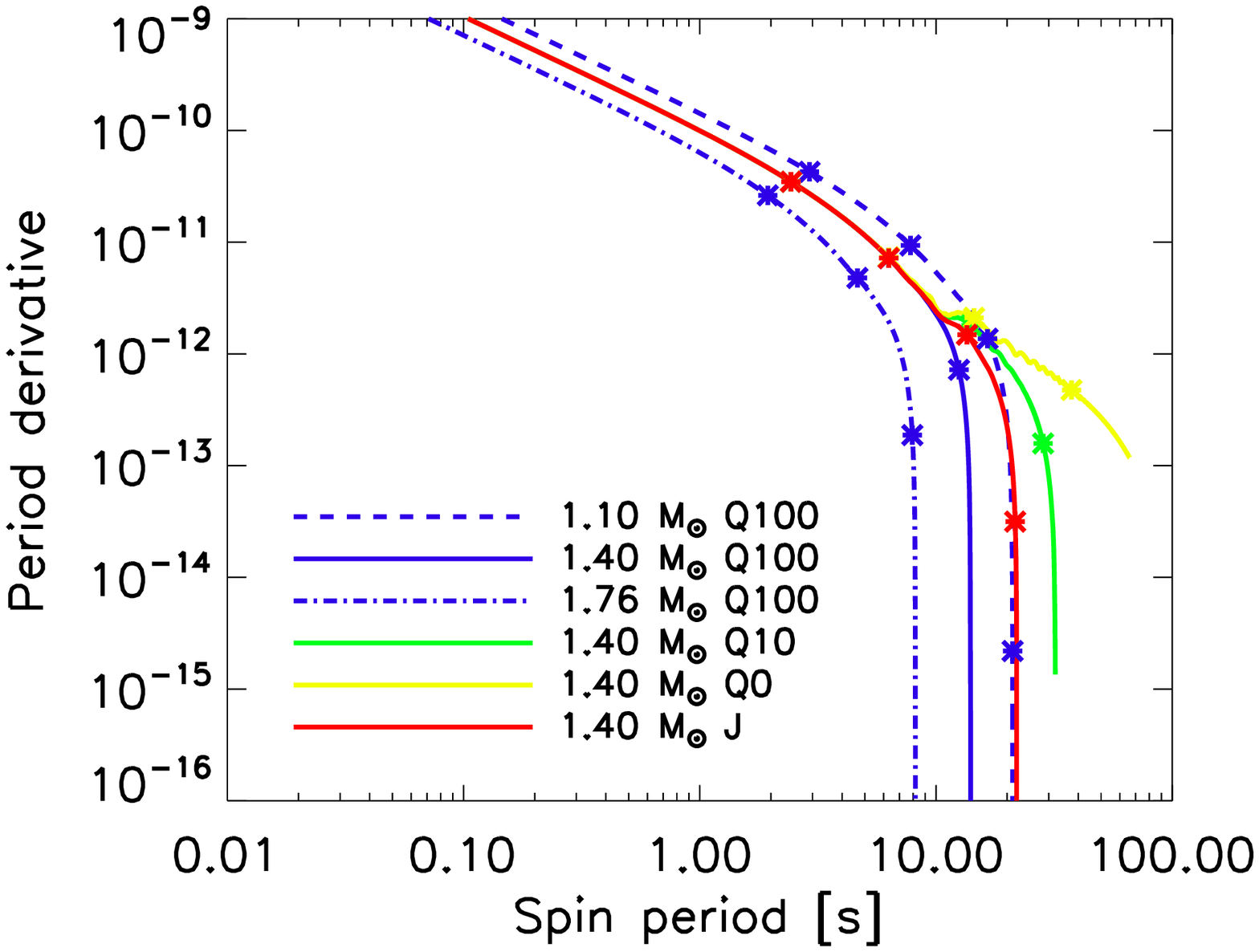}
\caption{Evolution of a model A with $B_p^0=3\times10^{14}$ G, but varying the neutron star mass and the impurity parameter as indicated in Fig.~\ref{fig:qimp}. We show the evolution of $B_p$ (top left panel) and luminosity (top right) during $5$ Myr, with the corresponding evolutionary tracks in the $P$-$\dot{P}$ diagram, with $f_\chi=1.5$ (aligned force-free rotator, bottom left), and $f_\chi=3$ (orthogonal force-free rotator, bottom right). Asterisks indicate the ages $t=10^3,10^4,10^5,10^6$ yr.}
 \label{fig:micro}
\end{figure}

In the top left panel of Fig.~\ref{fig:micro} we show the evolution of $B_p$ for the different models. During the first $\sim 30$ kyr, the magnetic field is dissipated by a factor of $\sim 2$ for all of the models. In this regime, the neutron star crust is still warm and electron scattering off impurities is not the dominant process, at least in the regions where currents are circulating. Thereafter, as the star cools and the currents are dragged inward, the evolution strongly depends on the impurity content of the inner crust. For low values of $Q_{imp}$ (model Q0), the magnetic field almost stops to dissipate and remains high, with some oscillations due to the Hall term. In contrast, a large value of $Q_{imp}$ (model Q100, blue lines) results in the dissipation of the magnetic field by one or even two orders of magnitude between 0.1 and 1 Myr. Models Q10 and J, both with moderate values of $Q_{imp}$, show an intermediate dissipation rate. Another consequence of a low impurity parameter is that, since the magnetic field dissipation time-scale is much longer than the star age, a similar asymptotic value of the magnetic field is reached for all neutron stars born as magnetars \citep{pons09}. On the contrary, in models with a high $Q_{imp}$ in the pasta region, the magnetic field dissipation is maintained for millions of years. The small quantitative differences between models with different masses (three blue lines) are caused by the different thickness of the crust (see Table~\ref{tab:nsmasses}).

%%%%%%%%%%%%%%%%%%%%%%%%%%%%%%%%%%%%%%%%%%%%%%%
%%%%%%%%%%%%%%%%%%%%%%%%%%%%%%%%%%%%%%%%%%%%%%%

Another important effect is that, because of the extra heat provided by the dissipation of magnetic energy, the dependence of the luminosity on the mass is strongly reduced, compared with non magnetized neutron star cooling models. The cooling curves for all models are similar during at least $10^4~$yr. We also note that varying the superfluid gaps has a visible effect only when the effect of Joule heating is negligible, i.e., for low values of $Q_{imp}$ in the pasta region or weak magnetic fields. Furthermore, if the resistivity is dominated by the electron-impurity scattering (large $Q_{imp}$), the resistivity is temperature-independent and the effect of varying the superfluid gap (or other parameters that may modify the local temperature) is negligible. 

In the bottom panels of Fig.~\ref{fig:micro}, we show the evolutionary tracks in the $P$--$\dot{P}$ diagram (integrating, as usual, eq.~\ref{eq:ppdot_spindown}, with $P_0=0.01$ s), up to an age of 5 Myr. We show the results for $f_\chi=1.5$ in eq.~(\ref{eq:k_spindown}) (bottom left) and $f_\chi=3$ (right panel). For models with high $Q_{imp}$, tracks bend downwards due to the enhanced dissipation of the magnetic field under the combined action of the Hall effect with the large resistivity in the innermost part of the crust. Therefore, an asymptotic value of $P$ is reached. The age at which this vertical drop begins depends on the initial field, being roughly $\sim 10^5$ yr for strong fields ($B_p^0\gtrsim 10^{14}$ G), and up to several Myr for weaker fields. The specific magnitude of the asymptotic value of $P$ depends on the initial magnetic field, the exact value of $Q_{imp}$ in the innermost part of the crust, the mass of the star (see the torque factors $C_{sd}$ in Table~\ref{tab:nsmasses}) and the value of $f_\chi$.

In \S~\ref{sec:period_clustering}, we will discuss the comparison with the observed distribution of periods, and how this constrains the properties of the inner crust. We will use the model Q100 and a neutron star mass of 1.4 $M_\odot$ for the rest of the section.

%%%%%%%%%%%%%%
\subsection{Expected magnetar outburst rates.}\label{sec:bursts}

As the magnetic field evolves, the lattice experiences Lorentz forces, expressed by the magnetic stress tensor, having components:

\begin{equation}
 M_{ij}(r,\theta)= \frac{B_i(r,\theta)B_j(r,\theta)}{4\pi}~,
\end{equation}
where $i,j=r,\theta,\varphi$. The lattice responds with an elastic stress $\sigma_b$, which, at equilibrium, compensates $M_{ij}$. However, during the evolution, local magnetic stresses can occasionally become too strong to be balanced by the elastic restoring forces of the crust, which hence breaks, and the extra stored magnetic/elastic energy becomes available for powering the observed bursts and flares \citep{thompson95,thompson96}. The maximum sustainable stress has been estimated with analytical arguments and through molecular dynamics simulations. In particular, \cite{chugunov10} obtained the fit 
\begin{equation}
\sigma_b^{\rm max} = \left(0.0195-\frac{1.27}{\Gamma_{coul}-71}\right) n_i \frac{Z^2 e^2}{a_i}~,
\label{eq:sigma}
\end{equation}
where $\Gamma_{coul}$ is the Coulomb parameter defined in eq.~(\ref{eq:coulomb_parameter}). This allowed \cite{perna11} to estimate the frequency and the location (depth and latitude) of {\em starquakes}, i.e. neutron star crust failures, without performing any simulation of the fast (seconds) fracture processes. The released energies were also estimated from the stored elastic energy.

Their work showed that the classification of objects as AXPs, or SGRs, or high-$B$ neutron stars, or standard radio pulsars, etc. does not correspond to an underlying intrinsic physical difference: outbursts can occur also in objects outside of the traditional magnetar range, albeit with a lower probability. The follow-up study by \cite{pons11} further highlighted the importance that the toroidal magnetic field has on the neutron star observed phenomenology: two objects with similar inferred dipolar $B$-field (as measured by $P$ and $\dot{P}$) can display a very different behaviour depending on the strength of the (unmeasured) toroidal magnetic field. The stronger the latter, the higher the luminosity of the object, and the more likely {\em starquakes} are to occur. In order to follow numerically the long-term evolution of neutron stars, these works used an approximation to treat the Hall term, which was only included when studying the short term evolution of the magnetic field. In the current work, the above approximation has been released, and the effect of the Hall term is fully implemented. Hence, in the following we update their results by using a similar formalism to compute the outburst statistics from the results of our last simulations.

%%%%%%%%%%%%%%%%%%%%%%%%%%%%%%%%%%%%
\begin{figure}
 \centering
\includegraphics[width=.6\textwidth]{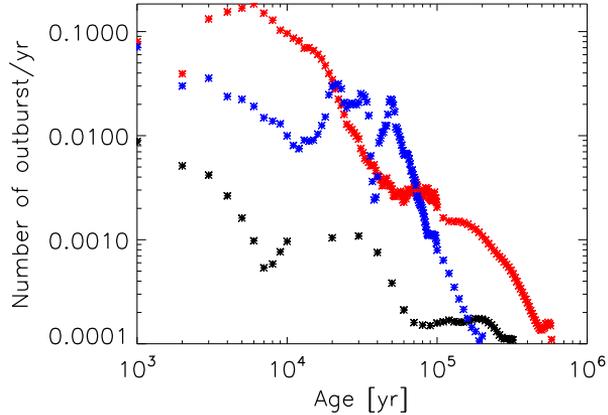}
\caption{Expected outburst rate for model A with $B_p^0=3\times 10^{14}$ G (black), $B_p^0=10^{15}$ G (red) and $B_p^0=10^{14}$ G with initial strong toroidal magnetic field $B_t^0=5\times 10^{15}$ G (blue).} 
 \label{fig:outburst}
\end{figure}
%%%%%%%%%%%%%%%%%%%%%%%%%%%%%%%%%%%%

In Fig.~\ref{fig:outburst} we show the evolution of the predicted outburst rate for models A15 (red), a model of type A with $B_p^0=3\times 10^{14}$ (black), and the model A14T (blue). While young neutron stars may undergo energetic outbursts every few years (for model A15) or tens of years (for the model with a weaker magnetic field), middle age sources are expected to show transient activity only every few thousands of years, and with less energy involved. Other models with initial lower magnetic fields, like model A14, do not give an appreciable event rate (less than $10^{-3}$ events per year even for young neutron stars). However, in presence of strong toroidal magnetic field, model A14T, the outburst rate is strongly enhanced.

Assuming a neutron star birth rate of $10^{-2}$ per year, there must be $\approx 10^4$ neutron stars in the Galaxy with ages $\lesssim 1$ Myr. If 10\% of them are born with $B\gtrsim 10^{14}$ G, a naive extrapolation of the estimated event rate at that age ($\lesssim 10^{-4}$ per year) leads to an outburst rate by old magnetars of one every few years. Therefore, we expect that more and more objects of this class will be discovered in the upcoming years. It is interesting to note that such old magnetars are expected to be detected only via their outburst activation, being too faint in the quiescent state even for very deep X-ray surveys. Since the launch of the {\em Swift} satellites in 2004, our capability of detecting and monitoring Galactic transients has largely increased, and 5 new magnetars have been discovered in the past years, two of those being relatively old and low magnetic field magnetars \citep{rea10,rea13,rea12,scholz12}.

On the other hand, the observational detection of an outburst event depends on the quiescent flux. The brightest sources, with $L\gtrsim 10^{35}$ erg/s, will experience a barely detectable flux enhancement, while for dimmer magnetars ($L\sim 10^{33}-10^{34}$ erg/s) the flux increase by 2 or 3 orders of magnitude during the outburst and are much easier to detect \citep{pons12b}.

\chapter{Unification of the isolated neutron stars diversity}\label{ch:unification}

Isolated neutron stars have been divided into different observational classes for historical reasons. However, a unifying vision considers them as different manifestations of the same underlying physics \citep{kaspi10}. In this context, one of the main theoretical tasks is to explain the varied phenomenology of their X-ray emission. X-ray spectra carry precious information about the surface temperature and the physics driving the cooling of the neutron stars. The detected X-ray flux, if accompanied by a reliable distance measurement, leads to an estimate of the bolometric luminosity, which can be confronted with neutron star cooling models to infer properties of dense matter in the neutron star interior. In addition, timing properties provide us with information about the rotational energy loss, which is believed to be regulated mainly by the dipolar component of the external magnetic field. The aim of this chapter is to contribute to unify the isolated neutron star population by establishing the evolutionary links between different sub-classes.

\section{The neutron star zoo.}

\subsection{Rotation-powered pulsars.}

Rotation-powered pulsars (RPPs) constitute the bulk of the detected neutron stars, with more than 2000 detected sources. They are usually seen in radio, and their emission is powered by the conversion of a fraction of their rotational energy into non-thermal radiation. Their rotation periods are in the range 1.3 ms--8.5 s, with inferred magnetic fields ranging from $\sim 10^8$ G to $\sim 10^{14}$ G. Pulsars with periods of the order of milliseconds are thought to be recycled in binary systems. They have undergone accretion phases during which they have spun up. We will not discuss them, since we are interested in isolated neutron stars.

A handful of RPPs constitutes the sub-class of the {\it high-B pulsars} \citep{ng11}. They have inferred magnetic fields $\sim 10^{13}$--$10^{14}$ G. At least one case has shown X-ray bursts (PSR~1846--0258, \citealt{gavriil08}). These properties point to high-B pulsars as the connection between RPPs and magnetars. Rotating radio transients (RRATs; \citealt{lyne09}) are a minority ($\sim$ 12) of RPPs showing sporadic and brief radio bursts, besides usual timing irregularities. They are thought to be a peculiar manifestation of RPPs, with extreme modulation of radio emission.

More than 100 radio-loud RPPs are also detected in X-rays \citep{becker09}, including one RRAT and several high-B pulsars. Most of them show a power-law component, powered by the rotation; however, in a few cases, they also show thermal emission. Among X-ray pulsars, several nearby, old ones ($\tau_c\gg~$Myr) show X-ray emission \citep{zavlin04}, which may be related to the particle bombardment of the polar cap by returning magnetospheric currents.

\subsection{Magnetars.}

Up to now, more than twenty X-ray pulsating sources are identified as {\it magnetars} (MAG, \citealt{duncan92,thompson95,thompson96}): young ($\sim 10^2$--$10^5$ yr) neutron stars with very strong magnetic fields $B \sim 10^{14}$--$10^{15}$~G, slow rotation ($P\sim 2$--$12~$s), and whose large X-ray quiescent luminosity cannot usually be explained in terms of rotational energy (see \citealt{mereghetti08} for an observational review), at variance with rotation-powered pulsars (RPPs). Historically, they are classified as Anomalous X-ray Pulsars (AXPs) or Soft Gamma Repeaters (SGRs). However, as more and better data have become available in the last decade, the separation between these two traditional classes has become thinner, and they are no longer considered different objects.  Magnetars show sporadic bursts in the X-ray and $\gamma$-ray bands, with episodes of outbursts lasting months to years \citep{rea11}. The decay and evolution of magnetic field is the dominant energy sources, and crustal deformations produced by the huge internal magnetic stresses are the likely cause of bursts and episodic giant flares.

\subsection{Nearby X-ray isolated neutron stars.}

The radio-quiet X-ray isolated neutron stars (XINSs), also known as {\it the magnificent seven} (see \citealt{haberl07,turolla09,kaplan09b} for recent reviews) are relatively old, nearby neutron stars, with the cleanest detected thermal emission and a relatively large magnetic field and long periods (3-11 s). Their proximity ($d< 1$ kpc) makes them the most promising candidates to study neutron star thermal spectra. In fact, the detected radiation is supposed to come only from residual cooling, with no magnetospheric contamination. Given their long periods, it is likely that their viewing angle is off their narrow radio beam.

\subsection{Central compact objects.}\label{sec:cco}

The handful of reported central compact objects (CCOs) forms a class of X-ray sources (see \citealt{gotthelf13} for a recent review), located close to the center of $\sim$ kyr old supernova remnants (SNRs). CCOs are supposed to be young, isolated, radio-quiet neutron stars. They show very stable, thermal-like spectra, with hints of temperature anisotropies, in the form of large pulsed fraction, or small emitting regions characterized by hot ($0.2$--$0.4$ keV) blackbody components. The period and period derivative are known for only three cases.

Applying the classical dipole-braking formula gives an estimate for the dipolar component of the \textit{external} magnetic field of $B_p\sim 10^{10}$--$10^{11}$~G. This implies that their spin period is basically the same as the natal one. For this reason, their characteristic ages are much longer than their real ages (see eq.~\ref{eq:treal_chage}), estimated by the study of the associated SNRs.

%%%%%%%%%%
\section{Data on cooling neutron stars.}\label{sec:observations}

To confront theoretical cooling models with observational data, we need to know simultaneously the age and some quantity related to the thermal emission from the neutron star surface (luminosity or temperature). The number of sources for which both measures are available is limited, and in most cases subject to large uncertainties. In this section, we will discuss in detail the sample of selected sources, which includes the following objects:

\noindent
$\bullet$ 4 CCO candidates, including the three with measured timing properties, and the very young neutron star in SNR Cassiopeia A. We have ignored the other CCOs candidates since they have spectral information with poor statistics and/or a very uncertain age of the associated SNR. \\

\noindent
$\bullet$ 8 RPPs, including the Vela pulsar and the so-called three Musketeers (PSR~B0656+14, PSR~B1055--52 and the $\gamma$-ray-loud, radio-quiet Geminga; \citealt{deluca05}). We have excluded most of the young pulsars, many of which are associated with pulsar wind nebulae, since in those cases data are compatible with non-thermal emission powered by the rotational energy loss, which is orders of magnitude larger than their X-ray luminosity (i.e. Crab pulsar and RX J0007.0+7303 in SNR CTA1; \citealt{caraveo10}). We also exclude several old pulsars due to the unclear nature of their X-ray emission (see above).\\

\noindent
$\bullet$ 7 XINSs. All of them have good spectra, and in most cases well known timing properties and good distance determinations (often with direct parallax measurements). \\

\noindent
$\bullet$ 4 high-B pulsars with good quality spectra. We have included the only RRAT detected so far in X-ray (PSR~J1819--1458, \citealt{mclaughlin07}). We have excluded the magnetar-like pulsar PSR~J1846--0258 because during quiescence its X-ray emission does not show a significant thermal component \citep{ng08,livingstone11b}, and it is orders of magnitude smaller than its rotational energy loss.\\

\noindent
$\bullet$ 17 magnetars with good spectra {\it in quiescence}, i.e., not during transient outbursts. Among the five most recently discovered magnetars with measured timing properties, we have included the last available observations after the outburst decay of Swift J1822.3--1606 \citep{rea12,scholz12} and SGR~0418+5729 \citep{rea10,rea13}, which are supposedly close to quiescence. Instead, we have excluded SGR~1833--0832 \citep{esposito11}, Swift J1834.9--0846 \citep{kargaltsev12} and SGR~1745--2900 \citep{mori13}, all discovered during an outburst and which have not yet been detected in quiescence.\\

We now discuss separately the timing and spectral properties of our sample. All the data presented in the following subsections, with links to abundant references, can also be found in our website.\footnote{\bf\tt http://www.neutronstarcooling.info/} We plan to update and extend periodically this website.

%%%%%%%%%%%%%%%%%%%%%%%%%%%%%%%%%%
\subsection{Timing properties and age estimates.}

%%%%%%%%%%%%%%%%%  TABLE 1  %%%%%%%%%%%%%%%%%%%%%%%%%%%%%
\setlength{\tabcolsep}{4pt}
\begin{table}[t]
\tiny
\begin{center}
\centering
\begin{tabular}{l l c c c c c c c l}
\hline
\hline
Source & assoc./nick & Class & $P$ & $\log(\dot{P})$ & $\log(\dot{E}_{rot})$ & $\log(B_p)$ & $\log(\tau_c)$ & $\log(\tau_k)$ & method\\
& & & [s] & & [erg/s] & [G] &  [yr] &  [yr] &\\
\hline
 CXOU J185238.6+004020 &      SNR         Kes79 &  CCO &  0.105 &  -17.1 &   32.5 &   10.8 &    8.3 &    3.7--3.9 &      SNR\\
        1E 1207.4-5209 &      SNR   G296.5+10.0 &  CCO &  0.424 &  -16.7 &   31.1 &   11.3 &    8.5 &    3.4--4.3 &      SNR\\
         RX J0822-4300 &      SNR       PuppisA &  CCO &  0.112 &  -17.0 &   32.4 &   10.8 &    8.3 &    3.5--3.6 &      SNR\\
  CXO J232327.9+584842 &      SNR          CasA &  CCO &     -  &     -  &     -  &     -  &     -  &        2.5 &H/SNR\\
        PSR J0538+2817 &      SNR          S147 &  RPP &  0.143 &  -14.4 &   34.7 &   12.2 &    5.8 &  $\sim$4.6 &      SNR\\
          PSR B1055-52 &                        &  RPP &  0.197 &  -14.2 &   34.5 &   12.3 &    5.7 &         -  &          -\\
        PSR J0633+1746 &      aka       Geminga &  RPP &  0.237 &  -14.0 &   34.5 &   12.5 &    5.5 &         -  &          -\\
          PSR B1706-44 &                        &  RPP &  0.102 &  -13.0 &   36.5 &   12.8 &    4.2 &         -  &          -\\
          PSR B0833-45 &      SNR          Vela &  RPP &  0.089 &  -12.9 &   36.8 &   12.8 &    4.1 &    3.7--4.2 &      SNR\\
          PSR B0656+14 &      SNR       Monogem &  RPP &  0.385 &  -13.3 &   34.6 &   13.0 &    5.0 &  $\sim$4.9 &      SNR\\
          PSR B2334+61 &      SNR    G114.3+0.3 &  RPP &  0.495 &  -12.7 &   34.8 &   13.3 &    4.6 &  $\sim$4.0 &      SNR\\
        PSR J1740+1000 &                        &  RPP &  0.154 &  -11.7 &   37.4 &   13.6 &    3.1 &         -  &          -\\
        PSR J0726-2612 &                        &   HB &  3.440 &  -12.5 &   32.4 &   13.8 &    5.3 &         -  &          -\\
        PSR J1119-6127 &      SNR    G292.2-0.5 &   HB &  0.408 &  -11.4 &   36.4 &   13.9 &    3.2 &    3.6--3.9 &      SNR\\
        PSR J1819-1458 &                   RRAT &   HB &  4.263 &  -12.2 &   32.5 &   14.0 &    5.1 &         -  &          -\\
        PSR J1718-3718 &                        &   HB &  3.378 &  -11.8 &   33.2 &   14.2 &    4.5 &         -  &          -\\
       RX J0420.0-5022 &                        & XINS &  3.450 &  -13.6 &   31.4 &   13.3 &    6.3 &         -  &          -\\
       RX J1856.5-3754 &                        & XINS &  7.055 &  -13.5 &   30.5 &   13.5 &    6.6 &    5.5--5.7 &  PM\\
       RX J2143.0+0654 &      aka       RBS1774 & XINS &  9.428 &  -13.4 &   30.3 &   13.6 &    6.6 &         -  &          -\\
       RX J0720.4-3125 &                        & XINS &  8.391 &  -13.2 &   30.7 &   13.7 &    6.3 &    5.8--6.0 &  PM\\
       RX J0806.4-4123 &                        & XINS & 11.370 &  -13.3 &   30.2 &   13.7 &    6.5 &         -  &          -\\
       RX J1308.6+2127 &      aka       RBS1223 & XINS & 10.310 &  -13.0 &   30.6 &   13.8 &    6.2 &    5.9--6.1 &  PM\\
       RX J1605.3+3249 &                        & XINS &     -  &     -  &     -  &     -  &     -  &  $\sim$5.7 &  PM\\
           1E 2259+586 &      SNR        CTB109 &  MAG &  6.979 &  -12.3 &   31.7 &   14.1 &    5.4 &    4.0--4.3 &      SNR\\
           4U 0142+614 &                        &  MAG &  8.689 &  -11.7 &   32.1 &   14.4 &    4.8 &         -  &          -\\
  CXO J164710.2-455216$^*$ &  cluster           Wd1 &  MAG & 10.611 &  -12.0 &   31.5 &   14.3 &    5.2 &         -  &          -\\
         XTE J1810-197$^*$ &                        &  MAG &  5.540 &  -11.1 &   33.3 &   14.6 &    4.1 &         -  &          -\\
        1E 1547.0-5408$^*$ &      SNR  G327.24-0.13 &  MAG &  2.072 &  -10.3 &   35.3 &   14.8 &    2.8 &         -  &          -\\
        1E 1048.1-5937$^*$ &                        &  MAG &  6.458 &  -10.6 &   33.5 &   14.9 &    3.7 &         -  &          -\\
    CXOU J010043.1-721 &                    SMC &  MAG &  8.020 &  -10.7 &   33.2 &   14.9 &    3.8 &         -  &          -\\
 1RXS J170849.0-400910 &                        &  MAG & 11.003 &  -10.7 &   32.7 &   15.0 &    4.0 &         -  &          -\\
 CXOU J171405.7-381031$^*$ &      SNR        CTB37B &  MAG &  3.825 &  -10.2 &   34.6 &   15.0 &    3.0 &  $\sim$3.7 &      SNR\\
           1E 1841-045 &      SNR         Kes73 &  MAG & 11.782 &  -10.4 &   33.0 &   15.1 &    3.7 &    2.7--3.0 &      SNR\\
         SGR 0501+4516 &      SNR           HB9 &  MAG &  5.762 &  -11.2 &   33.1 &   14.6 &    4.2 &  $\sim$ 4  &          -\\
           SGR 1627-41 &      SNR    G337.0-0.1 &  MAG &  2.595 &  -10.7 &   34.6 &   14.7 &    3.3 &  $\sim$3.7 &      SNR\\
           SGR 0526-66 &      SNR      N49(LMC) &  MAG &  8.054 &  -10.4 &   33.5 &   15.0 &    3.5 &  $\sim$3.7 &      SNR\\
           SGR 1900+14$^*$ &  cluster               &  MAG &  5.200 &  -10.0 &   34.4 &   15.1 &    3.0 &    3.6--3.9 &  PM\\
           SGR 1806-20$^*$ &  cluster       W31 &  MAG &  7.602 &   -9.6 &   34.4 &   15.5 &    2.6 &    2.8--3.0 &  PM\\
         SGR 0418+5729 &                        &  MAG &  9.078 &  -14.4 &   29.3 &   13.1 &    7.6 &         -  &          -\\
    Swift J1822.3-1606$^*$ &                        &  MAG &  8.438 &  -13.1 &   30.7 &   13.7 &    6.2 &         -  &          -\\
\hline
\hline
\end{tabular}
\end{center}
\caption{Timing properties and age estimates of our sample of isolated neutron stars. Here $\dot{E}_{rot}$ and $B_p$ are the rotational energy losses and the magnetic field strength at the pole, as given by eqs.~(\ref{eq:erot}) and (\ref{eq:inferred_bpole}) (with the widely used values $R_\star=10$ km, $I=10^{45}$ erg s$^{-1}$, $f_\chi=1$), and assuming that rotational energy losses are dominated by dipolar magnetic torques. Sources with multiple/variable $\dot{P}$ values in the literature are labelled with an asterisk ($^*$). The method used to estimate the real age has been indicated in the last column, as found in literature: SNR kinematic study (SNR), proper motion (PM), or historical records (H). For references, see our online catalogue$^1$, the ATNF pulsar catalogue$^2$ \citep{manchester05}, and the McGill magnetar catalogue$^3$.
In the text, we denote individual sources by short names or nicknames.}
\label{tab:timing}
\end{table} 
%%%%%%%%%%%%%%%%%%%%%%%%%%%%%%%%%%%%%%%%%%

If both the spin period and the period derivative of the source are known, the characteristic age $\tau_c=P/2 \dot{P}$ can be used as an approximation to the real age, with which it coincides only if the initial period was much shorter than the current value and the pre-factor $KB_p^2$, entering in the magnetic torque, eq.~(\ref{eq:ppdot_spindown}), has been constant during the entire neutron star life (see \S~\ref{sec:chage}). Unfortunately, this is not the most common situation, and usually, for middle-aged and old objects, $\tau_c$ is found to be larger than the real age, when the latter has been obtained by other methods. When the object is located in a SNR, a kinematic age can also be inferred by studying the expansion of the nebula (see \citealt{allen04} for a review with particular attention to the magnetar associations). For a few other nearby sources, e.g., some XINSs and few magnetars, the proper motion, with an association to a birth place, can give an alternative estimate of the age \citep{tetzlaff11,tendulkar12}. We have collected the most updated and/or reliable available information on timing properties and kinematic age from the literature, the ATNF catalogue\footnote{\tt http://www.atnf.csiro.au/people/pulsar/psrcat/} \citep{manchester05}, and the McGill online magnetar catalogue\footnote{\tt http://www.physics.mcgill.ca/$\sim$pulsar/magnetar/main.html}. We present in Table~\ref{tab:timing} all the sources in our sample, with their well-established associations \citep{gaensler01}, the known timing properties with the characteristic age and the alternative estimate for the age, when available. The inferred value of the surface, dipolar magnetic field at the pole $B_p$, assuming the standard dipole-braking formula, is shown as well.

We note that for RPPs timing properties are stable over a time-span of tens of years, and $P$ and $\dot{P}$ are precisely measured, but for magnetars the timing noise is much larger. In some cases, different values of $\dot{P}$ have been reported, differing even by one order of magnitude (see e.g. Table~2 of the online McGill catalogue). Consequently, we should take these values with caution, especially for the objects with the largest values of $\dot{P}$ (see \S~\ref{sec:unification} for further discussion).

\subsection{On luminosities and temperatures from spectral analysis.}

Luminosities and temperatures can be obtained by spectral analysis, but it is usually difficult to determine them accurately. The luminosity is always subject to the uncertainty in the distance measurement, while the inferred effective temperature depends on the choice of the emission model (blackbody versus atmosphere models, composition, condensed surface, etc.), which is subject to large theoretical uncertainties in the case of strong magnetic fields.  We often find that more than one model can fit equally well the data, without any clear, physically motivated preference for one of them, with inferred effective temperatures differing up to a factor of two. Photoelectric absorption from interstellar medium further constitutes a source of error in temperature measurements, since the value of the hydrogen column density $n_H$ is correlated to the temperature value obtained in spectral fits. Different choices for the absorption model and the metal abundances can also yield different results for the temperature. In the very common case of the presence of inhomogeneous surface temperature distributions, an approximation with two or three regions at different temperatures is usually employed. 

Finally, in the case of data with few photons and/or strong absorption features, the temperature is poorly constrained by the fit, adding a large statistical error to the systematic one. For all of these reasons, the temperatures inferred by spectral fits can hardly be directly compared to the {\it physical surface temperatures} extracted from cooling codes.

Because of the above considerations, the luminosity constitutes a better choice to compare data and theoretical models. Since it is an integrated quantity, it averages effects of anisotropy and the choice of spectral model. The main uncertainty on the luminosity is often due to the poor knowledge of the source distance. In the worst cases, the distance is known within an error of a few, resulting in up to one order of magnitude of uncertainty in the luminosity. In addition, the interstellar absorption acts predominantly in the energy band in which most of the middle age neutron stars emit ($E\lesssim 1$ keV). For this reason, hottest (magnetars) or closest (XINSs) sources are easier to detect. Similarly to the case of the temperature, the choice of different models of absorption and chemical abundances can yield additional systematic errors on the luminosity. However, for the worst cases, the relative error is about $30\%$, making it a secondary source of error compared with the distance.

\subsection{Data reduction.}

%%%%%%%%%%% TABLE 2 %%%%%%%%%%%%%%%%%%%%%%%%%%
\begin{table}[t]
\tiny
\begin{center}
\begin{tabular}{l c c c c}
\hline
\hline
Source & Date obs. & Obs.ID (sat.) & Exposure & Cts. \\
 & & & [ks] & [$10^3$] \\
\hline
  CXOU J185238.6+004020 &    2008-10-11 &    0550670601     (XMM) &  25.7 &     2.6\\
         1E 1207.4-5209 &    2002-08-04 &    0155960301     (XMM) &  74.6 &    90.2\\
          RX J0822-4300 &    2009-12-18 &    0606280101     (XMM) &  24.1 &    30.1\\
   CXO J232327.9+584842 &    2006-10-19 &          6690 (Chandra) &  61.7 &     9.1\\
         PSR J0538+2817 &    2002-03-08 &    0112200401     (XMM) &  10.0 &     4.3\\
           PSR B1055-52 &    2000-12-15 &    0113050201     (XMM) &  51.8 &    28.4\\
         PSR J0633+1746 &    2002-04-04 &    0111170101     (XMM) &  56.7 &    44.8\\
           PSR B1706-44 &    2002-03-13 &    0112200701     (XMM) &  28.4 &     5.1\\
           PSR B0833-45 &    2006-04-27 &    0153951401     (XMM) &  71.6 &  1150.0\\
           PSR B0656+14 &    2001-10-23 &    0112200101     (XMM) &   6.0 &    39.8\\
           PSR B2334+61 &    2004-03-12 &    0204070201     (XMM) &  26.8 &     0.3\\
         PSR J1740+1000 &    2006-09-28 &    0403570101     (XMM) &  28.6 &     2.4\\
         PSR J0726-2612 &    2011-06-15 &         12558 (Chandra) &  17.9 &     1.0\\
         PSR J1119-6127 &    2003-06-26 &    0150790101     (XMM) &  41.9 &     0.4\\
         PSR J1718-3718 &    2010-08-10 &         10766 (Chandra) &    10 &     0.0\\
         PSR J1819-1458 &    2008-03-31 &    0505240101     (XMM) &  59.2 &     6.8\\
        RX J0420.0-5022 &    2003-01-01 &    0141751001     (XMM) &  18.0 &     0.7\\
        RX J1856.5-3754 &    2011-10-05 &    0412601501     (XMM) &  18.2 &    46.7\\
        RX J2143.0+0654 &    2004-05-31 &    0201150101     (XMM) &  15.2 &    21.9\\
        RX J0720.4-3125 &    2003-05-02 &    0158360201     (XMM) &  39.7 &   105.7\\
        RX J0806.4-4123 &    2003-04-24 &    0141750501     (XMM) &  14.3 &    13.7\\
        RX J1308.6+2127 &    2003-12-30 &    0163560101     (XMM) &  17.3 &    33.0\\
        RX J1605.3+3249 &    2003-01-17 &    0671620101     (XMM) &  22.3 &    54.9\\
            1E 2259+586 &    2002-06-11 &    0038140101     (XMM) &  34.6 &   321.2\\
            4U 0142+614 &    2004-03-01 &    0206670101     (XMM) &  36.8 &  1780.0\\
   CXO J164710.2-455216 &    2006-09-16 &    0404340101     (XMM) &  40.4 &     1.7\\
          XTE J1810-197 & 2009-09-05/23 &06059902-3-401     (XMM) &  45.2 &    21.1\\
         1E 1547.0-5408 &    2004-02-08 &    0203910101     (XMM) &   6.4 &     0.6\\
         1E 1048.1-5937 &    2005-05-25 &    0147860101     (XMM) &  42.7 &   142.6\\
     CXOU J010043.1-721 &    2001-11-21 &    0018540101     (XMM) &  58.1 &     8.7\\
  1RXS J170849.0-400910 &    2003-08-28 &    0148690101     (XMM) &  31.1 &   212.5\\
  CXOU J171405.7-381031 &    2010-03-17 &    0606020101     (XMM) &  50.9 &    12.1\\
            1E 1841-045 &    2002-10-07 &    0013340201     (XMM) &   4.4 &    14.1\\
          SGR 0501+4516 &    2009-08-30 &    0604220101     (XMM) &  37.8 &    32.3\\
            SGR 1627-41 &    2008-09-25 &    0560180401     (XMM) & 105.0 &     3.1\\
            SGR 0526-66 &    2009-07-31 &         10808 (Chandra) &  28.7 &     5.1\\
            SGR 1900+14 &    2004-04-08 &    0506430101     (XMM) &  45.3 &    25.0\\
            SGR 1806-20 &    2005-10-04 &    0164561401     (XMM) &  22.8 &    27.5\\
          SGR 0418+5729 &    2012-08-25 &    0693100101     (XMM) &  63.1 &     0.5\\
     Swift J1822.3-1606 &    2012-09-08 &    0672283001     (XMM) &  20.2 &    17.4\\
\hline
\hline
\end{tabular}
\caption{Log of the observations by {\em XMM--Newton/EPIC--pn} and {\em Chandra/ACIS} used in this work.}
\label{tab:log}
\end{center}
\end{table}
%%%%%%%%%%%%%%%%%%%%%%%%%%%%%%%%%%%%%%%%%%%%%%%%%%%%%

In order to control the systematic errors in the luminosity and spectral parameters, we have homogeneously re-analysed all the data, extracted directly from the best available observations from {\em Chandra} or {\em XMM--Newton}. In Table~\ref{tab:log} we list the log of all the observations we used. These observations have been selected with the following criteria.
\begin{enumerate}
 \item For comparable exposure times of a given source, we always preferred the {\em XMM--Newton} observation, given the larger collecting area with respect to {\em Chandra}, hence resulting in a more detailed spectrum. Instead we used {\em Chandra} data for sources which have bright nebulae.
 \item For variable sources we took the longest available observation during the quiescent state of the neutron star.
 \item We excluded those objects for which the spectrum was fitted equally well without the addition of a thermal component.
\end{enumerate}

We processed all {\em XMM--Newton}  observations \citep{jansen01} listed in Table~\ref{tab:log} using {\tt SAS} version 11, and employing the most updated calibration files available at the time the reduction was performed (November 2012). Standard data screening criteria are applied in the extraction of scientific products, and we only used the {\em EPIC-pn} camera data. We extracted the source photons from a circular region of 30 arcsec radius, and a region of the same size was chosen for the background in the same CCD but as far as possible from the source position.\footnote{The only exception is PSR~B1055, for which the observation was performed in timing mode, and the extraction region is a box.} We restricted our analysis to photons having PATTERN$\leq$4 and FLAG=0.

The {\em Chandra} data we used in this work were all taken with the Advanced CCD Imaging Spectrometer ({\em ACIS-S}; \citealt{garmire03}). Data were analysed using standard cleaning procedures\footnote{{\tt http://asc.harvard.edu/ciao/threads/index.html}} and {\tt CIAO} version 4.4. Photons were extracted from a circular region with a radius of 3 arcsec around the source position, including more than $90\%$ of the source photons, and the background was extracted from a region of the same size, far from the source position.

\subsection{Data analysis.}

%%%%%%%%%%%%%%%%% TABLE 3 %%%%%%%%%%%%%%%%%%%%%%%%%
\begin{table}
\tiny
\begin{center}
\begin{tabular}{l c c c c c c c c c c}
\hline
\hline
Source & $\log(f_X)$ & $d$  & $k_bT_{bb}$ & $R_{bb}$ & $k_bT_2$ & $R_a$ & $\log(L)$ & best fit & $k_bT_c$ & $\log(L_c)$ \\
 & $\left[\frac{{\rm erg}}{{\rm cm}^2~{\rm s}}\right]$ & [kpc] & [eV] & [km] & [eV] & [km]  & [erg/s] & model & [eV] & [erg/s] \\
\hline
      Kes79 &  -12.3 &                   7.1 &  440 &  0.9 &  290 &  3.0 &   33.5--33.7 &      BB/nsa &  $<$ 100 &  $<$33.1\\
    1E 1207 &  -11.8 & 2.1$^{+ 1.8}_{- 0.8}$ &  190 &  9.6 &  145 &  7.4 &   33.0--34.0 &    BB*/nsa* &  $<$  60 &  $<$32.2\\ 
    PuppisA &  -11.3 &          2.2$\pm 0.3$ &  400 &  1.7 &  204 &  6.4 &   33.5--33.7 &      BB/nsa &  $<$  90 &  $<$32.9\\
       CasA &  -11.8 & 3.4$^{+ 0.3}_{- 0.1}$ &  450 &  1.7 &  288 &  2.7 &   33.4--33.6 &      BB/nsa &  $<$ 110 &  $<$33.3\\
  PSR J0538 &  -12.1 &          1.3$\pm 0.2$ &  160 &  2.6 &   -  &   -  &   32.7--32.9 &       BB+PL &  $<$  50 &  $<$31.9\\
  PSR B1055 &  -13.4 &         0.73$\pm0.15$ &  190 &  0.3 &   -  &   -  &   32.2--32.6 &      2BB+PL &     70 &      -\\
    Geminga &  -12.5 &0.25$^{+0.22}_{-0.08}$ &  140 &  0.1 &   -  &   -  &   31.6--32.5 &      2BB+PL &     42 &      -\\
  PSR B1706 &  -12.1 & 2.6$^{+ 0.5}_{- 0.6}$ &  160 &  3.3 &   -  &   -  &   31.7--32.1 &       BB+PL &  $<$  60 &  $<$32.2\\
       Vela &  -10.5 &         0.28$\pm0.02$ &  120 &  5.0 &   80 &  9.4 &   32.1--32.3 & (BB/nsa)+PL &  $<$  40 &  $<$31.5\\
  PSR B0656 &  -12.6 &         0.28$\pm0.03$ &  100 &  2.4 &   -  &   -  &   32.7--32.8 &      2BB+PL &     50 &      -\\
  PSR B2334 &  -14.0 & 3.1$^{+ 0.2}_{- 2.4}$ &  160 &  1.1 &   86 &  7.9 &   30.7--32.1 &      BB/nsa &  $<$  50 &  $<$31.9\\
  PSR J1740 &  -13.8 &                   1.4 &  170 &  0.4 &   68 &  7.8 &   32.1--32.2 &     2BB/nsa &     78 &      -\\
  PSR J0726 &  -14.0 &                   1.0 &   90 &  4.6 &   -  &   -  &   32.1--32.5 &          BB &  $<$  40 &  $<$31.5\\
  PSR J1119 &  -13.0 &          8.4$\pm 0.4$ &  270 &  1.5 &   -  &   -  &   33.1--33.4 &          BB &  $<$ 120 &  $<$32.9\\
  PSR J1819 &  -12.6 &                   3.6 &  130 & 12.3 &   -  &   -  &   33.6--33.9 &          BB &     -  &      -\\
  PSR J1718 &  -13.2 & 4.5$^{+ 5.5}_{- 0.0}$ &  190 &  2.0 &   -  &   -  &   32.8--33.5 &          BB &  $<$  90 &  $<$32.9\\
   RX J0420 &  -17.8 &                  0.34 &   50 &  3.4 &   -  &   -  &   30.9--31.0 &          BB &     -  &      -\\
   RX J1856 &  -14.4 &         0.12$\pm0.01$ &   63 &  4.1 &   -  &   -  &   31.5--31.7 &          BB &     -  &      -\\
   RX J2143 &  -13.1 &                  0.43 &  107 &  2.3 &   -  &   -  &   31.8--31.9 &          BB &     -  &      -\\
   RX J0720 &  -13.3 &0.29$^{+0.03}_{-0.02}$ &   84 &  5.7 &   -  &   -  &   32.2--32.4 &          BB &     -  &      -\\
   RX J0806 &  -13.4 &                  0.25 &  101 &  1.2 &   54 &  8.2 &   31.2--31.4 &    BB*/nsa* &     -  &      -\\
   RX J1308 &  -12.1 &                  0.50 &   94 &  5.0 &   -  &   -  &   32.1--32.2 &         BB* &     -  &      -\\
   RX J1605 &  -13.0 &                  0.10 &   99 &  0.9 &   56 &  5.3 &   30.9--31.0 &      BB/nsa &     -  &      -\\
1E 2259$^\dagger$ &  -10.3 &          3.2$\pm 0.2$ &  400 &  2.9 &  120 &   -  &   35.0--35.4 &      RCS+PL &  $<$ 120 &  $<$33.4\\
4U 0142$^\dagger$ &   -9.8 &          3.6$\pm 0.5$ &  400 &  6.5 &  290 &   -  &   35.4--35.8 &      RCS+PL &  $<$ 150 &  $<$33.8\\
  CXO J1647 &  -12.2 & 4.0$^{+ 1.5}_{- 1.0}$ &  330 &  0.6 &  150 &   -  &   33.1--33.6 &         RCS &  $<$ 120 &  $<$33.4\\
  XTE J1810 &  -11.7 &          3.6$\pm 0.5$ &  260 &  1.9 &   -  &   -  &   34.0--34.4 &         2BB &    116 &      -\\
    1E 1547 &  -11.5 &          4.5$\pm 0.5$ &  520 &  0.3 &  100 &   -  &   34.3--34.7 &         RCS &  $<$ 150 &  $<$33.8\\
    1E 1048 &  -10.8 &          2.7$\pm 1.0$ &  640 &  0.6 &  370 &   -  &   33.8--34.5 &         RCS &  $<$ 100 &  $<$33.1\\
 CXOU J0100 &  -12.5 &         60.6$\pm 3.8$ &  350 &  9.2 &  300 &   -  &   35.2--35.5 &         RCS &     -  &      -\\
 1RXS J1708$^\dagger$ &  -10.4 &          3.8$\pm 0.5$ &  450 &  2.1 &  320 &   -  &   34.8--35.1 &      RCS+PL &  $<$ 130 &  $<$33.6\\
 CXOU J1714 &  -11.4 &         13.2$\pm 0.2$ &  540 &  1.6 &  340 &   -  &   34.9--35.2 &         RCS &  $<$ 180 &  $<$34.1\\
1E 1841$^\dagger$ &  -10.4 & 9.6$^{+ 0.6}_{- 1.4}$ &  480 &  5.0 &  270 &   -  &   35.2--35.5 &      RCS+PL &  $<$ 200 &  $<$34.3\\
   SGR 0501 &  -11.3 & 1.5$^{+ 1.0}_{- 0.5}$ &  570 &  0.2 &  110 &   -  &   33.2--34.0 &         RCS &  $<$ 100 &  $<$32.9\\
   SGR 1627 &  -11.6 &         11.0$\pm 0.2$ &  450 &  2.0 &  280 &   -  &   34.4--34.8 &      RCS+PL &  $<$ 300 &  $<$34.9\\
   SGR 0526 &  -12.0 &         49.7$\pm 1.5$ &  480 &  3.6 &  320 &   -  &   35.4--35.8 &         RCS &  $<$ 200 &  $<$34.3\\
SGR 1900$^\dagger$ &  -11.1 &         12.5$\pm 1.7$ &  390 &  4.4 &  330 &   -  &   35.0--35.4 &      RCS+PL &  $<$ 150 &  $<$33.8\\
SGR 1806$^\dagger$ &  -10.6 &13.0$^{+ 4.0}_{- 3.0}$ &  690 &  2.0 &  390 &   -  &   35.1--35.5 &      RCS+PL &  $<$ 250 &  $<$34.7\\
SGR 0418$^\circ$ &  -14.0 &                   2.0 &  320 &  0.1 &   -  &   -  &   30.7--31.1 &          BB &  $<$  40 &  $<$31.5\\
Swift J1822$^\circ$ &  -11.5 &          1.6$\pm 0.3$ &  540 &  0.2 &  300 &   -  &   32.9--33.2 &         RCS &  $<$  70 &  $<$32.5\\
\hline
\hline
\end{tabular}
\end{center}
$^\circ$ The source has been recently discovered in outburst and it could have not reached the quiescence yet at the time of the latest available observation.\\
$^\star$ Absorption line(s) {\tt gabs} included in the fit.\\
$^\dagger$ Hard tail detected in quiescence hard X-ray spectra ($E\gtrsim 20~$keV).
\caption{Emission properties of the thermally emitting neutron stars. $f_X$ is the 1--10\,keV band unabsorbed flux according to the indicated best-fitting model (the {\tt bbodyrad} model flux is chosen whenever multiple models are equally compatible with the data). $k_bT_{bb}$ and $R_{bb}$ are the temperature and radius inferred by the {\tt bbodyrad} model (BB). $k_bT_2$ is the temperature inferred by the {\tt nsa} model for the weakly magnetized cases ($B\lesssim 5\times 10^{13}$ G) when an acceptable associated radius $R_a$ is found, also indicated. If $R_a$ is not indicated, $k_bT_2$ represents the temperature inferred by the {\tt RCS} model for strongly magnetized sources. $L$ is the bolometric luminosity from the thermal component(s) of the fit, and assuming the indicated distance (whose references are listed in the online table). The range of $L$ includes both statistical and distance errors; for strongly absorbed sources (i.e., most magnetars) a minimum arbitrary factor of $50\%$ uncertainty is assumed to account for systematical model-dependent uncertainty. $k_bT_c$ is either the lower temperature for models including 2 BBs, compatible with emission from the entire surface, or the upper limit for cases showing emission from a small spot $R_{bb}\sim$ few km. In the latter case, $L_c$ is the associated upper limit to the hidden thermal luminosity. See text for details on the spectral models. All radii, temperatures and luminosities are the values as measured by a distant observer.}
\label{tab:spectral}
\end{table}

\setlength{\tabcolsep}{10pt}

All spectra were grouped in order to optimize the signal-to-noise ratio in each spectral bin, but even for the fainter objects we grouped the spectra in order to have at least 25 counts per spectral bin, and reliably use the chi-squared statistic to assess the goodness of the fits. The response matrices were built using ad hoc bad-pixel files built for each observation.

\begin{figure}
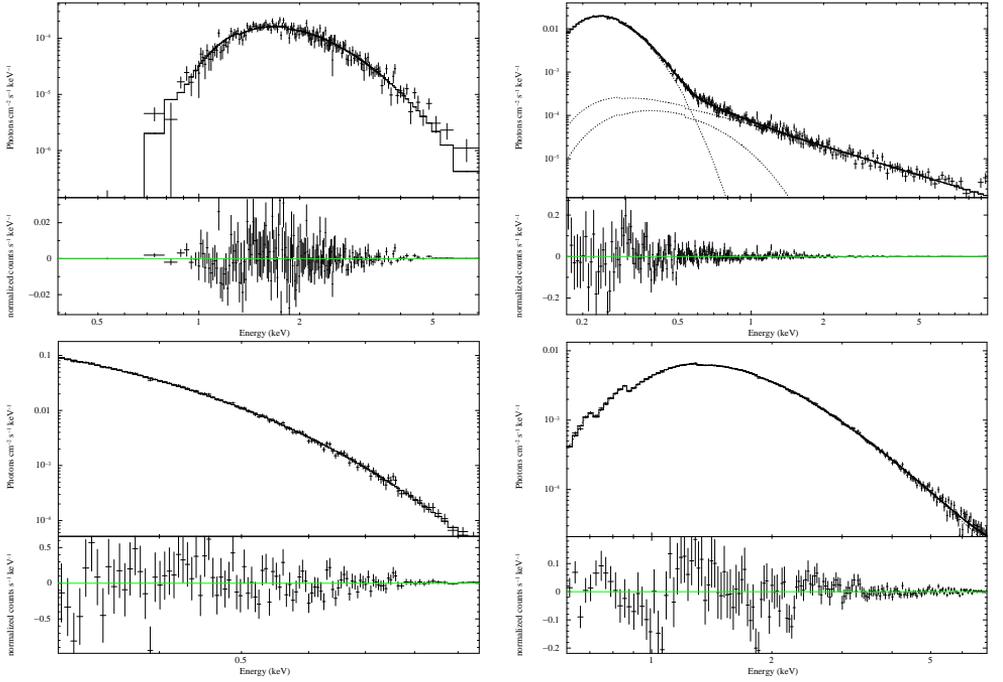

\centering
\includegraphics[width=.32\textwidth,angle=270]{images/spectrum_casA.ps}
\includegraphics[width=.32\textwidth,angle=270]{images/spectrum_geminga.ps}\\
\includegraphics[width=.32\textwidth,angle=270]{images/spectrum_j1856.ps}
\includegraphics[width=.32\textwidth,angle=270]{images/spectrum_1e2259.ps}
\caption{Unfolded spectra of four representative sources, with the fits reported in Table~\ref{tab:spectral}: the CCO in CasA (top left), the Geminga pulsar (top right), the XINS J1856 (bottom left) and the magnetar 1E2259 (bottom right).}
\label{fig:spectra}
\end{figure}

For the spectral analysis we used the {\tt XSPEC} package (version 12.4) for all fittings, and the {\tt phabs} photoelectric absorption model with the \cite{anders89} abundances, and the \cite{balucinska92} photoelectric cross-sections. We usually restricted our spectral modelling to the 0.3--10~keV energy band, unless the source was such that a smaller energy range was needed for the spectral analysis. We always excluded bad channels when needed. In Table~\ref{tab:spectral} we report the results of our analyses, and in Fig.~\ref{fig:spectra} we show the unfolded spectra of four representative sources.

Depending on the source, we either used a blackbody model ({\tt bbodyrad}) alone, or added a power law ({\tt power}) and/or a second blackbody component if statistically significant. When physically motivated, we also fitted data with an neutron star atmosphere model ({\tt nsa}: \citealt{zavlin96,pavlov95}), always fixing the neutron star mass to $M=1.4~M_\odot$, the radius to $R_\star=10$ km and $B$ to the closest value to the inferred surface value $B_p$ (see Table~\ref{tab:timing}).

We find that for CCOs the atmosphere models provide slightly better fits and radii closer to the typical neutron star values. In the case of RPPs, they can all be described by thermal component(s), with an additional power law. In particular, the three Musketeers (Geminga, PSR~B0656 and PSR~B1055) are well fitted by a double blackbody plus a power law. An atmosphere model (plus a power law) fits Vela better than a blackbody plus power law, and it is compatible with emission from the entire surface. However, in several other cases atmosphere models provide unphysical emitting radii. On the other hand, XINSs X-ray spectra are well fit by a single blackbody model, with absorption features in a few cases; in a couple of cases an atmosphere model can fit the data as well.

For magnetars we have also used a resonant Compton scattering model ({\tt RCS}: \citealt{rea08,lyutikov06}), adding a power law component when needed. In \S~\ref{sec:rcs_code}, we have briefly illustrated different models accounting for the resonant Compton scattering in the magnetosphere, with {\tt RCS} being the first (and naive) model. The main differences between these models are the physical parameters inferred from the fits (twist, temperature, viewing and magnetic angles), due to different hypotheses and approximations used. Note that we are interested in the unabsorbed flux, which is an integrated quantity, weakly dependent on the model. For this reason, we use the public {\tt RCS} model, implemented in the {\tt XSPEC} package, to account for the soft X-ray tail. In the {\tt RCS} model, the seed photons originate from the surface. However, part of the energy of the detected flux ultimately comes from the magnetospheric plasma, sustained by the large magnetic energy and dominated by pair-production. How to disentangle the surface and magnetospheric components is an open question and may be clarified only by a physically consistent model following the line of research of \cite{beloborodov13}. Lacking any better choice, we assume a thermal origin for the flux estimated with the {\tt RCS} model, discarding the flux of a second power law component, should the latter be needed to reproduce the spectra at very high energy.

%%%%%%%%%%%%%%%%%%%%%
%%%%%%%%%%%%%%%%%%%%%

Since our spectral fitting is aimed at a more reliable and homogeneous comparison with the theoretical models, we report in Table~\ref{tab:spectral} both the total 1--10\,keV unabsorbed flux, and the bolometric luminosity ($L$) of the thermal component(s) of the spectral fit. Note that, for rotational powered pulsars for which a strong non-thermal component is present, the luminosity we quote includes the blackbody component(s) only. Another quite common feature is the small size of the emitting region, typically a few km. Since spectra of most sources are strongly absorbed, this would hide possible cool components from an extended region. For this reason, we also report in Table~\ref{tab:spectral} the maximum temperature $kT_c$, and associated luminosity $L_c$, that a $\approx 10$ km radius neutron star would have while still being compatible with the lack of detection. These estimates rely on the particular spectral model we have chosen, and on the distance. They are indicative of the amount of possible hidden flux. For some strongly absorbed magnetars, this contribution could be in principle of the same order of magnitude as the flux detected from the visible hot spot.

Last, note that luminosities derived for the whole sample of objects span about five orders of magnitude, making the relative errors on luminosity much less substantial than those on the temperature. This further justifies our choice of taking into account luminosity instead of temperature to compare our cooling models to observations.

\section{The period clustering: constraining internal properties.}\label{sec:period_clustering}

%%%%%%%%%%%%%%%%%%%%%
\begin{figure}
\centering
\includegraphics[width=.6\textwidth]{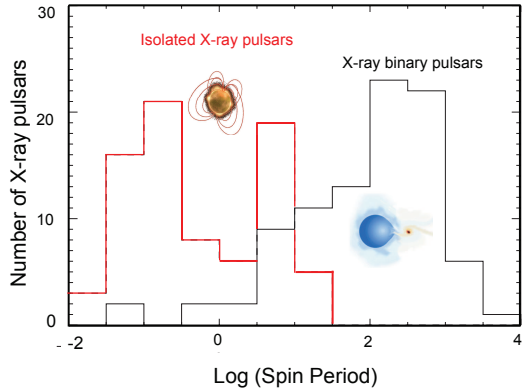}
\caption{Period distribution of isolated neutron stars (red) and neutron stars in binary systems (black).}
\label{fig:histogram_pulsars}
\end{figure}
%%%%%%%%%%%%%%%%%%%%%

Including all the known pulsars detected in any energy band (see ATNF catalogue), one realizes that there is an apparently sharp limit of $\sim 12$ s in the distribution of spin periods. Looking at the timing properties of our sample in Table~\ref{tab:timing}, we note that periods of most magnetars lie in a narrow range around $\sim 10$ s. In particular, some sources (e.g. 1E\,1841, SGR\,0526, SGR\,1806 among others) are known to be young and they already have periods from 7 to 12\,s. With their current estimated dipolar fields $B_p$, they should easily reach periods of 30 or 40\,s in a few thousand more years. This seems in contradiction with a steady pulsar spin down rate: where is the population of evolved high magnetic field neutron stars with long periods? Why none of the middle age magnetars, or the older X-ray pulsars have longer periods?

For RPPs, the strong dependence of the radio luminosity and the beaming angle with the rotation period, and the selection effect of several radio surveys for long spin periods, result in the lack of observed radio pulsars with periods longer than a few seconds. For X-ray pulsars, however, there is no reason to expect any selection effect. We plot in Fig.~\ref{fig:histogram_pulsars} the spin period distribution of isolated X-ray pulsars and X-ray binary pulsars, showing that there are no observational limitations to the detection of slow periods in X-ray binaries. When other torques are present (accretion), X-ray pulsars with rotation periods of hundreds or even thousand of seconds are clearly observed. The fact that no X-ray emitting isolated neutron star has been discovered so far with a period $>12$ s must therefore be a consequence of a real physical limit and not simply a statistical fluctuation \citep{psaltis02}. The easiest and long-standing answer is that the magnetic field decays as the neutron star gets older \citep{colpi00} in such a way that its spin-down rate becomes too slow to lead to longer rotation periods during the time it is still bright enough to be visible as an X-ray pulsar. In this scenario, {\it low-field magnetars} and XINSs are simply old magnetars whose external dipolar magnetic field has decayed to normal values \citep{turolla11,rea10}. However, no detailed quantitative predictions supported by realistic simulations had been able to reproduce the observational limits.

In \S~\ref{sec:geo} and \S~\ref{sec:magnetic_micro}, we discussed how the $P$--$\dot{P}$ evolutionary tracks obtained from the magneto-thermal simulations depend on the geometry of the initial magnetic field and on the impurity parameter in the pasta region (see Figs.~\ref{fig:geometry} and \ref{fig:micro}). In models with currents confined to the crust, and a high value of $Q_{imp}$, the vertical drop of the track gives a natural explanation to the observed upper limit to the rotation period of isolated X-ray pulsars, and the observed distribution with objects of different classes clustering in the range $P=2$--$12$ seconds while $\dot{P}$ varies over six orders of magnitude. On the other hand, in models with low impurity in the pasta region, the magnetic field decay is not fast. As a consequence, the period keeps increasing due to the slower dissipation of the magnetic field, which in principle predicts that pulsars of longer periods (20-100 s) should be visible. The slow release of magnetic energy through Joule heating keeps the neutron star bright and visible much longer than for models with large impurity in the pasta phase.

Other possible torque mechanisms, that would enter as extra-terms in equation (\ref{eq:ppdot_spindown}), such as stellar wind or accretion from a fallback disk, could act only in the early stages of a neutron star life and may contribute to explain the observed large values of $\dot{P}$ in some objects. Note, however, that our main conclusion is not affected: if there is a highly disordered inner crust, either due to the existence of the pasta phase or because the whole inner crust is amorphous, evolutionary tracks in the $P$--$\dot{P}$ diagram will bend down after $\sim 10^5$ yr regardless of the particular model. For this reason, we take $Q_{imp}=100$ in the pasta region in the baseline model (model Q100 of Fig.~\ref{fig:qimp}).

%%%%%%%%%%%%%%%%%%%%%%%%%%%%%%%%%%%%%%%%%%
\begin{figure}[t]
 \centering
\includegraphics[width=.9\textwidth]{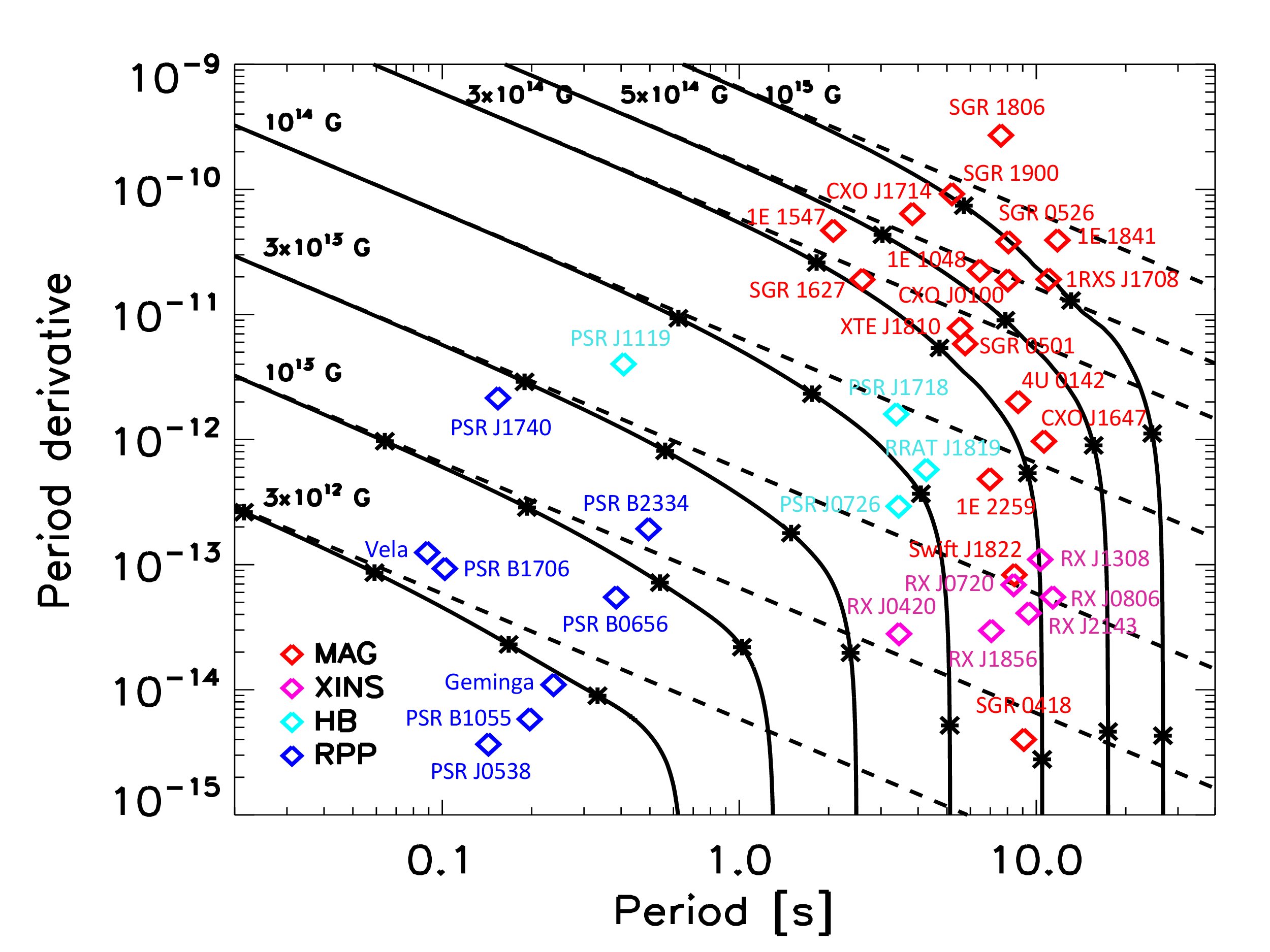}
\caption{Evolutionary tracks in the $P$--$\dot{P}$ diagram with mass and radius of our baseline model ($M=1.4~M_\odot$), with $B_p^0= 3\times10^{12}, 10^{13}, 3\times10^{13}, 10^{14}, 3\times10^{14}, 10^{15}$ G. Asterisks mark the real ages $t=10^3,10^4,10^5,5\times10^5$ yr, while dashed lines show the tracks followed in absence of field decay.} 
 \label{fig:ppdot_m140}
\end{figure}
%%%%%%%%%%%%%%%%%%%%%%%%%%%%%%%%%%%%%%%%%%%

In Fig.~\ref{fig:ppdot_m140} we show the evolutionary tracks in the $P$--$\dot{P}$ diagram for different initial values of the magnetic field strength, compared to the measured timing properties of X-ray pulsars. Asterisks mark different ages ($t=10^3,10^4,10^5,5\times10^5$ yr), while dashed lines show evolutionary tracks that the star would follow if there were no magnetic field decay. $B_p$ is almost constant during an initial epoch, $t\lesssim 10^3-10^5$ yr, which depends on the initial $B_p^0$: stronger initial fields decay before weaker ones.

An inspection of the data reveals that there may be common evolutionary links among groups of sources. One of such links roughly includes nine magnetars (1E~1547, SGR~1627, SGR~0501, XTE~J1810, CXO~J1647, 1E~2259, 4U~0142 and the low B magnetars Swift~1822, SGR~0418) and five of the XINSs (excluding only the faint source RX~J0420 and RX~J1605, that has no measure of $P$ and $\dot{P}$). For our particular model, this track corresponds to a neutron star born with $B_p^0=3 \times 10^{14}$ G (the exact value can change by a factor $\sim 2$ if the uncertainties in the inclination angle are considered in the spin-down formula).  In the upper right corner, we identify a second group of eight {\it extreme magnetars}, characterized by a larger $\dot{P}$ (CXOU~J1714, SGR~1900, 1E~1048, SGR~0526, CXOU~J0100, 1RXS~J1708, 1E~1841, SGR~1806). These would be consistent with neutron stars born with a higher initial field $B_p^0 \sim 10^{15}$ G.

%%%%%%%%%%%%%%
\section{The unification of the neutron star zoo.}\label{sec:unification}

%%%%%%%%%%%%%%%%%%%%%%%%%%%%%%%%%%%%%%%%%%
\begin{figure}[t]
 \centering
\includegraphics[width=.47\textwidth]{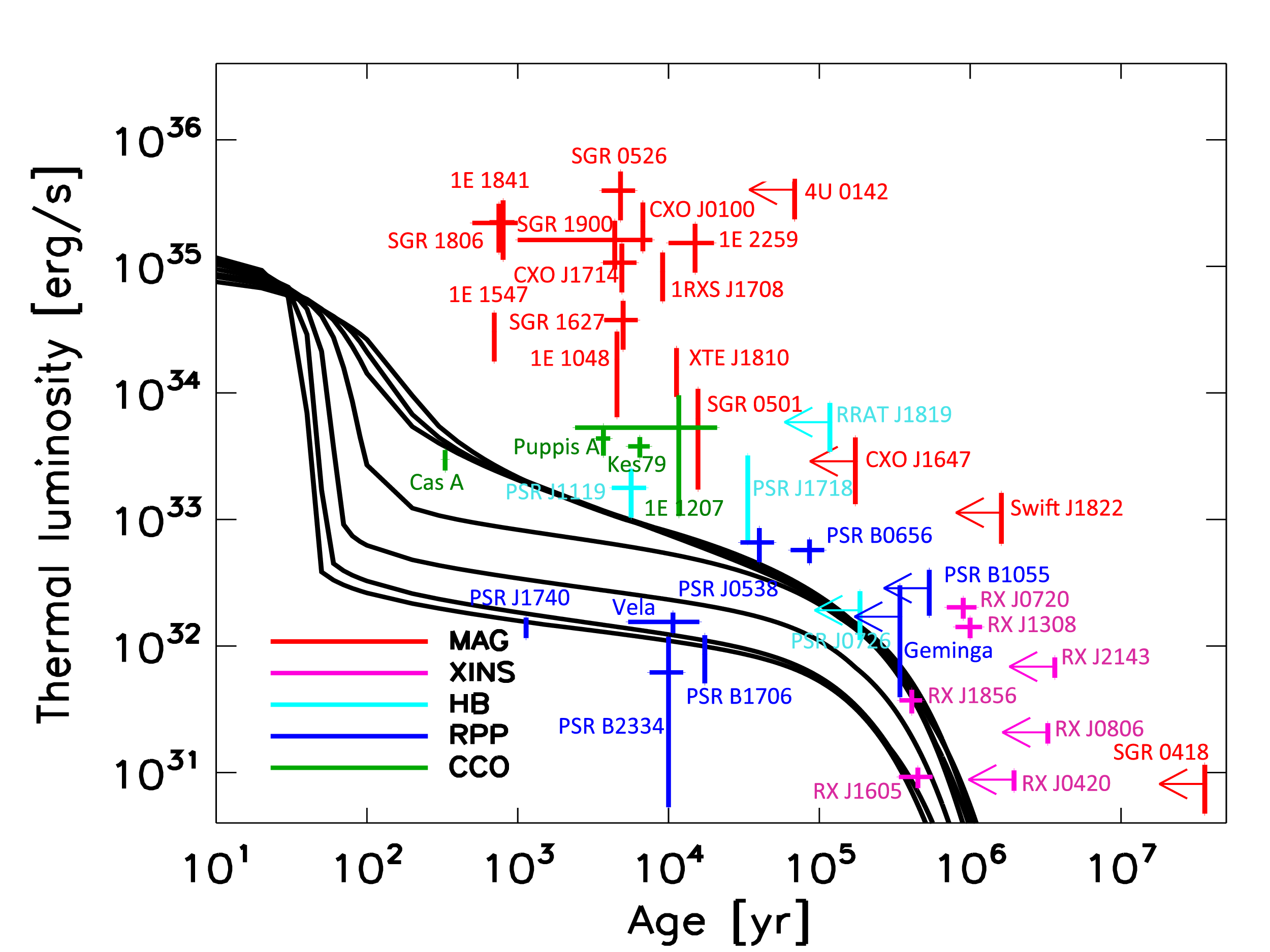}
\includegraphics[width=.47\textwidth]{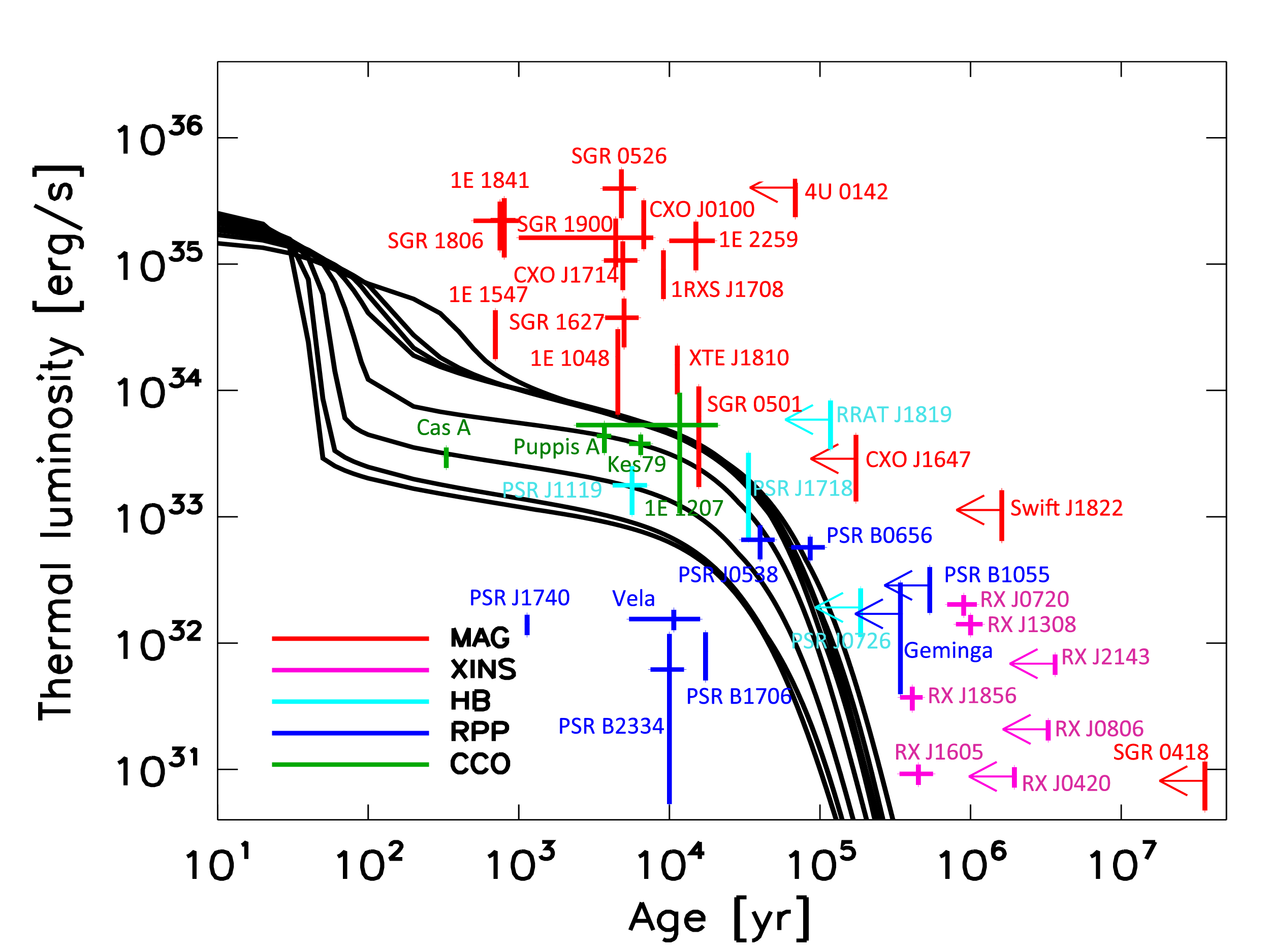}
\caption{Luminosity versus age for non-magnetized neutron star models. We show the same cooling curves as the bottom panels of Fig.~\ref{fig:b0} (increasing mass from top to bottom) compared with data presented in Tables \ref{tab:timing} and \ref{tab:spectral}. The left panel corresponds to models with iron envelopes and the right panel to models with light-element envelopes. Arrows label sources for which $\tau_c\gtrsim 50$ kyr, and no kinematic age is available, so that the real age is expected to be shorter. An uncertainty of $50\%$ has been arbitrarily taken for the kinematic age when error estimates have not been found in the literature.}
 \label{fig:cooling_data_b0}
\end{figure}
%%%%%%%%%%%%%%%%%%%%%%%%%%%%%%%%%%%%%%%%%%

In Fig.~\ref{fig:cooling_data_b0} we show the cooling curves for non-magnetized neutron stars, already shown in Fig.~\ref{fig:b0}, together with the luminosities extracted from observational data ($L$ in Table~\ref{tab:spectral}). The left/right panels correspond to models with heavy element (iron) envelopes and light element (hydrogen) envelopes, respectively. Sources with estimates of the kinematic age are shown with the associated error bar on the age. We indicate with arrows pointing towards the left the sources with $\tau_c\gtrsim 50$ kyr and no kinematic age, assuming that the characteristic age overestimates of the real age. If the decay of the magnetic field is strong, like in the model we use, eq.~(\ref{eq:treal_chage}), such an age overestimation could even be of more than one order of magnitude for the oldest objects (like SGR~0418+5729).

From a quick glance to the figure, the most interesting fact is that objects with larger inferred magnetic field are systematically hotter than what theoretical non-magnetized cooling curves predict. This provides a strong evidence in favor of the scenario in which magnetic field decay powers their larger luminosity. Even considering the (more than likely) overestimate of the neutron star age by using its characteristic age, which can reconcile some of the objects with standard cooling curves, it is clear that most magnetars and some high B neutron stars cannot be explained.

\begin{figure}[t]
 \centering
\includegraphics[width=.9\textwidth]{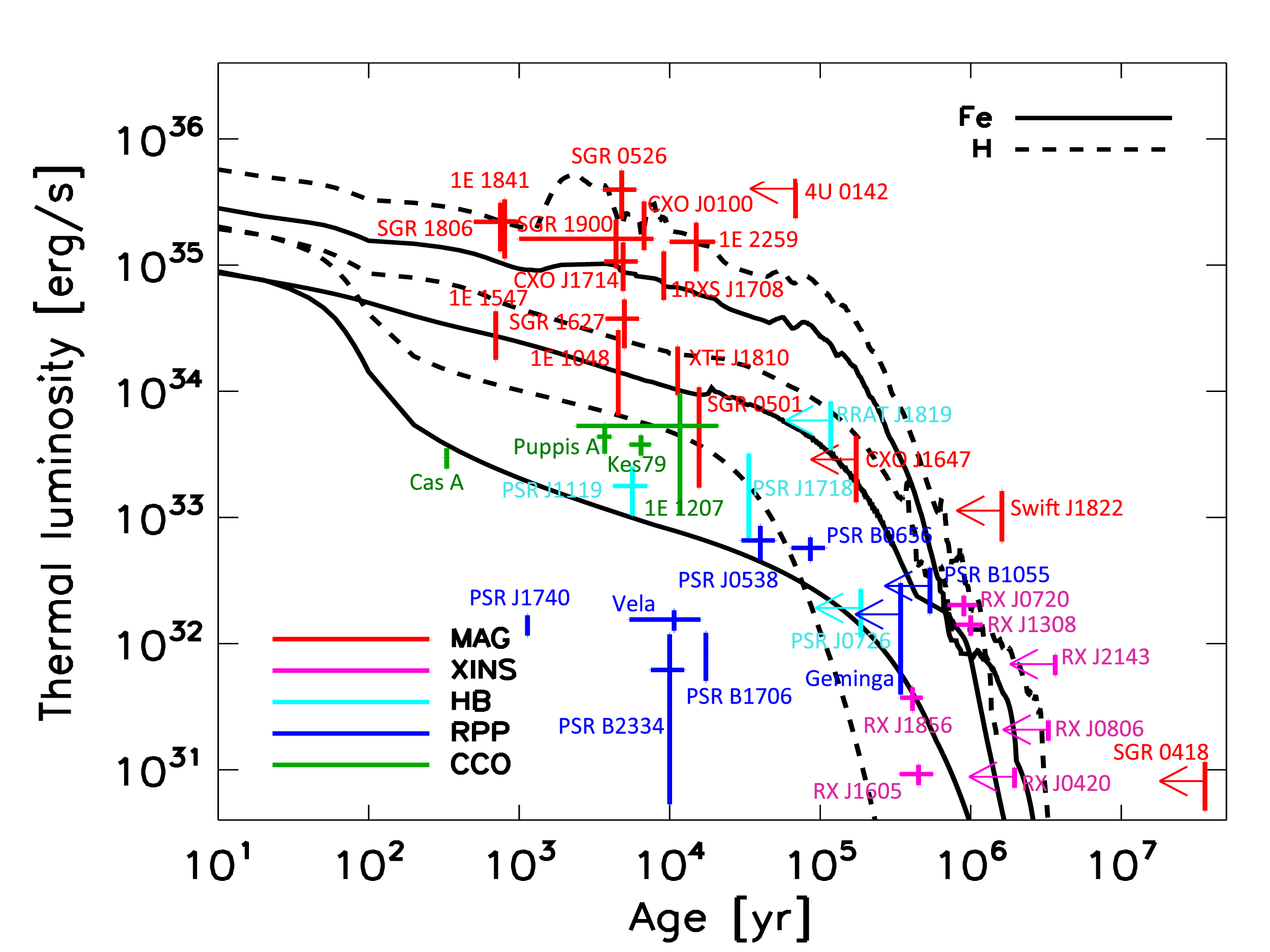} \\
\caption{Comparison between observational data and theoretical cooling curves. Models A with $B_p^0=0, 3\times 10^{14}, 3\times 10^{15}$ G are shown for Fe envelopes (solid) and light-element envelopes (dashed).} 
 \label{fig:cooling_data_all}
\end{figure}
%%%%%%%%%%%%%%%%%%%%%%%%%%%%%%%%%%%%%%%%%%%

In light of these results, we can repeat the classical comparison between cooling curves and data with the predictions of the full magneto-thermal evolution, for a range of magnetic field strengths. In Fig.~\ref{fig:cooling_data_all} we show the luminosity as a function of time for different values of the initial magnetic field up to $B_p^0=3\times 10^{15}$ G, compared with the same observational data.

Compared with the non-magnetized cooling curves (Fig.~\ref{fig:cooling_data_b0}), the most relevant difference is that the inclusion of the magnetic field allows us to generally explain objects with high luminosities. Several authors already pointed out the observed correlation between inferred magnetic field and luminosity or temperature in magnetars (e.g., \citealt{pons07a,an12}). We are now able to confirm and quantify the trend. Magnetic fields above $B\gtrsim 10^{14}$ G are strong enough to noticeably heat up the crust and power the observed X-ray radiation. The cooling time-scale for strongly magnetized objects is $\sim$ one order of magnitude larger than for weakly magnetized neutron stars.

For clarity of discussion, we now comment separately on the sources classified in three groups.
\begin{itemize}
\item {\it Neutron stars with initial field $B_p^0\lesssim 10^{14}$ G.} This group includes all RPPs, two high-$B$ PSRs (PSR~J0726 and PSR~J1119), and two of the XINSs (RX~J0420 and RX~J1605). For these relatively low fields, the luminosity is not expected to be significantly affected by the presence of the magnetic field.
\item {\it Neutron stars with initial field $B_p^0\sim 1$--$5\times 10^{14}$ G.} This group includes two high B pulsars (PSR~J1819 and PSR~J1718), 11 magnetars and 5 XINSs.
\item {\it Neutron stars with initial field $B_p^0\gtrsim 5\times 10^{14}$ G.} The 8 magnetars in the upper-right corner of Fig. \ref{fig:ppdot_m140}. 
\end{itemize}                                                                                                                                                                                         
We emphasize that this classification does not reflect intrinsic differences between groups; it is simply an arbitrary grouping that helps to highlight the evolutionary paths as a function of the initial magnetic field strength.

\subsection{Neutron stars with initial field $B_p^0\lesssim 10^{14}$ G.}

The standard cooling curves of Fig.~\ref{fig:cooling_data_b0} can account for all the weakly magnetized sources by simply varying the star mass. We note that there is actually more freedom in the theoretical models than in the particular neutron star model shown in Fig.~\ref{fig:cooling_data_all} , in which the microphysics input (e.g. gaps) and the neutron star mass have been fixed. Considering the uncertainties in the inferred ages and luminosities, the cooling curves for weakly magnetized neutron stars are consistent with all the observational data, with a few particular cases which are worth discussing. 

First, CCOs have very low $\dot{P}$ (below the shown range), implying a weak external magnetic field. Since we know that they are young (they are surrounded by SNRs), the small spin-down rate implies that their periods have not changed appreciably since birth, and no information about their real ages can be inferred from timing properties. The weak inferred magnetic field apparently contrasts with the inferred surface anisotropies and high luminosities (significantly higher than for standard RPPs) of the sources in the SNRs Kes 79 and Puppis A. Light-element envelope models (see Fig.~\ref{fig:b0}) can reconcile their age and large luminosity, but not the temperature anisotropies. The latter can be explained by the hidden magnetic field scenario (see chapter~\ref{ch:timing}).

Second, a few sources (Vela pulsar, PSR~B2334 and PSR~J1740) show evidence for enhanced cooling by direct URCA processes (or alternative exotic fast cooling neutrino processes), as already discussed by many authors.

\subsection{Neutron stars with initial field $B_p^0\sim 1$--$5\times10^{14}$ G.}

Comparing the cooling curves to the group of neutron stars compatible with initial magnetic fields $B_p^0\sim 1$--$5\times10^{14}$ G, we see that, although a few objects could still be compatible with non-magnetized curves, many others are clearly too luminous and need an additional heating source (see Fig.~\ref{fig:cooling_data_all}). Our results show that most of them can be reconciled with the theoretical models for a narrow range of initial magnetic fields, between $B_p^0\sim1$--$5 \times 10^{14}$ G. Note also that, for the few cases where we have both kinematic and characteristic ages (e.g., some XINSs), the latter overestimates the real age. For this reason, it is likely that the two XINSs and SGR 0418, placed on the right hand side of the cooling curves, have in fact real ages $\approx$0.5--1 Myr.

For the most luminous object that would belong to this group, 4U~0142, we do not have any alternative estimate of the age. However note that it is quite similar in both timing properties and luminosity to 1E~2259, whose kinematic age inferred for the SNR~CTB109 associated with 1E~2259 is about $10^4$ yr \citep{castro12}, more than one order of magnitude smaller than the characteristic age. We note that for these two objects it is difficult to reconcile the observed timing properties and luminosity, even with more extreme models (very strong toroidal magnetic field). While their luminosity is compatible with very high magnetic fields, the timing properties are more consistent with initial $B_p^0 \sim 3\times 10^{14}$ G.
                                                                                                                                           
Interestingly, these are also the two cases in which some weak evidence for the presence of fallback disks has been reported (see \citealt{wang06} for 4U~0142, and \citealt{kaplan09a} for 1E~2259). In both cases, the IR measurements are consistent with passive disks, i.e. disks in which the viscosity has been substantially reduced when they become neutral, as expected after $\sim 10^3$--$10^4$~yr \citep{menou01}. These disks are no longer interacting with the pulsar magnetosphere, and hence the measured characteristic age would again overestimate the actual neutron star age, since in the past the the torque pre-factor $KB_p^2$, eq.~(\ref{eq:ppdot_spindown}), was higher than today (see discussion in \S~\ref{sec:chage}).

\subsection{Neutron stars with initial field $B_p^0\gtrsim 5\times10^{14}$ G.}

For these sources, the cooling curves with iron envelopes for $B_p^0=10^{15}$~G are barely reaching their large luminosities, which may be an indication that these, still young, sources possess light element envelopes, or perhaps these objects are simply born with even higher magnetic fields, $\sim$~a few~$\times 10^{15}$~G.

We need to make some more considerations about these extreme magnetars. Firstly, we note from the evolutionary tracks in the $P$--$\dot{P}$ diagram that this group does not appear to have older descendants, in contrast to the first group of magnetars, which evolves towards the XINSs and SGR~0418+5729. The expected descendants of the extreme magnetars should have periods of few tens of seconds and would be bright enough to be seen. No selection effect in X-ray observations would prevent us from observing such sources.

Secondly, four of them (SGR~1900, SGR~1806, 1E~1841 and 1RXS~J1708) show a strong, non-thermal component in the hard X-ray band ($20-100$ keV), whose contribution to the luminosity is up to $10^{36}$ erg/s \citep{kuiper04,kuiper06,gotz06}. Together with 4U~0142 and 1E~2259, they are the only magnetars showing this hard X-ray emission persistently (i.e., not connected to outburst activity as in SGR~0501 and 1E~1547; \citealt{rea09,enoto10}). We stress again that the tails seen in soft and hard X-ray are ultimately due to the magnetospheric plasma, which can provide a significant amount of energy to the seed thermal spectrum emerging from the surface.

Thirdly, in most of the objects for which the kinematic age is available, a reverse, unusual mismatch with characteristic age, $\tau_k>\tau_c$, is observed (see Table~\ref{tab:timing}). This implies that the current value of $\dot{P}$ is larger than in the past. Note also that SGR~1900 and SGR~1806 are the magnetars with the most variable timing properties (see Table~2 of McGill catalogue).

These facts, i.e. the possible overestimate of both magnetic torque and thermal luminosity, and the absence of descendants, could point to some additional torque, apart from the dipole braking, acting temporarily during the early stages. One possibility discussed in the literature is wind braking \citep{tong13}, which can be effective during some epochs of the neutron star life, explaining the anomalous high values of $\dot{P}$ of some magnetars. In such a scenario, this group of most extreme magnetars would be born with $B_p^0\lesssim 10^{15}$ G, but they would be experiencing an extra torque due to temporary effects \citep{harding99}.

\chapter{Implications for timing properties}\label{ch:timing}

In this chapter we study the impact of magneto-thermal evolution on some specific aspects of timing properties observed in some neutron stars. First, we discuss the puzzling properties of CCOs (see \S~\ref{sec:cco}). The low value of their period derivative has led to the interpretation of CCOs as anti-magnetars, that is, neutron stars born with very low magnetic fields, which have not been amplified by dynamo action due to their slow rotation at birth. An alternative scenario is to consider how magnetic field gets screened by the fallback of the debris of the supernova explosion onto the newborn neutron star. In the first part of this chapter, we explore the viability of this {\em hidden magnetic field scenario} with the magneto-thermal evolution code described in the previous chapters.

A partially related topic is the braking index of pulsars. The spin-down by the torque of a constant dipolar field in vacuum predicts a value of 3. However, all measured braking indices deviate considerably from this value. In literature, different explanations have been proposed, mostly regarding the dynamics of the magnetosphere. In the second part of the chapter we quantify the expected imprints of the long-term magnetic field evolution on the braking index.

\section{CCOs and the hidden magnetic field scenario.}\label{sec:cco_hmf}

During a time interval of few months after the supernova explosion, a neutron star accretes material from the reversed shock, at a rate higher than the Eddington limit \citep{colgate71,blondin86,chevalier89,houck91,colpi96}. This episode of hypercritical accretion could bury the magnetic field into the neutron star crust, resulting in an external magnetic field (responsible for the spin-down of the star) much lower than the internal hidden magnetic field. In this phase, the value of $\dot{P}$ is very low.

When accretion stops, the screening currents localized in the outer crust are dissipated on Ohmic time-scales and the magnetic field eventually reemerges. The process of reemergence has been explored in past pioneering works \citep{young95,muslimov95,geppert99} with simplified 1D models and always restricted to dipolar fields. It was found that, depending on the depth at which the magnetic field is submerged (the submergence depth), it diffuses back to the surface on radically different time-scales $10^3$--$10^{10}$ yr. Thus, the key issue is to understand the submergence process and how deep can one expect to push the magnetic field during the accretion stage. This latter issue has been studied (also in 1D and for dipolar fields) by \cite{geppert99}, in the context of SN~1987A. They conclude that the submergence depth depends essentially on the total accreted mass. More recently, \cite{ho11} has revisited the same scenario in the context of CCOs, using a 1D cooling code and studying the reemergence phase of a buried, purely dipolar field, with similar conclusions to previous works. The hidden magnetic field scenario has also been proposed by \cite{shabaltas12} for the CCO in Kes 79. They can explain the observed high pulsed fraction ($f_p\simeq 60\%$) of the X-ray emission with a sub-surface magnetic field of $\simeq 10^{14}$~G, that causes the required temperature anisotropy. 

\subsection{Submergence of the magnetic field.}\label{sec:submergence}

In order to model the effect of the post-collapse accretion, we introduce an advective term $\vec{v}_a$ to obtain a modified induction equation, eq.~(\ref{eq:induction_hall_adv}), following \citep{geppert99}. This new term represents the sinking velocity of the accreted matter
\begin{equation}\label{eq:v_accr}
 \vec{v}_a=-\frac{\dot{m}(\theta)}{\rho}e^{-\nu}\hat{r}~,
\end{equation}
where $\dot{m}$ is the mass accretion rate per unit normal area, which in general depends on the latitude ($\theta$), and $\hat{r}$ is the unit vector in the radial direction. This expression is a result of the continuity equation (conservation of mass) together with the assumption that the accreted mass steadily piles up pushing the crust toward the interior without spreading (pure radial displacements, thus with $\dot{m}$ constant at any depth within a vertical column). 

We study the accretion phase by exploring the sensitivity of the results to two key parameters: the total accreted mass, $M_a$, and the angular dependence of $\dot{m}(\theta)$ in eq.~(\ref{eq:v_accr}). Neither the final geometry nor the submergence depth at the end of the accretion stage depend on the duration and time dependence of the fallback accretion period that follows a supernova explosion, because of its short duration ($\sim$ months), much shorter than the other relevant time-scales.

It should be noted that our simulations follow the submergence of the magnetic field in the solid crust, not the dynamical process of accretion outside. This would require MHD simulations of the accretion process in the exterior, as done for small columns above by the surface by \cite{bernal10,bernal13}, which is out of the scope of this work. Thus, we assume that any details about the dynamics of the accretion process is included in the phenomenological form of $\dot{m}(\theta)$, that describes the rate at which matter reaches the top of the crust at each latitude. How the complex external dynamics leads to a particular form of $\dot{m}(\theta)$ should be the purpose of separate studies. We also note that at the relatively high temperatures ($2-3 \times 10^9$ K) reached during the early accretion stage in our simulations, fast thermonuclear burning is likely to bring the composition close to the ground state, unlike what happens in accreting, old, cold neutron stars in binary systems. Therefore we assume that our crust is composed by matter in the ground state.

%%%%%%%%%%%%%%%%%%%%%%%%%%%%%
\begin{figure}[t]
 \centering
\includegraphics[width=.4\textwidth]{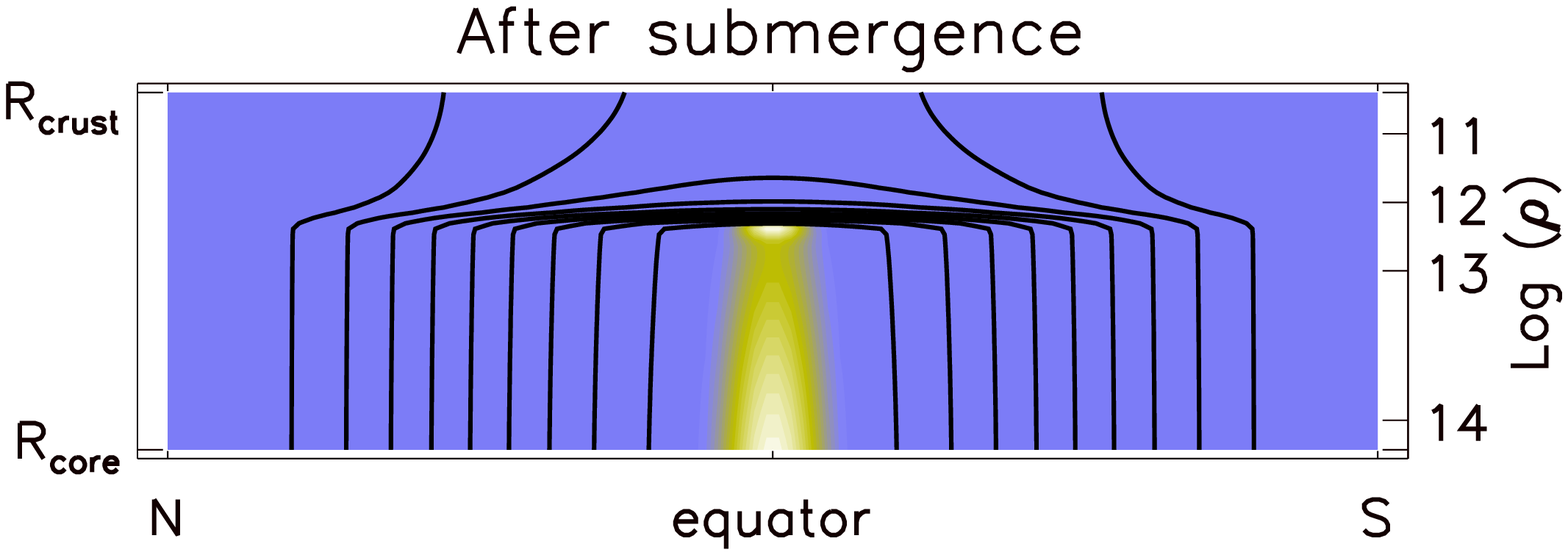}
\includegraphics[width=.4\textwidth]{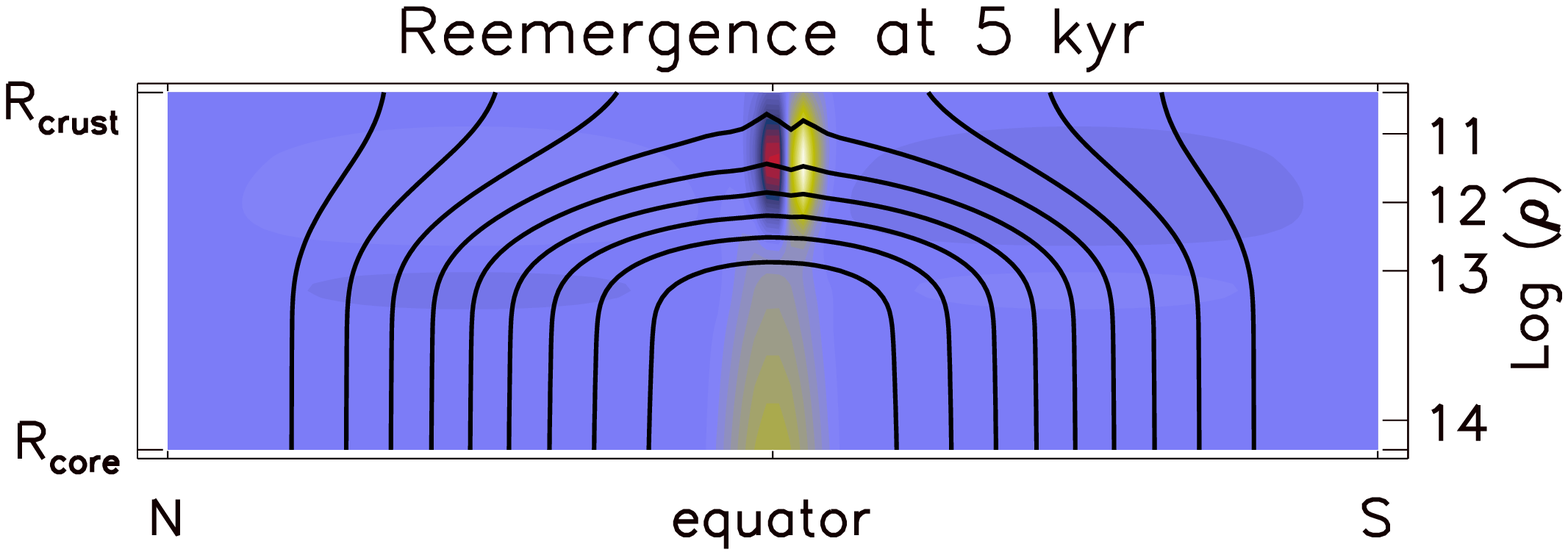}\\
\includegraphics[width=.4\textwidth]{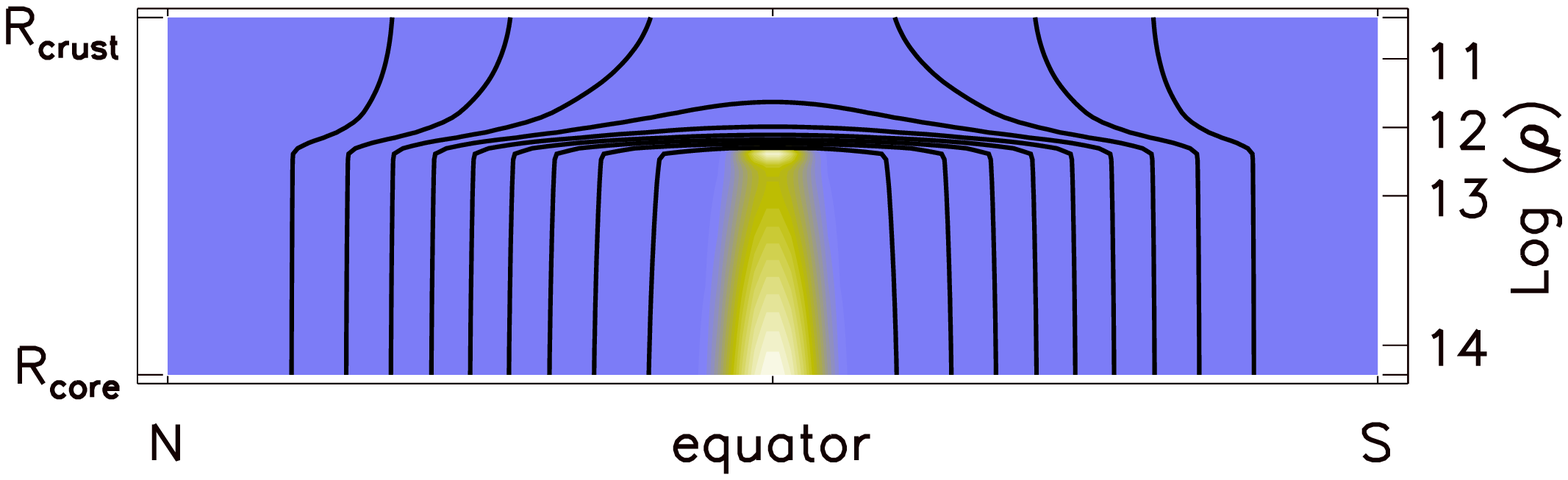}
\includegraphics[width=.4\textwidth]{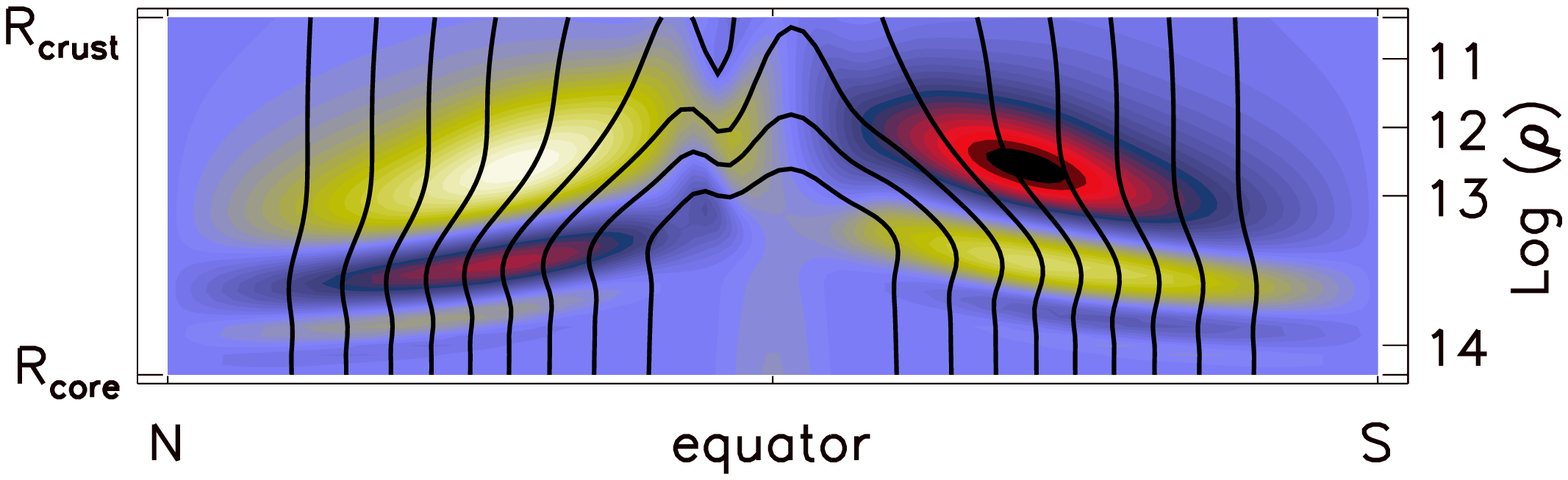}\\
\includegraphics[width=.4\textwidth]{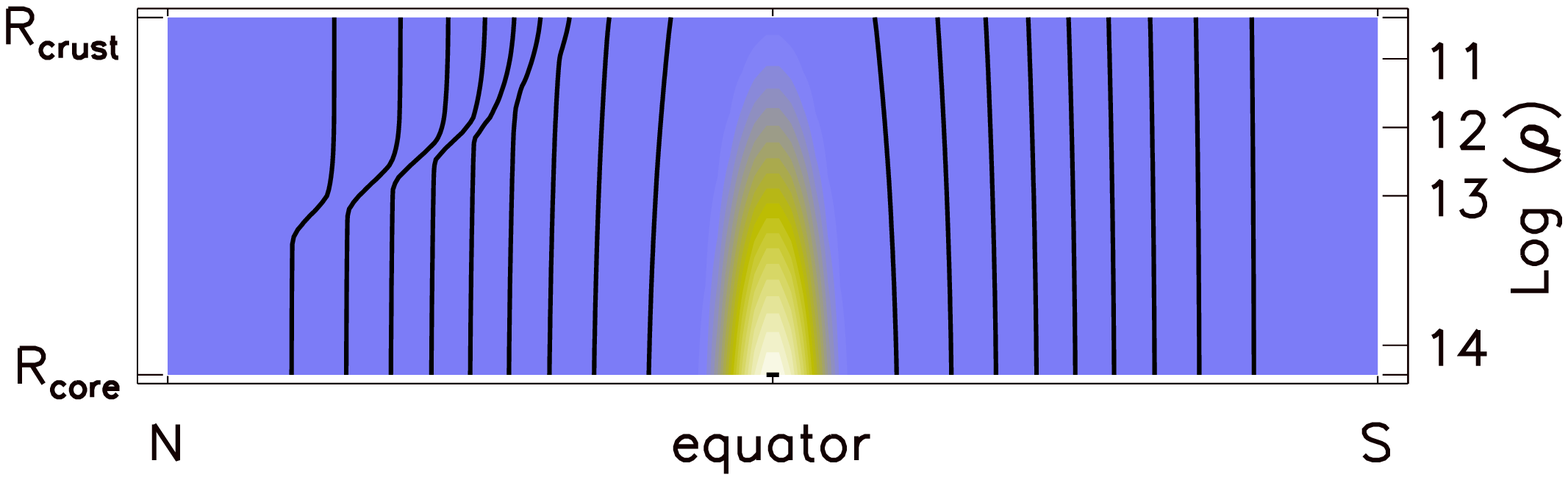}
\includegraphics[width=.4\textwidth]{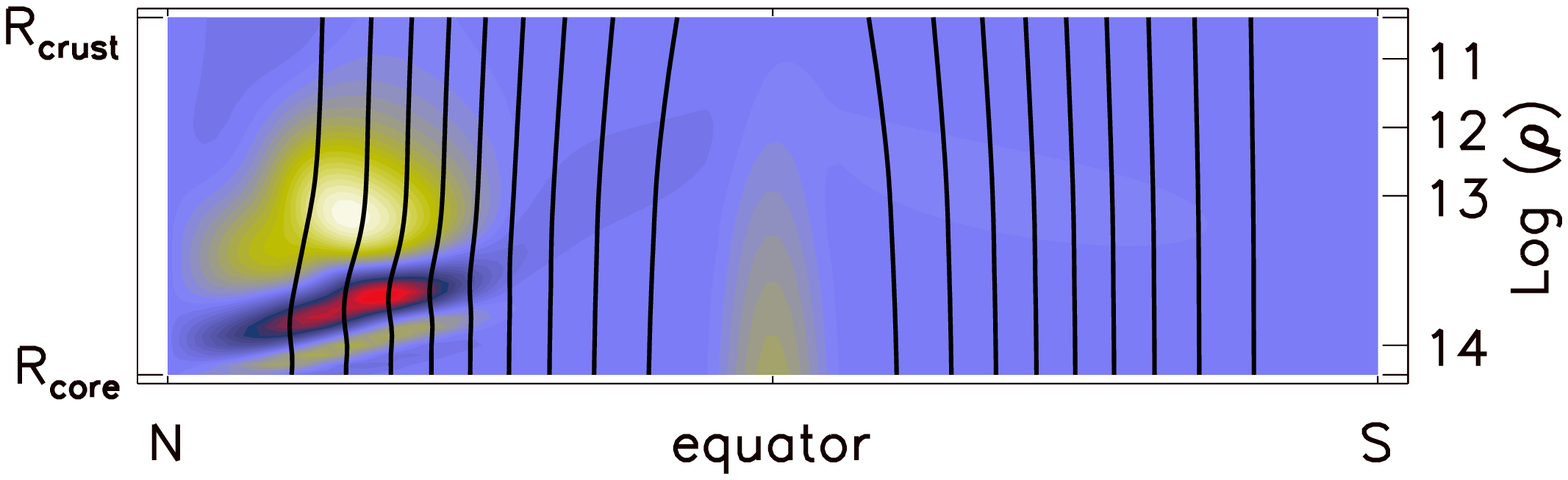}\\
\includegraphics[width=.4\textwidth]{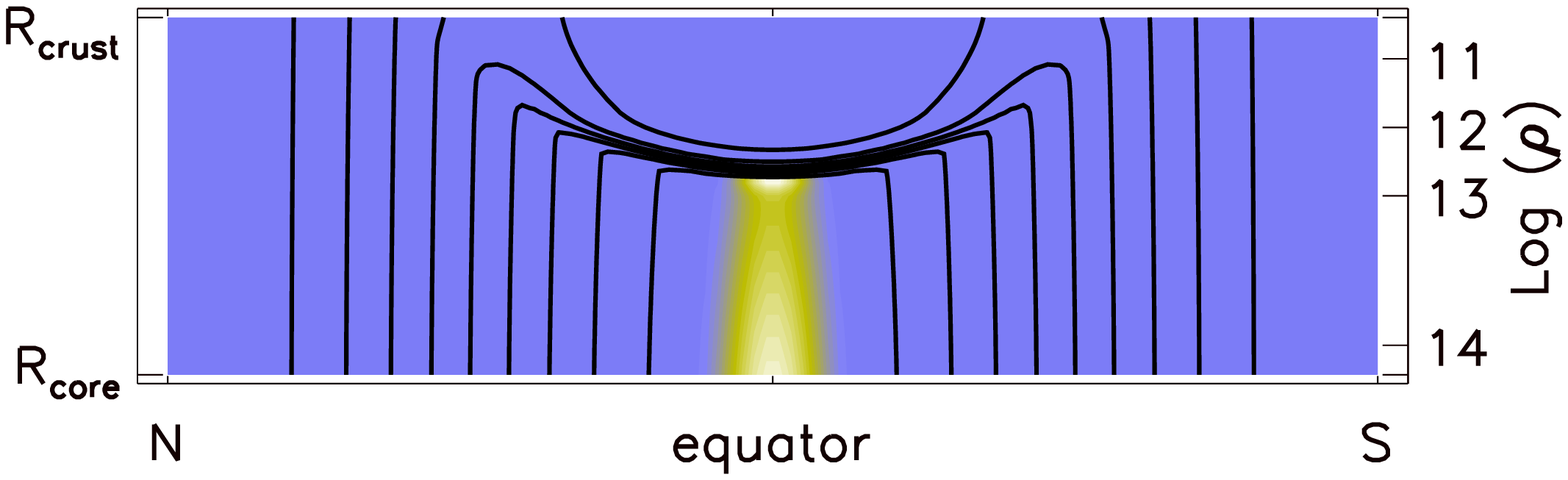}
\includegraphics[width=.4\textwidth]{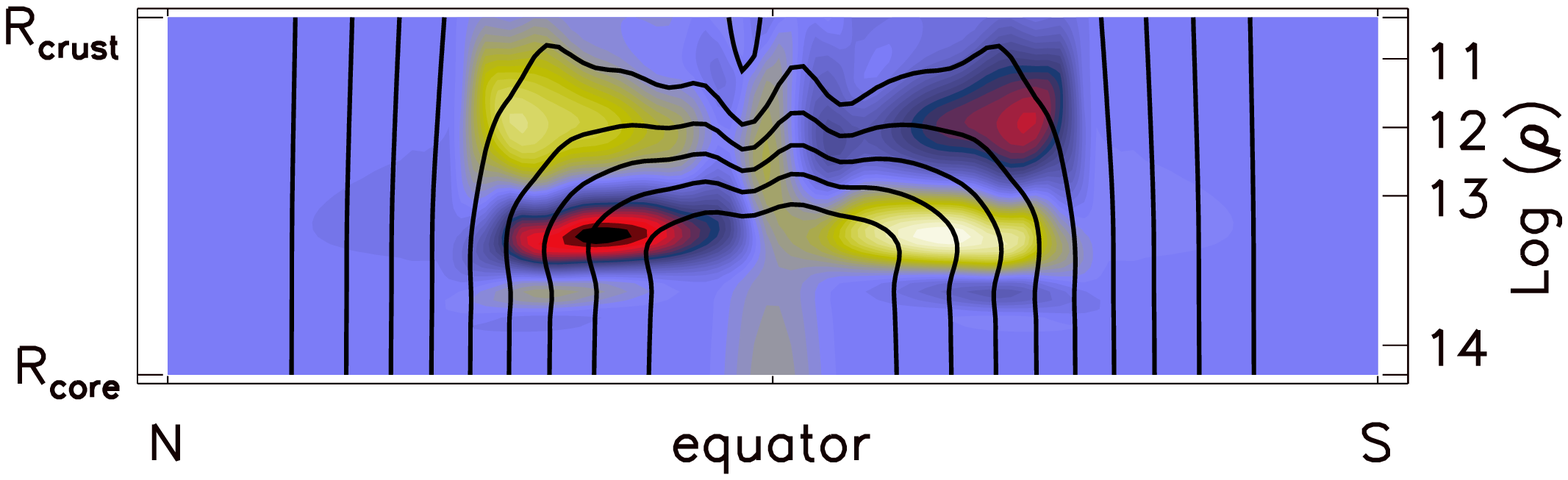}\\
\includegraphics[width=.4\textwidth]{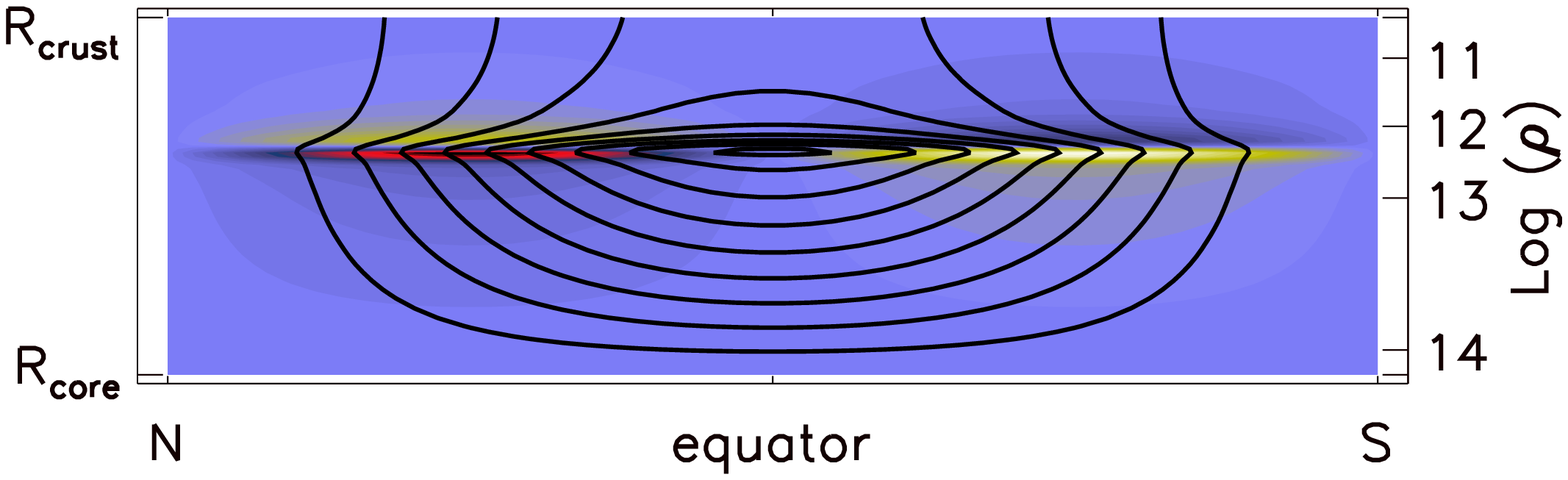}
\includegraphics[width=.4\textwidth]{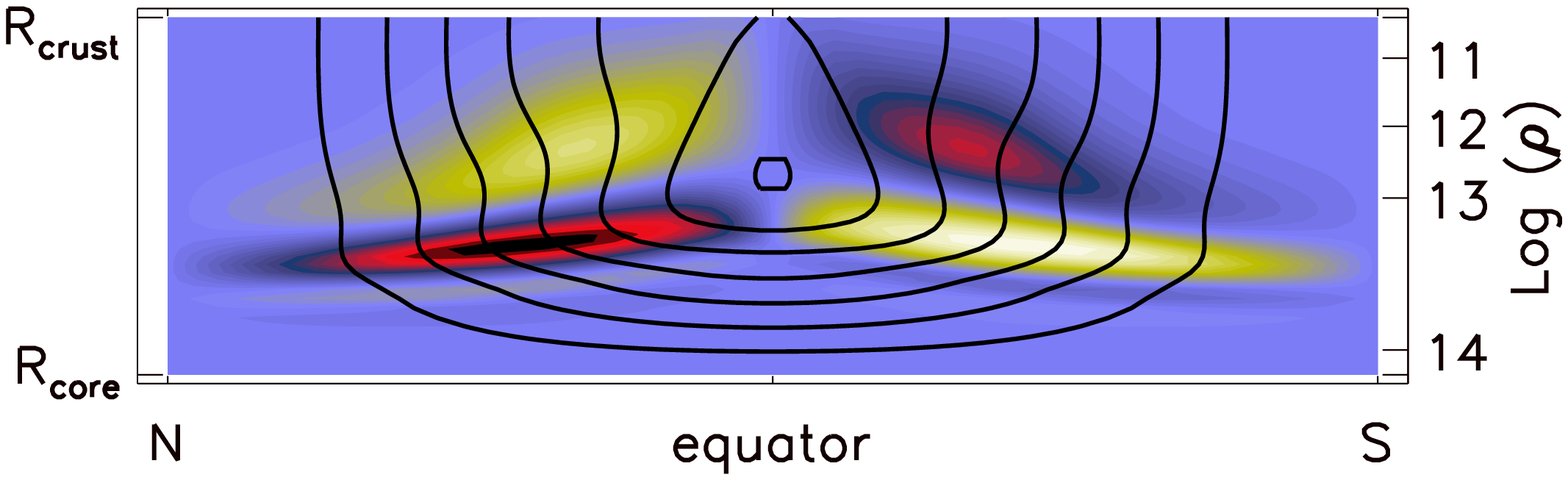}
\caption{Magnetic field evolution in the crust for different models, all with $M_a=10^{-4}~M_\odot$. Top row: model B12, spherical accretion. Second, third, fourth rows: model B14, with (s), (p) and (e) accretion rate geometries, respectively. Bottom row: model A14, spherical accretion. Left panels: submerged magnetic configurations after accretion stage. Right panels: reemergence stage, at $t=5$ kyr. The crust is represented in a planar map, with the radius (and the corresponding density) on the vertical axis (the crust/core interface is the bottom boundary), and the polar angle on the horizontal axis (north, equator and south are indicated). Solid lines represent the poloidal field, while the color scale indicates the toroidal field intensity (yellow: positive, red: negative, blue: zero).} 
\label{fig:bury_reemergence}
\end{figure}
%%%%%%%%%%%%%%

We begin by discussing how our results depend on the value of $M_a$ for three representative cases:\\
($s$) spherically symmetric, $\dot{m}(\theta)=k$;\\
($p$) channeled onto a polar cap, $\dot{m}(\theta)=ke^{-(\theta/\theta_w)^2}$;\\
($e$) channeled onto the equator, $\dot{m}(\theta)=ke^{-(\frac{\theta-\pi/2}{\theta_w})^2}$;\\
where $k$ is the normalization that fixes $M_a$. Hereafter we discuss our results with $\theta_w=0.4$ rad. As initial configuration, we will consider core-extended (models B) or crust-confined (models A) magnetic fields, already introduced in \S~\ref{sec:initial_b}. In the left panels of Fig.~\ref{fig:bury_reemergence} we compare the magnetic field in the crustal region immediately after accretion stops, for models with $M_a=10^{-4} M_\odot$. We look first at the simple case of spherical accretion (upper two panels) for models B12 and B14, i.e., with initial configuration B (see Fig.~\ref{fig:initial_b}), and $B_p^0=10^{12}$~G, $B_p^0=10^{14}$~G, respectively. During the accretion stage, magnetic field lines are advected toward the interior and screening currents are developed in the outer layers. The reduction in the surface magnetic field strength is compensated by the local amplification of $B_\theta$ at the submergence depth, where it can reach values up to few $10^{16}$~G. The geometry at the end of the submergence phase is not affected by the initial magnetic field strength (it simply scales with $B_p^0$), since the evolution is governed by the advective term $\vec{v}_a$, eq.~(\ref{eq:v_accr}). However, anisotropic accretion flows cause the irregular submergence of the magnetic field (middle and bottom panels). Strong screening currents also appear, but now they are localized in the regions where $\dot{m}(\theta)$ is larger. This has an important consequence: the external, large-scale, dipolar component  is not reduced as much as in the spherically symmetric case.

%%%%%%%%%%%%%%%%%%%%%%%%%%%%%
\begin{table}
\begin{center}
\footnotesize
 \begin{tabular}[ht!]{c c c c c c}
\hline
\hline
 geom. & $M_a$ 		& $\rho_d$	& $B_p^s$      & $B_p^r$  \\
       & [$M_\odot$]		& [g cm$^{-3}$] & [10$^{14}$G] & [10$^{14}$G]  \\
\hline 
 s & $2\times 10^{-3}$ & $ 3.0\times 10^{13}$ & $\sim 0$  & 0.76  \\
 s & $1\times 10^{-3}$ & $ 1.6\times 10^{13}$ & $10^{-5}$ & 0.84  \\
 s & $5\times 10^{-4}$ & $ 9.0\times 10^{12}$ & 0.005     & 0.90  \\
 s & $2\times 10^{-4}$ & $ 4.5\times 10^{12}$ & 0.12      & 0.93  \\
 s & $1\times 10^{-4}$ & $ 2.5\times 10^{12}$ & 0.34      & 0.95  \\
 e & $1\times 10^{-4}$ & $ 6.0\times 10^{12}$ & 0.52      & 0.92  \\
 p & $1\times 10^{-4}$ & $ 3.5\times 10^{13}$ & 0.96      & 0.98  \\
 s & $1\times 10^{-5}$ & $ 1.5\times 10^{11}$ & 0.86   	  & 0.98 \\
\hline
\hline
 \end{tabular}
\caption{Properties for different accretion geometries and total accreted mass ($M_a$) for model B14. The density at which the maximum compression of lines is reached is $\rho_d$. $B_p^s$ and $B_p^r$ stand for the values of $B_p$ at the end of the submergence stage and the maximum value reached during reemergence, respectively. }
\label{tab:sub_reem}
\end{center}
\end{table} 

In Table~\ref{tab:sub_reem} we summarize our results for model B14, for different values of $M_a$ and varying the accretion geometry, as listed above. Here $\rho_d$ denotes the density at the submergence depth and $B_p^s$ the final strength of the dipolar component of the magnetic field. We point out that the exact value of $B_p^s$ is very sensitive to the boundary conditions, i.e. the magnetic flux exchange with the exterior. We match with a general vacuum solution, while a more consistent approach should consider again the MHD accretion process in the exterior. On the contrary, we found that our results for $\rho_d$, $B_p^r$ and reemergence time-scales are not sensitive to the external boundary conditions. For spherical accretion ($s$), we found that if $M_a \lesssim 10^{-5} M_\odot$, the submergence depth is shallow. We observe a sharp transition from $M_a<10^{-4} M_\odot$, with at most a factor 3 decrease in the surface strength, to $M_a>10^{-3} M_\odot$, characterized by a very deep submergence, well within the inner crust.

Model A14 (fifth panel, model A with $B_p^0=10^{14}$ G) shows a similar behaviour: for a fixed $M_a$, the submergence depth is the same as in model B. No differences in the submergence process arise from the geometry or intensity of the magnetic field: the details of the $M_a-\rho_d$ relation depend mainly on the particular structure and mass of the neutron star. As a matter of fact, in agreement with the induction and continuity equations, the enclosed mass between $\rho_d$ and the surface is $M_a$, so that for $M_a\gtrsim 10^{-2} M_\odot$ the accreted mass has completely replaced the outer crust.

The rotational properties of the star would also be radically different, because the external dipole (third column) regulates the spin-down of the star. For $M_a \gtrsim 10^{-3} M_\odot$, and spherical (or nearly isotropic) accumulation of matter on top of the crust, $B_p$ is largely reduced and therefore the spin-down rate of the star becomes unusually small. Although estimates of the magnetic field based on timing properties suggest a very low magnetic field, the internal magnetic field could be orders of magnitude larger,  since it is simply screened by currents in the crust that will be dissipated on longer time-scales. Note also that equatorial accretion ($e$) produces a similar reduction of the external dipolar magnetic field as for spherical accretion, but the submergence is rather anisotropic (deeper in the equator). On the contrary, accretion concentrated in the polar cap ($p$) barely modifies the magnetic field strength or geometry for $M_a=10^{-4} M_\odot$.

We have also found that adding an azimuthal component of velocity $v_{\varphi}\neq 0$ in eq.~(\ref{eq:v_accr}) does not affect the submergence process, driven solely by the radial inflow. Instead, its effect is the twisting of magnetic field lines and the creation of toroidal magnetic field, which in turn enhances the Hall activity on longer time-scales, during the reemergence process.

\subsection{Reemergence of the magnetic field.}

After accretion stops, the reemergence process begins. It is mainly driven by the Ohmic dissipation of the screening currents, but also by the Hall term when the magnetic field strength locally exceeds $\sim 10^{14}$~G. In this case, there may be significant changes in the geometry of the magnetic field, including the generation of higher order multipoles in both poloidal and toroidal components (see chapter~\ref{ch:magnetic}).

In the right panels of Fig.~\ref{fig:bury_reemergence} we show the magnetic field configuration at $t=5$ kyr. If the magnetic field is weak (top panel), it the simply reemerges by diffusion. It shows very little Hall activity, seen as generation of toroidal magnetic field (white/black spots) and deformation of poloidal magnetic field in a small equatorial region of the outer crust. In the other four cases ($B_p^0=10^{14}$~G), the local generation of strong toroidal magnetic field and the creation of non-trivial structures are clearly seen, but these small scale features are mainly localized in the inner crust, which makes difficult to predict possible observational consequences. At the same time, the global, large-scale field at the surface is being restored to a shape closer to the original geometry, with additional small scale multipoles near the surface.

\begin{figure}
 \centering
\includegraphics[height=.35\textwidth]{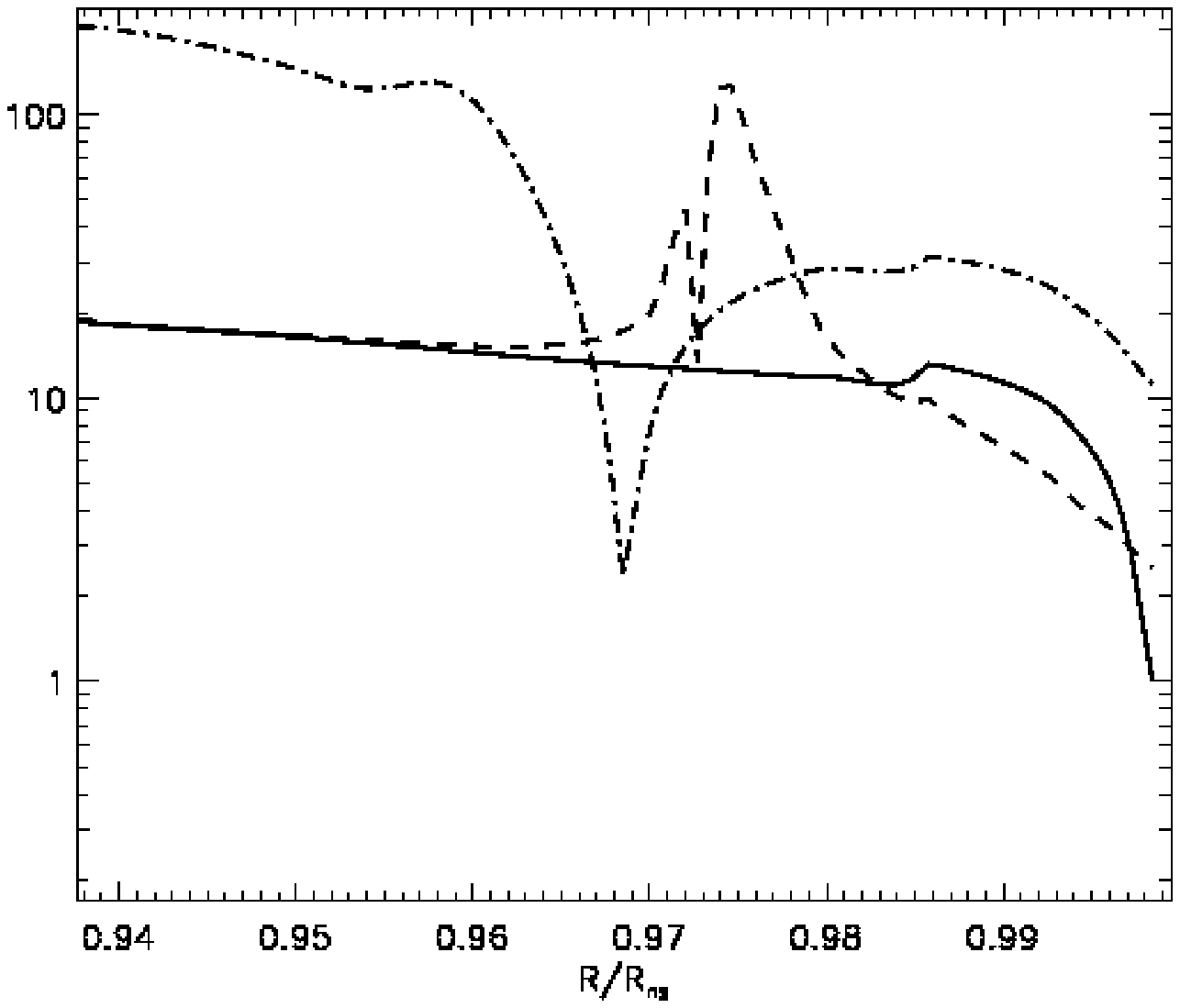}
\includegraphics[height=.35\textwidth]{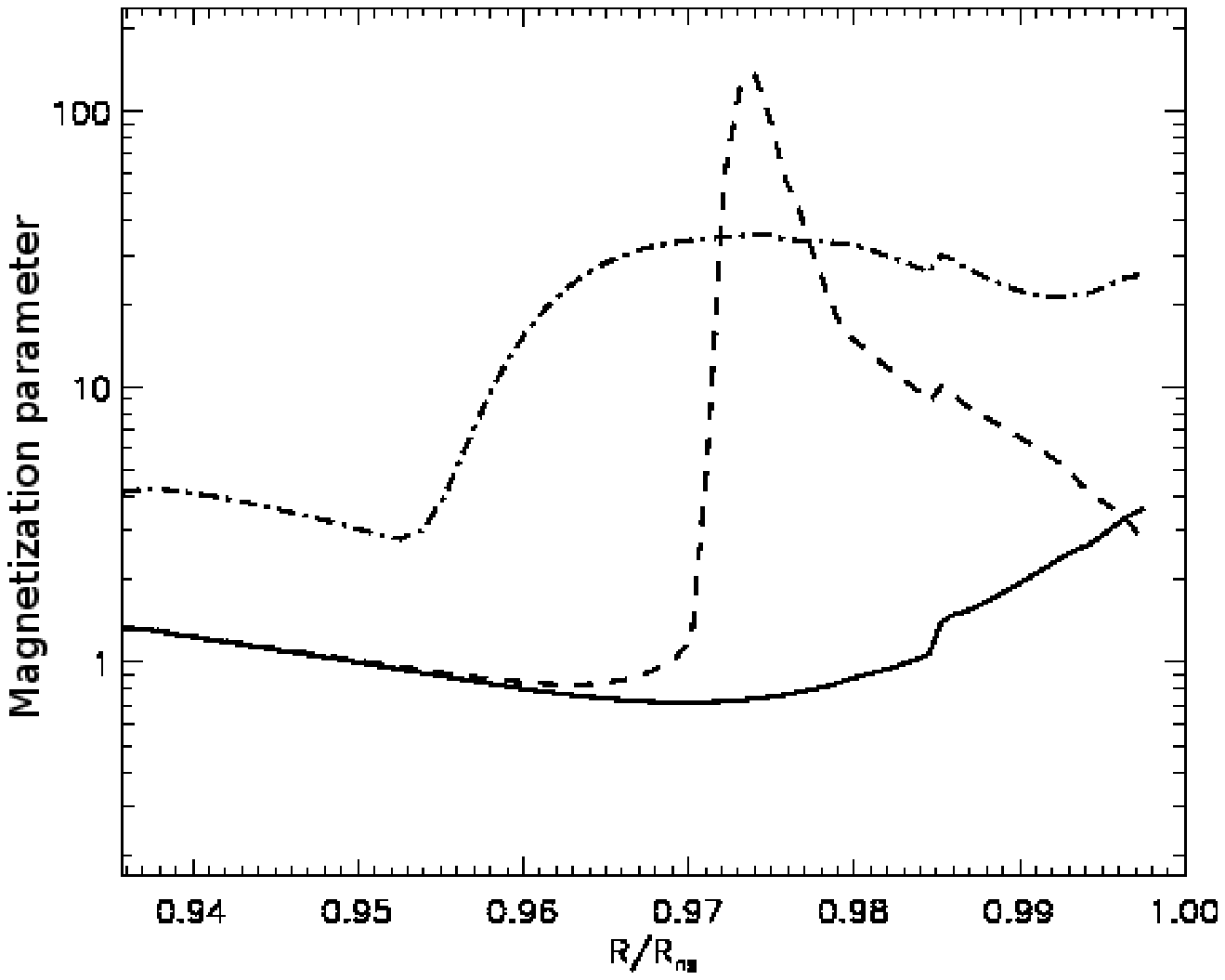}
\caption{Radial profile at the equator of $\omega_B \tau_e$ for initial model A14 (left) and model B14 (right), at three times: at the beginning (solid), after spherical accretion of $10^{-4} M_\odot$ (dashed) and at $t=5$ kyr, during reemergence (dot-dash).} 
 \label{fig:reynolds}
\end{figure}
%%%%%%%%%%%%%%%

The importance of Hall activity is shown in Fig.~\ref{fig:reynolds}. Here we plot the radial profile of the magnetization parameter, eq.~(\ref{eq:omtau}), at the equator, at different stages for model A14 (left panel) or model B14 (right panel). The solid line corresponds to the initial configuration, for which $\omega_B\tau_e$ is initially of order 1 for model B14, while it is ten times larger for model A14, due to the larger mean value of magnetic field. After the submergence (dashed), in both models $\omega_B\tau_e$ reaches much larger values in the region where the magnetic field has been compressed. After 5 kyr, the diffusion of the magnetic field and the lower temperature make $\omega_B\tau_e>1$ in the whole crust. As $\omega_B\tau_e$ scales with $B$, the same models, but with $B_p^0=10^{12}$~G, give a rescaling of $\omega_B\tau_e$ by two orders of magnitude, providing $\omega_B\tau_e\lesssim 1$ always. Thus, for such initially low fields almost no Hall activity is expected.

\begin{figure}[t]
 \centering
\includegraphics[width=.45\textwidth]{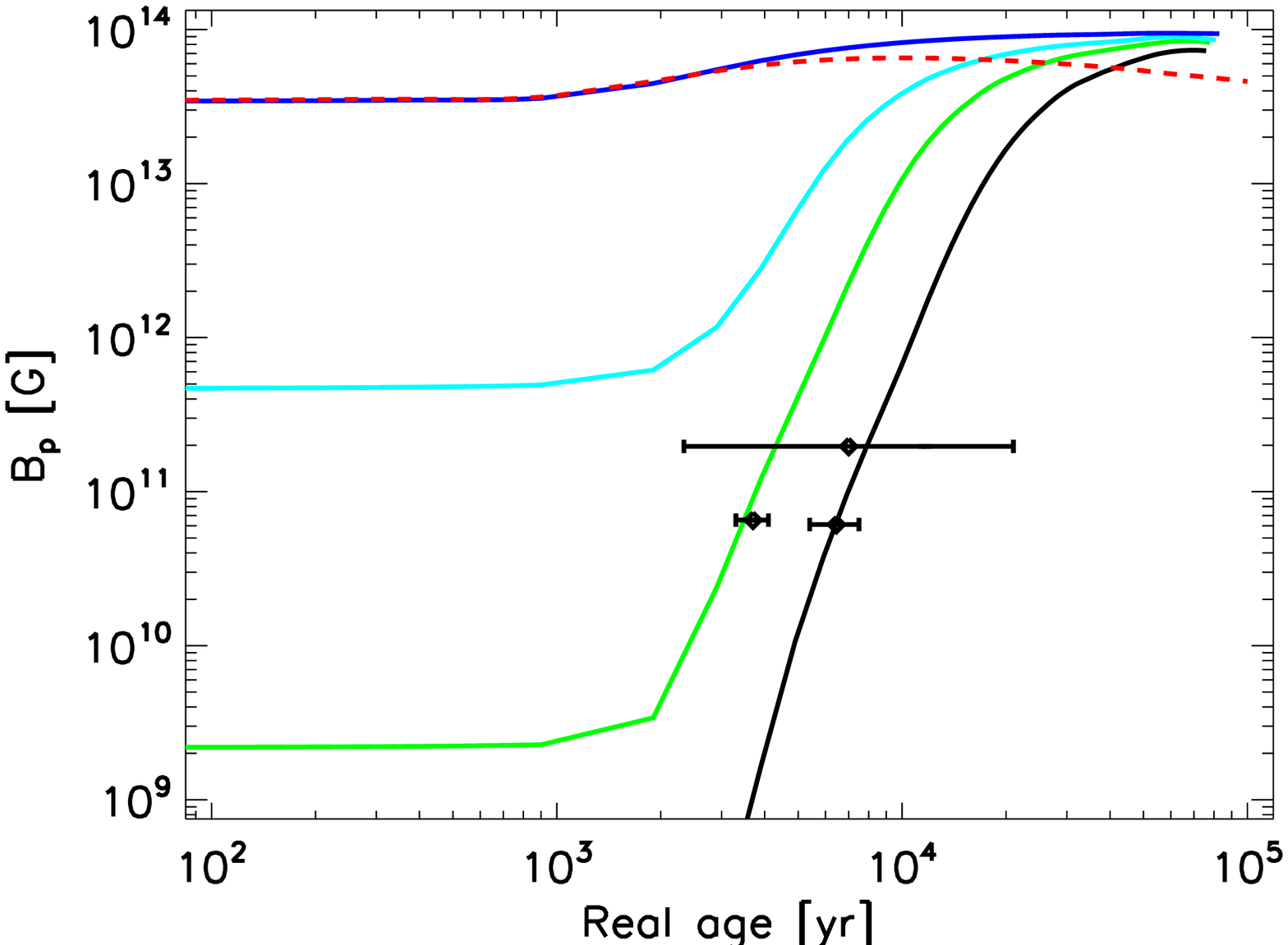}\\
\includegraphics[width=.45\textwidth]{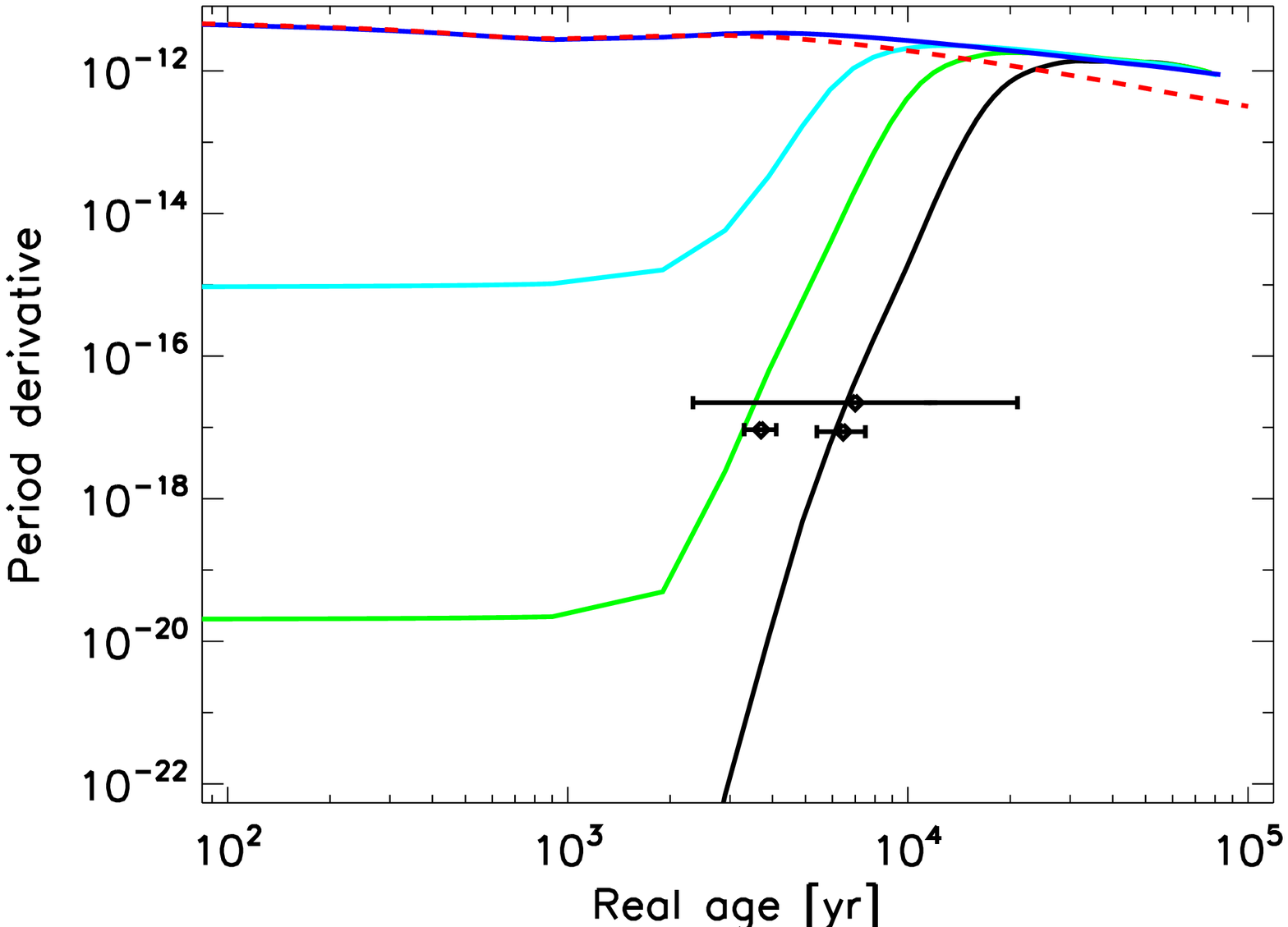}
\includegraphics[width=.45\textwidth]{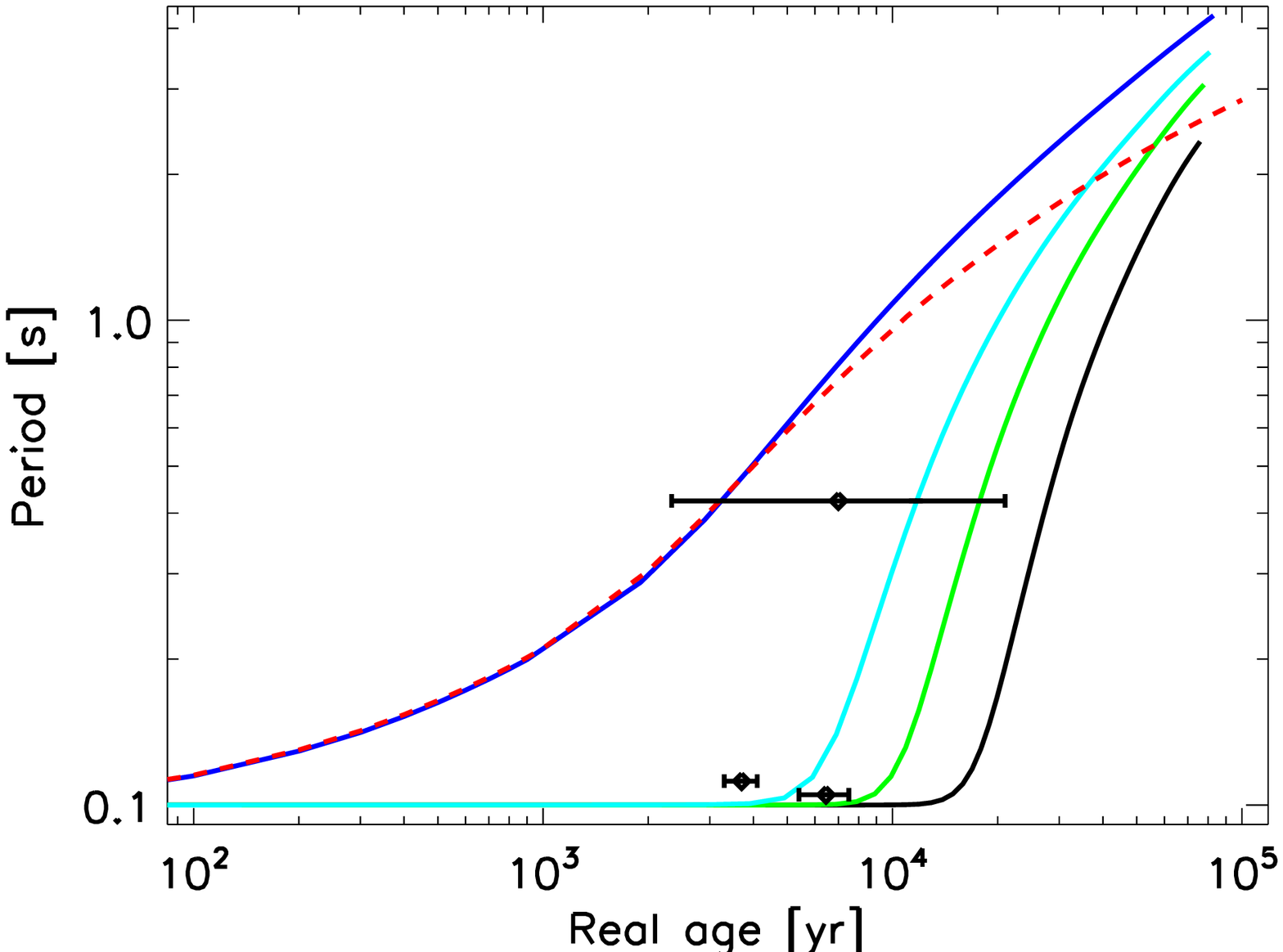}
\caption{Evolution of $B_p$ (top panel), $\dot{P}$ (bottom left) and $P$ (bottom right) during the reemergence phase after spherical accretion, assuming an initial period $P_0=0.1$ s. Different cases are shown: model A14, $M_a=10^{-4}M_\odot$ (red dashed); model B14, $M_a=[1,5,10,20]\times 10^{-4} M_\odot$ (blue, cyan, green and black lines, respectively). Observational data are shown with rhomboids; estimates and errors of the ages are taken as in Table~\ref{tab:timing}.}
 \label{fig:timing}
\end{figure}

In the top panel of Fig.~\ref{fig:timing} we show the evolution of $B_p$ as a function of time after a spherical fallback episode, with different $M_a$, for model A (red dashes) or model B (solid lines), together with the three inferred values for CCOs (see Table~\ref{tab:timing}). The {\it reemergence time-scale}, on which the surface magnetic field grows, depends basically on $\rho_d$. For total accreted matter in the range of interest, there is an initial delay of about $10^3$ yr before appreciable reemergence is observed. The reemergence process takes between $10^3$ and few $10^5$ yr, and it is determined by the local conditions (in particular, the resistivity) where the screening currents are located. The magnetic field will never reach the original strength, since some Joule dissipation is always expected. Generally speaking, $B_p$ is restored to close to its initial value for the core-extended configurations (model B). Only for the extreme case of $M_a \gtrsim 10^{-2} M_\odot$, when the reemergence process lasts a long time, $B_p^r$ is significantly reduced (see Table~\ref{tab:sub_reem}). In model A14, crustal currents are much stronger than in model B14, therefore both Ohmic and Hall time-scales are shorter. This implies that the reemerged field has been more dissipated, in agreement with 1D studies \citep{geppert99,ho11}. Furthermore, as the magnetic field is more compressed, the Hall activity is more intense (see dot-dash line in Fig.~\ref{fig:reynolds}), and the transfer of magnetic energy from the dipolar component to higher multipoles (i.e., small scale structures) is faster.

The reemergence stage has a strong imprint on timing properties. Assuming a period at the end of the accretion stage of $P_0=0.1$ s, we follow the evolution of the timing properties integrating, as usual, the spin-down equation (\ref{eq:ppdot_spindown}), arbitrarily fixing $f_\chi=1$. The bottom panels of Fig.~\ref{fig:timing} shows the evolution of $\dot{P}$ (left) and $P$ (right) as a function of the real age for several values of $M_a$ in the sensitivity range $[10^{-4}-10^{-3}] M_\odot$. Variations of a factor of a few in $M_a$ within this range result in an extremely low value of $\dot{P}$ during the first thousands of years of a neutron star life, with $P$ very close to the natal period. In terms of the spin-down age, $\tau_c=P/2\dot{P}$, this means that it overestimates the real age also by the same factor. Observational data reported for CCOs are consistent with $M_a \gtrsim10^{-3} M_\odot$.

\begin{figure}[t]
 \centering
\includegraphics[width=.5\textwidth]{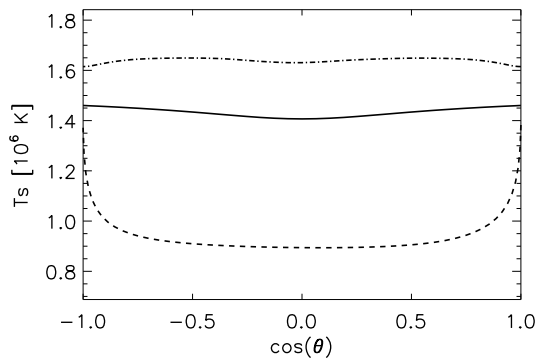}
\caption{Comparison between surface temperature distributions of an anti-magnetar (solid), and two models of hidden magnetars (dashed and dot-dashed) after accreting $M_a=10^{-3} M_\odot$. All three models are shown in an evolutionary phase when the external dipolar field is $B_p=10^{10}$~G.} 
 \label{fig:ts}
\end{figure}

A possible observable to distinguish between an anti-magnetar and the hidden magnetic field scenario is the surface temperature distribution, $T_s(\theta)$, that determines the average luminosity and light-curve in the X-ray band. Fig.~\ref{fig:ts} shows $T_s(\theta)$ for two high field neutron stars during the reemergence phase after a $10^{-3} M_\odot$ accretion episode and an anti-magnetar. The dot-dashed line refers to model B14 at $t_1=3$ kyr, while the dashed line corresponds to a model A13T, with a large dipolar toroidal component ($B_p^0=10^{13}$~G, $B_t^0=10^{15}$~G), at $t_2=6$ kyr. In both cases, the magnetic field has partially reemerged and has same value as for the  anti-magnetar model (solid lines), $B_p=10^{10}$~G. Model B14 has a slightly warmer surface than the anti-magnetar model, but both are almost isotropic. On the contrary, the crustal confined model shows a distinctive feature consisting in a lower average $T_s$ with hot polar caps. This is produced by the buried toroidal magnetic field that keeps the equatorial surface region insulated from the warmer core.

The thermal evolution during the reemergence phase in the hidden magnetic field scenario can yield to totally different degrees of anisotropy in the surface temperature, depending on the location of currents. A hidden magnetar with a large, extended toroidal magnetic field buried in the crust could qualitatively explain the surface temperatures anisotropies inferred in several CCOs: the large pulsed fraction observed in Kes 79 \citep{halpern10,shabaltas12}, the antipodal hot spots seen in Puppis A \citep{deluca12}, and the small emitting region of the blackbody components needed to fit the spectra of 1E 1207 \citep{deluca04}. Instead, weaker (or less extended) hidden toroidal magnetic field provide low pulsed fractions.

\section{Braking index of pulsars.}

In \S~\ref{sec:em_torque} we have summarized the theoretical description of pulsar spin-down by electromagnetic torques. We report again eq.~(\ref{eq:ppdot_spindown})
\begin{equation}\label{eq:ppdot_spindown_rep}
 P \dot{P} = K B_p^2~,
\end{equation}
where the proportionality constant $K$, eq.~(\ref{eq:k_spindown}),
\begin{equation}\label{eq:k_spindown_rep}
  K=f_\chi\frac{2\pi^2}{3}\frac{R_\star^6}{Ic^3}~,
\end{equation}
includes different dependences on the star radius, moment of inertia, magnetic field strength, and angle between rotation and magnetic axis. If all these quantities are constant in time, the magneto-dipole spin-down mechanism predicts a braking index $n=3$ (see \S~\ref{sec:braking_index_def} for the definition), but variations in time of any of these quantities may cause departures from this canonical value. Unfortunately, the accurate determination of the second derivative of the frequency, needed to estimate the  braking index, is not always possible because it is affected by glitches and other short-term timing irregularities.

At present, eight young pulsars have sufficiently steady rotations that stable values of their braking index are generally accepted \citep{lyne93,lyne96,middletich06,livingstone07,livingstone11b,welteverde11,espinoza11}, among which the most recent case is PSR J1734-3333, which has $n=0.9\pm0.2$ significantly below 3 \citep{espinoza11}. All these cases show the same trend: they are all young pulsars (Vela pulsar is the oldest pulsar of this sample) and they all have $n<3$. We note that gravitational wave emission predicts $n=5$, but, as already mentioned in \S~\ref{sec:em_torque}, it certainly does not contribute significantly to the torque of pulsars older than $100$ yr. 

Selecting a sample of 127 pulsars from the ATNF Pulsar Catalogue \citep{manchester05}, for which the quoted errors in  the second derivative of the spin frequency ($\ddot{\nu}$) are less than $10\%$, \cite{urama06} found a strong correlation of $\ddot{\nu}$ with $\dot{\nu}$, independent of the sign of $\ddot{\nu}$. They suggested that this trend can be accounted for by small stochastic deviations in the spin-down torque that are directly proportional (in magnitude) to the spin-down torque itself. Another point discussed in the literature is that some of the old pulsars ($>10^6$ yr) have braking indices with absolute values exceeding $|n|=10^4$. The occurrence of very high braking indices of both signs has been considered in the context of internal frictional instabilities occurring between the crust and the superfluid, almost independently of the evolution of the neutron star magnetic field \citep{shibazaki95}. However, this applies only for old neutron stars ($\tau \gtrsim 2 \times 10^7$ yr) and appears as extremely short term events oscillating about the canonical value $n=3$. \cite{barkusov10} proposed another explanation for the observed distribution with very high positive and negative braking indices by studying the effect of non-dipolar magnetic field components and neutron star precession on magnetospheric electric current losses. These large $n$ should be observable over relatively long periods of $10^3$-$10^4$ yr. Another possibility that can explain the observed variability of braking indices is the time-evolution of conductivity in the magnetosphere \citep{li12a}, which also has implications for the spin-down of intermittent pulsars and sub-pulse drift phenomena \citep{lyne09}.

Recalling the definition of braking index, eq.~(\ref{eq:BIstandard}), in a general case of a time-dependent magnetic field the braking index can be simply expressed as follows
\begin{equation}\label{eq:hatn} 
n=3-4 \frac{\dot{B_p}}{B_p} \tau_c \equiv 3-4 \frac{\tau_c}{\tau_B}~,
\end{equation}
where $\tau_c = P/2 \dot P$ is the characteristic age, eq.~(\ref{eq:chage}), and we defined the magnetic field evolution time-scale
\begin{equation}
 \tau_B \equiv \frac{B_p}{\dot{B_p}} = \frac{4}{n-3}\tau_c~.
\end{equation}
Eq.~(\ref{eq:hatn}) shows that any variation of $B_p$ results in a deviation from the $n=3$ standard value, which is obviously recovered for a constant magnetic field ($\dot{B_p} = 0$). For an increasing $B_p$ we will always obtain ${n} < 3$, while ${n} > 3$ is the signature of a decreasing $B_p$. Before showing results from our magneto-thermal simulations, we qualitatively analyse the effects of the most important physical process at different ages, considering three possible scenarios.

%%%%%%%%%%%%%%%%%%%%%%%%%%%%%%%%%%%%%%%%%%%%%%%%
\subsubsection{Amplification or reemergence of the dipolar surface magnetic field.}

The generic $n<3$ observed without exception for young pulsars can be caused by the re-diffusion of the magnetic field submerged in the crust during the supernova fallback episode as seen in \S~\ref{sec:cco_hmf}. Alternative mechanisms are time variations in the angle between the rotational and magnetic axes \citep{link97,ruderman98,barkusov09}, or the thermoelectric field generation that may proceed in the crust and envelope of young pulsars if a sufficiently strong temperature gradient is present  \citep{urpin86}. This process is limited by the condition that the surface temperature of the neutron star should not be lower than $3\times 10^6$ K \citep{wiebicke96}, which corresponds to about 1 kyr in the standard cooling scenario. In our simulations we included the reemergence of a screened magnetic field, but our present version of the code does not include magnetic field generation by thermoelectric effect, or other possible mechanisms such as magnetic flux expulsion from the superconducting core.

%%%%%%%%%%%%%%%%%%%%%%%%%%%%%%%%%%%%%%%%%%%%%%%%
\subsubsection{Ohmic decay.}

We can expect that for low-field middle-aged neutron stars ($10^4$--$10^5$ yr) the Ohmic decay is the dominant process, thus resulting in a negative $\tau_B$ in eq.~(\ref{eq:hatn}). Therefore, the braking index is expected to be $n \gg 3$, especially when the Ohmic diffusion time-scale is shorter than $\tau_c$.

%%%%%%%%%%%%%%%%%%%%%%%%%%%%%%%%%%%%%%%%%%%%%%%%
\subsubsection{Hall drift oscillatory modes.}
The Hall drift, through its nonlinear dependence on the magnetic field, has an influence on the magnetic field evolution either for magnetar conditions ($B>10^{14}$ G), or alternatively when the electrical resistivity becomes very low. In situations of quasi-equilibrium, oscillatory modes with magnetic energy exchange between the different crustal field modes may appear (see model with $Q_{imp}=0$ in Fig.~\ref{fig:micro}). The temporal evolution of the polar surface magnetic field can then be approximated by
\begin{equation}\label{eq:tauHall2}
B_p(t)= \bar{B}_p + \delta B\sin \left( \frac{2\pi t}{\tau_h}\right)~,
\end{equation}
and from eq.~(\ref{eq:hatn}) we have
\begin{equation}\label{eq:tauHall3}
n=3 -4\frac{\dot{B_p}}{B_p} \tau_c \approx 3-8\pi \frac{\tau_c}{\tau_h} \frac{\delta B}{\bar{B}_p} \cos\left( \frac{2\pi t}{\tau_h}\right)~,
\end{equation}                                                                                                                                                                                                                                                                                                                                                                                                                                                                                                                                                                                                                                                                                                                                                                                                                                                                                                                                                                                                                                                                     
where $\tau_h$ is the typical time-scale of the Hall dynamics (see eqs.~\ref{eq:hall_time-scale_pol} and \ref{eq:hall_time-scale}). These oscillations would be expected in young magnetars with $B_p>10^{14}$ G, if the Ohmic decay was not dominant (low $Q_{imp}$ in the pasta phase). It causes corrections to the canonical braking index $n=3$ of either positive or negative sign, according to eq.~(\ref{eq:tauHall3}). These corrections are expected to be small because of their short $\tau_c$, and to be increasingly important for objects with smaller $\dot{P}$ (old characteristic ages). In addition, as the star cools down, the drop in electrical resistivity may activate the Hall term even for normal pulsars $B_p \gtrsim 10^{12}$ and the occurrence of oscillatory modes during a {\it second Hall stage} at late times is a natural outcome. First estimates suggest that $\delta B/B_p$ could be as large as $\sim 10^{-1}$. When $\tau_c \gg \tau_h$, the second term in eq.~(\ref{eq:tauHall3}) dominates and the magnetic field oscillatory modes should result in equally probable positive and negative braking indices with high absolute values.

%%%%%%%%%%%%%%
\begin{figure}
\centering
\includegraphics[width=.5\textwidth]{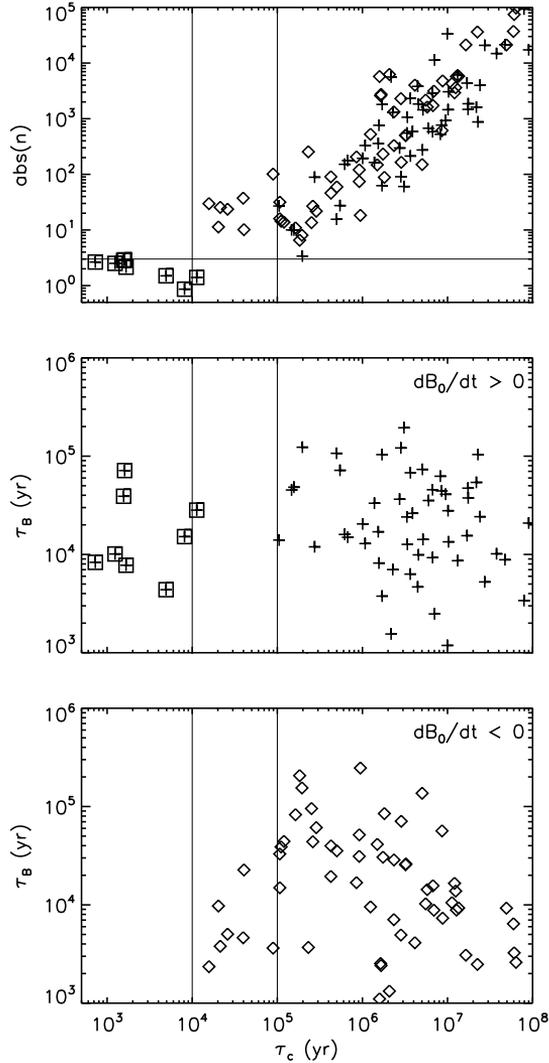}
\caption{Top: braking indices (absolute value) as a function of $\tau_c$ for our sample of 118 pulsars. We represent pulsars with $n<3$ with crosses and pulsars with $n>3$ with diamonds. Note that all objects marked with crosses and $\tau_c>10^5$ yr have actually $n<0$. The 8 youngest objects discussed in \cite{espinoza11} are marked with squares. Middle:  magnetic field evolution time-scale, $\tau_B=4 \tau_c/(3-n)$ for pulsars with $n<3$.  Bottom: absolute value of $\tau_B$ (they are all negative) for pulsars with $n>3$.  }
\label{fig:bi_taub}
\end{figure}
%%%%%%%%%%%%%%%%

\subsection{Magnetic field evolution in the pulsar population?}

From the whole population in the ATNF Pulsar Catalogue, we extracted a sample of pulsars for which $\ddot{\nu}$ has a quoted error smaller than 10\%. From this preselected sample, we excluded all pulsars in binaries and those with very short periods ($P < 15$ ms), likely to be recycled, and we doubled checked our list with the more recent review by \cite{hobbs10}, the first large-scale analysis of pulsar timing noise over time-scales $>10$ yr, which led us to remove several more pulsars whose revised values were inconsistent with the ATNF data, or with larger errors. After this selection, our sample contains 118 radio-pulsars, about half of which have negative braking indices.

We show in the top panel of Fig.~\ref{fig:bi_taub} the observed distribution of braking indices as a function of $\tau_c$. We considered the characteristic magnetic field evolution time-scale $\tau_B$ as given by eq.~(\ref{eq:hatn}), and separated the sample into two groups, those with positive (middle panel) and negative (bottom panel) $\tau_B$. The eight pulsars discussed in \cite{espinoza11} are marked with squares. With all due cautions regarding the uncertainties associated to these measurements, it is worth mentioning some interesting trends visible in this plot:

\begin{itemize}
\item
all young objects seem to have always $n<3$, which can be a hint of an increasing dipolar magnetic field;\footnote{We use the values collected in \cite{espinoza11} for this plot.}
\item
all middle age objects ($10^4$--$10^5$ yr), except Vela, that survived to our conservative selection criteria have negative time derivatives of $B_p$;\footnote{We did not include seven more objects, that fall in this region, with a quoted error of $\ddot{\nu}$ smaller than 10\% in the ATNF pulsar database. They are not considered in \cite{hobbs10}, probably because they did not have 10 yr of accumulated data, although their values have not been reported to change.}
\item
for old objects, there is no correlation at all, and there are similar numbers of objects with positive and negative derivatives of the magnetic field; the typical evolution time-scales are in the range $10^3$--$10^5$ yr.
\end{itemize}
The strong correlation between $|n|$ and $\tau_c$ seen in the old objects of the top panel simply reflects the definition of $n$, eq.~(\ref{eq:hatn}), with the additional piece of information that $\tau_B$ does not seem to be correlated with $\tau_c$ (see middle and bottom panels).

Note that our criteria automatically select the objects with a dominant contribution of $\ddot{\nu}$ in the timing phase residuals (i.e., the cleanest cubic lines in Fig.~3 of \cite{hobbs10}, e.g. B0114+58). This introduces a bias toward objects with high braking indices, and against pulsars with residuals dominated by higher order terms (e.g. B0136+57) or quasi-periodic terms (e.g. B1642-03 or B1826-17) in the time-dependent phase \citep{lyne10}. In the latter cases, the estimates of $\ddot{\nu}$ are subject to larger uncertainties, and $n$ strongly depends on the analysed time interval \citep{hobbs10}. However, we have checked that the general trends do not change if the sample is enlarged by including pulsars with quoted errors in $\ddot{\nu}$ up to 80\%. This simply increases the statistics ($>300$ sources) and includes some objects with lower value of $|n-3|$ (shorter $|\tau_B|$).

%%%%%%%%%%%%%%%%%
\begin{figure}
\centering
\includegraphics[width=0.45\textwidth]{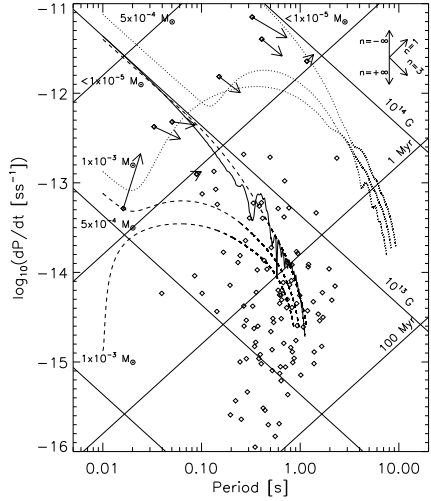}
\caption{Evolutionary tracks in the $P$-$\dot{P}$ diagram during the first 3 Myr for: model A13 (three dashed lines) and model A14 (three dotted lines), both for different values of 
accreted mass $M_a=[0.1,5,10] \times 10^{-4} M_\odot$; model A13O (solid) without accretion.}
\label{fig:bi_trajectories}
\end{figure}
%%%%%%%%%%%%%%%%%

\subsection{Expected evolution from theoretical models.}

We now present some numerical simulations including an episode of hypercritical accretion during their formation, as illustrated in \S~\ref{sec:cco_hmf}. In Fig.~\ref{fig:bi_trajectories} we plot the expected evolutionary tracks in the $P$-$\dot{P}$ diagram, compared to the pulsars of our sample. We show results for three initial magnetic field configurations, all of type A (crust-confined field, see \S~\ref{sec:initial_b}). Two of them are purely dipolar, although during the evolution other multipoles and toroidal magnetic field are naturally created. They differ by the initial magnetic field strengths: $B_p^0=10^{13}$ G (model A13, dashed lines) and $B_p^0=10^{14}$ G (model A14, dotted lines). In each case we compare results with three values of the total accreted mass. The third initial configuration (model A13O, solid line) has an initial dipolar field of $B_p^0=10^{13}$ G, but with an additional octupolar component of strength $B_o^0=3.5\times10^{14}$ G (at the pole). This model serves to make explicit the effect of the complex initial geometries.

For the eight youngest pulsars, we indicate their predicted movement for the next 2.5 kyr with arrows, assuming the present value of $n$, from \cite{espinoza11}, remains constant. The direction of the tangent vector to a given track is related to the braking index. For reference, we indicate in the legend on the upper right corner that $n=3$ and $n=1$ imply constant inferred $B_p$, and constant characteristic age, respectively.

The initial period $P_0$, which was assumed to be 0.01 s in all cases, only affects the early stage, while $\dot{P} \lesssim P_0/t$ (with $t$ being the real age), and $P\simeq P_0$. Models with deep submergence of the magnetic field but with different initial periods also have vertical trajectories shifted to the left/right depending on $P_0$, and quickly cross the range of $\dot{P}$ where the bulk of pulsars lie. At late times, tracks coming from the same model but with different $P_0$ are indistinguishable, typically converging after 

\begin{equation}
t \sim 30 \left(\frac{P_0}{0.01 s}\frac{10^{13}G}{B_p^0}\right)^2 \mbox{yr}~. 
\end{equation}
Therefore, $P_0$ has an appreciable long-term effect only in the deep submergence case ($M_a \gtrsim 10^{-3} M_\odot$), for which $B_p^0$ is strongly reduced after submergence: in these models, the first few $10^4$ yr are spent in the vertical trajectories, with $n\ll 0$. Note that in this reemergence phase there is no correlation between the real and characteristic ages.

When reemergence of the magnetic field has almost been completed, the trajectories reach the high-$\dot{P}$ region (i.e., largest $B_p$) and progressively bend. The extreme braking indices of PSR J1734-3333 ($n=0.9\pm0.2$; \citealt{espinoza11}) and PSR J0537-6910 ($n\sim -1.5$; \citealt{middletich06}) would be consistent with the last stage of the reemergence after a deep submergence into the inner crust. On the other hand, in the shallow submergence models ($M_a \sim 10^{-5}$--$10^{-4} M_\odot$), after accretion stage $B_p\simeq B_p^0$. These tracks initially run almost along the iso-magnetic lines: pulsars with $n$ slightly less than 3 are compatible with this scenario.

Independently of the early reemergence phase (if any), tracks with the same $B_p^0$ converge at middle-age, and have slopes corresponding to $n>3$,  characteristic of the slow Ohmic dissipation. During this phase, there is a correlation between real and characteristic ages, with typically $\tau_c$ being a factor of few longer than the real age. Some tracks show visible oscillations produced by the Hall activity when the star is cold enough ($t\gtrsim 10^5$ yr); in particular, model A13O (solid line) clearly shows that any complex initial geometry may have a distinct signature on the timing properties of pulsars.

We emphasize that we did not attempt to fit individual objects: our purpose with this sample of models is simply to show that, with reasonable assumptions, it is possible to explain the variability in the observed range of braking indices of young pulsars and to predict their evolutionary paths in the $P$-$\dot{P}$ diagram.

%%%%%%%%%%%%%%%%%%%%%%%%
\subsection{Braking index and evolution time-scale for realistic magnetic field evolution models.}

%%%%%%%%%%%%%%%%%
\begin{figure}[t]
\centering
\includegraphics[width=0.45\textwidth]{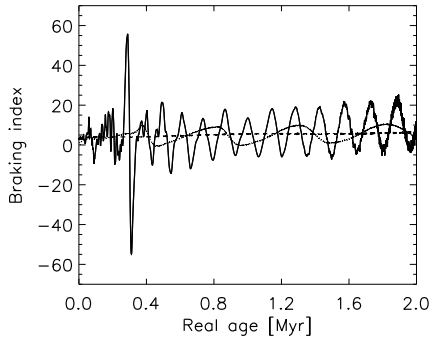}
\caption{Braking index as a function of age for model A13 (dashed line), model A14 with $M_a = 10^{-3} M_\odot$ (thin solid line), and model A13O (thick solid line).} 
\label{fig:bi_evo}
\end{figure}
%%%%%%%%%%%%%%%%%

In Fig.~\ref{fig:bi_evo} we plot the braking index evolution for three representative models. Note that the horizontal axis in Fig.~\ref{fig:bi_taub} represents the characteristic age $\tau_c$, while in Fig.~\ref{fig:bi_evo} we show our results as a function of the real age of each model, so that a direct comparison is not possible. We also used a linear scale in this plot to show the quasi-periodic oscillations during the long-term evolution more clearly. For fields $\lesssim 10^{13}$ G and simple dipolar geometries (dashed line) the braking index at late times is $n>3$, but its absolute value is low. In contrast, for strong dipolar fields, or for weak dipolar components but with strong higher order multipoles, Hall-drift induced oscillations appear sooner or later and,  in some situations, have large amplitudes that result in very high absolute values of the braking index. It is particularly interesting to compare models A13 and A13O (dashed and solid line),  which have the same initial dipolar component, representative of a typical pulsar (at $10^6$ yr the dipolar field is about $3$--$5\times 10^{12}$ G). However, the presence of a strong octupolar component at birth results in a radically different braking index behaviour, even if $P$ and $\dot{P}$ are similar. The amplitude of the oscillations and whether the modes are damped or excited depends on particular details of the small-scale structure of the magnetic field, which is unknown. The exact age at which these oscillatory modes are excited is connected to the temperature of the star, and therefore to its cooling history and internal physics (neutrino emission processes, superfluid gaps, etc.). For the standard cooling scenario this happens at $\approx 10^5$ yr. We can also observe that as the star evolves and the magnetic diffusivity decreases, the frequency of the dominant mode may vary (see solid line). For example, at early times in model A13O we observe variability on shorter time-scales (few kyr), while later we see longer oscillation periods ($10^5$ yr). After 1.5 Myr higher frequency modes seem to be growing again.

We also note that the strong oscillations, with positive and negative values of $n$, are seen only if the magnetic field evolution is dominated by the Hall term. However, the comparison with observational data in chapter~\ref{ch:unification}, indicates that strong magnetic field dissipation is taking place in the inner crust, with a consequent very large, but always positive, braking index.

%%%%%%%%%%%%%%%%%
\begin{figure}[t]
\centering
\includegraphics[width=0.45\textwidth]{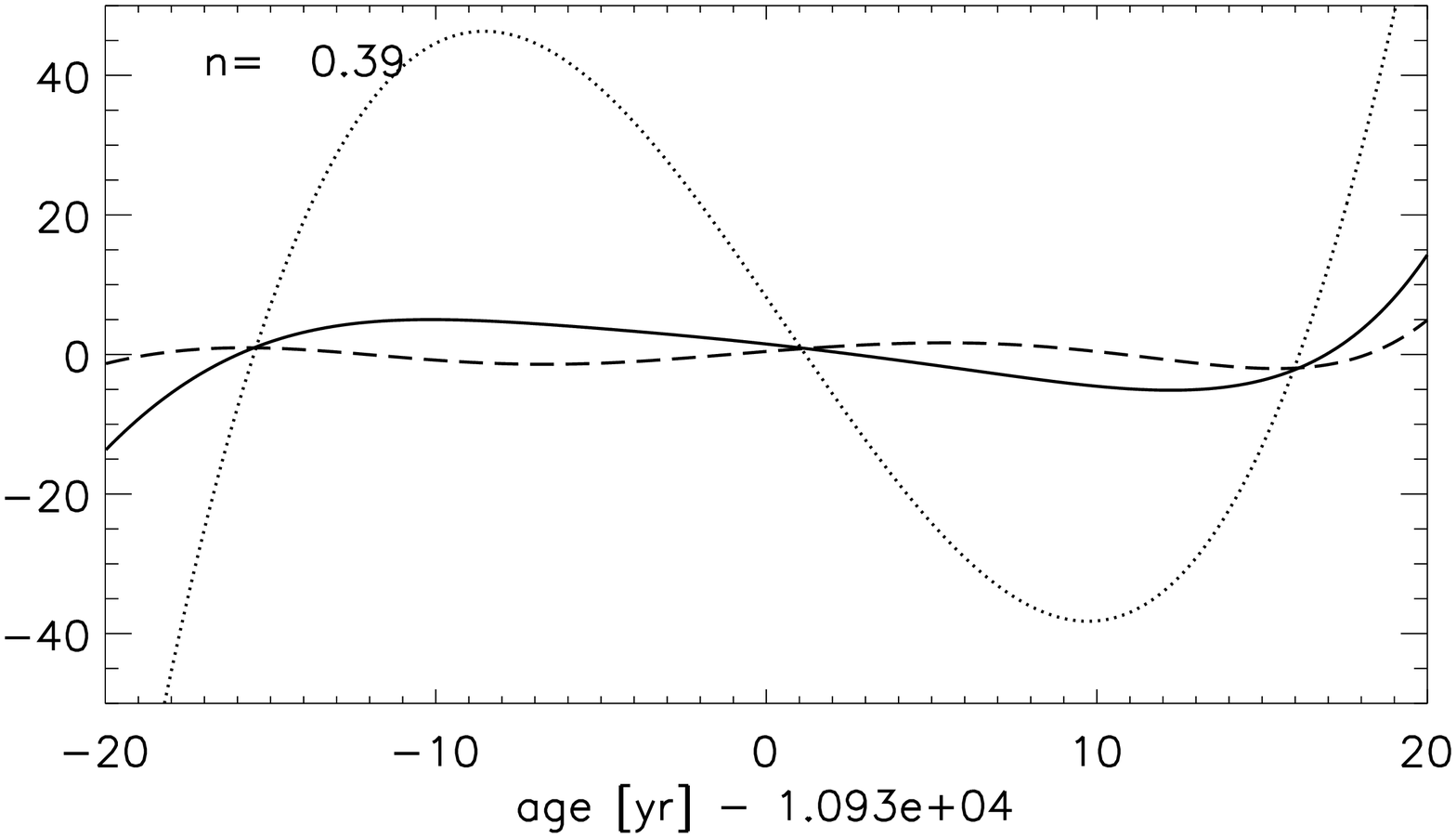}
\includegraphics[width=0.45\textwidth]{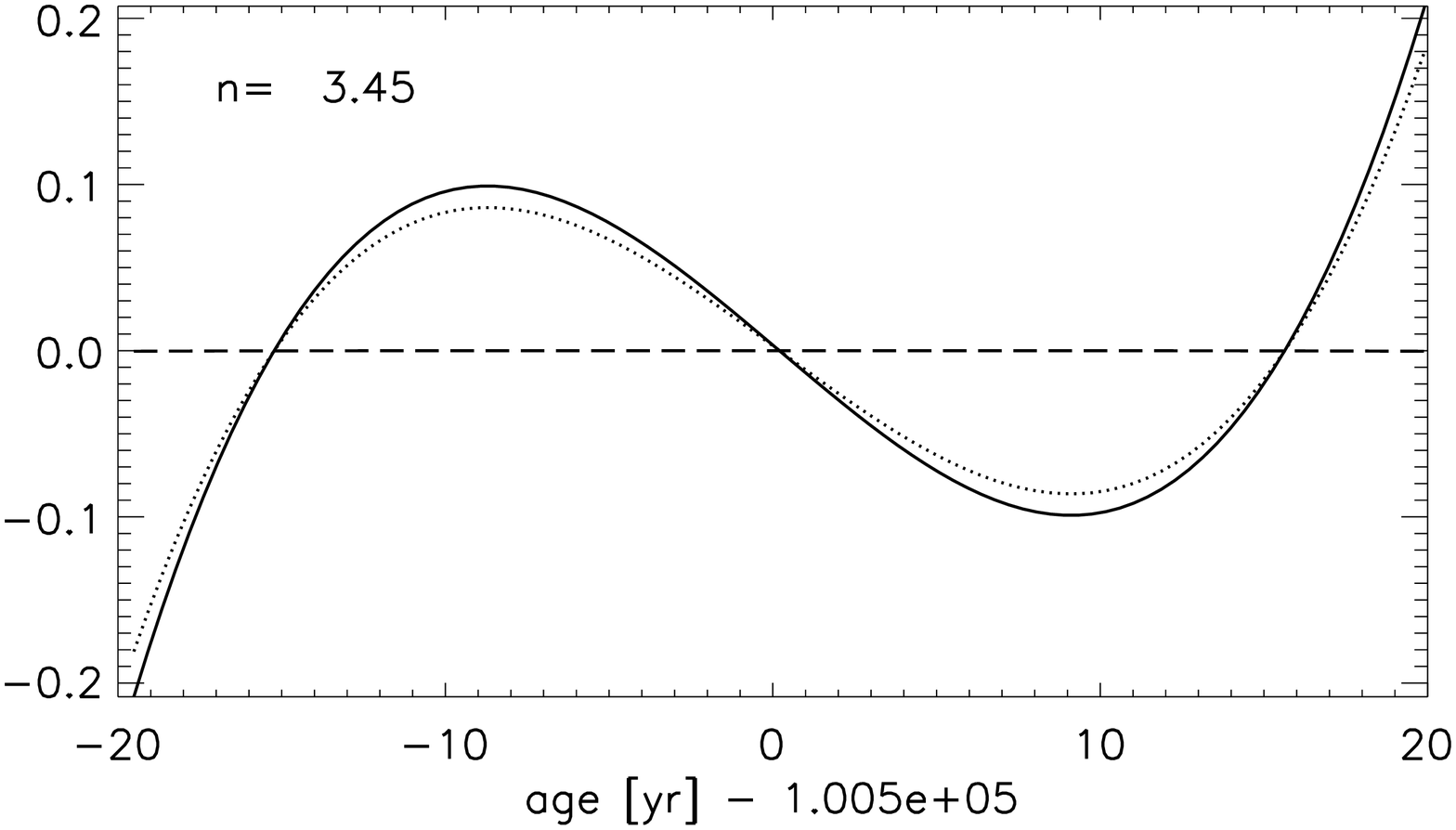}
\caption{Phase residuals for model A13O around $t_0=10.93$ kyr (left panel) and $t_0=100.5$ kyr (right panel), with the corresponding value of $n$ indicated in each case. We show residuals after removing the best-fit solution including quadratic terms (solid) or including cubic terms (long dashes); dotted lines show the cubic residuals obtained  assuming a constant $B_p$.}
\label{fig:bi_residuals}
\end{figure}
%%%%%%%%%%%%%%%%%

%%%%%%%%%%%%%%%%%%%%%%%%%%%%%
\subsection{Timing residuals.}

In timing analysis of radio pulsars, an important piece of information is the study of the residuals, i.e, the phase differences between the observed signal and the best-fit model, including the frequency and frequency derivative. As shown in detail in \cite{hobbs10}, there is a rich variety of shapes in the residuals of RPPs. As we have shown, the strong Hall-induced interplay between different multipoles and the toroidal magnetic field produces a complex evolution of $B_p$. To compare with observational timing analysis, we proceeded as follows: first, we choose a short interval of 40 yr in our simulations, centered on a fixed time denoted by $t_0$. This interval is similar to the longest periods for which phase-coherent timing analysis for radio pulsars can be performed. We obtain $\nu(t)=1/P(t)$ by integration of the period, and the phase $\Phi(t)=\int_{t_0}^{t} \nu(t')dt'$. Finally, we fit our synthetic time-dependent phase with a quadratic function
\begin{equation}
\Phi(t) = \Phi_0 + \nu_f (t-t_0) + \frac{1}{2}\dot{\nu}_f (t-t_0)^2 ~.
\end{equation}
The results of the fit, $\nu_f$ and $\dot{\nu}_f$, are of course in the range of values
of the {\it real} quantities in the time interval.

In Fig.~\ref{fig:bi_residuals} we show the phase residuals for model A13O (solid lines) at $t_0=10.93$ kyr (left panel) and $t_0=100.5$ kyr (right). During the two analysed intervals, the mean values of the dipolar magnetic field are $\bar{B_p}=8.1\times 10^{12}$ G and $\bar{B_p}=5.3\times 10^{12}$ G. The cubic shape indicates that the residuals are dominated by the next term in the Taylor series (i.e., red noise). Including in the fitting function a cubic term allows one to measure $\ddot{\nu}$ and the braking index ($n=0.39$ and $n=3.45$, respectively). With dashes we show the fourth-order residuals after subtracting the third-order term in the fitting function. For comparison, we also show with dotted lines the third-order residuals obtained assuming a constant value $B_p(t)=\bar{B_p}$ (that leads to $n=3$). 

In the first time interval, the change of magnetic field, $\delta B_p=1.3\times 10^{10}$ G $=1.6\times 10^{-3} \bar{B_p}$, is strong enough for the cubic residuals to be visibly different from the constant field case. The value of $|\ddot{\nu}|$ (and of $|n|$) is low, and the residuals of fourth order are significant. In the second time interval, $\delta B_p=1.4\times 10^8$ G $=2.6\times 10^{-5} \bar{B_p}$, and $n$ is close to 3. As a consequence, the deviation from the constant field spin-down behavior is slow and fourth-order residuals are orders of magnitude weaker (dashed line).

\begin{figure}[t]
\centering
\includegraphics[width=0.42\textwidth]{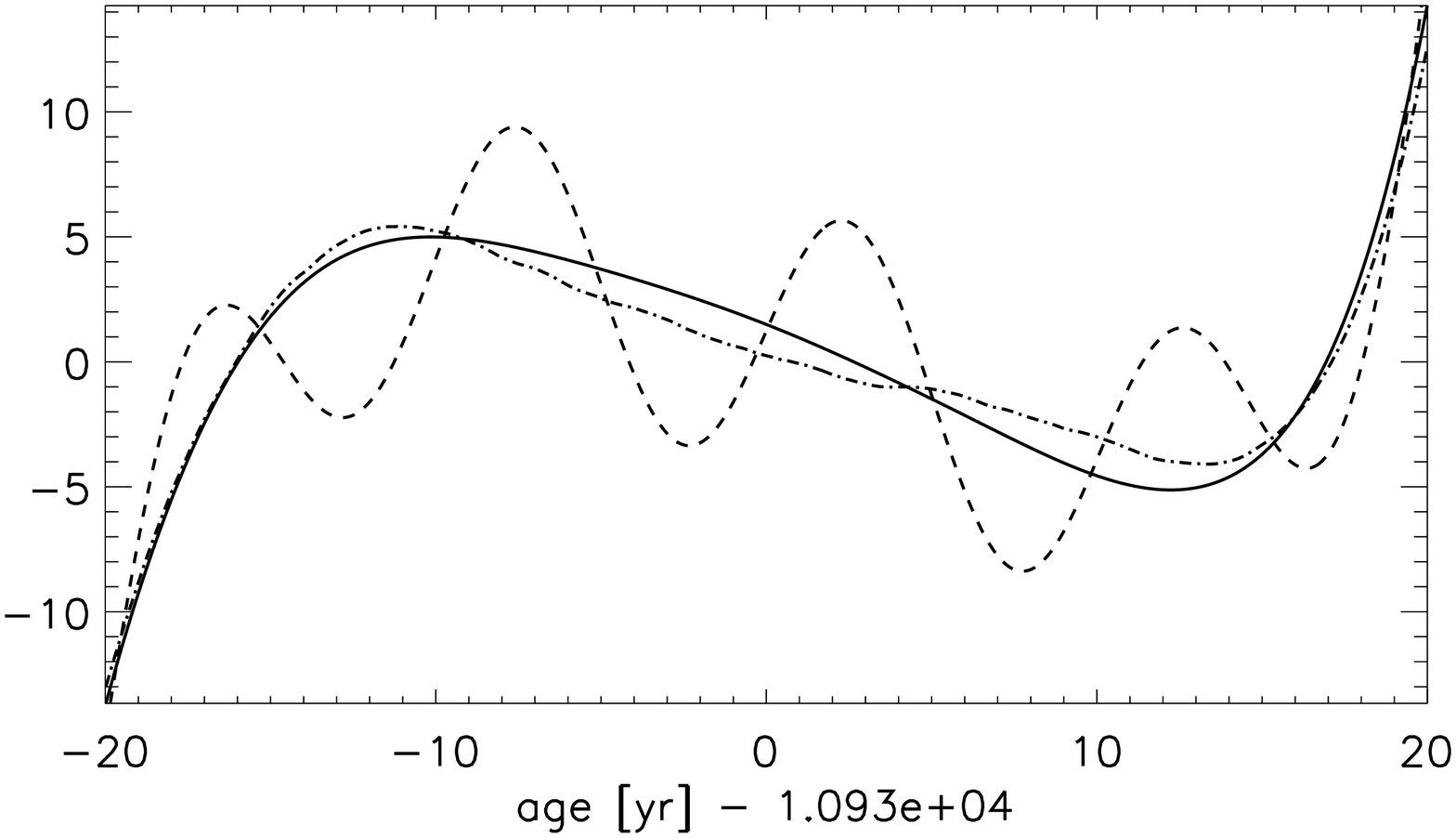}
\caption{Comparison of the cubic residuals of the same theoretical model (solid line) shown in the left panel of Fig.~\ref{fig:bi_residuals}, with the result after adding a sinusoidal perturbation with frequency $0.1$ yr$^{-1}$ (dashed) or a random perturbation (dot-dashed). For both perturbation forms we considered a maximum amplitude of $(\delta B_p/2)=6.5\times 10^9$ G.}
\label{fig:bi_residuals_c}
\end{figure}

To investigate the effect of short-time irregularities, we repeated the process but artificially added two additional sources of noise to our theoretical values of $B_p(t)$ of the first time interval, with amplitude of the same order as the theoretical variations $\delta B_p$. In Fig.~\ref{fig:bi_residuals_c} we show the residuals obtained by adding a sinusoidal perturbation, $B_1=(\delta B_p/2)\sin(2\pi t/T)$, with $T=10$ yr (dashed lines), or a random perturbation of maximum absolute value $\delta B_p/2$ (dash-dotted lines). In our analysis, we sampled values every $\sim 10^{-2}$ yr, comparable with the typical integration time in observational data. We checked that increasing the time interval strongly reduces the contribution of the random noise at a fixed amplitude, as expected. Similarly, the periodic short-term noise is reduced when the integration interval becomes much longer than the perturbation period.

%%%%%%%%%%%%%%%%%
%%%%%%%%%%%%%%%%%

Note that small perturbations of $B_p$ produce visible effects, but the cubic residual still dominates. In contrast, in many pulsars a smooth cubic phase residual is clearly seen, which means that the spin-down is very stable, short-term irregularities are completely negligible (or averaged out in the long observation period), and the measure of a braking index (even a high value) dominated by the secular evolution is robust. Glitches should appear as cusps in these plots (e.g. see data for PSR B0154+61 in \citealt{hobbs10}), but a more detailed investigation in this direction is beyond the scope of this work.

\section{Discussion.}

We have explored the influence of the long-term magnetic field evolution on the timing properties of CCOs and young pulsars, taking into account the effects of having an initial hyper-critical accretion phase, lasting weeks or months. CCOs properties are compatible with the hidden magnetic field scenario. In this framework, the lack of detection of orphaned CCOs (i.e., ''anti-magnetars`` whose SNRs have faded away) is explained by the evolution of CCOs into normal pulsars, after the re-emergence phase. However, we note that selection effects both in radio and X-ray thermal radiation would play against the detection of these objects, thus any claim should be supported by population synthesis study. Furthermore, the number of CCOs with timing properties (three) is much smaller than the number of unconstrained parameters of the model (initial magnetic field, amount of mass and geometry of accretion, initial period...), thus a fit to individual objects is not physically constraining.

The large values of pulsar braking indices, up to $|n|=10^8$, are usually addressed to magnetospheric effects. Here we have explored the contribution of the long-term evolution. For the young objects, the reemergence of magnetic field is a viable explanation for the commonly observed values $n<3$. For the old objects, the models we have explored show large oscillations of positive and negative values, but limited to $|n|\lesssim 100$ in models with large, multipolar magnetic fields, and low impurity parameter everywhere. Considering also that the upper limit of $\sim 12$ s indicates a highly resistive inner crust, we conclude that the long-term evolution is not the dominant driver of the observed values of braking indices in old pulsars. This leaves the magnetosphere as the most promising candidate to influence the pulsar timing noise (see, e.g., \citealt{bucciantini06,kramer06,lyne10}).

\chapter{Conclusions}\label{ch:conclusions}

In this thesis we have studied in detail different theoretical and observational aspects related to magnetic fields in neutron stars, with special attention to the magnetosphere and the evolution of the internal field. The understanding of neutron star physics involves fundamental physics, due to the extreme conditions met in these compact stars. In this sense, it is evident that magnetic field is a main actor determining the observational properties, and its evolution and features are worth studying. The diversity of observational manifestations has historically led to dub different classes and sub-classes of neutron stars. Thousands of isolated neutron stars have been detected in different energy bands, but only in the recent years, thanks to the increasing data, a unified picture has begun to emerge.

The modeling of force-free magnetospheres is important to understand the magnetar phenomenology. From an analytical or semi-analytical point of view, practical results are limited to the few cases in which restrictive mathematical assumptions can be made. Therefore, we have faced the problem by constructing and thoroughly testing a numerical code that relaxes an arbitrary initial model to a force-free solution with given boundary conditions. With our numerical simulations we can find general solutions of twisted magnetospheres with complex geometries.

Our main conclusion is that the possible family of relaxed configurations depends on the initial conditions: the solution is not unique. This reflects the freedom we have in choosing the pattern of currents, for a given radial magnetic field at the surface. Starting with two different configurations with the same value of volume-integrated helicity does not necessarily lead to the same final configuration. Thus we were able to find a variety of solutions, which can be qualitatively different from the self-similar models. Some of the new magnetosphere models that we found are characterized by a high degree of nonlinearity, twists up to a few radians, and a nontrivial functional form of the threading current.

One of the most testable imprints of the magnetosphere is present in the X-ray spectra of magnetars. In these objects, the magnetosphere is supposed to be responding to the strong magnetic activity in the interior. The magnetospheric plasma is very dense compared with standard pulsars and distorts the thermal spectrum emerging from the surface. We have studied spectral changes in a few of our force-free models, finding that the magnetic field geometry has a visible effect, but it is very hard to disentangle it from similar effects due to variations of other parameters, like the velocity distribution of plasma particles. A fully consistent approach to the problem, considering both the plasma dynamics and the large-scale magnetic configuration, is still missing, although promising studies have been recently published. With further advances, and enlarging the wealth of observational data, it will be possible to constrain the geometric and dynamical properties of the magnetosphere.

We have also presented a comprehensive study of the magneto-thermal evolution of isolated neutron stars, exploring the influence of the initial magnetic field strength and geometry,  neutron star mass, envelope composition, and relevant microphysical parameters such as the impurity content of the innermost part of the crust (the pasta region). Our state-of-the-art, 2D magneto-thermal code, is the first able to work with large magnetic field strengths, while consistently including the Hall term throughout the full evolution. We have reviewed the basic ingredients regulating the cooling: thermal and electrical conductivities, heat capacity, and superfluidity. At these extreme conditions of pressure, magnetic field and density, physical processes are not always well understood, and often only theoretical calculations are available. In particular, the effect of superfluidity on the cooling is thought to be very important, and many works have been dedicated to its study. However, the aim of this thesis was to explore the effects of the magnetic field evolution, advancing in our understanding of its imprint on observations.

The Hall term plays a very important role in the long-term evolution, strongly enhancing the dissipation of energy over the first $\sim 10^6$~yr of neutron star life, with respect to the purely resistive case. This is due to two main reasons: the generation of small structures and currents sheets, and the gradual compression of currents and toroidal magnetic field towards the crust/core interface. Hence, the rate of field dissipation strongly depends on the resistivity given by the amount of impurities in the innermost region of the crust.

A main result of our work is that there is a direct correlation between the maximum observed spin period of isolated X-ray pulsars and the existence of a highly resistive layer in the inner crust of neutron stars. This is a plausible evidence for the existence of nuclear pasta (or a disordered phase) near the crust/core interface. The precise density at which the pasta phase appears, the geometry of the nuclear clusters, and their charge distribution is strongly dependent on the symmetry energy of the nuclear interaction and its density dependence. In any case, the imprint of the poorly known nuclear properties of the inner crust are present in the evolution of the magnetic field, and our results open the possibility of constraining these parameters using a new astrophysical observable. In particular, as present and future space missions, such as {\em LOFT} (Large Observatory For X-ray Timing; \citealt{feroci12}) and {\em ATHENA+} \citep{nandra13}, keep increasing the statistics of X-ray pulsars, and realistic theoretical models are used as input for neutron star population synthesis studies, we will be able to accurately constrain the properties of the inner crust of neutron stars and, therefore, the equation of state of dense matter.

For weakly magnetized objects, low mass stars are systematically brighter than high mass stars. Our results also show that this separation is smeared out for highly magnetized stars. We also confirm that light-element envelopes are able to maintain a higher luminosity (up to an order of magnitude) than iron envelopes for a long period of time, $\gtrsim 10^4$~yr, regardless of the magnetic field strength. 

In order to compare theoretical models with observations, we have revised the observational data for all currently known isolated neutron stars with clearly detected surface thermal emission. The sample includes magnetars, high-$B$ pulsars, standard radio pulsars, XINSs, and CCOs. For all of them we have performed a homogeneous, systematic analysis of the X-ray spectra, inferring the thermal luminosity, and trying to reduce at the minimum the systematic errors due to different data analysis and modeling. We published on--line the results of our analysis, including detailed references for every source.

A comparison between a number of theoretical models and the observations (both timing and luminosities) has shown that, for the objects with relatively low fields ($B_p\lesssim 10^{14}$~G), the magnetic field has little effect on the luminosity. These objects, of which the RPPs are the most notable representatives, have luminosities compatible with the predictions by standard cooling models, with the dimmest sources (Vela pulsar, PSR~B2334 and PSR~J1740) requiring an iron envelope and the activation of fast cooling processes.

Most magnetars, with inferred magnetic fields of a few times $10^{14}$~G, have luminosities too high to be compatible with standard cooling. The magneto-thermal evolutionary models with initial magnetic fields $B_p^0\sim 3$--$5 \times 10^{14}$~G can account for their range of luminosities at the corresponding timing properties. As these objects evolve and their magnetic fields dissipate, their observational properties become compatible with those of the XINSs.

Finally, the most extreme magnetars, endowed with magnetic fields of strength $\gtrsim 10^{15}$~G as inferred from their timing parameters, are characterized by the highest luminosities, barely compatible with the $B_p^0=10^{15}$~G cooling curve with an iron envelope. However, for the same initial magnetic field, a light-element envelope is able to account for the luminosity of even the brightest objects. Note that most of these objects show non-thermal emission in the  hard X-ray band, and the estimate of their dipolar magnetic fields from timing properties is unreliable, due to the large timing noise. We suggest that, at least for some of them, the magnetosphere plays an important role in influencing timing properties and their magnetic fields may not be higher than in other magnetars.

We have found that the initial magnetic field configuration also plays an important role in the observational properties of the highly magnetized neutron stars. The presence of strong currents circulating in the crust is required to explain the timing and spectral properties of isolated neutron stars. As a matter of fact, if the currents sustaining the magnetic field are mainly circulating in the core, the very high conductivity of the core results in a much slower dissipation (time-scales of Gyr compared to $10^4-10^5$ yr or less for the crust), and neither the observed enhanced luminosity nor the maximum spin period can be explained. This does not imply that the magnetic field cannot penetrate in the core, but only that at least a large fraction of the total magnetic energy has to be stored in the crust. The physical origin of this confinement is unclear, and further studies regarding the initial configuration in MHD equilibrium, and the evolution of magnetic field in the superconducting core could help to clarify this point.

We have also studied the {\em hidden magnetic field scenario} for CCOs, proving that it is not in contradiction with any observational data. An initial phase of magnetic field submergence, caused by the post-supernova fallback of debris, can explain the puzzling behaviour of these objects, that seem to have low external magnetic fields together with hints for more complex and intense magnetic fields below the surface. We also consider the possibility that not all the CCO candidates are of the same nature: large, magnetar-like hidden magnetic fields are particularly suitable for those CCOs showing temperature anisotropies. On the other hand, some of the young RPPs could be in their final stage of reemergence, representing the connection between CCOs and RPPs or magnetars. This is in agreement with the braking indices observed in pulsars: we have quantified the contribution expected from the long-term magnetic evolution to the observed timing noise of pulsars.

In summary, our magneto-thermal simulations help to paint a unified picture of the variety of observational properties of isolated neutron stars, and their evolutionary paths. Our results can account for the overall phenomenology, whereas in-depth testing of each specific magnetic field configuration will require detailed spectral and timing modeling of the thermal component of each source. Further, deeper observations and discoveries of magnetars will allow in the next years to refine the comparison with data and constrain the theoretical models. In this sense, the apparently puzzling variety in the neutron star zoo can be understood in terms of only a few intrinsic differences, such as mass, age, inclination angle and initial magnetic field, and some observational biases affecting the detectability in different energy bands, like distance, position in the sky, and viewing angle.

\chapter*{Conclusiones}

En esta tesis se han estudiado en detalle los diferentes aspectos te\'oricos y observacionales relacionados con los campos magn\'eticos de las estrellas de neutrones, con especial atenci\'on a la magnetosfera y la evoluci\'on del campo interno. Los procesos f\'isicos en las estrellas de neutrones implican muchas ramas de la f\'isica, algunas de f\'isica fundamental, como la estructura de la materia en las condiciones extremas que se encuentran en estos objetos compactas. En este sentido, es evidente que el campo magn\'etico es un actor principal en la determinaci\'on de las propiedades observacionales, y su
evoluci\'on es un aspecto importante a estudiar. La gran diversidad de manifestaciones observacionales ha llevado hist\'oricamente a agrupar las estrellas de neutrones aisladas en clases y subclases diferentes. Entre las miles de estrellas de neutrones detectadas en diferentes bandas de energ\'ia, un panorama unificado ha comenzado a surgir en los \'ultimos a\~{n}os, debido a la mejora en cantidad y calidad de los datos.

El modelado de magnetosferas en la aproximaci\'on {\it force-free} es importante para entender la fenomenolog\'ia de los magnetars. Desde un punto de vista anal\'itico o semi-anal\'itico, los resultados pr\'acticos se limitan a unos pocos casos en los que se pueden hacer suposiciones matem\'aticas restrictivas. Por lo tanto, nos hemos enfrentado al problema construyendo un c\'odigo num\'erico que usa un m\'todo de relajaci\'on para llevar un modelo inicial arbitrario a una soluci\'on {\it force-free}, con condiciones de contorno dadas. Con nuestras simulaciones num\'ericas podemos encontrar soluciones generales de magnetosferas con geometr\'ias complejas.

Nuestra conclusi\'on principal de esta parte de la tesis es que la posible familia de soluciones depende de las condiciones iniciales: la soluci\'on no es \'unica. Esto refleja la libertad que tenemos en la elecci\'on de la configuraci\'on de corrientes, para un mismo campo magn\'etico radial conocido en la superficie.
Partiendo de dos configuraciones diferentes con el mismo valor de helicidad total, no se llega necesariamente a la misma configuraci\'on final. As\'i, hemos sido capaces de encontrar una variedad de soluciones que pueden ser cualitativamente diferentes de los anteriores modelos auto-semejantes. Algunos de los nuevos modelos de magnetosferas que hemos encontrado se caracterizan por un alto grado de no linealidad, con lineas de campo retorcidas, y una forma funcional no trivial para las corrientes.

Uno de las huellas m\'as visibles de la geometr\'{\i}a de la magnetosfera est\'a presente en el espectro de rayos X de los magnetars. En estos objetos, se supone que la magnetosfera es compleja y din\'amica, debido a su fuerte acoplamiento con la actividad magn\'etica en el interior. El plasma magnetosf\'erico es muy denso en comparaci\'on con los p\'ulsares est\'andar y distorsiona el espectro t\'ermico que emerge de la superficie. Hemos estudiado los cambios espectrales en algunos de nuestros modelos, encontrando que la geometr\'ia del campo magn\'etico tiene un efecto visible, pero que es muy dif\'icil de separar de efectos similares debidos a la variaci\'on de otros par\'ametros, como la distribuci\'on de la velocidad de las part\'iculas del plasma. Un enfoque totalmente coherente del problema, teniendo en cuenta tanto la din\'amica del plasma como la configuraci\'on magn\'etica a gran escala, todav\'ia no existe, aunque algunos estudios prometedores han sido publicados recientemente. Con los nuevos avances, y el aumento de datos de observaci\'on, en un futuro pr\'oximo
ser\'a posible restringir las propiedades geom\'etricas y din\'amicas de la magnetosfera.

En la parte central de la tesis, hemos presentado un estudio exhaustivo de la evoluci\'on magneto-t\'ermica de estrellas de neutrones aisladas, explorando la influencia de  la intensidad y geometr\'ia del campo magn\'etico inicial, la masa de la estrella, la composici\'on de la envoltura, y algunos par\'ametros 
microf\'isicos, como el contenido de impurezas de la parte m\'as interna de la corteza (la regi\'on de pasta nuclear). Nuestro c\'odigo magneto-t\'ermico 2D de vanguardia es el primero capaz de calcular con campo magn\'eticos intensos, incluyendo el t\'ermino Hall a lo largo de toda la evoluci\'on. Hemos revisado los principios b\'asicos que regulan el enfriamiento: conductividad t\'ermica y el\'ectrica, capacidad de calor y superfluidez. En estas 
condiciones extremas de presi\'on, campo magn\'etico y densidad, los procesos f\'isicos no son siempre bien comprendidos, no hay apenas datos experimentales y a menudo s\'olo est\'an disponibles  c\'alculos te\'oricos . En particular, se cree que el efecto de la superfluidez en el enfriamiento es muy importante, y muchos trabajos se han dedicado a su estudio. El objetivo de esta tesis ha sido estudiar los efectos de la evoluci\'on del campo magn\'etico y avanzar en nuestra comprensi\'on de su huella en las observaciones.

El t\'ermino Hall juega un papel muy importante en la evoluci\'on a largo plazo, aumentando fuertemente la disipaci\'on de la energ\'ia en los primeros $\sim 10^6$ a\~{n}os de vida, con respecto al caso puramente resistivo. Esto se debe a dos razones principales: la generaci\'on de estructuras peque\~{n}as y discontinuidades magn\'eticas, y la compresi\'on gradual de las corrientes y del campo magn\'etico toroidal hacia la interfaz de corteza / n\'ucleo. 
Por lo tanto, la tasa de disipaci\'on de campo depende en gran medida de la resistividad el\'ectrica causada por la cantidad de impurezas en la regi\'on m\'as interna de la corteza. 

En este sentido, uno de los resultados principales de nuestro trabajo es que existe una correlaci\'on directa entre el per\'iodo de rotaci\'on  m\'aximo de los 
p\'ulsares aislados de rayos X y la existencia de una capa de conductividad el\'ectrica baja en la corteza interna de las estrellas de neutrones. Esta es una evidencia convincente de la existencia de la pasta nuclear (o de una fase desordenada similar) cerca de la interfaz de corteza/n\'ucleo. 
La densidad precisa a la que aparece la fase de la pasta, la geometr\'ia de los grumos nucleares y su distribuci\'on de carga, es fuertemente dependiente de la energ\'ia de simetr\'ia de la interacci\'on nuclear y de su variaci\'on con la densidad. En cualquier caso, la huella de las propiedades nucleares de la corteza interior est\'an presentes en la evoluci\'on del campo magn\'etico, y nuestros resultados abren la posibilidad de limitar algunos par\'ametros utilizando un nuevo observable astrof\'isico. A medida que las misiones espaciales actuales y futuras, como {\em LOFT} \citep{feroci12} y {\em ATHENA+} \citep{nandra13}, vayan aumentando la estad\'istica de p\'ulsares de rayos X, y los modelos te\'oricos realistas se refinen y utilicen en estudios de s\'intesis de poblaci\'on de estrellas de neutrones, seremos capaces de limitar con precisi\'on las propiedades de la corteza interior de las estrellas de neutrones y, por lo tanto, la ecuaci\'on de estado de la materia densa.

Nuestros resultados tambi\'en muestran que, mientras que para los objetos d\'ebilmente magnetizados, las estrellas de baja masa ($M\lesssim 1.4 \, M_\odot$) son sistem\'aticamente m\'as brillantes que las estrellas de masa alta, esta separaci\'on se suaviza para las estrellas altamente magnetizadas. Adem\'as, se confirma que las envolturas de elementos ligeros son capaces de mantener una luminosidad superior (hasta un orden de magnitud) que las envolturas de hierro durante un periodo de tiempo considerable, $\gtrsim 10^4$ a\~{n}os, independientemente de la intensidad del campo magn\'etico.

Con el fin de comparar los modelos te\'oricos con las observaciones, hemos revisado los datos observacionales de todas las estrellas de neutrones aisladas conocidas en las cuales se haya detectado inequ\'{\i}vocamente emisi\'on termica superficial. La muestra incluye los magnetares, los p\'ulsares de alto campo
magn\'etico, radio-p\'ulsares normales, XINSs y CCOs. Para todos ellos hemos realizado un an\'alisis homog\'eneo y sistem\'atico de los espectros de rayos X, llegando a obtener la luminosidad t\'ermica, y tratando de reducir al m\'inimo los errores sistem\'aticos. 
Hemos publicado en un cat\'aloco online los resultados de nuestro an\'alisis, incluyendo referencias detalladas para cada fuente.

La comparaci\'on entre una serie de modelos te\'oricos y las observaciones ha demostrado que, para los objetos con campos relativamente bajos ($B_p \lesssim 10^{14}$ G), el campo magn\'etico tiene poco efecto sobre la luminosidad. Estos objetos, de los cuales los radio-p\'ulsares son los representantes m\'as conocidos, tienen luminosidades compatibles con las predicciones de los modelos de enfriamiento est\'andar, excepto para las fuentes m\'as d\'ebiles (Vela p\'ulsar, PSR~B2334  y PSR~J1740), que requieren una envoltura de hierro y la activaci\'on de los procesos de enfriamiento r\'apido.

La mayor\'ia de los magnetars, con campos magn\'eticos inferidos de pocas veces $10^{14}$ G, son demasiado luminosos para ser compatibles con el modelo de enfriamiento est\'andar (sin efectos de campo). En cambio, nuestros modelos evolutivos con campos magn\'eticos iniciales $B_p^0 \sim 3$--$5 \times 10^{14}$ G pueden dar cuenta de su rango de luminosidades a las correspondientes edades. A medida que evolucionan y sus campos magn\'eticos se disipan, las propiedades observacionales de los magnetars se vuelven compatibles con las de las XINSs.

Por \'ultimo, los magnetars m\'as extremos, dotados de campos magn\'eticos $\gtrsim 10^{15}$~G, como se deduce de sus par\'ametros de rotaci\'on, se caracterizan por las luminosidades m\'as altas, apenas compatibles con la curva de enfriamiento de campo $B_p^0=10^{15}$ G con una envoltura de hierro. Sin embargo, para el mismo campo magn\'etico inicial, una envoltura de elementos ligeros es capaz de dar cuenta incluso de la luminosidad de los objetos 
m\'as brillantes. Teniendo en cuenta que la mayor\'ia de estos objetos muestran emisi\'on no t\'ermica en la banda de rayos X duros, y la estimaci\'on de sus campos magn\'eticos dipolares es poco fiable, sugerimos que, al menos para algunos de ellos, la magnetosfera juega un papel importante alterando las propiedades de rotaci\'on (variaci\'on del periodo) y sus campos magn\'eticos pueden ser en realidad similares a otros magnetars.

Hemos encontrado que la configuraci\'on inicial del campo magn\'etico tambi\'en juega un papel importante y que se requiere la presencia de fuertes corrientes que circulan en la corteza para explicar las propiedades observadas de las estrellas de neutrones altamente magnetizadas.
De hecho, si las corrientes que sustentan el campo magn\'etico est\'an circulando principalmente en el n\'ucleo, la muy alta conductividad lleva a 
una disipaci\'on mucho m\'as lenta (escalas de tiempo de Gyr en comparaci\'on con $10^4$--$10^5$ a\~{n}os de la corteza), y ni la alta luminosidad observada ni la existencia de un periodo m\'aximo de rotaci\'on se pueden explicar. Esto no implica que el campo magn\'etico no puede penetrar en el n\'ucleo, pero s\'i
que al menos una gran parte de la energ\'ia magn\'etica total tiene que estar almacenada en la corteza. El origen f\'isico de este confinamiento no es claro, y 
m\'as estudios sobre el equilibrio MHD, y la evoluci\'on del campo magn\'etico en un n\'ucleo superconductor podr\'{\i}an ayudar a aclarar este punto.

Tambi\'en hemos estudiado el escenario de {\em campo magn\'etico oculto} en los CCOs, demostrando que no est\'a en contradicci\'on con los datos observacionales. Una fase inicial de sumergimiento del campo magn\'etico, causada por la ca\'ida de restos de la supernova, puede explicar el comportamiento desconcertante de estos objetos, que parecen tener campos externos bajos pero campos m\'as intensos por debajo de la superficie. 
Nuestros resultados apuntan a que no todos los CCOs son de la misma naturaleza: los campos magn\'eticos intensos ocultos son m\'as probables en
aquellos CCOs que muestran anisotrop\'ias de temperatura. Por otro lado, algunos de los p\'ulsares est\'andar j\'ovenes, podr\'ian estar en su etapa final de reemergencia, y representan la conexi\'on entre CCOs y magnetares. Esto est\'a de acuerdo con los \'indices de frenado observados en los p\'ulsares, para los cuales hemos cuantificado la contribuci\'on debida a la evoluci\'on magn\'etica a largo t\'ermino.

En resumen, nuestras simulaciones ayudan a dibujar una imagen unificada de la variedad de propiedades observacionales de las estrellas de neutrones aisladas, y sus distintas rutas evolutivas. Nuestros resultados pueden explicar de la fenomenolog\'ia general, pero se requieren estudios particulares m\'as 
detallados para entender cada fuente individual.
Observaciones m\'as profundas y el descubrimiento de nuevos magnetares nos permitir\'an en los pr\'oximos a\~{n}os refinar la comparaci\'on con los datos y restringir los modelos te\'oricos. En este sentido, la variedad aparentemente desconcertante en el zoo de las estrellas de neutrones se podr\'a entender en 
t\'erminos de s\'olo unas pocas diferencias intr\'insecas, como la masa, la edad, el \'angulo de inclinaci\'on y el campo magn\'etico inicial, adem\'as de
algunos sesgos observacionales que afectan a la detecci\'on en diferentes bandas de energ\'ia, como la distancia,
la posici\'on en el cielo, y el \'angulo de visi\'on.

\appendix
\chapter{Mathematical notes for the magnetic field formalism}\label{app:mf_formalism}

\section{Poloidal-toroidal decomposition.}\label{app:poloidal-toroidal}

In MHD different formalisms can describe the magnetic field. Here we describe the most common notations found in the literature. For any three-dimensional, solenoidal vector field $\vec{B}$, like the magnetic field, we can always introduce the {\em vector potential} $\vec{A}$ so that
\begin{equation}
  \vec{B}=\vec{\nabla}\times\vec{A}~.
\end{equation}
$\vec{B}$ can also be expressed by two scalar functions $\Phi(\vec{x})$, $\Psi(\vec{x})$, that define its {\em poloidal} and {\em toroidal} components as follows:

\begin{eqnarray}\label{eq:def_decomposition_app}
  && \vec{B}_{pol}=\vec{\nabla}\times(\vec{\nabla}\times\Phi\vec{k})~,\\
  && \vec{B}_{tor}=\vec{\nabla}\times\Psi\vec{k}~,
\end{eqnarray}
where $\vec{k}$ is an arbitrary vector. This decomposition is useful in problems where $\vec{k}$ can be taken to be normal to the physical boundaries, and the boundary conditions in the toroidal direction are periodic. Therefore, for a spherical domain, and using the spherical coordinates $(r,\theta,\varphi)$, the suitable choice is $\vec{k}=\vec{r}$. In this case, $\vec{\nabla}\times\vec{r}=0$, therefore

\begin{eqnarray}\label{eq:def_pot_vect_app}
  && \vec{B}_{pol}=\vec{\nabla}\times(\vec{\nabla}\Phi\times\vec{r})~,\\
  && \vec{B}_{tor}=\vec{\nabla}\Psi\times\vec{r}~.
\end{eqnarray}
In axial symmetry, $\Phi=\Phi(r,\theta)$, $\Psi=\Psi(r,\theta)$, and the toroidal magnetic field is directed along the azimuthal direction $\hat{\varphi}$.

\begin{table}[t]
 \begin{center}
 \begin{tabular}{l r r}
\hline
\hline
Poloidal function 	& $\Gamma(r,\theta)$					& $\Phi(r,\theta)$ \\
Toroidal function 	& $\alpha(\Gamma)$					& $\Psi(r,\theta)$ \\
Magnetic flux		& $2\pi\Gamma$						& $-2\pi r\sin\theta\partial_\theta \Phi$\\
Enclosed current	& $(c/2)\int\alpha\de\Gamma$			    	& $-(cr\sin\theta/2)\partial_\theta\Psi$\\
$A_\varphi$ 		& $\Gamma(r,\theta)/r \sin\theta$		    	& $-\partial_\theta\Phi$ \\
$\vec{B}_{pol}$ 	& $(\vec{\nabla}\Gamma\times\hat{\varphi})/r\sin\theta$  	& $\vec{\nabla}\times(\vec{\nabla}\Phi\times\vec{r})$ \\
$\vec{B}_{tor}$ 	& $[\int\alpha \de\Gamma]\hat{\varphi}/r\sin\theta$    	& $\vec{\nabla}\Psi\times\vec{r} $\\
$B_r$			& $(\partial_\theta\Gamma)/r^2\sin\theta$      		& $-[\partial_\theta(\sin\theta\partial_\theta\Phi)]/r\sin\theta$\\
$B_\theta$		& $-(\partial_r\Gamma)/r\sin\theta$ 			& $[\partial_r(r \partial_\theta\Phi)]/r$ \\
$B_\varphi$		& $[\int\alpha\de\Gamma]/r\sin\theta$ 			& $-\partial_\theta\Psi$ \\
 \hline
 \hline
 \label{tab:formalism}
 \end{tabular}
 \caption{Comparison between different notations. In our work, we use the formalism of the left column, valid for force-free magnetic fields (see \S\ref{sec:pulsar_eq}).}
 \end{center}
\end{table}

Alternatively, the magnetic field can be expressed in terms of two other scalar functions:
\begin{equation}\label{eq:mf_clebsch_app}
  \vec{B}=\vec{\nabla} \Gamma \times \vec{\nabla} \Theta~.
\end{equation}
In axial symmetry, and with the choice $\Theta=\varphi-\xi (r,\theta)$, the \textit{magnetic flux function} $\Gamma(r,\theta)$ is related to the $\varphi-$component of the vector potential by
\begin{equation}\label{eq:gamma_aphi_app}
  \Gamma(r,\theta)= A_\varphi(r,\theta) r\sin\theta~.
\end{equation}
The poloidal and toroidal components are expressed as

\begin{eqnarray}\label{eq:def_poloidal_app}
  && \vec{B}_{pol}=\frac{\vec{\nabla} \Gamma(r,\theta) \times \hat{\varphi}}{r \sin\theta}~,\\
  && \vec{B}_{tor}=(\vec{\nabla} \xi)_{pol} \times (\vec{\nabla}\Gamma)_{pol}~.
\end{eqnarray}
This formalism is useful to rewrite the induction equation as follows. Consider a general electric field and velocity related each other by $\vec{E}=-(\vec{v}\times \vec{B})/c$, so that

\begin{equation}
 \partial_t\vec{B}=-c\vec{\nabla}\times \vec{E} = \vec{\nabla}\times(\vec{v}\times \vec{B}) ~.
\end{equation}
where $\partial_t$ indicates the time derivative. If we consider eq.~(\ref{eq:mf_clebsch_app}) and exploit the orthogonality between $\vec{B}$ and both $\vec{\nabla}\Theta$ and $\vec{\nabla}\Gamma$, then the induction equation, multiplied by $\vec{\nabla}\Theta$, is manipulated as follows:
\begin{eqnarray}
 && \vec{\nabla}\Theta\cdot[\partial_t (\vec{\nabla}\Gamma\times\vec{\nabla}\Theta)]=\vec{\nabla}\Theta\cdot[\vec{\nabla}\times(\vec{v}\times \vec{B})] ~,\nonumber\\
 && \vec{\nabla}\Theta\cdot[\vec{\nabla}\Gamma\times\vec{\nabla}\partial_t\Theta]=\vec{\nabla}\cdot[(\vec{v}\times \vec{B})\times \vec{\nabla}\Theta]+(\vec{v}\times \vec{B})\cdot(\vec{\nabla}\times\vec{\nabla}\Theta)~,\nonumber\\
 && (\nabla\Theta\times\nabla\Gamma)\cdot\nabla\partial_t\Theta=\nabla\cdot[\vec{B}(\vec{v}\cdot\vec{\nabla}\Theta)-\vec{v}(\vec{B}\cdot\vec{\nabla}\Theta)]~,\nonumber\\
 && -\vec{B}\cdot\vec{\nabla}\partial_t\Theta=\vec{B}\cdot\vec{\nabla}(\vec{v}\cdot\vec{\nabla}\Theta)~.
\end{eqnarray}
If we repeat the same algebra, multiplying the induction equation by $\vec{\nabla}\Gamma$ instead of $\vec{\nabla}\Theta$, then we obtain an advection equation for each scalar function:
\begin{eqnarray}
 && \partial_t \Gamma+\vec{v}\cdot\vec{\nabla} \Gamma=0~, \label{eq:roum1_app}\\
 && \partial_t \Theta+\vec{v}\cdot\vec{\nabla} \Theta=0~. \label{eq:roum2_app}
\end{eqnarray}
A static solution is achieved if and only if $\vec{v}=0$, since the velocity field, $\vec{\nabla}\Gamma$, and $\vec{\nabla}\Theta$ are all orthogonal to the magnetic field by definition.

In chapter~\ref{ch:magnetosphere}, we adopt the formalism based on the magnetic flux function $\Gamma(r,\theta)$. Additionally, for force-free fields, as explained in detail in \S~\ref{sec:pulsar_eq}, the toroidal field is easily expressed in terms of the enclosed current function:

\begin{equation}\label{eq:bphi_definition_app}
 B_\varphi=\frac{2}{cr\sin\theta}I(\Gamma)~.
\end{equation}
The conversion between the two formalisms is shown in Table~\ref{tab:formalism}.

\section{Twist of a magnetic field line.}\label{app:twist}

We define the twist of a magnetic field line as the azimuthal displacement between its surface footprints (for a dipolar-like configuration, one footprint in each hemisphere). The equation for a magnetic field line is:
\begin{equation}\label{eq:line_equation_app}
 \frac{\delta r}{B_r}=\frac{r\delta\theta}{B_\theta}=\frac{r\sin\theta\delta\varphi}{B_\varphi}~.
\end{equation}
The first equality is easily obtained from the differential of $\Gamma$ in axial symmetry:

\begin{equation}
\de\Gamma=\frac{\partial(A_\varphi r\sin\theta)}{\partial r}\de r + \frac{\partial(A_\varphi r\sin\theta)}{\partial \theta}\de\theta=0~. 
\end{equation}
The line twist is defined as the integral of $\delta\varphi$ along the magnetic field line $l_\Gamma$, the surface footprints of which lie at latitudes $\theta_{1,2}$:

\begin{equation}\label{eq:def_twist_app}
  \Delta \varphi_{tw}(\Gamma)\equiv\int_{\theta_1}^{\theta_2} \frac{B_\varphi(r(\theta,\Gamma),\theta)}{B_\theta(r(\theta,\Gamma),\theta)\sin\theta}\de\theta ~,
\end{equation}
where the dependence $r(\theta,\Gamma)$ can be found by solving the field line equations~(\ref{eq:line_equation_app}). The constraint $\Delta \varphi_{tw}(0)=0$ means no toroidal magnetic field at the poles.

\section{Magnetic helicity.}\label{app:helicity}

The standard definition of magnetic helicity,
\begin{equation}\label{eq:helicity_gen}
 {\cal H}\equiv\int_V (\vec{A}\cdot \vec{B}) \de V~,
\end{equation}
is gauge-dependent: any transformation of the vector potential $\vec{A}\rightarrow \vec{A}+\vec{\nabla}\Xi$, with $\Xi$ a scalar function, gives the same magnetic field $\vec{B}=\vec{\nabla}\times \vec{A}$, but the helicity changes: ${\cal H} \rightarrow {\cal H} + \int \vec{B}\cdot\vec{\nabla}\Xi$.

The helicity conservation theorem is obtained considering the following identity:

\begin{eqnarray}
 \partial_t {\cal H} &=& -\int_V \vec{A}\cdot (\vec{\nabla}\times \vec{E})\de V + \int_V \partial_t \vec{A}\cdot \vec{B}\de V = \nonumber\\
	      &=& -\int_V \vec{E}\cdot (\vec{\nabla}\times \vec{A})\de V + \int_V \vec{\nabla}\cdot(\vec{A}\times \vec{E})\de V - \int_V \vec{E}\cdot \vec{B}\de V - \int_V \vec{B}\cdot \vec{\nabla}\Xi\de V = \nonumber\\
	      &=& -2\int_V \vec{E}\cdot \vec{B}\de V + \int_V \vec{\nabla}\cdot(\vec{A}\times \vec{E})\de V - \int_V \Xi\vec{B}\de V~,
\end{eqnarray}
where in the second passage we have used $\partial_t \vec{A}=-\vec{E}-\vec{\nabla}\Xi$, coming from the homogeneous Maxwell equations (Faraday's law and $\vec{\nabla}\cdot\vec{B}=0$). The helicity conservation theorem is then

\begin{equation}\label{eq:helicity_cons_gen}
 \partial_t {\cal H} = -2\int_V \vec{E}\cdot \vec{B}\de V + \int_S (\vec{A}\times \vec{E} - \Xi\vec{B})\cdot \hat{n}\de S~,
\end{equation}
where $S$ is the surface enclosing the volume $V$, and $\hat{n}$ its normal unit vector. Eq.~(\ref{eq:helicity_cons_gen}) resembles the Poynting theorem for the electromagnetic energy conservation, with $\vec{E}\cdot \vec{B}$ playing the dissipative role of $\vec{J}\cdot \vec{B}$ and $\vec{A}\times \vec{E}$ representing the helicity flux at surface. In the trivial case of a stationary solution, $\vec{E}\cdot\vec{B}=\partial_t \vec{A}=0$. For configurations changing with time, we have a volume integral, which is zero if the force-free condition is satisfied, plus a surface term. 

In axial symmetry, a suitable choice of gauge is $\vec{A}_{pol}\cdot\vec{B}_{pol}=0$, therefore the helicity we use in this work is:
\begin{equation}\label{eq:helicity_tor}
 {\cal H}\equiv\int_V (A_\varphi B_\varphi)\de V~.
\end{equation}
Taking the time derivative and using the induction equation, we obtain the helicity conservation theorem:

\begin{eqnarray}\label{eq:helicity_cons_tor}
 \partial_t {\cal H} &=& -\int A_\varphi \hat{\varphi}\cdot(\vec{\nabla}\times \vec{E})\de V - \int E_\varphi B_\varphi\de V = \nonumber\\
 &=& \int \left[ \vec{\nabla} \cdot (A_\varphi \hat{\varphi} \times \vec{E} ) - \vec{E} \cdot \vec{\nabla}\times (A_\varphi \hat{\varphi}) \right]\de V - \int E_\varphi B_\varphi\de V = \nonumber\\
 &=&  \int_{\partial V} (A_\varphi\hat{\varphi}\times \vec{E})\cdot \de\vec{S} - \int (\vec{E}\cdot\vec{B})\de V ~.
\end{eqnarray}
If the poloidal electric field vanishes at the boundaries, the total helicity is conserved.

\section{Legendre polynomials.}\label{app:legendre}

The Legendre polynomials $P_l(\mu)$ are a complete orthonormal basis in the domain $\mu\in[-1,1]$ and satisfy the {\it Legendre's differential equation}
\begin{equation}\label{eq:def_legendre_app}
  \frac{d}{d\mu}\left[(1-\mu^2)\frac{d P_l(\mu)}{d\mu}\right]+l(l+1)P_l(\mu)=0~.
\end{equation}
Each Legendre polynomial $P_l(\nu)$ is an $l^{th}$-degree polynomial. With the normalization $P_l(1)=1$, they may be expressed using the Rodrigues' formula:

\begin{equation}
    P_l(\mu) = \frac{1}{2^l l!}\frac{d^l [(\mu^2 -1)^l]}{d\mu^l}~. 
\end{equation}
The first polynomials are:

\begin{eqnarray}
 && P_0(\mu)=1~, \nonumber \\
 && P_1(\mu)=\mu~, \nonumber \\
 && P_2(\mu)=\frac{1}{2}(3\mu^2-1)~, \nonumber \\
 && P_3(\mu)=\frac{1}{2}(5\mu^3-3\mu)~.
\end{eqnarray}
The following orthogonality relations hold:

\begin{eqnarray}
  && \int_{-1}^1 P_l(\mu)P_m(\mu)\de\mu=\frac{2}{2l+1}\delta_{lm}~,\label{eq:leg_on_app}\\
  && \int_{-1}^1 \frac{d P_l(\mu)}{d\mu}\frac{d P_m(\mu)}{d\mu}(1-\mu^2)\de\mu=\frac{2l(l+1)}{2l+1}\delta_{lm}~.\label{eq:leg_on2_app}
\end{eqnarray}
The recurrence relation between polynomials is:

\begin{equation}\label{eq:legendre_recurrence_app}
(1-\mu^2)\frac{d P_m}{d\mu}=\frac{m(m+1)}{2m+1}(P_{m-1}-P_{m+1})~.
\end{equation}
The {\it associated Legendre polynomials} are defined as

\begin{equation}\label{eq:def_legendre_ass_app}
  P_l^t(\mu)=(-)^t(1-\mu^2)^{t/2}\frac{d^t}{d\mu^t}P_l(\mu)~,
\end{equation}
where $t$ is the order of the associated polynomial. Their recurrence relation is given by

\begin{equation}\label{eq:def_legendre_ass_recurrence_app}
-(2l+1)\sqrt{1-\mu^2}P_l^t=P_{l+1}^{t+1}-P_{l-1}^{t+1}~. 
\end{equation}

\section{Potential magnetic field.}\label{app:vacuum_bc}

If the magnetic field is potential (i.e. no current, $\vec{\nabla}\times\vec{B}=0$), we can express the magnetic field as a gradient of the {\it magnetostatic potential} $\chi_m$. The $\vec{\nabla}\cdot\vec{B}=0$ condition implies that $\chi_m$ satisfies the Laplace equation:

\begin{eqnarray}
 && \vec{B}=\vec{\nabla}\chi_m~, \label{eq:magnetostatic_potential}\\ 
 && \nabla^2 \chi_m=0~.\label{eq:laplace_potential}
\end{eqnarray}
The Legendre expansion of the potential $\chi_m$ reads:

\begin{equation}\label{eq:bl_def_potential}
 \chi_m = B_0R_\star \sum_l \left[ b_lP_l(\mu)\left(\frac{R_\star}{r}\right)^{(l+1)} + d_l \left(\frac{r}{R_\star}\right)^l\right]~,
\end{equation}
where $\mu\equiv\cos\theta$, $B_0$ is a normalization, and the sum is over $l\ge 1$ ($l=0$ would correspond to the magnetic monopole, violating $\vec{\nabla}\cdot\vec{B}=0$). The dimensionless weigths $b_l$ and $d_l$ are associated to the $l$-multipole of two branches of solutions. The second branch, $\propto (r/R_\star)^l$, diverges for a domain extending to $r\rightarrow \infty$, therefore we set $d_l=0,~ \forall l$ if we are in a region that formally extends to infinity, like the magnetosphere. The corresponding magnetic field components are

\begin{eqnarray}
  && B_r      = - B_0 \sum_l b_l (l+1)P_l(\mu)\left(\frac{R_\star}{r}\right)^{(l+2)}~, \label{eq:br_vacuum}\\
  && B_\theta = - B_0 \sum_l b_l \sqrt{1-\mu^2}\frac{d P_l(\mu)}{d\mu}\left(\frac{R_\star}{r}\right)^{(l+2)}~, \label{eq:bth_vacuum}\\
  && B_\varphi   = 0~. \label{eq:bphi_vacuum}
\end{eqnarray}
Note that when general relativity is considered, eq.~(\ref{eq:bth_vacuum}) is modified by non-trivial relativistic factors $f_l^{rel}$ \citep{pons09}:

\begin{equation}
 B_\theta = - B_0 \sum_l f_l^{rel} b_l \sqrt{1-\mu^2}\frac{d P_l(\mu)}{d\mu}\left(\frac{R_\star}{r}\right)^{(l+2)}~. \label{eq:bth_vacuum_rel}
\end{equation}
We neglect these corrections for simplicity. The intensity at the north pole is
\begin{equation}
 B_N = - B_0\sum_l b_l(l+1)~.
\end{equation}
For instance, a purely dipolar field (with $-b_l(l+1)=\delta_{l1}$) is given by

\begin{eqnarray}
 && A_\varphi   = B_0 \frac{R_\star^3}{r^2} \frac{\sin\theta}{2}~, \\
 && B_r      = B_0 \left(\frac{R_\star}{r}\right)^3 \cos\theta~, \\
 && B_\theta = B_0 \left(\frac{R_\star}{r}\right)^3 \frac{\sin\theta}{2}~,
\end{eqnarray}
while a purely quadrupolar field (with $-b_l(l+1)=\delta_{l2}$) is given by

\begin{eqnarray}
 && A_\varphi   = B_0 \frac{R_\star^4}{r^3}\frac{\sin\theta\cos\theta}{2}~, \\
 && B_r      = B_0 \left(\frac{R_\star}{r}\right)^4 \frac{3\cos^2\theta-1}{2}~,\\
 && B_\theta = B_0 \left(\frac{R_\star}{r}\right)^4 \sin\theta\cos\theta~.
\end{eqnarray}

Now we give the prescription to match, at some radius $r=R$, a general magnetic field in the region $r<R$, with a general potential solution outside, as given by eqs.~(\ref{eq:br_vacuum})-(\ref{eq:bphi_vacuum}). Across the radial direction, the radial magnetic field has to be continuous, while $B_\theta$ can in principle have a discontinuity supported by a toroidal current $J_\varphi$. Similarly, $B_\varphi$ can be discontinuous if sustained by a non-zero current component $J_\theta$. If the matching is smooth, i.e. no surface current sheets are allowed, then $B_\varphi\rightarrow 0$ for $r\rightarrow R$.

We present two techniques: the first relies on the Legendre spectral decomposition presented above, while the second is a finite difference method. In both cases, given a $B_r$ at the boundary $r=R$, we find the form of $B_\theta$ corresponding to the potential solution. 

\subsection{Spectral method.}\label{app:vacuum_spectral} 

Multiplying the radial component of the potential magnetic field, eq.~(\ref{eq:br_vacuum}) by $P_m(\mu)$, and using the identity (\ref{eq:leg_on_app}), the integral in $\mu\in[-1,1]$ gives:

\begin{equation}\label{eq:bl_def}
 \int_{-1}^1 B_r(R,\mu)P_m(\mu)\de\mu= - B_0b_l \frac{2(l+1)}{2l+1}~.
\end{equation}
Given a $B_r(R,\mu)$, the numerical integration of the left-hand side allows the extraction of the weights $b_l$. Then, $B_\theta(r,\mu)$ is reconstructed following (\ref{eq:bth_vacuum}). In a numerical simulation with a grid of $n_r$ radial points, we decompose $B_r^{(i,n_r)}$, where $i$ is the angular index of the grid, and obtain the values of $b_l$ up to a maximum multipole $l_{max}$. The latter cannot exceed $n_\theta/2$, where $n_\theta$ is the number of angular points of the grid. In the simulations presented in our work, we typically use $l_{max}=30$. We are often interested to the external dipolar component, defined by eq.~(\ref{eq:br_vacuum}) with $l=1$: $B_p=-2B_0b_1$. For a given $B_r(R,\mu)$, it is obtained by setting $m=1$ in the Legendre decomposition, eq.~(\ref{eq:bl_def}):

\begin{equation}\label{eq:def_bpole_app}
  B_p = \frac{3}{2}\int_{-1}^1  B_r(R,\mu) \mu~\de\mu~.
\end{equation}

\subsection{Green's method.}

The spectral method is very accurate for smooth functions $B_r(R,\mu)$. In case of sharp features in $B_r$, like the step-like profile created by the Hall evolution of the radial field in the crust (see chapter~\ref{ch:magnetic}), the largest multipoles acquire a non-negligible weight, and, if $l_{max}$ is not large enough, it gives fake oscillations in the reconstructed $B_\theta$ (Gibbs phenomenon).

For this reason, we employ an alternative method, employing a formalism often used in electrostatic problems. We express the potential magnetic field in terms of the magnetostatic potential $\chi_m$, eq.~(\ref{eq:magnetostatic_potential}). The second Green's identity, applied to a volume enclosed by a surface $S$, relates $\chi_m$ with a {\em Green's function} $G$ (see eq.~1.42 of \citealt{jackson91}):

\begin{equation}\label{eq:green_psi1}
 2\pi \chi_m(\vec{r}) = -\int_S \frac{\partial G}{\partial n'}(\vec{r},\vec{r}') \chi_m(\vec{r}') {\rm d}S' + \int_S G(\vec{r},\vec{r}')\frac{\partial \chi_m}{\partial n'}(\vec{r}'){\rm d}S'~,
\end{equation}
where $\hat{n}'$ is the normal to the surface. Comparing with the electrostatic problem, we see that no volume integral is present, because $\dive\vec{B}\equiv \nabla^2\chi_m=0$. Note also that the factor $2\pi$ appears instead of the canonical $4\pi$, because inside the star eq.~(\ref{eq:laplace_potential}) does not hold, thus $2\pi$ is the solid angle seen from the surface. The Green's function has to satisfy $\nabla'^2 G(\vec{r},\vec{r}')=-2\pi \delta(\vec{r}-\vec{r}')$. The functional form of $G$ is gauge dependent: given a Green's function $G$, any function $F(\vec{r},\vec{r}')$ which satisfied $\nabla'^2 F=0$ can be used to build a new Green's function $\tilde{G}=G+F$. The boundary conditions determine which gauge is more appropriate for a specific problem.

In our case the volume is the outer space, $S$ is a spherical boundary of radius $R$ (e.g., the surface of the star), and $\hat{n}'=-\hat{r}'$. We face a {\em von Neumann boundary condition problem}, because we know the form of the radial magnetic field
\begin{equation}
 B_r(R,\theta) \equiv \frac{\partial\chi_m}{\partial r}(R,\theta)~.
\end{equation}
In order to reconstruct the form of
\begin{equation}
 B_\theta(R,\theta)\equiv \frac{1}{R}\frac{\partial\chi_m}{\partial\theta}(R,\theta)~,
\end{equation}
we have to solve the following integral equation for $\chi_m$:

\begin{eqnarray}\label{eq:green_psi2}
  2\pi \chi_m(\vec{r}) & = & R^2\left\{\int_0^\pi\int_0^{2\pi}  \frac{\partial G}{\partial r'}(\vec{r},\vec{r}') \chi_m(R,\theta')\sin\theta' \de \varphi' \de \theta' + \right.\nonumber\\
  && \left. - \int_0^\pi\int_0^{2\pi} G(\vec{r},\vec{r}') B_r(\theta')\sin\theta' \de \varphi' \de \theta'\right\}~.
\end{eqnarray}
So far, we have not specified the Green's function. In our case, the simplest Green's function is:

\begin{eqnarray}\label{eq:green0}
 && G(\vec{r},\vec{r}')=\frac{1}{|\vec{r}-\vec{r}'|}= [(r\sin\theta\cos\varphi - r'\sin\theta'\cos\varphi')^2+\nonumber\\
 && + (r\sin\theta\sin\varphi - r'\sin\theta'\sin\varphi')^2+(r\cos\theta - r'\cos\theta')^2]^{-1/2}~.
\end{eqnarray}
In axial symmetry, we can set $\varphi=0$, to obtain

\begin{equation}\label{eq:green1}
 G(\vec{r},\vec{r}')=[(r\sin\theta - r'\sin\theta'\cos\varphi')^2+(r'\sin\theta'\sin\varphi')^2+(r\cos\theta - r'\cos\theta')^2]^{-1/2}~.
\end{equation}
Its derivative with respect to $r'$ is:

\begin{eqnarray}\label{dergreen1}
 && \derparn{r'}{G}(\vec{r},\vec{r}')=\nonumber\\
 && = -G^3[(r'\sin\theta'\cos\varphi'-r\sin\theta)\sin\theta'\cos\varphi'+r'\sin^2\theta'\sin^2\varphi' + (r'\cos\theta'-\cos\theta)\cos\theta'] \nonumber\\
 && = -G^3[r'(\sin^2\theta'\cos^2\varphi'+\sin^2\theta'\sin^2\varphi'+\cos^2\theta')-r(\sin\theta\sin\theta'\cos\varphi'+\cos\theta\cos\theta')] \nonumber\\
 && = G^3[r(\sin\theta\sin\theta'\cos\varphi'+\cos\theta\cos\theta')-r']~.
\end{eqnarray}
We evaluate $G$ and its radial derivative at $r=r'=R$
\begin{eqnarray}
 && G(R,\theta,\theta',\varphi')=\frac{1}{\sqrt{2}R}\left[1-\cos(\theta-\theta')+2\sin\theta\sin\theta'\sin^2\left(\frac{\varphi'}{2}\right)\right]^{-1/2}~, \label{green_eps0} \\
 && \derparn{r'}{G}(R,\theta,\theta',\varphi') \rightarrow -\frac{G}{2R}~. \label{dergreen_eps0}
\end{eqnarray}
Casting the two formulas above in eq.~(\ref{eq:green_psi2}), we note that the following integral appears in the two right-hand side terms:

\begin{equation}
 f(\theta,\theta')\equiv \sin\theta'\int_0^{2\pi}RG(R,\theta,\theta',\varphi')\de \varphi'~.
\end{equation}
As $G$ depends on $\varphi'$ via $\sin^2(\varphi'/2)$, we can change the integration limits to $[0,\pi/2]$, and $\varphi'\rightarrow 2\varphi'$, therefore

\begin{equation}\label{eq:green_f}
 f(\theta,\theta')=\sqrt{8}\sin\theta'\int_0^{\pi/2} [1-\cos(\theta-\theta')+2\sin\theta\sin\theta'\sin^2\varphi']^{-1/2}\de \varphi'~.
\end{equation}
Casting eq.~(\ref{eq:green_f}) in eq.~(\ref{eq:green_psi2}), and substituting $\chi_m(\theta)=R\int_0^\theta B_\theta(R,\theta') \de\theta'$, we have

\begin{equation}\label{eq:greeneps0_final}
 4\pi\int_0^\theta B_\theta(\theta') \de \theta' + \int_0^\pi  B_\theta(\theta')\left[\int_{\theta'}^{\pi}f(\theta,\theta'')\de\theta''\right]\de\theta' = -2\int_0^\pi B_r(\theta')f(\theta,\theta')\de\theta'~.
\end{equation}
In eq.~(\ref{eq:green_f}), if $\theta=\theta'$, then $f(\theta,\theta')\rightarrow 2\int_0^{\pi/2}(\sin\varphi')^{-1}\de \varphi'$, which is not integrable because of the singularity in $\varphi'=0$ (corresponding to $\vec{r}=\vec{r}'$). However, in both terms where it appears, the function $f(\theta,\theta')$ is integrated in $\theta'$, and both terms of the equation are integrable.

For numerical purposes, we can express eq.~(\ref{eq:greeneps0_final}) in matrix form, introducing $f_{ij}=f(\theta_i,\theta'_j)$ evaluated on two grids with vectors $\theta_i,\theta'_j$, with $m$ steps $\Delta\theta$. The coefficients of the matrix $f_{ij}$ are purely geometrical, therefore they are evaluated only once, at the beginning. The grid $\theta_i$ coincides with the locations of $B_r(R,\theta)$, while the resolution of the grid $\theta'_j$ is $M$ times the resolution of the grid $\theta_i$ ($M\gtrsim 5$) to improve the accuracy of the integral function $f_{ij}$ near the singularities $\theta_i\rightarrow\theta_j$. The resolution of the grid of $\varphi'_k$ barely affects the result, provided that it avoids the singularities $\varphi'=0,\pi/2$. We typically use $M=10$ and $n_\varphi'=1000$. The calculation of the factors $f_{ij}$ is performed just once and stored. The matrix form is:

\begin{equation}\label{eq:greeneps0_matrix}
 \sum_{j=1}^m [4\pi\delta_{ij} + f_{ij}\Delta\theta] \chi_m(\theta_j)= \sum_{j=1}^m [-2f_{ij}\Delta\theta]B_r(\theta_j), \qquad i=1,m~.
\end{equation}
From this, we obtain $B_\theta$ by taking the finite difference derivative of $\chi_m(\theta)$.

\chapter{Force-free magnetic fields}\label{app:forcefree}

A generic magnetic field has $\vec{\nabla}\times\vec{B}\neq 0$ and it cannot be described by the magnetostatic potential, eq.~(\ref{eq:magnetostatic_potential}). In \S~\ref{app:poloidal-toroidal} we have introduced the scalar potentials $\Gamma, \Phi, A_\varphi$ related to the poloidal magnetic field (see Table~\ref{tab:formalism} for their mutual relations). In terms of Legendre polynomials, they are conveniently written as follows:

\begin{eqnarray}\label{eq:potentials_general_app}
  && \Phi(r,\mu)  =\frac{B_0R_\star}{2}\sum_l a_l(r)P_l(\mu)~,\\
  && \Gamma(r,\mu)=\frac{B_0R_\star}{2}\sum_l ra_l(r)(1-\mu^2)\frac{d P_l(\mu)}{d\mu}~, \label{eq:potential_gen_app}\\
  && A_\varphi(r,\mu)=\frac{B_0R_\star}{2}\sum_l a_l(r)\sqrt{1-\mu^2}\frac{d P_l(\mu)}{d\mu}~,
\end{eqnarray}
where $a_l(r)$ is a dimensionless radial function giving the weight of the $l$-pole, the normalization $B_0$ is the magnetic field strength at the pole in the case of purely dipolar field with $a_1=1$, and $R_\star$ is the radius of the star. Considering eq.~(\ref{eq:def_poloidal_app}) and the Legendre's differential equation (\ref{eq:def_legendre_app}), the poloidal magnetic field components are given by

\begin{eqnarray}\label{eq:mf_leg_app}
  && B_r      = \frac{B_0}{2}\frac{R_\star}{r}\sum_l l(l+1)P_l(\mu)a_l(r) ~, \\
  && B_\theta = - \frac{B_0}{2}\frac{R_\star}{r}\sum_l \sqrt{1-\mu^2}\frac{d P_l(\mu)}{d\mu}\frac{d (ra_l(r))}{dr}~.
\end{eqnarray}
For construction, the conditions $B_\theta(\mu=\pm 1)=0$ are guaranteed. The intensity at north pole is $B_N=B_0\sum_{l=1}^\infty l(l+1)a_l(R_\star)$. The potential field, eqs.~(\ref{eq:br_vacuum}), (\ref{eq:bth_vacuum}), is recovered if

\begin{equation}\label{eq:al_bl_app}
  a_l(r) = -\frac{2b_l}{l} \left(\frac{R_\star}{r}\right)^{(l+1)} ~.\\
\end{equation}
In \S~\ref{sec:semianalytical}, we search solutions to the pulsar equation in the non-rotating limit, eq.~(\ref{eq:rotB_nonrotating}):

\begin{equation}\label{eq:rotB_nonrotating_app}
\curlB=\alpha(\Gamma)\vec{B}~,
\end{equation}
which can now be written as

\begin{equation}\label{eq:gs_ode_leg_app}
  B_0R_\star\frac{l(l+1)}{2l+1}\left[\frac{d^2 (ra_l(r))}{dr^2}-l(l+1)\frac{a_l(r)}{r}\right]= -\int_{-1}^1\frac{d P_l(\mu)}{d\mu}\left[\alpha\int \alpha(\Gamma)\de\Gamma\right]\de \mu~.
\end{equation}

\section{Force-free spherical Bessel solutions.}\label{app:bessel}

The force-free condition $\curlB=\alpha \vec{B}$, with $\alpha=k/R_\star$ constant, reduces to a Bessel differential equation for each function $a_l(r)$:
\begin{equation}
  r^2\frac{d^2 a_l(r)}{dr^2}+2r\frac{d a_l(r)}{dr}+\left[\left(k\frac{r}{R_\star}\right)^2 - l(l+1)\right]a_l(r)=0~.
\end{equation}
The analytical solutions to this equation can be written in terms of the variable $x=k r/R_\star+\delta$, where $\delta$ is a constant phase that hereafter we set it to zero without loss of generality. The general solution reads

\begin{equation}\label{eq:al_bl_blapp}
 a_l(r)=m_lJ_l(x) + c_lY_l(x)~,
\end{equation}
where the coefficients $m_l, c_l$ are the weights of the $l$-polar spherical Bessel function of the first and second kind, respectively given by the following Rayleigh's formulas:

\begin{eqnarray}\label{spher_bessel}
  && J_l(x)=(-x)^l\left(\frac{1}{x}\frac{\partial}{\partial x}\right)^l\frac{\sin x}{x}~,\\
  && Y_l(x)=-(-x)^l\left(\frac{1}{x}\frac{\partial}{\partial x}\right)^l\frac{\cos x}{x}~.
\end{eqnarray}
The spherical Bessel functions have the following properties:

\begin{eqnarray}
  && \lim_{x\rightarrow 0}J_l(x)\rightarrow x^l~,\label{eq:bessel_I_x0}\\
  && \lim_{x\rightarrow 0}Y_l(x)\rightarrow -x^{-(l+1)}~, \label{eq:bessel_II_x0}\\
  && \lim_{x\rightarrow \infty} J_l(x),Y_l(x),\frac{d^n J_l(x)}{dx^n},\frac{d^n Y_l(x)}{dx^n}\rightarrow \frac{1}{x}~.
\end{eqnarray}
Furthermore, for large $x$, the following properties hold:

\begin{eqnarray}
 && \lim_{x\rightarrow \infty}Y_{l}(x)=J_l(x-\pi/2)~,\\
 && \lim_{x\rightarrow \infty}Y_{l+1}(x+\pi/2)=Y_l(x)~.
\end{eqnarray}
For large $x$, the zeros of spherical Bessel functions form a grid of step $\sim\pi$, with a shift for different multipoles $l$ and $m$ given by

\begin{equation}
  \{x_{l,0}\}=\{x_{m,0}\}+(l-m)\pi/2 \qquad x\rightarrow \infty~.
\end{equation}
The magnetic field components are given by

\begin{eqnarray}\label{eq:sol_bessel_app}
  && B_r=\frac{B_0}{2}\frac{R_\star}{r}\sum_l l(l+1)P_l(\mu)(m_lJ_l(x)+c_lY_l(x))~,\\
  && B_\theta=-\frac{B_0}{2}\frac{R_\star}{r}\sum_l \sqrt{1-\mu^2}\frac{d P_l(\mu)}{d\mu}\frac{d [x(m_lJ_l(x)+c_lY_l(x))]}{d x}~,\\
  && B_\varphi=k\frac{B_0}{2}{\sin\theta}\sum_l \frac{d P_l(\mu)}{d\mu}(m_lJ_l(x)+c_lY_l(x)) ~.
\end{eqnarray}
Note that a smooth matching with the vacuum is not possible because it would require that, at the same radius $R$, $B_\varphi(R,\theta)=0$ and $B_r(R,\theta)\neq 0$, a condition that cannot be satisfied because those two components have the same radial dependence $m_lJ_l(x)+c_lY_l(x)$, implying having the same zeros.

\subsubsection{Solution for the magnetosphere.}
The $Y_l$ functions are appropriate to describe the magnetic field in regions extending to infinity (magnetosphere, \S~\ref{sec:bessel}). In this case, the potential solution is recovered in the limit $k\rightarrow 0$.

\subsubsection{Solution for the interior.}
On the other hand, the $J_l$ function are suitable for domains with $r\rightarrow 0$ (interior of the star, \S~\ref{sec:pure_ohmic_mode}). In particular, the only solution which is regular in the center, with nonzero magnetic field, is the $l=1$ first kind spherical Bessel function, for which:

\begin{eqnarray}
  && J_1(x)=\frac{\sin x}{x^2} - \frac{\cos x}{x}~,\\
  && A_\varphi=B_0R_\star\frac{\sin\theta}{2x} \left(\frac{\sin x}{x}-\cos x\right)~,\\
  && B_r=\frac{B_0R_\star}{r}\cos\theta \left(\frac{\sin x}{x^2} - \frac{\cos x}{x}\right)~,\\
  && B_\theta=-\frac{B_0R_\star}{2r}\sin\theta\left(\frac{\sin x}{x^2}-\frac{\cos x}{x}-\sin x\right)~,\\
  && B_\varphi=\frac{kB_0R_\star}{2}\sin\theta \left(\frac{\sin x}{x^2}-\frac{\cos x}{x}\right) ~.
\end{eqnarray}
In the limit $x\rightarrow 0$, we recover the solution corresponding to a homogeneous field aligned with the magnetic axis

\begin{equation}
\vec{B} \rightarrow \frac{k}{3}B_0R_\star~\hat{z}~. 
\end{equation}

\subsubsection{Crust-confined configuration.}\label{app:bessel_bc}

Spherical Bessel solutions can be used to build crustal-confined poloidal magnetic fields with a smooth matching with the potential dipole outside. The vacuum condition for the poloidal magnetic field, $\vec{\nabla}\times\vec{B}=0$, is satisfied if the toroidal magnetic field is neglected, which implies that the configuration is not force-free. It requires a mixture of functions of both (first and second) kind, with $m_1=\cos \alpha$, $c_1 =\sin \alpha$ for the dipolar case \citep{aguilera08b}. If $B_r=0$ at some inner interface $R_c$, as in superconducting boundary condition (see \S~\ref{sec:magnetic_bc}) the value of $\alpha$ is found by solving the condition

\begin{equation}\label{eq:build_crustal_confined}
\tan{[\alpha(R_c-R_\star)]}=\alpha R_\star~.
\end{equation}
On the other hand, it is not possible in general to satisfy both the vacuum boundary condition and the regularity at the center together, since the latter requires $c_l=0$. 

\subsection{A diverging solution.}\label{sec:leg_diverging}

Choosing $\alpha=(k/R_\star) |\Gamma/\Gamma_0|^{-1/2}$, the toroidal magnetic field is given by
\begin{equation}\label{bphi_alpha1}
  B_\varphi=2k\sqrt{\frac{|B_0\sum_l a_l(r)\frac{d P_l(\mu)}{d\mu}|}{r}}\mbox{sgn}(\Gamma)~.
\end{equation}
Assuming $\Gamma$ is positive everywhere, eq.~(\ref{eq:gs_ode_leg}) becomes

\begin{equation}\label{ode_alpha1}
  R_\star\frac{l(l+1)}{2l+1}\left[\frac{d^2 (ra_l(r))}{dr^2}-l(l+1)\frac{ra_l(r)}{r^2}\right]=-k^2\int_{-1}^1\frac{d P_l(\mu)}{d\mu}\de \mu~.
\end{equation}
For even multipoles the right hand side is null, and the potential solutions are recovered. For odd multipoles, the right hand side is always $-2k^2$, therefore the general solution is the homogeneous one (vacuum solution) plus a particular one, which can be found analytically by means of the Euler-Cauchy method. For $l=1$ the particular solution is given by $a_1^{part}(r)=-k^2r\log r$. For odd $l\ge 3$, this particular solution is $a_l^{part}(r)=\eta_{part} r$, where
\begin{equation}
  \eta_{part}=\frac{k^2}{R_\star}\frac{2(2l+1)}{l(l+1)[l(l+1)-2]}~.
\end{equation}
Finally we have

\begin{eqnarray}\label{sol_alpha1}
  && a_l(r)=\left\{
  \begin{array}{ll}
  c_1\left(\frac{R_\star}{r}\right)^2 - k^2\frac{r}{R_\star}\log\left(\frac{R_\star}{r}\right)	& l=1\\
  c_l\left(\frac{R_\star}{r}\right)^{(l+1)}+\eta \frac{r}{R_\star}				& l\mbox{ odd}\ge 3\\
  c_l\left(\frac{R_\star}{r}\right)^{(l+1)}							& l\mbox{ even}
  \end{array}\right.~,\\
  && B_r=B_0\sum_ll(l+1)P_l(\mu)\left\{
  \begin{array}{ll}
  c_1\left(\frac{R_\star}{r}\right)^3-k^2\log \left(\frac{R_\star}{r}\right)	& l=1\\
  c_l\left(\frac{R_\star}{r}\right)^{(l+2)}+\eta				& l\mbox{ odd}\ge 3\\
  c_l\left(\frac{R_\star}{r}\right)^{(l+2)}					& l\mbox{ even}
  \end{array}\right.~,\\
  && B_\theta=B_0\sqrt{1-\mu^2}\sum_l\left\{
  \begin{array}{ll}
  c_1\left(\frac{R_\star}{r}\right)^3+k^2\left[2\log \left(\frac{R_\star}{r}\right)+1\right]	& l=1\\
  lc_l\frac{d P_l(\mu)}{d\mu}\left(\frac{R_\star}{r}\right)^{(l+2)}-2\eta			& l\mbox{ odd}\ge 3\\
  lc_l\frac{d P_l(\mu)}{d\mu}\left(\frac{R_\star}{r}\right)^{(l+2)}				& l\mbox{ even}
  \end{array}\right.~,\\
  && B_\varphi=2k\sqrt{\frac{B_0\sum_l a_l(r)\frac{d P_l(\mu)}{d\mu}}{r}}~.
\end{eqnarray}
We see that, if $r\gg R_\star$, the magnetic field of odd modes grows ($l=1$) or approaches a constant ($l\ge 3$), that is the magnetic flux across a spherical shell $\int B_r r^2 d\Omega$ grows with increasing $r$. We conclude that, except the trivial case in which all the odd multipoles are null, this solution is unphysical.

\subsection{Gaunt solution.}\label{app:gaunt}

If $\alpha=(k/R_\star)|\Gamma/\Gamma_0|^{1/2}$, eq.~(\ref{eq:gs_ode_leg_app}) becomes

\begin{equation}\label{ode_alpha3}
  \frac{l(l+1)}{2l+1}\left[\frac{d^2 (ra_l(r))}{dr^2}-l(l+1)\frac{a_l(r)}{r}\right]=-\frac{k^2}{3}\frac{1}{R_\star}\int_{-1}^1\frac{d P_l(\mu)}{d\mu}\left(\frac{\Gamma}{\Gamma_0}\right)^2\mbox{sgn}(\Gamma)\de \mu ~.
\end{equation}
The integral on the right-hand side can be written as
\begin{eqnarray}\label{alpha3_int_part}
 {\cal I} &=& \frac{1}{\Gamma_0^2}\int_{-1}^1\frac{d P_l(\mu)}{d\mu}\Gamma^2\mbox{sgn}(\Gamma)\de \mu=\frac{1}{\Gamma_0^2}[\Gamma^2 \mbox{sgn}(\Gamma)P_l(\mu)]^1_{-1} + \nonumber\\
   && - \frac{2}{\Gamma_0^2}\left[\int_{-1}^1\Gamma\frac{d \Gamma}{d\mu}\mbox{sgn}(\Gamma)P_l(\mu)\de \mu  + \int_{-1}^1\Gamma^2\delta(\Gamma)\frac{d \Gamma}{d\mu}P_l(\mu)\de \mu \right] ~,
\end{eqnarray}
where the third term comes from $\mbox{sgn}'(x)=2\delta(x)$. The first and third terms in this equation are zero. Next, we assume by simplicity that $\Gamma \ge 0$ in the whole domain, so that $\mbox{sgn}(\Gamma)\equiv 1$, and we define the dimensionless radial functions
\begin{equation}
 f_l(r)\equiv \frac{r}{R_\star}a_l(r)~.
\end{equation}
Hereafter we drop the dependences on $\mu$ and $r$ for conciseness. Using the Legendre's differential equation (\ref{eq:def_legendre_app}), we obtain from eq.~(\ref{eq:potential_gen_app})
\begin{eqnarray}\label{alpha3_int_part2}
 \frac{d \Gamma}{d\mu} &=& -\Gamma_0\sum_n n(n+1)f_nP_n~,
 \end{eqnarray}
and we can express the integral ${\cal I}$ in a more compact form
\begin{eqnarray}\label{eq:integral_compact}
 {\cal I} &=& 2 \sum_{n=1}^{\infty} n(n+1) f_n\sum_{m=1}^\infty f_m \int_{-1}^1 (1-\mu^2) P_l \frac{d P_m}{d\mu} P_n \de\mu=\\
   &=& \sum_{m,n=1}^{\infty}\frac{2m(m+1)}{2m+1}n(n+1)f_m f_n \left[\int_{-1}^1 P_lP_{m-1}P_n\de\mu - \int_{-1}^1 P_lP_{m+1}P_n\de\mu\right]\nonumber ~,
\end{eqnarray}
where we have also used the recurrence relations between Legendre polynomials, eq.~(\ref{eq:legendre_recurrence_app}). For each multiple $l$ eq.~(\ref{eq:gs_ode_leg}) reads
\begin{equation}\label{ode_sol3}
  \frac{d^2 f_l}{d r^2}-l(l+1)\frac{f_l}{r^2}=-\left(\frac{k}{R_\star}\right)^2 \sum_{m,n=1}^\infty f_mf_n g_{lmn}~,
\end{equation}
where
\begin{equation} 
g_{lmn}=\frac{2}{3}\frac{2l+1}{2m+1}\frac{m(m+1)}{l(l+1)}n(n+1) \left[\int_{-1}^1 P_lP_{m-1}P_n\de\mu - \int_{-1}^1 P_lP_{m+1}P_n\de\mu\right]~.\nonumber
\end{equation}
The integral of the product of three Legendre polynomials can be expressed by the {\em Wigner 3-$j$ symbols}:

\begin{equation}\label{wigner}
 \int_{-1}^1 P_aP_bP_c\de\mu=\left(
  \begin{array}{ccc}
  a & b & c\\
  0 & 0 & 0
  \end{array}\right)^2~.
\end{equation}
Alternatively, we can express it in terms of associated Legendre polynomials, defined in eq.~(\ref{eq:def_legendre_ass_app}). Eq.~(\ref{eq:integral_compact}) becomes
\begin{eqnarray}\label{eq:integral_compact_ass}
 {\cal I} &=& 2 \sum_{n=1}^{\infty} n(n+1) f_n\sum_{m=1}^\infty f_m \int_{-1}^1 \sqrt{1-\mu^2} P_l^0 P_m^1 P_n^0\de\mu~.
\end{eqnarray}
We use the recurrence relation, eq.~(\ref{eq:def_legendre_ass_recurrence_app}), applied to $\sqrt{1-\mu^2}P_l^0$ and obtain:

\begin{equation}\label{glmn}
  g_{lmn}=\frac{2}{3}\frac{n(n+1)}{l(l+1)}\left[\int_{-1}^1 P_{l+1}^1P_m^1P_n^0\de\mu - \int_{-1}^1 P_{l-1}^1P_m^1P_n^0\de\mu\right]~.
\end{equation}
These integrals can be evaluated analytically using the following Gaunt's formula \citep{gaunt30}:

\begin{eqnarray}\label{gaunt}
  && {\cal I}_3=\int_{-1}^1 P_a^u P_b^v P_c^w\de\mu=\nonumber\\
  && 2\,(-)^{(s-b-w)}\frac{(b+v)!(c+w)!(2s-2c)!s!}{(b-v)!(s-a)!(s-b)!(s-c)!(2s+1)!}\times\\
  && \times\sum_{t=p}^q (-)^t \frac{(a+u+t)!(b+c-u-t)!}{t!(a-u-t)!(b-c+u+t)!(c-w-t)!}~,\nonumber
\end{eqnarray}
where 
\begin{eqnarray}
  && s\equiv\frac{a+b+c}{2}~,\nonumber\\
  && p\equiv{\rm max}\{0,c-b-u\}~,\\
  && q\equiv{\rm min}\{b+c-u,a-u,c-w\}~.\nonumber
\end{eqnarray}
The integral ${\cal I}_3$ is non-zero if and only if both the following relations are satisfied:
\begin{itemize}
 \item $b-c\le a\le b+c$, a triangular condition;
 \item $a+b+c$ is even.
\end{itemize}
In our case, $u=v=1$, $w=0$, thus for each fixed $l$ in eq.~(\ref{glmn}), the conditions above are satisfied for an infinite number of pairs $(m,n)$ (see the values of ${\cal I}_3$ in Table~\ref{tab_gaunt}), with $g_{lmn}$ having numerical values of $\sim O(1)$, as illustrated in Table~\ref{tab_coupling}. Therefore, the $g_{lmn}$ factors in eq.~(\ref{ode_sol3}) couple each $l$-multipole with an infinite number of other multipoles. In order to find a solution, we have to solve eq. (\ref{ode_sol3}) with boundary conditions at the surface, following these steps:
\begin{itemize}
  \item we fix a value of $k$ and the maximum order of multipoles we consider, $l_{max}$: $f_l\equiv 0 \,\forall l>l_{max}$; fix the amplitude $f_l(r=R_\star)\, \forall l\le l_{max}$, with the requirement that  $\Gamma$ has to be positive-definite in all its domain; for instance, we can arbitrarily choose $f_l=1/l$;
  \item we guess an initial value of derivatives at the surface: the first attempt can be the relation valid for the vacuum solution: $z_l\equiv f_l'(R_\star)=-lf_l(R_\star)$;
  \item we solve numerically the equation up to a maximum radius $r_{max}$ to be chosen as large as possible;
  \item we check the boundary conditions at the external boundary $r=r_{max}$: in order to match to the vacuum solution, for each $l$ the following check-function has to approach zero $F_l(z_1,z_2,...z_{l_{max}})= f_l'(r_{max})+lf_l(r_{max})/r_{max}~;$
  \item we use an iterative Newton-Raphson method until $\sum_l |F_l|\le \epsilon$.
\end{itemize}

We can find solutions for small values of twist, $k\lesssim 0.1$, close to be potential. A strong limitation of the model is the condition $\Gamma \ge 0$ in its whole domain $(r,\mu)\in[1,r_{max}]\times[-1,1]$. However, a non positive $\Gamma$ would make the calculation of Gaunt's factors much harder. In order to relax the $\Gamma \ge 0$ constrain, we could consider the following form $\alpha=(k/R_\star)(\Gamma/\Gamma_0+A)^{1/2}$, with $A$ a dimensionless constant such that $\Gamma/\Gamma_0 + A \ge 0$ everywhere. The resulting differential equation is:
\begin{equation}
  \frac{d^2 f_l}{d r^2}-l(l+1)\frac{f_l}{r^2}=-\frac{k^2}{R_\star^2} \left(\sum_{m,n=1}^\infty f_mf_n g_{lmn} + \frac{4}{3}Af_l + \frac{2}{3}A^2\int_{-1}^1\frac{dP_l}{d\mu}\de \mu\right)~.
\end{equation}
The problem is that for $r\gg R_\star$ the right hand side is dominated by the constant term $A\Gamma_0$, thus the solutions will diverge as that in \S~\ref{sec:leg_diverging}.

% find '  0  '
% replace '  0  '
% file gaunt_lmn_print.dat from gaunt_lmn_print.f90
\begin{table}
\begin{center}
 \begin{tabular}{|l|c c c c c c|}
 \hline
a=1  & c=1 & c=2 & c=3 & c=4 & c=5 & c=6 \\
 \hline
b=1 &    0   & -0.267 &    0   &    0   &    0   &    0   \\
b=2 &  0.800 &    0   & -0.343 &    0   &    0   &    0   \\
b=3 &    0   &  0.686 &    0   & -0.381 &    0   &    0   \\
b=4 &    0   &    0   &  0.635 &    0   & -0.404 &    0   \\
b=5 &    0   &    0   &    0   &  0.606 &    0   & -0.420 \\
b=6 &    0   &    0   &    0   &    0   &  0.587 &    0   \\
 \hline
a=2  & c=1 & c=2 & c=3 & c=4 & c=5 & c=6 \\
 \hline
b=1 &  0.800 &    0   & -0.343 &    0   &    0   &    0   \\
b=2 &    0   &  0.343 &    0   & -0.457 &    0   &    0   \\
b=3 &  1.371 &    0   &  0.229 &    0   & -0.519 &    0   \\
b=4 &    0   &  1.143 &    0   &  0.173 &    0   & -0.559 \\
b=5 &    0   &    0   &  1.039 &    0   &  0.140 &    0   \\
b=6 &    0   &    0   &    0   &  0.979 &    0   &  0.117 \\
 \hline
a=3  & c=1 & c=2 & c=3 & c=4 & c=5 & c=6 \\
 \hline
b=1 &    0   &  0.686 &    0   & -0.381 &    0   &    0   \\
b=2 &  1.371 &    0   &  0.229 &    0   & -0.519 &    0   \\
b=3 &    0   &  0.686 &    0   &  0.104 &    0   & -0.599 \\
b=4 &  1.905 &    0   &  0.519 &    0   &  0.040 &    0   \\
b=5 &    0   &  1.558 &    0   &  0.440 &    0   &    0   \\
b=6 &    0   &    0   &  1.399 &    0   &  0.392 &    0   \\
 \hline
a=4  & c=1 & c=2 & c=3 & c=4 & c=5 & c=6 \\
 \hline
b=1 &    0   &    0   &  0.635 &    0   & -0.404 &    0   \\
b=2 &    0   &  1.143 &    0   &  0.173 &    0   & -0.559 \\
b=3 &  1.905 &    0   &  0.519 &    0   &  0.040 &    0   \\
b=4 &    0   &  0.981 &    0   &  0.360 &    0   & -0.031 \\
b=5 &  2.424 &    0   &  0.759 &    0   &  0.280 &    0   \\
b=6 &    0   &  1.958 &    0   &  0.653 &    0   &  0.230 \\
 \hline
a=5  & c=1 & c=2 & c=3 & c=4 & c=5 & c=6 \\
 \hline
b=1 &    0   &    0   &    0   &  0.606 &    0   & -0.420 \\
b=2 &    0   &    0   &  1.039 &    0   &  0.140 &    0   \\
b=3 &    0   &  1.558 &    0   &  0.440 &    0   &    0   \\
b=4 &  2.424 &    0   &  0.759 &    0   &  0.280 &    0   \\
b=5 &    0   &  1.259 &    0   &  0.559 &    0   &  0.197 \\
b=6 &  2.937 &    0   &  0.979 &    0   &  0.461 &    0   \\
 \hline
a=6  & c=1 & c=2 & c=3 & c=4 & c=5 & c=6 \\
 \hline
b=1 &    0   &    0   &    0   &    0   &  0.587 &    0   \\
b=2 &    0   &    0   &    0   &  0.979 &    0   &  0.117 \\
b=3 &    0   &    0   &  1.399 &    0   &  0.392 &    0   \\
b=4 &    0   &  1.958 &    0   &  0.653 &    0   &  0.230 \\
b=5 &  2.937 &    0   &  0.979 &    0   &  0.461 &    0   \\
b=6 &    0   &  1.527 &    0   &  0.737 &    0   &  0.364 \\
 \hline

\end{tabular}
\caption{Values of the integrals ${\cal I}_3$ with $u=v=1$ and $w=0$, calculated with the Gaunt's formula~(\ref{gaunt}).}
\label{tab_gaunt}
\end{center}
\end{table}

% file coupling_lmn_print.dat from gaunt_lmn_print.f90

\begin{table}
 \begin{center}
 \begin{tabular}{|l|c c c c c c|}
 \hline
l=1  & n=1 & n=2 & n=3 & n=4 & n=5 & n=6 \\
 \hline
m=1 &  0.533 &    0   & -1.371 &    0   &    0   &    0   \\
m=2 &    0   &  0.686 &    0   & -3.048 &    0   &    0   \\
m=3 &  0.914 &    0   &  0.914 &    0   & -5.195 &    0   \\
m=4 &    0   &  2.286 &    0   &  1.154 &    0   & -7.832 \\
m=5 &    0   &    0   &  4.156 &    0   &  1.399 &    0   \\
m=6 &    0   &    0   &    0   &  6.527 &    0   &  1.645 \\
 \hline
l=2  & n=1 & n=2 & n=3 & n=4 & n=5 & n=6 \\
 \hline
m=1 &    0   &  0.635 &    0   & -0.847 &    0   &    0   \\
m=2 &  0.127 &    0   &  0.762 &    0   & -1.732 &    0   \\
m=3 &    0   &    0   &    0   &  1.077 &    0   & -2.797 \\
m=4 &  0.423 &    0   & -0.154 &    0   &  1.480 &    0   \\
m=5 &    0   &  1.039 &    0   & -0.370 &    0   &  1.958 \\
m=6 &    0   &    0   &  1.865 &    0   & -0.653 &    0   \\
 \hline
l=3  & n=1 & n=2 & n=3 & n=4 & n=5 & n=6 \\
 \hline
m=1 & -0.089 &    0   &  0.652 &    0   & -0.673 &    0   \\
m=2 &    0   &  0.267 &    0   &  0.700 &    0   & -1.305 \\
m=3 &  0.059 &    0   &  0.194 &    0   &  0.932 &    0   \\
m=4 &    0   & -0.054 &    0   &  0.207 &    0   &  1.233 \\
m=5 &  0.269 &    0   & -0.186 &    0   &  0.233 &    0   \\
m=6 &    0   &  0.653 &    0   & -0.363 &    0   &  0.263 \\
 \hline
l=4  & n=1 & n=2 & n=3 & n=4 & n=5 & n=6 \\
 \hline
m=1 &    0   & -0.137 &    0   &  0.658 &    0   & -0.587 \\
m=2 & -0.091 &    0   &  0.324 &    0   &  0.659 &    0   \\
m=3 &    0   &  0.175 &    0   &  0.224 &    0   &  0.839 \\
m=4 &  0.035 &    0   &  0.096 &    0   &  0.240 &    0   \\
m=5 &    0   & -0.060 &    0   &  0.080 &    0   &  0.276 \\
m=6 &  0.196 &    0   & -0.168 &    0   &  0.069 &    0   \\
 \hline
l=5  & n=1 & n=2 & n=3 & n=4 & n=5 & n=6 \\
 \hline
m=1 &    0   &    0   & -0.169 &    0   &  0.661 &    0   \\
m=2 &    0   & -0.152 &    0   &  0.358 &    0   &  0.632 \\
m=3 & -0.085 &    0   &  0.234 &    0   &  0.234 &    0   \\
m=4 &    0   &  0.130 &    0   &  0.130 &    0   &  0.244 \\
m=5 &  0.023 &    0   &  0.059 &    0   &  0.121 &    0   \\
m=6 &    0   & -0.057 &    0   &  0.038 &    0   &  0.124 \\
 \hline
l=6  & n=1 & n=2 & n=3 & n=4 & n=5 & n=6 \\
 \hline
m=1 &    0   &    0   &    0   & -0.192 &    0   &  0.663 \\
m=2 &    0   &    0   & -0.198 &    0   &  0.381 &    0   \\
m=3 &    0   & -0.148 &    0   &  0.275 &    0   &  0.240 \\
m=4 & -0.077 &    0   &  0.187 &    0   &  0.147 &    0   \\
m=5 &    0   &  0.104 &    0   &  0.091 &    0   &  0.135 \\
m=6 &  0.016 &    0   &  0.040 &    0   &  0.075 &    0   \\
 \hline
 \end{tabular}
 \caption{Values of the coupling factor $g_{lmn}$ defined in eq.~(\ref{glmn}).}
 \label{tab_coupling}
 \end{center}
\end{table}

%%%%%%%%%%%%%%%%%%%%

\chapter{Notes for finite difference time domain methods}\label{app:fdtd}

\section{Grid and geometrical elements.}\label{app:grid}

In this thesis we have extensively used the induction equation
\begin{equation}\label{eq:induction_app}
  \frac{1}{c}\frac{\partial \vec{B}}{\partial t} = - \vec{\nabla}\times(e^\nu\vec{E})~,
\end{equation}
discretized in spherical coordinates. The redshift correction $e^\nu$ and the space curvature factor $e^\lambda$ are $1$ for flat metric (like in the magnetospheric code, \S~\ref{sec:numerical_magnetosphere_code}). If, instead, the relativistic metric is considered, eq.~(\ref{eq:metric}), they depend on radius $r$.

In order to numerically integrate eq.~(\ref{eq:induction_app}), we define the geometrical elements, fluxes and circulations. Consider the three surfaces $S_n^{(i,j)}$ centered on a node $(i,j)$ and normal to the unit vectors $\hat{n}=\hat{r},\hat{\theta},\hat{\varphi}$. The elementary areas are calculated by the following exact formulas:

\begin{eqnarray}\label{surfaces}
 && S_r^{(i,j)}      = \int_0^{2\pi} \int_{\theta_{i-1}}^{\theta_{i+1}} r_j^2 \sin\theta\,\de \varphi\,\de \theta 	= 2\pi r_j^2(\cos\theta_{i-1}-\cos\theta_{i+1})~, \\
 && S_\theta^{(i,j)} = \int_0^{2\pi} \int_{r_{j-1}}^{r_{j+1}} r\sin\theta_i \,\de \varphi\,\de r        	= \pi(r_{j+1}^2-r_{j-1}^2)e^{\lambda_j}\sin\theta_i~,     \\
 && S_\varphi^{(i,j)}   = \int_{\theta_{i-1}}^{\theta_{i+1}} \int_{r_{j-1}}^{r_{j+1}} r\,\de \theta\,\de r = (\theta_{i+1}-\theta_{i-1})\frac{(r^2_{j+1}-r^2_{j-1})}{2}e^{\lambda_j}~.
\end{eqnarray}
Since the right-hand side of eq.~(\ref{eq:induction_app}) contains $\vec{\nabla}\times\vec{E}$, it is useful to apply Stokes' theorem as follows. The magnetic flux across $S_n^{(i,j)}$ is approximated by

\begin{equation}\label{fluxes}
\Phi_n^{(i,j)}=B_n^{(i,j)}S_n^{(i,j)}~,
\end{equation}
and the magnetic fluxes are advanced in time using

\begin{equation}\label{derfluxes}
\Phi_n^{(i,j)}(t+\Delta t)=\Phi_n^{(i,j)}(t) - c{\Delta}t\oint_{\partial S_n^{(i,j)}} e^\nu\vec{E} \cdot \de \vec{l}~,
\end{equation}
where the numerical circulation of the electric field along the edges of the surface $S_n$ is approximated by using the values of $\vec{E}$ in the middle of the edges, whose lengths are
\begin{eqnarray}\label{eq:lengths}
 && l_r^{(i,j)}=(r_{j+1}-r_{j-1})e^{\lambda_j}~,\\
 && l_\theta^{(i,j)}=r_j(\theta_{i+1}-\theta_{i-1})~,\\
 && l_\varphi^{(i,j)}=2\pi r_j\sin\theta_i~.
\end{eqnarray}
Explicitly, the circulation of $\vec{E}$ along the edge of each surface $S_n^{(i,j)}$ is
\begin{eqnarray}
  \oint_{\partial S_r^{(i,j)}} e^\nu\vec{E} \cdot \de \vec{l}=&& 
e^{\nu_j}E_\varphi^{(i+1,j)}l_\varphi^{(i+1,j)}-e^{\nu_j}E_\varphi^{(i-1,j)}l_\varphi^{(i-1,j)}~, \label{eq:circulations1}\\
  \oint_{\partial S_\theta^{(i,j)}} e^\nu\vec{E} \cdot \de \vec{l}=&& 
-e^{\nu_{j+1}}E_\varphi^{(i,j+1)}l_\varphi^{(i,j+1)}+e^{\nu_{j-1}}E_\varphi^{(i,j-1)}l_\varphi^{(i,j-1)}~, \label{eq:circulations2}\\
  \oint_{\partial S_\varphi^{(i,j)}} e^\nu\vec{E} \cdot \de \vec{l}=&& 
e^{\nu_{j+1}}E_\theta^{(i,j+1)}l_\theta^{(i,j+1)}-e^{\nu_{j-1}}E_\theta^{(i,j-1)}l_\theta^{(i,j-1)} \nonumber\\
&& -e^{\nu_j}E_r^{(i+1,j)}l_r^{(i+1,j)}+e^{\nu_j}E_r^{(i-1,j)}l_r^{(i-1,j)}~. \label{eq:circulations3}
\end{eqnarray}
We make use of Stokes' theorem to calculate the current $n$-component ($n=r,\theta,\varphi$) at each node:

\begin{equation}
 J_n^{(i,j)}=\frac{c}{4\pi}\frac{e^{-\nu_j}}{S_n^{(i,j)}}\oint_{\partial S_n^{(i,j)}} e^\nu\vec{B} \cdot \de \vec{l}~.
\end{equation}
For each cell $(i,j)$ the local divergence of $\vec{B}$ is defined as the net magnetic flux flowing across the surfaces divided by the cell volume (fluxes through the toroidal surfaces $S_\varphi$ cancel due to axial symmetry):
\begin{equation}\label{divb}
(\vec{\nabla}\cdot\vec{B})^{(i,j)}=\frac{1}{V^{(i,j)}}[\Phi_r^{(i,j+1)}-\Phi_r^{(i,j-1)}+\Phi_\theta^{(i+1,j)}-\Phi_\theta^{(i-1,j)}]~,
\end{equation}
with
\begin{equation}\label{volume}
V^{(i,j)} = \frac{2\pi}{3}(\cos\theta_{i-1}-\cos\theta_{i+1})(r^3_{j+1}-r^3_{j-1})e^{\lambda_j} ~.
\end{equation}
The numerical method ensures that the local divergence is conserved to machine accuracy by construction. As a matter of fact, eqs.~(\ref{derfluxes}) and (\ref{divb}) imply
\begin{eqnarray}
\frac{d}{dt}(\vec{\nabla}\cdot\vec{B})^{(i,j)} &=& c\left[-\oint_{\partial S_r^{(i,j+1)}}e^\nu\vec{E}\cdot\de \vec{l} + \oint_{\partial S_r^{(i,j-1)}}e^\nu\vec{E}\cdot\de \vec{l}\right. \nonumber\\
 && \left.-\oint_{\partial S_\theta^{(i+1,j)}}e^\nu\vec{E}\cdot\de \vec{l} + \oint_{\partial S_\theta^{(i-1,j)}}e^\nu\vec{E}\cdot\de \vec{l} \right]~.
\end{eqnarray}
According to eqs.~(\ref{eq:circulations1})-(\ref{eq:circulations3}), the last equation is written as a sum of toroidal elements $e^\nu E_\varphi l_\varphi$. They are evaluated twice at each of the four surrounding edges (with center $\theta_{i\pm1},r_{j\pm1}$), with opposite sign, thus they cancel exactly. Therefore, provided there is an initial divergence-less field, it is guaranteed that the numerical divergence defined by eq. (\ref{divb}) remains zero to round-off level at each time step.

\section{Numerical couplings.}\label{app:couplings}

\begin{figure}[t]
\centering
\includegraphics[width=14cm]{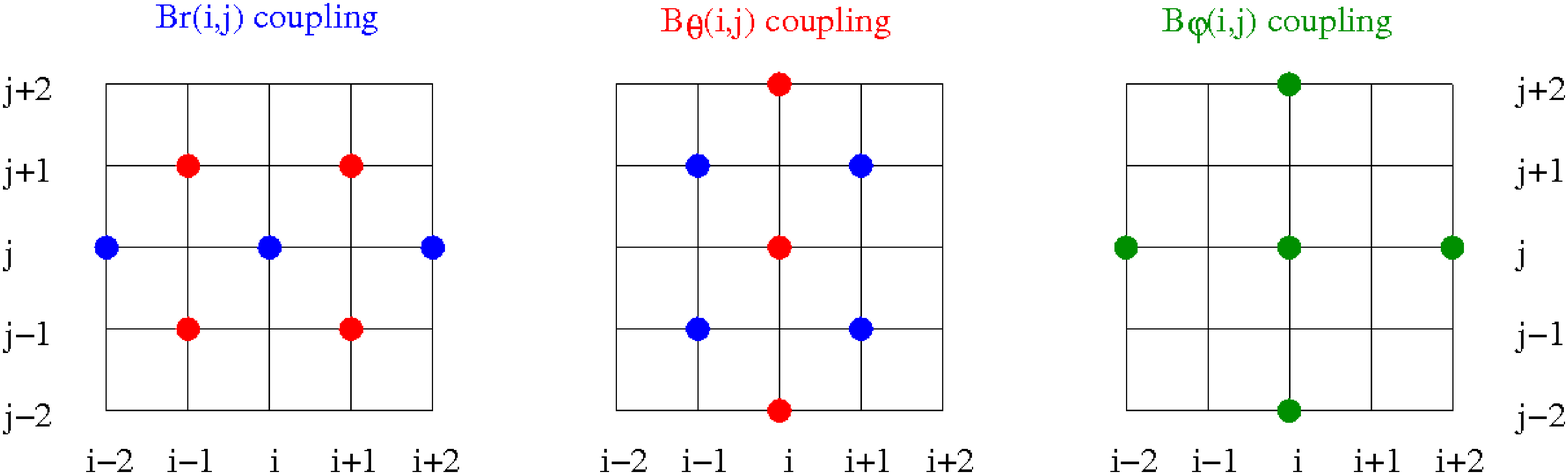}\\
\bigskip
\hrule
\bigskip
\includegraphics[width=14cm]{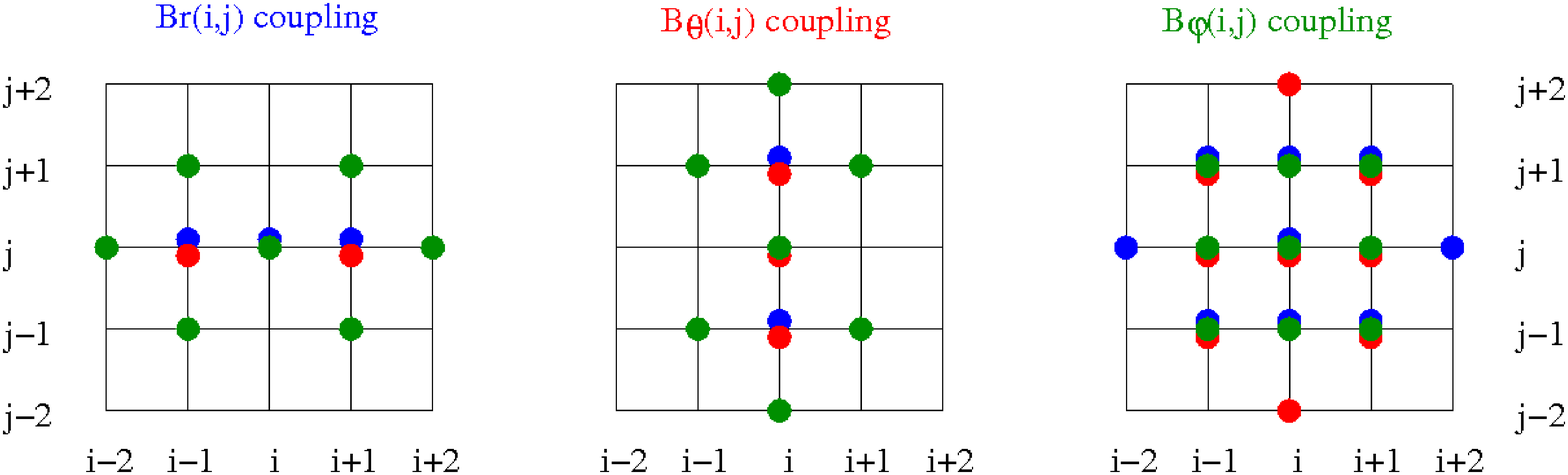}\\
\bigskip
\hrule
\bigskip
\includegraphics[width=14cm]{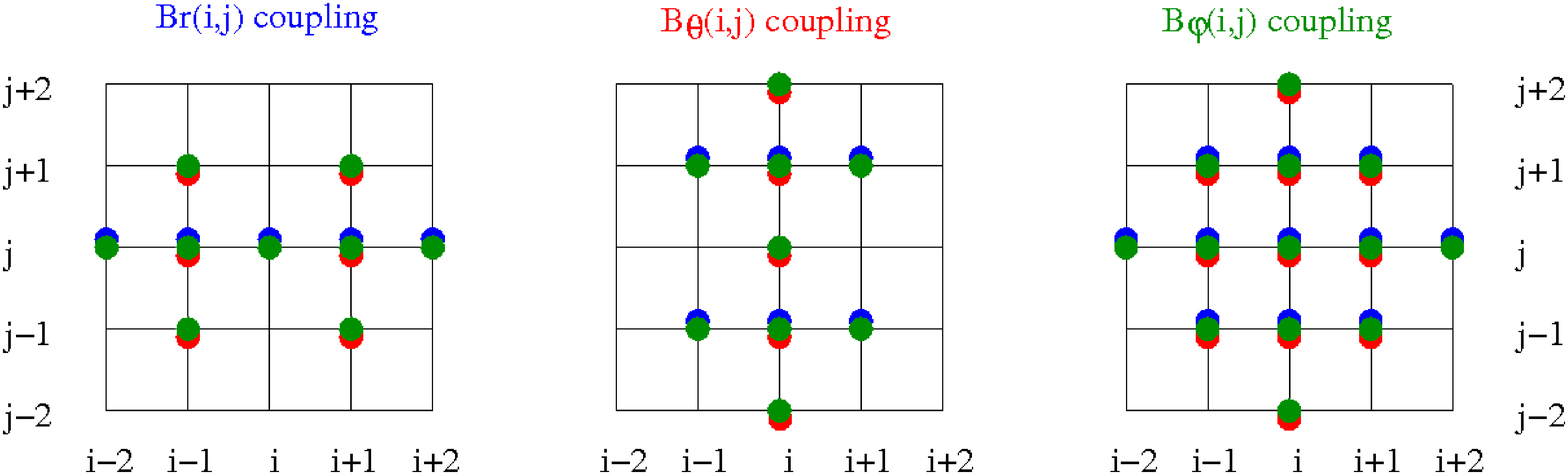}
\caption{Couplings between magnetic field components ($B_r^{(i,j)}$, left panels; $B_\theta^{(i,j)}$, center; $B_\varphi^{(i,j)}$, right) at each time step, for different forms of $\vec{E}$ in the induction equation: Ohmic (top panels), Hall (middle), and magnetofrictional (bottom). The coupled neighbors $B_r,B_\theta,B_\varphi$ are indicated respectively with blue, red and green points.} 
\label{fig:couplings}
\end{figure}

In the induction equation (\ref{eq:induction_app}) we have considered three possible forms of $\vec{E}$:
\begin{itemize}
 \item $\vec{E}=\eta \vec{J}$, Ohmic induction equation (chapter~\ref{ch:magnetic});
 \item $\vec{E}=f_h \vec{J}\times\vec{B}$, Hall induction equation (chapter~\ref{ch:magnetic});
 \item $\vec{E}=\nu(\vec{J}\times\vec{B})\times\vec{B}/cB^2$, magnetofrictional induction equation (chapter~\ref{ch:magnetosphere}).
\end{itemize}
We want to study which points of the numerical grid are coupled at each time-step. In Fig.~\ref{fig:couplings} we schematically show, for each one of the three choices of $\vec{E}$, the couplings between the value of each magnetic field component at $(i,j)$ and the other neighbors components around. Colors mark the different components.

In the Ohmic induction equation (top panel) the poloidal and toroidal parts are completely decoupled. Note that in this case a staggered grid works well, because one can define the three components coherently, picking out from our figure just the positions that are coupled each other. In the Hall induction equation (middle), the coupling is complicated, and in fact values on a staggered grid are not enough and we need to obtain by interpolation some components on different grid points. The magnetofrictional induction equation (bottom) requires even more interpolations of variables at different grid points. The coefficients of the couplings are not trivial, because they include geometrical terms, microphysical parameters ($\eta$, $f_h$), or because the terms themselves are non-linear.

\def \aj {Astron. J.}
\def \ATel {Astron. Tel.}
\def \apj {ApJ}
\def \aplett {ApL}
\def \apjl {ApJL}
\def \apjs {ApJS}
\def \aap {A\&A}
\def \aaps {A\&AS}
\def \aapr {A\&ARv.}
\def \aara {A\&A Rev.}
\def \apss {AP\&SS}
\def \araa {Annu. Rev. Astro. Astrophys.}
\def \cjaa {Chin. J. Astron. Astrophys.} 
\def \gca {Geochim. Cosmochim. Acta}
\def \jcp {J. Comput. Phys.}
\def \mnras {MNRAS}
\def \nat {Nature} 
\def \npa {Nuc. Phys. A}
\def \physrep {Phys. Rep.}
\def \pasj {PASJ}
\def \pr {Phys. Rev.}
\def \pra {Phys. Rev. A}
\def \prc {Phys. Rev. C}
\def \prd {Phys. Rev. D}
\def \pre {Phys. Rev. E}
\def \prl {Phys. Rev. Lett.}
\def \rmxaa {RMAA}
\def \solphys {Sol. Phys.}
\def \sovast {Soviet Ast.}

\begin{spacing}{0}
\bibliographystyle{pers}
\bibliography{biblio}
\end{spacing}

% ACKS
\newpage\null
\pagestyle{empty}
\chapter*{Agradecimientos}

Mi primer y enorme agradecimiento es para Marta. No se puede resumir en pocas palabras lo que es pasar ocho a{\~n}os juntos, de los cuales la mayor\'ia han sido en la distancia. Sei la mia stella. I nostri viaggi per il mondo, di un giorno o un mese, hanno cancellato ogni volta la distanza: i puffins, le palme amazzoniche, le escursioni alpine (magari con una bella porta di legno come coperta), la tenda alla Coveta Fum\'a o in un campo coltivato delle F{\ae}r {\O}er. E la nostra (troppo breve) convivenza \`e stata la cosa pi\`u bella.

En estos cuarenta meses en Alicante, lo que m\'as he aprendido no viene en ning\'un libro, c\'odigo o conocimiento cient\'ifico: hay personas que, sin saberlo, ense{\~n}an c\'omo vivir y c\'omo ser una persona sensible, libre, sencilla, coerente y, sobre todo, tranquilamente feliz. Gracias Luis, no hace falta que diga m\'as. Y hay peronas que no se dan cuenta de sus grandes cualidades: la curiosidad hacia el mundo, la apertura de mente, la inteligencia, la sensibilidad y la energ\'ia que puede transmitir. Tama, eres un hermano.

Sembra banale, ma l'appoggio costante \`e fondamentale per andare avanti: so di avere una famiglia eccezionale, che ha sempre favorito questa mia esperienza fantastica. Mamma e pap\`a, grazie per capirmi, rispettare ed appoggiare le mie scelte e il mio modo di pensare, so che spesso non \`e stato facile. Cri, tu sei come una lampadina in una giornata estiva alicantina: in piena luce, durante l'attivit\`a diurna, non ci penso, ma so che se cala il sole, o, di rado, nuvole scure coprono il bagliore, in qualsiasi momento puoi illuminarmi e farmi vedere tutto pi\`u chiaro e semplice.

Taner, Miguel y Marcel, ?`qu\'e puedo decir? Simplemente que ha sido una gran suerte encontraros. ?`C\'omo voy a olvidar los mil discursos y payasadas, y la m\'itica furgo con fuego de Marcel? Cuenca, Cabo de Gata, Sierra Nevada, se quedan entre los mejores recuerdos.

Un pensiero particolarmente affettuoso va a chi mi \`e vicino anche se a distanza, indipendentemente da quanto spesso ci si senta: Annalisa, Marta E., Serena, Ludo, parlare con voi mi fa sempre bene.

Al voley alicantino, que llevar\'e siempre en mi coraz\'on. Las ma{\~n}anas de invierno en el Postiguet con Tama, desafiando t\'imidamente a los reyes callejeros de la playa. Y luego San Juan, y los cursos. Un Mikasa, una red y la playa han sido durante tres a{\~n}os una v\'alvula de escape y diversi\'on, lucha y lugar de encuentros, est\'imulos y risas. Rebozarse en arena y pegarse un ba{\~n}o en cualquier fecha del a{\~n}o con vistas al Puig Campana, antes de volverse pedaleando los diez kil\'ometros hasta ``Sanvi'', con el calor de verano o la niebla de invierno, no tiene precio. Quiero agradecer much\'isimo a los entrenadores que he tenido: Lambe, sus explicaciones, su i-Lamb de arena, y su gran paciencia; Nadia, con sus cabreos y frases inolvidables (por ejemplo, ``?`!`Pero esto qu\'e es, la asociaci\'on de los mancos?!''); Carol, con su sonrisa y su disponibilidad; Pablo, que nunca se cansa de repetir cien veces la misma cosa. Sois unos cracks, ha sido incre\'ible poder entrenar con vosotros, aunque probablemente no haya aprendido ni la d\'ecima parte de lo que sab\'eis.

Gracias a esas amigas como Mar\'ia (telepizza, caballos y chilenas, pero sobre todo charlas y amistad), Ceci y Irene (con vuestros cotilleos), Cris, Montse con sus ``cortitas'', Juan (``!`No hay! !`Pegaleeeeeeeee!''), y las cientas de personas con los cuales he jugado uno o veinte partidos.

No podr\'e olvidarme de las horas pasadas disfrutando del clima alicantino con H\'ector ``hay-que-darle-a-todo'', el multicampe\'on Jorge y nuestros trofeos (gran maestro de p\'adel y michirones, pero no de pocha), Al, \'Angel (queda pendiente el desaf\'io de tomates), Samy (la reina del Xorret de Cat\'i), Gon, y decenas m\'as.

A Jos\'e y Juan Antonio: si durante mi doctorado, al fin y al cabo, he podido vivir muy a gusto, es en gran parte m\'erito vuestro: m\'as que cient\'ificamente, mi agradecimiento va al terreno personal, porque desde el primer d\'ia me he sentido muy c\'omodo. Creo que no es nada com\'un tener directores de tesis tan humanos con los cuales se pueda tener, adem\'as, tanta confianza. He de a{\~n}adir que, aunque sea por \'osmosis, con vosotros se aprende un mont\'on, y lo mismo vale para Nanda. Rosalba, grazie per l'ospitalit\`a a Boulder e NY, le corse lungo Boulder Creek e Mesa Trail, e soprattutto le parole incoraggianti e sagge che ben ricordo.

Dani, t\'u eres una de las pocas personas en las cuales instintivamente conf\'io, una simple conversaci\'on de vez en cuando me aporta mucho: ?`ser\'a que tienes pinta de fil\'osofo sabio? V\'ictor y Mar\'ia Jos\'e, s\'e que con vostros siempre podr\'e hablar de todo, y que sab\'eis aguantar mi falta de diplomacia (a ver si alg\'un d\'ia, despu\'es de tantas estrellas, aprendo algo de tu libro, MJ). Vull donar les gr\`acies al club de la gespa (Bernat, Carmen, Miguel, \`Angela, Marta), a les desenes de dinars, discussions i migdiades. Estranyar\'e els \`anecs, les tortugues, les preguntes ing\`enues, aix\'i com el cor d'or de Bernat i les q\"uestions morals de Carmen.

Me d\'a mucha pena tener que abandonar el piso y el barrio Santa Isabel: he agradecido mucho la convivencia serena y alegre (bueno, excepto cuando las sillas colapsan) a Luis, Quique y su tranquilidad contagiadora tan importante en los \'ultimos meses (todav\'ia no me creo que tengas solo 19 a{\~n}os, si supieras como me portaba yo con tu edad...), y el tr\'io de acogida: Gon (!`Viva la polenta!), Violeta y Andrea.

A Carlitos, por las infinitas met\'aforas rebuscadas y nobles. Pablo, Fernando y Giovanni, a\'un recuerdo c\'omo, cada uno con propuestas diferentes, me acogisteis las primeras semanas: p\'adel, teter\'ia, bici por los puertos alicantinos, tequilas, !`gracias por todo! Gracias tambien a la ``italiana'' Mar\'ia Jos\'e y sus charlas por las tardes, a Fanny, Marisol y \'Oscar, por vuestra ayuda con los papeleos, y un saludo al reci\'en llegado Jes\'us.

Para acabar, una promesa: !`volver\'e a vivir por aqu\'i!

\bigskip
\begin{flushright}
{\it Daniele (con una ``l'')} 
\end{flushright}

\newpage\null

% BACK COVER
\newpage\null
% \newpage\null
% \pagecolor{yellow}
% \newpage

\end{document}